\magnification = \magstep0
\hsize=15.0truecm  \hoffset=0.0truecm  \vsize=20.1truecm  \voffset=1.5truecm
\font\sc=cmr8   
\font\sit=cmti8
\font\sbf=cmbx8
\font\ssbf=cmbx9

\def\scbaselines{\baselineskip=8pt    \lineskip=0pt   \lineskiplimit=0pt}
\def\smedbaselines{\baselineskip=9pt    \lineskip=0pt   \lineskiplimit=0pt}

\def\refbaselines{\baselineskip=11pt    \lineskip=0pt   \lineskiplimit=0pt}

\def\dblbaselines{\baselineskip=12pt    \lineskip=0pt   \lineskiplimit=0pt}
\def\widebaselines{\baselineskip=13pt    \lineskip=0pt   \lineskiplimit=0pt}
\def\semiwidebaselines{\baselineskip=10pt    \lineskip=0pt   \lineskiplimit=0pt}
\def\vsl{\vskip\baselineskip}   \def\vs{\vskip 6pt} \def\vsss{\vskip 3pt}
\parindent=10pt \nopagenumbers

\def\omit#1{\empty}

\input colordvi
%
%
\def\endtable{\endgroup}
\def\tableheight{\vrule width 0pt height 8.5pt depth 3.5pt}
{\catcode`|=\active \catcode`&=\active 
    \gdef\tabledelim{\catcode`|=\active \let|=\vbar
                     \catcode`&=\active \let&=\nobar} }
\def\table{\begingroup
    \def\twidth{\hsize}
    \def\tablewidth##1{\def\twidth{##1}}
    \def\defaultheight{\vrule width 0pt height 8.5pt depth 3.5pt}
    \def\heightdepth##1{\dimen0=##1
        \ifdim\dimen0>5pt 
            \divide\dimen0 by 2 \advance\dimen0 by 2.5pt
            \dimen1=\dimen0 \advance\dimen1 by -5pt
            \vrule width 0pt height \the\dimen0  depth \the\dimen1
        \else  \divide\dimen0 by 2
            \vrule width 0pt height \the\dimen0  depth \the\dimen0 \fi}
    \def\spacing##1{\def\defaultheight{\heightdepth{##1}}}
    \def\nextheight##1{\noalign{\gdef\tableheight{\heightdepth{##1}}}}
    \def\end{\cr\noalign{\gdef\tableheight{\defaultheight}}}
    \def\zerowidth##1{\omit\hidewidth ##1 \hidewidth}    
    \def\hline{\noalign{\hrule}}
    \def\skip##1{\noalign{\vskip##1}}
    \def\bskip##1{\noalign{\hbox to \twidth{\vrule height##1 depth 0pt \hfil
        \vrule height##1 depth 0pt}}}
    \def\header##1{\noalign{\hbox to \twidth{\hfil ##1 \unskip\hfil}}}
    \def\bheader##1{\noalign{\hbox to \twidth{\vrule\hfil ##1 
        \unskip\hfil\vrule}}}
    \def\spanloop{\span\omit \advance\mscount by -1}
    \def\extend##1##2{\omit
        \mscount=##1 \multiply\mscount by 2 \advance\mscount by -1
        \loop\ifnum\mscount>1 \spanloop\repeat \ \hfil ##2 \unskip\hfil}
    \def\vbar{&\vrule&}
    \def\nobar{&&}
    \def\hdash##1{ \noalign{ \relax \gdef\tableheight{\heightdepth{0pt}}
        \toks0={} \count0=1 \count1=0 \putout##1\end 
        \toks0=\expandafter{\the\toks0 &\end} \xdef\piggy{\the\toks0} }
        \piggy}
    \let\e=\expandafter
    \def\putspace{\ifnum\count0>1 \advance\count0 by -1
        \toks0=\e\e\e{\the\e\toks0\e&\e\multispan\e{\the\count0}\hfill} 
        \fi \count0=0 }
    \def\putrule{\ifnum\count1>0 \advance\count1 by 1
        \toks0=\e\e\e{\the\e\toks0\e&\e\multispan\e{\the\count1}\leaders\hrule\hfill}
        \fi \count1=0 }
    \def\putout##1{\ifx##1\end \putspace \putrule \let\next=\relax 
        \else \let\next=\putout
            \ifx##1- \advance\count1 by 2 \putspace
            \else    \advance\count0 by 2 \putrule \fi \fi \next}   }
\def\tablespec#1{
    \def\vdimens{\noexpand\tableheight}
    \def\tabby{\tabskip=0pt plus100pt minus100pt}
    \def\r{&################\tabby&\hfil################\unskip}
    \def\c{&################\tabby&\hfil################\unskip\hfil}
    \def\l{&################\tabby&################\unskip\hfil}
    \edef\templ{\noexpand\vdimens ########\unskip  #1 
         \unskip&########\tabskip=0pt&########\cr}
    \tabledelim
    \edef\body##1{ \vbox{
        \tabskip=0pt \offinterlineskip
        \halign to \twidth {\templ ##1}}} }

\def\t#1{#1} 
\def\t#1{\empty}
    
  \def\ba{\kern -1pt}
\parskip = 0pt 
\def\ts{\thinspace} \def\cl{\centerline}
\def\ni{\noindent}

\def\nhi{\noindent  \hangindent=1.0truecm}

\def\nhi2{\noindent \hangindent=2.79cm}
  
\def\nhhj{\noindent \hangindent=9.33truecm}  
\def\nhi1{\indent \hangindent=0.8truecm}
\def\nhii{\indent \hangindent=0.95truecm}
\def\nhij{\indent \hangindent=1.13truecm}

\def\makeheadline{\vbox to 0pt{\vskip-30pt\line{\vbox to8.5pt{}\the
                               \headline}\vss}\nointerlineskip}
\def\toppageno{\headline={\hss\tenrm\folio\hss}}
\def\footnoterule{\kern-3pt \hrule width \hsize \kern 2.6pt \vskip 3pt}
\output={\plainoutput}    \pretolerance=10000   \tolerance=10000

\def\sup1{$^{\rm 1}$} \def\sup2{$^{\rm 2}$}
\def\r0{$\rho_0$}   
 \def\0{\phantom{0}} \def\bb{\kern -2pt}
\def\1{\phantom{1}}         
\def\etal{{et~al.\ }}
\def\gapprox{$_>\atop{^\sim}$} \def\lapprox{$_<\atop{^\sim}$}


\def\spose#1{\hbox to 0pt{#1\hss}}
\def\lesssim{\mathrel{\spose{\lower 3pt\hbox{$\mathchar"218$}}
     \raise 2.0pt\hbox{$\mathchar"13C$}}}
\def\gtrsim{\mathrel{\spose{\lower 3pt\hbox{$\mathchar"218$}}
     \raise 2.0pt\hbox{$\mathchar"13E$}}}


\def\kms{km~s$^{-1}$}          
\newdimen\sa  \def\sd{\sa=.1em \ifmmode $\rlap{.}$''$\kern -\sa$
                               \else     \rlap{.}$''$\kern -\sa\fi}
\newdimen\sb\def\degd{\sa=.15em\ifmmode $\rlap{.}$^\circ$\kern -\sa$
                               \else     \rlap{.}$^\circ$\kern -\sa\fi}
\def\ss{\ifmmode ^{\prime\prime}$\kern-\sa$ \else $^{\prime\prime}$\kern-\sa\fi}
\def\mm{\ifmmode ^{\prime}$\kern-\sa$ \else $^{\prime}$\kern-\sa \fi}
  \def\msund{M$_{\odot}$}
\def\mbh{$M_{\bullet}$~}  
\def\m31{M{\ts}31} \def\mm32{M{\ts}32} \def\mmm33{M{\ts}33} \def\M87{M{\ts}87}

\def\lax{{$\mathrel{\hbox{\rlap{\hbox{\lower4pt\hbox{$\sim$}}}\hbox{$<$}}}$}}
\def\gax{{$\mathrel{\hbox{\rlap{\hbox{\lower4pt\hbox{$\sim$}}}\hbox{$>$}}}$}}
\input psfig.tex

\font\almostbig=cmbx12 scaled 850

\font\big=cmbx12 scaled 1100
\font\bigau=cmr12 scaled 1200
\font\bigbig=cmr12 scaled 2000
\font\bigit=cmti10 scaled 1200


\def\bigpoint{
  \font\fiverm=cmr5
  \font\sevenrm=cmr7
  \font\bigrm=cmr12 scaled 1200
  \font\fivei=cmmi5 scaled 1200
  \font\seveni=cmmi7 scaled 1200
  \font\bigi=cmmi12 scaled 1200
  \font\fivesy=cmsy5  scaled 1200
  \font\sevensy=cmsy7 scaled 1500
  \font\bigsy=cmsy12 scaled 1200
  \font\bigit=cmti12 scaled 1200
  \font\bigbf=cmbx12 scaled 1200
  \font\bigsl=cmsl12 scaled 1200
  \textfont0=\bigrm \scriptfont0=\sevenrm     
    \scriptscriptfont0=\fiverm                  
  \def\rm{\fam0 \bigrm}   
  \textfont1=\bigi  \scriptfont1=\seveni  
    \scriptscriptfont1=\seveni                  
  \def\mit{\fam1 } \def\oldstyle{\fam1 \bigi}
  \textfont2=\bigsy \scriptfont2=\sevensy 
    \scriptscriptfont2=\sevensy                  
}

\vfill\eject

\def\ARRed{\textColor{.23 1. 1. .17}}
\def\ARGreen{\textColor{.85 0.22 0.99 .10}}
\def\ARDiscard{\textColor{1 0.1 0.1 0.1}}

\cl{\null} \vskip 0pt  
 
\cl{\bigbig Coevolution (Or Not) of} \vsss

\cl{\bigbig Supermassive Black Holes and Host Galaxies}

\vsl
 
\cl{\bigau John Kormendy$^1$ and Luis C.~Ho$^2$}
\vsl
\cl{$^1$Department of Astronomy, University of Texas at Austin,}
\cl{2515 Speedway C1400, Austin, TX 78712-1205; email: kormendy@astro.as.utexas.edu}

\vsl
\cl{$^2$The Observatories of the Carnegie Institution for Science,}
\cl{813 Santa Barbara Street, Pasadena, CA 91101; email: lho@obs.carnegiescience.edu}
 
\vsl

\dblbaselines

\ni {\big\ARRed Abstract}\textBlack
\vs


      Supermassive black holes (BHs) have been found in 87 galaxies by dynamical modeling of spatially resolved kinematics.
The {\it Hubble Space Telescope} revolutionized~BH~research by advancing the subject from its proof-of-concept phase into 
quantitative studies of BH demographics.  Most influential was the discovery of a tight correlation between BH{\ts}mass{\ts}$M_\bullet$ and the 
velocity dispersion{\ts}$\sigma$ of the bulge component of the host galaxy.  Together with similar correlations with bulge luminosity and 
mass, this led to the widespread belief that BHs and bulges coevolve by regulating each other's growth.  Conclusions based on one set of
correlations from $M_\bullet \sim 10^{9.5}$ $M_\odot$ in brightest cluster ellipticals to \hbox{$M_\bullet \sim 10^6$\ts$M_\odot$} in 
the smallest galaxies dominated BH work for more than a decade. \vs

      New results are now replacing this simple story with a richer and more plausible picture in which 
BHs correlate differently with different galaxy components.  A reasonable aim is to use~this~progress 
to refine our understanding of BH{\ts}--{\ts}galaxy coevolution.  BHs with masses of $10^5$\ts--\ts$10^6$ $M_\odot$ 
are found in many bulgeless galaxies.  Therefore, classical (elliptical-galaxy-like) bulges are not necessary for BH formation.  
On the other hand, while they live in galaxy disks, BHs do not correlate with galaxy disks.  Also, any $M_\bullet$ correlations 
with the properties of disk-grown pseudobulges and dark matter halos are weak enough to imply no close coevolution.  \vs

      The above and other correlations of host galaxy parameters with each other and with $M_\bullet$ suggest that there are four
regimes of BH feedback. (1) Local, secular, episodic, and stochastic feeding of small BHs in largely bulgeless galaxies
involves too little energy to result~in~coevolution.  (2) Global feeding in major, wet galaxy mergers rapidly grows giant BHs in 
short-duration, \hbox{quasar-like} events whose energy feedback does affect galaxy evolution.  The resulting hosts are classical bulges
and coreless-rotating-disky ellipticals.~(3) After these AGN phases and at the highest galaxy~masses, maintenance-mode BH feedback 
into X-ray-emitting gas has the primarily negative effect~of~helping to keep baryons locked up in hot gas and thereby keeping 
galaxy formation from going to completion.  This happens in giant, core-nonrotating-boxy ellipticals.  Their properties, including 
their tight correlations between $M_\bullet$ and core parameters, support the conclusion that core ellipticals form by 
dissipationless major mergers.  They inherit coevolution effects from smaller progenitor galaxies.  Also, (4) independent of any
feedback physics, in BH growth~modes~(2)~and~(3), the averaging that results from successive mergers plays a major role in decreasing
the scatter in $M_\bullet$ correlations from the large values observed in bulgeless and pseudobulge galaxies to the small values 
observed in giant elliptical galaxies. 

\vfill\eject

\ni {\big\ARRed 1. INTRODUCTION}\textBlack
\vsss

\pageno=2 \toppageno

     Coevolution (or not) of supermassive black holes (BHs) and host galaxies is the central theme of this review. 
The heyday of this activity was the first half of the history of the universe, 7\ts--\ts12 Gyr ago. Conditions were different 
then than they~are~now.~~Lema\^\i tre~(1931) captured this challenge elegantly and accurately:~``The evolution of the
universe can~be~likened~to~a display of fireworks that has just ended: some few red wisps, ashes, and smoke.  Standing on a 
well-chilled cinder, we~see the fading of the suns and try to recall the vanished brilliance of the origin of the worlds.''  

      This paper reviews the archaeology of supermassive cinders.  For more than a decade, the most important tool 
for this research was the {\it Hubble Space Telescope\/} (HST).  Now, as HST nears the end of its life and begins to be 
replaced by other tools that have different capabilities, BH work is branching out in new directions.  It is an appropriate
time to review what we have learned.

      If an iconic event can be said to have started work on BHs, it was the discovery (Schmidt 1963) 
that the 13$^{\rm th}$ magnitude, starlike object identified with the powerful radio source 3C 273 had the shockingly
high redshift (for such a bright object) of $z = 0.158$.  Assuming that the redshift was due to the Hubble expansion 
of the Universe, 3C 273 was the second-most-distant object~then~known.  3C 273 is $\sim$\ts10 times
more luminous than~the~brightest~galaxies.~The~first~quasar~discovery was quickly followed by many others.~Rapid 
variability~implied~tiny~sizes.~After a Darwinian struggle between competing theories, the idea that quasars and 
other, less luminous active galactic nuclei (AGNs) are powered by accretion onto supermassive BHs 
(Hoyle \& Fowler 1963; 
Salpeter 1964; 
Zel'dovich 1964; 
Lynden-Bell 1969, 1978; 
Lynden-Bell \& Rees 1971) 
quickly gained acceptance.  Decades of productive work based on this idea followed.  However, there was no dynamical evidence 
that BHs with the required masses, $M_\bullet \sim 10^6$\ts--\ts$10^{9}$\ts\msund, actually exist.  At the same time, 
it was clear that quasars were much more numerous at $z$ \gapprox \ts2 than they~are~now.  After they stop accreting, 
dark BH remnants should live essentially forever.  So dead quasar engines should hide in many nearby galaxies.  Finding them 
became one of the ``holy grails'' of astronomy.

\includegraphics{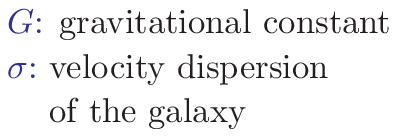}

       However, the above masses are only $\sim$\ts0.1\ts\% of the stellar masses of the host galaxies.~So~BHs~are expected 
to dominate the local gravity and to lead to observable dynamical consequences only inside 
a sphere-of-influence radius $r_{\rm infl} \equiv G M_\bullet / \sigma^2 \sim 1$ to 100 pc.  At distances $D = 1$~to~20~Mpc
for nearby galaxies, this corresponds to $\sim 0\sd1$ to $1^{\prime\prime}$.  Thus it was inevitable that dynamical evidence for 
BHs would be hard to find.  In fact, the first stellar-dynamical BH detections followed as soon as they became feasible, i.{\ts}e., 
in the mid- to late-1980s, when CCDs became available on spectrographs.  A ``proof-of-concept'' phase followed, when~the~emphasis 
was on finding out whether ground-based observations could provide convincing evidence for BHs.  
Given our expectations about $r_{\rm infl}$, work at optical wavelengths required the excellent seeing of observatories
like Palomar and Mauna Kea.  This period is reviewed~in Kormendy~\& Richstone (1995, hereafter KR95).  The foundations of the 
BH search are laid~out~there.  Here, we bring the story up-to-date through 2013 February.

\vfill
\ni {\big\ARRed 1.1 Road Map}\textBlack
\vsss

      This paper has two parts.  Sections 2 and 3 review the history of dynamical~BH~detections.
We concentrate on improvements in spatial resolution, modeling machinery, and numbers of detections.  
Our confidence that we are finding supermassive black holes is discussed here. 
On the other hand, readers who are mainly interested in BH--host-galaxy 
correlations and the physics of coevolution can skip directly to 
Sections 4 (the distinction between classical and pseudo bulges), 
         5 (BH database), 
         6~and~7 (BH demographics), and 
         8 (BH--host-galaxy coevolution).
Section  9 sums up.

      Our picture of BH demographics divides itself naturally into two phases, one dominated~by~many HST BH detections 
and a post-HST phase when qualitatively new results emerge largely from ground-based, AO-assisted and maser BH detections 
that target a broader range of galaxy types. We begin (Section 1.2) by introducing these phases.

\eject
 
\vs
\ni {\big\ARRed 1.2 HST and the Two Phases of BH Demographic Studies}\textBlack
\vs

      HST helped to complete the proof-of-concept phase by confirming ground-based BH detections with higher 
spatial resolution.  Also, by delivering five-times-better resolution than ground-based, optical spectroscopy, HST made
it possible to find BHs in many more galaxies.  Thus, it could search statistically fair galaxy samples.  This led to the convincing 
conclusion that BHs are present in essentially every galaxy that has a bulge component.  By opening the floodgates of 
discovery, HST advanced the study of supermassive BHs into a new era in which we could study BH demographics.  

      Some results have been in place almost from the beginning of this subject.  
Most important is the agreement between the global volume density of BH mass that is predicted by AGN energetics 
and what we observe
(So\l tan 1982;
Yu \& Tremaine 2002;
Marconi \etal 2004).
But most work in this subject has concentrated on correlations between $M_\bullet$ and properties of host galaxies.

      We differentiate between two phases of this work:

      {\ARRed\bf Sections 6.1 and 6.2:}\textBlack~From the early 1990s until recently, we believed that BH masses satisfy one 
correlation each with the luminosity, mass, and velocity dispersion~of~the~host~galaxy.  Even during this period, we knew that 
these correlations apply to bulges and ellipticals but not to disks.  Still, work to explain the correlations ignored component 
structure.  The emphasis was on using the the potential well depth of the galaxy as measured by $\sigma$\ts--{\ts}and nothing 
else\ts--{\ts}to engineer coevolution.

      {\ARRed\bf Sections 6.3\ts--\ts6.14 and 7:}\textBlack~In contrast, we are now starting to realize that BHs correlate in different 
ways with different components of galaxies.  This is ``low-hanging fruit'': it reveals new connections between galaxy and BH growth.  
This phase of BH work is still in its early stages.

\vs
\ni {\big\ARRed 1.3 The Scope of this Review}\textBlack
\vs

      Sections 2\ts--\ts3 discuss the robustness of BH detections and $M_\bullet$ estimates via optical absorption- and 
emission-line spectroscopy and radio-wavelength maser spectroscopy.  AGN $M_\bullet$ estimates via reverberation mapping 
and single-epoch spectroscopy are discussed in Supplemental Information. 

      Section 4 briefly reviews the difference between classical and pseudo bulges of disk galaxies.

      Section 5 is the $M_\bullet$ and host galaxy database for BHs detected via spatially resolved dynamics.

      Sections 6\ts--\ts7 discuss correlations (or not) of $M_\bullet$ with host galaxy bulges, pseudobulges, disks, 
nuclear star clusters, core properties, globular cluster systems, and dark matter halos.   Section\ts6.15 lists but does not 
discuss possible correlations that we do not confirm.~Section\ts8 reviews coevolution.

      BHs are now too big a subject for a single ARA\&A review.  We must omit many topics, including ones that are of 
great current interest.  
In particular, the theory of BH{\ts}--{\ts}host-galaxy coevolution is too large a topic for this paper.  It has been the subject 
of whole conferences and deserves a review of its own.  We concentrate on observed BH correlations and their implications.  
This includes a preliminary look at the evolution of these correlations with cosmological lookback time and what we learn
from that evolution (Section 8.6).

      Many reviews preceed ours.  
Rees (1984) and Begelman, Blandford \& Rees (1984) review early theory work;
KR95,
Richstone \etal (1998),
Ho (1999a), and 
Kormendy \& Gebhardt (2001) review observations; 
Melia \& Falcke (2001), Melia (2007), and Genzel, Eisenhauer \& Gillessen (2010) discuss the Galactic center;
Ferrarese \& Ford (2005), Peterson (2008), and Marziani \& Sulentic (2012) review the connection with AGN research;
Ho (2008) reviews BHs in low-luminosity AGNs;
Miller \& Colbert (2004) is about intermediate-mass BHs;
Volonteri (2010) and
Greene (2012) are on BH seeds and early growth;
Merritt \& Milosavljevi\'c (2005) discuss the evolution of BH binaries;
Centrella \etal (2010) is on general relativistic effects in BH binaries;
Ho (2004a) and
Cattaneo \etal (2009) 
summarize conferences on BH-galaxy coevolution;
McNamara \& Nulsen (2007; 2012),
Alexander \& Hickox (2012), 
Schawinski (2012), and
Fabian (2012) review AGN feedback, and
Fabian (2013) reviews BH spins.
A short summary of the present paper is Kormendy (2013).

\vfill\eject

\vs
\ni {\big\ARRed 1.4 Terminology}\textBlack
\vs
  
      To avoid confusion, we explain some terminology used in this review:

      {\bf\ARRed Supermassive black holes (BHs)}\textBlack\ are associated with the centers of galaxies~that~contain bulges 
or pseudobulges.  ``Supermassive'' as opposed to~what? Answer: supermassive as opposed to 
ordinary-mass BHs of $\sim$\ts10\ts$M_\bullet$ that are end products of stellar evolution. A ``gray area'' involves 
the several-hundred-solar-mass BHs that may be the end products of the evolution of zero-metallicity stars 
(Bond \etal 1984;
Fryer, Woosley \& Heger 2001;
Heger \etal 2003).
They are plausible seeds of supermassive BHs
(Larson 2000;
Madau \& Rees 2001;
Volonteri, Haardt \& Madau 2003;
Volonteri \& Rees 2005;
Volonteri \& Natarajan 2009;
Volonteri 2010).  
We call BHs that are remnants of stellar evolution without additional growth
\hbox{\bf\ARRed stellar-mass~BHs}\textBlack.  If some BHs get ejected from galaxies as a consequence of 
the evolution of BH triples, we do not invent a new name~for~them.

      {\bf\ARRed  Intermediate-mass black holes (IMBHs)}\textBlack\ here refers only to BHs that may live in the centers 
of globular clusters and in the nuclear star clusters ({\bf\ARRed nuclei}\textBlack) of bulgeless, late-type, and spheroidal 
galaxies.  The mass functions of BHs and IMBHs overlap (Section 7).

      {\bf\ARRed Classical bulges}\textBlack\ are defined purely by observational criteria: they are indistinguishable from 
elliptical galaxies, except that they are embedded in disks (Renzini 1999).  Classification criteria~are summarized in the 
Supplementary Information.  E.{\ts}g., classical bulges satisfy the fundamental plane correlations for ellipticals
(Kormendy \etal 2009, hereafter KFCB;
Fisher \& Drory 2010;
Kormendy \& Bender 2011).
Underlying the practical definition is the assumption that we use in interpretations, which is that
classical bulges form as ellipticals do, in major galaxy mergers. 

      {\bf\ARRed Pseudobulges:}\textBlack\ Kormendy \& Kennicutt (2004) and Kormendy (2012) review observations which show that 
some central galaxy components that we used to identify as classical bulges have properties~that~are more 
disk-like than those of classical bulges and are, we infer, made not by galaxy mergers but by slow (``secular'') 
evolution internal to isolated galaxy disks.  Pseudobulges may be augmented by minor mergers; this
is not an issue here.  The difference between classical~and pseudo bulges is important, because we find that they 
do not correlate in the same~way~with~BHs.  The above papers, Kormendy \& Bender (2013b), and the Supplementary Information here summarize
classification criteria.  We emphasize that they are purely observational; they do not depend on interpretation.  For reliable 
classification, it is important to use as many classification criteria as possible.  All classifications in {\bf Table 3\/} are based 
on at least two and in some cases as many as five independent criteria. Section 4 discusses this subject in more detail.

      {\bf\ARRed (Pseudo)bulge}\textBlack\ refers either to a classical bulge or to a pseudobulge, without prejudice. 

      \hbox{\kern 10pt{\bf\ARRed Spheroidal galaxies (Sphs)}\textBlack\ are morphologically like ellipticals and are called
``dwarf~ellipticals''} \ni by many authors (e.{\ts}g., Binggeli, Sandage, \& Tarenghi 1984; 
Binggeli, Sandage, \& Tammann 1985, 1988; 
Binggeli, Tammann, \& Sandage 1987; 
Sandage \& Binggeli 1984; Sandage, Binggeli, \& Tammann 1985).
But they are robustly different from ellipticals when examined quantitatively in terms of effective~radii~$r_e$ that contain 
half of the total light and effective brightnesses~$\mu_e$~at~$r_e$.  Whereas ellipticals satisfy fundamental plane parameter 
correlations such that less luminous ellipticals have higher effective brightnesses, Sphs define a parameter sequence 
that is roughly perpendicular to that of ellipticals such that less luminous Sphs have lower effective brightnesses
(Wirth \& Gallagher 1984; 
Kormendy 1985, 1987, 2009;
Binggeli \& Cameron 1991;
Bender,{\ts}Burstein \&{\ts}Faber\ts1992;
KFCB;
Kormendy \& Bender 2012).~The sequences overlap over a factor of at least~10 in luminosity.  It is important to note that 
Sandage \& Binggeli (1984) and Sandage, Binggeli, \& Tammann (1985)
distinguish between elliptical and dwarf elliptical galaxies over this whole luminosity range.  
This is testament to their different properties. 
Kormendy (1985, 1987, 2009),
Kormendy \etal (2009), and
Kormendy \& Bender (2012)
show that the $\mu_e$\ts--\ts$r_e$\ts--{\ts}luminosity sequence of Sphs is closely similar
to that of Sd{\ts}--{\ts}Im galaxies.  \hbox{They conclude that Sphs are defunct late-type} galaxies
that have been transformed by a variety of internal and environmental gas removal processes.  Physically,
spheroidal galaxies are equivalent to bulgeless S0 galaxies (Kormendy \& Bender 2012).  Local Group
dwarf spheroidals include Draco, Sculptor, and Fornax; brighter examples are NGC 147, NGC 185, and NGC 205.  The Virgo cluster 
contains spheroidals up to $M_V \simeq -18$.  At still higher luminosities, Sph-like galaxies contain bulges and so are called S0s.

      The importance of distinguishing E and Sph galaxies is this:~So-far scanty data~imply~that Sphs and pure-disk galaxies 
play similar roles in hosting BHs (Section 7.2).~The luminosity function of Sph galaxies rises steeply at low luminosities;
in the Virgo~cluster,~$\phi$\ts$\propto$\ts$L^{-1.35}$ (Binggeli, Sandage, \& Tammann 1988).
In contrast, the luminosity function of ellipticals has a maximum at $M_V \sim -18 \pm 2$ and then drops
steeply at higher and lower $L$ (Binggeli, Sandage,~\&~Tammann~1988,~Fig.\ts1).  Classical bulges are similar.
So the luminosity and mass functions of the components that correlate closely enough with $M_\bullet$ to imply coevolution 
are bounded at low masses.  Understanding coevolution is easier~if~we~do~not~have to face
the prospect that it still happens in dwarf galaxies.

      {\bf\ARRed AGNs}\textBlack~are active galactic nuclei that emit largely nonthermal radiation from BH accretion disks.

      {\bf\ARRed Nuclei}\textBlack\ have a specific meaning, different from any generic use to refer to the centers of galaxies.   
Nuclei are compact central clusters of stars that are distinct from -- i.{\ts}e., smaller and denser than -- the galaxy components
 in which they are embedded (see 
Kormendy \& McClure 1993; 
Lauer \etal 1998 
for the Local Group example~in~M{\ts}33,
B\"oker \etal 2002, 2004;
C\^ot\'e \etal 2006 
for properties of more distant examples, and
B\"oker 2010 for a brief review).
They are similar to globular clusters; their structural parameters $r_e$ and $\mu_e$ 
extend the correlations shown by globular clusters to higher luminosities $L$ (see the above papers;
Carollo 1999;
Hopkins~et~al.~2009a).
Therefore nuclei are not small bulges. 
Some ``globular clusters'' in our Galaxy and others may be defunct nuclei of galaxies whose main bodies have been
stripped away by tidal forces (Section~7.4).  Nuclei are found at virtually all Hubble types but are especially
common in late-type and spheroidal galaxies.  The relationship between nuclei and BHs is discussed in Section 6.11.

      {\bf\ARRed Black holes correlate with}\textBlack\ [some component in galaxies] means more than just that 
this component frequently contains a BH.  By using this phrase, we mean that $M_\bullet$ is observed to correlate 
with parameters of the host component, usually mass, luminosity, and velocity dispersion.

      {\bf\ARRed Tight correlations}\textBlack\ are ones in which the observed scatter is closely similar to the
parameter measurement errors.  These correlations are particularly important because they are the ones that are
suggestive of coevolution.  We will see (Section 6) that $M_\bullet$ correlates tightly with the velocity dispersions,
$K$-band luminosities, and stellar masses of classical bulges and ellipticals but little or not at all with any 
property of pseudobulges and disks.

      {\bf\ARRed Coevolution of BHs and galaxies}\textBlack\ happens differently in different galaxies. At its weakest, it must
happen to all supermassive black holes that live in galaxies.  They must evolve together.~But they do not have to influence 
each other.  A weak form of coevolution is (1) that galaxies affect how BHs grow because they control BH feeding and merging 
via global, galaxy-wide processes.  A stronger form is (2) that BHs may control galaxy properties via energy and momentum feedback 
into the galaxy evolution process.  When people interpret tight BH--host-galaxy correlations as implying coevolution, 
they usually mean the strong version when both of the above happen.  This is what we mean when we use ``coevolution'' without
qualification.  We argue in Section 8 that this strong coevolution is less prevalent than the galaxy formation literature 
suggests.  When only (1) applies but largely not (2), we still use the term ``coevolution'', but we make the distinction clear. 
In contrast, when BH feeding involves local (not galaxy-wide) processes and when the resulting AGNs have little noticeable 
effect on their hosts, then we say that no coevolution takes place, even though the BHs are, of course, still growing inside galaxies. 
Thus Section 8 discusses four regimes of AGN feedback that range from no coevolution to strong coevolution in the above senses.

\vfill\eject

\vs
\ni {\big\ARRed 2. PROGRESS IN BH DETECTION TECHNOLOGY. I.}
\vsss

\ni {\big\ARRed \quad{\kern 4pt}IMPROVEMENTS IN SPATIAL RESOLUTION}\textBlack
\vs

      {\bf Figure 1} shows the history of $M_\bullet$ measurements for all galaxies that have BH detections based 
on observations of spatially resolved dynamics.  Multiple measurements for each galaxy are joined by straight line 
segments to show how the available spatial resolution has improved with time.  For BHs found with HST, only the
discovery observations are shown; these have not been superseded.  The individual measurements for our Galaxy, M{\ts}31, 
and M{\ts}32 are listed in {\bf Table 1}.  Spatial resolution is parametrized by the ratio of the radius $r_{\rm infl}$ 
of the sphere of influence of the BH to the effective Gaussian dispersion radius $\sigma_*$ of the point-spread function 
(see notes to {\bf Table 1}).

\vs
\ni {\big\ARRed 2.1 Early Ground-Based BH Discoveries}\textBlack
\vs

      KR95 reviews the seven ground-based BH detections (and one based on HST) available in 1995.

      The first dynamical BH discovery was in M{\ts}32 (Tonry 1984, 1987).  It was made with barely 
enough resolution to be reasonably secure (Kormendy 2004; \S\S\ts2.2.1 here).  
It is typical of a subject with urgent expectations and much at stake that the first~discovery is made 
as soon as it becomes barely feasible and at a time when the result still has only~modest~significance.~These 
early papers are successful if the discovery gets more secure as it gets tested with better technology.  
This has happened for M{\ts}31, M{\ts}32, NGC 3115, NGC 3377, and NGC 4594.  M{\ts}32 has the most accurate
$M_\bullet$ measurement based on absorption-line spectra of unresolved stellar populations.

      Other pre-HST BH detections based on optical, absorption-line spectroscopy were those~in~M{\ts}31 
(Dressler \& Richstone 1988; Kormendy 1988a),
NGC 4594~(Sombrero galaxy, Kormendy 1988b), 
NGC 3115 (Kormendy \& Richstone 1992), and
NGC 3377 (Kormendy 1992a, b; KR95; Kormendy \etal 1998).
A ground-based BH detection in NGC 4486B was reported in the HST era (Kormendy \etal 1997).
All these have been confirmed with HST except NGC 4486B (HST data are available but not yet modeled).
These results are reviewed in Section 2.2.

\vfill




 \includegraphics{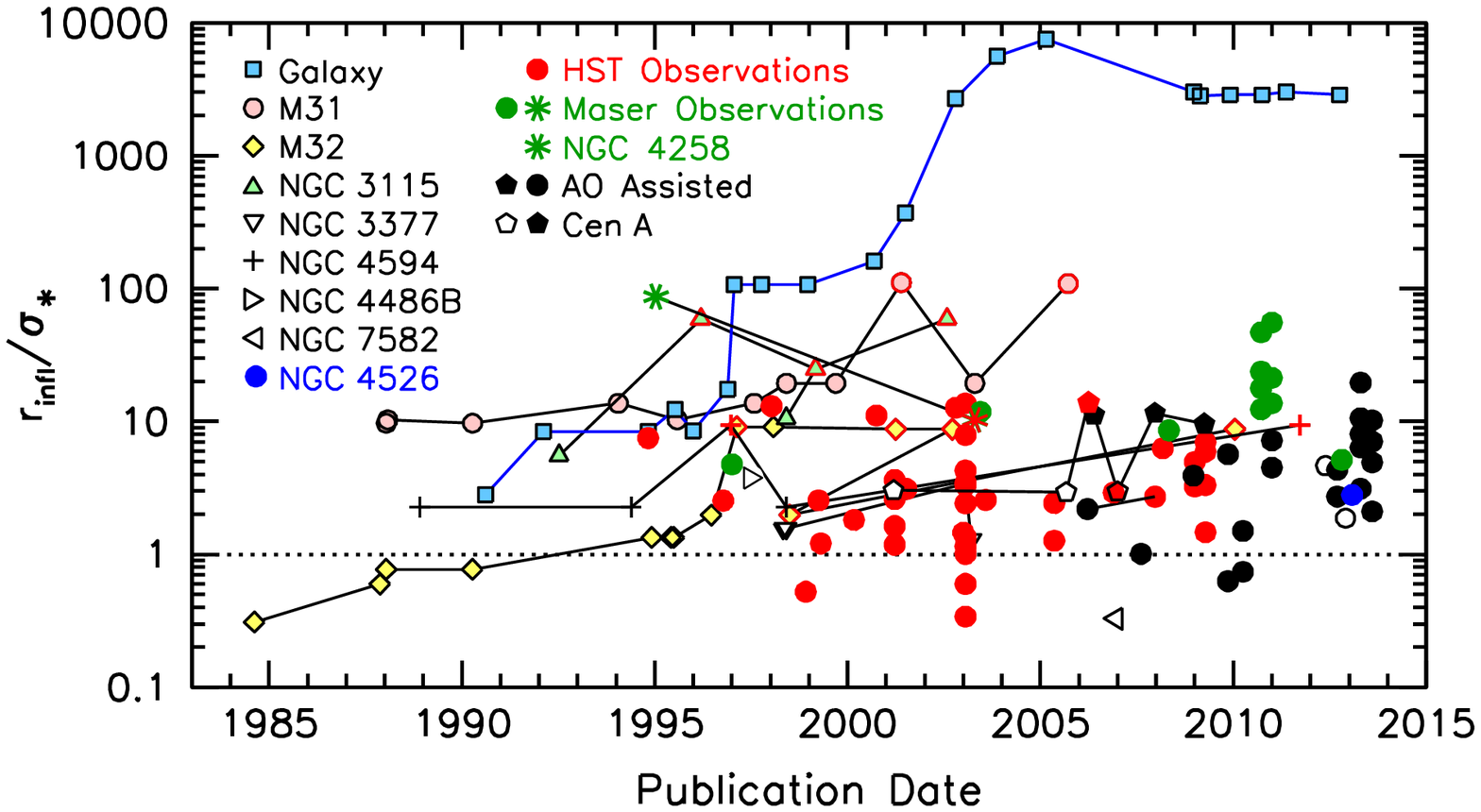}

\ni {\bf \textBlue Figure 1}\textBlack 

\vskip 1pt
\hrule width \hsize
\vskip 2pt

\ni Spectroscopic spatial resolutions for BH discoveries ({\bf Tables 1 -- 3\/}) and
    histories of improvements in resolution.  AO-assisted resolutions are approximate.  The maser discoveries at 2010.9\ts$\pm$\ts0.2 
    and the AO discoveries at 2013.4\ts$\pm$\ts0.2
    have been spread out enough so that they can be distinguished.~Note that most
    pre-HST, ground-based BH discoveries were made at similar or higher $r_{\rm infl}/\sigma_*$~than most HST BH discoveries.
    But HST has 5 -- 10 times better spatial resolution (absent adaptive optics), 
    so it is used to discover smaller BHs in more distant galaxies.  Updated from Kormendy\ts(2004).

\eject

\hsize=17truecm  \hoffset=-0.8truecm  \vsize=25.3truecm  \voffset=-0.5truecm

\cl{\null}
\vskip -63pt
\cl{\null}

\hfuzz=20pt
\def\5{\phantom{.}}

$$
\table
\scbaselines
\hskip -10pt
\tablewidth{15.3truecm}
\tablewidth{16.9truecm} 
\tablewidth{15.0truecm}
\tablewidth{17.0truecm}
\tablespec{\l\c\c\c\c\c\c\l}
\body{
\header{\bf \Blue{Table 1 \quad  Mass measurements of supermassive black holes in our Galaxy, M{\ts}31, and M{\ts}32}}
\skip{3pt}
\hline
\skip{2pt}
&{Galaxy}&   {$D$}  & {$\sigma_e$}           & {$M_\bullet$ ($M_{\rm low},M_{\rm high}$)} & {r$_{\rm infl}$}\0  & {$\sigma_*$} & {r$_{\rm infl}/\sigma_*$} & {Reference} \end
&        &  {(Mpc)} & {\llap{(}km s$^{-1}$\rlap{)}} & {($M_\odot$)}                              & {\llap{(}arcsec)} & {(arcsec)}   &                           &             \end
&{(1)}   &  {(2)}   & {(3)}                  & {(4)}                                      & {(5)~~}             & {(6)}        & {(7)}                       & {(8)}     \end
\skip{2pt}
\hline
\skip{2pt}
&Galaxy   &      &     & 4.41(3.98\ts--\ts4.84)    e6       &             &\00.0146 &\llap{2}868.\0 & Meyer et al. 2012       \end
&Galaxy   &      &     & 4.2\0(3.9\0\ts--\ts4.6\0) e6       &             &\00.0139 &\llap{3}013.\0 & Yelda et al. 2011       \end
&Galaxy  &0.00828& 105 & 4.30(3.94\ts--\ts4.66)    e6       &\llap{4}1.9\0&\00.0146 &\llap{2}868.\0 & Genzel, Eisenhauer \& Gillessen 2010      \end
&Galaxy  &0.00828& 105 & 4.30(3.94\ts--\ts4.66)    e6       &\llap{4}1.9\0&\00.0146 &\llap{2}868.\0 & Gillessen et al. 2009a  \end
&Galaxy   &      &     & 4.09(3.74\ts--\ts4.43)    e6       &             &\00.0148 &\llap{2}829.\0 & Gillessen et al. 2009b  \end
&Galaxy   &      &     & 4.25(3.44\ts--\ts4.79)    e6       &             &\00.0139 &\llap{3}013.\0 & Ghez et al. 2008        \end
&Galaxy   &      &     & 3.80(3.60\ts--\ts4.00)    e6       &             &\00.0056 &\llap{7}478.\0 & Ghez et al. 2005        \end
&Galaxy   &      &     & 3.7\0(3.3\0\ts--\ts4.1\0) e6       &             &\00.0075 &\llap{5}583.\0 & Ghez et al. 2003        \end
&Galaxy   &      &     & 3.8\0(2.3\0\ts--\ts5.4\0) e6       &             &\00.0155 &\llap{2}702.\0 & Sch\"odel et al.~2002   \end
&Galaxy   &      &     & 2.1\0(1.3\0\ts--\ts2.8\0) e6       &             & 0.113   &        371.\0 & Chakrabarty \& Saha~2001\end
&Galaxy   &      &     & 3.1\0(2.6\0\ts--\ts3.6\0) e6       &             & 0.26\0  &        161.\0 & Genzel et al.~2000      \end
&Galaxy   &      &     & 2.7\0(2.5\0\ts--\ts2.9\0) e6       &             & 0.39\0  &        107.\0 & Ghez et al.~1998        \end
&Galaxy   &      &     & 2.70(2.31\ts--\ts3.09)    e6       &             & 0.39\0  &        107.\0 & Genzel et al.~1997      \end
&Galaxy   &      &     & 2.55(2.12\ts--\ts2.95)    e6       &             & 0.39\0  &        107.\0 & Eckart \& Genzel 1997      \end
&Galaxy   &      &     & 2.8\0(2.5\0\ts--\ts3.1\0) e6       &             & 2.4\0\0 &        \017.4 & Genzel et al.~1996      \end
&Galaxy   &      &     & 2.0\0(0.9\0\ts--\ts2.9\0) e6       &             & 4.9\0\0 &       \0\08.5 & Haller et al.~1996      \end
&Galaxy   &      &     & 2.9\0(2.0\0\ts--\ts3.9\0) e6       &             & 3.4\0\0 &        \012.3 & Krabbe et al.~1995      \end
&Galaxy   &      &     & 2.\0\0\phantom{(2.6\0\ts--\ts3.7\0)} e6 &        & 5\5\0\0\0 &     \0\08.4 & Evans \& de Zeeuw 1994\end
&Galaxy   &      &     & 3.\0\0\phantom{(2.6\0\ts--\ts3.7\0)} e6 &        & 5\5\0\0\0 &     \0\08.4 & Kent 1992             \end
&Galaxy   &      &     & 5.4\0(3.9\0\ts--\ts6.8\0) e6       &             &15\5\0\0\0\0&    \0\02.8 & Sellgren et al.~1990  \end
\skip{2pt}
\hline
\skip{2pt} \Red{
&M{\ts}31 & 0.774& 169 &  1.4 (1.1\ts--\ts2.3) e8           & 5.75        & 0.053   &  109.\0       & Bender et al.~2005      \end}\textBlack
&M{\ts}31 &      &     &  1.0 \phantom{(0.0\ts--\ts0.0)} e8 &             & 0.297   & \019.4        & Peiris \& Tremaine 2003 \end\Red{
&M{\ts}31 &      &     &  6.1 (3.6\ts--\ts8.7) e7           &             & 0.052   &  111.\0       & Bacon et al.~2001       \end}\textBlack
&M{\ts}31 &      &     &  3.3 (1.5\ts--\ts4.5) e7           &             & 0.297   & \019.4        & Kormendy \& Bender 1999 \end
&M{\ts}31 &      &     &  6.0 (5.8\ts--\ts6.2) e7           &             & 0.297   & \019.4        & Magorrian et al.~1998   \end
&M{\ts}31 &      &     &  9.5 (7\5\0\ts--\5\ts10) e7        &             & 0.42\0  & \013.7        & Emsellem \& Combes 1997 \end
&M{\ts}31 &      &     &  7.5 \phantom{(0.0\ts--\ts0.0)} e7 &             & 0.56\0  & \010.3        & Tremaine 1995           \end
&M{\ts}31 &      &     &  8.0 \phantom{(0.0\ts--\ts0.0)} e7 &             & 0.42\0  & \013.7        & Bacon et al.~1994       \end
&M{\ts}31 &      &     & 5\5\0 (4.5\ts--\ts5.6) e7          &             & 0.59\0  &\0\09.7        & Richstone, Bower \& Dressler 1990   \end
&M{\ts}31 &      &     &  3.8 (1.1\ts--\5\ts11) e7          &             & 0.56\0  & \010.3        & Kormendy 1988a          \end
&M{\ts}31 &      &     &  5.6 (3.4\ts--\ts7.8)  e7          &             & 0.59\0  &\0\09.7        & Dressler \& Richstone 1988    \end
\skip{2pt}
\hline
\skip{2pt} \Red{
&M{\ts}32 &0.805 &\077 &  2.45(1.4\ts--\ts3.5) e6           & 0.46        & 0.052   &  \18.76       & van den Bosch \& de Zeeuw 2010\end
&M{\ts}32 &      &     &  2.9 (2.7\ts--\ts3.1) e6           &             & 0.052   &  \18.76       & Verolme et al.~2002     \end
&M{\ts}32 &      &     &  3.5 (2.3\ts--\ts4.6) e6           &             & 0.052   &  \18.76       & Joseph et al.~2001      \end}\textBlack
&M{\ts}32 &      &     &  2.4 (2.2\ts--\ts2.6) e6           &             & 0.23\0  &  \11.98       & Magorrian et al.~1998   \end\Red{
&M{\ts}32 &      &     &  3.9 (3.1\ts--\ts4.7) e6           &             & 0.050   &  \19.11       & van der Marel et al.\ts1998a \end
&M{\ts}32 &      &     &  3.9 (3.3\ts--\ts4.5) e6           &             & 0.050   &  \19.11       & van der Marel et al.\ts1997a, 1997b\end}\textBlack
&M{\ts}32 &      &     &  3.2 (2.6\ts--\ts3.7) e6           &             & 0.23\0  &  \11.98       & Bender, Kormendy \& Dehnen 1996      \end
&M{\ts}32 &      &     &  2.1 (1.8\ts--\ts2.3) e6           &             & 0.34\0  &  \11.34       & Dehnen 1995             \end
&M{\ts}32 &      &     &  2.1 \phantom{(2.6\ts--\ts3.7)} e6 &             & 0.34\0  &  \11.34       & Qian et al.~1995        \end
&M{\ts}32 &      &     &  2.1 (1.7\ts--\ts2.4) e6           &             & 0.34\0  &  \11.34       & van der Marel et al.\ts1994a \end
&M{\ts}32 &      &     &  2.2 (0.8\ts--\ts3.5) e6           &             & 0.59\0  &  \10.77       & Richstone, Bower \& Dressler 1990   \end
&M{\ts}32 &      &     &  9.3 \phantom{(4.7--18.9)}      e6 &             & 0.59\0  &  \10.77       & Dressler \& Richstone 1988    \end
&M{\ts}32 &      &     &  7.5 (3.5--11.5)      e6           &             & 0.76\0  &  \10.60       & Tonry 1987              \end
&M{\ts}32 &      &     &  5.8 \phantom{(2.6\ts--\ts3.7)} e6 &             & 1.49\0  &  \10.31       & Tonry 1984              \end
\skip{2pt}
\hline
}
\endtable
$$

\scbaselines

\null
\vskip -22pt
\null

\semiwidebaselines

{\sc\noindent\null\vskip -4pt
\vbox{\Red{Lines based on HST spectroscopy are in red.}\textBlack\
Column 2 is the assumed distance.  
Column 3 is the stellar velocity dispersion inside the ``effective radius'' that encompasses half of the light of the bulge.
Column 4 is the measured BH mass with the one-sigma range that includes 68\ts\% of the probability in parentheses.  Only the top four $M_\bullet$
         values for the Galaxy include distance uncertainties in the error bars. 
Column 5 is the radius of the sphere of influence of the BH; the line that lists $r_{\rm infl}$ contains the adopted $M_\bullet$.
Column 6 is the effective resolution of the spectroscopy, estimated as in Kormendy (2004).  It is a \underbar{radius}
that measures the blurring effects of the telescope point-spread function or ``PSF,'' the slit
width or aperture size, and the pixel size.  The contribution of the telescope is estimated 
by the dispersion $\sigma_{*\rm tel}$ of a Gaussian fitted to the core of the average radial brightness profile
of the PSF.  In particular, the HST PSF has $\sigma_{*\rm tel} \simeq$ 0\sd036 from a single-Gaussian fit
to the PSF model in van der Marel, de Zeeuw \& Rix (1997a).  Then the resolutions in the directions
parallel and perpendicular to the slit are $\sigma_{*\parallel}$, the sum in quadrature of $\sigma_{*\rm tel}$ 
and 1/2 of the pixel size, and $\sigma_{*\bot}$, the sum in quadrature of $\sigma_{*\rm tel}$ and 1/2 of the
slit width.  Finally, the effective $\sigma_*$ is the geometric mean of $\sigma_{*\parallel}$ and $\sigma_{*\bot}$.   
The top 7 lines for the Galaxy use a proxy $\sigma_*$ equal to the smallest pericenter radius for the stellar orbits 
that were included in the derivation; usually this is for star S2, but for lines 6 and 7, it is star S16.
Note: $\sigma_*$ measures~a~PSF~radius; for readers who prefer a PSF diameter, FWHM $\equiv 2.35\ts\sigma_*$.
\hbox{Column 7 is the effective measure of the degree to which the observations} reach inside the sphere of influence
            and reliably see the BH.
Column 8 lists the reference.  Updated from Kormendy (2004).}
\lineskip=-20pt \lineskiplimit=-20pt
}

\dblbaselines

\vfill\eject

\hsize=15.0truecm  \hoffset=0.0truecm  \vsize=20.1truecm  \voffset=1.5truecm

      Some early authors worried that $r_{\rm infl}$ is too small for ground-based BH detection (e.{\ts}g., 
Rix 1993; 
Emsellem, Bacon, \& Monnet 1995).  
This was never the problem: the initial, ground-based $r_{\rm infl}/\sigma_*$ was equal to or better than the median for HST discoveries 
({\it red points} in {\bf Figure 1}) for our Galaxy, M{\ts}31, NGC 4594, NGC 3115, and NGC 4486B.  That is, the easiest cases were 
found from the ground with $r_{\rm infl}/\sigma_* \sim 5$ to 10 before HST had a chance to discover them at higher $r_{\rm infl}/\sigma_*$.

      Instead, there were two dangers:  

      First was the danger that $M_\bullet$ would be overestimated if $r_{\rm infl}$ was not well enough resolved.  
This happened slightly in Magorrian \etal (1998, see Kormendy 2004), which mostly used ground-based data that were not obtained with 
high spatial resolution for the BH search.~Magorrian \etal (1998) derived a BH-to-bulge mass ratio of 
$0.0052^{+0.0014}_{-0.0011}$, higher than previous and subsequent values based on the same galaxies and physical assumptions.
Ironically, the present, improved BH masses that give greater emphasis to giant, core ellipticals give (Section 6.6.1) the same BH mass 
fraction that Magorrian got.  In any case, the ground-based BH detections reviewed in KR95 were obtained with sufficient resolution, 
and their $M_\bullet$ estimates ({\bf Table\ts1}) were not systematically too high.

      Second was the danger that ground-based work would succeed only for a biased sample~with larger-than-average 
$M_\bullet$.  But KR95 already got a mean BH-to-bulge mass ratio of $0.0022^{+0.0016}_{-0.0009}$.  Later values 
based on larger HST samples range from $0.0013^{+0.0023}_{-0.0008}$
(Merritt \& Ferrarese 2001;~cf.
McLure \& Dunlop 2002)
to $0.0023^{+0.0020}_{-0.0011}$
(Marconi \& Hunt 2003).  Modern searches with improved modeling machinery can measure $M_\bullet$ properly even when
$r_{\rm infl}/\sigma_* \sim 0.3$\ts--\ts0.5 (\S\ts3.1.1).~Concerns about sample biases have not vanished.  But other effects 
prove to be more important.~As discussed in Sections 3 and 6, improvements in $M_\bullet$ measurement technology (e.{\ts}g.,
omission of $M_\bullet$ values based on gas dynamics when broad emission-line widths were not taken into account and inclusion 
of dark matter in stellar dynamical models) have revised typical BH masses upward and now lead (Section 6.61) to a substantially 
larger BH-to-bulge mass fraction.

      The strongest BH cases---the ones in our Galaxy (\S\S\ts2.3) and in NGC 4258 
(\S\ts3.3)~-- come entirely from ground-based work.  The maser galaxy NGC 4258 has been observed 
with HST, but only to check our modeling machinery (Pastorini \etal 2007; Siopis \etal 2009).

\vs
\ni {\big\ARRed 2.2 HST confirmed ground-based BH detections at high resolution}\textBlack
\vs

      Conceptually (but not chronologically), the first important contribution from HST was to confirm the ground-based
BH detections at $\sim$\ts5 times higher spatial resolution.  Early detections were viewed with varying 
degrees of skepticism.  The subject had a shaky~history.  KR95 emphasized the problem:  ``It is easy to believe 
that we have proved what we expect~to~find.''~Some kinds of evidence, e.{\ts}g., finding cuspy brightness profiles 
(Young \etal 1978;
Lauer \etal 1992, 1993, 1995;
Crane \etal 1993)
that look similar to the cusps expected from BHs 
(Peebles 1972;
Bahcall\ts\&{\ts}Wolf 1976;
Young 1980) 
were interpreted as revealing BHs (Young \etal 1978) 
but proved not to be evidence for BHs at all (KR95).  The authors of early stellar dynamical BH papers 
did their best to be careful, but detections were bound to be more secure -- and mass estimates were
bound to be more accurate -- when verified with higher spatial resolution.   
Improvements in modeling machinery were under way in parallel.  Early papers made simplifying assumptions (e.{\ts}g.,
spherical symmetry, isotropic velocity dispersion tensors) that were quickly superseded.  We discuss spatial 
resolution in the present section and improvements in modeling in \S\ts3.1.

      Our Galaxy, M{\ts}32, and M{\ts}31 have the most detailed histories of observational and modeling improvements in 
measurements of $M_\bullet$.  {\bf Table 1} lists all measurements of $M_\bullet$ in these galaxies, using
$r_{\rm infl}/\sigma_*$ as a measure of how well they resolve the BH sphere of influence.  They are ordered by 
publication date (updated from Kormendy~2004, which also lists such data for other galaxies).  
{\bf Figure 2} illustrates this record for M{\ts}32.

\vfill\eject

\vs

\ni {\bf\ARRed 2.2.1.~M{\ts}32.}\textBlack {\bf ~Figure 2} illustrates our point that the BH in M{\ts}32 was discovered as early as 
our technology allowed.  With $r_{\rm infl}/\sigma_* < 1$, Tonry (1984) overestimated~$M_\bullet$,~although~not~significantly with 
respect to the uncertainties.~Ten years were needed to achieve \hbox{$r_{\rm infl}/\sigma_* > 1$}~(van~der~Marel~et~al. 1994b; 
Dehnen 1995; Qian \etal 1995).  By then, $\sigma_*$ had improved by a factor~of~4 from Tonry's data, and $M_\bullet$ had converged
 to its present value.  Observations with the Canada-France-Hawaii telescope improved $\sigma_*$ by an additional factor of 1.5 and further 
confirmed $M_\bullet$ (Bender, Kormendy \& Dehnen 1996).  Soon afterward, HST spectra with 5 times better resolution became available and 
resoundingly confirmed that stellar dynamical $M_\bullet$ measurements were working.~The total 
improvement in resolution since Tonry (1984) is a factor of 30.  But $M_\bullet$ estimates have stayed stable through 20 years of 
improvements in resolution and modeling.  This is an example of a general result: constrained by formation physics, galaxies do not use 
their freedom to indulge in perverse orbit structures that would cause simple models to provide wrong masses. 
The above successful history has been critically important to our confidence in~BH~detections.

\vs

\vfill


 \includegraphics{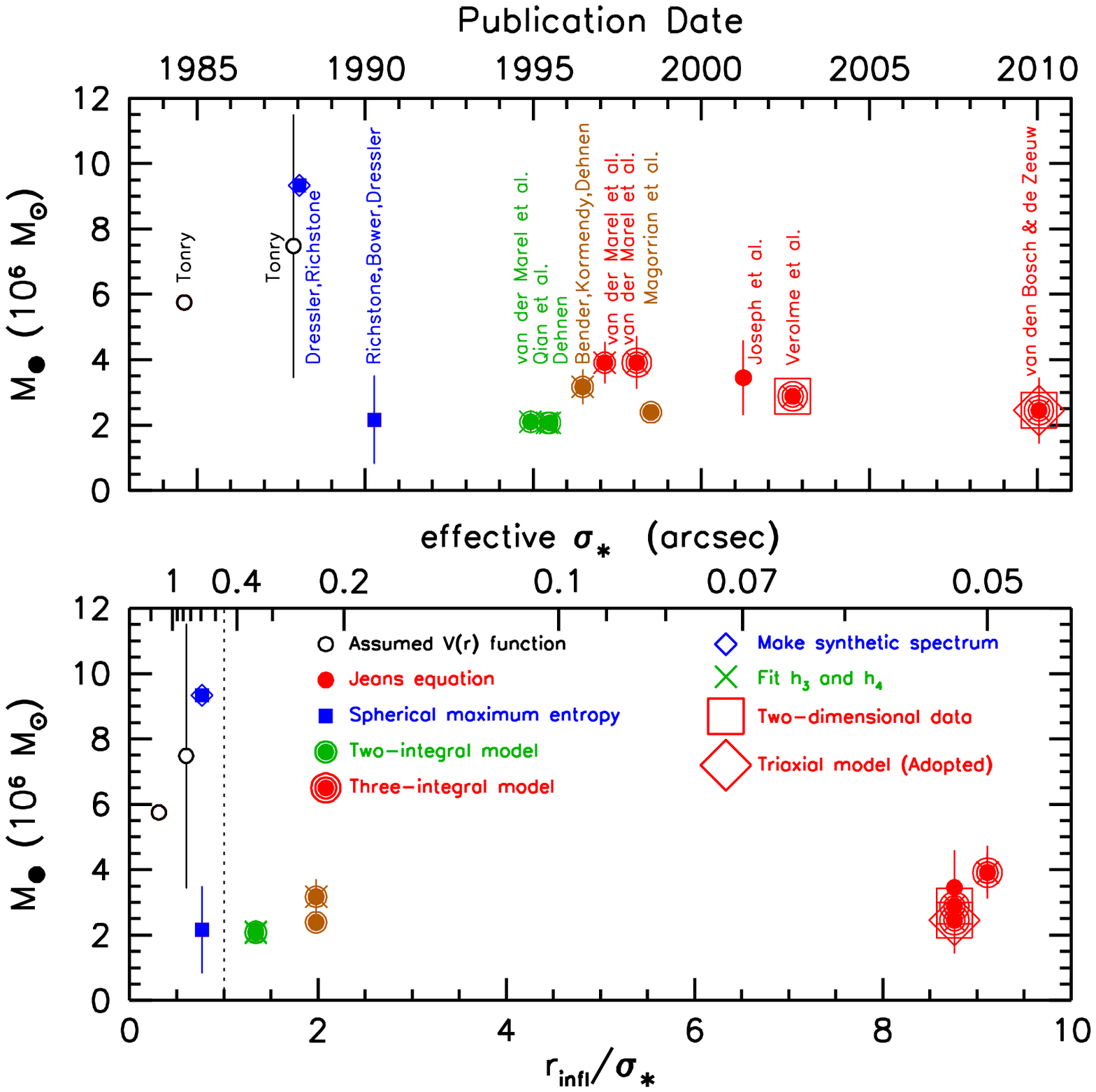}


\ni {\bf \textBlue Figure 2}\textBlack 

\vskip 1pt
\hrule width \hsize
\vskip 2pt

\ni History of stellar-dynamical measurements of $M_\bullet$ in M{\ts}32 (updated from Kormendy 2004).  Derived BH mass is shown 
as a function of ({\it top\/}) publication date and ({\it bottom\/}) spectroscopic~spatial resolution.  Resolution is measured by the effective
Gaussian dispersion radius $\sigma_*$ of the PSF.  More relevant physically is the ratio of the radius of the BH 
sphere of influence, $r_{\rm infl} = GM_\bullet/\sigma^2$,~to~$\sigma_*$.  
\hbox{If $r_{\rm infl}/\sigma_*$ \lapprox \ts1,} velocities are dominated by the mass distribution of the stars even in the central~pixel  
and BH detection is difficult.  If $r_{\rm infl}/\sigma_* \gg 1$, we resolve the region~where~velocities 
are controlled by the BH.  Symbol shapes encode improvements in observations, in kinematic analysis, and in dynamical modeling techniques
(see \S\ts3.1).  HST measurements are shown in red.

\eject

\ni {\bf\ARRed 2.2.2. NGC 3115, NGC 3377, and NGC 4594.}\textBlack ~Similar improvements in  $r_{\rm infl}/\sigma_*$ 
of a factor of 11 for NGC 3115 (from Kormendy \& Richstone 1992 to Kormendy \etal 1996b; Emsellem, Dejonghe \& Bacon 1999),
   a factor of 4.1 for NGC 4594 (Kormendy 1988b to Kormendy \etal 1996a; Jardel \etal 2011), and
   a factor of 2.2 for NGC 3377 (Kormendy \etal 1998 to Gebhardt \etal 2003) 
also improved confidence as the earlier $M_\bullet$ estimates were confirmed~(Table~1.1~in~Kormendy~2004).  

\vfill

\includegraphics{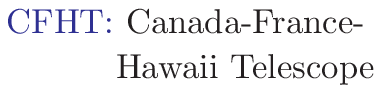}

\ni {\bf\ARRed 2.2.3. M{\ts}31.}\textBlack ~Improvements in $r_{\rm infl}/\sigma_*$ of a factor of 11 from Dressler 
\& Richstone (1988) and Kormendy (1988a) to Bender \etal (2005) have been especially important to our confidence in BH detection.  
M{\ts}31 was in some ways easy, because $r_{\rm infl}/\sigma_*$\ts$\simeq$\ts10 already in the 1988 papers.  But Lauer \etal (1993) discovered
that the galaxy has an asymmetric double nucleus of red stars, P1 and P2 ({\bf Figure~3}).  The good news was that P1$+$P2 are rotation-dominated; 
this reduces uncertainties caused by poorly known velocity anisotropy.  The bad news was that the asymmetry was at first not understood.  
It caused some authors to worry that the nucleus is not in equilibrium and~that~$M_\bullet$ can not be derived 
(Bacon \etal 1994;
van der Marel 1995; 
Ferrarese \& Ford 2005).  
The explanation of the double nucleus as an eccentric disk (Tremaine 1995; Peiris \& Tremaine 2003) whose signature kinematic 
asymmetries are well confirmed at CFHT and HST resolution (Kormendy \& Bender 1999; Bender \etal 2005) removes this uncertainty.  
\hbox{In any case, these complications are now moot}.  CFHT observations by Kormendy \& Bender (1999) showed that the BH lives 
in a third, tinier nucleus (``P3'') of blue stars embedded in P2 ({\bf Figure 3}).  The highest-resolution BH mass measurement is 
now based entirely on the dynamics of P3, as follows (Bender \etal 2005).

      {\bf Figure 4} shows the rotation curve and velocity dispersion profile of P3 as measured~with~HST. The Keplerian rotation curve 
of a cold, razor-thin disk ({\it top panel\/}) fits the measurements well after PSF convolution and pixel sampling ({\it bottom two panels 
of plots\/}).  PSF-convolved surface brightnesses, line-of-sight velocity dispersions, and mean velocities are 
shown at right in {\bf Figure~4}.  Note that the apparent dispersion is entirely caused by rotational line broadening.  
Orbit-superposition, maximum entropy models also were fitted to the data; the conclusion is that 
$M_\bullet = 1.4 \times 10^8$ $M_\odot$ with a 1 sigma confidence range of (1.1 -- 2.3) $\times 10^8$ $M_\odot$.  
This larger $M_\bullet$ estimate is consistent with the error bars in earlier measurements from dynamical models of P1 and P2 but is more reliable. 
Thus the BH mass in M{\ts}31 has now been confirmed in a series of 
studies that greatly improved both the spatial resolution and the accuracy of modeling assumptions.

\vs
\vfill

\includegraphics{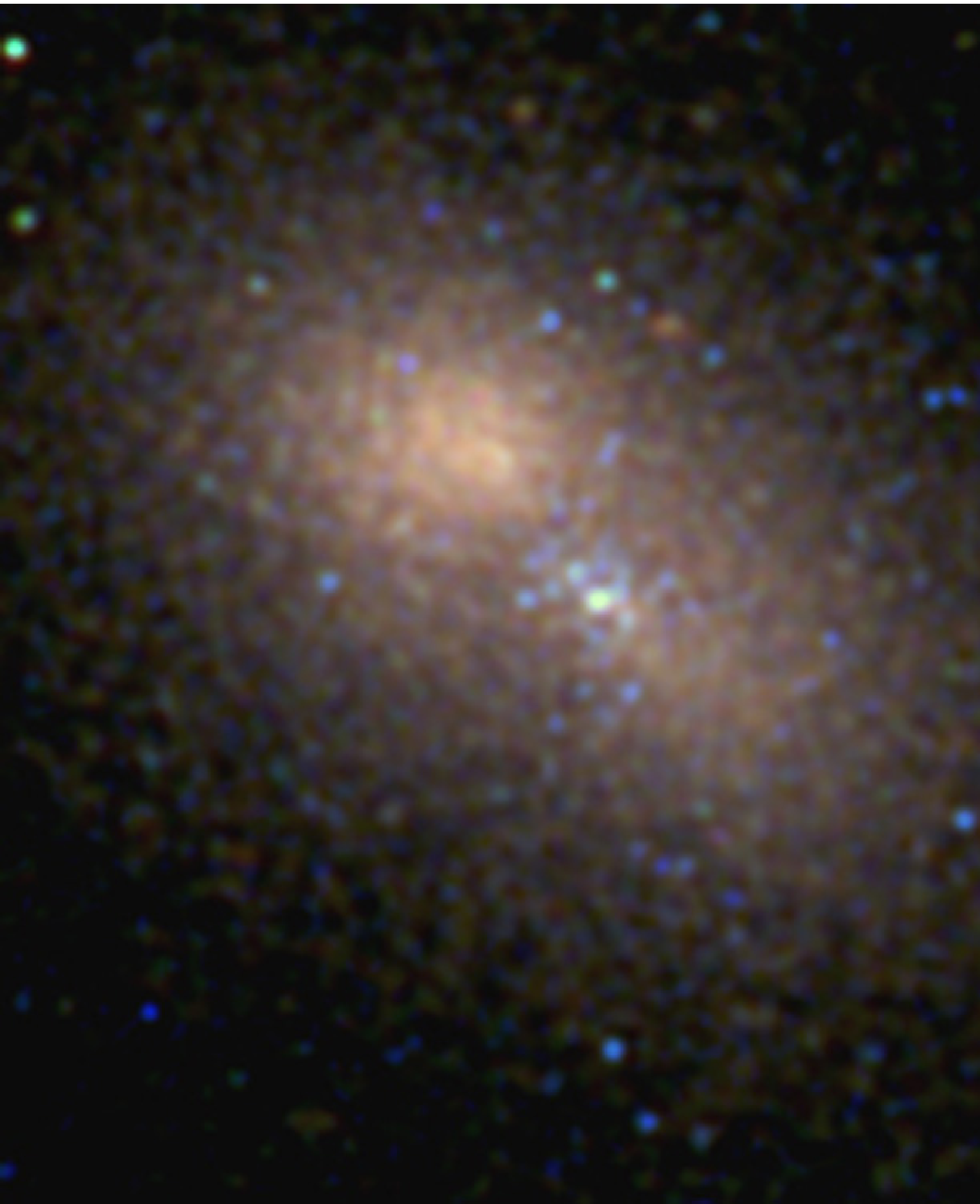}

\def\figindent{\indent \hangindent=3.25truein}

{\parindent=3.25truein
\figindent \ni {\bf \textBlue Figure 3}\textBlack 

\vskip 3pt
\nointerlineskip \moveright 3.25truein\vbox{\hrule width 2.655truein}\textBlack\nointerlineskip
\vskip 3pt

\figindent Nyquist-sampled color image of the triple nucleus of M{\ts}31 made from $V$-, $B$-, and 3000-\AA-band images
obtained with the HST Advanced Camera for Surveys (Lauer~et~al. 2012).  The High Resolution Camera scale, 0\sd028 $\times$ 0\sd025, 
does not adequately sample the PSF.  Dithered exposures were therefore obtained in a 2$\times$2 square pattern of 0.5-pixel steps 
and combined using the Fourier image reconstruction technique of Lauer (1999).  The final, PSF-deconvolved image has a scale of 
0\sd0114 pixel$^{-1}$, a resolution FWHM of 0\sd030 = 0.11 pc,  and a field of view of $3^{\prime\prime} \times 3^{\prime\prime}$.  
North is up and east is at left.  The upper-left brightness peak is P1; the lower-right peak is P2.  The brightest blue source 
at the center is P3.  The~BH~lives~in~P3.  

}

\eject

\cl{\null} \vskip 3.65truein

\includegraphics{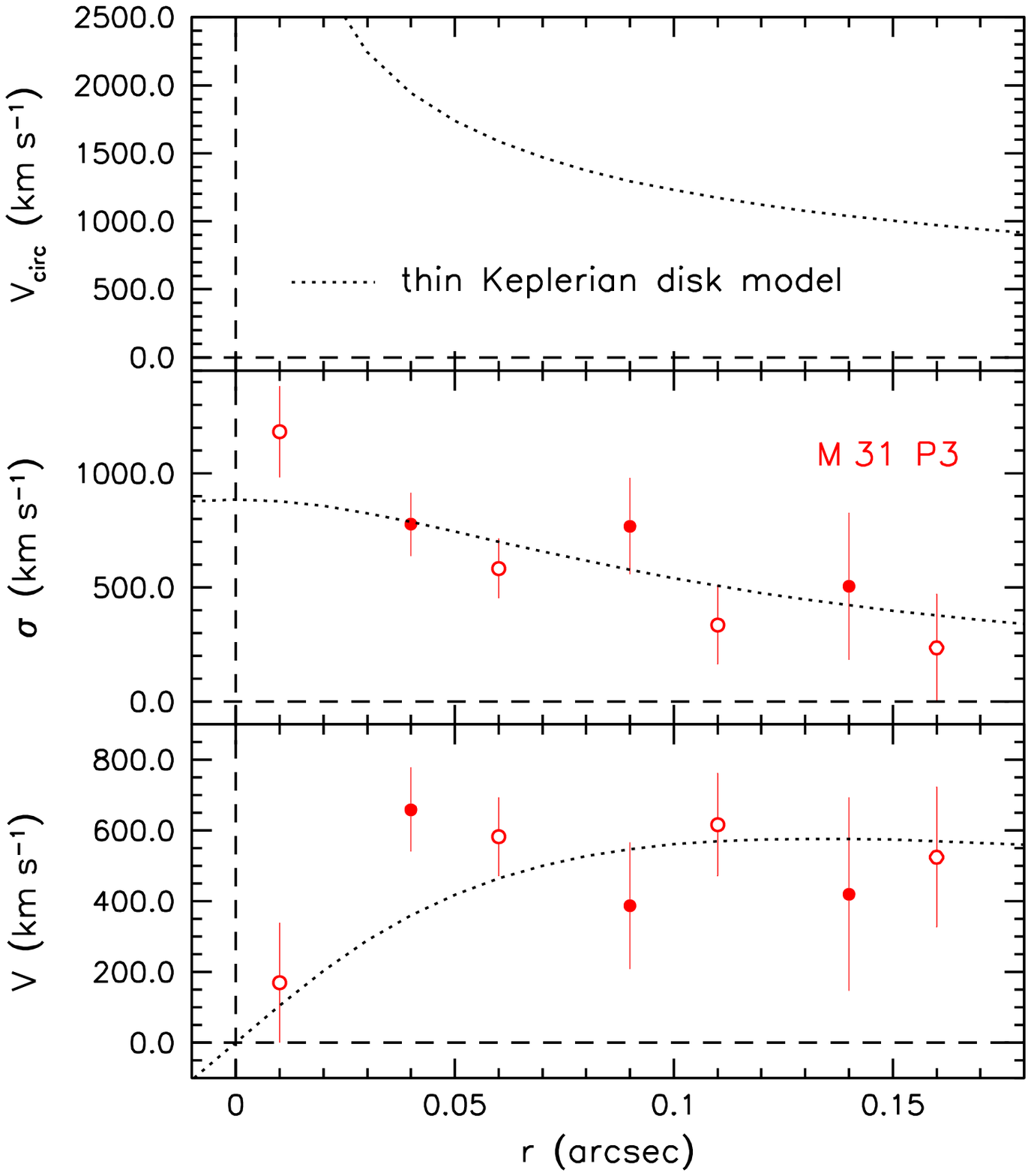}




 \includegraphics{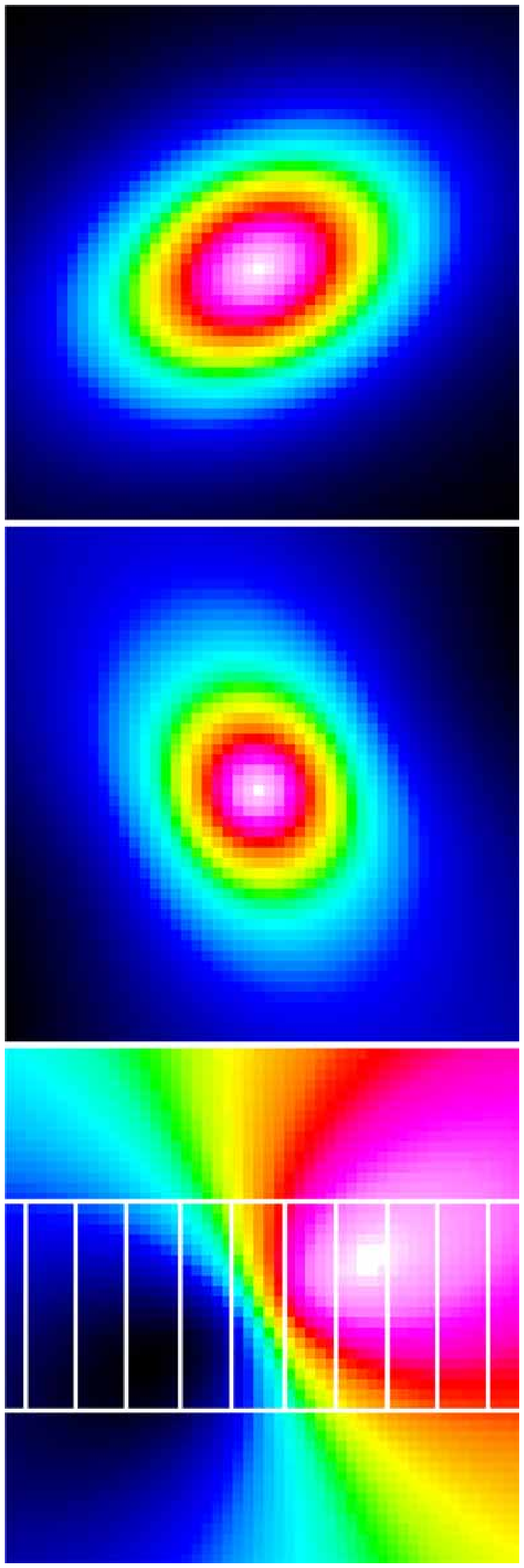}

\ni {\bf \textBlue Figure 4}\textBlack 

\vskip 1pt
\hrule width \hsize
\vskip 2pt

\ni The bottom two plots show the measured rotation velocity $V$ and velocity dispersion $\sigma$ along
    the major axis of P3 with light from P1 and P2 subtracted.  Spectral decomposition is easy, because P3 is dominated
    by Balmer absorption lines that are essentially absent in the P1~and~P2.  Open and filled circles are from opposite sides
    of the center.  Dotted curves show the best-fit Keplerian rotation curve of an infinitely thin disk at the top and, 
    after PSF convolution and pixel sampling, in the bottom panels.  The 0\sd5 $\times$ 0\sd5 color images show the
    model surface brightness ({\it top}, linear scale from 0 = black to 1 = white), apparent velocity dispersion ({\it middle}, 
    black to white corresponds to 150 to 1000 km s$^{-1}$), and rotation velocity ({\it bottom}, black to white corresponds 
    to $-700$ to $+700$ km s$^{-1}$).  All three panels show PSF-convolved values; the bottom panel in addition superposes 
    the spectrograph pixels on the rotation velocity field.  Note that the apparent velocity dispersion is all due to the
    velocity range seen by each pixel. Figure adapted from Bender \etal (2005).

\vs

      HST has taken us on a remarkable journey in M{\ts}31.~At the beginning, we knew only that~the nucleus is asymmetric 
in surface brightness (Light, Danielson \& Schwarzschild 1974) and kinematics (Dressler \& Richstone 1988; Kormendy 1988a).  
The BH was found by the latter papers, but asymmetries were ignored.  HST revealed the double nucleus; this was interesting 
as a signature of stellar dynamics near BHs, but it raised concerns about the $M_\bullet$ determination.  
Higher-resolution STIS spectroscopy now takes us far inside P1 and P2 into the blue cluster that surrounds the BH.  The 
double nucleus has become irrelevant in the same way that the bulge and disk are irrelevant to $M_\bullet$ measurement.
HST reveals a flat apparent rotation curve from $r = 0\sd05 \pm 0\sd01$ to $0\sd15 \pm 0\sd01$; all of this radius
range is well inside ground-based PSFs.  The corresponding, intrinsic rotation velocities are 1000 -- 2000 km s$^{-1}$, 
i.{\ts}e., $0.003 c$ to $0.007 c$, still far from relativistic but among the\rlap{\hbox{~~\Blue{\bf c:}\textBlack\rm ~speed of light}}
largest velocities seen in any galaxy.  The above is exactly the kind of journey through several generations of improved observations 
that HST has so importantly made possible.

\eject

\vs
\ni {\big\ARRed 2.3 Our Galaxy is the strongest BH case}\textBlack
\vs

      By far the most remarkable series of improvements in BH observations are the ones in our Galaxy.  It
is decisively our best case for a supermassive BH.   All of this work is ground-based.

      The Galactic center is so close ($D$\ts$\simeq$\ts$8.28 \pm 0.33${\ts}kpc; Genzel,{\ts}Eisenhauer{\ts}\&{\ts}Gillessen\ts2010:{\ts}GEG10)
that individual stars can be resolved and followed through their orbits ({\bf Figure 5}).  The shortest orbit periods observed 
so far are 15.8 yr (star S2) and 11.5 yr (star S102).  Each star will eventually provide an independent measure of $M_\bullet$, 
but robust results are available so far only for star S2.  

      The early history of BH mass measurements in our Galaxy is reviewed in Genzel, Hollenbach~\& Townes (1994) and in KR95.  
This work was based on spectroscopy of gas or stellar subpopulations, like that discussed in Sections 3.1 and 3.2, with the 
advantage of better spatial resolution but the disadvantage of discreteness and systematic effects connected with the 
types of stars observed.  The $M_\bullet$ values in {\bf Table 1} start with the first ones based on dynamical models
(Genzel \& Townes~1987 review earlier observations).  Early $M_\bullet$ measurements in {\bf Table 1} were based on the 
kinematics of individual stars or on integrated, $K$-band spectra of CO absorption bands.  Their best resolution was a few arcsec, 
but even so, they attained values of $r_{\rm infl}/\sigma_* \simeq 8$ to 12.~Thus our Galaxy enters {\bf Table\ts1} 
at $r_{\rm infl}/\sigma_*$ values that are comparable to those of the best HST BH discoveries ({\bf Figure\ts1}).

      At that time, our Galaxy was a solid BH case, but it was not yet one of the best. 
Since then, spectacular advances in instrumentation culminating in adaptive optics (AO) work with effective
resolution $\sim 0\sd0146 \simeq 0.00059$ pc (see below) has transformed our Galaxy into the poster child for 
supermassive BHs.  This progress is reviewed in GEG10.  Our discussion is therefore brief.

\includegraphics{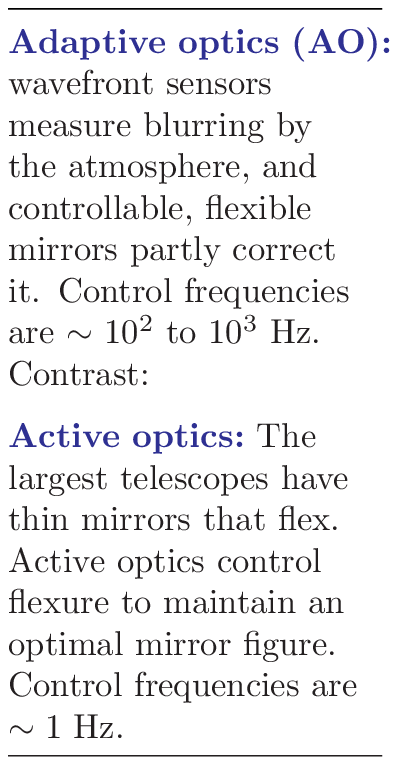}

      Tour-de-force, independent observing programs by Reinhard Genzel and Andrea Ghez and their groups have dramatically
improved the effective resolution.  The first improvements were achieved with speckle observations 
that shrunk effective PSF sizes enough to resolve the nuclear cluster of stars at $r \sim 1^{\prime\prime}$.  Both groups then
participated in the development of AO, and now dozens of stars are seen in the central arcsec of our Galaxy.  
At $D = 8.3$ kpc, $1^{\prime\prime}$ corresponds to 0.04 pc or 8280 AU.  The Galactic BH has a mass of 4 $\times 10^6$ $M_\odot$; 
at 0.04~pc, the circular-orbit rotation velocity is $\sim 700$ km s$^{-1}$, similar to that of the P3 disk in 
M{\ts}31.  But the Galactic center is much closer.~At radii that are accessible to ground-based~AO, 
we enter a new realm of galactic observations.   Normally, we see only snapshots of stellar positions and galactic structures 
whose changes are too slow to be observed in human lifetimes.  But a circular orbit at $r = 0.04$ pc around the Galactic center has an 
orbital time of $\sim 400$ years.  Farther in, it is possible to measure proper motions and watch stars orbit around the center.  
More than one complete orbit of the star S2 has already~been~observed ({\bf Figure 5}).  It gives
$M_\bullet = 4.30 \pm 0.20\ts({\rm stat}) \pm 0.30\ts({\rm sys}) \times 10^6$\ts$M_\odot$~(GEG10), the gold standard of available BH masses.
Note that the estimated error is dominated by uncertainties in the distance. The same is true of many extragalactic BHs, even
though most authors (and our {\bf Tables 2} and {\bf 3}) do not include distance errors in $M_\bullet$ uncertainties.

      The observation that the orbit of star S2 is (within errors) closed means that essentially all of the attracting mass is located inside 
its pericenter radius (Ghez \etal 2008; GEG10; and {\bf Figure 5} give the constraint on any extended mass $M_{\rm extended}$;
see Ghez \etal 2009 for a review).  
S2's pericenter radius is 0\sd0146 $\simeq$ 0.00059$~{\rm pc} = 122$ AU $\simeq$ $1400 r_S$.  This is the effective spatial resolution of the 
BH mass measurement in our Galaxy.  No journey through a sequence of improving observations and astrophysical arguments for 
supermassive BHs is as persuasive as the one in our Galaxy.  The pericenter velocity of S2 is $>$ 6000 km s$^{-1} \simeq 0.02 c$.  
Star S16 is even more extreme.  Its pericenter radius is $45 \pm 16$ AU = $(529 \pm 193) r_S$ and its pericenter velocity is $\sim 12,000$
km s$^{-1}$ = $0.04 c$; it has been observed through pericenter but not yet through a complete $36 \pm 17$ yr orbit (Ghez \etal 2005).  
These observations set the standard for how close to the Schwarzschild radius $r_S \equiv 2 G M_\bullet / c^2$ we have come with 
observations that provide stellar-dynamical BH detections and masses.

\cl{\null} \vskip 2.35truein

\includegraphics{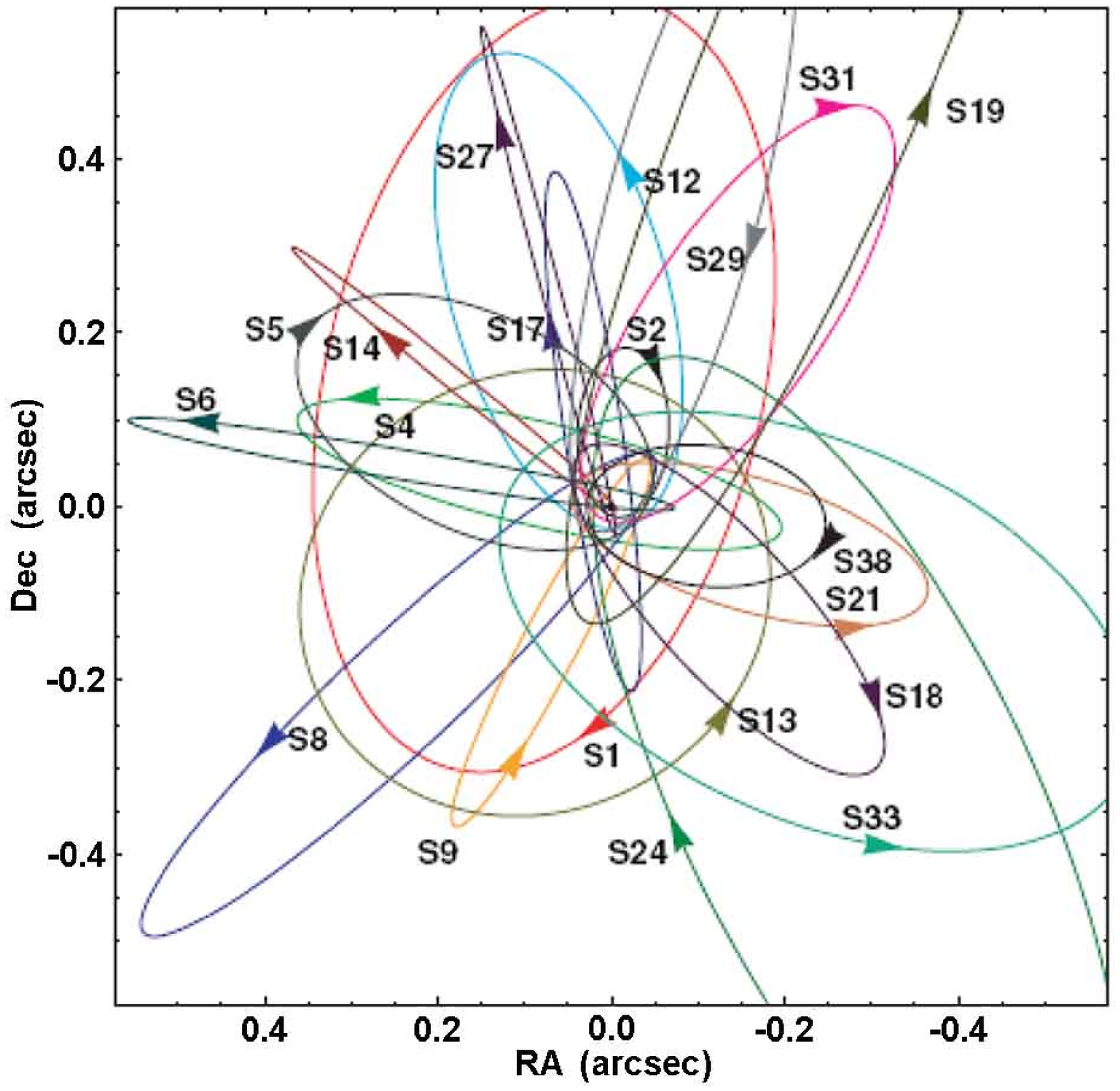}

\includegraphics{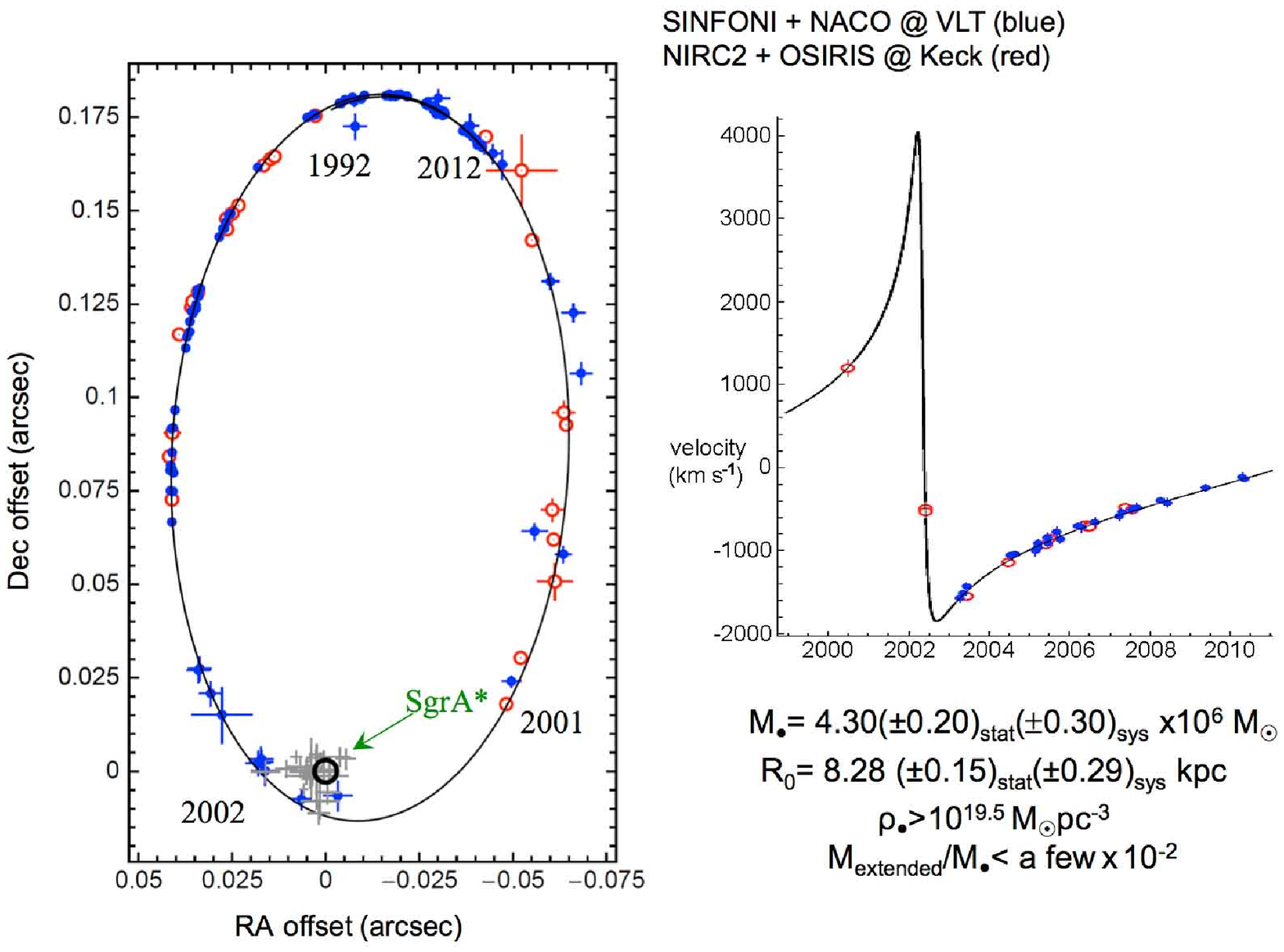}

\ni {\bf \textBlue Figure 5}\textBlack 

\vskip 1pt
\hrule width \hsize
\vskip 2pt

\ni ({\it left\/}) Orbits of individual stars near the Galactic center.  ({\it right\/}) Orbit of star S2 around the BH and {\ARGreen associated
           radio source Sgr A$^*$}\textBlack~based on observations of its position from 1992 to 2012.  Results 
           from the {\textRed Ghez group using the Keck telescope}\textBlack~and 
           from the {\textBlue Genzel group using the Europen Very Large Telescope (VLT)}\textBlack~are combined.  
           This figure is updated from Genzel, Eisenhauer \& Gillessen (2010) and is kindly provided by Reinhard Genzel.

\vs

      These results establish the existence and mass of the central dark object beyond any reasonable doubt.~They also eliminate 
astrophysical plausible alternatives to a BH.~These include brown dwarfs and stellar remnants (e.{\ts}g., 
Maoz 1995, 1998;
Genzel \etal 1997, 2000;
Ghez \etal 1998,~2005)
and even fermion balls
(Ghez \etal 2005;
GEG10).
Boson balls
(Torres \etal 2000;
Schunck \& Mielke 2003;
Liebling~\& Palenzuela 2012)
are harder to exclude; they are highly relativistic, they do not have hard surfaces, and they are consistent with 
dynamical mass and size constraints.~But~a~boson ball is like the proverbial elephant in a tree: it is OK where it is, 
but how did it ever get there?  
GEG10 argue that boson balls are inconsistent with astrophysical constraints based on AGN radiation.  Also, the So\l tan (1982) 
argument implies that at least most of the central dark mass observed in galaxies grew by accretion in AGN phases, and this quickly 
makes highly relativistic objects collapse into BHs.  Finally (Fabian 2013), X-ray AGN observations imply that we see, in some objects, 
material interior to the innermost stable circular orbit of a non-rotating BH; this implies that these BHs are rotating rapidly and 
excludes boson balls as alternatives to all central dark objects.   
Arguments against the most plausible BH alternatives -- failed stars and dead stars~-- are also made for other galaxies in
Maoz (1995, 1998) and in Bender \etal (2005).  Exotica such as sterile neutrinos or dark matter WIMPs could still have detectable
(small) effects, but we conclude that they no longer threaten the conclusion that we are detecting supermassive black holes.

      KR95 was titled ``Inward Bound{\ts}--{\ts}The Search for Supermassive Black Holes in Galactic~Nuclei.''
HST has taken us essentially one order of magnitude inward in radius.~A few other telescopes~take~us closer. 
But mostly, we are still working at $10^4$ to $10^5$ Schwarzschild radii.  In our Galaxy, we have observed individual
stars in to $\sim 500$ Schwarzschild radii.  Only the velocity profiles of relativistically broadened Fe K$\alpha$ lines 
(e.{\ts}g., Tanaka et al.~1995; Fabian 2013) probe radii that are comparable to the Schwarzschild radius.  So we are still 
inward bound.  Joining up our measurements made at thousands of $r_S$ with those probed by Fe K$\alpha$
emission requires that we robustly integrate into our story the rich and complicated details of AGN physics; that is,
the narrow-- and broad--emission-line regions.  That journey still has far to go.

\vs
\ni {\big\ARRed 2.4 HST Discovered Most BHs and Enabled Demographic Studies}\textBlack
\vs
  
      This brings us to what is decisively the most important contribution that HST has made: Because it reliably 
delivers high spatial resolution, HST made it feasible to find BHs in many more galaxies.  By opening the floodgates of 
discovery, it put work on BH demographics into high gear.  As we write this paper, most dynamically detected
BHs ({\bf Tables 2, 3}) were discovered with HST.

      The burst of BH discoveries centered in 2001 $\pm$ 2 is well illustrated in {\bf Figure 1}.  It fueled the
first period of demographic work discussed in the Introduction, when all bulges and ellipticals -- and, indeed,
perhaps even objects as different as globular clusters -- appeared to satisfy a single set of correlations between
$M_\bullet$ and various properties of the host galaxies.

\vsss\vsss\vsss
\hfuzz=100pt
\ni {\hbox{\big\ARRed 2.5 The Post-HST Era of Adaptive Optics and Maser BH Detections}}\textBlack
\vs

      Recently, this subject has emerged from the 15-year plateau of the above demographic picture.  One reason is that 
AO enables new discoveries using ground-based telescopes
(Beckers 1993; Davies \& Kasper 2012).  
AO delivers an important advantage for late-type galaxies: it works best in the infrared, where problems with young stars 
and dust are minimized.  Second, after a spectacular beginning with NGC 4258 (Miyoshi \etal 1995) and a dry period that 
followed, maser BH searches are starting to have substantial success again (e.{\ts}g.,
Kuo \etal 2011).
Both of these trends are visible in {\bf Figure\ts1}. 
Third, $M_\bullet$ estimates via reverberation mapping of AGNs have become convincing.  This adds new
observations that are not included in {\bf Figure 1}.  All three techniques tend to work best for late-type
galaxies, so the range of galaxy types with BH demographic information has grown broader.~Fourth, dynamical modeling 
has made a quantum jump in sophistication~(\S\ts3.1).
Sections 6 and 7 review the resulting observations that are taking us beyond the demographic plateau of the past 
decade.  We now find that $M_\bullet$ correlates (or fails to correlate) differently with different structural components
in galaxies.  This is rich and only partly explored territory.

\vs\vs
\ni {\big\ARRed 3. PROGRESS IN BH DETECTION TECHNOLOGY. II.}
\vsss

\ni {\big\ARRed \quad{\kern 4pt}SPECTROSCOPY \& DYNAMICAL MODELING}\textBlack
\vs

      Early on in this subject, it was commonly believed that BHs could be
detected photometrically.  Based on experience with globular clusters, it was thought
that brightness~profiles of elliptical galaxies would flatten near the center into 
nearly-constant-density cores.   This expectation~may~also have been influenced
by analogy with stars, in which the density, temperature, and other properties are 
continuous, differentiable functions of radius near the center.  If the luminosity density $j(r)$
is non-singular at the center of a spherical galaxy, then the surface brightness must vary 
with radius~$R$ as $I(R)=I_0 + aR^2 +\hbox{O}(R^4)$ (Tremaine 1997).  Adding a BH converts
the volume brightness profile into a cusp with a singular central luminosity density.  
A power-law cusp, $j(r)\propto r^{1-\gamma}$, gives rise to a cusp in the surface density, 
$I(R)\propto R^{-\gamma}$, for $\gamma > 1$.  The theory of such cusps was described in seminal papers by
Peebles (1972),
Bahcall\ts\&{\ts}Wolf (1976), and
Young (1980). 

      We now realize that brightness profiles by themselves provide no evidence for or against~BHs.  
Kormendy~(1993a) and KR95 showed that higher $M_\bullet$ is associated with shallower, not steeper, 
central brightness profiles.   Also, it is straightforward to construct equilibrium stellar systems with 
power-law cusps that have no central point mass (e.{\ts}g., 
Dehnen 1993;
Tremaine \etal 1994).
And $N$-body simulations of hierarchical clustering produce cuspy dark halos in the absence of any baryons 
(Navarro, Frenk \& White 1997). 
Ironically, the cluster of old stars near the center of the Milky Way{\ts}--{\ts}the galaxy with the strongest 
evidence for a BH{\ts}--{\ts}appears to have a constant-density core with a core radius of $\sim$\ts0.4 pc 
(Genzel, Eisenhauer, \& Gillessen 2010).  Some giant elliptical galaxies even contain partly hollow cores 
(Lauer \etal 2002).~There is physics in this (Section\ts6.13).

      Therefore detection of BHs in nearby galaxies requires both photometry to measure the density distribution of the stars 
and spectroscopy to measure their kinematics.  For photometry, HST provides a PSF whose Gaussian core has a dispersion radius 
of $\sigma_{*\rm tel} \simeq 0\sd036$ at visible wavelengths (van der Marel \etal 1997a; see {\bf Table 1\/} here).  For spectroscopy, 
taking pixel sizes and telescope PSF into account, HST provides a best resolution Gaussian dispersion radius $\sigma_* \simeq 0\sd05$
({\bf Table\ts1\/}).  Higher-resolution observations can be provided by AO, but the fraction of the light that is in the PSF core is much 
smaller for AO than it is for HST.  Still higher resolution is available from radio interferomtery of water maser disks and for our Galaxy.

\vs\vsss
\ni {\big\ARRed 3.1 Stellar dynamics}\textBlack
\vs

      Stellar-kinematic observations and dynamical models used to interpret them have improved dramatically during the $\sim$\ts30 years 
of the BH search.  This is important, because most~BH~detections in {\bf Tables\ts2\/} and {\bf 3\/} are based on stellar dynamics.  
Also, the technique covers almost~the~whole~range of observed BH masses.  Therefore, improvements in modeling reliability have been central 
to our understanding of BH demographics and coevolution physics.  
Kormendy \& Richstone~(1995),
Kormendy (2004), and 
Ferrarese \& Ford (2005)
review the history of modeling machinery from the first claimed BH detection (Sargent{\ts}et{\ts}al.{\ts}1978, who used spherical, isotropic models 
on observations of M{\kern0.09ex}87) to mass measurements based only on Equation 1, below, to spherical anisotropic models, and to oblate-spheroidal, 
two-integral models with distribution functions $f(E, L_z)$ based on energy and the axial component of angular momentum.~Meanwhile, spectroscopy
improved~in~two~ways,~(1) spatial resolution improved from $\sim$\ts1$^{\prime\prime}$ ground-based work to $\sim$\ts0\sd05 for HST, and 
(2) the derived data improved from $V$ and $\sigma$ measurements using long-slit spectroscopy to measurement of full line-of-sight
velocity distribtions (LOSVDs) to ground-based, two-dimensional spectroscopy.  The history of these improvements is summarized here in Section\ts2, 
especially in {\bf Table\ts1\/} and {\bf Figure\ts2\/}.  We concentrate here on the state of the art since $\sim$\ts2000, i.{\ts}e., three-integral models 
based on Schwarzschild's (1979, 1993) orbit superposition method. 

      The big difficulty in stellar dynamics is the unknown anisotropy in the velocity~distribution.  It is worth abstracting 
from Kormendy \& Richstone (1995) the arguments about how and when this is important.  A heuristic understanding is provided by the
idealized case of spherical symmetry and a velocity ellipsoid that everywhere points at the center.  Then the first velocity moment of the 
collisionless Boltzmann equation gives the mass $M(r)$ within radius $r$,
$$
M(r) = {{V^2r}\over G} + {{\sigma_r^2r}\over G}~\biggl[- \ts{{d\ln{\nu}}\over{d\ln{r}}} - {{d\ln{\sigma_r^2}}\over{d\ln{r}}} 
                       - \biggl(1 - {\sigma_{\theta}^2 \over \sigma_r^2}\biggr) - \biggl(1 - {\sigma_{\phi}^2 \over \sigma_r^2}\biggr)\biggr]\ts, \eqno{(1)}
$$
where $V$ is the rotation velocity, $\sigma_r$, $\sigma_{\theta}$, and $\sigma_{\phi}$ are the radial and azimuthal components of the velocity
dispersion, and $\nu$ is the density of the tracer population whose kinematics we measure (not the total density).  All quantities except $M$ are 
unprojected.  Thus~$V$~and~$\sigma$ contribute similarly to $M(r)$, but the $\sigma^2_r r / G$ term is multiplied by the sum of four terms.
The first two are almost always positive, but the anisotropy terms can be positive or negative.  Anisotropy is most important when
$V$ is small and when the anisotropy terms threaten to cancel the first two terms in the bracket.  Under what circumstances can this happen?

      Sections 6.7, 6.13, and 8.4 discuss the distinction between elliptical galaxies that do or do not contain cores; i.{\ts}e., breaks at $\sim$\ts$10^2$ pc
in $\nu(r)$ from steep outer profiles to shallow inner power~laws.  This distinction matters here in two ways.  First, core galaxies have shallow
inner profiles with $0.5$\ts\lapprox\ts$-d \ln{\nu}/d \ln{r}$\ts\lapprox\ts$1$ (Kormendy \etal 1996c; Gebhardt \etal 1996).  Coreless galaxies 
typically have $-d \ln{\nu}/d \ln{r} \simeq 1.9 \pm 0.3$.  The $-d\ln{\sigma_r^2} / d\ln{r}$ term cannot easily be $>$\ts1 and is larger in coreless galaxies.
So the first two terms add up to 1\ts--\ts1.5 for core galaxies and almost~3 for coreless galaxies.   Secondly, coreless 
galaxies rotate more rapidly and are less anisotropic than galaxies with cores.  So, for coreless galaxies (including classical bulges), the 
$V^2 r / G$ term is more important and the velocity anisotropy terms are both smaller and in competition with larger values of the first two terms.  
Velocity anisotropy is not negligible, but its effects tend to be small.  In contrast, if a core elliptical has $\sigma_\theta$ and $\sigma_\phi$ 
as small as 60\ts--\ts70\ts\%~of~$\sigma_r$, the anisotropy terms essentially cancel the first two terms.  And $V^2 r / G \sim 0$.  Similarly, if $\sigma_r$ 
is smaller than the other two components, the derived mass can easily double.  So velocity anisotropy is a big issue for core ellipticals.  And it is
not safe to ignore it even for coreless ellipticals and bulges.

      Dealing with anisotropy has controlled the history of stellar-dynamical BH work (e.{\ts}g.,~{\bf Figure\ts2}).
Early BH detections did not involve dynamical models; they used Equation 1 to measure~masses, much as H{\ts}{\sc I} rotation curves $V(r)$ give
mass distributions $M(r) \simeq V^2 r / G$ except that~we~also have to deal with $\sigma$, projection of the model light distribution and kinematics, 
and~PSF~convolution.  These papers
(Tonry 1984, 1987:~M{\ts}32; 
Kormendy 1988a,{\ts}b:~M{\ts}31, NGC\ts4594; 
Kormendy \& Richstone 1992:~NGC\ts3115) 
succeeded because the galaxies were picked to have almost-edge-on bulges in which velocity anisotropy was known to be small 
(Kormendy\ts\&{\ts}Illingworth\ts1982; Jarvis\ts\&{\ts}Freeman\ts1985). 

      The first papers to incorporate anisotropy pointed out that, by suitably tuning \hbox{$\sigma_r$\ts$>$\ts$\sigma_{\rm tangential}$} near the center, 
the Sargent \etal (1978) observations of M{\ts}87 can be fitted without a BH 
        (Duncan \& Wheeler 1980;
         Binney \& Mamon 1982;  
         Richstone \& Tremaine 1985; 
         Dressler \& Richstone 1990).  
In the context of other work which showed that giant (but not lower-luminosity) ellipticals rotate 
         slowly and are triaxial (e.{\ts}g.,
         Illingworth 1977;
         Binney 1978a, b; 
         Davies \etal 1983),
these papers firmly established that velocity anisotropy is the central problem in stellar-dynamical BH searches.  

      The broadest-impact advance in stellar-dynamical BH search technology{\ts}--{\ts}one whose descendants define the state of the art 
today{\ts}--{\ts}was application of Schwarzschild's (1979, 1993) orbit superposition method to model galaxies.  The gravitational potential is 
defined as the sum of a central point mass $M_\bullet$ (to be determined) and stellar densities equal to the light distribution 
times the mass-to-light ratio $M/L$ of the stellar population~(to~be~determined) assumed to be constant with radius.  Then ``all possible orbits'' in
this mass distribution are calculated as functions of energy and angular momentum and integrated long enough to give three-dimensional distributions of 
time-averaged densities, velocities, and velocity dispersion components.  Finally, an optimum linear combination of these orbital distributions is 
calculated to fit the observed light and velocity distributions after projection and PSF convolution.  This method has the major advantage that an 
explicit distribution function $f$ which is a solution to the collisionless~Boltzmann~equation~need~not~be~defined. Instead, all valid distribution 
functions can be calculated numerically, provided that all possible orbits are explored (see below).  Uniqueness is a problem.  But it is easy to minimize 
a ``cost function'' or maximize a ``profit function'' to accomplish specific goals that circumvent the uniqueness problem by finding extreme solutions 
that are of astrophysical interest.  In particular, if the algorithm is required to minimize $M_\bullet$ at all cost to the other variables and if it fails, 
then a central dark mass has robustly been detected.  A common profit function is maximization of an entropy $\simeq$ the integral of $-f\ln{{\kern-0.1ex}f}$ 
over position and velocity space.  Doing this makes the distribution function smoother but more specialized.  Modern data are so detailed that smoothness 
is not an issue, so entropy maximization is given low weight.  The earliest code made spherical models (Richstone \& Tremaine 1984, 1985, 1988).
Applications to the BH search were 
Dressler \& Richstone (1988);  
Richstone, Bower, \& Dressler (1990), and
Kormendy \& Richstone (1992).
This code evolved into one of three extant modeling machines, i.{\ts}e., the Nuker code 
(Gebhardt \etal 2000d, 2003;
 Richstone~et~al.~2004;
 Thomas \etal 2004;
 Siopis \etal 2009;
 Thomas 2010).
The others are the Leiden code
(van der Marel \etal 1998a;
 Cretton       \etal 1999a;
 Cappellari    \etal 2002, 2006; 
 Verolme       \etal 2002; 
 Shapiro       \etal 2006), 
and the Valluri code
(Valluri, Merritt \& Emsellem 2004; 
 Valluri~et~al.~2005). 

      These codes are widely applied in BH searches and in other galaxy studies.  Much science{\ts}--{\ts}including most 
of the results in this paper{\ts}--{\ts}depend on their success.  Nevertheless, they involve nontrivial limitations and concerns that 
are not fully addressed in the literature.  For example: \vs

\nhi1 1. Until Gebhardt \& Thomas (2009), BH modeling papers did not include halo dark matter.  This problem
         is now essentially solved (see below). \vsss\vskip 2pt

\nhi1 2. Core ellipticals are known to be moderately triaxial, but the above codes make axisymmetric models.
         First explorations with triaxial models are discussed below.  This remains an issue. \vsss\vskip 2pt

\nhi1 3. In many Nuker papers (e.{\ts}g., Gebhardt \etal 2003; G\"ultekin \etal 2009b), galaxies are assumed to be edge-on.  
         Limited tests show little systematic effect of varying the inclination $i$, but the main reason is that $i$ couples with $M_\bullet$
         and $M/L$ strongly enough so that constraints on $i$ are weak.  This is a problem especially
         for round ellipticals, since studies of observed axial ratio distributions
        (Sandage, Freeman, \& Stokes 1970; Binney \& de Vaucouleurs 1981; Tremblay \& Merritt 1996) 
         show that intrinsically spherical ellipticals are rare.  If galaxies are more face-on than we think, then we can
         underestimate $M_\bullet$.  An example is NGC 3379, discussed below.  \vsss\vskip 2pt

\nhi1 4. We highlight the need to sample ``all possible orbits'', because this has been a contentious~issue.
         Valluri, Merritt, \& Emsellem (2004) argue the danger 
         that, given indeterminacies caused by unknown triaxiality, velocity anisotropy, inclination, and 
         projection, many parameters (including $M_\bullet$ and $M/L$) are strongly coupled, so different
         distribution functions distributed along tilted valleys in multidimensional $\chi^2$ space
         fit the data almost~equally~well.  They warn us that, if too few orbits are used, any Schwarzschild code 
         will fail to discover the freedom to explore this valley and will give a spuriously narrow valley in marginalized $\chi^2$ 
         around a minimum that may be biased.  And they argue that, once the number of orbits is sufficiently increased, the problem 
         will reveal itself as a flat bottom to the $\chi^2$ valley that favors no particular minimum because the intrinsic, 
         multidimensional $\chi^2$ valley is also flat-bottomed.  \vsss\vskip 2pt

\nhi1 \phantom{4.\ts~}Richstone \etal (2004) test this problem for NGC 821, which has the second-most-poorly- resolved BH sphere of
         influence in Gebhardt \etal (2003): $r_{\rm infl}/\sigma_* \simeq 0.6$.  They argue that the 7,000 orbits used in 
         Gebhardt \etal (2003) were sufficient (althought the $\chi^2$ minimum was rather jagged) by showing that 10,000,
         15,000 and 30,000 orbits give the same BH mass.  The $\chi^2$ valley is moderately broad over a factor of
         slightly more than 2 in $M_\bullet$, but it is not flat-bottomed, and 1-$\sigma$ error bars are well
         defined.  Nevertheless, we now realize that sufficient sampling of ``all possible orbits'' is
         not easy.  Shen \& Gebhardt (2010) found that: ``Our new $M_\bullet$ is about 2 times larger
         than the previous published value [for NGC 4649]; the earlier model did  not adequately sample the orbits
         required to match the large tangential anisotropy in the galaxy center.''  Shen and Gebhardt used 30,000 orbits each for 16,000
         models, compared with $\sim$\ts7000 orbits used by Gebhardt \etal (2003).  In recent work, it seems unlikely that
         orbit undersampling is still a problem, as authors have increased the sampling very substantially.  For example, 
         Gebhardt \& Thomas (2009) use 25,000 orbits each for 25,000 orbit libraries calculated for M{\ts}87; 
         Gebhardt \etal (2011) use 40,000 orbits per library; 
         Rusli \etal (2013) use 24,000 orbits per library, and 
         Cappellari \etal (2006) and Shapiro \etal (2006) use 444,528 orbits per library to fit 15,552 observables in NGC 3379. \vs

      Nevertheless, issues (1) and (2) remain.  Two further developments are crucial to our confidence.  First, we discuss intercomparisons,
e.{\ts}g., of $M_\bullet$ measurements made with the above codes on the same data or made with one of the above codes versus an entirely 
independent BH measurement.  Second, we discuss two further astrophysical improvements, the addition of dark matter to the models and 
an initial exploration of the effects of triaxiality in NGC 3379 and in M{\ts}32.

      Siopis \etal (2009) tested the Nuker code by comparing their stellar dynamical measurement of
$M_\bullet = (3.3 \pm 0.2) \times 10^7$\ts$M_\odot$ in NGC\ts4258 with the known, accurate~mass 
$M_\bullet$\ts=\ts$(3.81 \pm 0.04)$\ts$\times$\ts$10^7$\ts$M_\odot$ 
obtained from the kinematics of a maser disk at $r$\ts=\ts0.1{\ts}--{\ts}0.3{\ts}pc (Herrnstein{\ts}et{\ts}al.{\ts}2005;~\S\ts3.3.1{\ts}here).  
Their HST STIS and ground-based spectroscopy were similar to that used in other BH measures.  The spatial resolution was good,
$r_{\rm infl}/\sigma_* \simeq 5.8$.  The HST-determined, stellar-dynamical mass is lower by (13\ts$\pm$\ts5)\ts\%; i.{\ts}e., by 3$\sigma$.  
But if the above were typical of RMS errors in stellar-dynamical mass measurements, then these would be negligibly small for most purposes. 
Moreover, NGC\ts4258 has complications that do not occur for most galaxies:
(1) The light distribution used in the stellar-dynamical measurement is compromised at $r < 0\sd2$ because AGN continuum radiation 
    could not accurately be subtracted.
(2) Dust in the bulge and especially near the center affects the light distribution.
(3) The ellipticity varies from 0.26 at $r$\ts$\sim$\ts0\sd5 to 0.4{\ts}--{\ts0.45 in much of the bulge to 0.6 in the inclined disk.
    However, the model ellipticity was fixed at 0.35.  Assuming $\epsilon = 0.45$ increased the mass estimate to 
    $M_\bullet = (3.5 \pm 0.4) \times 10^7$\ts$M_\odot$, closer to and formally consistent with the accepted value.
(4) NGC 4258 is weakly barred and strongly oval; the kinematic signatures of both features (see Bosma 1981 for ovals) are clear in the H{\ts}{\sc I} 
    velocity field (van Albada 1980).  Van Albada derives an inclination of 72$^\circ$ based on the assumption that the outer H{\ts}I is in circular 
    motion.  However, if the nested ovals in the light distribution are elongated normally, i.{\ts}e., perpendicular to the bar at large $r$ 
    (Kormendy 1982, 2012), then the galaxy could be more face-on.  For $i = 62^\circ$, ``a value close to the lowest inclination angles that we
    found in the literature'', $M_\bullet = (3.6 \pm 0.4) \times 10^7$\ts$M_\odot$ (Siopis \etal 2009).  
So it is appears that the stellar-dynamical and maser masses~are~not~formally~inconsistent.  
In any case, underestimates of BH masses are less worrying than overestimates, because overestimates could cause us to believe that
a BH has been detected when none is present.

      A test of independent observations and model codes is available for the core elliptical NGC\ts3379.  From HST Faint Object
Spectrograph (FOS) and ground-based, long-slit spectroscopy, Gebhardt \etal (2000d, see G\"ultekin \etal 2009c) got
$M_\bullet$\ts=\ts$1.2^{+0.8}_{-0.6}$\ts$\times$\null$10^8$\ts$M_\odot$ (1-$\sigma$ errors) for \kern-0.3ex$D$\ts=\ts11.7{\ts}Mpc.  
Based on two-dimensional spectroscopy from SAURON at large radii and CFHT OASIS at small radii, Shapiro \etal (2006) got
$M_\bullet$\ts=\ts$1.6^{+3.0}_{-1.1}$\ts$\times$\null$10^8$\ts$M_\odot$ (3-$\sigma$ errors).  Both groups assumed~that the galaxy is edge-on,
although Shapiro includes results at $i = 50^\circ$ in the~error~bars.  NGC\ts3379 is a difficult
galaxy; the Gebhardt BH mass was based largely on the asymmetry~in~the~central~LOSVD.  With HST, Gebhardt
had $r_{\rm infl}/\sigma_*$\ts$\sim$\ts2.2, whereas Shapiro had $r_{\rm infl}/\sigma_*$\ts$\sim$\ts0.7 for the~CFHT~data.
However, the Shapiro data are two-dimensional and of very high $S/N$.  The resulting $M_\bullet$ is substantially more robust than
the HST value.~This emphasizes a problem that is usually unstated: given the difficulty of getting HST time, observers ask for
approximately the minimum $S/N$ that they believe will be adequate.  This is unfortunate.  Still, the above comparison
is reassuring, because the measurements and dynamical models are~entirely~independent.

      A less reassuring test of independent observations and analysis codes is available for the core elliptical NGC 1399, the 
brightest galaxy in the Fornax cluster.~Houghton~et~al.~(2006) measure $M_\bullet = 1.26(0.63-1.78) \times 10^9$ $M_\odot$ based on NAOS-CONICA 
AO spectroscopy with the European Southern Observatory's Very Large Telescope.  The effective resolution is 
$\sigma_* = 0\sd086$ as defined in the notes to {\bf Table 1}.  The models are a variant on the Leiden code.  In contrst, Gebhardt \etal 
(2007) measure $M_\bullet = 0.50(0.43-0.57) \times 10^9$ $M_\odot$ based on HST STIS spectra with $\sigma_* = 0\sd070$.  The number of 
orbits used by the two studies is comparable ($\sim 10^4$).  Their 1-$\sigma$ error bars almost overlap.~Nevertheless, the comparison is 
sobering.~We adopt the mean of the two results in Section\ts5. \vsss\vskip 2pt

      Two additional modeling improvements have proven to be very important: \vfill\eject

      Since 2009, halo dark matter has been included in stellar-dynamical $M_\bullet$ measurements.  Omitting it was dangerous, mostly
for ellipticals with cores.  Because these are more anisotropic than coreless ellipticals, stars visit larger ranges of radii.  Even if 
dark matter densities near the center are small compared to stellar densities (and this is not guaranteed), the stars in the radius range in 
which we have kinematic data can visit large radii where dark matter dominates the potential.  Then the brightness profile gives the wrong 
potential and hence the wrong orbit structure.  Including dark matter almost inevitably increases $M_\bullet$ estimates.  This is because we
assume that stellar population $M/L$ ratios are independent of radius.~If we add dark matter to the models and therefore attribute some of the 
mass density at large radii to that dark matter, then $M/L$ must be decreased there.~Therefore we decrease $M/L$ near the center,~too. 
As a result, $M_\bullet$ must be increased to 
maintain a good fit to the kinematics.  In their first study of M{\ts}87, Gebhardt and Thomas (2009) found that 
$M_\bullet = (6.1 \pm 0.5) \times 10^9$\ts$M_\odot$ for $D = 17$~Mpc, essentially double the mass
$M_\bullet = (3.6 \pm 1.0) \times 10^9$\ts$M_\odot$ 
given by the emission-line rotation curve (Macchetto \etal 1997).
This review was delayed for several years because the same problem existed for other core galaxies, too, but we had no revised BH masses.
Now, this problem is moot.  All core galaxies with stellar-dynamical BH masses except NGC 3607 and NGC 5576 have been (re)analyzed using 
models that include dark matter (see {\bf Table 2\/}).~This includes~new~BH~mass~measurements.  We are grateful to S.~Rusli and R.~Saglia for 
communicating nine new BH masses before publication. Figure 9 in Rusli \etal (2013) shows how the correction factor for $M_\bullet$
derived without using dark matter depends on the spatial resolution of the spectroscopy.  We also have an independent calibration of the
corrections from Schulze \& Gebhardt (2011).  However, as noted in the notes to {\bf Table 2} (see Supplemental Information), the two
calibrations agree poorly at the low effective resolution for NGC 3607, so we cannot reliably correct $M_\bullet$.  We
therefore consider the BH detection to be valid, but we omit the galaxy from correlation fits (see {\bf Figure 12}).  NGC 5576 is the
only other galaxy that required a significant correction to $M_\bullet$.  That correction is robustly determined and has been applied in 
{\bf Table 2}.

      The second improvement is a first exploration of triaxial models.  Developed by van den Bosch \etal (2008), these
have been applied to the core galaxy NGC 3379 and the coreless galaxy M{\ts}32 by van den Bosch \& de Zeeuw (2010).  Triaxial models are
expensive in complexity and in the computing resources needed for an exploration of the (much enlarged) parameter space.  But
the improvement in $M_\bullet$ is important for core galaxies.  For NGC\ts3379, axisymmetric models produced a well-defined 
estimate of $M_\bullet = 1.6^{+0.3}_{-0.9} \times 10^8$ $M_\odot$ (1-$\sigma$ errors) (Shapiro \etal 2006).
Van den Bosch \& de Zeeuw (2010) confirm this result when they force their models to be axisymmetric.  But their triaxial models give
a three-times-higher mass, $M_\bullet = (4.6 \pm 1.1) \times 10^8$ $M_\odot$.
Triaxiality is more important than usual for NGC 3379, because it is a core elliptical that looks nearly round.
We noted above that intrinsically spherical ellipticals are rare.  So NGC{\ts}3379 is likely to be nearly face-on, and 
a wide range of near-face-on inclinations are allowed by triaxial models that are excluded by axisymmetric models.  
We use van den Bosch's BH mass in {\bf Table 2\/}, but we note with concern that other core ellipticals  have not been analyzed with triaxial models.  

      In contrast, for the coreless elliptical M{\ts}32, the triaxial models of van den Bosch\ts\&{\ts}de{\ts}Zeeuw\ts(2010) give a BH mass 
$M_\bullet = (2.45 \pm 1.05) \times 10^6$\ts$M_\odot$ that is consistent with results from axisymmetric models.  It is less likely that
triaxiality is a problem for $M_\bullet$ measurements in coreless galaxies.

      For $\sim$\ts25 years, there has been a disconnect between the good agreement of various authors' $M_\bullet$ measurements
and worst-case scenarios motivated by the worries in our discussion of Equation 1, allowed formally by the collisionless Boltzmann equation,
and exemplified by Valluri~et~al.~(2004).  Results almost always turn out more benign than the worst-case scenarios.  
Galaxies do not use their freedom to indulge in perverse orbit structures.  In practice, velocity distributions are nearly isotropic
near galaxy centers; in core galaxies, $\sigma_\theta > \sigma_r$ inside the core radius, and $\sigma_r > \sigma_\theta$ and $\sigma_\phi$ 
at large radii.  There is physics in this.  Near-isotropy at small radii results from the destruction of box orbits that pass close
enough to BHs to scatter off of them 
(\hbox{Norman, May, \& van Albada 1985;}
Gerhard \& Binney 1985;
Merritt \& Quinlan 1998).  
A bias toward circular orbits can result~(1)~from embedded~nuclear~disks, 
(2) from slow BH accretion (Young~1980; Goodman \& Binney 1984), and 
(3) from core scouring by binary~BHs~(Gebhardt~2004).~Orbits at large radii tend to be radial because stars get flung there from small radii 
    during violent~relaxation.~Thus, 
the physics of galaxy formation limits the distribution functions that galaxies can have.  Despite this,
and despite the reassuring tests discussed in this section, nagging doubts remain about stellar-dynamical BH
masses.~As data and models improve, statistical error bars shrink.  Almost certainly, they are already smaller~than systematic errors
that we do not know how~to~evaluate.  Systematics that cause concern include 
(1) triaxiality in core ellipticals, 
(2) radial mass-to-light ratio variations caused by radial gradients in stellar populations (a very solvable problem), and
(3) whether we have the correct density profiles of dark matter after realistic merger histories and baryonic pulling.  
For these reasons, we do not concentrate on the scatter in BH--host-galaxy correlations.  Rather, we focus on more qualitative 
results about the differences in such correlations for different kinds of galaxy components.

\vs
\noindent{\bf \ARRed 3.1.1 What Spatial Resolution Do We Need To Measure M$_\bullet$ With Stellar~Dynamics?}\textBlack
\vs

      Gebhardt et al.~(2003) answer this question for the three-integral, orbit-superposition models constructed with the
Nuker code.  They derive $M_\bullet$ using HST and ground-based spectroscopy.   They also measure how $M_\bullet$ estimates 
change when they use only the ground-based spectroscopy.  The HST spectra are higher in resolution by a mean factor of 11.2\ts$\pm$\ts1.2.
We assume here that masses based on these spectra are accurate.  Then results based only on the ground-based spectra can be used to map
out how $M_\bullet$ estimates deteriorate at $r_{\rm cusp}/\sigma_*$\ts\lapprox{\kern0.06ex}1.  Gebhardt \etal (2003) conclude that 
$M_\bullet$ error bars are larger when the HST data are omitted but that systematic errors in $M_\bullet$ are small.  
In particular, results from the three-integral models are reliable provided that $r_{\rm cusp}/\sigma_*$ \gapprox \ts0.3.  All 
stellar-dynamical BH detections in {\bf Tables 2\/} and {\bf 3\/} satisfy this criterion.  At lower resolution, the 
error bars grow, but there is no sign that they are unrealistic or that $M_\bullet$ is systematically wrong.

      All of the ground-based BH discoveries listed in KR95 had $r_{\rm cusp}/\sigma_* > 1$ except the earliest papers on M{\ts}32.
{\bf Figure 2\/} shows that they overestimated $M_\bullet$.  But HST BH discoveries that are made with $r_{\rm cusp}/\sigma_* \simeq 0.3$ to 1 
are more secure than these early results on M{\ts}32, because we now fit LOSVDs and because three-integral models 
are more reliable than simpler models.

\vfill

\eject

\vs
\ni {\big\ARRed 3.2 Ionized gas dynamics}\textBlack
\vs

Gas-dynamical mass measurements in principle offer several major
advantages compared to stellar-dynamical modeling.  The central few hundred
parsecs of practically all spiral galaxies and $>$\ts50\% of S0 and
elliptical galaxies have detectable optical nebular line emission (Ho,
Filippenko, \& Sargent 1997a).  The inferred amount of warm ($10^4$ K) ionized
gas is quite modest, typically only $\sim$\ts$10^4$\ts--\ts$10^5{\ts}M_\odot$ (Ho, 
Filippenko, \& Sargent 2003), but it is readily detectable at \hbox{ground-based} and
(Hughes \etal 2005; Shields \etal 2007) HST resolution.  From a practical point
of view, nebular emission lines are much easier to measure than stellar
absorption lines.  They not only have larger equivalent widths, but their
relatively simple line profiles make measurement of velocities and velocity
dispersions straightforward.  By contrast, the stellar-based approach relies on
measurement of higher-order moments of the LOSVDs of weaker spectral
features, which is often tremendously challenging for more massive, lower
surface brightness galaxies.  For example, stellar-dynamical analysis
of brightest cluster galaxies has only recently become feasible (e.{\ts}g.,
McConnell{\ts}et{\ts}al.\ts2011a, 2012), but only with the use of 8\ts--\ts10-meter-class
telescopes equipped with AO; the 2.4-meter HST is simply not up to this
task.

Another appeal of using gas instead of stars lies in the conceptual simplicity 
of the dynamical modeling.  If the gas is in Keplerian rotation in a dynamically 
cold disk, the analysis is vastly less intricate and computationally cheaper than 
the orbit-based
machinery needed to treat the stars.  Moreover, complexities such as orbital 
anisotropy, triaxiality, or the influence of the dark matter halo can be 
neglected.   The basic steps involved in the modeling process were outlined in 
Macchetto \etal (1997) and further refined in van{\ts}der{\ts}Marel \& van{\ts}den{\ts}Bosch 
(1998b), Barth\ts\etal (2001), Maciejewski \& Binney (2001), and Marconi \etal 
(2001, 2003).  Starting with the assumption that the gas is in a thin disk 
that rotates in circular orbits in the principal plane of the galaxy potential, 
the goal is to compute a model velocity field that simultaneously matches
the observed velocities, velocity dispersions, and surface brightness 
distribution of the line emission.  The gravitational potential of the galaxy 
consists of the contribution from stars as measured by the projected 
stellar surface brightness and an assumed stellar $M/L$ plus
an additional dark mass (i.{\ts}e.,~the BH).  To compare the model with the data, 
one must take into consideration the instrumental PSF, the size of the spectroscopic 
aperture, and the intrinsic spatial distribution of the line emission.  An additional 
problem that commonly accompanies emission-line gas is dust that complicates the 
analysis of the light distribution.

Early HST BH searches capitalized on the perceived advantages of
gas measurements.  The~first iconic observation by Harms \etal (1994)
targeted the 100-pc-scale circumnuclear disk in M{\ts}87 detected in H$\alpha$
emission by Ford \etal (1994).  Although the single-aperture FOS provided 
limited spatial sampling, the observations demonstrated that ionized gas in 
two positions at radii 0\sd25 on either side of the center has 
velocities $\pm 500$ \kms\ relative to the systemic velocity of the galaxy.  
Assuming that the gas is in Keplerian rotation, the implied enclosed mass is 
$\sim (2.4\pm0.7) \times 10^9\,M_\odot$.  The corresponding central mass-to-light
ratio, $(M/L)_I = 170 \,(M_\odot/L_\odot)$, implies that most of the mass is 
dark and presumably attributable to the much-sought-after supermassive BH.  
Similar cases quickly followed: 
NGC~4261 (Ferrarese, Ford, \& Jaffe 1996), 
NGC~7052 (van{\ts}der{\ts}Marel \& van{\ts}den{\ts}Bosch 1998b), 
NGC~6251 (Ferrarese \& Ford 1999), and 
IC~1459 (Verdoes{\ts}Kleijn{\ts}et{\ts}al.\ts2000).  
These studies adopted the same strategy, focusing on massive ellipticals 
known to host radio jets, which, as expected, often turn out to be accompanied 
by a nuclear disk of dust and gas aligned perpendicular to the jet axis.  
A supermassive BH was discovered in every case.  The modeling 
analysis used in some of these early studies did not take into account all the 
effects described above.  Moreover, the line widths, while clearly observed to 
be quite substantial, were either ignored or treated in a highly 
simplified manner.  The disks were largely assumed to be dynamically cold.

Macchetto \etal (1997) reobserved the circumnuclear disk of M{\ts}87 with the
long-slit mode of the Faint Object Camera (FOC), using a
0\sd063$\times$13\sd5 slit positioned on the center and in two
parallel, flanking positions.  Apart from the higher spatial resolution of the
FOC, the new data provided much improved spatial coverage to better map the
distribution of the gas.  This is essential to establish the detailed
kinematics of the nuclear disk.  Moreover, these authors demonstrated the
importance of considering the instrumental effects of the telescope PSF and
finite slit size, as well as folding into the analysis the spatial
distribution of the line emission.  As with the FOS observations, the line
widths were broader than the instrumental resolution, but apparently they could
be fit self-consistently with the radial velocities using a thin disk in
Keplerian rotation about a central mass of $(3.2\pm0.9) \times 10^9\,M_\odot$.

However, gas dynamics have worse complications than we naively imagine.
Requirements to apply it successfully are more stringent than we usually believe,
and the probability that galaxies cooperate to meet these requirements turns out in hindsight
to be disappointingly low.~At~a~most basic level, gas must be distributed in radius to properly 
sample the BH sphere of influence.  At the same time, the kinematics
of the gas need to be ordered enough to be interpretable.  Unlike stars, gas is a collisional 
fluid, and it responds to nongravitational perturbations such as turbulence, shocks, radiation 
pressure, and magnetic fields.  Gas is easy to push around.  Every galaxy must be scrutinized 
on a case-by-case basis to check {\it  a posteriori} whether the gas has sufficiently settled 
into an equilibrium configuration whose kinematics are dominated by gravity.  And there is a
problem that is not often considered: dust absorption may make the
gas distribution opaque enough so that we cannot assume that we see through it in projection.

Unfortunately, the majority of galaxies, especially disk systems, {\it do not\/} fulfill these 
requirements.  Beginning with the seminal observation of M{\ts}84\ts={\ts}NGC\ts4374 by Bower \etal (1998), 
the \hbox{long-slit} capability of the Space Telescope Imaging Spectrograph (STIS) has been used to 
map the circumnuclear regions of relatively large numbers of nearby galaxies.
Most efforts have focused on the bulges of early- to mid-type S0 and spiral galaxies (e.{\ts}g., 
Hughes \etal 2005; 
Shields \etal 2007), 
but ellipticals have also been observed 
(Noel-Storr \etal 2003).  
While these efforts have yielded a number of BH discoveries and masses
(Barth \etal 2001;
Sarzi  \etal 2001; 
Verdoes~Kleijn \etal 2002; 
Devereux \etal 2003; 
Marconi \etal 2003;  
Atkinson \etal 2005; 
Coccato \etal 2006; 
De~Francesco, Capetti, \& Marconi 2006, 2008; 
Pastorini \etal 2007; 
Dalla~Bont\`a \etal 2009; 
Walsh, Barth, \& Sarzi 2010), 
the success rate of these programs has been much~lower~than~was~anticipated.  
Fewer than 20\% of randomly selected nearby bulges have regular, symmetric velocity fields amenable to dynamical
analysis.  Preselection by narrow-band imaging or dust-lane morphology boosts the chances of finding promising 
candidates (Ho \etal 2002), but morphological regularity does not guarantee that velocities are well behaved.  
E.{\ts}g., NGC~3379 has a regular-looking~nuclear~gas~disk,
but it shows strong non-circular motions that are difficult to interpret (Shapiro \etal 2006).  
Another challenge is the determination of disk orientations.  The inclination angles of disks 
are often poorly constrained using just their kinematics.  Then we resort to the morphological 
appearance of the gas or dust distribution on resolvable scales.  The assumption 
that the gas is coplanar is unlikely always to hold, especially in cases where 
the disk appears warped.  Patchy emission and dust obscuration complicate 
matters further.  Because $M_\bullet \propto 1/{\rm sin}^2 i$, a significant 
fraction of the final error budget in the mass estimate comes from the 
uncertainty in the disk orientation.

As a consolation prize, the low success rate of gas-kinematic programs 
has yielded a collection of more than 100 $M_\bullet$ upper limits 
(Sarzi \etal 2002; 
Verdoes~Kleijn, van~der~Marel, \& Noel-Storr 2006; 
Beifiori \etal 2009).  
With a typical detection sensitivity of a few$\times 10^6\,M_\odot$, these limits 
provide useful constraints on the BH--host-galaxy scaling relations, especially for 
galaxy types that are sparsely populated by current detections (Beifiori \etal 2012).

Even when emission-line gas clearly rotates about the center, it almost always shows
line widths that are substantially larger than instrumental and thermal broadening.  
These velocity dispersions usually increase toward the center, and they can be comparable
to the rotation velocities, ranging from $\sigma/v$ \lax\ 0.3 to $\sigma/v \gg\ 1$ (e.{\ts}g., 
Verdoes~Kleijn \etal 2000;
Barth \etal 2001;
Dalla~Bont\`a \etal 2009).  
In some galaxies, the observed line widths can be attributed to unresolved rotation (e.{\ts}g.,
M{\ts}87:~Macchetto \etal 1997; 
NGC~3998:~De~Francesco, Capetti \& Marconi 2006), 
but excess dispersion is definitely present in others (e.{\ts}g., 
NGC~3245:~Barth \etal 2001; 
NGC\ts4374:~Walsh, Barth, \& Sarzi 2010).  
The nature of this intrinsic gas dispersion is not understood.  It is the biggest 
systematic uncertainty that plagues gas-based BH mass estimates.  There is no 
universal agreement about whether gas velocity dispersions are dynamically important or
about how to deal with them.  The problem lies in their physical interpretation.  
If the line widths arise from nongravitational processes such as local turbulence 
(e.{\ts}g., van~der~Marel \& van~den~Bosch 1998b), then they can be \phantom{0000}

\vfill


 \includegraphics{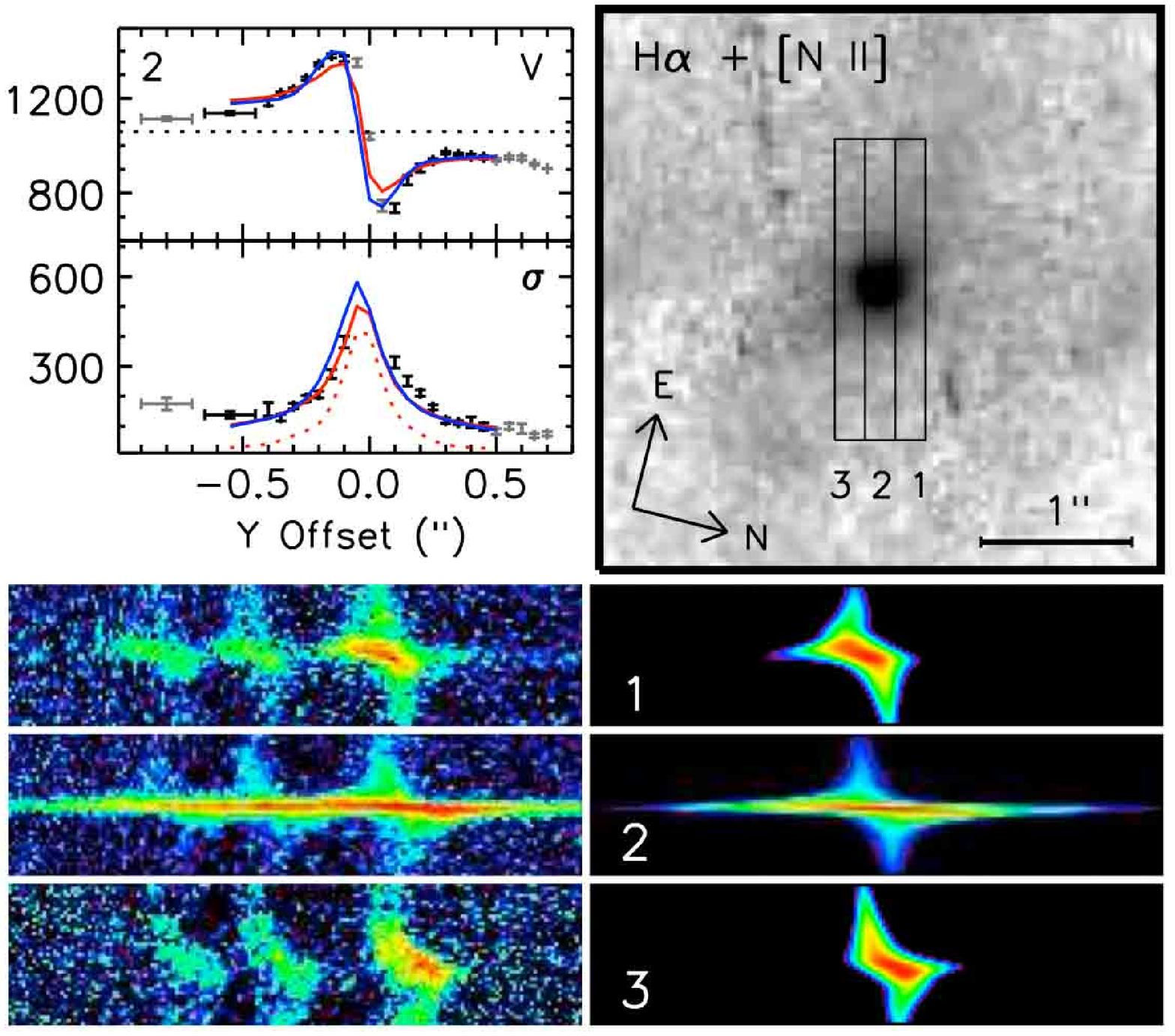}

\ni {\bf \textBlue Figure 6}\textBlack

\vskip 1pt
\hrule width \hsize
\vskip 2pt
\ni STIS observations of NGC{\ts}4374 (Bower \etal 1998) as analyzed by Walsh, Barth, \& Sarzi (2010). The 
top-right panel is a WFPC2 continuum-subtracted H$\alpha$+[N{\ts}II] image showing the 
nuclear~disk of ionized gas.~The footprints of the three slits are overlaid.~The top-left panels
show the radial profiles of [N{\ts}II] $\lambda$6583 mean velocity and velocity dispersion (in km s$^{-1}$) 
along the central~slit~position.  Superposed are predictions of the best BH model with ({\it blue curve}) and 
without ({\it red solid curve}) correction for asymmetric drift.  The red dotted curve shows the contribution 
from rotational line broadening; because these velocity dispersions are smaller than the ones observed, 
the intrinsic velocity dispersion must be significant.  The bottom panels show the continuum-subtracted,
two-dimensional STIS spectra of the H$\alpha$+[N{\ts}II] 
region ({\it left\/}) and the synthetic spectra of [N{\ts}II] $\lambda$6583 ({\it right\/}).  The 
vertical axis is the spatial direction, and the horizontal axis shows wavelength increasing toward the right.  
The three slit positions correspond to the locations labeled in the top grey-scale image.

\eject

\noindent  neglected in the mass models.  Alternatively, the line widths may measure the internal
velocity structure of a dynamically hot distribution of clouds.  Then the dispersion contributes to the hydrostatic 
support of the gas against gravity, and the observed mean rotation velocity is less than the circular velocity.  
Not accounting for this ``asymmetric drift'' then causes us to underestimate the central mass.

Walsh, Barth, \& Sarzi (2010) reanalized the Bower \etal (1998) STIS observations of NGC\ts4374 following the techniques 
of Barth \etal (2001).  Their model fully accounts for the propagation of line profiles through the HST and
STIS optics.  It reproduces the distinctive rise toward the center in the rotation curve and also the large and 
complicated line widths ({\bf Figure 6}).  Note that, among all galaxies whose gas kinematics have been studied with HST,
NGC{\ts}4374 is the only object whose BH sphere of influence is well enough resolved to show Keplerian rotation,
$V(r) \propto r^{-1/2}$.  The best-fit model requires an intrinsic velocity dispersion, which the authors 
model as asymmetric drift, using analysis machinery that is familiar from stellar dynamics (Barth \etal 2001).  
The resulting BH mass, $M_\bullet$\ts=\ts$9.3^{+1.0}_{-0.9} \times 10^8\,M_\odot$ ({\bf Table\ts2}), is significantly smaller
than Bower's original value of $1.6^{+1.2}_{-0.7}\times10^9\,M_\odot$, but it is a factor of two larger than the 
result obtained when asymmetric drift is neglected. 

Strictly speaking, the formalism for the asymmetric drift correction is only 
valid in the limit $\sigma/v \ll 1$ (Binney \& Tremaine 2008).  NGC\ts4374 has
$\sigma/v \approx 0.4-0.6$ and so approximately fulfills this condition.  So does
NGC~3245, whose disk has $\sigma/v < 0.35$ (Barth \etal 2001).  Its applicability, however, 
is more suspect in some giant ellipticals, which have $\sigma/v > 1$.  
Where do we go from here?  Our options are quite limited under these circumstances.  
As a limiting{\ts}--{\ts}if unrealistic{\ts}--{\ts}case, we can ascribe all of the 
observed velocity dispersion to gravity and treat the gas as a spherical, isotropic distribution 
of collisionless cloudlets governed by the Jeans equation.  This approach typically yields 
masses that are factors of $\sim 3-4$ larger than the opposite limiting case of a pure thin disk (e.{\ts}g., 
Verdoes~Kleijn \etal 2000; 
Cappellari \etal 2002).  
The truth lies in between.  A physically better motivated but still tractable alternative is to consider 
a kinematically hot disk with an isotropic pressure.  This is the model employed by 
H\"aring-Neumayer \etal (2006) and Neumayer \etal (2007) for NGC~5128. 

Given the many problems that cloud the reliability of gas-dynamical models, it is crucial that BH masses 
derived using this technique be verified independently.  Stellar-dynamical masses provide a cross-check.  
Eight galaxies (not counting the Milky Way) have been studied using both methods.  We do not 
include NGC~3227 and NGC~4151 because their bright active nuclei make the dynamical modeling especially
challenging for both gas (Hicks \& Malkan 2008) and stars (Davies \etal 2006; Onken \etal 2007).  
{\bf Figure 7} compares the BH mass measurements for the remaining objects.  At first glance, the results 
do not look encouraging.  The scatter is large and most of the gas-based masses are systematically smaller
than the star-based masses.  To be fair, three of the objects (plotted as open symbols) do not provide stringent 
tests because of difficulties involving one or both of the masses estimates 
(IC~1459:~Cappellari \etal 2002; 
NGC~3379:~Shapiro \etal 2006; 
NGC 4335:~Verdoes~Kleijn \etal 2002).  
Removing these three helps to bring the two sets of measurements into better agreement at the expense of 
decimating an already small sample to the point that we cannot conclude much that is statistically meaningful.

Still, three points deserve notice.   The two spiral galaxies (M{\ts}81 and
NGC~4258) lie reasonably close to the 1:1 line, although the relatively large
uncertainty of the mass based on ionized gas kinematics for NGC~4258
(Pastorini \etal 2007) precludes a rigorous comparison with the
stellar-dynamical mass from Siopis \etal (2009).  The excellent agreement for
M{\ts}81 comes as a bit of a surprise, considering the preliminary
nature of the stellar-dynamical mass (Bower et al. 2000) and the highly
simplified analysis applied to the gas data (Devereux \etal 2003).
The gas kinematics of NGC~5128 (Cen~A) have been revisited many times
using both ground-based and HST observations 
(Marconi \etal 2001, 2006;
H\"aring-Neumayer \etal 2006; 
Krajnovi\'c, Sharp \& Thatte 2007).  
The latest efforts by Neumayer \etal (2007) give a gas-dynamical mass that is
a factor of 4 smaller than the stellar-dynamical mass from Silge \etal (2005).
But it agrees well with the mass from Cappellari \etal (2009), which is
based on the same set of AO data used for the gas analysis.  These results
are listed and discussed further in the notes to {\bf Table 2}.

The two most worrisome points are those for M{\ts}87 and NGC~3998.  As illustrated in {\bf Figure\ts12}, 
their gas-based BH masses are almost factors of 2 and 4 smaller than their 
star-based masses, respectively (Gebhardt \etal 2011; Walsh \etal 2012).  The ionized gas disks of
both of these objects display very large line widths, up to FWHM \gapprox \ts1000 km s$^{-1}$
near the centers.  Is asymmetric drift to blame for the systematically smaller gas-based masses?  
Upward $M_\bullet$ corrections by factors of 2\ts--\ts4 are certainly plausible for some galaxies
(Verdoes~Kleijn \etal 2000; 
H\"aring-Neumayer \etal 2006;  
Walsh, Barth \& Sarzi 2010), 
although, according to Macchetto \etal (1997) and De~Francesco, Capetti, \& Marconi (2006), rotational 
broadening alone can reproduce the observed line widths in these two objects.  Gas-dynamical analyses
in general urgently need to be checked. 

In view of all this, it is legitimate to question the accuracy of BH masses that are based on ionized gas kinematics
when broad emission lines were not explicitly taken into account.  {\bf Tables 2\/} and {\bf 3\/} list 16 galaxies 
analyzed using ionized gas techniques.  The most massive ellipticals are 
of special concern.  Their nuclear spectra not only show large line widths, 
but their kinematics often suggest the presence of substantial random motions 
(Noel-Storr \etal 2003; 
Noel-Storr, Baum, \& O'Dea 2007)
presumably agitated by nongravitational forces associated with radio jets (Verdoes~Kleijn, 
van~der~Marel, \& Noel-Storr 2006).  Section 6.3 compares the gas-based BH masses systematically
against the BH mass correlations determined using other techniques.  We conclude there that $M_\bullet$
values determined from ionized gas kinematics that do not include corrections for large line widths
are systematically underestimated.  We regard the BH discoveries in these objects as legitimate, and
we list them in {\bf Tables 2} and {\bf 3}, but we list them in turquoise color.  We omit them
from all correlation diagrams subsequent to {\bf Figure 12}, and we omit them from all correlation fits.

\vfill

\includegraphics{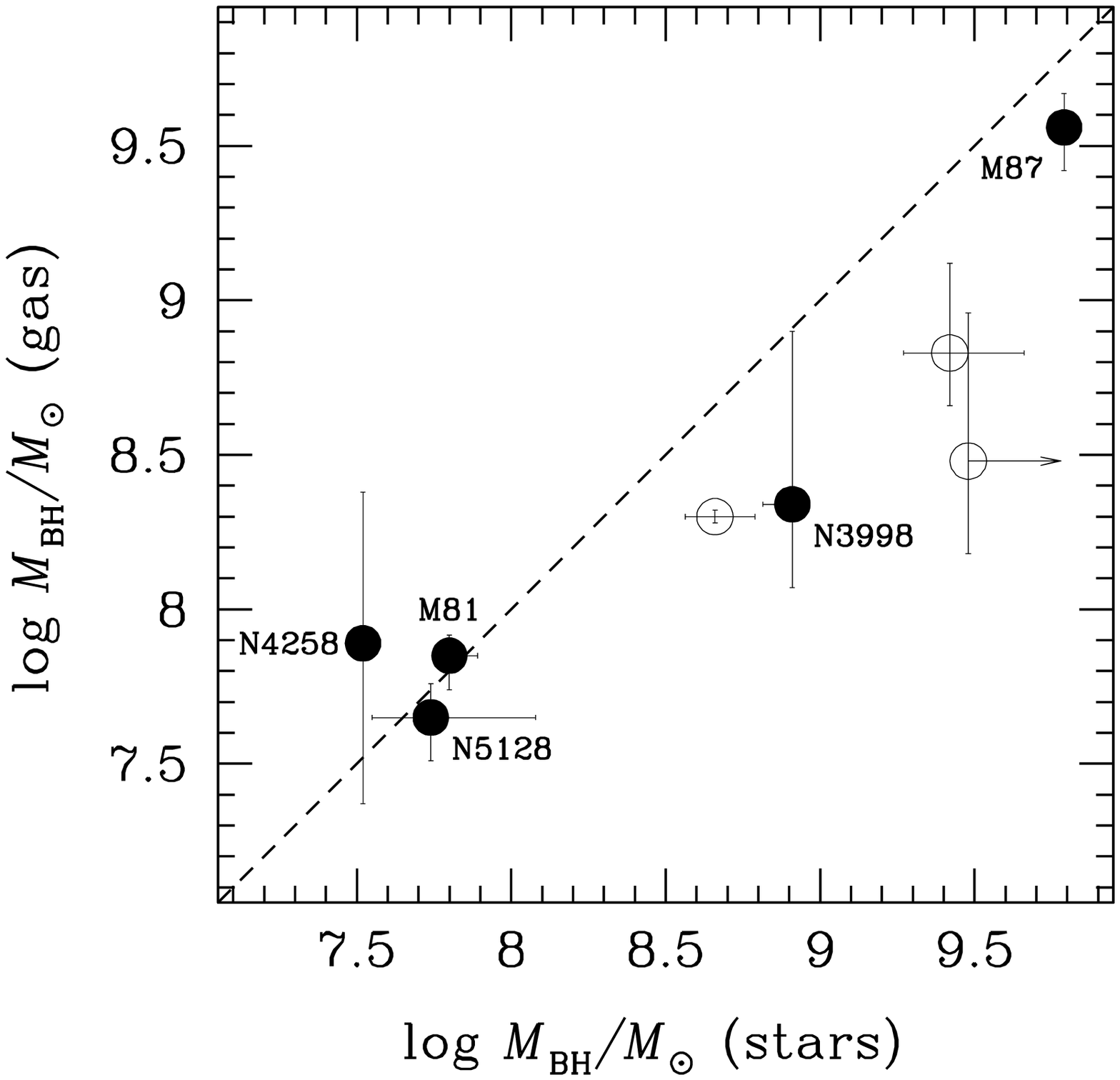}

\ni {\bf \textBlue Figure 7}\textBlack

\vskip 1pt
\hrule width \hsize
\vskip 2pt
\ni Comparison between BH masses derived from gas and stellar dynamics.~Three objects ({\it open~circles\/}) 
do not have sufficiently accurate masses from one or both of the methods.  

\eject

\vs
\ni {\big\ARRed 3.3 Maser dynamics of molecular gas}\textBlack
\vs

\hsize=15.truecm  \hoffset=0.0truecm  \vsize=20.5truecm  \voffset=1.2truecm

      The Miyoshi \etal (1995) study of the Keplerian central rotation curve of NGC\ts4258 established 
radio interferometry of water maser emission from circum-BH molecular~gas~disks as one of the 
most powerful techniques of measuring BH masses.  In favorable cases, it is reliable; it reaches angular 
radii of a few milliarcsec (two dex better than almost all other~techniques), and it is most useful in 
just the galaxies{\ts}--{\ts}gas-rich, optically obscured, star-forming, and often with AGN emission{\ts}--{\ts}that 
cannot be studied well by other techniques.  Excellent angular resolution is 
particularly important because these galaxies tend to have small BHs.  Maser dynamics greatly strengthen 
both the accuracy and the dynamic range of our $M_\bullet$ measurements.

      The early success in NGC 4258 was reviewed in KR95 and updated in Moran (2008).~We review this story as 
the prototypical example of two independent ways of using masers to measure $M_\bullet$,
(\S\ts3.3.1) using the Keplerian rotation curve of the non-systemic-velocity components and (\S\ts3.3.2) using the 
drift and centripetal acceleration of the near-systemic components.  Section 3.3.3 reviews the main complication:
masing molecular disks often have masses that are comparable to $M_\bullet$.

      NGC 4258 is a normal SABbc ({\bf Figure 8}), one of the latest-type galaxies that has a 
classical bulge (Siopis~et~al.~2009).  Kormendy \& Bender (2013b) derive $B/T = 0.12 \pm 0.02$.  
NGC 4258 is a Seyfert~1.9 (Ho, Filippenko, \& Sargent 1997a) with curved radio jets 
(Cecil \etal 2000; cf.~van der Kruit, Oort, \& Mathewson 1972) shown in the middle panel of {\bf Figure 8}.
They are believed to be responsible for the similarly-curved, ``anomalous'' H$\alpha$ spiral~arms (e.{\ts}g., 
Cecil, Wilson, \& De Pree 1995)
and the nearly-coincident X-ray emission (Yang \etal 2007; right panel of {\bf Figure 8}).  
How the jets heat the disk is complicated, because they curve out of the disk plane.  The important point 
here is that, at $r$\ts\lapprox\ts0.1 pc, the jets have PA $\simeq -3^\circ$ and are perpendicular to the masing
gas disk (Cecil \etal 2000).  Its orientation, structure, and rotation curve are illustrated in {\bf Figure 9}.

\vfill


 \includegraphics{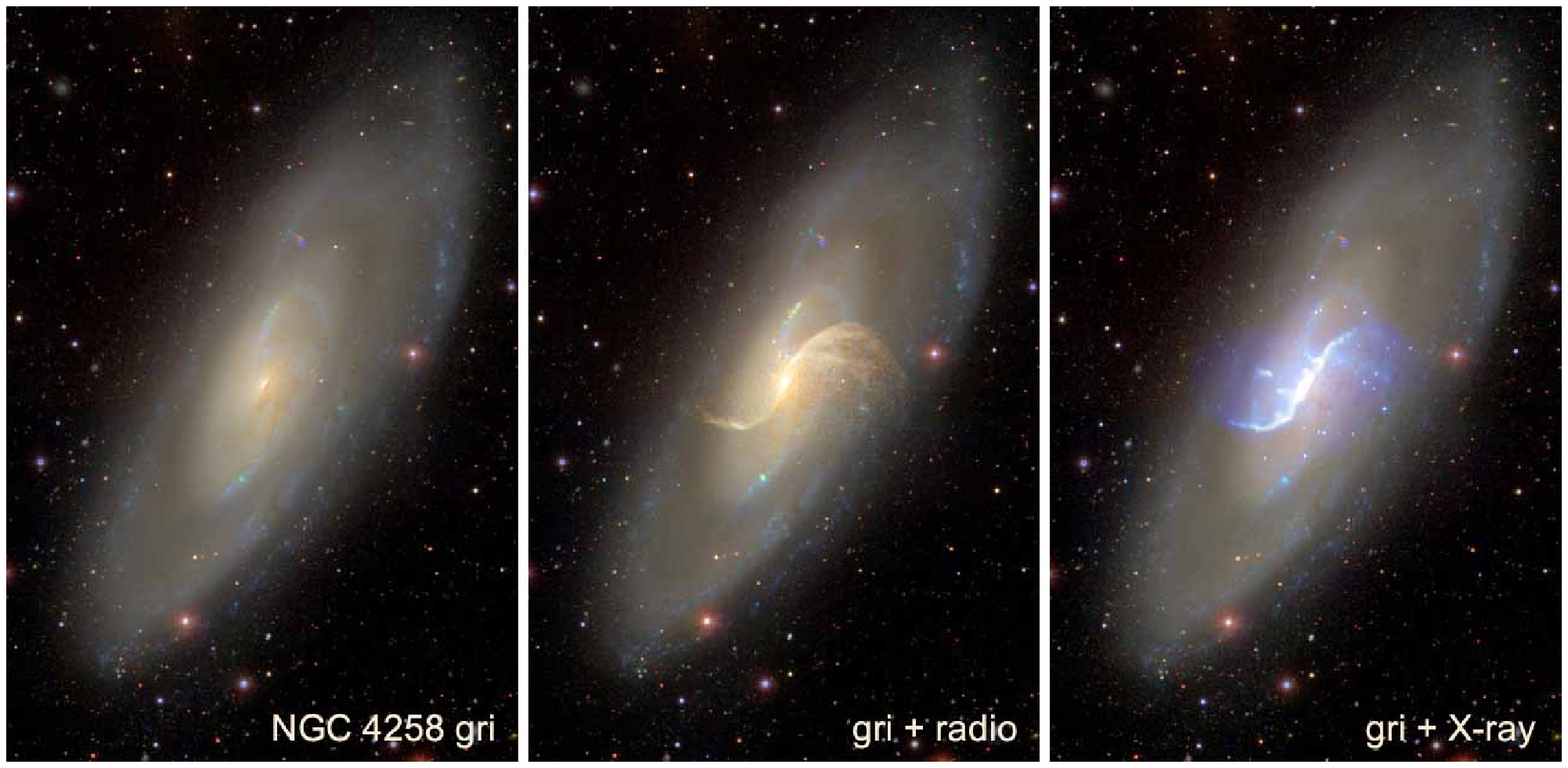}

\ni {\bf \textBlue Figure 8}\textBlack 

\vskip 1pt
\hrule width \hsize
\vskip 2pt

\ni NGC 4258 
({\it left\/}) in an Sloan Digital Sky Survey (SDSS) $gri$ color image from NED, 
({\it center\/}) with the addition in orange of the image of the radio jet (Cecil \etal 2000), and
({\it right\/}) with the addition instead and in blue of the X-ray image from Yang \etal (2007).  
The radio and X-ray images are from {\bf http://chandra.harvard.edu/photo/2007/ngc4258}.
The right side of the galaxy is the near side; the north = top side is receding.  The maser disk
has PA $\simeq 84^\circ$, and the east = left side is receding.  The maser disk
therefore counter-rotates with respect to the main galaxy disk.

\eject

\cl{\null}

\hsize=15.0truecm  \hoffset=-0.3truecm  \vsize=20.1truecm  \voffset=1.5truecm

\vfill


 \includegraphics{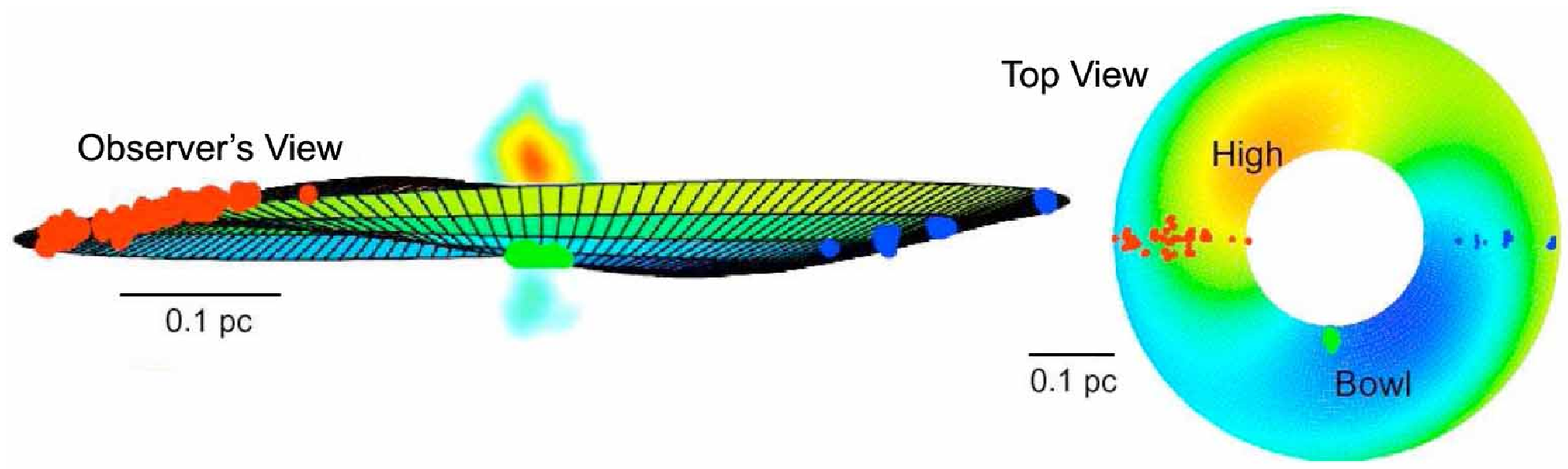}


 \includegraphics{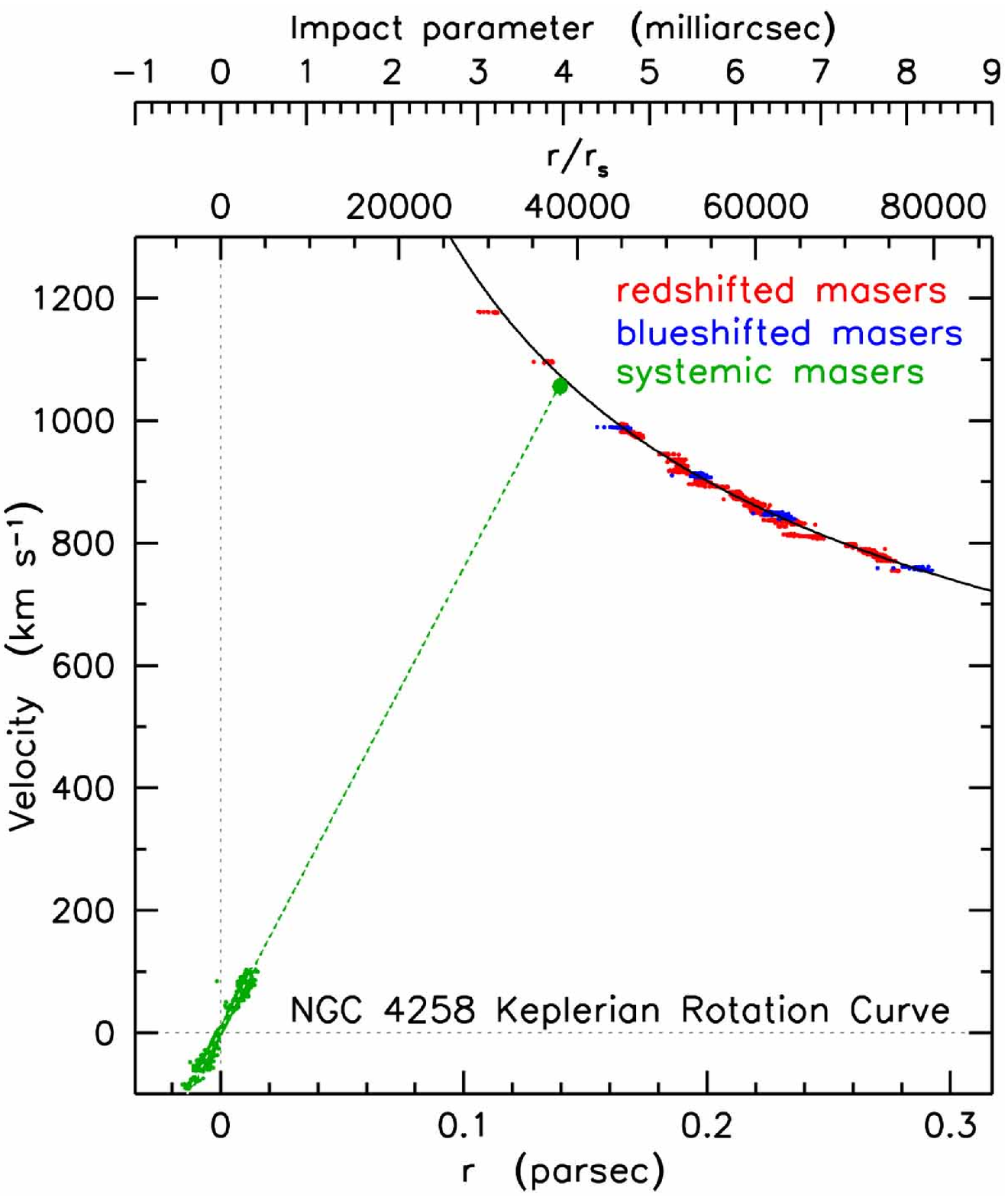}

\ni {\bf \textBlue Figure 9}\textBlack 

\vskip 1pt
\hrule width \hsize
\vskip 2pt

\ni ({\it top\/}) Schematic views of the almost-edge-on, warped maser disk of NGC 4258 (from Moran 2008)
with warp parameters from Herrnstein \etal (2005) and including the inner contours of the radio jet. 
The relative positions of the receding, near-systemic, and approaching masers are indicated by red, 
green, and blue spots, respectively.  Differences in line-of-sight projection corrections to the slightly 
tilted maser velocities account for the departures in the high-$|V|$ masers from exact Keplerian rotation.  
The near-systemic masers are seen tangent to the bottom of the maser disk bowl along the line of sight.  
They drift from right to left in $\sim 12$ years across the green patch where amplification
is sufficient for detection; this patch subtends $\pm 4^\circ$ as seen from the center~(Moran~2008). 

\ni ({\it bottom\/}) NGC 4258 rotation curve $V(r)$ versus radius in units of pc ({\it bottom axis\/}),
Schwarzschild radii ({\it top axis}), and milliarcsec ({\it extra axis\/}).  The black curve
is a Keplerian fit to 4255 velocities of red- and blue-shifted masers ({\it red and~blue~dots}). 
The small green points and line show 10036 velocities of near-systemic masers and a linear fit to them.
The green filled circle is the corresponding mean $V(r)$ point (\S\ts3.3.2).  The maser data are taken 
from Argon \etal (2007).

\eject

\ni {\bf\ARRed 3.3.1. Keplerian Rotation Curve of the Non-Systemic Masers.}\textBlack ~The high-velocity 
masers in NGC 4258 ({\it red and blue points} in {\bf Figure 9}) almost exactly have $V \propto r^{-1/2}$,
so it is likely (but not quite guaranteed:~\S\ts3.3.3) that the disk mass is negligible and that 
$M_\bullet = V^2 r/G$.  Here, $V(r)$ was constructed by assuming that
all masers move exactly toward or away from us.  The illustrated fit is 
$V(r) = 2083/\sqrt{r} + 27$ km s$^{-1}$, where $r$ is in
milliarcsec.  The fit is not quite Keplerian, and the innermost masers fall slightly
below the curve.  This simplest fit gives $M_\bullet = 3.65 \times 10^7$ $M_\odot$.

      Herrnstein \etal (2005) ``{\it find $\sim 2 \sigma$ evidence for deviation from Keplerian rotation in
the NGC 4258 disk, with a preferred rotation law of $r^{-0.48 \pm 0.01}$}'' (their emphasis).  ``This
amounts to a total flattening in the high-velocity rotation curve of about 9.0 km s$^{-1}$ across the 
high-velocity~masers.''  This is the inward flattening in {\bf Figure 9}.  Herrnstein and collaborators 
investigate several possible explanations, including the effects of a central dark cluster with finite radius 
or alternatively a massive accretion disk.  Their preferred explanation is that the true rotation curve is 
exactly Keplerian but that the 9 km s$^{-1}$ total apparent departure from a Keplerian is accounted for by 
projection corrections applied to differently tilted velocities of masers in a thin disk that is differently 
almost edge-on at different radii. 
Herrnstein \etal (2005) fit a warped-disk geometry to the maser positions ({\bf Figure 9, {\it top}})
and find that the disk inclination varies from $81\degd4$ at $r = 3.9$ milliarcsec to $\sim 91^\circ$ at 
$r = 9.1$ milliarcsec.  Then the true rotation curve is Keplerian and the central mass is 
$M_\bullet = (3.81 \pm 0.01) \times 10^7$ $M_\odot$ for our adopted distance of 7.27 Mpc ({\bf Table~3}). 
However, the error bars depend on the assumptions that the adopted geometry is correct and that the angles
between the galactocentric radius vectors of the masers and our lines of sight are all $90^\circ$.  In view 
of the uncertainties in the assumptions, we adopt the above BH mass but use as the error bar the RMS of the 
mass determinations from the pure Keplerian fit, the central cluster model, the massive disk model, and the 
preferred warp model.  The result is $M_\bullet = (3.81 \pm 0.04) \times 10^7$ $M_\odot$.

\vfill

\ni {\bf\ARRed 3.3.2. Rotation Curve of the Near-Systemic Masers.}\textBlack ~We can also derive 
$M_\bullet$ independently of the non-systemic-velocity components using only the near-systemic masers.  To do
this, we need measurements of the centripetal acceleration $V(r)^2/r$ and the linear velocity gradient along 
the line of central masers.  The latter is measured in km s$^{-1}$ milliarcsec$^{-1}$
but is easily converted to km s$^{-1}$ pc$^{-1}$ using the assumed distance.  It is then $V/r$.  This can
be done for every individual maser source that provides both an acceleration and a drift measurement, so it 
could be used to measure $V(r)$.  Here, we derive $M_\bullet$ from the ensemble of 10036 systemic masers 
plotted in {\bf Figure\ts9}.  We derive their mean $V$ = (acceleration)/(drift) and hence $r$ and the single 
$V(r)$ point that is plotted as the filled green circle in the figure.  Given that the masers are projected 
nearly in the direction to the center, the only assumption is that they are in circular motion around the center.  

      This method was used in early papers written before the high-$|V|$ masers in NGC 4258 were discovered 
(Haschick, Baan, \& Peng 1994; Watson \& Wallin 1994).  However, it has been overpowered by the appeal of the Keplerian 
rotation curve of the high-$|V|$ components, and it does not seem to have been used since their discovery.  
For NGC 4258, this method provides important verification that the maser orbits are circular.  In other galaxies
in which high-velocity masers are not discovered, it nevertheless can provide $M_\bullet$ (caveat \S\ts3.3.3),
especially if $V(r)$ can be measured. Mass measurements using near-systemic maser components, although less accurate than 
ones based on well-defined high-$|V|$ rotation curves, are nevertheless often more accurate than ones based on 
(say) stellar dynamics.  These measurements deserve a place in our BH sample.

      For the systemic masers, we measure a mean drift of $V = (266.5 \pm 0.3)$ km s$^{-1}$ 
milliarcsec$^{-1}$ ({\it green dashed line} in {\bf Figure 9}).  We adopt the mean of 12 measurements of the
centripetal accelerations listed in Table 4 of Humphreys \etal (2008), $V^2/r = (8.17 \pm 0.12)$ km s$^{-1}$ yr$^{-1}$.
Including a 1\ts\% correction (to $r$) for the $8^\circ$ tilt of the maser disk, $M_\bullet = (3.65 \pm 0.09) \times 10^7$ $M_\odot$.

      We adopt the weighted mean of the two independent results, $M_\bullet = (3.78 \pm 0.04) \times 10^7$ $M_\odot$.

\eject

\cl{\null} \vskip 1.85truein


 \includegraphics{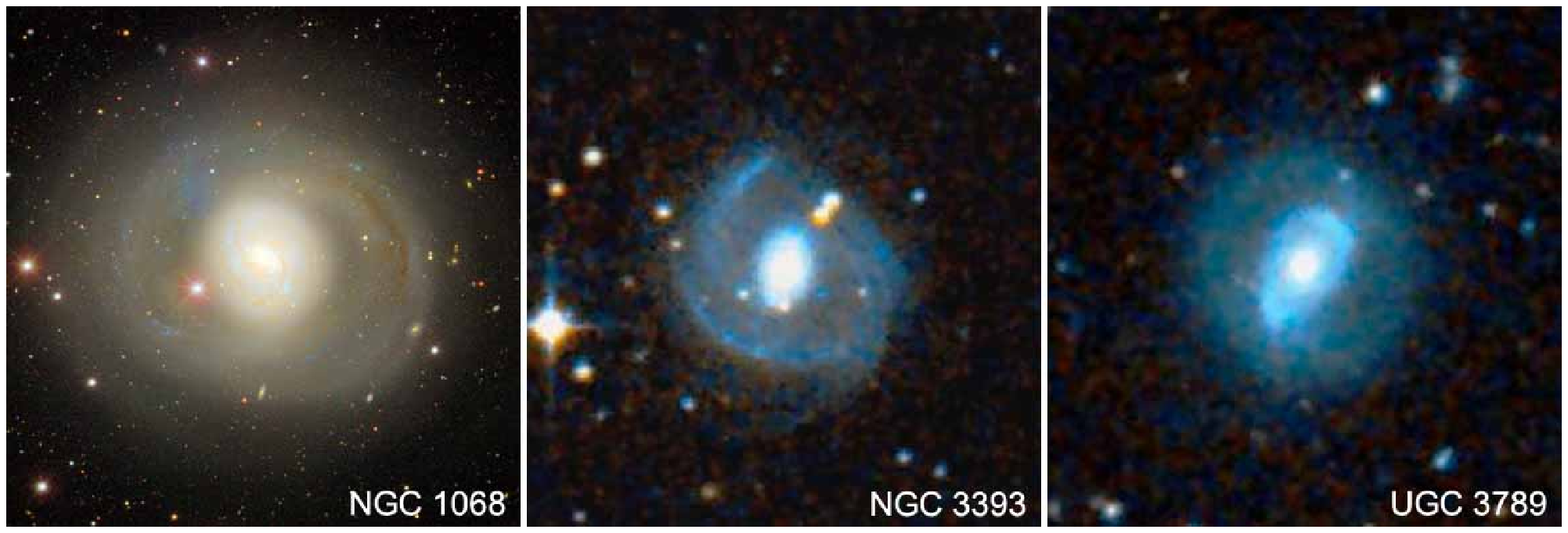}

\ni {\bf \textBlue Figure 10}\textBlack 

\vskip 1pt
\hrule width \hsize
\vskip 2pt

\ni Three almost-face-on galaxies that contain edge-on maser disks.~All three show the two nested~ovals of 
different surface brightness, axial ratio, and position angle that identify oval disks (Kormendy 1982; Kormendy \& 
Kennicutt 2004).  Their axial ratios (face-on $b/a \sim 0.8$) and PAs
(perpendicular to each other) imply that all three galaxies are not far from face-on.  All three 
contain pseudobulges; $PB/T \simeq 0.30$, $0.27 \pm 0.06$, and $0.32 \pm 0.03$ in NGC 1068, NGC 3393, and 
UGC 3789, respectively.  NGC 1068 and NGC 3393 are well-known Seyfert~2 
galaxies, and the high-excitation narrow-line spectrum of UGC\ts3789 implies that 
it also contains a Seyfert\ts2 nucleus.~NGC\ts1068 is from the SDSS.  The others are
from the Digital Sky Survey via {\bf http://www.wikisky.org}; these have much bluer color
balances than the SDSS.  NGC 1068 is an Sb; the other galaxies are Sab in Hubble type.
\vs

      Within errors, the rotation curve of NGC 4258 from the high-velocity masers agrees with the single point, 
$V(0.140 \pm 0.002~{\rm pc}) = (1055 \pm 15)$ km s$^{-1}$ ({\it filled green circle} in {\bf Figure~9}),
derived from the systemic masers.  This confirms that the masers are in circular orbits around the BH. 
The check is relevant, because eccentric disks are easily possible at $r \ll r_{\rm infl}$ (Tremaine 1995).

      Published data allow a similar comparison of a Keplerian rotation curve from high-$|V|$ masers
with a single $V(r)$ measurement from near-systemic masers for only one other galaxy.  This is the face-on oval
galaxy UGC 3789 shown in {\bf Figure 10}.  Reid \etal (2009), Braatz \etal (2010), and Kuo \etal (2011) show that
the galaxy contains an edge-on masing disk with a rotation curve that is essentially as clean and Keplerian
as that of NGC 4258.  From these data, Braatz \etal (2010) derive a distance of $49.9 \pm 7.0$ Mpc.  Then
$V(r) = 440.4 \pm 3.0$ km s$^{-1}$ mas$^{1/2}$ and the gravitating mass is $M = (1.09 \pm 0.15) \times 10^7$ 
$M_\odot$ (Braatz \etal 2010).  Kuo \etal (2011) reanalyze the same data with a different dynamical center
and obtain $(1.12 \pm 0.05) \times 10^7$ $M_\odot$.

      Braatz \etal (2010) show that the near-systemic masers are divided into two groups with different 
velocities and accelerations.  They therefore lie in rings at different galactocentric radii.~For ``ring\ts1'', 
they find an acceleration of $1.50 \pm 0.13$ km s$^{-1}$ yr$^{-1}$ and a velocity drift of 
$695 \pm 65$ km s$^{-1}$ mas$^{-1}$.  Ring 2 shows acceleration of $4.96 \pm 0.34$ km s$^{-1}$ yr$^{-1}$ 
and velocity drift of $1654 \pm 440$~km~s$^{-1}$~mas$^{-1}$.  From these, we derive two $V(r)$ and $M$ values
that are independent of each other~and~of~the results from the high-$|V|$ masers.~Ring 1 implies that 
$V(0.179 \pm 0.028\ts{\rm pc}) = 513 \pm 59$ km s$^{-1}$, compared with $513 \pm 3$
km s$^{-1}$ expected from the \hbox{high-$|V|$} masers. 
Ring 2 implies that $V(0.11 \pm 0.04~{\rm pc}) = 712 \pm 196$ km s$^{-1}$, compared with $671 \pm 5$
km s$^{-1}$ from the \hbox{high-$|V|$} masers.  
The ring 1 value confirms that these masers are accurately in circular
motion around the center.  Rings 1 and 2 imply that the gravitating mass is $M = (1.09 \pm 0.26) \times 10^7$
$M_\odot$ and $M = (1.23 \pm 0.67) \times 10^7$ $M_\odot$, respectively.  Ring 1 accurately confirms the mass 
implied by the high-$|V|$ masers.  For NGC 4258, which also has a Keplerian rotation curve, we identified this
gravitating mass as $M_\bullet$.~This turns out~to~be~valid (Hur\'e \etal 2011).  Here, we consider a caveat---finite 
maser disk mass---before we decide on $M_\bullet$.

\vs
\ni {\bf\ARRed 3.3.3. Molecular Disks With Masses M$_{\bf disk}$ $\sim$ M$_\bullet$}\textBlack ~prove to be the major
complication~of maser BH measurements.  For example, NGC 1068 \hbox{({\bf Figure 10}) has a sub-Keplerian rotation curve},
$V(r) \propto r^{-0.31 \pm 0.02}$ (Greenhill \etal 1996).~Possible explanations include massive molecular disks 
(Hur\'e 2002;
Lodato \& Bertin 2003;
Hur\'e \etal 2011),
nuclear star clusters with radii comparable to those of the gas disks (Kumar~1999), and disk warps.  In many cases, 
warps are disfavored, because the masers 
are \hbox{observed with VLBI to lie along a thin straight line.}  Nuclear clusters are more difficult to exclude. 
Masers are almost necessarily associated with high obscuration.  Many of these galaxies also contain AGNs.
So we do not generally have photometric constraints on nuclear star clusters.  We only have indirect hints
from observations of nearby, non-masing galaxies.  But in large, late-type galaxies, nuclei with masses 
of $10^6$ to $10^{7.5}$ $M_\odot$ and effective radii of 0.3 to 10 pc similar to the radii of maser disks are common 
(Walcher \etal 2005; Kormendy \etal 2010).  Still, it is astrophysically plausible that maser disks have 
significant mass, and most attention has been give to this alternative.  The differences in mass contributions by 
disks and by spherical clusters are several tens of percents.  It would be prudent to keep in mind that quoted 
$M_\bullet$ values may be overestimated and that quoted errors based on analyses of massive 
disks may be underestimated.  With this caveat, we review $M_\bullet$ measurements based on the approximation that
departures from Keplerian rotation are due to massive molecular disks.  

      Hur\'e \etal (2011) emphasize that the rotation curve $V^2 = G M_\bullet / r + r d \Phi_{\rm disk}/d r$ of a BH
plus accretion disk is in general not a power law in radius.  Also, the disk potential $\Phi_{\rm disk}$ is complicated 
(Binney \& Tremaine 1987, 2008).  
For practical reasons, different authors make different simplifying 
assumptions.  Lodato \& Bertin (2003) assume that the disk is self-regulated at the threshold of axisymmetric Jeans
instability in the presence of rotation, $Q \equiv c_s \kappa / \pi G \Sigma = 1$, where $c_s$ is the sound speed, $\kappa$
is the epicyclic frequency, and $\Sigma$ is the disk surface density.  This involves poorly known physics.
In contrast, Hur\'e \etal (2011) develop a formalism to deal with massive accretion disks that surround
point masses on the basis of the mathematically tractable assumption that the disk surface density is $\Sigma 
\propto r^{-s}$, where 0 \lapprox \ts$s$ \lapprox 3 is astrophysically plausible (see their discussion) and
adequate to fit real data.   NGC 1068 provides an example.  For distance $D = 15.9$ Mpc, Lodato \& Bertin (2003) 
find that $M_{\rm disk} = (9.5 \pm 0.6) \times 10^6$ $M_\odot$ and that $M_\bullet = (8.9 \pm 0.3) \times 10^6$ $M_\odot$. 
Similarly, Hur\'e \etal (2011) get $M_{\rm disk} = 12.1 \times 10^6$ $M_\odot$ and 
$M_\bullet = (8.0 \pm 0.3) \times 10^6$ $M_\odot$.  Hur\'e (2002) also finds similar results with a more specialized
disk model.  We adopt the mean of the Lodato and Hur\'e estimates, $M_\bullet = (8.39 \pm 0.44) \times 10^6$ $M_\odot$.

      NGC 1068 and similar galaxies below raise a concern about $M_\bullet$ as measured with lower (e.{\ts}g., HST)
spatial resolution in late-type galaxies that could contain substantial nuclear gas.  Even if the measurement
of the contained mass is correct, any BH could have a significantly smaller mass.  For example, we reject the
$M_\bullet$ measurement in the Sc galaxy NGC 3079 because the maser rotation curve is flat (Kondratko, Greenhill \& Moran 2005).
Better constrained maser $M_\bullet$ measurements add important dynamic range to our demographic results, but we 
retain them with due caution.

      Returning to UGC 3789: Despite the convincingly Keplerian rotation curve (Reid \etal 2009; Kuo \etal 
2011), Hur\'e \etal (2011) find a clean solution in which $M_{\rm disk}/M_\bullet \simeq 0.77$.  They
therefore suggest that $M_\bullet = 0.81 \times 10^7$ $M_\odot$.  The difference between this value and the total
gravitating mass $M$ derived in \S\ts3.3.2 affects no conclusions in this paper.  We will be conservative and adopt the
mean, using the difference as an error estimate.  Thus we adopt $M_\bullet = (9.65 \pm 1.55) \times 10^6$ $M_\odot$.

      Two objects analyzed by Hur\'e \etal (2011) have $M_{\rm disk} > M_\bullet$.~IC\ts1481 has $M_{\rm disk} = 3.55
\times 10^7$\ts$M_\odot$ and $M_\bullet = 1.30 \times 10^7$ $M_\odot$.  It proves to be a merger in progress, so the
galaxy is listed in {\bf Table\ts2} and discussed in the Table notes for elliptical galaxies.  NGC 3393 ({\it middle panel}
of {\bf Figure 10}) has  $M_{\rm disk} \sim 6 M_\bullet$ according to Hur\'e \etal (2011).
Particulars are discussed in the {\bf Table 3\/} notes for galaxies with pseudobulges.

      The galaxies illustrated in {\bf Figure 10} are remarkably similar and serve to emphasize an
important puzzle about BHs.  Each BH is embedded in an edge-on accretion disk that is nearly perpendicular
to the disk of the host galaxy.  Each host galaxy is oval and contains a pseudobulge that is growing
secularly.  It is easy to identify a global mechanism that feeds gas toward the center.  
So why are the angular momentum vectors of this gas and the gas in the accretion disk almost 
perpendicular?  Maser disks (Greenhill \etal 2009) and, more generally, AGN jets (Kinney \etal 2000)
are oriented randomly with respect to their host galaxies.  In NGC 4258, the maser disk even counter-rotates with
respect to the host galaxy ({\bf Figure 8}).  

      Note that, if the mass of the accretion disk is comparable to or larger than the mass of the BH, the answer is 
not that the spinning BH tips the accretion disk until it is in the BH's equatorial plane.

\vs
\ni {\bf\ARRed 3.3.4. Other Maser BH Cases.}\textBlack ~Greenhill (2007) reviews the early history of maser 
detections in galaxies.  Finding maser disks is difficult, not least because \hbox{they must be edge-on,} and the orientation 
of the host galaxy gives no clue about when this is the case.  After the spectacular start with NGC 4258, further
progress was slow.  Other early detections not reviewed above include Circinus and NGC 4945.  They are
discussed in the notes to {\bf Table 3\/}.

      Now, progress on this subject is accelerating, in part because of the advent of the highly sensitive 
Robert C.~Byrd Green Bank Telescope of the National Radio Astronomy Observatory (Greene \etal 2010).  Kuo \etal
(2011) report maser disk detections and BH mass measurements in seven galaxies:
NGC 1194, an edge-on S0 with a prominent classical bulge;
NGC 2273, an unusual (RR)SB(rs)a galaxy with a small pseudobulge;
UGC 3789 (\S\ts3.3.3 and {\bf Figure 10});
NGC 2960, a merger in progress, hence listed in {\bf Table 2\/} with ellipticals;
NGC 4388 (edge-on SBbc);
NGC 6264 (SBb), and
NGC 6323 (SBab).
The host galaxies are discussed in Greene \etal (2010) and Kormendy \& Bender (2013b); parameters adopted here are from
the latter paper.  When necessary, summaries are given in the notes to {\bf Table 3\/}.  In all cases, the masers are 
distributed along lines indicative of edge-on disks with little or no warp.  The radius range is large enough
in all cases to reveal outward-decreasing rotation curves, and mostly, the evidence for a Keplerian disk is strong.  
The most uncertain case is NGC 4388.  These galaxies are very important to BH demographic results~at low
$M_\bullet$ and especially to our understanding of BHs in pseudobulges (Greene \etal 2010; \S\ts6.8 here).

      Several more galaxies arguably have maser-based BH discoveries but $M_\bullet$ measurements that are
not accurate enough for our sample.  Promising objects are discussed by
Wilson, Braatz \& Henkel (1995),
Kondratko, Greenhill \& Moran (2006), and
Greenhill \etal (2009).
It is reasonable to expect rapid progress in this subject. 

\vs
\ni {\big\ARRed 3.4 BH mass measurements for active galactic nuclei (AGNs)}\textBlack
\vs

      Reverberation mapping and single-epoch spectroscopy of AGN broad emission lines now provide large numbers of BH masses 
for objects in which we cannot resolve $r_{\rm infl}$.   They greatly extend~the range in Hubble~types, distances, and AGN 
luminosities of galaxies with BH demographic data.  However, these methods require a dedicated review. Also, we use them only 
in a limited way in Section 7.  Therefore we include only a short summary in the Supplementary Information.

\hsize=15.0truecm  \hoffset=0.0truecm  \vsize=20.1truecm  \voffset=1.5truecm

\vfill\eject

\vsss
\ni {\big\ARRed 4. CLASSICAL BULGES VERSUS PSEUDOBULGES}\textBlack
\vsss

      The morphological similarity between elliptical galaxies and giant bulges in disk galaxies (e.{\ts}g., the Sombrero galaxy: 
{\bf http://heritage.stsci.edu/2003/28/big.html}) is a well-known part of the Hubble classification scheme 
(Hubble 1930; 
Sandage 1961; 
de Vaucouleurs \etal 1991;
Buta, Corwin, \& Odewahn 2007).~That similarity is quantitative: it includes fundamental plane and other parameter
correlations (e.{\ts}g., 
Bender, Burstein, \& Faber 1992;
Kormendy \& Bender 2012).  
One of the biggest success stories in galaxy formation is the demonstration that these bulges and elliptical
galaxies formed in the same way, when the dynamical violence of major galaxy mergers scrambles disks into
ellipsoids. Extensive evidence is based on theory 
(Toomre\ts\&{\ts}Toomre\ts1972; 
Toomre\ts1977), 
observations (e.{\ts}g.,
Schweizer 1990), 
and numerical modeling (e.{\ts}g., 
Barnes 1989, 1992).
``By the time of the reviews of Schweizer (1990, 1998), Barnes \& Hernquist (1992), Kennicutt (1998b), and Barnes (1998),
the merger revolution in our understanding of elliptical galaxies was a `done~deal'\ts'' (Kormendy 2012).  Since then, our
picture of the formation of ellipticals has grown still richer, both observationally and through detailed modeling
(see KFCB and Kormendy  2012 for reviews).  When ellipticals accrete cold gas and grow new disks around themselves 
(Steinmetz \& Navarro~2002), we stop calling them ``ellipticals'' and instead call them ``bulges'' (Renzini 1999).
Thus merger-built bulges are well understood within our picture of galaxy formation.  And they can be identified purely 
observationally.  We refer to them as ``classical bulges.''

      Beginning in the 1980s (e.{\ts}g., Kormendy 1982), it became apparent that elliptical-galaxy-like classical
bulges are not the only kinds of dense stellar systems found at the centers of disk galaxies.  Another kind of ``bulge''
is distinguished observationally as having diskier properties than do classical bulges.  Concurrent work on 
the slow (``secular'') internal evolution of isolated disks showed that outward angular momentum transport by bars and other 
nonaxisymmetries dumps large amounts of gas into galaxy centers.  The Schmidt (1959)\ts--{\ts}Kennicutt (1989,\ts1998a)~observation 
that star formation rates increase rapidly with gas density then leads us to expect that gas infall feeds starbursts.  Meanwhile, 
intense nuclear starbursts were observed preferentially in barred and oval galaxies.  And they have timescales that are 
reasonable to grow disky bulge-like components similar to those seen in barred and oval galaxies.  That is, we see the growth
of these high-density, disky centers of galaxies in action.  Gradually, the observed disky bulges and the theoretical evolution 
scenario were connected into a now-robust picture of internal galaxy evolution that complements hierarchical clustering.  
These often-disky central components are now known as ``pseudobulges'' to take account of their superficial similarity to 
classical bulges (with which they were often confused) while recognizing that they formed mainly by slow processes that do not 
involve major mergers.  This~subject~is~reviewed~in~Kormendy~\& ~Kennicutt (2004)~and~in~Kormendy~(1993b, 2012).  

      The distinction between classical and pseudo bulges is important because (Section 6.8) we find that classical bulges and 
ellipticals correlate closely with $M_\bullet$, but pseudobulges hardly correlate with BHs at all.  Observational criteria that 
have nothing to do with BHs divide central components into two kinds that correlate differently with BHs.  This is a substantial 
success of the secular evolution picture. 

      In {\bf Tables 2\/} and {\bf 3\/} in the next section, (pseudo)bulge classifications and (pseudo)bulge-to-total luminosity ratios, 
$B/T$ and $PB/T$, are taken from Kormendy \& Bender (2013b).  The observational criteria used to distinguish classical bulges from 
pseudobulges are listed in Kormendy \& Kennicutt (2004).  They have since been refined slightly via new observations; the criteria 
as used in Kormendy \& Bender (2013b) and in this paper are listed in the Supplementary Information.

      Classifications are more robust if they are based on many criteria.   All of our classifications are based on at least two and 
sometimes as many as five criteria.

\vfill\eject

\vs
\ni {\big\ARRed 5. BH DATABASE}\textBlack
\vs

      This section is an inventory of galaxies that have BH detections and $M_\bullet$ measurements based on stellar dynamics, 
ionized gas dynamics, CO molecular gas disk dynamics, or maser disk dynamics.  {\bf Table 2\/} lists ellipticals, including
mergers-in-progress that have not yet relaxed into equilibrium.  {\bf Table 3\/} lists disk galaxies with classical bulges
(upper part of table) and pseudobulges (lower part of table).  The demographic results discussed in the following sections 
are based on these tables.   Both tables are provided in machine-readable form in the electronic edition of this paper.

      We reviewed the $M_\bullet$ measurements in Sections 2 and 3.~Notes on individual objects~in~{\bf Tables\ts2} and {\bf 3}
follow the tables and 
provide more detail.
%
Derivation of host galaxy properties is relatively straightforward for ellipticals as discussed in the notes 
and in Supplementary Information.  These derivations are more complicated for disk-galaxy hosts, because 
(pseudo)bulge classification is crucial and because (pseudo)bulge{\ts}--{\ts}disk photometric decomposition is necessary.  This work 
is too long to fit here; it is published in a satellite paper written in parallel with this review (Kormendy \& Bender 2013b).  
Some details are repeated here in the table notes for the convenience of readers.  

      Implicit in the tables are decisions about which published $M_\bullet$ measurements are reliable enough for inclusion.  
No clearcut, objective dividing line separates reliable and questionable measurements.  Our decisions are personal judgments.  
Our criteria are similar to those in G\"ultekin \etal (2009c); when we made a different decision, this is explained in the notes 
on individual objects.~We try to be conservative.~With a few exceptions that are not included in correlation fits, 
stellar-dynamical masses are retained only if they are based on three-integral models.  Nevertheless (Section 3), it is likely that 
systematic errors -- e.{\ts}g., due to the neglect of triaxiality in giant ellipticals -- are still present in some data.  
For this reason, we do not discuss correlation scatter in much detail.  We~do, however, derive the most accurate correlations 
that we can with present data (Section 6.6).

      The sources of the adopted $M_\bullet$ measurements are given in the last column of each table, and 
earlier measurements are discussed in the notes on individual objects.  The $M_\bullet$ error bars 
present a problem, because different authors present error bars with different confidence intervals.  For
consistency, we use approximate one-sigma standard deviations, i.{\ts}e., 68\ts\% confidence intervals.
When authors quote two-sigma or three-sigma errors, we follow Krajnovi\'c \etal (2009) and estimate that
one-sigma errors are $N$ times smaller than $N$-sigma errors.  Flags in Column 12 of {\bf  Table\ts2} and Column 19
of {\bf Table 3} encode the method used to determine $M_\bullet$, whether the galaxy has a core (ellipticals
only), and whether $M_\bullet$ was derived with models that include triaxiality or dark matter and large orbit libraries.  
Only the models of M{\ts}32, NGC\ts1277, and NGC\ts3998 include~all~3.

      We use a distance scale (Column 3) based mainly on surface brightness fluctuation measurements at small distances
and on the WMAP 5-year cosmology at  large distances ($H_0$\ts=\ts$70.5${\ts}km{\ts}s$^{-1}${\ts}Mpc$^{-1}$; Komatsu \etal 2009). 
Details are in the notes that follow {\bf Tables 2} and {\bf 3}.  Velocity dispersions~$\sigma_e$ (Colums 11 and 17 of
{\bf Tables 2} and {\bf 3}) are problematic; they are discussed in the table notes.

      We discuss luminosity correlations only in terms of $K_s$-band absolute magnitudes.  However,
we also provide $V$ magnitudes for the convenience of readers and because we use them to check~the~$K_s$
magnitudes.  Readers should view $(B - V)_0$ as a galaxy color that contains physical 
information but $(V$\null$-$\null$K)_0$ mainly as a sanity check of the independent $V$ and $K_s$ magnitude systems.  
Our $K_s$ magnitudes are on the photometric system of the Two Micron All Sky Survey (2MASS: Skrutskie \etal 2006); 
the effective wavelength is $\sim 2.16$ $\mu$m.  To good approximation, $K_s = K - 0.044$
(Carpenter 2001; Bessell 2005), where $K$ is Johnson's (1962) 2.2 $\mu$m bandpass.  
Except in this paragraph and in the tables, we abbreviate $K_s$ as $K$ for convenience.~{\bf Tables 2}~and~{\bf 3}
list $K_s$ apparent magnitudes of the galaxies from the 2MASS Large Galaxy Atlas (Jarrett \etal 2003) or from the online 
Extended Source Catalog.  Corrections (usually a few tenths of a mag) have been made for some of the brightest or
angularly largest galaxies as discussed in the table notes.

\vfill\eject


\hsize=19.truecm  \hoffset=-1.5truecm  \vsize=25.5truecm  \voffset=-0.5truecm

\magnification = \magstep0



\font\sc=cmr8
\def\scbaselines{\baselineskip=8pt    \lineskip=0pt   \lineskiplimit=0pt}

\def\dblbaselines{\baselineskip=12pt    \lineskip=0pt   \lineskiplimit=0pt}
\def\vsl{\vskip\baselineskip}   \def\vs{\vskip 6pt} \def\vsss{\vskip 3pt}
\parindent=10pt \nopagenumbers

\input colordvi

\def\t#1{#1} 
\def\t#1{\empty}
    
  \def\ba{\kern -1pt}
\parskip = 0pt 
\def\ts{\thinspace} \def\cl{\centerline}
\def\ni{\noindent}  

\def\nhi{\noindent \hangindent=1.0truecm}
  
\def\ihi{\indent\hangindent=1.85truecm}
\def\ihij{\indent\null{\kern -4.7pt}\hangindent=1.85truecm}

\def\makeheadline{\vbox to 0pt{\vskip-30pt\line{\vbox to8.5pt{}\the
                               \headline}\vss}\nointerlineskip}
\def\toppageno{\headline={\hss\tenrm\folio\hss}}
\def\footnoterule{\kern-3pt \hrule width \hsize \kern 2.6pt \vskip 3pt}
\output={\plainoutput}    \pretolerance=10000   \tolerance=10000

\def\sup1{$^{\rm 1}$} \def\sup2{$^{\rm 2}$}
\def\r0{$\rho_0$}   
 \def\0{\phantom{0}} \def\bb{\kern -2pt}
\def\1{\phantom{1}}         
\def\etal{{et~al.\ }}
\def\gapprox{$_>\atop{^\sim}$} \def\lapprox{$_<\atop{^\sim}$}
\def\kms{km~s$^{-1}$}          
\newdimen\sa  \def\sd{\sa=.1em \ifmmode $\rlap{.}$''$\kern -\sa$
                               \else \rlap{.}$''$\kern -\sa\fi}
\def\ss{\ifmmode ^{\prime\prime}$\kern-\sa$ \else $^{\prime\prime}$\kern-\sa\fi}
\def\mm{\ifmmode ^{\prime}$\kern-\sa$ \else $^{\prime}$\kern-\sa \fi}
  \def\msund{M$_{\odot}$}
\def\mbh{$M_{\bullet}$~}  
\def\m31{M{\ts}31} \def\mm32{M{\ts}32} \def\mmm33{M{\ts}33} \def\M87{M{\ts}87} 
\def\mbh{$M_\bullet$}

\def\nhi2{\noindent \hangindent=2.79cm}

\font\big=cmbx12 scaled 1100
\font\bigau=cmr12 scaled 1200
\font\bigbig=cmr12 scaled 2000
\font\bigit=cmti10 scaled 1200

\font\small=cmr8


\def\tenpoint{
  \font\fiverm=cmr5
  \font\sevenrm=cmr7
  \font\tenrm=cmr10
  \font\fivei=cmmi5
  \font\seveni=cmmi7
  \font\teni=cmmi10
  \font\fivesy=cmsy5
  \font\sevensy=cmsy7  
  \font\tensy=cmsy10
  \font\it=cmti10
  \font\bf=cmbx10
  \font\sl=cmsl10
  \textfont0=\tenrm \scriptfont0=\sevenrm     
    \scriptscriptfont0=\fiverm                 
  \def\rm{\fam0 \tenrm}   
  \textfont1=\teni  \scriptfont1=\seveni  
    \scriptscriptfont1=\fivei                  
  \def\mit{\fam1 } \def\oldstyle{\fam1 \teni}
  \textfont2=\tensy \scriptfont2=\sevensy 
    \scriptscriptfont2=\fivesy                 
}


\def\ninepoint{
  \font\fiverm=cmr5
  \font\sevenrm=cmr7
  \font\ninerm=cmr9
  \font\fivei=cmmi5
  \font\seveni=cmmi7
  \font\ninei=cmmi9
  \font\fivesy=cmsy5
  \font\sevensy=cmsy7  
  \font\ninesy=cmsy9
  \font\it=cmti9
  \font\bf=cmbx9
  \font\sl=cmsl9
  \textfont0=\ninerm \scriptfont0=\sevenrm     
    \scriptscriptfont0=\fiverm                 
  \def\rm{\fam0 \ninerm}   
  \textfont1=\ninei  \scriptfont1=\seveni  
    \scriptscriptfont1=\fivei                  
  \def\mit{\fam1 } \def\oldstyle{\fam1 \ninei}
  \textfont2=\ninesy \scriptfont2=\sevensy 
    \scriptscriptfont2=\fivesy                 
}


\def\eightpoint{
  \font\fiverm=cmr5
  \font\sevenrm=cmr7
  \font\eightrm=cmr8
  \font\fivei=cmmi5
  \font\seveni=cmmi7
  \font\eighti=cmmi8
  \font\fivesy=cmsy5
  \font\sevensy=cmsy7  
  \font\eightsy=cmsy8
  \font\rm=cmr8
  \font\it=cmti8
  \font\bf=cmbx8
  \font\sl=cmsl8
  \textfont0=\eightrm \scriptfont0=\sevenrm     
    \scriptscriptfont0=\fiverm                 
  \def\rm{\fam0 \eightrm}   
  \textfont1=\eighti  \scriptfont1=\seveni  
    \scriptscriptfont1=\fivei                  
  \def\mit{\fam1 } \def\oldstyle{\fam1 \eighti}
  \textfont2=\eightsy \scriptfont2=\sevensy 
    \scriptscriptfont2=\fivesy                 
}

\ninepoint
\small

\def\tblbaselines{\baselineskip=10pt    \lineskip=0pt   \lineskiplimit=0pt}
\scbaselines

\hfuzz=100pt


\hsize=19.5truecm  \hoffset=-1.8truecm  \vsize=25.8truecm  \voffset=-1.1truecm

\def\x{\textBlack}
\def\y{\textPurple}
\def\y{\ARDiscard}
\def\g{\textColor{1. .1 1. .1}}

\cl{\null}

\vskip -40pt

\cl{\null}

\table
\tblbaselines
\hskip -10pt
\tablewidth{20truecm} 
\tablespec{\l\l\c\c\c\c\c\c\c\c\c\c\l}
\body{
\header{\bf \Blue{Table 2 \quad  Supermassive black holes detected dynamically in 45 elliptical galaxies (December 2012)}\textBlack}
\skip{5pt}
\hline
\skip{3pt}
&   Galaxy &\llap{T}ype& Distance &$  K_s   $&$\0M_{KsT}$ &$ M_{VT}$ & ($V$\null$-$\null$K_s)_0\0$&($B$\null$-$\null$V)_0 $&\0$\log{M_{\rm bulge}}$&  \mbh (low \mbh\ -- high \mbh)     &  $\sigma_e$  &       Flags       &  Source                \end %
&          &        &     (Mpc)   &          &            &          &             &   & ($M_\odot$)&       ($M_\odot$)                & (km s$^{-1}$)&~M{\ts}C{\ts}$M_\bullet$ &                  \end %
& (1)      & (2)    &      (3)    &   (4)    &     (5)    &  (6)     &     (7)     &    (8)    &        (9)       &              (10)                    &     (11)     &       (12)        &  (13)                  \end %
\skip{3pt}	        	     	                	   				        
\hline		        	     	                	   				        
\skip{3pt}	        	     	                	   				        
&   M{\ts}32 &  E2  & \0\00.805 7 &$ 5.10   $&$  -19.45  $&$ -16.64 $&$  2.816    $&$ 0.895   $&$\09.05 \pm 0.10 $&$  2.45(   1.43-\03.46 ) \times 10^6 $&$\077\pm\03  $&   1     0     1   &  van{\ts}den{\ts}Bosch{\ts}+{\ts}2010  \end \skip{-2pt} %
&\g NGC 1316 &  E4  &  20.95 1    &$ 5.32   $&$  -26.29  $&$ -23.38 $&$  2.910    $&$ 0.871   $&$ 11.84 \pm 0.09 $&$  1.69(   1.39-\01.97 ) \times 10^8 $&$ 226\pm\09  $&   1     0     0   &  Nowak + 2008        \x\end \skip{-2pt} 
&   NGC 1332 &  E6  &  22.66 2    &$ 7.05   $&$  -24.73  $&$ -21.58 $&$  3.159    $&$ 0.931   $&$ 11.27 \pm 0.09 $&$  1.47(   1.27-\01.68 ) \times 10^9 $&$ 328\pm\09  $&   1     0     0   &  Rusli + 2011          \end \skip{-2pt} 
&   NGC 1374 &  E0  &  19.57 1    &$ 8.16   $&$  -23.30  $&$ -20.43 $&$  2.874    $&$ 0.908   $&$ 10.65 \pm 0.09 $&$  5.90(   5.39-\06.51 ) \times 10^8 $&$ 167\pm03   $&   1     0     1   &  Rusli+2013            \end \skip{-2pt} 
&   NGC 1399 &  E1  &  20.85 1    &$ 6.31   $&$  -25.29  $&$ -22.43 $&$  2.863    $&$ 0.948   $&$ 11.50 \pm 0.09 $&$  8.81(   4.35-17.81  ) \times 10^8 $&$ 315\pm03   $&   1     1     0   &  see notes             \end \skip{-2pt} %
&   NGC 1407 &  E0  &  29.00 2    &$ 6.46   $&$  -25.87  $&$ -22.89 $&$  2.980    $&$ 0.969   $&$ 11.74 \pm 0.09 $&$  4.65(   4.24-\05.38 ) \times 10^9 $&$ 276\pm\02  $&   1     1     1   &  Rusli+2013            \end \skip{-2pt} 
&   NGC 1550 &  E1  &  52.50 9    &$ 8.77   $&$  -24.87  $&$ -21.89 $&$  2.974    $&$ 0.963   $&$ 11.33 \pm 0.09 $&$  3.87(   3.16-\04.48 ) \times 10^9 $&$ 270\pm 10  $&   1     1     1   &  Rusli+2013            \end \skip{-2pt} 
&\y NGC 2778 &  E2  &  23.44 2    &$ 9.51   $&$  -22.34  $&$ -19.39 $&$  2.955    $&$ 0.911   $&$ 10.26 \pm 0.09 $&$  1.45(   0.00-\02.91 ) \times 10^7 $&$ 175\pm\08  $&   1     0     1   &  Schulze + 2011      \x\end \skip{-2pt} %
&\g NGC 2960 &  E2  &  67.1\0 9   &$ 9.78   $&$  -24.36  $&$ -21.30 $&$  3.068    $&$ 0.880   $&$ 11.06 \pm 0.09 $&$  1.08(   1.03-\01.12 ) \times 10^7 $&$ 166\pm 16  $&   3     0     0   &  Kuo + 2011          \x\end \skip{-2pt} %
&   NGC 3091 &  E3  &  53.02 9    &$ 8.09   $&$  -25.54  $&$ -22.56 $&$  2.980    $&$ 0.962   $&$ 11.61 \pm 0.09 $&$  3.72(   3.21-\03.83 ) \times 10^9 $&$ 297\pm 12  $&   1     1     1   &  Rusli+2013            \end \skip{-2pt} 
&   NGC 3377 &  E5  &  10.99 2    &$ 7.16   $&$  -23.06  $&$ -20.08 $&$  2.980    $&$ 0.830   $&$ 10.50 \pm 0.09 $&$  1.78(   0.85-\02.72 ) \times 10^8 $&$ 145\pm\07  $&   1     0     1   &  Schulze + 2011        \end \skip{-2pt} 
&   NGC 3379 &  E1  &  10.70 2    &$ 6.27   $&$  -23.88  $&$ -21.01 $&$  2.867    $&$ 0.939   $&$ 10.91 \pm 0.09 $&$  4.16(   3.12-\05.20 ) \times 10^8 $&$ 206\pm 10  $&   1     1     1   &  van{\ts}den{\ts}Bosch{\ts}+{\ts}2010  \end \skip{-2pt} 
&\y NGC 3607 &  E1  &  22.65 2    &$ 6.99   $&$  -24.79  $&$ -21.92 $&$  2.872    $&$ 0.911   $&$ 11.26 \pm 0.09 $&$  1.37(   0.90-\01.82 ) \times 10^8 $&$ 229\pm 11  $&   1     1     0   &  G\"ultekin + 2009b  \x\end \skip{-2pt} %
&   NGC 3608 &  E1  &  22.75 2    &$ 7.62   $&$  -24.17  $&$ -21.19 $&$  2.980    $&$ 0.921   $&$ 11.01 \pm 0.09 $&$  4.65(   3.66-\05.64 ) \times 10^8 $&$ 182\pm\09  $&   1     1     1   &  Schulze + 2011        \end \skip{-2pt} 
&   NGC 3842 &  E1  &  92.2\0 9   &$ 8.84   $&$  -25.99  $&$ -23.01 $&$  2.980    $&$ 0.941   $&$ 11.77 \pm 0.09 $&$\y9.09(   6.28-11.43  ) \times 10^9$\x&$270\pm 27  $&   1     1     1   &  McConnell + 2012      \end \skip{-2pt} 
&\y NGC 4261 &  E2  &  32.36 2    &$ 6.94   $&$  -25.62  $&$ -22.64 $&$  2.980    $&$ 0.974   $&$ 11.65 \pm 0.09 $&$  5.29(   4.21-\06.36 ) \times 10^8 $&$ 315\pm 15  $&   2     1     0   &  Ferrarese + 1996    \x\end \skip{-2pt} 
&   NGC 4291 &  E2  &  26.58 2    &$ 8.42   $&$  -23.72  $&$ -20.76 $&$  2.954    $&$ 0.927   $&$ 10.85 \pm 0.09 $&$  9.78(   6.70-12.86  ) \times 10^8 $&$ 242\pm 12  $&   1     1     1   &  Schulze + 2011        \end \skip{-2pt} %
&   NGC 4374 &  E1  &  18.51 1    &$ 5.75   $&$  -25.60  $&$ -22.62 $&$  2.980    $&$ 0.945   $&$ 11.62 \pm 0.09 $&$  9.25(   8.38-10.23  ) \times 10^8 $&$ 296\pm 14  $&   2     1     0   &  Walsh + 2010          \end \skip{-2pt} 
&\g NGC 4382 &  E2  &  17.88 1    &$ 5.76   $&$  -25.51  $&$ -22.53 $&$  2.980    $&$ 0.863   $&$ 11.51 \pm 0.09 $&$  1.30(   0.00-22.4\0 ) \times 10^7 $&$ 182\pm\05  $&   1     1     0   &  G\"ultekin + 2011   \x\end \skip{-2pt} 
&\y NGC 4459 &  E2  &  16.01 1    &$ 7.15   $&$  -23.88  $&$ -20.91 $&$  2.975    $&$ 0.909   $&$ 10.88 \pm 0.09 $&$\y6.96(   5.62-\08.29 ) \times 10^7$\x&$167\pm\08  $&   2     0     0   &  Sarzi + 2001          \end \skip{-2pt} 
&   NGC 4472 &  E2  &  16.72 1    &$ 4.97   $&$  -26.16  $&$ -23.18 $&$  2.980    $&$ 0.940   $&$ 11.84 \pm 0.09 $&$  2.54(   2.44-\03.12 ) \times 10^9 $&$ 300\pm\07  $&   1     1     1   &  Rusli + 2013          \end \skip{-2pt} 
&   NGC 4473 &  E5  &  15.25 1    &$ 7.16   $&$  -23.77  $&$ -20.89 $&$  2.874    $&$ 0.935   $&$ 10.85 \pm 0.09 $&$  0.90(   0.45-\01.35 ) \times 10^8 $&$ 190\pm\09  $&   1     0     1   &  Schulze + 2011        \end \skip{-2pt} %
&   M{\ts}87 &  E1  &  16.68 1    &$ 5.27   $&$  -25.85  $&$ -22.87 $&$  2.980    $&$ 0.940   $&$ 11.72 \pm 0.09 $&$  6.15(   5.78-\06.53 ) \times 10^9 $&$ 324^{+28}_{-12}\0$& 1 1     1   &  Gebhardt + 2011       \end \skip{-2pt} 
&   NGC 4486A&  E2  &  18.36 1    &$ 9.49   $&$  -21.83  $&$ -18.85 $&$  2.980    $&$ \dots   $&$ 10.04 \pm 0.09 $&$  1.44(   0.92-\01.97 ) \times 10^7 $&$ 111\pm\05  $&   1     0     0   &  Nowak + 2007          \end \skip{-2pt} 
&   NGC 4486B&  E0  &  16.26 1    &$10.39\0 $&$  -20.67  $&$ -17.69 $&$  2.980    $&$ 0.991   $&$\09.64 \pm 0.10 $&$\y6.\0\0(4.\0\0-\09.\0\0)\times 10^8 $&\x$185\pm\09$&   1     0     0   &  Kormendy + 1997       \end \skip{-2pt} 
&   NGC 4649 &  E2  &  16.46 1    &$ 5.49   $&$  -25.61  $&$ -22.63 $&$  2.980    $&$ 0.947   $&$ 11.64 \pm 0.09 $&$  4.72(   3.67-\05.76 ) \times 10^9 $&$ 380\pm 19  $&   1     1     1   &  Shen+Gebhardt\ts2010  \end \skip{-2pt} 
&   NGC 4697 &  E5  &  12.54 1    &$ 6.37   $&$  -24.13  $&$ -21.33 $&$  2.799    $&$ 0.883   $&$ 10.97 \pm 0.09 $&$  2.02(   1.52-\02.53 ) \times 10^8 $&$ 177\pm\08  $&   1     0     1   &  Schulze + 2011        \end \skip{-2pt} %
&   NGC 4751 &  E6  &  32.81 2    &$ 8.24   $&$  -24.38  $&$ -21.22 $&$  3.158    $&$ 0.983   $&$ 11.16 \pm 0.09 $&$  2.44(   2.07-\02.56 ) \times 10^9 $&$ 355\pm 14  $&   1     0     1   &  Rusli + 2013          \end \skip{-2pt} 
&   NGC 4889 &  E4  & 102.0\0~9\1 &$ 8.41   $&$  -26.64  $&$ -23.63 $&$  3.007    $&$ 1.031   $&$ 12.09 \pm 0.09$&\y$\02.08( 0.49-\03.66  ) \times 10^{10}$\x&$347\pm 5$&   1     1     1   &  McConnell + 2012      \end \skip{-2pt} 
&   NGC 5077 &  E3  &  38.7\0 9   &$ 8.22   $&$  -24.74  $&$ -21.66 $&$  2.949    $&$ 0.987   $&$ 11.28 \pm 0.09 $&$  8.55(   4.07-12.93  ) \times 10^8 $&$ 222\pm 11  $&   2     1     0   &  De{\ts}Francesco{\ts}+{\ts}2008 \end \skip{-2pt} %
&\g NGC 5128 &  E   & \03.62 6    &$ 3.49   $&$  -24.34  $&$ -21.36 $&$  2.980    $&$ 0.898   $&$ 11.05 \pm 0.09 $&$  5.69(   4.65-\06.73 ) \times 10^7 $&$ 150\pm\07  $&   1     1     0   &  Cappellari + 2009   \x\end \skip{-2pt} 
&   NGC 5516 &  E3  &  55.3~\ts~9 &$ 8.31   $&$  -25.47  $&$ -22.50 $&$  2.970    $&$ 0.993   $&$ 11.60 \pm 0.09 $&$  3.69(   2.65-\03.79 ) \times 10^9 $&$ 328\pm 11  $&   1     1     1   &  Rusli + 2013          \end \skip{-2pt} 
&   NGC 5576 &  E3  &  25.68 2    &$ 7.83   $&$  -24.23  $&$ -21.29 $&$  2.939    $&$ 0.862   $&$ 11.00 \pm 0.09 $&$  2.73(   1.94-\03.41 ) \times 10^8 $&$ 183\pm\09  $&   1     1     0   &  G\"ultekin + 2009b    \end \skip{-2pt} %
&   NGC 5845 &  E3  &  25.87 2    &$ 9.11   $&$  -22.97  $&$ -19.73 $&$  3.238    $&$ 0.973   $&$ 10.57 \pm 0.09 $&$  4.87(   3.34-\06.40 ) \times 10^8 $&$ 239\pm 11  $&   1     0     1   &  Schulze + 2011        \end \skip{-2pt} %
&   NGC 6086 &  E   & 138.0\0~9\1 &$ 9.97   $&$  -25.74  $&$ -22.84 $&$  2.884    $&$ 0.965   $&$ 11.69 \pm 0.09 $&$  3.74(   2.59-\05.50 ) \times 10^9 $&$ 318\pm\02  $&   1     1     1   &  McConnell + 2011b     \end \skip{-2pt} 
&\y NGC 6251 &  E1  & 108.4\0~9\1 &$ 9.03   $&$  -26.18  $&$ -23.18 $&$  2.998    $&$ \dots   $&$ 11.88 \pm 0.09 $&$  6.14(   4.09-\08.18 ) \times 10^8 $&$ 290\pm 14  $&   2     1     0   &  Ferrarese + 1999    \x\end \skip{-2pt} 
&   NGC 6861 &  E4  &  28.71 2    &$ 7.71   $&$  -24.60  $&$ -21.42 $&$  3.179    $&$ 0.962   $&$ 11.25 \pm 0.09 $&$  2.10(   2.00-\02.73 ) \times 10^9 $&$ 389\pm\03  $&   1     0     1   &  Rusli + 2013          \end \skip{-2pt}
&\y NGC 7052 &  E3  &  70.4~\ts~9 &$ 8.57   $&$  -25.70  $&$ -22.86 $&$  2.841    $&$ 0.86\0  $&$ 11.61 \pm 0.10 $&$  3.96(   2.40-\06.72 ) \times 10^8 $&$ 266\pm 13  $&   2     1     0   &  van{\ts}der{\ts}Marel{\ts}+{\ts}1998b   \x\end \skip{-2pt} %
&   NGC 7619 &  E3  &  53.85 2    &$ 8.03   $&$  -25.65  $&$ -22.83 $&$  2.821    $&$ 0.969   $&$ 11.65 \pm 0.09 $&$  2.30(   2.19-\03.45 ) \times 10^9 $&$ 292\pm\05  $&   1     1     1   &  Rusli + 2013          \end \skip{-2pt} 
&   NGC 7768 &  E4  & 116.0\0~9\1 &$ 9.34   $&$  -26.00  $&$ -23.19 $&$  2.811    $&$ 0.906   $&$ 11.75 \pm 0.09 $&$  1.34(   0.93-\01.85 ) \times 10^9 $&$ 257\pm 26  $&   1     1     1   &  McConnell + 2012      \end \skip{-2pt} 
&   IC 1459  &  E4  &  28.92 2    &$ 6.81   $&$  -25.51  $&$ -22.42 $&$  3.081    $&$ 0.966   $&$ 11.60 \pm 0.09 $&$  2.48(   2.29-\02.96 ) \times 10^9 $&$ 331\pm\05  $&   1     0     0   &  Cappellari + 2002     \end \skip{-2pt} %
&\g IC 1481  &  E1.5&  89.9\0 9   &$10.62\0 $&$  -24.17  $&$  \dots $&$  \dots    $&$ \dots   $&$       \dots    $&$  1.49(   1.04-\01.93 ) \times 10^7 $&$ \dots      $&   3     0     0   &  Hur\'e + 2011       \x\end \skip{-2pt} %
&\y A1836 BCG&  E   & 152.4\0~9\1 &$ 9.99   $&$  -25.95  $&$ -22.64 $&$  3.310    $&$ 1.043   $&$ 11.81 \pm 0.10 $&$  3.74(   3.22-\04.16 ) \times 10^9 $&$ 288\pm 14  $&   2     1     0   &  Dalla{\ts}Bont\'a{\ts}+\ts2009\x\end \skip{-2pt} 
&\y A3565 BCG&  E   &  49.2~\ts~9 &$ 7.50   $&$  -25.98  $&$ -23.03 $&$  2.948    $&$ 0.956   $&$ 11.78 \pm 0.09 $&$  1.30(   1.11-\01.50 ) \times 10^9 $&$ 322\pm 16  $&   2     1     0   &  Dalla{\ts}Bont\'a{\ts}+\ts2009\x\end \skip{-2pt} 
&\y Cygnus A &  E   & 242.7\0~9\0 &$10.28\0 $&$  -26.77  $&$ -23.23 $&$  3.54\0   $&$ \dots   $&$       \dots    $&$  2.66(   1.91-\03.40 ) \times 10^9 $&$ 270\pm 90  $&   2     1     0   &  Tadhunter + 2003    \x\end \skip{-2pt} 
\skip{3pt}											    
\hline												    
}												    
\endtable											    
												    

\scbaselines

\null
\vskip -15pt
\null

{\sc\parindent=39pt\eightpoint
\def\nhi{\noindent \hangindent=1.4truecm}
\lineskip=-20pt \lineskiplimit=-20pt
\noindent Column 1 is the galaxy name; BCGs are brightest cluster galaxies in the Abell clusters named. \y Cyan listings are not included in the fits (Section 6.3).\par\x
\nhi Column 2 is Hubble type (mostly RC3). \g Green lines are for mergers in progress (Section 6.4). \hfill \y If only $M_\bullet$ is cyan, we accept it but do not include it\x\par
\nhi Column 3 is the assumed distance from the following sources, starting with the highest-priority sources: \hfill \y in the correlation fits (see Section 6.6).\x        \par
\ihi            (1) Blakeslee \etal (2009) surface-brightness fluctuation (SBF) distances for individual galaxies in the Virgo and Fornax clusters; \par
\ihi            (2) Tonry \etal (2001) SBF corrected via Equation A1 in Blakeslee \etal (2010);                                                     \par
\ihi            (3) Mei \etal (2007) SBF mean distance to the Virgo W$^\prime$ cloud (de Vaucouleurs 1961);                                         \par
\ihi            (4) Mei \etal (2007) SBF mean distance to the Virgo cluster (no W$^\prime$ cloud);                                                  \par
\ihi            (5) Thomsen \etal (1997) SBF distance to NGC 4881 in the Coma cluster;                                                              \par
\ihi            (6) mean of distance determinations adopted in Kormendy \etal (2010); sources are given there;                                      \par
\ihi            (7) Monachesi \etal (2011); agrees with (8);                                                                                        \par
\ihi            (8) mean of many determinations listed on NED, using mainly Cepheids, SBF, TRGB, and RR Lyrae stars.                                \par
\ihi            (9) As a last resort, we adopt D (Local Group) given by NED for the recession velocity of the galaxy 
                    (if isolated) or its group (if in a group or cluster) and for the WMAP 5-year cosmology parameters (Komatsu \etal 2009).        \par
\ihij          (10) van den Bosch \etal (2012).                                                                                                     \par
\nhi Column \04 is the 2MASS $K_s$ total magnitude.  When $(V - K_s)_0 = 2.980$ in Column 7, $K_s$ has been corrected as discussed in 
                   {\ARRed\sbf Apparent
                    Magnitude Corrections}~\textBlack in the Table notes  in the Supplemental Information.                   \par
\nhi Columns 5 and 6 are the $K_s$- and $V$-band absolute magnitudes based on the adopted distances and Galactic absorption corrections from Schlegel
                \etal (1998) as recalibrated by Schlafly \& Finkbeiner (2011).  The $V$-band magnitudes are taken, in order of preference, from KFCB, from RC3, or from Hyperleda 
                (usually ``integrated photometry'' but sometimes the main table if it implies a more realistic $(V - K_s)_0$ color).        \par
\nhi Columns \ts7 and 8 are the $V - K_s$ and $B - V$ colors of the galaxy corrected for Galactic reddening.                                             \par
\nhi Column \09 is the base-10 logarithm of the bulge mass (Section 6.6.1).\par
\def\nhi{\noindent \hangindent=1.55truecm}
\nhi Column 10 is the measured BH mass with 1-sigma range in parentheses from sources in Column 13.\par
\nhi Column 11 is the stellar velocity dispersion $\sigma_e$.  We adopt the usual convention that $\sigma_e^2$ is the intensity-weighted mean of $V^2 + \sigma^2$ 
               out to a fixed fraction of the effective radius $r_e$ that contains half of the light of the galaxy.  As discussed in {\ARRed\sbf Corrections to effective
               velocity dispersions}~\textBlack in the table notes, 
               we adopt $r_e/2$ when we calculate $\sigma_e$ from photometry and published kinematics (see notes on individual objects).  When no note is given,
               $\sigma_e$ is from the $M_\bullet$ source paper or from G\"ultekin \etal (2009c).  \par
\nhi Column 12 lists three flags: ``M'' encodes the method used to measure $M_\bullet$, using 1 for stellar dynamics, 2 for ionized gas dynamics, and 3 for maser dynamics.
                 ``C'' = 1 implies that the galaxy has a core (e.{\ts}g., Lauer \etal 1995).
                 ``$M_\bullet$'' = 1 implies that the BH mass has been ``corrected'' by making dynamical models that include large orbit libraries and triaxiality (M{\ts}32 and NGC 3379) 
                 or dark matter halos.\par
}

\parindent=10pt

\vfill\eject 

\hsize=15.0truecm  \hoffset=0.0truecm  \vsize=20.1truecm  \voffset=1.5truecm

\vfill\eject

\hsize=19.truecm  \hoffset=-1.5truecm  \vsize=24.7truecm  \voffset=-0.5truecm

\cl{\null}

\vskip 1truecm
\includegraphics{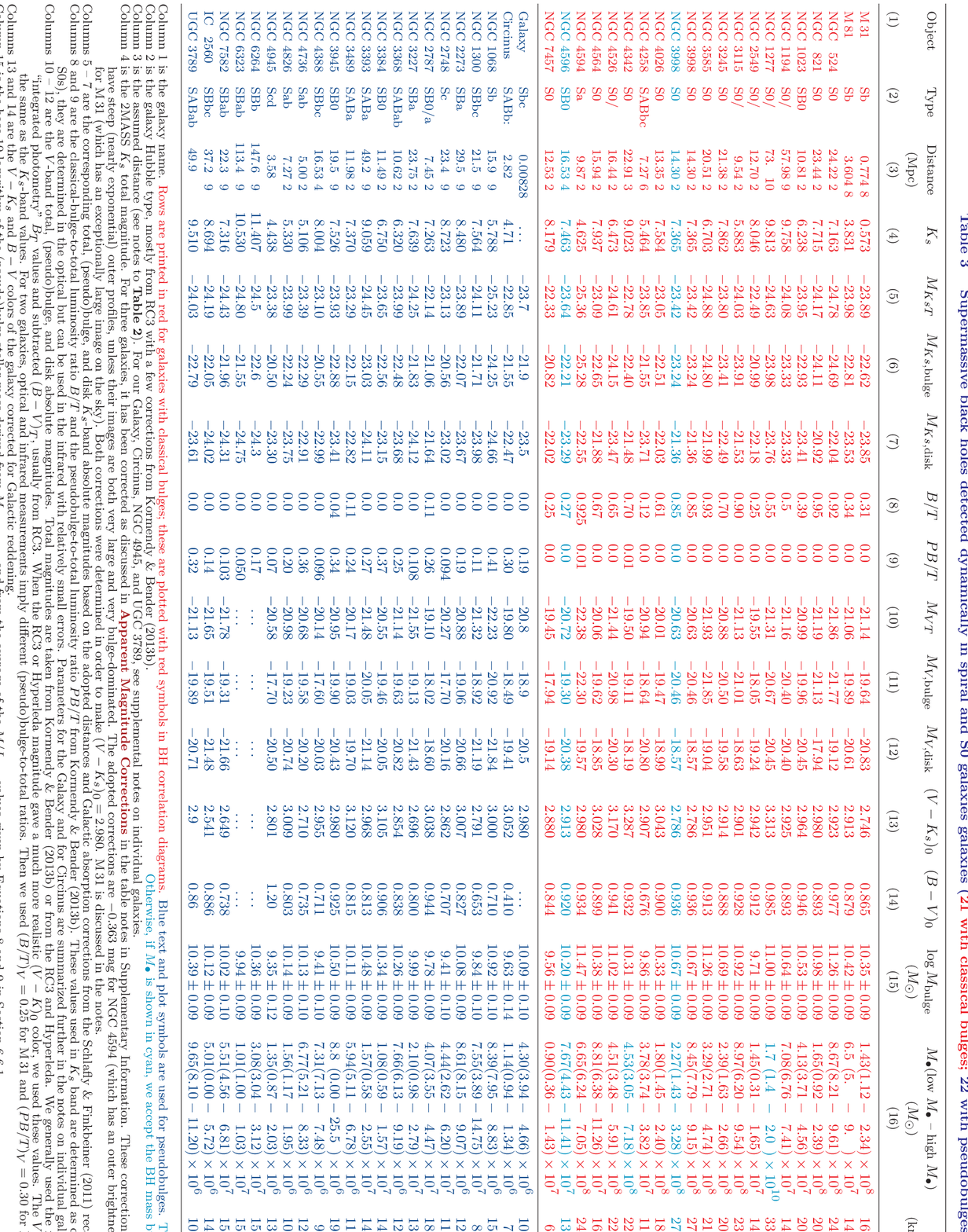}

\cl{\null}
\vfill\eject

{\eightpoint\small\smedbaselines

\def\vsn{\vskip 4pt}

\lineskip=-30pt \lineskiplimit=-30pt

\vs
{\bf\ARRed Notes on Technical Problems with the Parameters in Tables 2 and 3:}\textBlack
\vs

{\bf\ARRed Apparent Magnitude Corrections:}~\textBlack Section S1 in the Supplemental Information discusses the reasons for and 
derives corrections to 2MASS $K$ magnitudes for the biggest and brightest galaxies.  A summary is provided here.  

      An important consistency check on $V$- and $K$-band apparent magnitudes is provided by the observation that galactic-absorption-corrected 
$(V - K)_0$ colors are exceedingly well behaved for almost all galaxies.  Exceptions are the most internally absorbed or starbursting galaxies, 
but they are not relevant here.~For other galaxies, we find a tight correlation between $(V - K)_0$ and $(B - V)_0$ for 
0.6 \lapprox \ts$(B - V)_0$ \lapprox \ts1.2.  Classical bulges and ellipticals have $(V - K)_0 \simeq 2.980$ with a total scatter (not a dispersion) 
of about $\pm 0.15$.  Both colors are tabulated in {\bf Tables\ts2}~and~{\bf 3}.  We use them to check the apparent magnitudes.  For a few, 
usually faint galaxies, discrepant colors suggest that the $V$ magnitudes in NED and Hyperleda are wrong, usually by a few tenths of a magnitude.  
For these, 2MASS is more accurate, and we correct the $V$ magnitudes to make $(V - K)_0 = 2.980$.  More often -- usually for the largest galaxies 
on the sky -- the 2MASS magnitudes are the problem.  We correct them as follows.

      The 2MASS photometric system is very accurate; Jarrett \etal (2003) state 
that the photometric zero-point calibration is accurate to 2\ts\% -- 3\ts\% across the sky.  To the extent
that we have been able to check them (Kormendy \& Bender 2013b), their integrated magnitudes are correspondingly
accurate at least within the radii out to which they have data.  The survey is somewhat shallow; the 1-sigma
sky noise is 20.0 mag arcsec$^{-2}$ in $K$, although profiles can be derived somewhat fainter than this by
averaging over many pixels.  When there is a problem, it is with the extrapolation to total magnitudes, called
$K_{\rm tot}$ in Jarrett \etal (2003) and {\tt k\_m\_ext} in the online catalog.  The extrapolation is made by
fitting a S\'ersic function to the parts of the profile that are relatively safe from noise and from the PSF
and then integrating the extrapolated function out to ``about four disk scale lengths.''  This procedure works best 
for disk galaxies, i.{\ts}e., ones that have nearly exponential outer profiles.  For ellipticals, the radial range
is too small to yield an accurate S\'ersic index (Section A2 in KFCB), and their tabulated S\'ersic 
indices are too small.  The result is to underestimate the total brightnesses of ellipticals, particularly 
giant ellipticals that have $n > 4$.  The derivation of this conclusion and how we correct for it are the subjects 
of Section S1.  In brief, the $(V - K)_0$ colors of our BH hosts are well enough behaved so that we can use them and 
the $V$ magnitudes to correct the $K$ magnitudes in problem cases.  We are conservative in making this correction -- 
we  make it only when we observe a color $(V - K)_0 < 2.75$ that is \underbar{much} too blue.  That is, if the
correction is less than 0.230 mag, we do not use it.  Also, in some marginal cases, it is not clear whether the
$V$ or the $K$ magnitude is the problem. In all these cases, we use the total magnitude from 2MASS, even for ellipticals.  
In fact, so little is known about some of the more distant BH hosts that the 2MASS magnitudes are much more accurate 
than any available $V$-band measurements.  Finally, in a few cases, our own photometry yields a composite, $K$-band 
profile with 2MASS zeropoint whose integral is more accurate than the 2MASS result because our measurements reach
 out to much larger radii.  We adopt these total magnitudes also.

      The largest correction, $\Delta K = -0.411$, is derived for M{\ts}31.  Jarrett \etal (2003) warn us that
``The Andromeda result should be viewed with caution, as we are not fully confident that the total flux 
of M{\ts}31 has been captured, because of the extreme angular extent of the galaxy and the associated difficulty
with removing the infrared background.''  This problem happens in $V$ band, too.  We adopt $V_T = 3.47 \pm 0.03$ 
from Walterbos \& Kennicutt (1987;
the error estimate is optimistic).  Then the 2MASS magnitude $K = 0.984$ implies that $(V - K)_0 = 2.335$.
This is much too blue for an Sb galaxy with $(B - V)_0 = 0.87$ ({\bf Figure S1}).  As in other such cases, 
we correct $K \rightarrow K - 0.411$ to make $(V - K)_0 = 2.980$ for the bulge only.  The $V$-band
bulge magnitude is determined from the $V$-band $B/T = 0.25 \pm 0.01$.  The correction gives us a $K$-band
magnitude for the bulge only.  The infrared $B/T = 0.31 \pm 0.01$ then gives us the total observed magnitude of the
galaxy, $K_T = 0.573$.  Including both the red bulge and the bluer disk, the total colors of the galaxy are
$(B - V)_0 = 0.87$ and $(V - K)_0 = 2.75$ (cf.~{\bf Figure S1}).

      In summary, 2MASS quoted errors on $K$-band magnitudes, usually 0.02\ts--\ts0.04~mag, are most reliable for
disk galaxies but can be too optimistic for giant Es.  We believe that our corrected magnitudes are generally 
accurate to $\sim 0.1$ mag, although the values for M{\ts}31 and for galaxies with $(V - K)_0$ very different from 
2.980 are more uncertain.  We quote $K_s$ to higher precision because we do not wish to lose precision in arithmetic.  

\vs

{\bf\ARRed Corrections to stellar-dynamical M$_\bullet$ for core galaxies:}~\textBlack 
Gebhardt \& Thomas (2009),
Shen \& Gebhardt (2010),
Schulze \& Gebhardt (2011), and
Rusli \etal (2013)
demonstrate convincingly that stellar dynamical mass models that do not include dark matter halos underestimate
$M_\bullet$ by factors of 2 or more when the BH sphere of influence is not extremely well resolved.  
The effect is small for coreless galaxies; we neglect it when models that include dark matter are not available.
Fortunately, almost all stellar-dynamical $M_\bullet$ estimates for core galaxies are now based on models that include
dark matter.  For one exception, NGC 5576, we correct $M_\bullet$ as discussed in the notes on individual objects.  
For NGC 3607, the correction is too uncertain and we omit the object (see table notes and orange point in {\bf Figure 12}).

\vs

{\bf\ARRed Corrections to effective velocity dispersions:}~\textBlack The velocity dispersion $\sigma_e$ that we correlate with 
$M_\bullet$ is more heterogeneously defined in different papers and less consistently measured within these definitions than we suppose. 
This stealth ``can of worms'' could be a bigger problem than the more obvious uncertainties in measuring $M_\bullet$ that preoccupy authors.  
We check values when we can and fix a few problems.  But the necessary data are not available for all objects.  Fortunately, we can show 
that this problem is not severe.

      The worries are these: 
(1) {\it A priori}, we do not know how best to define $\sigma_e$ so that we learn important physics from the $M_\bullet$\ts--\ts$\sigma_e$ 
    correlation.  Clearly we should not include data at $r$ \lapprox \ts$r_{\rm infl}$ in the average.  But inside what fraction
    of $r_e$ should we average $\sigma(r)$?  We usually claim that we average inside $r_e$ and call the result $\sigma_e$.  However: 
(2) Accurate values of $r_e$ are known for few galaxies.  KFCB demonstrate via high-dynamic-range photometry that brightness 
    profiles of giant ellipticals extend farther out than we have thought.  The $r_e$ values derived in KFCB are more accurate
    than previous results, and they are larger than previous values for almost all giant ellipticals.  We use them here.  But we do
    not have such data for most BH hosts.  It is safe to assume that the $r_e$ values in common use are too small.
    For bulges, the situation is worse.  Accurate decompositions are available for a few galaxies, but they are being derived
    for most galaxies in parallel with the writing of this paper (Kormendy \& Bender 2013b).  
(3) Different fractions of $r_e$ are used by different authors. For example, Ferrarese \& Merritt (2000) use $r_e/8$, whereas the 
    Nuker team uses $r_e$.  
(4) Different authors perform the radial averaging differently.  We follow the Nuker team practice (e.{\ts}g., 
    Pinkney \etal 2003; G\"ultekin~et~al.~2009c) and use $\sigma_e^2$ = average of $V(r)^2 + \sigma(r)^2$, weighting by $I(r) dr$.  
    In most cases, we adopt $\sigma_e$ from the $M_\bullet$ source paper or from G\"ultekin~et~al.~(2009c).  When we calculate it, 
    we perform the average inside $r_e/2$ and include a comment in the notes on individual objects.  However, the above practice
    can be contrasted with $\sigma_e$ values from the SAURON team: They add spectra that sample the galaxy in two dimensions 
    inside $r_e$ or inside the SAURON field, whichever is smaller (Emsellem \etal 2007).  Because $\sigma$
    and $V$ rather than $\sigma^2 + V^2$ are averaged and because the weighting of different radii is essentially by $2 \pi r I(r) dr$ 
    rather than by $I(r) dr$, the resulting $\sigma_e$ values are smaller than the ones that we use.
    No study has proved that the Nuker definition produces a tighter $M_\bullet$\ts--\ts$\sigma_e$ correlation or that a
    tighter correlation is more physically meaningful.  However, because it is most commonly used and therefore most widely
    available, we use the Nuker definition here.  

      In view of these concerns, it is prudent to check how much our results might depend on the definition of $\sigma_e$ or on
how well its measurement follows that definition.  {\sbf Figure 11} compares our $\sigma_e$ values, adopted after much private
wailing and gnashing of teeth, with central velocity dispersions tabulated in HyperLeda and with $\sigma_e$ values calculated as
described above by the SAURON/ATLAS3D team.  Clearly, adopting either alternative would have minimal effect on the scatter in the
$M_\bullet$\ts--\ts$\sigma_e$ correlation and essentially no effect on qualitative conclusions.

      HyperLeda includes a few bad measurements; these are particularly expected for pseudobulges, which can have small
dispersions that are undersampled by poor wavelength resultion and serious problems with dust and star formation.  But it is relatively
easy to find and discard these problems and get an improved central $\sigma$ that would agree with our $\sigma_e$ as well as classical bulges 
and ellipticals do.

      SAURON/ATLAS3D $\sigma_e$ values mostly agree very well with ours.  Some SAURON values are smaller, as expected.
But conclusions would not be changed if we had SAURON $\sigma_e$ values for all galaxies. 
The average shift $\Delta \log{\sigma_e} = \log ({\rm Nuker~\sigma_e}) - \log ({\rm SAURON~\sigma_e}) = 0.0299$ or a factor of 1.07
will be relevant in Section 8 when we compare the $z \simeq 0$ $M_\bullet$\ts--\ts$\sigma_e$ relation with ones derived for galaxies
at large redshifts.  High-$z$ observations necessarily add spectra (not $\sigma^2$) inside apertures that are large in kpc.
SAURON $\sigma_e$ values are the closest match at $z \simeq 0$.  We therefore use the above factor to correct our least-squares fit to
the local $M_\bullet$\ts--\ts$\sigma_e$ correlation for comparison with high-$z$ objects.

\cl{\null}

\vfill

\includegraphics{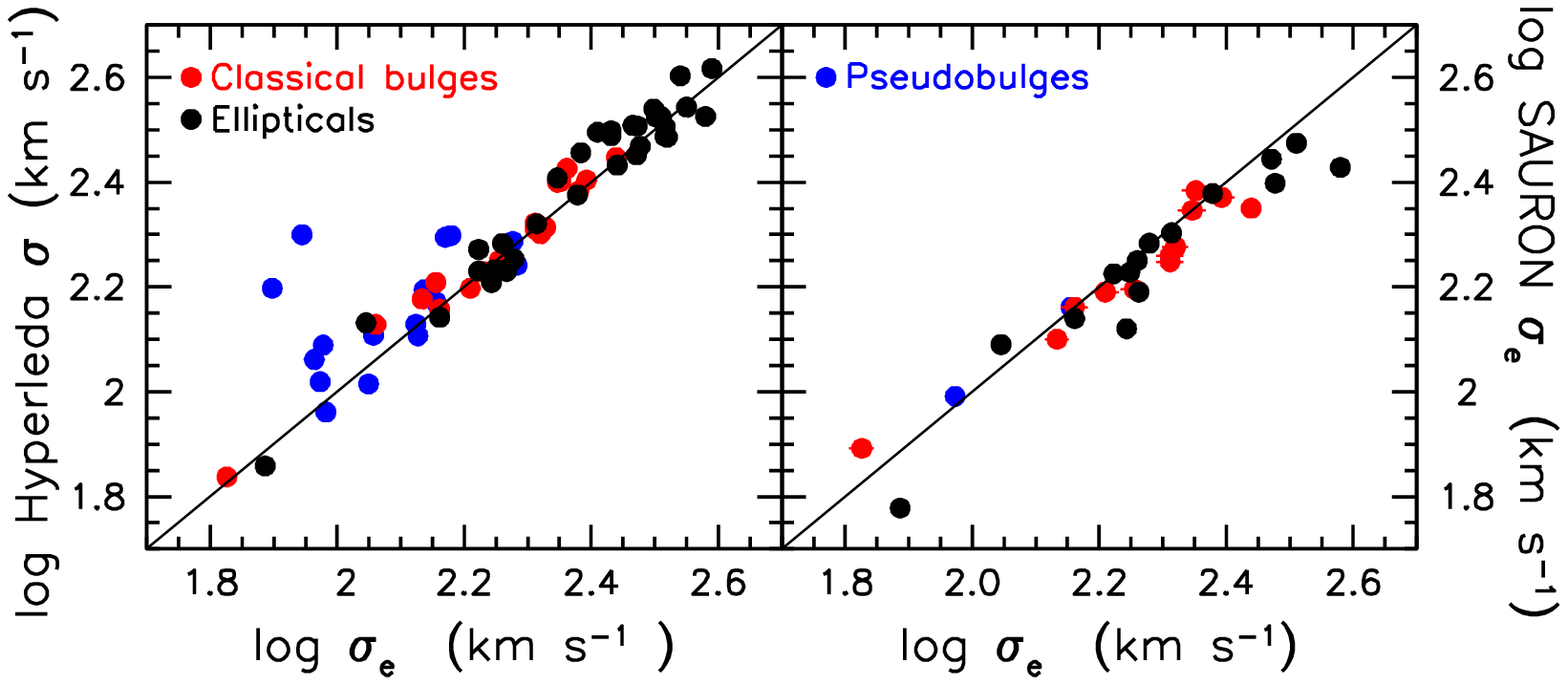}

\ni {\bf \textBlue Figure 11}\textBlack 

\vskip 1pt
\hrule width \hsize
\vskip 2pt

\ni Comparison of our adopted $\sigma_e$ ({\sbf Tables 2} and {\sbf 3}) to ({\it left}) the central velocity dispersion as tabulated in HyperLeda
    and ({\it right\/}) the $\sigma_e$ value tabulated by the SAURON/ATLAS3D teams (Emsellem \etal 2007; Cappellari \etal 2013).  The straight
    lines are not fits; they indicate equality.  Given our definition, it is reassuring that our $\sigma_e$ is approximately the geometric mean
    of HyperLeda central and SAURON/ATLAS3D $\sigma_e$.

\vs
{\bf\ARRed Notes on Individual Elliptical Galaxies:}\textBlack
\vs

M{\ts}32: The BH discovery papers and history of $M_\bullet$ measurements are discussed in \S\ts2.2.1. We adopt $M_\bullet$
          from the triaxial models of van den Bosch \& de Zeeuw (2010).   We calculate $\sigma_e = 77 \pm 3$ km s$^{-1}$ from 
          our photometry ($r_e/2 = 38^{\prime\prime}/2$) and kinematics in Simien \& Prugniel (2002).  The value averaged inside 
          $r_e$ is 1 km s$^{-1}$ smaller.  This is in good agreement with $\sigma_e = 75 \pm 3$ km s$^{-1}$ in 
          Tremaine \etal (2002),
          G\"ultekin \etal (2009c), and
          McConnell \& Ma (2013).
          In contrast, Graham \& Scott (2013) adopt $\sigma_e = 55$ km s$^{-1}$, presumably by not including $V(r)$.

\vsn

NGC 2778, NGC 3608, NGC 4291, NGC 4473, NGC 4649, NGC 4697, and NGC 5845:~The BH discovery is by Gebhardt \etal (2003). \vsn

NGC 1316: This galaxy is bluer than the typical giant E, with $(B - V)_0 = 0.87$ (NED).  Based on our experience with
          published total magnitudes of bright galaxies (e.{\ts}g., KFCB), we adopt $B_T = 9.17$ from the Hyperleda 
          integrated photometry table.  This yields $M_{VT} = -23.38$.  The 2MASS $K = 5.587$ magnitude then implies
          that $(V - K)_0 = 2.64$.  This is implausible for the $(B - V)_0$ color, indicating that the $K$ luminosity
          is underestimated.  We use the $(V - K)_0$ versus $(B - V)_0$ correlation to derive $(V - K)_0 = 2.91$ and 
          hence to correct the $K$ magnitude by $-0.268$ magnitudes to $K = 5.319$.  
          This is a typical correction for a nearby giant elliptical. \vsn

NGC 1332 and NGC 4751: These galaxies initially presented us with an interpretation dilemma.  We believe that it is solved and
          that both galaxies are best interpreted as (rather extreme) ellipticals.  However, realizing this required us to learn
          somthing new about elliptical galaxies.  This note explains our conclusions and summarizes the consequeunces
          of the more canonical alternative that these are S0 galaxies. 

          Both galaxies are highly flattened (E6) and contain prominent, almost-edge-on nuclear dust disks.  It is remarkable how
          many BH host ellipticals contain nuclear dust disks: NGC 1332, NGC 3379 (faintly), NGC 3607, NGC 4261,
          NGC 4374, NGC 4459, NGC 4486A and NGC 5845 (in which much of the nuclear gas disk has formed stars), NGC 4697, NGC 4751, 
          NGC 6251, NGC 6861, NGC 7052, NGC 7768, A1836 BCG, A3565 BCG, and probably IC 1459.  Several more contain nuclear disks
          of stars that plausibly formed out of gas-and-dust disks in the manner illustrated by NGC 4486A (Kormendy \etal 2005) and
          NGC 5845 (Lauer \etal 1995).  Of course, this does not prove that all these objects -- especially NGC 1332 and NGC 4751 --
          are ellipticals; bona fide S0$_3$ galaxies contain nuclear dust disks, too (e.{\ts}g., NGC 5866 in the {\sit Hubble Atlas},
          Sandage 1961).  One reason why dust-lane E and S0 galaxies are preferentially found among BH hosts is that 
          seeing a dust lane motivates authors to measure an emission-line rotation curve.  The prevalence of central gas
          disks in BH hosts is interesting from a BH feeding point of view, but their importance in this note is that they
          tell us that NGC 1332 and NGC 4751 are almost edge-on.
 
          Images of NGC 1332 and NGC 4751 suggest that both galaxies contain two components, a central one that is relatively round
          and that has a steep brightness gradient and an outer one that looks flatter and that has a shallower brightness gradient.
          These are defining features of S0 galaxies.  If the central component is interpreted as a bulge and the outer one as a disk,
          then plausible decompositions are possible and give $B/T = 0.43$ for NGC 1332 (Rusli \etal 2011) and $B/T = 0.55 \pm 0.05$
          for NGC 4751 (Kormendy \& Bender 2013b).  We emphasize:  {\sit If these interpretations are correct, then NGC 1332 closely
          resembles the S0 galaxies NGC 1277 and MGC 4342 in having a bulge that contains an abnormally high-mass~BH (see {\bf Figure 14}
          for illustration).  NGC 4751 is similar but less extreme.  Like NGC 1277 and NGC 4486B, both galaxies then also have high velocity
          dispersions that are well outside the scatter in the Faber-Jackson (1976) correlation between E or bulge luminosity and velocity
          dispersion.  That is, if NGC 1332 and NGC 4751 are S0s, then they are further examples of the high-$M_\bullet$
          deviant galaxies discussed in Section 6.5.}

          However, a compelling argument suggests that these galaxies are extremely flattened extra-light ellipticals:

          KFCB present and review evidence that Virgo cluster ellipticals are naturally divided into two kinds, $M_{VT} < -21.6$ galaxies that 
          have cores and $M_{VT} \geq -21.5$ galaxies that have central extra light above the inward extrapolation of the outer $n \simeq 3 \pm 1$
          S\'ersic-function main body (see \textRed Sections 6.X\textBlack~and 6.9 here).  They suggest that and (e.{\ts}g.)~Hopkins
          \etal (2009b) model how extra-light ellipticals form in wet mergers, such that the main body of the galaxy is the scrambled-up
          remnant of the pre-merger disk and bulge stars and the extra-light component was manufactured by a starburst during the merger.
          In Virgo ellipticals, the fraction of the stellar mass that is in the extra-light component is $\sim$\ts5\ts\% (if just the 
          extra light is counted: KFCB) or as much as a few 10s of percents (if a standard S\'ersic-S\'ersic decomposition is applied:
          Hopkins \etal 2009b).  Both kinds of ellipticals are represented among our BH hosts; NGC 3377 and NGC 4459 are typical extra 
          light ellipticals.

          Huang \etal (2013a) make a similar study of ellipticals in field environments.
          They show that field ellipticals are different from cluster ellipticals in three ways that are relevant here: (1) extra light 
          (their ``two inner components'') makes up a larger fraction $\sim$\ts20\ts\% to 40\ts\% of the galaxies, (2) extra-light 
          ellipticals extend to higher luminosity in the field than in the Virgo cluster, and (3) field ellipticals
          can be as flat as E6 (their Figure\ts1).  In the context of these results, NGC 1332 and NGC 4751 are more plausible interpreted
          as extra-light ellipticals, not S0s.  In fact, even though NGC 1332 is exceedingly close to edge-on, its isophotes are more
          rectangular than those of an edge-on thin disk; this motivated Sandage \& Bedke (1994) to emphasize (their italics) that the
          galaxy contains a ``{\sit thick disk}''.  

          We now believe that there may be almost a continuum in the properties (although not a seamless overlap in numbers) of outer bodies
          of these galaxies from E5 ellipticals with a modest amount of extra light (NGC 3377) to E6 ellipticals that are roughly half extra
          light (NGC 4751 and NGC 6861) to E6 galaxies whose outer parts resemble thickened disks (NGC 1332) to true S0s with thin disks
          (NGC 5866).  Objects like NGC 4751 and NGC 1332 may be rare, and it appears that they are confined to the field, perhaps because 
          this environment favors formation by a small number of gentle mergers that involve progenitors with large gas fractions.  This is 
          the interpretation of NGC 1332 and NGC 4751 that we adopt when we construct BH correlation diagrams and least-squares fits.
          {\sbf Figure 15} illustrates both interpretations for NGC 1332. \vsn

NGC 1399: As discussed in Section 3.1, we adopt the mean $M_\bullet$ measured by Houghton \etal (2006) and by Gebhardt \etal (2007).
          Conservatively, we adopt 1-$\sigma$ errors that span the complete range obtained in both measurements.  Also,
          $\sigma_e = 315$ km s$^{-1}$ is calculated using kinematic data from Graham \etal (1998), intensity-weighting $V^2 + \sigma^2$
          out to $0.5 r_e = 56^{\prime\prime}$ using our photometry.

\eject

NGC 2778 had a BH detection in Gebhardt \etal (2003) but only a $M_\bullet$ upper limit in Schulze \& Gebhardt (2011). However,  
         $M/L_K = 3.3$ is too big for an old stellar population minus dark matter, implying that the $M_\bullet$
          limit is too small.  We illustrate it in {\bf Figure 12} and then omit it. \vsn

NGC 2960:  The BH discovery paper is Henkel \etal (2002); a reliable BH mass was determined by Kuo \etal (2011).
           NGC 2960 has frequently been classified as Sa?~(RC3, UGC, NED), but Kormendy \& Bender (2013b) show that it is a merger in
           progress.  The galaxy is therefore listed here, with the ellipticals. The dispersion is from Greene \etal 2010. \vsn

NGC 3377: The BH was found by Kormendy \etal (1998), whose measurements $M_\bullet = (2.1 \pm 0.9) \times 10^8$ $M_\odot$ and 
          $M/L_V = 2.0 \pm 0.2$ agree well with $M_\bullet = (1.9 \pm 1.0) \times 10^8$ $M_\odot$ and $M/L_V = 2.3 \pm 0.4$ in 
          Schultze \& Gebhardt (2011).  The reasons are (1) that the resolution of the CFHT spectroscopy was very good,
          (2) that the assumption by Kormendy of an isotropic velocity distribution in this low-luminosity, 
          extra-light (Kormendy 1999), and rapidly rotating (Emsellem \etal 2004) elliptical was close enough to the truth, and (3) that 
          this E5 galaxy is essentially guaranteed to be edge-on.  The BH mass in NGC 3377 was also measured in Gebhardt \etal (2003).  
          The 2MASS $K$-band magnitude gives an implausible color of $(V - K)_0 = 2.69$.  We have corrected $K_s$ in {\bf Table 2} 
          to give the mean color for giant ellipticals, $(V - K)_0 = 2.980$.  \vsn

NGC 3379 is a core elliptical with dynamical models that include triaxiality (the only core galaxy that has such models)
         but not dark matter (van den Bosch \& de Zeeuw 2010).  However, the HST FOS spectra (Gebhardt \etal 2000d) 
         resolve the BH sphere of influence with $r_{\rm infl}/\sigma_* \simeq 7.1$.  Schulze \& Gebhardt (2011)
         and Rusli \etal (2013) show that $M_\bullet$ does not require a significant correction for dark matter 
         under these circumstances.   \vsn

NGC 4374 (M{\ts}84) is the fifth-brightest elliptical in the Virgo cluster, but it is only 0.61 mag fainter than NGC\ts4472~(KFCB).  
         Like many radio galaxies, it (3C 272.1) has a nuclear gas and dust disk (Bower \etal 1997), which makes it feasible 
         to search for a BH relatively independently of the stellar mass distribution (Walsh, Barth \& Sarzi 2010) by measuring 
         the emission-line gas rotation curve.  NGC 4374 is the first galaxy in which a BH discovery was made using HST STIS 
         (Bower \etal 1998).  STIS's long-slit capability made it possible to see the prominent zig-zag in the emission lines 
         that is the signature of the BH.  

         However, as discussed in Section 3.2, the line profile is complicated by a two-component structure, and this has led to 
         some~uncertainty~in~$M_\bullet$.  Bower{\ts}et{\ts}al.{\ts}(1998) decompose the line profiles into slowly- and rapidly-rotating components 
         and use the latter to get $M_\bullet$\ts=\ts$1.63(0.99$\null$-$\null2.83)$\times$\kern -0.4pt$10^9$\ts$M_\odot$. 
         Maciejewski \& Binney (2001) suggest that the complicated line profile is caused by integrating the light of
         the nuclear disk inside a spectrograph slit that is broader (0\sd2 wide) than the telescope PSF.  They estimate
         $M_\bullet = 4.4 \times 10^8$ $M_\odot$.  The large difference between these two determinations has been a cause of
         concern, not only for NGC 4374 but also for other $M_\bullet$ determinations based on emission lines.  Recently, 
         Walsh, Barth \& Sarzi (2010) model the observations in much greater detail and -- importantly -- include the effects of the 
         velocity dispersion in the gas.  After application of this ``asymmetric drift'' correction, they get 
         $M_\bullet = 9.25(8.38~-~10.23) \times 10^8$ $M_\odot$.  We adopt their value but note (\S\ts3.2) that there are
         still significant uncertainties in emission-line $M_\bullet$ measurements.  Certainly the folklore that gas rotation curves 
         give $M_\bullet$ easily and without the complications and uncertainties in stellar-dynamical modeling is too optimistic. \vsn

NGC 4382:~The G\"ultekin{\ts}et{\ts}al.{\ts}(2011) $M_\bullet$ limit is based on stellar-dynamical models that do not include dark matter.~For 
          $M_\bullet$\ts$\leq$\ts$1.3^{+5.2}_{-1.2}\ts\times\null10^8$\ts$M_\odot$, $r_{\rm infl}/\sigma_* \simeq 0.3$ is not well resolved.  
          Since this is a core galaxy, we need to ask whether an upward correction to the $M_\bullet$ limit is required.  
          Examination of G\"ultekin's analysis persuades us that no correction is needed: (1) The core is unusually 
          small (break radius $r_b = 81${\ts}pc, Lauer \etal 2005) for the high luminosity of the galaxy.  Our only BH host E that has a smaller 
          $r_b = 54$ pc in Lauer \etal (2005) is NGC 3608, for which Schulze \& Gebhardt (2011) find the same $M_\bullet$ with and without dark matter. 
          (2) The kinematic measurements analyzed by G\"ultekin only reach 0.28\ts$r_e$, i.{\ts}e., the inner part of the galaxy that is most
          dominated by visible matter.  Large $M_\bullet$ corrections in Schulze \etal (2011) happen when the ground-based observations reach larger
          fractions of $r_e$.  In confirmation, (3) G\"ultekin \etal (2011) remark: ``Within $\sim$\ts2 kpc, about the outer 
          extent of our data, [Nagino \& Matsushita 2009, who] study the gravitational potential as revealed by X-ray emission from the
          interstellar medium, \dots~find a constant $B$-band mass-to-light ratio consistent with a potential dominated by stellar mass.''
          We therefore use G\"ultekin's BH mass limit as published.  \vsn

NGC 4459 is the second-brightest extra-light elliptical in the Virgo cluster and the brightest one that has S\'ersic $n < 4$
          ($n = 3.2 \pm 0.3$: KFCB).  Therefore -- as indicated by the normal color, $(V - K)_0 = 2.975$ -- the 2MASS $K_T$ 
          magnitude is accurate.   The galaxy has a prominent nuclear dust disk; in this sense, it closely resembles NGC 1332,
          NGC 4751, and many other BH host ellipticals discussed in the notes to those objects. \vsn

M{\ts}87: The early history of BH searches is reviewed in KR95.  We consider the BH discovery paper to be the HST gas-dynamical 
          study by Harms \etal (1994).~They derived $M_\bullet = (2.7 \pm 0.8) \times 10^9$ $M_\odot$ (all masses are corrected 
          to the SBF distance of 16.68 Mpc:~Blakeslee \etal 2009).   For many years, the definitive mass measurement -- also based 
          on HST gas kinematics -- was $M_\bullet = (3.6 \pm 1.0) \times 10^9$ $M_\odot$ (Macchetto \etal (1997).  Stellar-dynamical 
          measurement of $M_\bullet$ is difficult, because the central brightness profile is shallow inside the break radius $r_b = 5\sd66$ 
          that defines the ``core'' (Lauer \etal 1992, 2007b).  The result (Kormendy 1992a, b) is unfavorably small luminosity weighting 
          inside the BH sphere of influence, $r_{\rm infl} \simeq 3\sd1$.  Even a steep central increase in $\sigma(r)$ is strongly 
          diluted by projection.  Measuring LOSVDs helps if one can detect the resulting high-velocity wings (van der Marel 1994b).  
          M{\ts}87 remains too expensive for HST absorption-line spectroscopy, but high-$S/N$ ground-based spectroscopy is successful.  
          Gebhardt \& Thomas (2009) fit a variety of kinematic measurements, including two-dimensional spectroscopy~from~SAURON 
          (Emsellem \etal 2004) and higher-resolution, long-slit spectroscopy from van der Marel \etal (1994a:~seeing FWHM = 0\sd6; 
          slit width = 1$^{\prime\prime}$).  They for the first time include dark matter in the dynamical models; this is important because 
          the tradeoff in mass between dark and visible matter inevitably decreases the measured stellar mass-to-light ratio at large radii.  
          Analysis machinery is still based on the assumption that $M/L$ is independent of radius, so the consequence is to reduce $M/L$ near
          the center, too.  To maintain a good fit to the kinematics, $M_\bullet$ must be increased.  
          Gebhardt \& Thomas (2009) derive $M/L_V = 10.9 \pm 0.4$ and $M_\bullet = (2.1 \pm 0.6) \times 10^9$ $M_\odot$ without 
          including dark matter and        $M/L_V =  6.8 \pm 0.9$ and $M_\bullet = (6.0 \pm 0.5) \times 10^9$ $M_\odot$ 
          including dark matter.  The change is in the expected sense. 
          Recently, Gebhardt \etal (2011) add Gemini telescope integral-field spectroscopy aided by laser-guided~AO; the resulting PSF 
          has a narrow core (FWHM = 0\sd06) that contributes 14\ts--\ts45\ts\% of the PSF light.
          Such good resolution allows a substantial improvement in the reliability of the BH mass measurement.
          They get and we adopt $M_\bullet = 6.15(5.78~-~6.53) \times 10^9$ $M_\odot$. 

          As in many other galaxies in which stellar- and gas-dynamical $M_\bullet$ measurements can be compared, the stellar-dynamical mass
          is substantially larger.  In M{\ts}87, it is a factor of $1.74 \pm 0.50$ larger than the Macchetto \etal (1997) value.  Such comparisons 
          are discussed further in Section 3.2. \vsn

NGC 4486A: This galaxy has a bright star 2\sd5 from its center that affects most published magnitude measurements.  KFCB measure the total
           $V_T = 12.53$ magnitude without this star, and this provides $M_{VT} = -18.85$.  The 2MASS magnitude appears to include the star, so we do
           not~use~it.  Instead, we adopt the well determined $(V - K)_0 = 2.980$ color for old elliptical galaxies and derive $M_{KT} = -21.83$
           from $M_{VT}$.  For $M_\bullet$, we read 1-$\sigma$ error bars from the $\chi^2$ contour diagram in Figure 6 of Nowak \etal (2007). \vsn

NGC 4486B is one of the lowest-luminosity, normal ellipticals known.  The most accurate photometry (KFCB) gives $V_T = 13.42$ and, with the present 
          distance, $M_{VT} = -17.69$.  M{\ts}32 is about 1 mag fainter.  Again, we adopt $(V - K)_0 = 2.980$ in preference to the 2MASS $K$ magnitude 
          to get $M_{KT} = -20.67$.  The BH mass $M_\bullet \simeq 6.1^{+3.0}_{-2.0} \times 10^8$ $M_\odot$ is the only one based on spherical, 
          isotropic stellar-dynamical models that we use in this paper.  We use it (1) because the central velocity dispersion $\sigma = 291 \pm 25$
          km s$^{-1}$ is much larger than the upper envelope $\sigma \sim 160$ km s$^{-1}$ (for the galaxy's luminosity) of the scatter in the
          Faber-Jackson (1976) correlation.  As in the case of NGC 1277 (q.{\ts}v.),~this is a strong indicator of unusually high masses, even though anisotropic 
          models can formally fit the data without a BH.  However, we have already noted that low-luminosity, coreless ellipticals (e.{\ts}g.,
          M{\ts}32) are not generally very anisotropic.  NGC 4486B is a rapid rotator.~(2) We do not use $M_\bullet$ in
          any correlation fits.  Instead, we include NGC 4486B as the earliest discovery of a compact, early-type galaxy which deviates from
          the $M_\bullet$--host-galaxy correlations in the direction of abnormally high $M_\bullet$.  The most extreme such galaxy is
          NGC 1277 ({\bf Table 3} and {\bf Figure 15}). \vsn

NGC 5077: We use the BH mass that was calculated including emission-line widths in the analysis (De Francesco \etal 2008, see p.~361) and
          adopt the same high-$M_\bullet$ error bar.  The low-$M_\bullet$ error bar is from the analysis that does not include line widths. 

\vfill\eject

NGC 5128: At a distance of 3.62 Mpc, NGC 5128 = Centaurus A is the second-nearest giant elliptical (after Maffei 1 at 2.85 Mpc) and the nearest 
          radio galaxy.  It is therefore very important for the BH search.  It is also a merger-in-progress, although that merger may be 
          relatively minor.  It will prove to be important to our conclusion in Section 6 that mergers-in-progress often have BH masses
          that are small compared to expectations from the $M_\bullet$--host-galaxy correlations.

          The problem is that the many published $M_\bullet$ measurements show disquieting disagreements (Section 3.2).
          Five measurements based on the rotation of a nuclear gas disk are available.  In recent years, as measurements and modeling
          have improved, the mass measurements have converged reasonably well:  We have \par\noindent
               $M_\bullet = 2.07(0.62~-~5.18) \times 10^8$ $M_\odot$ (Marconi \etal 2001)  based on ESO VLT observations without AO; \par\noindent
\nhhj          $M_\bullet = 1.14(0.60~-~1.24) \times 10^8$ $M_\odot$ (Marconi \etal 2006)  based on HST STIS spectroscopy and including
                                                                                           uncertainties from the poorly constrained
                                                                                           inclination of the gas disk in the error bars; \par\noindent
               $M_\bullet = 0.63(0.55~-~0.69) \times 10^8$ $M_\odot$ (H\"aring-Neumayer \etal 2006) based on VLT with AO and including the
                                                                                           gas velocity dispersion in the estimate; \par\noindent
\nhhj          $M_\bullet = 0.85(0.71~-~0.93) \times 10^8$ $M_\odot$ (Krajnovi\'c, Sharp, \& Thatte 2007) based on Gemini telescope spectroscopy without AO;
                                                                                           the Pa$\beta$ rotation curve is consistent
                                                                                           with zero velocity dispersion; and \par\noindent
\nhhj          $M_\bullet = 0.47(0.43~-~0.52) \times 10^8$ $M_\odot$ (Neumayer \etal 2007) based on VLT SINFONI spectroscopy with AO; the gas
                                                                                           dispersion is taken~into~account. 
          The convergence of $M_\bullet$ measurements based on gas dynamics is reassuring.  However, two stellar-dynamical measurements
          agree poorly: \par
\nhhj          $M_\bullet = 2.49(2.28- 2.80) \times 10^8$ $M_\odot$ (Silge \etal 2005) based on Gemini observations without AO, and \par\noindent
\nhhj          $M_\bullet = 0.57(0.47- 0.67) \times 10^8$ $M_\odot$ (Cappellari \etal 2009) based on VLT SINFONI spectroscopy with AO. \par\noindent
          The last set of measurements has the highest resolution and $S/N$, and it agrees with the latest gas-dynamical results.  We adopt 
          this $M_\bullet$. \vsn

NGC 5576: This is a core elliptical (Lauer \etal 2007b) with a BH detection and $M_\bullet$ measurement in G\"ultekin \etal (2009b).
          Halo dark matter was not included in the dynamical models, but the resolution $r_{\rm infl}/\sigma_* = 3.3$ is good
          enough so that we can apply a correction calibrated by Schulze \& Gebhardt (2011) and by Rusli \etal (2013).  The
          mean of the two corrections is a factor of $1.6 \pm 0.3$.  We applied this correction and added the uncertainty 
          in the correction to the uncertainty in G\"ultekin's $M_\bullet$ measurement in quadrature to give the estimated 1-$\sigma$
          error quoted in the table. \vsn

NGC 6861 is classified as an S0$_3$ galaxy in Sandage \& Tammann (1981) because of its prominent nuclear dust disk.
          M\'endez-Abreu \etal (2008) estimate that $B/T \simeq 0.64$.  However, our preliminary photometry (Kormendy \&
          Bender 2013b) shows little or no significant departure from an $n \simeq 2$ S\'ersic-function main body with central extra light.  That is,
          the galaxy is similar to NGC 4459, another extra-light elliptical (KFCB) with a central dust disk that motivated an S0$_3$ 
          classification in Sandage \& Tammann (1981).  There could be a faint disk component in NGC 6861 such as the one in NGC 3115; if so,
          it will make no difference to any conclusions in this paper.  We therefore classify the galaxy as an extra-light elliptical.\vsn

IC 1459: The BH discovery by Verdoes Kleijn \etal (2000) is based on HST WFPC2 photometry, HST FOS spectroscopy through
         six apertures to measure the emission-line rotation curve, and ground-based (CTIO 4 m telescope) spectroscopy with FWHM
         $\sim 1\sd9$ seeing to measure the stellar kinematics.  They get $M_\bullet \sim (4~\rm to~6) \times 10^9$ $M_\odot$ from
         stellar dynamical modeling but $M_\bullet \sim (0.2~\rm to~0.6) \times 10^9$ $M_\odot$ from the HST gas dynamics.
         In contrast, Cappellari \etal (2002) combine HST STIS spectroscopy with high-$S/N$ ground-based spectroscopy (CTIO 4 m 
         telescope; seeing FWHM $\approx 1\sd5$; slit width = 1\sd5; CCD scale = 0\sd73 pixel$^{-1}$) that provide both emission-line
         and absorption-line kinematics.  Again, the agreement between gas and star measurements is not superb: stellar dynamics
         give $M_\bullet = 2.65(2.28~-~3.03) \times 10^9$ $M_\odot$ (which we adopt), whereas gas dynamics give
         $M_\bullet \approx 3.6 \times 10^9$ $M_\odot$.  As noted in \S\ts3.2, the comparison between gas and stellar dynamics
         is not reassuring. 

         At $r \leq 43^{\prime\prime}/2 = r_e/2$, we measure $\sigma_e = 329$ km s$^{-1}$ from the kinematic data in Cappellari \etal (2002)
         and $\sigma_e = 332$ km s$^{-1}$ from the kinematic data in Samurovi\'c \& Danziger (2005).  We adopt $\sigma_e = 331 \pm 5$ km s$^{-1}$.
         The data reach far enough out to measure a value inside $r_e$; this would be only 2.6 km s$^{-1}$ smaller than the value that we adopt.
         Our photometry shows that this is an extra-light elliptical with $n \simeq 3.1^{+0.4}_{-0.3}$.  \vsn

IC 1481: The BH discovery paper is Mamyoda \etal (2009). They detect maser sources distributed along a line indicative of an edge-on 
         molecular disk, and they see a symmetrical rotation curve.  But $V(r) \propto r^{-0.19 \pm 0.04}$ is substantially sub-Keplerian.  
         They conclude that the maser disk is more massive than the BH.  Hur\'e \etal (2011) present an analysis method that is suitable 
         for a wide range of disk-to-BH mass ratios.  They measure $M_\bullet = (1.59 \pm 0.45) \times 10^7$ $M_\odot$ and a maser disk mass
         of about $4.1 \times 10^7$ $M_\odot$ (see \S\ts3.3.3).  We adopt these values.

         The host galaxy is discussed in Kormendy \& Bender (2013b).  SDSS images show loops, shells, and dust lanes characteristic
         of a major merger in progress.  The overall light distribution is that of a normal extra-light elliptical with S\'ersic
         $n = 2.5^{+0.25}_{-0.2}$ (KFCB).  Consistent with this, the central $2^{\prime\prime} \times 2^{\prime\prime}$
         of the galaxy has an A{\ts}--{\ts}F, post-starburst spectrum (Bennert, Schulz \& Henkel 2004).  The ellipticity profile shows 
         that this object is turning into an E1.5 elliptical, so we list it in {\bf Table 2}.  In discussions in the next section, we include 
         it among mergers in progress. 

\vsn
{\bf\ARRed Notes on Disk Galaxies With Classical Bulges:}\textBlack
\vsn

M{\ts}31: The BH mass measurements are discussed in \S\ts2.  We adopt $M_\bullet$ determined
          from P3, the blue cluster part of the triple nucleus (Bender \etal 2005).  We recomputed $\sigma_e = 169 \pm 8$ km s$^{-1}$
          from our photometry and kinematic data in Saglia \etal (2010).  Integrating to $r_e$ or $r_e/2$ gives
          the same result.  Agreement with $\sigma_e = 160 \pm 8$ km s$^{-1}$ in G\"ultekin \etal (2009c) is good. 
          Chemin, Cargnan \& Foster (2009) provide $V_{\rm circ}$.

          Photometry is difficult, because the galaxy is large.  The $V$- and $K$-band magnitudes are discussed in the table notes 
          on {\bf\ARRed Apparent Magnitude Corrections}~\textBlack as an example of the correction of 2MASS magnitudes.  Since our BH 
          correlations are derived in $K$ band, we use an infrared measurement of the bulge-to-disk ratio.  This is the mean
          $B/T = 0.31 \pm 0.01$ of four, $L$-band measurenets by 
          Seigar, Barth \& Bullock (2008),
          Tempel, Tamm \& Tenjes (2010),
          Kormendy \etal (2010), and
          Courteau \etal (2011). 
          We list this value in {\bf Table 3}.  However, $B/T = 0.25 \pm 0.01$ is smaller in $V$ band, and
          we use this smaller value in deriving the $V$-band bulge and disk magnitudes. \vsn

M{\ts}81: The BH discovery and $M_\bullet$\ts=\ts$6\ts(\pm\ts20\ts\%)$\null$\times$\kern -0.6pt$10^7$\ts$M_\odot$ measurement are 
          reported in Bower \etal (2000).  There are two concerns: this result is based on axisymmetric, two-integral stellar-dynamical
          models, and it has never been published in a refereed journal.  Also, Devereux \etal (2003) measure $M_\bullet$
          using HST STIS spectroscopy to get the ionized gas rotation curve; the 
          problems here are that the [N\ts\sc II] emission lines are blended with and had 
          to be extracted from broad H$\alpha$ emission and that the width of the [N\ts\sc II] emission~lines~is~not~discussed.  
          But, whereas the danger is that the emission-line rotation curve will lead us to 
          underestimate $M_\bullet$, Devereux \etal (2003) get $M_\bullet = 7.0(6.0-8.9) \times 10^7$\ts$M_\odot$, 
          larger than Bower's value.  Both measurements are problematic, but they agree.
          Also, Bower's measurement of $M_\bullet$ in NGC 3998 in the same abstract agrees
          with a reliable stellar-dynamical measurement in Walsh \etal (2012).  So we
          adopt the average of the Bower and Devereux $M_\bullet$ measurements.

          Available $B/T$ measurements in the visible and infrared agree within errors.~We
          adopt the mean $B/T = 0.34 \pm 0.02$ (Kormendy \& Bender~2013b). \vsn

NGC 524: Krajnovi\'c \etal (2009) use $\sigma_e = 235$ km s$^{-1}$ from Emsellem \etal (2007), but this is from a 
          luminosity-weighted sum of spectra inside $r_e$.  It is therefore not consistent with the G\"ultekin \etal
          (2009c) definition.  For consistency, we computed $\sigma_e = 247$ km s$^{-1}$ from kinematic data in
          Simien \& Prugniel (2000) and our photometry.  \vsn

NGC 821, NGC 3384, NGC 4564, and NGC 7457: The BH discovery is by Gebhardt \etal (2003). \vsn

NGC 821 is usually considered to be an elliptical galaxy, but the shapes of the isophotes in 
        the image in the Carnegie Atlas of Galaxies (Sandage \& Bedke 1994) suggests that it is an almost-edge-on
        S0.  In fact, Scorza \& Bender (1995) did a bulge-disk decomposition and got $B/T = 0.943$.
        Kormendy \& Bender (2013b) collect $V$-band photometry; they get $V_T = 10.96$ and $B/T = 0.969$.  We adopt
        $B/T = 0.95$ here.  We emphasize that no conclusions depend on $B/T$ or on our reclassification of the galaxy
        as an S0.  The bulge S\'ersic index is $\sim$ 4.9; under these circumstances, it is commonly necessary 
        to correct the 2MASS $K_T$ magnitude slightly.  We determine a correction of $\Delta K_T = -0.185$ and apply 
        it to derive the photometric parameters listed in {\bf Table 3}. \vsn

NGC 1023: The asymptotic outer rotation velocity $V_{\rm circ} = 251 \pm 15$ km s$^{-1}$ is from Column (12) of
          Table 1 in Dressler \& Sandage (1983). \vsn

NGC 1277: We adopt the Perseus cluster distance and NGC 1277 BH mass from van den Bosch \etal (2012).
          However, our analysis of the host galaxy (Kormendy \& Bender 2013b) is different from that of van den Bosch,
          who decompose the light distribution into four radially overlapping components.  This is operationally 
          analogous to a multi-Gaussian expansion in the sense that it forces the S\'ersic indices of all components
          to be small.  Partly for this reason, they concluded that the bulge is not classical.  We find that the
          ellipticity at large radii is similar to the ellipticity near the center; this is a sign also seen in many
          edge-on S0s in the Virgo cluster (Kormendy \& Bender 2012) and indicates that the bulge dominates 
          at both small and large radii.  We decomposed the galaxy into two components such that the bulge
          dominates at both small and large radii.  The decomposition is robust, the bulge has a S\'ersic 
          index of $3.5 \pm 0.7$, and $B/T = 0.55 \pm 0.07$.  Both results imply that the bulge is classical. \vsn

NGC 2549: Krajnovi\'c \etal (2009) find that $M_\bullet = (1.4^{+0.2}_{-1.3}) \times 10^7$ $M_\odot$ for $D = 12.3$ Mpc,
          quoting 3-$\sigma$ errors.  In this case, dividing the 3-$\sigma$ error bars by 3 would obscure the 
          fact that this is an unusually weak BH detection.  We therefore read the 1-$\sigma$ errors directly from the
          $\chi^2$ contours shown in their paper.  The result is approximate, $M_\bullet = 1.45(0.31 - 1.65) \times 10^7$ $M_\odot$
          for our adopted $D = 12.70$ Mpc, but more realistic. \vsn

NGC 3115: Kormendy \& Richstone (1992) discovered the BH and got $M_\bullet$\ts=\ts1.0(0.3$-$3.3)$\times$\kern -0.5pt$10^9$\ts$M_\odot$ 
          from isotropic models and a smallest possible $M_\bullet$ = 1$\times$\null$10^8$\ts$M_\odot$ from the most extreme anisotropic 
          model that fit their CFHT kinematic data.  These are consistent with $M_\bullet$=0.90(0.62$-$0.95)$\times$\kern -0.5pt$10^9$\ts$M_\odot$
          adopted from Emsellem, Dejonghe \& Bacon (1999).  Additional measurements have ranged from 5$\times$\kern -0.5pt$10^8$\ts$M_\odot$ to 
          2$\times$\kern -0.5pt$10^9$\ts$M_\odot$ (Kormendy \etal 1996b; Magorrian \etal 1998). \vsn

NGC 3585: The outer rotation velocity $V_{\rm circ}$ for the embedded disk is from Scorza \& Bender (1995). \vsn

NGC 3998 is listed twice in {\bf Table 3}, once with the BH mass that we adopt from stellar-dynamical models (Walsh
          \etal 2012) and once with the smaller BH mass based on the emission-line rotation curve (De Francesco \etal 2006).
          We illustrate this in {\bf Figure 12} as an example of why we do not use $M_\bullet$ values determined from
          ionized gas rotation curves when line widths are not taken into account.

          We use $B/T = 0.85 \pm 0.02$ from bulge-disk decompositions in Kormendy \& Bender (2013b) and in
          S\'anchez-Portal \etal (2004).  With this $B/T$, NGC 3998 is the most significant bulge outlier to the 
          $M_\bullet$\ts--\ts$M_{K,\rm bulge}$ correlation, as Walsh \etal (2012) concluded.  There is a possibility that 
          a three-component, bulge-lens-disk decomposition is justified; if so, $B/T$ would be smaller, $\sim 0.66$.  
          Then NGC 3998 would be a more significant outlier, in the manner of NGC 4342 and the galaxies discussed in Section 6.5.
          For $\sigma_e$, we adopt the mean of $\sigma_e = 270$ km s$^{-1}$ found by Walsh for $r_e \simeq 18^{\prime\prime}$ 
          from our photometry and $\sigma_e = 280$ km s$^{-1}$ which we find using G\"ultekin's definition, our photometry,
          and kinematic data from Fisher (1997). \vsn

NGC 4526: This is the brightest S0 galaxy in the Virgo cluster and the first galaxy to have a BH discovered using
          the central CO rotation curve (Davis \etal 2013).  We use $B/T = 0.65 \pm 0.05$ from Kormendy \& Bender (2013b) 
          and $\sigma_e = 222 \pm 11$ km s$^{-1}$ from Davis \etal (2013), but we checked that $\sigma_e$ is consistent 
          with our definition of how to average $V^2(r) + \sigma^2(r)$.
          The asymptotic circular velocity is from Pellegrini, Held, \& Ciotti (1997), but it is uncertain whether
          the measured rotation curve reaches far enough out in this and almost any bulge-dominated S0. \vsn

NGC 4258: The spectacular H$_2$O maser disk and consequent accurate BH  mass measurement were discovered by Miyoshi \etal (1995).
          Herrnstein \etal (1999) uses the masers to measure a direct geometric distance $D = 7.2 \pm 0.3$ Mpc to NGC 4258.
          Herrnstein \etal (1999) interprets small departures from precise Keplerian rotation in terms of a warped gas disk
          and derives an improved BH mass.  Our adopted mass is based in large part on this result.  Section 3.3 provides the
          details.  Sources for $V_{\rm circ}$ are listed in Kormendy \etal (2010).

          Given a ``bomb-proof'' accurate BH mass in a conveniently inclined galaxy, NGC 4258 has been used to test both
          stellar-dynamical and ionized-gas-dynamical $M_\bullet$ measurement machinery (Sections 3.1 and 3.2, respectively). \vsn

NGC 4594 = M{\ts}104 = the Sombrero Galaxy: The BH discovery paper was Kormendy (1988b), who obtained
           $M_\bullet = 5.5(1.7~-~17) \times 10^8$ $M_\odot$.  The quoted error bar was conservative, 
           but the best-fitting mass was within 17\ts\% of the present adopted value,
           $M_\bullet = 6.65(6.24~-~7.05) \times 10^8$\ts$M_\odot$ (Jardel \etal 2011).   The BH
           detection was confirmed at HST resolution by Kormendy \etal (1996a), but the mass was
           estimated only by reobserving at HST resolution a set of models that were designed for ground-based data.
           As a result, $M_\bullet \sim 1.1 \times 10^9$ $M_\odot$ was not very accurate.  
           Emsellem \etal (1994) measured $M_\bullet \sim 5.3 \times 10^8$ $M_\odot$ and
           Magorrian \etal (1998) got $M_\bullet = 6.9(6.7~-~7.0) \times 10^8$ $M_\odot$ based
           on two-integral models.   The presently adopted BH mass is based on three-integral models.

           We adopt the total magnitude measurement $B_T = 8.71$ in Burkhead (1986) and correct the
           2MASS $K$ magnitude to give $(V - K)_0 = 2.980$.  Also, $V_{\rm circ}$ is from Faber \etal (1977) 
           and Bajaja \etal (1984). \vsn

NGC 4596: We adopt the $M_{\rm BH,fix}$ mass in Table 2 of Sarzi \etal (2001).  Also, $B/T = 0.27 \pm 0.04$
          comes from comparing Benedict's (1976) decomposition at surface brightnesses \lapprox \ts23.5 B mag
          arcsec$^{-2}$ with the adopted total magnitude $B_T = 11.37$, i.{\ts}e., the mean of values in RC3,
          the Hyperleda main table, and the Hyperleda integrated photometry table.  The rotaton velocity
          corrected for asymmetric drift is from Kent (1990). \vsn

NGC 7457: We confirm G\"ultekin's value of $\sigma_e = 67 \pm 3$ km s$^{-1}$ with our photometry and kinematic measurements. 
          The outer disk circular velocity is from our kinematic data and those of 
          Cherepashchuk \etal 2010, corrected for asymmetric drift by them but for our assumed
          inclination of the galaxy, $i = 59^\circ \pm 2^\circ$.

\vsn
{\bf\ARRed Notes on Disk Galaxies With Pseudobulges:}\textBlack
\vsn

Our Galaxy: Photometric parameters are discussed in Kormendy \& Bender (2013b).  Our Galaxy requires special procedures because 
we live inside it.  For the convenience of readers, we summarize the provenance of the photometric parameters here.  The Galaxy 
has a ``boxy bulge''
(Weiland \etal 1994;
Dwek \etal 1995)
that is generally understood as an almost-end-on bar
(Combes \& Sanders 1981;
Blitz \& Spergel 1991).
It is therefore a pseudobulge -- a component built out of the disk.  There is no photometric or kinematic sign of a classical bulge (see 
Freeman 2008,
Howard \etal 2009, 
Shen \etal 2010, and
Kormendy \etal 2010 
for reviews and for some of the evidence).  We average pseudobulge-to-total luminosity ratios from
Kent, Dame \& Fazio (1991) and
Dwek \etal (1995)
to get $PB/T = 0.19 \pm 0.02$.  To get $M_{KsT} = -23.7$, we adopt the total $K$-band luminosity
$L_K = 6.7 \times 10^{10}$ $L_{K\odot}$ from Kent, Dame \& Fazio (1991) and convert it from their assumed
distance of 8 kpc to our assumed distance of 8.28 kpc from Genzel \etal (2010).
The disk and pseudobulge absolute magnitudes follow from $PB/T$.
Finally, $V$-band magnitudes are derived from $K$-band magnitudes by assuming that $(V - K)_0 = 2.980$.
The bulge absolute magnitude is reasonably accurate, because the bulge is old; the main effect of 
estimating $M_{V,\rm bulge}$ from $M_{K,\rm bulge}$ is to implicitly correct for internal extinction.
The disk magnitude is much more uncertain, because the assumed color does not take young stars into account.
However, this has only minimal effects on our conclusions.

      The adopted BH mass is now securely derived from the orbits of individual stars.
The history of the remarkable improvement in $M_\bullet$ measurements is reviewed in Genzel, Eisenhauer \& Gillessen (2010);
early stages were covered in KR95.  The velocity dispersion $\sigma_e$ is from Tremaine \etal (2002).

\vsn

Circinus is like M{\ts}31 in structure and inclination, but it is a smaller galaxy with a gas-rich
pseudobulge, and it has a smaller BH than M{\ts}31.  It is a difficult case, because it is close to 
the Galactic plane.  The Galactic absorption is large, and our estimates of it are uncertain.  Kormendy \& Bender (2013b)
measure the galaxy's photometric parameters; the total apparent magnitude is $K_T = 4.71$.  The pseudobulge classification and 
$PB/T = 0.30 \pm 0.03$ are from the same paper and from Fisher \& Drory (2010).  We adopt $V_T = 10.60 \pm 0.04$ 
as the average of values tabulated in the RC3 (de~Vaucouleurs \etal 1991) and Hyperleda (Paturel \etal 2003).  
Comparing $K_T$ and $V_T$ in the context of various published estimates of the Galactic absorption,  we adopt $A_V = 3.15$ 
from Karachentsev \etal (2004), because it gives the most reasonable total color for the galaxy, $(V - K)_0 = 3.05$.  This 
then determines the other photometric parameters, including the distance $D = 2.82$ Mpc (Karachentsev \etal 2004).  
 
      Greenhill \etal (2003) measure masers both in outflowing gas and in a
well-defined, essentially edge-on accretion disk.  The latter masers show a well-defined Keplerian rotation
curve which implies that $M_\bullet = (1.14 \pm 0.20) \times 10^6$ $M_\odot$. We adopt this value, although
Hur\'e \etal (2011) find hints that $M_\bullet$ may be smaller.  We are uncomfortable about the conflicting published
velocity dispersion measurements: Oliva \etal (1995) measure 168 km s$^{-1}$ consistently (RMS = 10 km s$^{-1}$)
from four infrared CO bands; their instrumental resolutions ($\sigma_{\rm instr} \simeq 80$ and 51 km s$^{-1}$
should be sufficient.  But Maiolino \etal (1998) measure $\sigma \simeq 79 \pm 3$ km s$^{-1}$ at
$\sigma_{\rm instr} \simeq 64$ km s$^{-1}$ in the 2.3\ts--\ts2.4 $\mu$m CO bands using an integral-field 
spectrograph and AO; there is little gradient in the central 1\sd2 except that the nucleus has a
bulge-subtracted velocity dispersion of $\sigma = 55 \pm 15$ km s$^{-1}$.  More recently, M\"uller-S\'anchez 
\etal (2006) use SINFONI AO integral-field specrtoscopy on the VLT to measure $\sigma \simeq 80$
km s$^{-1}$ in the central 0\sd4 $\times$ 0\sd4.  We adopt $\sigma_e = 79 \pm 3$ km s$^{-1}$.  \vsn

NGC 1068 is a prototypical oval galaxy (Kormendy \& Kennicutt 2004) with an unusually massive pseudobulge that is more than
         a magnitude more luminous and a factor of $\sim$\ts4 more massive than the classical bulge of M{\ts}31.
         Kormendy \& Bender (2013b) find that the pseudobulge-to-total luminosity ratio is quite different in 
         the optical and infrared; $PB/T \simeq 0.41$ at $H$ but $\simeq 0.3$ at $r$ and $i$.  We use these values at $K$ and
         $V$, respectively.  We adopt $\sigma_e = 151 \pm 7$ km s$^{-1}$ from G\"ultekin \etal (2009c) and $V_{\rm circ} = 283 \pm 9$
         km s$^{-1}$ from Hyperleda but note that the latter value is uncertain.  We know of no two-dimensional analysis of the outer
         velocity field that takes the two differently oriented nested ovals into account; for a galaxy that is close to edge-on,
         this is very important.
 
         The BH discovery papers are Gallimore \etal (1996) and Greenhill \etal (1996) who found and
         measured positionally resolved H$_2$O maser emission with the VLA and with VLBA, respectively.  The
         case is not as clean as that in NGC 4258, because the rotation velocity in the non-systemic-velocity
         sources decreases with increasing radius more slowly than a Keplerian, $V(r) \propto r^{-0.31 \pm 0.02}$
         (Greenhill \etal 1996).  The simplest and most plausible explanation is that the mass of the masing
         disk is not negligible with respect to the BH.  Ignoring this, the above papers derive a first approximation
         to $M_\bullet$ of $1 \times 10^7$ $M_\odot$.  Greenhill \& Gwinn (1997) report additional VLBI observations
         and refine the total mass to $1.54 \times 10^7$ $M_\odot$.  Lodato \& Bertin (2003) confirm this: they get $M_\bullet =
         (1.60 \pm 0.02) \times 10^7$ $M_\odot$ using the approximation of a Keplerian rotation curve. 
         However, both Lodato \& Bertin (2003) and Hur\'e (2002; see also Hur\'e \etal 2011) derive models that account for the disk mass,
         and we adopt the average of their results, $M_\bullet = (8.39 \pm 0.44) \times 10^6$ $M_\odot$ (Section 3.3.3).

         Note again the extreme misalignment of the maser disk, which is essentially edge-on, and the rest of the
         galaxy, which is $\sim$\ts21$^\circ$ from face-on. \vsn

NGC 1300: We adopt $D$ (Local Group) = 21.5 Mpc, consistent with the distances to neighbors NGC 1297 and NGC 1232, all members
          of grouping 51 $-7$ $+4$ (Tully 1988).~However, Tonry \etal (2001) find $D$\ts=\ts28.5{\ts}Mpc for NGC\ts1297.
          We cannot tell whether there is a problem~with~one of the distances or whether NGC 1297 is fortuitously
          close to NGC\ts1300 in the sky but half of the distance from us to the Virgo cluster behind it.  This is one example of
          a general problem: Distances remain uncertain, and we do not fold these uncertainties into our error estimates.

          The effective radius of the pseudobulge is $r_e \simeq 4\sd5 \pm 0\sd1$ (Fisher \& Drory 2008; Weinzirl \etal 2009).
          For $\sigma_e$, we use the mean dispersion $88 \pm 3$ km s$^{-1}$ interior to 3\sd5 as shown in Figure 6 of
          Batcheldor \etal (2005). 
          \vsn

NGC 2273: We calculated $\sigma_e = 125 \pm 9$ km s$^{-1}$ from our photometry and from kinematic data in Barbosa \etal (2006).
          Also, $V_{\rm max} = 196 \pm 5$ km s$^{-1}$ is from H{\ts}I data in Noordermeer \etal (2007). \vsn

NGC 2787 is our only explicit example of a phenomenon that must be moderately common -- a galaxy that contains both a classical
         and a pseudo bulge.~Erwin \etal (2003) make a decomposition with $B/T = 0.11$ and $PB/T = 0.26$.  We adopt this
         decomposition to make the~above~point.  However, at our present level of understanding, trying to separate bulges from 
         pseudobulges is risky.  In other galaxies, we identify the dominant component and assign all of the (pseudo)bulge
         light to it.  Here, too, Columns 6 and 11 list the magnitudes of the bulge and pseudobulge together, and we treat
         this as a pseudobulge galaxy.  The outer rotation velocity $V_{\rm circ}$ is from Shostak (1987) and from van Driel
         \& van Woerden (1991). \vsn

NGC 3227: We use the corrected SBF $D = 23.75$ Mpc for companion galaxy NGC\ts3226 (Tonry \etal 2001). Mundell \etal (1995) provide $V_{\rm circ}$.\vsn

NGC 3368 is a pseudobulge-dominated S(oval)ab spiral galaxy with a central decrease in $\sigma$ at $r < 1^{\prime\prime}$ (Nowak \etal 2010).
          As emphasized by these authors, different definitions give different values of $\sigma_e$ and this affects whether or not the BH
          falls within the scatter of the $M_\bullet$\ts--\ts$\sigma_e$ relation.  Luminosity-weighted within the VLT SINFONI field of view
          of $3^{\prime\prime} \times 3^{\prime\prime}$, $\sigma = 98.5$ km s$^{-1}$ and the BH is consistent with $M_\bullet$\ts--\ts$\sigma_e$.
          However, we use the definition that $\sigma_e$ is the luminosity-weighted mean of $V^2 + \sigma^2$ within approximately $r_e$ (the
          exact radius makes little difference).  Because rotation contributes and because $r_e \simeq 11\sd2 \pm 2\sd7$ for the pseudobulge,
          we get a substantially larger value of $\sigma_e \simeq 125 \pm 6$ km s$^{-1}$.  This is based on our photometry and on kinematic
          data in H\'eraudeau \etal (1999) corrected inside $r_e/2$ to agree with Nowak \etal (2010).  Sarzi \etal (2002) derived
          $\sigma_e = 114 \pm 8$ km s$^{-1}$ using dispersions only; this shows approximately how much difference rotation makes to the
          definition.  Many pseudobulge galaxies are similar in that central velocity dispersions are much smaller than the $\sigma_e$
          that is obtained from the $V^2 + \sigma^2$ definition. 

          Nowak \etal (2010) suggest that NGC 3368 contains a small classical bulge in addition to the dominant pseudobulge.  We add them together.\vsn

NGC 3384: G\"ultekin \etal (2009c) used $\sigma_e = 143 \pm 7$ km s$^{-1}$.  We essentially confirm this: With our photometry and kinematic
          measurements, we get $\sigma_e = 150 \pm 8$ km s$^{-1}$.  We adopt the mean. \vsn

NGC 3393 is another prototypical oval galaxy with a large pseudobulge.  {\bf Figure 10} is included to emphasize its similarity to NGC 1068: 
         It is only $\sim$\ts$13^\circ$ from face-on (Cooke \etal 2000), but it contains an edge-on, masing accretion disk.
         
         All measurements of this galaxy are somewhat uncertain.  Kormendy \& Bender (2013b) find a preliminary $PB/T = 0.27 \pm 0.06$.
         The BH mass measurement is based on the rotation curve of a masing molecular disk (Kondratko, Greenhill \& Moran 2008).  
         The maser sources are well distributed along a line indicative of an edge-on disk, but they cover only a small radius range, 
         so they do not securely measure the rotation curve shape.  They are consistent with Keplerian; this gives an enclosed~mass of 
         $M_\bullet = (3.55 \pm 0.23) \times 10^7$ $M_\odot$ at $r \leq 0.41 \pm 0.02$ pc.  But there are signs that the rotation curve
         is slightly flatter than Keplerian.  For their best-fitting sub-Keplerian rotation curve, Kondratko \etal (2008) get
         $M_\bullet \simeq 2.97 \times 10^7$ $M_\odot$.  In contrast, Hur\'e \etal (2011) find a good solution with a maser disk that 
         is 6 times as massive as the BH.  Then $M_\bullet \simeq 0.67 \times 10^7$ $M_\odot$.  This is the only galaxy in our sample 
         in which two such analyses give substantially different results.  We adopt the mean of the two masses and half of the difference
         as our error estimate.

         The velocity dispersion $\sigma_e$ is securely measured by Greene \etal (2010), but $V_{\rm circ}$ is too
         uncertain, because the galaxy is too close to face-on. \vsn

NGC 3489 is a weakly barred S0 with a dominant (pseudo)bulge that contributes $\sim$\ts35\ts\% of the light of the galaxy (Nowak \etal 2010).
         These authors argue plausibly that about one-third of this component is a classical bulge.  We conservatively add them together.  
         Also, we derive $\sigma_e = 113 \pm 4$ km s$^{-1}$ from our photometry and kinematic data in McDermid \etal (2006). \vsn

NGC 4388: We have only central velocity dispersion data for this galaxy.  Greene \etal (2010) measure $\sigma = 107 \pm 7$ km s$^{-1}$;
          Ho \etal (2009) get $91.7 \pm 9.5$ km s$^{-1}$, and we adopt the average, $\sigma_e = 99 \pm 10$ km s$^{-1}$.  This is likely
          to be an underestimate of $\sigma_e$ as we define it, because it neglects rotation inside the half-light radius $r_e \simeq 3\sd0$
          of the pseuodbulge. \vsn

NGC 4736 and NGC 4826: We are most grateful to Karl Gebhardt for making $M_\bullet$ available before publication (Gebhardt \etal 2013).
         For NGC 4736, we calculated $\sigma_e$ from our photometry (Kormendy \& Bender 2013b) and kinematic data in M\"ollenhoff \etal (1995).
         For NGC 4826, $\sigma_e$ is from our photometry and kinematic data in Rix \etal (1995).  For both galaxies, the result 
         is not significantly different if we integrate inside the pseudobulge $r_e \simeq 9\sd7$ and $16\sd7$, respectively (Fisher \& Drory 2008)
         or inside $r_e/2$.  Sources for $V_{\rm max}$ are given in Kormendy \etal (2010). \vsn

NGC 4945 is an edge-on, dusty Scd with a small pseudobulge ($PB/T = 0.07$) that is heavily absorbed at optical wavelengths.  
         We use $D = 3.58$ Mpc, i.{\ts}e., the mean of two ``TRGB'' distances determined from the magnitude of the tip of the red giant
         branch in the stellar color-magnitude diagram (3.36 Mpc: Mouhcine \etal 2005 and 3.80 Mpc: Mould \& Sakai 2008).
         We adopt $K_T = 4.438$, i.{\ts}e., the integral of the surface brightness and ellipticity profiles measured in Kormendy \& Bender (2013b). 
         For comparison, 2MASS lists $K_s = 4.483$.  Our total magnitude implies a slightly more plausible color $(V - K)_0 = 2.712$.  

         The BH detection in Greenhill, Moran \& Herrnstein (1997) was rejected by G\"ultekin \etal (2009c) because the maser 
         rotation curve is asymmetric and because the maser disk inclination is only approximately constrained to be edge-on by its 
         linear distribution at PA $\approx 45^\circ$.  Still, the assumption that the disk is edge-on is at least as secure as
         many other assumptions that we routinely make.  And the rotation curve decreases cleanly with radius on one side of the center. 
         Thus $M_\bullet$ is not more uncertain than the most problematic cases based on stellar and ionized gas dynamics. 

         The maser disk and the galaxy disk have similar PA, are similarly edge-on, and rotate in the same direction (Greenhill \etal 1997). \vsn

NGC 6264 and NGC 6323 are the most distant disk galaxies with maser BH detections.  Kormendy \& Bender (2013b)
      measure $r$- and $K$-band brightness profiles, respectively.  No HST imaging is available for either galaxy, 
      although CFHT images with PSF dispersion radii of $\sigma_* = 0\sd23$ are available for NGC 6323.  Both galaxies
      have small pseudobulges; NGC 6264 has $PB/T \simeq 0.17 \pm 0.03$ and NGC 6323 has $PB/T \simeq 0.05 \pm 0.01$.    
      For both galaxies, we adopt $\sigma_e$ equal to the central velocity dispersion measured by Greene \etal (2010). \vsn

IC 2560: Evidence for a BH based on the dynamics of a H$_2$O maser disk was reported in Ishihara \etal (2001) and
         refined with further observations in Yamauchi \etal (2012).  We adopt $M_\bullet$ and its upper error bar
         from the latter paper.  However, only one point in the rotation curve is observed from high-$|V|$ masers,
         so we cannot tell whether the rotation curve is Keplerian.  Centripetal acceleration of the systemic masers
         is accurately measured, but their velocity gradient with position along the major axis of the disk
         is not accurately enough known to give a second meaningful ($V$, $r$) point as discussed in Section  3.3.2.
         Since some maser disks have gas masses that are a significant fraction of the BH mass, we must regard the
         BH mass determination for IC 2560 as an upper limit.  Nevertheless, $M_\bullet$ is clearly small enough   
         to support our conclusion that BHs do not correlate with pseudobulges in the same way as they do with
         classical bulges. 

         We have only a central velocity dispersion measurement (Greene \etal 2010).  The outer rotation velocity
         $V_{\rm circ}$ is from Hyperleda. \vsn

UGC 3789: We adopt the geometric distance $D = 49.9 \pm 7.0$ Mpc from Braatz \etal (2010).  It provides a
          check of our D (Local Group) distances ({\bf Table 2}, Column 3, source 9), which would have been 47.8 Mpc. 
          The $K$ magnitude is from 2MASS, but the $V$ magnitude is estimated from $K$ using a photoelectric measurement of
          $B - V = 0.92$ in a 50$^{\prime\prime}$ diameter aperture by Arkhipova \& Saveleva (1984).  The correlation
          of $(V - K)_0$ with $(B - V)_0$ then gives $(V - K)_0 \simeq 2.90$.

          UGC 3789 is an another example of a surprisingly common phenomenon for which we know no explanation: It is a 
          prototpical, almost-face-on oval galaxy with an edge-on molecular disk surrounding the BH ({\bf Figure 10}).
          The velocity dispersion is from Greene \etal (2010) and $V_{\rm circ}$ is from Hyperleda. \vsn

\vs
{\bf\ARRed Notes on Discarded Galaxies {\ARDiscard (Cyan Lines in Tables 2 and 3})\ARRed:}\textBlack
\vs

NGC 2778 is discarded (1) because it provides only an upper limit on $M_\bullet$ (Schulze \& Gebhardt 2011)
         and (2) because the implied mass-to-light ratio $M/L_K = 3.3$ is too large for an old stellar population
         from which the dark matter has been subtracted and modeled separately (Section 6.6).  We conclude that
         dark matter is still included and therefore that the $M_\bullet$ upper limit is substantially too small.
         This galaxy is a good illustration of the importance of adding $M/L$ constraints to our mass measurements,
         something that has not heretofore been done \vsn

      NGC 3607 is a core elliptical with a BH detection and $M_\bullet$ measurement in G\"ultekin \etal (2009b).
The modeling did not include dark matter.  Schulze \& Gebhardt (2011) and Rusli \etal (2013) show
that this is not a large problem if $r_{\rm infl}$ is very well resolved, but they show that $M_\bullet$ is
systematically underestimated if $r_{\rm infl}/\sigma_*$ \lapprox \ts5.  Both papers provide calibrations of
how the $M_\bullet$ correction factor depends on  $r_{\rm infl}/\sigma_*$, but the two calibrations disagree
even though they both use variants of the Nuker code.  The disagreement may result from technical details such 
as the number of orbits used in the modeling.  But the problem is severe for NGC 3607, for which the apparent
value of $r_{\rm infl}/\sigma_* \simeq 1.5$.  Based on the Schulze calibration, we should multiply the
G\"ultekin $M_\bullet = 1.4 \times 10^8$ $M_\odot$ ({\bf Figure 12}) by a factor of $\sim$2.~The Rusli calibration
gives a factor~of~$\sim$4.7.  We~conclude that the BH detection is reliable but that we do not
know the BH mass well enough to retain NGC 3607 in our sample.  \vsn

NGC 4261, NGC 6251, NGC 7052, A1836 BCG, and A3565 BCG: All of these galaxies have valid BH detections based on
          optical emission-line kinematic observations in the papers listed in Column 12.  However, the widths
          of the emission lines are comparable to the rotation velocities near the center, and these widths were
          not taken into account in estimating $M_\bullet$.  As discussed in the BH discovery papers and in Section 3.2
          here, it is not guaranteed that line widths imply a ``contribution'' to $M_\bullet$ as they would for
          absorption-line velocity dispersions.  But it is likely that they are not ignorable.  In Section 6.2, we
          compare these BH masses to the BH--host-galaxy correlations that we derive for the most reliable BH masses.
          We find that the above galaxies do indeed have anomalously small BH masses.  The conservative conclusion 
          therefore is that neglecting emission-line widths can result in underestimated BH masses.  We therefore 
          omit all such masses, even those for lower-luminosity bulges in which the measured $M_\bullet$ does not
          obviously deviate from the correlations. \vsn

Cygnus A: The BH discovery (Tadhunter \etal 2003) is based on the optical emission-line rotation curve measured with HST STIS.
          The authors note that the emission lines are very broad, but they do not include line widths in their $M_\bullet$
          determination.  Therefore the BH mass is probably underestimated.
          The velocity dispersion is from Thornton \etal (1999) and is very uncertain; it is based on the observed width of the
          Ca infrared triplet absorption lines but not on any of the standard methods of comparing the spectrum to a standard star.
          Finally, although we used the brightest $V$-band magnitude in the literature (from Paturel \etal 2000 as listed by Hyperleda),
          the $(V - K)_0 = 3.54$ color is implausibly large.  It may be affected by large internal absorption.  If so, the $K$-band 
          magnitude may be usable.  Nevertheless, for all of the above reasons, this galaxy is plotted in Section 6.2 and thereafter
          is omitted from all correlations and fits.

\vfill\eject


\hsize=15.0truecm  \hoffset=0.0truecm  \vsize=20.1truecm  \voffset=1.5truecm

\tenpoint
\tenrm
\dblbaselines

      Photometric decomposition of brightness distributions into (pseudo)bulge and disk components is carried out in the same way 
for classical and pseudo bulges.  The procedure (Kormendy 1977) begins by assuming a functional form for bulge brightness profiles 
that is motivated by pure-bulge galaxies (ellipticals) and a functional form for disk brightness profiles that is 
motivated by pure-disk galaxies.  Here, we fit bulges with S\'ersic (1968) functions, $\log {I(r)} \propto r^{1/n}$.  Many studies
have shown that the S\'ersic index $n$ is generally $> 4$ for giant ellipticals, $\sim 2$ to 3 for smaller ellipticals, 
\gapprox \ts2 for classical bulges, and \lapprox \ts2 for pseudobulges (see KFCB; Kormendy\ts2012 for reviews).  Disks are usually 
exponential, $I(r) \propto e^{-r/r_0}$, where $r_0$ is the scale length (e.{\ts}g., Freeman 1970; see van der Kruit \& Freeman 2010 
for a review).  We then decompose the observed, bulge\ts$+${\ts}disk profile into two components with these analytic forms that 
add up to the observed profile.   Sometimes a third component (usually a bar or a lens) is needed; such components are included in
disk light.

      Fifteen of the galaxies in {\bf Table 3} have $(P)B/T$ measurements from 2 -- 7 independent sources.  The rest have only
one reliable source; most of them are measured in Kormendy \& Bender\ts(2013b).  Composite brightness profiles are derived there
from our own images and from images available through online archives.  The HST archive is particularly important, because 
we need as much spatial resolution as possible to provide leverage on bulges that dominate over disks within only a small radius range 
near the center.  However, no $(P)B/T$ ratio is seriously compromised by the lack of HST data.  Various decomposition 
algorithms are used in published papers; many are carried out on two-dimensional images.  Kormendy \& Bender (2013b) construct
S\'ersic-exponential decompositions of major-axis profiles but take ellipticity profiles fully into account in calculating $(P)B/T$. 
This is more reliable than two-dimensional decomposition when, in constructing the latter, it is assumed that the flattening of each 
component is constant~with~radius.  This assumption is often wrong, especially for bulges. Then two-dimensional algorithms must resort 
to decompositions into many components that are not physically meaningful in order to fit the isophote ellipticities.
When we can, we average $(P)B/T$ from multiple sources; this helps to estimate uncertainties.  A galaxy can contain both a classical bulge 
and a pseudobulge (Section 4), so both $B/T$ and $PB/T$ can be $> 0$.  The resulting $B/T$ and $PB/T$ values are listed in {\bf Table 3}.

\vs
\ni {\big\ARRed 5.1 Comparison of Our Sample With Recent BH Studies} \textBlack
\vsss

      Our sample has 45 ellipticals, 20 classical bulges, and 22 pseudobulges.~Of these, we omit seven ellipticals
and one classical bulge from correlation fits, because $M_\bullet$ was determined using ionized gas kinematics without taking
broad emission-line widths into account.  We regard these as valid BH discoveries, but $M_\bullet$ may be underestimated, at least
for the bigger galaxies (Section 6.3). These galaxies are identified {\y with cyan lines in tables and with cyan symbols
in plots}\textBlack.
NGC\ts4382 and NGC\ts3945 have only $M_\bullet$ upper limits, but these will be important to the conclusions.  Almost all core ellipticals
with $M_\bullet$ measurements based on stellar dynamics have halo dark matter included in the models;~when~not, and when a correction 
to $M_\bullet$ is required, this is discussed in the Supplementary Information table notes on {\bf\ARRed Corrections to stellar-dynamical
BH masses for core galaxies}\textBlack\ and in notes to individual objects.

      It is useful to compare this study with three recent studies of BH--host-galaxy correlations:

      G\"ultekin \etal (2009c) had 25 ellipticals, 14 classical bulges, and 10 pseudobulges.~However,~they did not have $(P)B/T$ ratios
for 17 galaxies, so the $M_\bullet$\ts--\ts$M_{B,\rm bulge}$ sample is smaller than the $M_\bullet$\ts--\ts$\sigma$
sample.  G\"ultekin did not have enough information to conclude that $M_\bullet$ values based on emission lines that did not model 
line widths should be omitted; at that time, these galaxies still included M{\ts}87.  The stellar-dynamical models used did not include 
dark matter.  

      McConnell \& Ma (2013) have 36 elliptical galaxies and 34 bulges.  Classical and pseudo bulges are not distinguished,
although early- and late-type galaxies are distinguished.  This is almost but not quite the same thing.  Here, we argue that the
bulge-pseudobulge distinction is crucial, because classical bulges show $M_\bullet$--host-galaxy correlations whereas pseudobulges 
largely do not.  Almost all of the $M_\bullet$ determinations based on stellar dynamics do include dark matter in the models.  
Galaxies with $M_\bullet$ based on emission-line rotation curves for which line widths were ignored are not discarded from fits.
This paper is an extension of G\"ultekin \etal  (2009c) mostly in the classical tradition 
of exploring a single set of $M_\bullet$--host-parameter correlations but with clear signs of some of the present results (especially
the ``saturation'' of $M_\bullet$\ts--\ts$\sigma$ at high $M_\bullet$).

      Graham \& Scott (2013) have 30 ellipticals and ostensibly 48 disk galaxies, but 7 ``disk galaxies'' (M{\ts}32, NGC 1316, 
NGC 2960, NGC 3607, NGC 4459, NGC 5128, and Cygnus A) are ellipticals.  Graham and Scott do not distinguish between classical bulges and 
pseudobulges; instead, they compare barred and unbarred galaxies.  But many unbarred galaxies contain pseudobulges, and many barred galaxies 
contain classical bulges.  Some stellar-dynamical BH masses include dark matter in the models (the McConnell \etal 2011a,{\ts}b, 2012 galaxies), 
but many do not (the Gebhardt \etal 2003 galaxies) even though masses that include dark matter in the models are published (Schulze \& Gebhardt 2011).  
The sample also contains a number of questionable $M_\bullet$ values.   For example, NGC 2778 now has just an upper limit on $M_\bullet$
(Schulze \& Gebhardt 2011) not a BH detection (Gebhardt \etal 2003).  Also, Graham and Scott include seven BH masses read off of an unlabeled 
plot in Cappellari \etal (2008).  Michele Cappellari (private communication) recommends that we not use these preliminary values. 
We follow his suggestion. Graham also retains $M_\bullet$ values that were derived from emission-line rotation curves for which line widths were ignored.  

      Comparisons with results in the above papers are included in Section 6.

\vsss\vsss
\ni {\big\ARRed 6. BH CORRELATIONS -- OBSERVATIONS}\textBlack
\vsss

      Based on {\bf Tables\ts2} and {\bf 3}, this section rederives and brings up to date the correlations between BH masses and host galaxy components. 
In doing so, we review the two phases of BH demographic studies that we differentiated in Section\ts1.2. 

      {\bf\ARRed Sections 6.1 and 6.2}\textBlack~review the first phase, i.{\ts}e., the pre-HST and HST-era derivations of the correlations 
between $M_\bullet$ and bulge luminosity or mass and velocity dispersion, respectively.  These derivations mostly did not differentiate between
ellipticals and the bulges of disk galaxies or between bulges of different types.  A single set of correlations 
\hbox{($M_\bullet$\ts--\ts$L_{\rm bulge}$, $M_\bullet$\ts--\ts$M_{\rm bulge}$, and $M_\bullet$\ts--\ts$\sigma$)}
was found for all hot stellar parts of galaxies.  This period was a decade-long plateau~in~our
demographic picture in which the focus was on debates about correlation slopes, zeropoints, and scatter, based on different choices of analysis 
machinery, on different selection of objects (e.{\ts}g., inactive galaxies versus AGNs), and on different views of what constitutes a reliable BH sample.
Theoretical work on BH--host-galaxy coevolution focused on using a single parameter -- the gravitational potential well depth as measured by
bulge velocity dispersion -- to control the effects of AGN feedback.

      Now, we have moved beyond this plateau in our understanding as it has become possible to distinguish different correlations for different 
kinds of host-galaxy components.  The rest of Section\ts6 updates the BH correlations using our Section 5 sample, and in doing so, it reviews and
further develops this new phase in BH demographic studies.  Section 6 is long, so a road map is useful:

      {\bf\ARRed Section 6.3}\textBlack~is a ``lemma'' which discusses why we omit from correlation studies those BH masses that were derived
from optical, emission-line rotation curves without taking (usually large) emission line widths into account.  In much the same way that
stellar dynamics would cause us to underestimate $M_\bullet$ if we ignored large velocity dispersions, we now have clear signs that these
emission-line-based measurements underestimate $M_\bullet$.  Section 6.3 also identifies two other galaxies for which we believe that $M_\bullet$ 
has been underestimated.  

      {\bf\ARRed Section 6.4}\textBlack~presents a new result.  We and Bender \etal (2013) find (essentially independently) that major mergers in progress
deviate from the $M_\bullet$--host-galaxy correlations in having unusually small BHs for their galaxy luminosities, stellar masses, and velocity dispersions.

      {\bf\ARRed Section 6.5}\textBlack~discusses a small number of galaxies that are known to deviate in the opposite sense:~They have unusually
large BH masses, outside the scatter shown by BHs in other galaxies.  We discuss what these galaxies may be telling us.  Both kinds of 
deviators{\ts}--{\ts}large-$M_\bullet$ monsters and small-$M_\bullet$ BHs in mergers{\ts}--{\ts}are omitted from derivations of BH parameter correlations.

      {\bf\ARRed Section 6.6}\textBlack~derives, for ellipticals and for classical (elliptical-galaxy-like) bulges only, updated
correlations between $M_\bullet$ and host luminosity, stellar mass, and velocity dispersion.  We revise the derivation of the ratio of BH mass
to bulge mass.  The new ratios are substantially larger than ones in past papers, and the ratio increases slightly with increasing bulge mass.

      {\bf\ARRed Section 6.7}\textBlack~introduces the observed distinction between two kinds of elliptical galaxies that, we suggest, are remnants 
of major, dissipational (``wet'') and dissipationless (``dry'') mergers.  With our new and larger sample, we confirm the strong hints seen by 
applying~the Faber-Jackson (1976) $M_{\rm bulge} \propto \sigma^{(\sim 4)}$ relation (Lauer \etal 2007a) and more directly by recent papers on 
the biggest~BHs (McConnell \etal 2011a, b; 2012) that the $M_\bullet$\ts--\ts$\sigma$ correlation saturates at high $M_\bullet$ and that this happens 
only for core galaxies.~For these, $M_\bullet$ becomes essentially independent of $\sigma$ (but not~$M_{\rm bulge}$) at the highest BH masses.  
This in turn is evidence that the most recent evolution of core ellipticals was via dry mergers, because then BH and galaxy masses grow but $\sigma$
is essentially preserved.

      After this update of the tight BH correlations with classical bulges and ellipticals, we broaden the discussion by examining 
correlations with other kinds of galaxy components:

      {\bf\ARRed Section 6.8}\textBlack~shows that BHs do not correlate tightly enough with pseudobulges to imply close coevolution.
Pseudobulge BHs generally deviate by having smaller masses than the scatter of points in the correlations for classical bulges and ellipticals.
This motivates the suggestion in Sections 8.1 and 8.3 that there are two different modes of BH feeding for classical and pseudo bulges.

      {\bf\ARRed Section 6.9}\textBlack~confirms that $M_\bullet$ does not correlate with the properties of galaxy disks.

      {\bf\ARRed Section 6.10}\textBlack~reviews in more detail the arguments of Kormendy, \& Bender (2011) that BHs do not correlate
tightly enough with the properties of dark matter halos to imply that there is any special coevolution physics beyond what is already
implied by the BH\ts--{\ts}bulge correlations.  Of course, dark matter gravity ultimately controls hierarchical clustering and therefore
much of galaxy formation.  But the data require no conceptual leap beyond the physics of wet major mergers to the idea that additional
coevolution physics is controlled by dark matter in the absence of bulges.

      {\bf\ARRed Section 6.11}\textBlack~reviews the close but puzzling connection between BHs and nuclear star clusters in their relations
with host galaxies.

      {\bf\ARRed Section 6.12}\textBlack~discusses the even more remarkable correlation between BHs and globular cluster systems that surround
elliptical galaxies.

      {\bf\ARRed Section 6.13}\textBlack~~reviews correlations of $M_\bullet$ with the stellar light and mass whose ``absence'' defines the
shallow-density-gradient cores of the biggest elliptical galaxies.  The now-canonical explanation is that cores are scoured by black hole binaries
that are made in galaxy mergers.  The BHs in a binary then decay toward an eventual BH--BH merger by flinging stars away from the center.  The 
tightness of the observed correlations between $M_\bullet$ and core properties provides a ``smoking gun'' connection between BHs and cores that 
does much to support the above picture.

      {\bf\ARRed Sections 6.14 and 6.15}\textBlack~mention other BH correlations, e.{\ts}g., bivariate, ``fundamental plane''
correlations between $M_\bullet$ and any two of $\sigma$, effective radius, and effective brightness.  Finally:

      {\bf\ARRed Section 6.16}\textBlack~is our summary: BH masses correlate tightly enough~to~imply~strong coevolution
(Section\ts1.4) only with classical bulges and ellipticals and with no other galaxy component.

       {\bf\ARRed Section 7}\textBlack~then continues by discussing BHs in pure-disk galaxies; that is, ones that contain no classical bulges
and essentially no pseudobulges.  With all of our observational results in place, Section 8 then updates our understanding of what we learn
about the coevolution (or not) of BHs and their host galaxies.

\vs
\ni {\big\bigpoint\ARRed\kern -20pt 6.1 The M$_\bullet$ -- L\lower2pt\hbox{\almostbig bulge} and 
                        M$_\bullet$ -- M\lower2pt\hbox{\almostbig bulge} correlations}\textBlack
\vs

      The earliest BH demographic result was the correlation between $M_\bullet$ and the luminosity 
of the bulge component of the host galaxy.  Based on two objects, Dressler \& Richstone (1988) 
already noted that ``the [central dark] object in M{\ts}31 is 5\ts--\ts10 times more massive than the 
one in M{\ts}32, closer to the ratio of spheroid luminosities ($\sim$\ts15) than it is to the ratio of
total luminosities~($\sim$\ts70).''  This foreshadowed the correlation of $M_\bullet$ with bulge but not disk luminosity.
Dressler (1989) was the first to propose such a correlation, based on five objects.  Independently, Kormendy (1993a) was 
the first to illustrate the correlation, based on seven objects.  In 1993, it was still possible that we were discovering 
only the easiest cases, so the correlation could have been the upper envelope of a distribution that extended to lower 
$M_\bullet$.  Remarkably, the mean ratio of BH mass to bulge mass = \hbox{(bulge luminosity )\ts$*$\ts(mass-to-light ratio)}
was $\langle M_\bullet/M_{\rm bulge}\rangle = 0.0022^{+0.0017}_{-0.0009}$, consistent with more recently published values, 
$\sim$\ts0.13\ts\%, and smaller than the ratio that we find here in Section 6.6.1.  So the BH sample was not severely biased.  
An update of the $M_\bullet$\ts--\ts$L_{\rm bulge}$ correlation that includes the maser BH discovery in NGC 4258 is in KR95.

      Magorrian \etal (1998) made the first study of a large sample of galaxies that were not chosen to
favor easy BH detection.  Virtues were the large and unbiased sample of 32 galaxies and the uniform
analysis machinery of axisymmetric, two-integral dynamical models.  The ``down side'' was that the data were
very heterogeneous, ranging from high-spatial-resolution kinematic observations from Mauna Kea and Palomar
({\bf Table 1}; notes to {\bf Tables 2} and {\bf 3}) to ground-based spectroscopy with several arcsec resolution.  
BH masses proved to be accurate when based on high-resolution data but were overestimated for poorly 
resolved galaxies (e.{\ts}g., Merritt \& Ferrarese 2001; Kormendy 2004).  Nevertheless, 
this paper provided an important ``sanity check'' on the generality of published BH discoveries.  Also, BH detections 
in all but six of the 32 galaxies led to the belief that essentially all bulges contain BHs that satisfy the
$M_\bullet$\ts--\ts$L_{\rm bulge}$ correlation.

      Further elaboration of the $M_\bullet$\ts--\ts$L_{\rm bulge}$ correlation followed for larger and different kinds of 
galaxy samples and for different $M_\bullet$ quality cuts (e.{\ts}g.,
Ho 1999a;
Merritt \& Ferrarese~2001;  
Laor~2001;                   
Kormendy \& Gebhardt 2001;
McLure \& Dunlop 2002; 
Marconi \& Hunt 2003;
Ferrarese \& Ford 2005;
Graham 2007;
G\"ultekin \etal 2009c;
Sani \etal 2011;
Kormendy \etal 2011;
Vika \etal 2012;
Graham \& Scott 2013;
McConnell \& Ma 2013).
Particularly important was the change from visible-light \hbox{($B$- to $R$-band)}
absolute magnitudes to $K$ band in order to minimize effects of internal absorption and young stars;
this resulted~in significantly reduced scatter (Marconi\ts\&{\ts}Hunt\ts2003).

      The $M_\bullet$--$L_{\rm bulge}$ correlation was always implicitly about bulge mass, but it was important to derive masses directly 
and to express the correlation explicitly in terms of $M_{\rm bulge}$.  Bulge masses were derived using the virial theorem, 
$M_{\rm bulge} \propto r_e \sigma^2$, where $r_e$ is the effective radius and $\sigma$ is the velocity dispersion (Marconi \& Hunt 2003)
or from dynamical modeling (Magorrian \etal 1998; H\"aring \& Rix 2004) or from scaling relations implied by Fundamental Plane parameter 
correlations (McLure \& Dunlop 2002).  The results were generally consistent in finding a good correlation with a 
few outliers (e.{\ts}g., NGC 4342).  Most studies are consistent with a linear relation, $M_\bullet \propto M_{\rm bulge}$,
but a few studies have~found~hints (H\"aring \& Rix 2004) or more of~a~steeper-than-linear relationship, 
$M_\bullet \propto M_{\rm bulge}^{1.54 \pm 0.15}$ (Laor 2001 complete sample).  It is difficult to compare these studies fairly,
because some include BH masses from AGN emission-line widths (Laor 2001) and others do not.  Section 6.6 updates these
results with the present sample.

      There has been good agreement on the mean ratio of BH mass to bulge mass~and~its~scatter.
Merritt \& Ferrarese (2001) got $\langle M_\bullet/M_{\rm bulge} \rangle = 0.0013$ with a standard deviation of a factor~of~2.8;
Kormendy \& Gebhardt (2001) got $\langle M_\bullet/M_{\rm bulge} \rangle = 0.0013$; 
McLure \& Dunlop (2002) got $\langle M_\bullet/M_{\rm bulge} \rangle = 0.0012$; 
Marconi \& Hunt (2003) got $\langle M_\bullet/M_{\rm bulge} \rangle = 0.00234$ with an intrinsic dispersion of 0.27 dex, and
H\"aring \& Rix (2004) got $\langle M_\bullet/M_{\rm bulge}\rangle = 0.0014 \pm 0.0004$. 
Even the more indirect indications via the mutual correlations between BH mass, the ``mass~deficit'' that defines the cores of
elliptical galaxies and BH mass fraction all point to a typical ratio 
$\langle M_\bullet/M_{\rm bulge} \rangle \simeq 0.001$ with hints of a scatter to values 10 times smaller (Kormendy \& Bender~2009).
However, with recent upward corrections to BH masses in giant ellipticals, with the omission of emission-line-based BH masses that we
argue are underestimated, and with separate consideration for pseudobulges, we will find a larger ratio of $M_\bullet/M_{\rm bulge}$
for classical bulges and ellipticals listed in {\bf Tables 2} and {\bf 3}.

      The correlation between $M_\bullet$ and bulge mass or luminosity remains a well established, useful~result that, in coming 
sections, will open the door to new conclusions about which parts of galaxies coevolve with AGNs.  Certainly it kindled suspicions 
that galaxy and BH growth affect each~other.  However, the event that electrified the community and that --
to paraphrase Christopher Marlowe~--
``launched a thousand [papers]'' on BH--galaxy coevolution was the discovery of the tighter correlation between $M_\bullet$
and host galaxy velocity dispersion.

\vs
\ni {\big\bigpoint\ARRed\kern -20pt 6.2 The M$_\bullet$ -- $\sigma$ correlation}\textBlack 
\vs

      Avi Loeb suggested independently to Karl Gebhardt and to Laura Ferrarese that they should look for a correlation between BH 
mass and galaxy velocity dispersion.  It is remarkable in retrospect and a testament to the unpredictable way that research lurches 
forward that the correlation was~not found much earlier.  In any case, a tight $M_\bullet$\ts--\ts$\sigma$ correlation was discovered, 
announced at the Spring 2000 meeting of the American Astronomical Society by Gebhardt (see Kormendy \etal 2000) and by Ferrarese and 
published in Ferrarese \& Merritt (2000) and in Gebhardt \etal (2000a).

      The immediate reaction was that \underbar{the} fundamental relationship between BHs and host galaxies had been found.
The reason was that the scatter was ``only 0.30 dex \dots~over almost 3 orders of magnitude in [$M_\bullet$]'' and ``most of
this is due to observational errors'' (Gebhardt \etal 2000a).  Ferrarese \& Merritt (2000) similarly concluded that the ``scatter
[is] no larger than expected on the basis of measurement error alone.''  Both papers emphasized that the correlation is 
important at a practical level because it allows accurate prediction of $M_\bullet$ from an easy-to-make measurement and because
it implies that BH growth and bulge formation are closely linked.

      Many papers have expanded on this result with bigger samples and have explored its applicability to AGNs, 
to other kinds of objects (e.{\ts}g.,~globular clusters, \S\ts7.4 here), and to the distant Universe.  Papers that elaborate on the  
$M_\bullet$\ts--\ts$\sigma$ relation as revealed by the growing sample~of~BH detections include
Merritt \& Ferrarese (2001);
Kormendy \& Gebhardt (2001);
Tremaine \etal (2002);
Marconi \& Hunt (2003);
Ferrarese \& Ford~(2005);
Wyithe (2006a, b);
Graham (2007, 2008a, b);
Hu (2008);
G\"ultekin \etal (2009c);
Greene \etal (2010);
Graham \etal (2011);
Sani \etal (2011);
Kormendy et al.~(2011);
McConnell \etal (2011a);
Beifiori \etal (2012);
Graham \& Scott (2013); and
McConnell~\&~Ma~(2013).

      We update the $M_\bullet$ -- $\sigma$ correlation in Sections 6.6 and 6.7.  One conclusion will be that the
$M_\bullet$\ts--\ts$M_{\rm bulge}$ and $M_\bullet$\ts--\ts$\sigma$ correlations have the same intrinsic scatter.
This is important because we calculate $M_\bullet$ in part from $\sigma$.  Therefore, when we plot $M_\bullet$ 
against $\sigma$, we are to some degree plotting a function of $\sigma$ against $\sigma$ (See Section 6.5).  Seeing 
the same small scatter in the $M_\bullet$\ts--\ts$M_{\rm bulge}$ correlation is therefore compelling.

\vfill \eject

\vs
\ni {\big\bigpoint\ARRed\kern -20pt 6.3 We omit BH masses based on kinematics of ionized gas when}
\ni {\big\bigpoint\ARRed\kern -20pt ~\quad\quad~\ts~broad emission line widths are not taken into account.}\textBlack
\vs

      Section\ts3.2 showed why we believe that BH masses based on ionized gas rotation curves~are~often underestimated
when large emission-line widths are not taken into account.~It~is not {\it a priori} certain that large line widths imply that 
some dynamical support of the gas comes from random motions, so that an ``asymmetric drift correction'' (Mihalas \& Routly 1968; 
Binney \& Tremaine 1987) is required, as it would be for a stellar system in which $\sigma \sim V$.  However, when the same galaxy 
is analyzed using emission- and absorption-line kinematics and both show large line widths, we almost always get larger $M_\bullet$
estimates from absorption lines than from~emission~lines.~The~examples of M{\ts}87 and NGC\ts3998 are shown in {\bf Figure\ts12} with
pairs of points, a cyan point for the low-$M_\bullet$, emission-line result connected with a cyan line to a black point for the 
high-$M_\bullet$ result~from stellar dynamics.  The correction is particularly big for NGC 3998.  Thus, it is reasonable to suspect, 
even when we cannot check them, that $M_\bullet$ values are underestimated by ionized gas rotation curves when large line widths are not 
taken into~account.  These cases are listed in cyan in the tables and are shown by cyan points in {\bf Figure 12}.  Also shown in black 
with cyan centers are three BH masses that are based on emission-line measurements in which line widths were taken into account.

      These latter points agree with the $M_\bullet$ correlations determined from stellar and maser dynamics.  So do cyan 
points for galaxies with $\sigma_e$\ts$\sim$\ts160\ts$\pm$\ts20{\ts}km{\ts}s$^{-1}$.~But when $\sigma_e$\ts$>$\ts250{\ts}km{\ts}s$^{-1}$, 
many cyan points fall near the bottom of or below~the~scatter~for~black~points.    For M{\ts}87, 
both points fall within the scatter, but using the emission-line $M_\bullet$ would -- and, historically, did -- contribute to our missing 
a scientific result, i.{\ts}e., that $\sigma_e$ ``saturates'' at high $M_\bullet$ (Section 6.7).  Other conclusions are at stake, too.  
Because Graham \& Scott (2013) retain the cyan points but do not have the high-$M_\bullet$ BHs from Rusli \etal (2013), they incorrectly
conclude that the $M_\bullet$\ts--\ts$M_{K,\rm bulge}$ correlation has a kink to lower slope at high $M_\bullet$ whereas we conclude that it has
no kink, and they see little or no kink in $M_\bullet$\ts--\ts$\sigma_e$ whereas we conclude that $\sigma_e$ saturates at high $M_\bullet$.
From here on, we omit the all-cyan points from plots and fits.  We also omit NGC\ts2778 ($M/L_K$ is too big) and NGC\ts3607 (we cannot correct 
$M_\bullet$ for the omission of dark matter from the dynamical~models\ts--{\ts}see table supplementary notes on individual objects).

\vfill

\cl{\null} 

\includegraphics{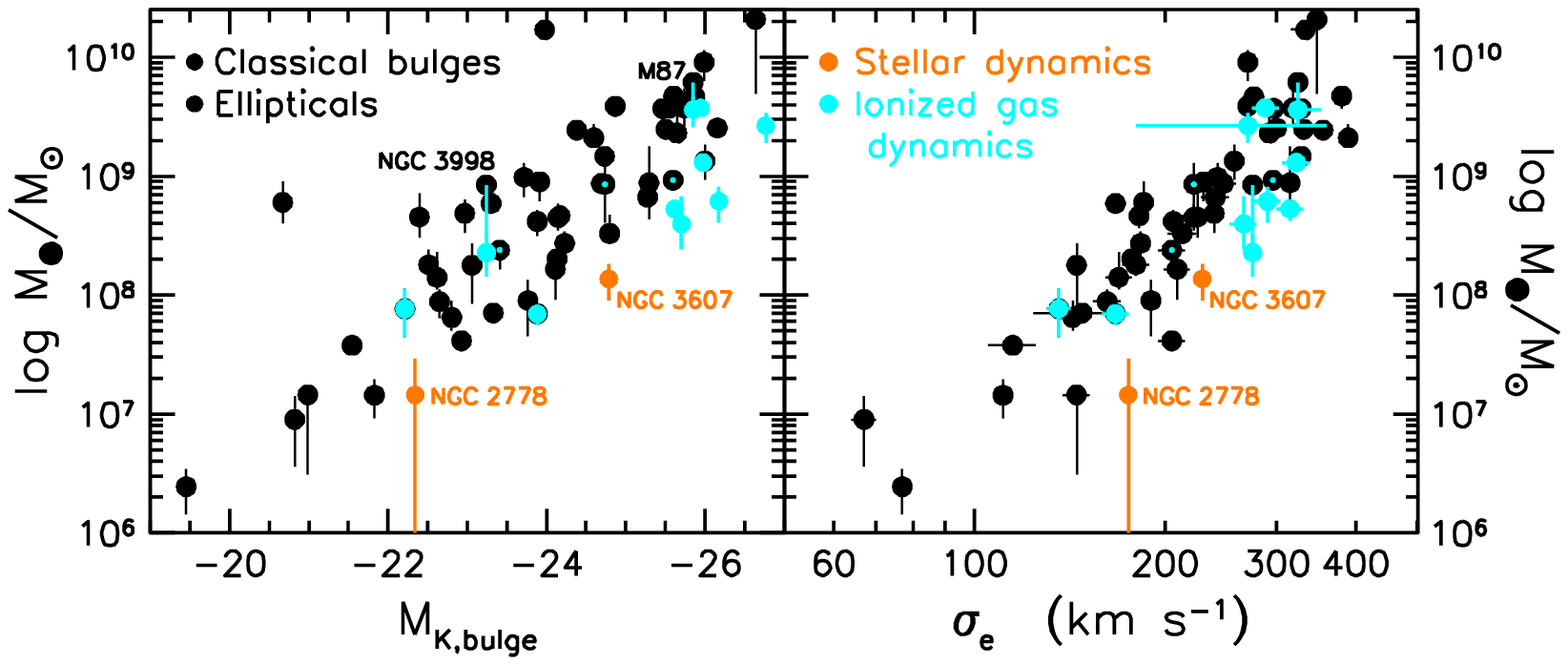}

\ni {\bf \textBlue Figure 12}\textBlack 

\vskip 1pt
\hrule width \hsize
\vskip 2pt
\ni Correlations between BH mass $M_\bullet$ and ({\it left\/}) the $K$-band absolute magnitude of the classical bulge or elliptical and
    ({\it right\/}) its effective velocity dispersion for the sample in {\bf Tables 2} and {\bf 3}.  Galaxies or $M_\bullet$
    measurements that we omit from further illustrations and fits are shown in orange and cyan.

\eject

\vs
\ni {\big\bigpoint\ARRed\kern -20pt 6.4 Mergers in progress have abnormally small BH masses.}\textBlack
\vs

      BHs are detected dynamically in 5 galaxies that are mergers in progress (green lines in {\bf Table\ts2}).  The~two that are 
least well known are illustrated in {\bf Figure 13}.  When we add these mergers to the BH correlations ({\bf Figure 14\/}), we find that at
least these five BHs have abnormally small masses for their (giant!)~galaxies' luminosities.  They have more marginally low BH masses for their
$\sigma_e$. 

      Hints of this result were noticed in some of the BH discovery papers, but a comprehensive picture has never emerged.  Misses are 
easy to understand.  NGC 2960 was misclassified as Sa?~in~the~RC3, and IC\ts1481 was misclassified~as~Sb:{\ts}pec in van den Bergh, Li \&
Filippenko (2002).  So authors have considered NGC 2960 to be one example among many of a pseudobulge (Greene \etal 2010) or a late-type
galaxy (McConnell \& Ma 2013) with a relatively small BH.  Neither McConnell \& Ma (2013) nor Graham \& Scott (2013) have IC 1481 in their sample.  
However, both galaxies clearly show tidal tails, loops, and shells that are characteristic of mergers in progress ({\bf Figure 13}).  Also,
Cappellari \etal (2009) correctly noted that NGC 5128 satisfies the $M_\bullet$\ts--\ts$\sigma_e$ correlation.

\vfill


\includegraphics{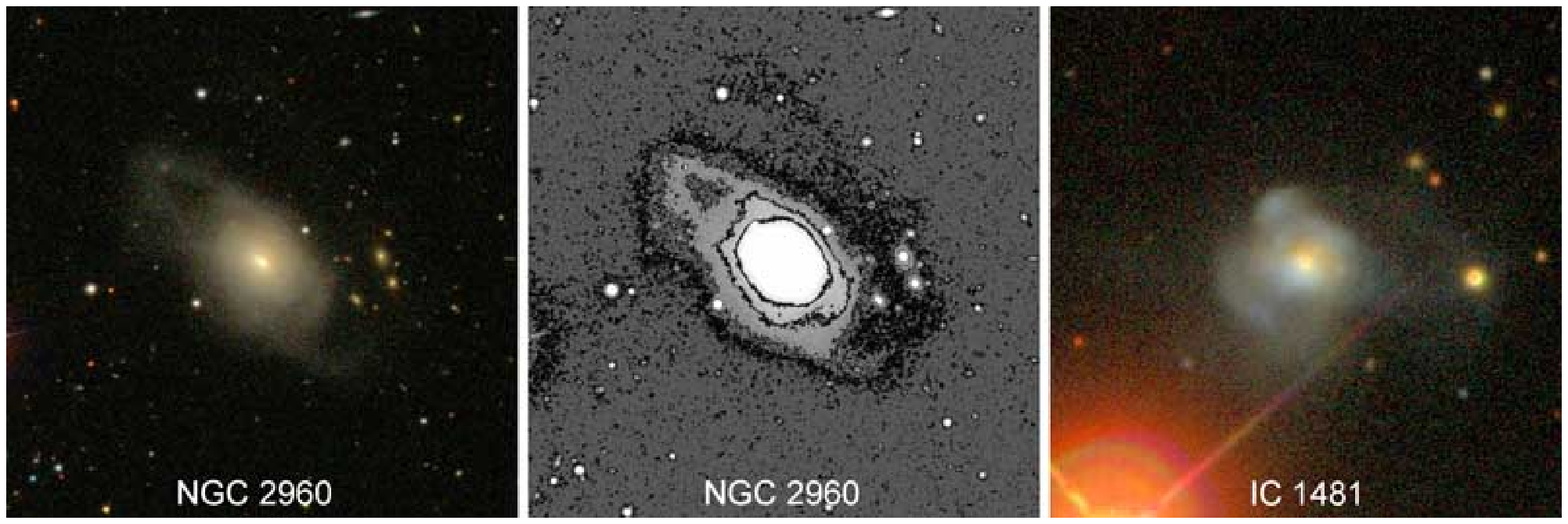}

\ni {\bf \textBlue Figure 13}\textBlack 

\vskip 1pt
\hrule width \hsize
\vskip 2pt
\ni ({\it left\/}) SDSS {\tt WIKISKY} image and ({\it center\/}) isophotes of NGC 2960 (Kormendy \& Bender 2013b).
    ({\it right\/}) SDSS {\tt WIKISKY} image of IC 1481.  The merger nature of NGC 4382 is illustrated in KFCB (their Figure 9).
     Additional mergers with abnormally small $M_\bullet$ are added by Bender \etal (2013).

\vskip 6.1truecm


\includegraphics{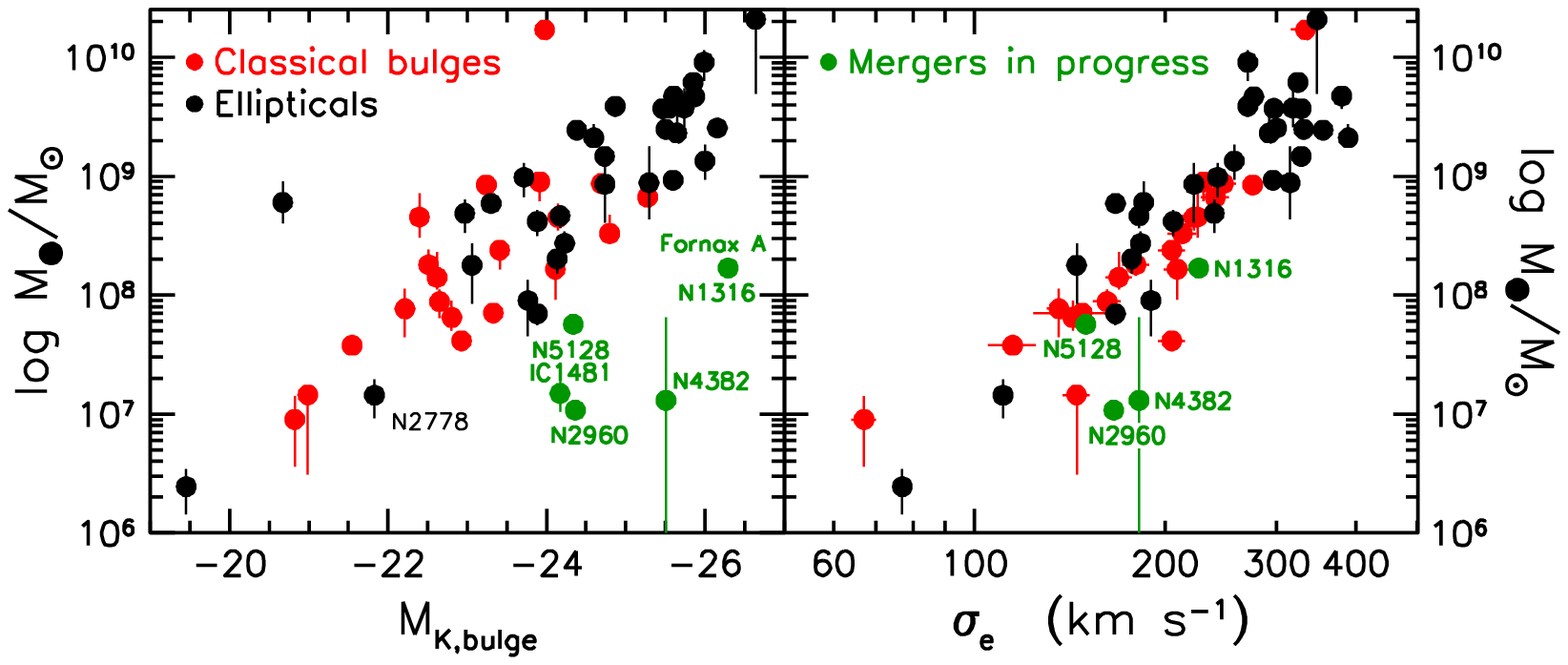}

\ni {\bf \textBlue Figure 14}\textBlack 

\vskip 1pt
\hrule width \hsize
\vskip 2pt
\ni Correlations between $M_\bullet$ and ({\it left\/}) the $K$-band absolute magnitude of the bulge or elliptical and
    ({\it right\/}) its effective velocity dispersion for classical bulges, ellipticals, and mergers in progress.

\eject

      However, Nowak \etal (2008) certainly realized that, while NGC 1316 = Fornax A satisfies the $M_\bullet$\ts--\ts$\sigma_e$ correlation,
its BH mass is ``a factor of $\sim$\ts4 smaller than expected from its bulge mass and the Marconi \& Hunt [2003] relation''.  The deviation is
even larger with the corrected BH masses in {\bf Figure\ts14}.  And G\"ultekin \etal (2009c) discussed whether $M_\bullet$ is anomalously
low in NGC\ts4382.  They concluded that the 1$\sigma$ upper bound on $M_\bullet$ is consistent with the 1$\sigma$ lower envelope of the scatter 
in the $M_\bullet$\ts--\ts$\sigma_e$ relation.  On the other hand, the 1$\sigma$ lower envelope of the $M_\bullet$\ts--\ts$L_V$ correlation
implies a BH mass that ``is more than a factor of 2 larger than our 3$\sigma$ upper limit for the black hole mass [in NGC 4382].  In this sense, 
the black hole mass is anomalously low''.

      We now see that this is generally true for the five mergers in progress that are in our sample.  It is not plausible that their 
stellar populations are so young that $K$-band luminosities are brightened over those of old stellar populations by enough to account for the 
horizontal deviations of the green points in {\bf Figure 14} ({\it left}).  G\"ultekin \etal (2009c) get $M/L_V = 3.74 \pm 0.10$\ts$M/L_{V\odot}$ 
for NGC\ts4382, a normal value (Schulze \& Gebhardt 2011).  Nowak \etal (2008) get $M/L_K = 0.58^{+0.07}_{-0.04}$~(3$\sigma$)~$M/L_{K\odot}$ 
for NGC\ts1316, and Cappellari \etal (2009) get $M/L_K = 0.63 \pm 0.14$ (3$\sigma$) $M/L_{K\odot}$ for NGC\ts5128.  These are only slightly smaller
than the normal values for old stellar populations that are given, e.{\ts}g., by Equations 8 and 9 in Section 6.6.1.  We use $K$-band magnitudes
in {\bf Figure 14} precisely so that stellar population effects are minimized.

      Bender \etal (2013) discuss the galaxies in {\bf Figure 13} in more detail and find a new example of an anomalously low-$M_\bullet$ BH 
in a merger in progress.

      Why does this happen?  We suggest several possibilities.  Some involve merger physics and others involve limitations in our
search techniques.  It is not clear that they add up to the whole story.

      The merger physics possibility is this:~In a major merger of two disk-dominated galaxies~at~$z$\ts$\sim$\ts0, all pre-merger
disk mass gets scrambled up into a new elliptical but -- unlike the situation at high~$z$~-- little gas is available
to feed up the BH mass in proportion to how much the bulge mass has grown.  This can easily account for a factor of $\sim$\ts2 or more
deviation in the $M_\bullet$\ts--\ts$M_{K,\rm bulge}$ relation.  It can account for much larger deviations if one or both progenitors had
pseudo or no bulges with (consequently) low-mass BHs (see Section 6.8).  However, NGC\ts1316 and NGC\ts4382 have such large stellar masses 
that they cannot be products of one major merger of two pure-disk galaxies.

      In terms of BH detection technology, we could imagine the ``dark horse'' possibility~that~one~of~the progenitor BHs has not yet
arrived at the center.  We suffer the problem of the drunkard looking for his keys under a lamppost:~the center is usually the only place 
where we look and perhaps the only place where we can succeed. Thus, it is reassuring that two of our five merger-in-progress~BHs 
were discovered by their maser emission and could have been discovered elsewhere in the galaxy provided that they were accreting water.
Moreover, the $M_\bullet$ measurement in NGC 2960 is relatively secure (the rotation curve is nicely Keplerian, Kuo \etal 2011), though the
measurement of IC\ts1481 is more uncertain (the molecular disk mass is larger than the BH mass, Section 3.3.3).  The measurements of NGC\ts1316 
and NGC\ts5128 are also very robust.~And $M/L_V$ is not particularly small in NGC\ts4382.  In fact, our biggest qualm in terms of BH detection  
is that all five BHs are so close to the centers of their galaxies.  Has there really been time for the BHs to settle to the center after all
the recent merger violence?  Also, to the extent that we can tell (maser disks and nonthermal AGNs help) none of these five BHs are doubles.   
Galaxies like those discussed in this section should be the most obvious place to look for the BH binaries that are expected to result from mergers.

      It is possible that the deviations in {\bf Figure 14} are the sum of several small effects; e.{\ts}g., mergers of low-$M_\bullet$ disks,
or systematic errors in $M_\bullet$ measures, or errant, non-central BHs.  But the apparently low BH masses in mergers are general enough 
to be compelling, and it is possible~that some new and interesting physics awaits discovery.  Or an unrecognized problem with our BH searches.

      For reasons of clarity, we omit the mergers in progress from further correlation diagrams.

\vfill\eject

\vs
\ni {\big\bigpoint\ARRed\kern -20pt 6.5 BH monsters in relatively small bulges and ellipticals}\textBlack
\vs

      Prominent in the late-2012 news was the discovery of a spectacularly overmassive BH in the normal-looking, almost-edge-on S0 NGC\ts1277 
in the Perseus cluster (van den Bosch \etal 2012).  At $M_\bullet$\ts=\ts(1.7\ts$\pm$\ts0.3)\ts$\times$\null$10^{10}$\ts$M_\odot$, it is 
essentially as massive as the biggest BH known.  But that BH lives in NGC\ts4889, one of the two giant, central galaxies in the Coma cluster,
whereas NGC\ts1277 is an obscure S0 galaxy that is two magnitudes fainter.  {\bf Figure 15} shows it in the $M_\bullet$ correlations.  At left, 
the high-luminosity end of the $M_{K,\rm bulge}$ error bar is at the total luminosity of the galaxy.  But BHs correlate with classical bulges, 
not (it will turn out) with disks, and the bulge constitutes~\lapprox1/2 of the galaxy. Van den Bosch \etal (2012) make a photometric decomposition 
into four components; the sum of the inner two gives $B/T = 0.27$, and this provides the faint end of the $M_{K,\rm bulge}$ error bar in 
{\bf Figure 15} ({\it left}).  Our photometry (Kormendy \& Bender 2013b) shows only two components, a normal, S\'ersic $n = 3.5 \pm 0.7$ 
classical bulge that dominates the light at both large and small radii and an exponential disk.  We get $B/T = 0.55 \pm 0.07$.  This gives the 
absolute magnitude of the point plotted in {\bf Figure 15}.  For any of these magnitudes, the BH is more massive than the upper envelope of the 
scatter in the $M_\bullet$--$M_{K,\rm bulge}$ correlation by at least an order of magnitude.

      A similarly overmassive BH was found in NGC 4486B by Kormendy \etal (1997).~This~is~a~tiny ($M_{VT} = -17.7$) but otherwise normal
elliptical (KFCB), a companion of M{\ts}87.  Spectroscopy~with the CFHT in FWHM = 0\sd52 seeing reveals an extraordinarily steep 
velocity dispersion gradient~to $\sigma = 291 \pm 25$ km s$^{-1}$ at the center.  This value is characteristic of giant ellipticals
that~are~\gapprox\ts4~mag brighter than NGC 4486B.  Not surprisingly, isotropic dynamical models imply a large MH mass of
$M_\bullet$\ts=\ts$6^{+3}_{-2}$\ts$\times$\null$10^8$\ts$M_\odot$.~Extreme anisotropic models can explain the kinematics without a
BH, so the BH detection has been regarded as weak.  But NGC\ts4486B and NGC\ts1277 are equally remarkable~in having steep inward velocity dispersion
gradients up to central $\sigma$ values that are far outside the scatter~in~the Faber-Jackson (1976) relation.  Both $M_\bullet$ determinations
have some uncertainties (we have trouble understanding how NGC\ts1277 can have a completely normal-looking bulge when all of it lives inside
the sphere-of-influence radius of its BH).  But it seems unlikely that we are wrong in concluding that both galaxies have unprecedentedly
large BH-to-bulge mass ratios.  For NGC\ts4486B, our estimate of $M/L_K$ gives $M_\bullet/M_{\rm bulge} = 0.14^{+0.05}_{-0.04}$.  For NGC 1277,
our measurement of $B/T$ and $M/L_K$ gives $M_\bullet/M_{\rm bulge} \simeq 0.18^{+0.06}_{-0.04}$ ({\bf Figure 18}).  Who ordered this?

\vfill


 \includegraphics{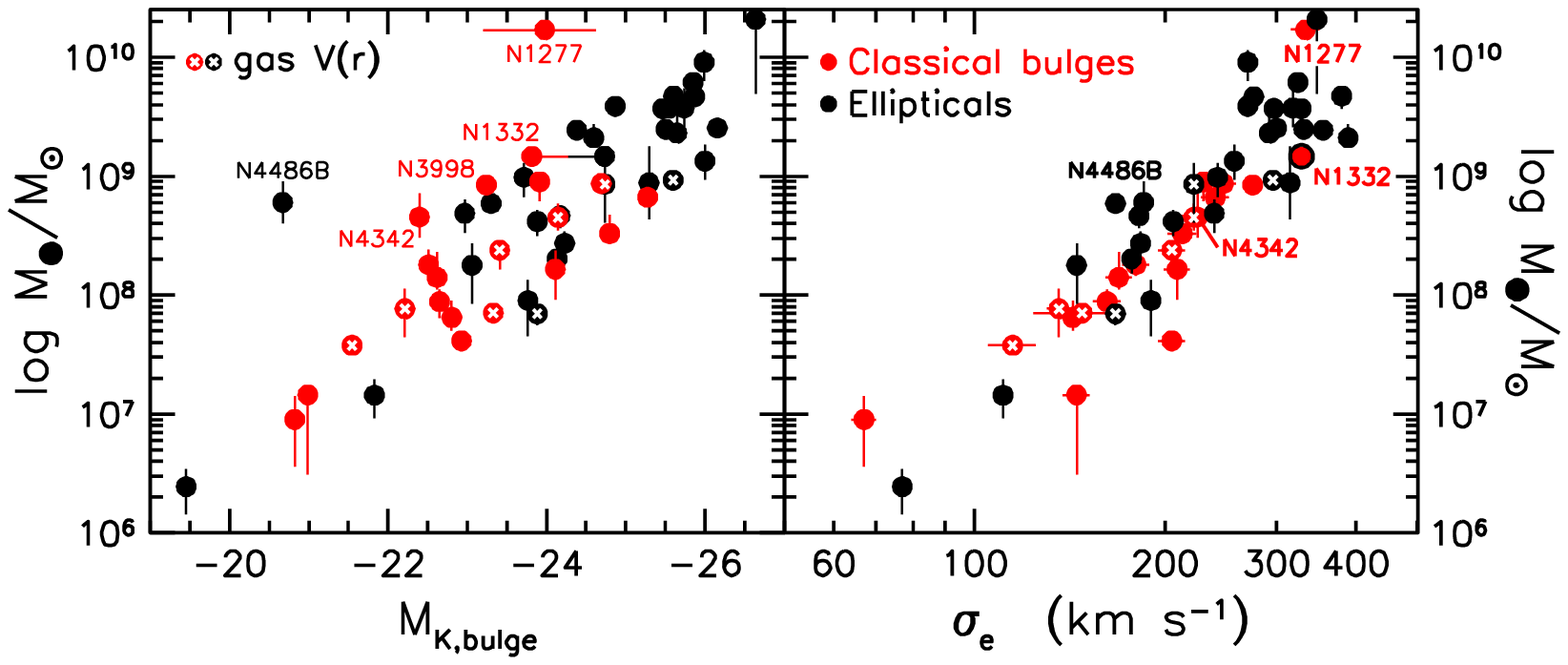}

\ni {\bf \textBlue Figure 15}\textBlack 

\vskip 1pt
\hrule width \hsize
\vskip 2pt
\ni The $M_\bullet$\ts--{\ts}$M_{K,\rm bulge}$ and $M_\bullet$\ts--{\ts}$\sigma_e$ correlations with abnormally high-mass BHs identified. 
NGC 1332 is plotted twice; the points are joined by a line.  If it is an S0, then it contains a marginal example 
of a BH monster.  If, as we suggest in the Table notes, it is an elliptical, then its BH is normal.

\eject

      These objects are rare in the $z$\ts$\sim$\ts0 universe, but they are not unique.  Van den Bosch \etal (2012) are working on additional 
compact galaxies with extraordinarily high central $\sigma$ (their Table\ts1). And we have had hints of more subtle examples.  The best known 
is NGC 4342, \hbox{another edge-on~S0}, this time in the Virgo W$^\prime$ cloud.  The BH mass determined by Cretton \& van den Bosch (1999b), 
together with our measurement of $B/T = 0.70$, makes it the biggest outlier among more normal classical bulges in the $M_\bullet$\ts--\ts$M_{K,\rm bulge}$ 
correlation.  It has $M_\bullet/M_{\rm bulge} \simeq 0.023^{+0.007}_{-0.006}$ ({\bf Figure 18}).  Another possible example is NGC 4350, a Virgo cluster, 
edge-on S0 with a bulge magnitude of $M_{K,\rm bulge} = -22.2$ and $M_\bullet \sim (7 \pm 2) \times 10^8$\ts$M_\odot$ (Pignatelli, Salucci \& Danese 2001).
We do not include it in {\bf Table 3} and {\bf Figure 15} partly because the BH  mass is based only on isotropic dynamical models but mostly because the 
velocity dispersion~is normal for the bulge luminosity.  However, if the mass is correct, it is a bigger outlier~in $M_\bullet$\ts--\ts$M_{K,\rm bulge}$
than NGC\ts4342.  Finally, in case we are wrong in the {\bf Table 2} notes and NGC\ts1332 should be classified as an S0 rather~than an elliptical, 
we note that it would be at the high-$M_\bullet$ extreme of the scatter in {\bf Figure 15} ({\it left}).  However, as explained in the notes, 
we henceforth regard it as an elliptical (black point connected to the red point~in~the~figure).~Note: NGC\ts3998, the next most extreme outlier,
is quite a different sort of galaxy; with $B/T$ = 0.85, it is almost an elliptical. 

      In summary, we have two BH monsters, hints of more to come, and signs that these are extremes of a
distribution of overmassive BHs that connects up with the scatter in the $M_\bullet$ correlations.  

      It is worth emphasizing that none of these galaxies are major outliers in the $M_\bullet$\ts--\ts$\sigma_e$ correlation.  
In Section 6.4, too, merger BHs were extreme outliers in $M_\bullet$\ts--\ts$M_{K,\rm bulge}$ but only marginal outliers 
in $M_\bullet$\ts--\ts$\sigma_e$. It is common to think of the small scatter in $M_\bullet$\ts--\ts$\sigma_e$ as astrophysically magic.~But~we 
see in these two sections that we learn rather more from the luminosity correlation.  The sobering lesson may be this:~$M_\bullet$ is measured 
by interpreting central changes (for example, a slight increase~$\delta\sigma$)~in~$\sigma$ (and, of course, rotation) 
from values $\simeq$\ts$\sigma_e$ that vary only slowly at large radii. 
But the same~$\delta\sigma$ gives a bigger BH mass for a bigger $\sigma_e$.  We may get a small scatter in $M_\bullet$\ts--\ts$\sigma_e$ for 
galaxies in general and for $M_{K,\rm bulge}$ outliers in particular in part because we are plotting a function of $\sigma_e$ against~$\sigma_e$.  
Note the astonishingly small scatter for classical bulges in {\bf Figure 15}.  When we resolve $r_{\rm infl}$ exceedingly well,
this concern should vanish.  But many $M_\bullet$ values are based on subtle features in the LOSVDs near galaxy centers.  
This caution applies throughout this review.

      At the same time, {\bf Figure 15} includes an important check.  If $M_\bullet$ was measured using an emission-line rotation curve of gas,
then the point is marked with a white cross.  This includes galaxies in which optical emission-line widths were taken into account
and even two low-$\sigma_e$ galaxies (NGC\ts4459 and NGC\ts4596) for which line widths were not taken into account.  The other points
are for maser measurements and for one molecular gas \hbox{measurement (NGC\ts4526:~Davis \etal 2013)}.  These points are immune from the above
worry.  But they are fully consistent with the small scatter in $M_\bullet$\ts--\ts$\sigma_e$, including that for bulges.  The weakest
stellar-dynamical BH measurements{\ts}--{\ts}including those of the monsters{\ts}--{\ts}should be checked.  But we are reassured by the gas points 
in {\bf Figure 15}.

      What do we learn from BH monsters?  We do not have a definitive  answer; we only float a speculation for further study.  It is interesting 
to note that the bulge of NGC 1277 
has an effective radius of $r_e \simeq 0.97$ kpc.  NGC 4486B is tinier, $r_e = 0.20 \pm 0.01$ kpc (KFCB).  But even the bulge of NGC 1277 
is similar to the compact ``red~nuggets'' that constitute most early-type galaxies at high redshifts (e.{\ts}g., Szomoru, Franx \& van Dokkum 2012 ).
We know from spectral energy distributions that these were already old then.  We also struggle to understand how at least some BHs grew supermassive 
very quickly in the early universe.  It will be interesting to investigate whether objects like NGC 1277 are remnants -- this one now cloaked in a disk -- of
early red nuggets that already had overmassive BHs at high $z$.  Is NGC 1277 left over from a bygone era before the present BH correlations
were engineered?

\vfill\eject

\cl{\null}

\vfill



 \includegraphics{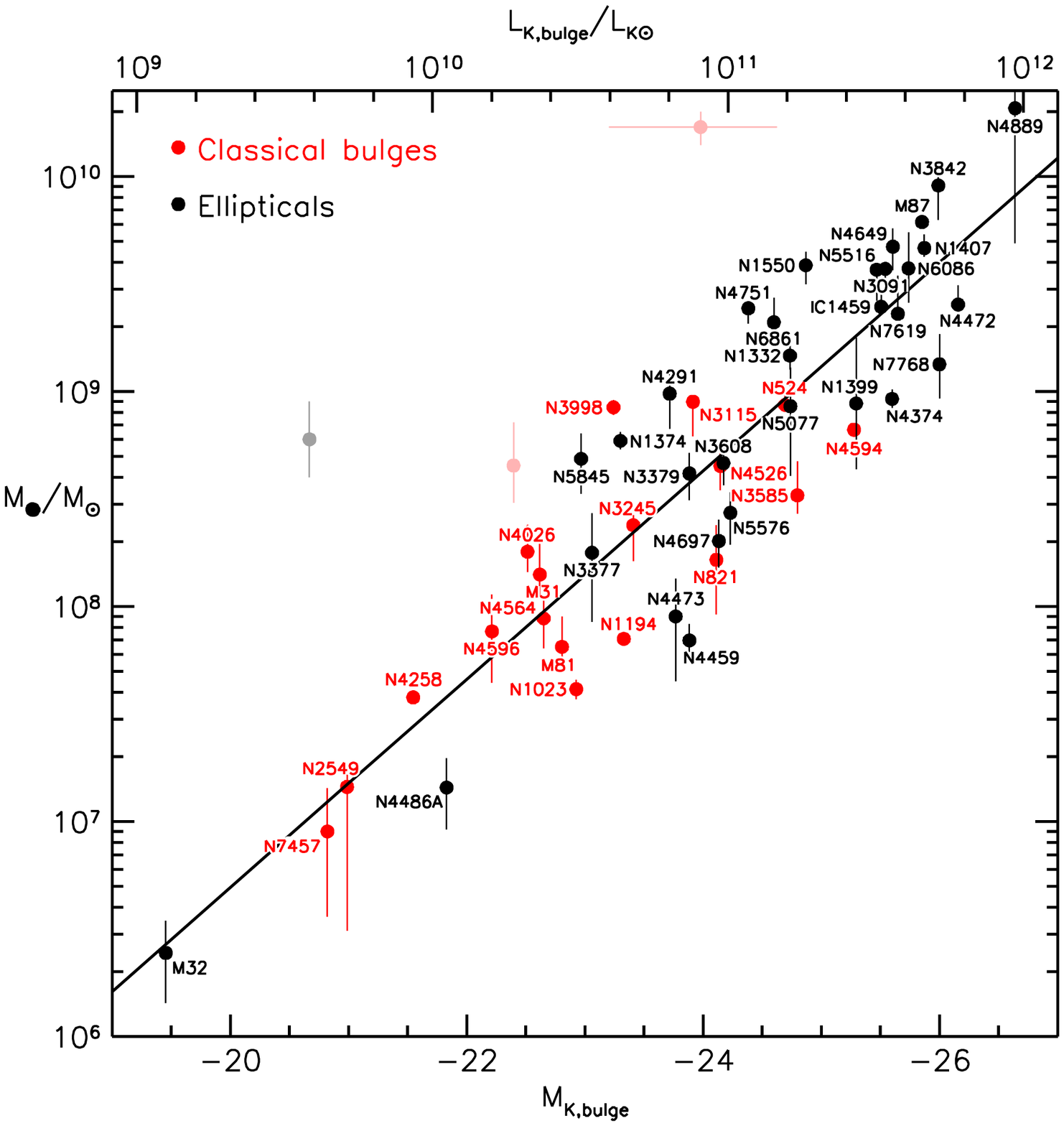}
 \includegraphics{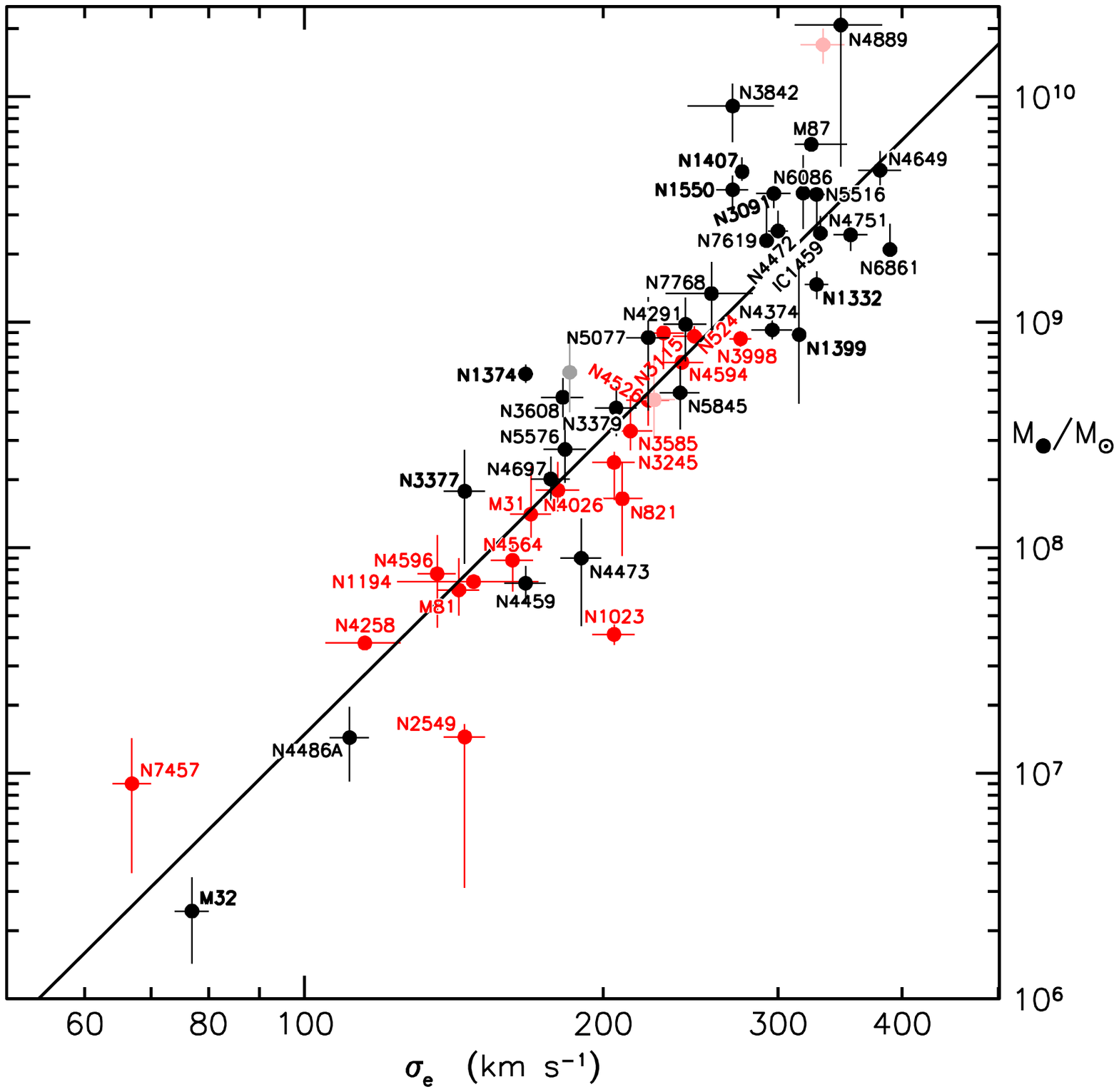}

\ni {\bf \textBlue Figure 16}\textBlack 

\vskip 1pt
\hrule width \hsize
\vskip 2pt
\ni Correlation of dynamically measured BH mass $M_\bullet$ with ({\it left}\/) $K$-band absolute magnitude 
    $M_{\rm K,bulge}$ and luminosity $L_{K,\rm bulge}$ and ({\it right\/}) velocity dispersion~$\sigma_e$ 
    for ({\it red\/}) classical bulges and ({\it black\/}) elliptical galaxies. The lines are symmetric 
    least-squares fits to all the points except the monsters ({\it points in light colors}), NGC 3842, and NGC 4889.  
    {\bf Figure 17} shows this fit with 1-$\sigma$ error bars.

\vs\vskip 2pt
\ni {\big\bigpoint\ARRed\kern -20pt 6.6 The M$_\bullet$\ts--{\ts}L\lower2pt\hbox{\almostbig bulge},
                                            M$_\bullet$\ts--{\ts}M\lower2pt\hbox{\almostbig bulge}, and 
                                            M$_\bullet$\ts--\ts$\sigma$\lower2pt\hbox{\almostbig\kern -1.2pt e} correlations for} \par
\vskip 2pt
\ni {\big\bigpoint\ARRed\kern 5pt classical bulges and elliptical galaxies}\textBlack
\vs

      {\bf Figure\ts16} shows the updated correlations of $M_\bullet$ with bulge luminosity and velocity dispersion.
Recent advances allow us to derive more robust correlations and to better understand the systematic effects in their scatter.  
First, we distinguish classical bulges that are structurally like ellipticals from pseudobulges that are
       structurally more disk-like than classical bulges.~There is now a strong case that classical bulges are made in major mergers,~like 
       ellipticals, whereas pseudobulges are grown secularly by the internal evolution of galaxy disks.~We show in Section\ts6.8 that 
       pseudobulges do not satisfy the same tight $M_\bullet$--host-galaxy correlations as classical~bulges~and ellipticals.  
       Therefore we omit them here.
Second, we now have bulge and pseudobulge data for all BH galaxies (Kormendy \& Bender 2013b).
Third (Section 3), we have more accurate BH masses, partly because of improvements in data \hbox{(ground-based AO and integral-field} spectroscopy),
                                        partly because of improvements in modeling (e.{\ts}g., three-integral models that include dark matter), and
                                        partly because we are now confident that emission-line rotation curves underestimate $M_\bullet$
                                        unless broad line widths are taken into account (Section 6.3).  We omit these masses. 
Fourth, we have reasons to omit BH monsters, mergers in progress, and (Section 6.7) the two largest BHs known in ellipticals. 
Finally, the sample of galaxies with dynamical BH detections is larger and broader in Hubble types. 
These developments lead to a significant recalibration of the ratio of BH mass to the mass of the host bulge and, 
as we have already begun to see, to qualitatively new conclusions. 

\eject

\cl{\null}

\vfill

 \includegraphics{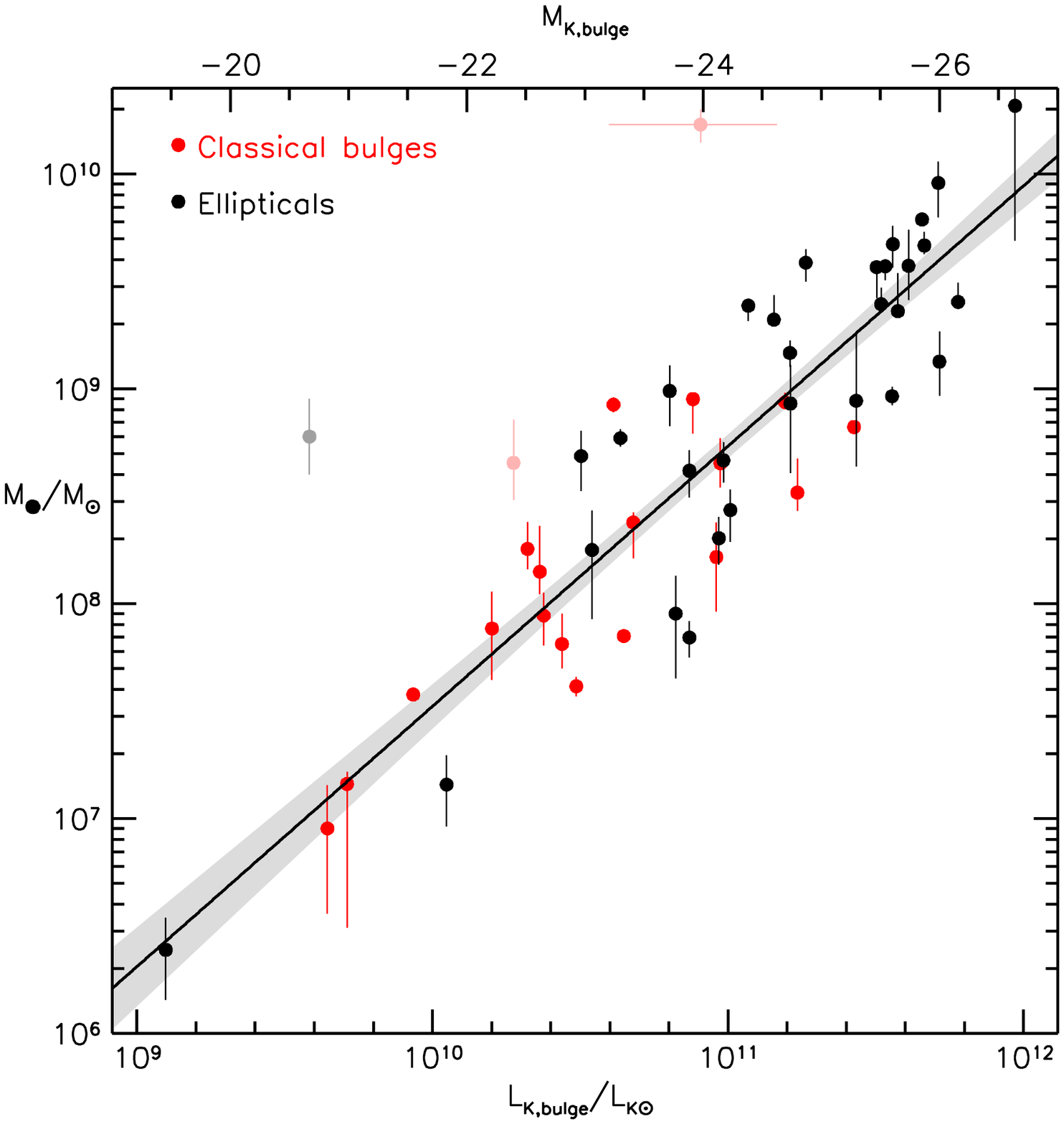}
 \includegraphics{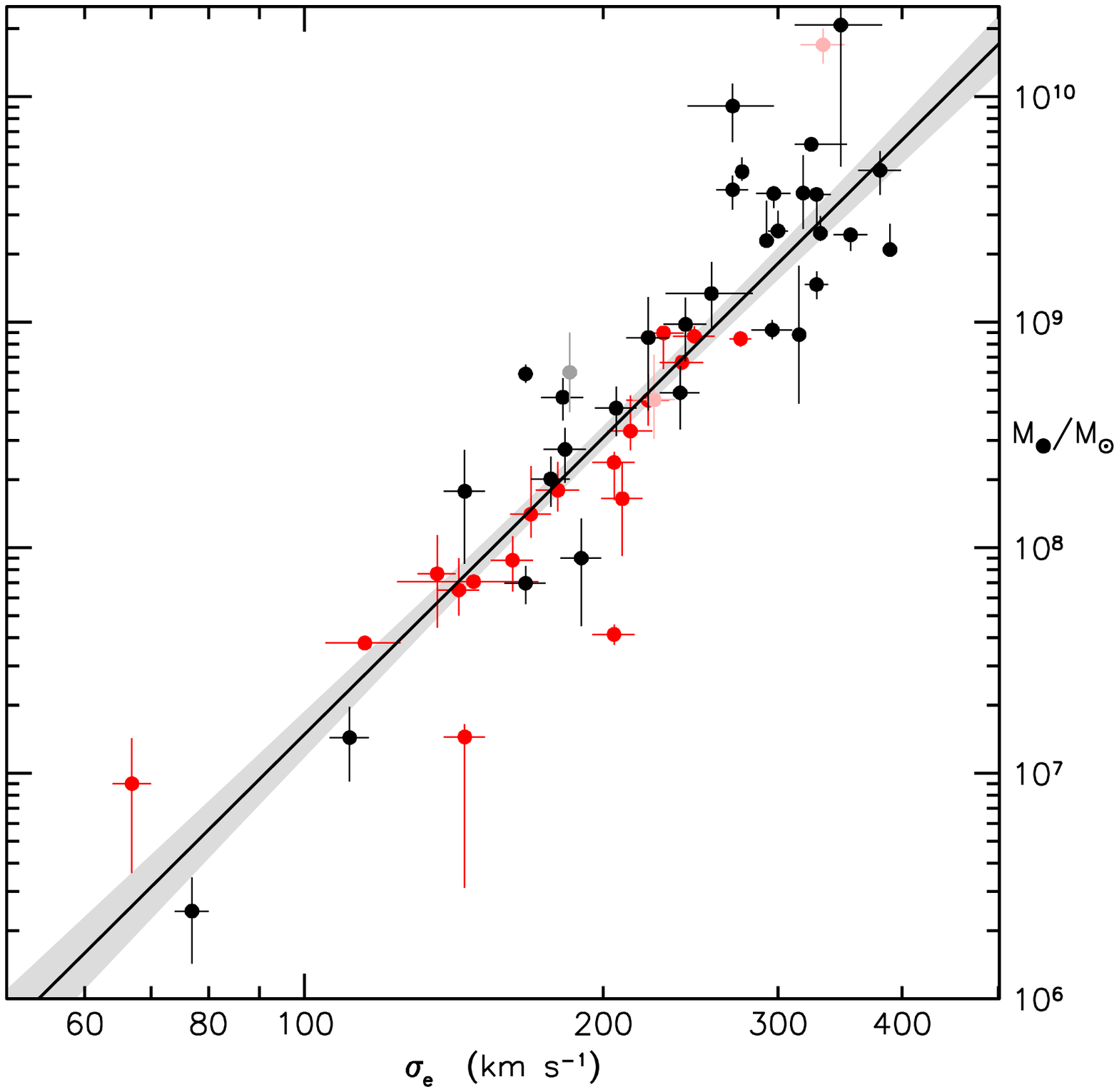}

\ni {\bf \textBlue Figure 17}\textBlack 

\vskip 1pt
\hrule width \hsize
\vskip 2pt
\ni The $M_\bullet$--$M_{\rm K,bulge}$ and $M_\bullet$--$\sigma_e$ correlations with symmetric (Tremaine \etal 2002) least-squares~fits 
    (Equations~2~and~3) and the 1$\sigma$ range of the fits ({\it gray shading\/}).  Here we give equal weight~to all the points.  
    Fits that use the individual $M_\bullet$
    measurement errors (Equations 4 and 5) are almost identical.  Among the plotted points, all fits omit the BH monsters ({\it points in 
    light colors\/}), $M_\bullet$ values determined from ionized gas rotation curves without taking line widths into account (NGC\ts4459 and NGC\ts4596),
    and the two highest-$M_\bullet$ ellipticals (NGC\ts3842 and NGC\ts4889, see text). The bottom axis at left shows $K$-band luminosity.
    Uncertainties in $M_{K,\rm bulge}$ are typically $\pm 0.2$, smaller for some ellipticals and significantly larger for a few bulges with 
    uncertain $B/T$ ratios.

\vs\vsss

      {\bf Figure\ts17} shows our adopted symmetric, least-squares fits calculated as in Tremaine{\ts}et{\ts}al.{\ts}(2002).
Symmetrizing around $L_{K,\rm bulge} = 10^{11}$\ts$L_{K\odot}$ and $\sigma_e = 200$ km s$^{-1}$, these are
\vskip -28pt
\null
$$
\log\biggl(\kern -2pt{{M_\bullet} \over {10^9{\ts}M_\odot}}\kern -2pt\biggr) 
              = -(0.266 \pm 0.052) - (0.484 \pm 0.034) (M_{K,\rm bulge} + 24.21);~{\rm intrinsic~scatter} = 0.31;  \eqno{(2)}
$$
\cl{\null} 
\vskip -42pt 
\cl{\null}
$$
\log\biggl(\kern -2pt{{M_\bullet} \over {10^9{\ts}M_\odot}}\kern -2pt\biggr) 
              = -(0.510 \pm 0.049) + (4.377 \pm 0.290) \log \biggl({{\sigma} \over {200~{\rm km~s}^{-1}}}\biggr);~{\rm intrinsic~scatter} = 0.29. \eqno{(3)}
$$                   
\null
\vskip -14pt
\noindent For the above version, we adopt equal errors of $\Delta M_{K,\rm bulge} = 0.2$ and $\Delta \log{M_\bullet} = 0.117$ = the mean 
for all fitted galaxies.  If, instead, we use individual errors in $M_{K,\rm bulge}$ ($\pm$\ts0.2) and $\log{\sigma_e}$ and add individual 
errors in $\log{M_\bullet}$ to the intrinsic scatter in quadrature and iterate the intrinsic scatter until the reduced $\chi^2 = 1$, then
\vskip -28pt
\null
$$
\log\biggl(\kern -2pt{{M_\bullet} \over {10^9{\ts}M_\odot}}\kern -2pt\biggr) 
                = -(0.253 \pm 0.052) - (0.484 \pm 0.036) (M_{K,\rm bulge} + 24.21);~{\rm intrinsic~scatter} = 0.31;  \eqno{(4)}
$$
\cl{\null} 
\vskip -43pt 
\cl{\null}
$$
\log\biggl(\kern -2pt{{M_\bullet} \over {10^9{\ts}M_\odot}}\kern -2pt\biggr) 
   = -(0.501 \pm 0.049) + (4.414 \pm 0.295) \log \biggl(\kern -2pt{{\sigma} \over {200~{\rm km~s}^{-1}}}\kern -2pt\biggr);~{\rm intrinsic~scatter} = 0.28.\eqno{(5)}
$$        

\null
\vskip -12pt
\noindent The difference between the two sets of fits is small.  Taking account of a variety of fits, we conclude 
\eject

\noindent that the intrinsic $\log{M_\bullet}$ scatter in $M_\bullet$--$M_{K,\rm bulge}$ is $0.31 \pm 0.02$, almost the same as the 
                                                       intrinsic scatter $0.29 \pm 0.03$ in $M_\bullet$--$\sigma_e$. 
This conclusion has also been reached by other authors who use infrared luminosities (e.{\ts}g., Marconi\ts\&{\ts}Hunt\ts2003;
Sani \etal 2011).

      We adopt the Equations 2 and 3 fits, because they do not give undue weight to a few points that fortuitously have tiny error bars 
but that could be anywhere in the intrinsic scatter.  When the intrinsic scatter is much larger than the measurement errors for a few but 
not all points, it is preferable to give all points equal weight (see Tremaine \etal 2002 for further discussion).   Note that all of the
intrinsic scatter is assumed to be in $\log{M_\bullet}$.

      Rewriting Equations 2 and 3 in physically more transparent forms,
\vskip -5pt
$$
{{M_\bullet} \over {10^9~M_\odot}} = \biggl(0.542^{+0.069}_{-0.061}\biggr)\ 
                                           \biggl({{L_{K,\rm bulge}} \over 
                                                  {10^{11}{\ts}L_{K\odot}}}\biggr)^{1.21 \pm 0.09} \eqno{(6)}
$$
\null
\vskip -26pt
\null
$$
{{M_\bullet} \over {10^9~M_\odot}} = \biggl(0.309^{+0.037}_{-0.033}\biggr)\ 
                                           \biggl({{\sigma} \over 
                                                  {200~{\rm km~s}^{-1}}}\biggr)^{4.38 \pm 0.29} \eqno{(7)}
$$

\vs
\ni \hbox{{\bf\ARRed 6.6.1.{\ts}The M$_\bullet$\ts--{\ts}M\lower2pt\hbox{\sbf bulge} correlation and the ratio of BH mass to bulge mass}}~\textBlack

      Galaxy formation work requires the mass equivalent of Equation 6, the $M_\bullet$\ts--\ts$M_{\rm bulge}$ correlation.
This is trickier to derive than it sounds.~It is not just a matter of multiplying the bulge luminosity~by a mass-to-light ratio that
is provided automatically by the stellar dynamical models that~give~us~$M_\bullet$.  Bulge mass is inherently less well defined than 
bulge luminosity. Mass-to-light ratios~of~old~stellar populations are uncertain; 
(1) the initial mass function (IMF) of star formation 
is poorly known; it may vary with radius in an individual galaxy or from galaxy to galaxy;
(2) stellar population age and metallicity distributions affect~$M/L$ and are famously difficult to disentangle;
    one consequence is that late stages of stellar evolution{\ts}--{\ts}especially asymptotic giant branch (AGB) stars{\ts}--{\ts}affect $M/L$~but
    but are poorly constrained observationally (Portinari \& Into 2011).  Most important,
(4) dark matter contributes differently at different radii and probably differently in different galaxies.

      Graves \& Faber (2010) discuss these problems.  They conclude~that all of the above are important, that stellar population age and 
metallicity account for $\sim$\ts1/4 of the variations in optical mass-to-light ratios, and that some combination of IMF and dark matter 
variations account for the rest.  However, this field is unsettled; extreme points of view are that even $K$-band mass-to-light ratios 
vary by factors of $\sim$\ts4~from~galaxy to galaxy and that~all of this range is due to IMF variations (Conroy \& van Dokkum 2012) or 
contrariwise that IMFs vary little from one place to another (Bastian, Covey, \& Meyer~2010).

      These problems are background worries that may yet hold unpleasant surprises, but mostly,~they are beyond the scope of this paper.  
The extensive work of the SAURON and ATLAS3D teams (Cappellari \etal 2006, 2013) shows that dynamically determined $I$- and $r$-band mass-to-light 
ratios are very well behaved.  For 260 ATLAS3D galaxies, $M/L_r \propto \sigma_e^{0.69 \pm 0.04}$ with an intrinsic~scatter of only 22\ts\%.  
Since $M/L_K$ is likely to vary less from galaxy to galaxy than $M/L_r$, this suggests that we proceed by finding a way to estimate
$M/L_K$.  In particular, we want an algorithm that does not involve the use of uncertain effective radii $r_e$.  Here's why:

      Published studies often derive $M_{\rm bulge}$ dynamically from $r_e$, $\sigma_e$, and a virial-theorem-like relation 
$M_{\rm bulge}$\ts=\ts$k \sigma_e^2 r_e / G$, where $k$ is, e.{\ts}g., 
3 (Marconi \& Hunt 2003) or
5 (Cappellari \etal 2006, 2010) or
8 (Wolf~et~al.~2010).
This situation is unsatisfactory; different assumptions about the density profile are one reason~why $k$ is uncertain.  Also, 
$r_e$ values are less well measured than we think.  An example of the problem can be found in Marconi \& Hunt (2003):~they 
derive mass-to-light ratios that range from $M/L_K$\ts=\ts0.31 for NGC\ts3384 to $M/L_K$\ts=\ts1.94 for NGC\ts4649.  The machinery
that we adopt below gives a much smaller difference in $M/L_K$ for these two galaxies that are both made of old stars. They may contain
different fractional contributions of dark matter within $r \sim r_e$.  But the difference could also signal a problem with the $r_e$ data
(see KFCB for a review).  Given heterogeneous $r_e$ measurements, we prefer to avoid using any relation of the
form $M_{\rm bulge} \propto \sigma_e^2 r_e$.

      Cappellari \etal (2006, 2013) and other papers on early-type galaxies (see Gerhard 2013 for a review) find that
dark matter contributes $\sim$\ts10\ts--\ts40\ts\% of the mass within~$r_e$, differently in different galaxies.~This is one reason why we choose
not to use $M/L$ ratios from the papers that derive~$M_\bullet$.  Observations reach differently far out into the dark-matter-dominated
parts of different galaxies, and different papers take dark matter into account (or not) in different ways.  We hoped that accounting separately 
for dark matter would provide mass-to-light ratios that are representative of pure, old stellar populations, but (for example) the $M/L_K$ 
ratios derived from the 12 galaxies in Schulze \& Gebhardt (2011) range over a factor of more than 10.  This is a bigger variation than we get 
below even from purely dynamical, total mass-to-light ratios that include dark matter. 

      We also have no guidance about whether we should include halo dark matter in $M_{\rm bulge}$ or not.
It seems silly even to try; bulge stars and dark matter are distributed differently in radius, so how we combine
them would depend (e.{\ts}g.)~on an arbitrary choice of radius.  Anyway, most dark matter is at large radii, and it distribution is unknown.
Moreover, we conclude in Section 6.10 that BH masses do not correlate with halo dark matter in any way that goes beyond the correlation
with bulge properties.  On the other hand, dark matter certainly contributes most of the gravity that makes hierarchical clustering
happen.  So it must affect BH accretion indirectly.  It seems prudent neither to ignore halo gravity completely nor to attempt to add 
all halo mass to the stellar mass.  We adopt an intermediate approach that essentially averages stellar population mass-to-light ratios
and ones that are determined from dynamics.  But we use a zeropoint based on stellar dynamics. 

      As a matter of principle, we prefer to derive mass-to-light ratios in a way that is as independent as possible 
of the $M_\bullet$ measurements.  We also need an algorithm that can be applied to galaxies without dynamical $M/L$ measurements; e.{\ts}g.,
the maser BH galaxies.  And we need an algorithm that can be applied to pseudobulges with poorly observed but young stellar populations.  
This is one reason why we base our results in part on mass-to-light ratios from stellar population models.  We use $M/L_K$ because it is
relatively insensitive to dust absorption and to stellar population~age.  But the contribution of AGB stars to $K$-band light is of paramount
importance and had not until recently been sufficiently taken into account. Fortunately, Into \& Portinari (2013) provide
a revised calibration of mass-to-light ratios against galaxy colors, both over a wide range of bandpasses.  In particular, they
find a much steeper dependence of $M/L_K$ on $(B - V)_0$ color than did (e.{\ts}g.) Bell \& de Jong (2001).  With their calibration, {\it
the dependence of stellar population mass-to-light ratio $M/L_K$ on $\sigma_e$ is completely consistent with that given by the dynamical 
models used below.}

      We therefore use two independent methods to estimate $M/L_K$, one from the galaxy's $(B - V)_0$ color and Into \& Portinari's
(2013) Table 3 and the other using a dynamically measured correlation between $M/L_K$ and $\sigma_e$.  The normalization of the
stellar population $M/L$ values is uncertain for reasons given above.  In particular, the population mass is dominated by
low-mass stars that we do not observe.  So an IMF must be assumed.  Into \& Portinari (2013) assume a Kroupa (2001)~IMF.  However,
we adopt a dynamically measured zeropoint, as follows.

       We begin with Williams, Bureau \& Cappellari (2009), who determine masses and $K$-band luminosities for 
14 S0 galaxies using axisymmetric dynamical models that have contant anisotropies in the meridional plane. 
Cappellari \etal (2006) provide mass-to-light ratios $M/L_I$ determined from detailed dynamical models for 
12 additional galaxies that either are known BH hosts or that have photometry available in KFCB.  We use our $V_T$, $(V - I)_e$ 
from Hyperleda, and Cappellari's $M/L_I$ values to calculate bulge masses and then $M_{KT}$ to determine $M/L_K$ values.
These are very consistent with the Williams \etal (2009) results.  All 26 galaxies together show a shallow correlation of $M/L_K$
with $\sigma_e$.  The four largest values, $M/L_K \geq 1.6$, are outliers and are omitted.  Presumably these galaxies
contain larger contributions of dark matter that we choose not to include.
The remaining 22 galaxies satisfy $\log{(M/L_K)} = 0.287 \log{\sigma_e} - 0.637$ with an RMS scatter of 0.088.  As expected, 
the relation is shallower than the one in $r$ band (above).  It has essentially the same scatter of $\sim$\ts23\ts\%.  
Dynamically, $M/L_K = 1$ at $\sigma_e = 166$ km s$^{-1}$, where the Into \& Portinari (2013)
calibration gives $M/L_K \simeq 0.76$.  Cappellari \etal (2006) argue that the difference may be due to 
the inclusion of some dark matter in the dynamical models.  We use the dynamical zeropoint.

      To shift the Into \& Portinari $\log M/L_K$ values to the above, dynamical zeropoint, we first use their Table 3 relation 
$\log {M/L_K} = 1.055 (B - V)_0 - 1.066$ to predict an initial, uncorrected $M/L_K$.  This correlates tightly with $\sigma_e$: 
$\log {M/L_K} = 0.239 \log{\sigma_e} - 0.649$ with an RMS scatter of only~0.030.
We then apply the shift $\Delta \log{M/L_K} = 0.1258$ or a factor of 1.34 that makes the corrected Into \& Portinari
mass-to-light ratio agree with the dynamic one, $M/L_K = 1.124$, at $\sigma_e = 250$ km s$^{-1}$.  

      We then have two ways to predict $M/L_K$ that are independent except for the above shift,
\vskip -5pt
$$
\log{M/L_K} = 0.2871 \log{\sigma_e} - 0.6375;\quad\quad\quad{\rm RMS} = 0.088; \eqno{(8)}
$$
\vskip -15pt
$$
\log{M/L_K} = 1.055 (B - V)_0 - 0.9402;\quad\quad{\rm RMS} = 0.030, \eqno{(9)}
$$
where we use the RMS scatter of the correlation with $\sigma_e$ to estimate errors for~the~latter~equation.
We adopt the mean of the mass-to-light ratios given by Equations 8 and 9.  For the error estimate, we
use $0.5\sqrt{0.088^2 + 0.030^2 + ({\rm half~of~the~difference~between~the~two}~\log{M/L_K}~{\rm values})^2}$.  We use the resulting
$M/L_K$ together with $M_{K,\rm bulge}$ to determine bulge masses.  For the $\log{M_{\rm bulge}}$ error estimate,  we add the
above in quadrature to $(0.2/2.5)^2$.  The results are listed in {\bf Tables 2}~and~{\bf 3}.

      {\bf Figure 18} shows the correlation of $M_\bullet$ with bulge mass $M_{\rm bulge}$.  A symmetric,
least-squares fit to the classical bulges and ellipticals omitting the monsters and (for consistency with 
$M_\bullet$\ts--\ts$\sigma_e$), the emission-lime $M_\bullet$ values for NGC 4459 and NGC 4596 plus NGC 3842 and NGC 4889 gives 
the mass equivalent of Equation 6,
\vskip -5pt
$$
\quad{{M_\bullet} \over {10^9~M_\odot}} = \biggl(0.49^{+0.06}_{-0.05}\biggr)\ 
                                           \biggl({{M_{\rm bulge}} \over 
                                                  {10^{11}{\ts}M_{\odot}}}\biggr)^{1.16 \pm 0.08};~
                           {\rm intrinsic~scatter} = 0.29~{\rm dex}. \eqno{(10)}
$$
Thus the canonical BH-to-bulge mass ratio is $M_\bullet/M_{\rm bulge} = 0.49^{+0.06}_{-0.05}$\ts\% at $M_{\rm bulge} = 10^{11}$\ts$M_\odot$.

      This BH mass ratio at $M_{\rm bulge}$\ts=\ts$10^{11}$\ts$M_\odot$ is 2--4 times larger than previous values, which range from 
$\sim$\ts0.1\ts\% (Sani \etal 2011),
0.12\ts\% (McLure \& Dunlop 2002), and
$0.13^{+0.23}_{-0.08}$\ts\% (Merritt \& Ferrarese 2001; Kormendy \& Gebhardt~2001) to
$0.23^{+0.20}_{-0.11}$\ts\% (Marconi \& Hunt 2003).  
The reasons are clear: (1) we omit pseudobulges; these do not satisfy the tight correlations in
Equations\ts2\ts--\ts7; (2) we omit galaxies with $M_\bullet$ measurements based on ionized gas dynamics that do not take broad
emission-line widths into account; (3) we omit mergers in progress.  All three of these tend to have smaller BH  masses 
than the objects that define the above correlations.  Also, the highest BH masses occur in core ellipticals (more on these
below), and these have been revised upward, sometimes by factors of $\sim 2$, by the addition of dark matter to dynamical
models.  Moreover, thanks to papers like Schulze \& Gebhardt (2011) and Rusli \etal (2013), we have~many~such~objects.

      The exponent in Equation\ts10 is slightly larger than 1, in reasonable agreement with H\"aring \& Rix (2004), who got 
$M_\bullet \propto M_{\rm bulge}^{1.12 \pm 0.06}$ and again a lower normalization, BH mass fraction 
$\simeq$\ts15\ts\% at $M_{\rm bulge} = 10^{11}$\ts$M_\odot$.  McConnell \& Ma (2013) get a similar range of exponents from
$1.05 \pm 0.11$ to $1.23 \pm 0.16$ depending on how the bulge mass is calculated (dynamics versus stellar populations).

\vfill \eject

      To better emphasize the slight variation in $M_\bullet/M_{\rm bulge}$, we plot this ratio expressed as a percent
against bulge mass in {\bf Figure 18}.  A direct fit to these points gives
\vskip -28pt
\null
$$
\null\quad\quad100\biggl({{M_\bullet} \over {M_{\rm bulge}}}\biggr) = \biggl(0.49^{+0.06}_{-0.05}\biggr)\ 
                                           \biggl({{M_{\rm bulge}} \over 
                                                  {10^{11}{\ts}M_\odot}}\biggr)^{0.14 \pm 0.08},~~
                        {\rm intrinsic~scatter} = 0.29~{\rm dex}. \eqno{(11)}
$$
\null
\vskip -12pt
\noindent BH mass ratios range from 0.1\ts\% to $\sim$\ts1.8\ts\%, with NGC 4486B and NGC 1277 standing out at 14\ts\% and 17\ts\%,
respectively.  The systematic variation in $M_\bullet/M_{\rm bulge}$ with $M_{\rm bulge}$  is one reason why AGN feedback 
has little effect on galaxy structure at low BH masses and instead becomes important at the largest BH masses (Section 8).
Note:~the RMS scatter $\Delta \log{M_\bullet} = 0.327$ in {\bf Figure\ts18} ({\it top}) is only marginally smaller than $\Delta \log{M_\bullet} =
0.341$ in the luminosity correlation ({\bf Figures 16} and {\bf 17}).  Conversion from $L_{K,\rm bulge}$ to $M_{\rm bulge}$ does 
not make much difference for old stellar populations.

\vfill


\includegraphics{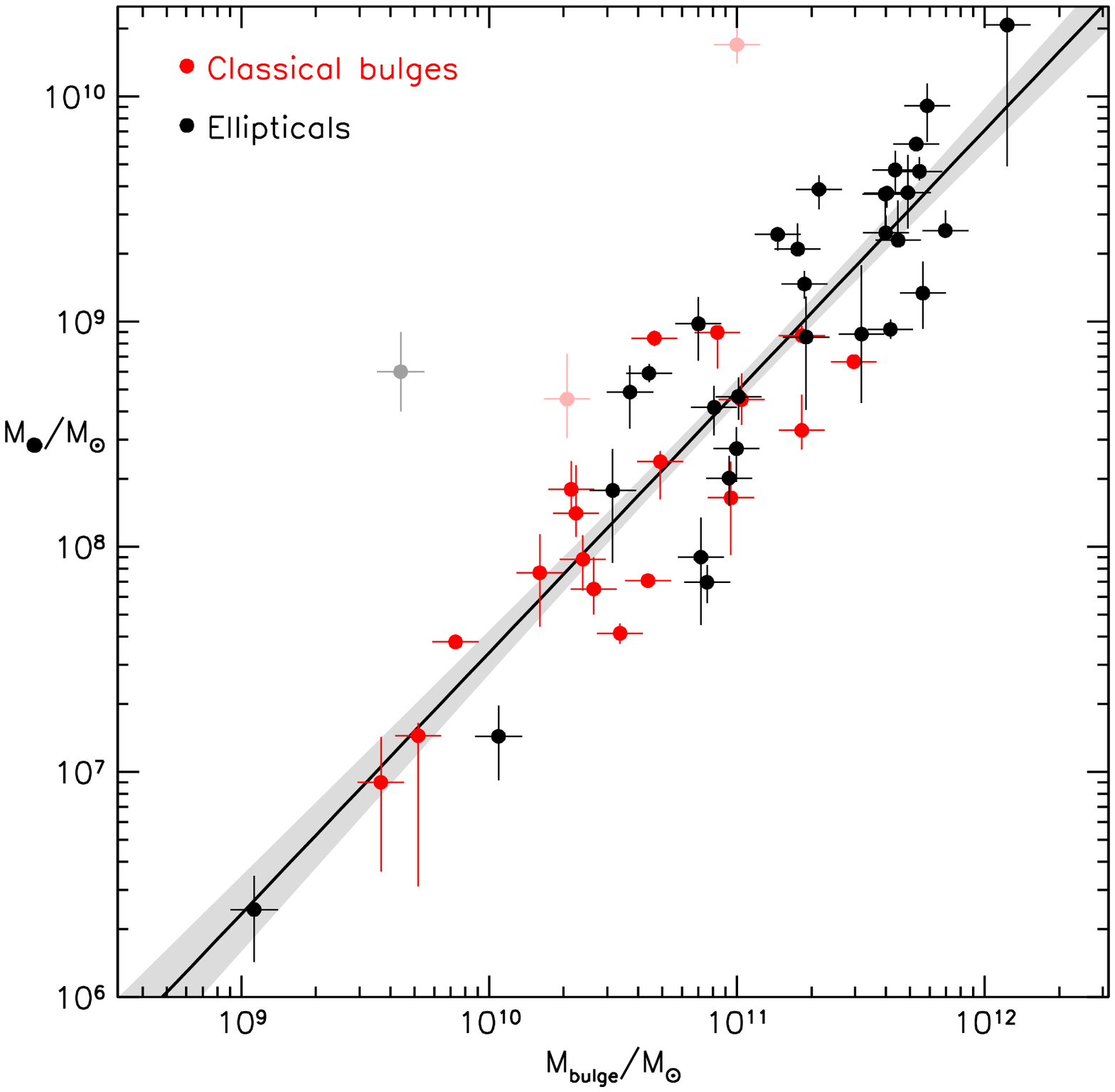}
\includegraphics{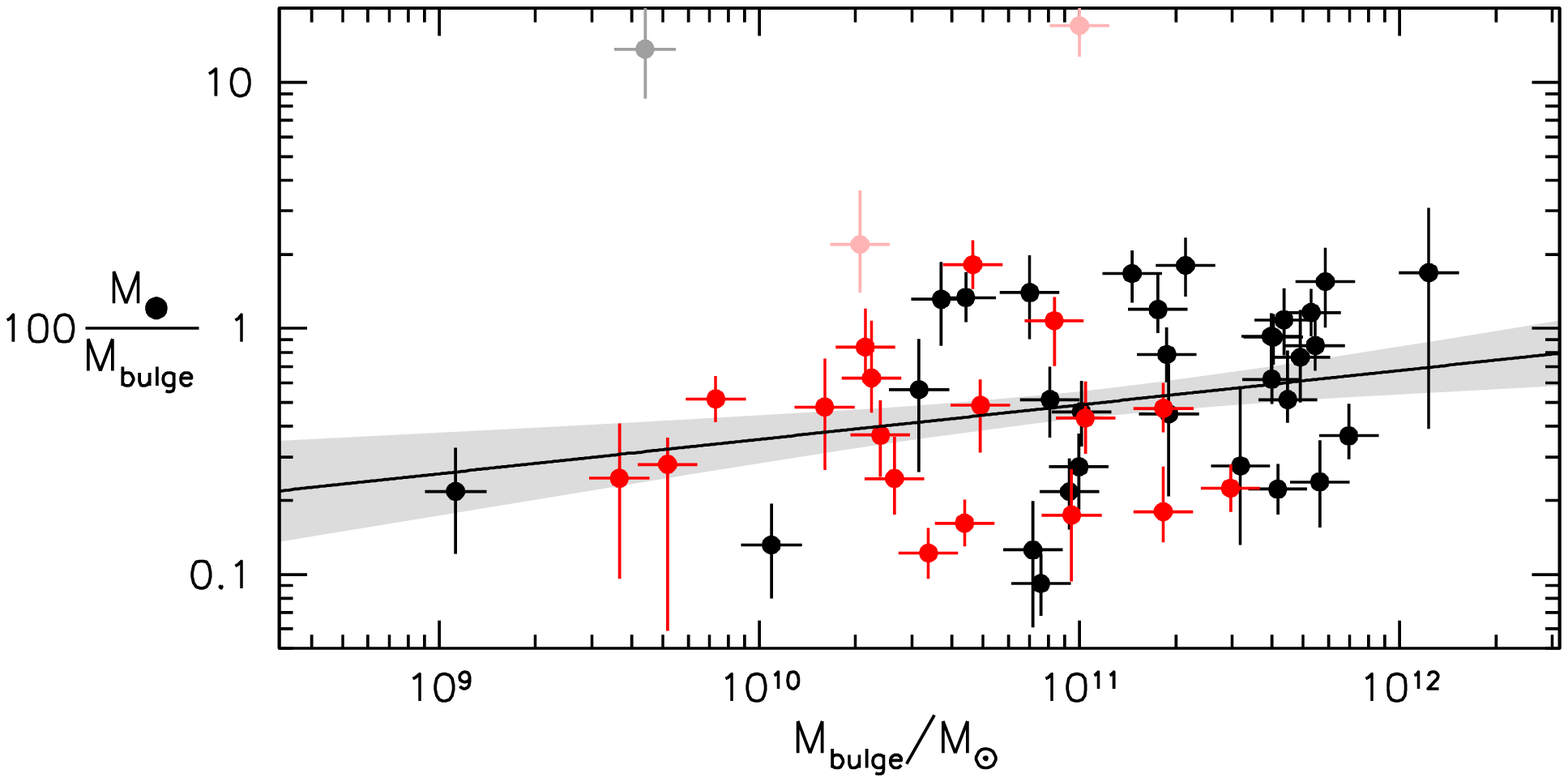}

\ni {\bf \textBlue Figure 18}\textBlack 

\vskip 1pt
\hrule width \hsize
\vskip 2pt
\ni ({\it top}) BH mass and ({\it bottom}) percent ratio of BH mass to bulge mass as functions of bulge mass.  The lines
    are Equations\ts10 and 11.   The scatter in $M_\bullet/M_{\rm bulge}$ is larger than the systematic variation. 

\eject    

      The history of our developing understanding of the upward revision in $M_\bullet/M_{\rm bulge}$ is an interesting
sanity check.  Early in the writing of this paper, when we first separated out pseudobulges and considered them separately from
the BH--host correlations, we were uncomfortable to find that classical bulges were offset toward high $M_\bullet$ with respect
to ellipticals.  We were, of course, not entitled to assume that this was a problem -- whether bulges and ellipticals have the 
same correlations is a scientific question that we need to ask.  But the hint made us uncomfortable, especially since Gebhardt
\& Thomas (2009) had recently revised $M_\bullet$ upward in M{\ts}87, and soon afterward, van den Bosch \& de Zeeuw (2010) revised
$M_\bullet$ upward in NGC 3379.  It was likely that $M_\bullet$ values in core galaxies would generally get revised upward, so we
delayed the completion of this paper for several years until this was done.  With the $M_\bullet$ revisions in Schulze \& Gebhardt (2011)
and the many new and large $M_\bullet$ values published by McConnell and Rusli and collaborators, the offset between classical bulges
and ellipticals vanished.  But then stellar-dynamical BH masses in classical bulges and ellipticals started convincingly to disagree 
with BH masses based on ionized gas kinematics, only, however, when the $M_\bullet$ derivation did not take broad emission-line widths into 
account ({\bf Figure 12}).  Our suspicions about those masses therefore crystallized and we began to omit them.  Concurrently, the 
mergers in progress ({\bf Figure 14\/}) began to disagree convincingly with the revised correlations, too.  In this way, we converged on the 
correlations in {\bf Figures 16} and {\bf 17} and the new, larger mean ratios of BH mass to bulge mass in {\bf Figure 18}.

      We emphasize that Equations 10 and 11 apply only to the classical bulges and ellipticals that participate in the tight BH--host-galaxy 
correlations.  These dominate the BH mass budget.  But small
BHs in pseudobulges and in bulgeless galaxies have systematically smaller BH  masses and BH-to-pseudobulge mass ratios, and these may 
be the most common BHs in the universe.

\vs
\ni \hbox{{\bf\ARRed 6.6.2.{\ts}Comparison with Other Studies of BH--Host-Galaxy Correlations}}~\textBlack Explorations of the 
$M_\bullet$\ts--\ts$L_{\rm bulge}$, 
$M_\bullet$\ts--\ts$M_{\rm bulge}$, and
$M_\bullet$\ts--\ts$\sigma_e$ 
correlations have become a major industry as BH samples have increased in size, as $M_\bullet$ estimates 
have become available for AGNs through methods such as reverberation mapping and single-epoch spectroscopy of broad emission
lines, and as opportunities have become available to do photometry in the infrared.  A partial list since 
Kormendy \& Gebhardt (2001) includes
Tremaine~et~al.~(2002);
McLure~\&~Dunlop (2002);
Marconi \& Hunt (2003);
H\"aring \& Rix (2004);
Ferrarese \& Ford (2005);
Greene \& Ho (2006b, 2007c);
Aller \& Richstone (2007);
Greene, Ho, \& Barth (2008);
G\"ultekin \etal (2009c);
Kormendy, Bender, \& Cornell (2011);
Graham \etal (2011);
Sani \etal (2011);
Vika \etal (2012);
Graham \& Scott (2013); and
McConnell \& Ma (2013).  Comparison of these papers with each other and with our results would take more space than we can afford
and would be less illuminating than a few general~comments. 

      With one main exception, published studies get shallower slopes than we do for all~3~correlations.
Reasons:~(1) 
    they include giant Es for which we believe that $M_\bullet$ has been underestimated from emission-line rotation curves;
(2) they include NGC 1316 or NGC 5128, which have low BH masses for their high bulge masses;
(3) studies before 2011 have few or no BH masses that have been corrected upward as a result of the inclusion of dark matter in dynamical models;
(4) even later studies do not generally have all the extraordinarily high-$M_\bullet$ objects published recently by McConnell,
    Gebhardt, Rusli, and collaborators.
McConnell \& Ma (2013) is closest to our study; they include the Rusli galaxies.    Their fits to early-type galaxies give slopes of 
$1.11 \pm 0.13$ (we get $1.21 \pm 0.09$) for $M_\bullet$\ts--\ts$L_{\rm bulge}$ and 
$5.08 \pm 0.38$ (we get $4.38 \pm 0.29$) for $M_\bullet$\ts--\ts$\sigma_e$.  For the above reasons and because we omit pseudobulges, published
studies also generally get lower $M_\bullet$ zeropoints.

      Published studies get steeper slopes than we do when they include \hbox{pseudobulges or the high-$M_\bullet$} BHs in NGC 3842 and NGC 4889
or both.  Pseudobulge BHs deviate from the correlations primarily to low $M_\bullet$, and they do so at small pseudobulge masses.  Including them
steepens the correlations.  Also, we will see in the next section what is already apparent in {\bf Figures 16} amd {\bf 17}:  The 
$M_\bullet$\ts--\ts$\sigma_e$ correlation ``saturates'' at the high end such that $M_\bullet$ becomes essentially independent of $\sigma_e$.  
We found it sufficient to account for this by excluding only NGC 3842 and NGC 4889 from the fits.  Including them steepens
$M_\bullet$\ts--\ts$\sigma_e$ for the complete sample of galaxies.  McConnell \& Ma (2013) find this when they
get an $M_\bullet$\ts--\ts$\sigma_e$ slope of $5.57 \pm 0.33$ and Graham and Scott (2013) find it when they get 
$6.08 \pm 0.31$ both for their complete galaxy samples.
Even when Graham \& Scott (2013) exclude barred galaxies, they get a slope of $5.14 \pm 0.31$.  As we have noted, some SB galaxies
contain classical bulges and many unbarred galaxies contain pseudobulges, so excluding barred galaxies does not more cleanly isolate
unique physics than does not excluding them.

     In this paper, we do not include fits that mix classical and pseudo bulges.  We strongly believe that we learn the most
physics when we make fits only to collections of objects that are sufficiently similar in formation and structure.  The fits
themselves tell us when this happens.  E.{\ts}g., when two samples that are distinguished on physical grounds
have different correlations (classical bulges here and pseduobulges in Section 6.8), then finding this out supports
the physical distinction.  Similarly, when a correlation that is essentially linear over most of its dynamic range suddenly
shows a kink at one end, this is plausibly a sign that new physics is involved.  This happens in the next section, and it
proves to be easy to diagnose the new physics.

\vs
\ni {\big\ARRed 6.7 The Distinction Between Core and Coreless Ellipticals:}\par
\ni {\big\ARRed \phantom{6.7}~Dry Mergers and the Saturation of the 
     M\lower.3ex\hbox{$\bullet$}\ts--\ts$\sigma$\lower2pt\hbox{\almostbig\kern -1.2pt e} Correlation}\textBlack
\vs

      Ellipticals are divided into two kinds that differ in physical properties (see Kormendy 2009 for a summary and KFCB 
for a detailed review).  This difference is important here because its origin explains --
we suggest -- an important feature in the $M_\bullet$\ts--\ts$\sigma_e$ correlation.   Also, it will turn out in Section 8 
that the two kinds of ellipticals have different forms of AGN feedback.

\vs
\noindent \ARRed {\bf 6.7.1 Two kinds of elliptical galaxies}\textBlack~ As illustrated in Section\ts6.13,
giant ellipticals~have~cores: their brightness profiles show a break toward small radii from steep outer 
S\'ersic functions to shallow inner cusps.  In contrast, Virgo ellipticals that lack cores have extra light 
at small radii above the inward extrapolation~of~the outer S\'ersic profile.  The extra light is interpreted 
as the remnant of the central starbursts that happen in wet mergers 
(Mihos \& Hernquist 1994; 
Kormendy 1999; 
C\^ot\'e \etal 2007;
KFCB; 
Hopkins \etal 2009a).  
Faber \etal (1997) show that core galaxies generally have boxy isophotes and rotate slowly, whereas coreless galaxies 
generally have disky isophotes and rotate rapidly.  Lauer (2012) confirms that the distinction between core and coreless 
galaxies is essentially the same as the distinction between nonrotating and rotating ellipticals made by the SAURON and 
ATLAS$^{\rm 3D}$ teams (Emsellem \etal 2007, 2011; Cappellari \etal 2007, 2011).  Classical bulges are essentially 
equivalent to coreless ellipticals.  

{\ARRed\bf Coreless-disky-rotating ellipticals}~\textBlack in the Virgo cluster have $V$-band absolute magnitudes 
$M_V$\ts\gapprox\ts$-21.5$ and stellar masses $M_*$\ts\lapprox\ts2\ts$\times$\null$10^{11}$\ts$M_\odot$.  KFCB summarize 
the evidence that they formed by dissipative mergers that included central starbursts.~Local prototypes are ULIRGs 
like Arp\ts220.

{\ARRed\bf Core-boxy-nonrotating ellipticals}~\textBlack have $M_V$\ts\lapprox
\ts$-21.5$.~Their most recent (one or more) major mergers were dissipationless.
The relative importance of major and minor mergers is much debated in recent literature, 
so we present one piece of evidence below for the importance of major mergers. 

      The transition between coreless and core galaxies happens at a stellar mass that corresponds
to a dark matter mass that is somewhat greater than $M_{\rm DM} \simeq 10^{12}$\ts$M_\odot$ $\equiv$ $M_{\rm crit}$ that is invoked to control
``$M_{\rm crit}$ quenching'' of star formation by hot, X-ray-emitting gas  
(Dekel \& Birnboim 2006;
Cattaneo \etal 2006, 2009;
Faber \etal 2007).  Core ellipticals contain large amounts of X-ray gas and coreless galaxies do not; KFCB suggest that this is
the reason for the difference between the two kinds.  Section 8.4 reviews
the evidence for this and discusses implications for AGN feedback.  Here, we are interested in the implications of the
two kinds of ellipticals for BH--host-galaxy correlations.

\vfill\eject

\cl{\null}

\vskip 2.1truein

 \includegraphics{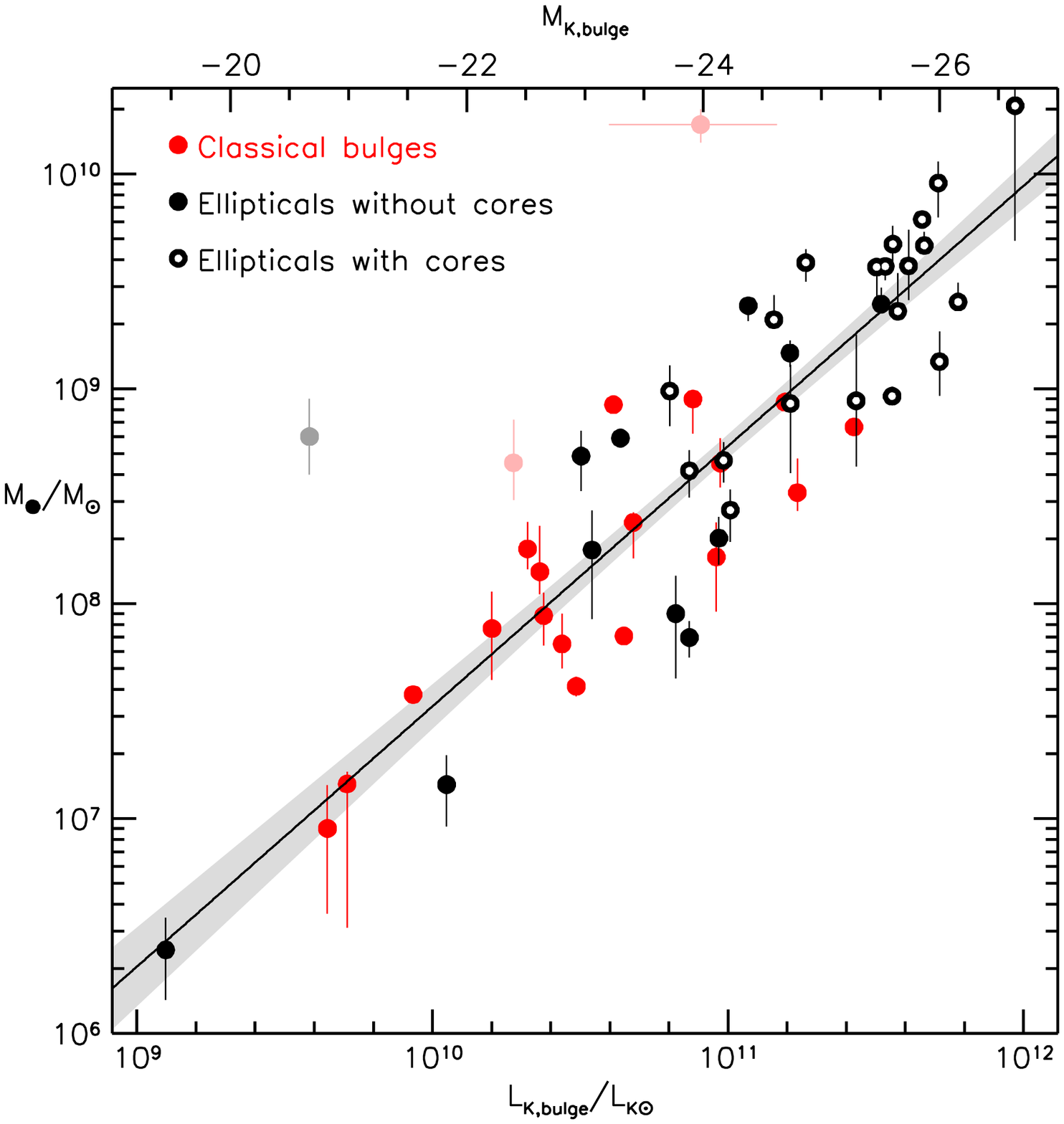}
 \includegraphics{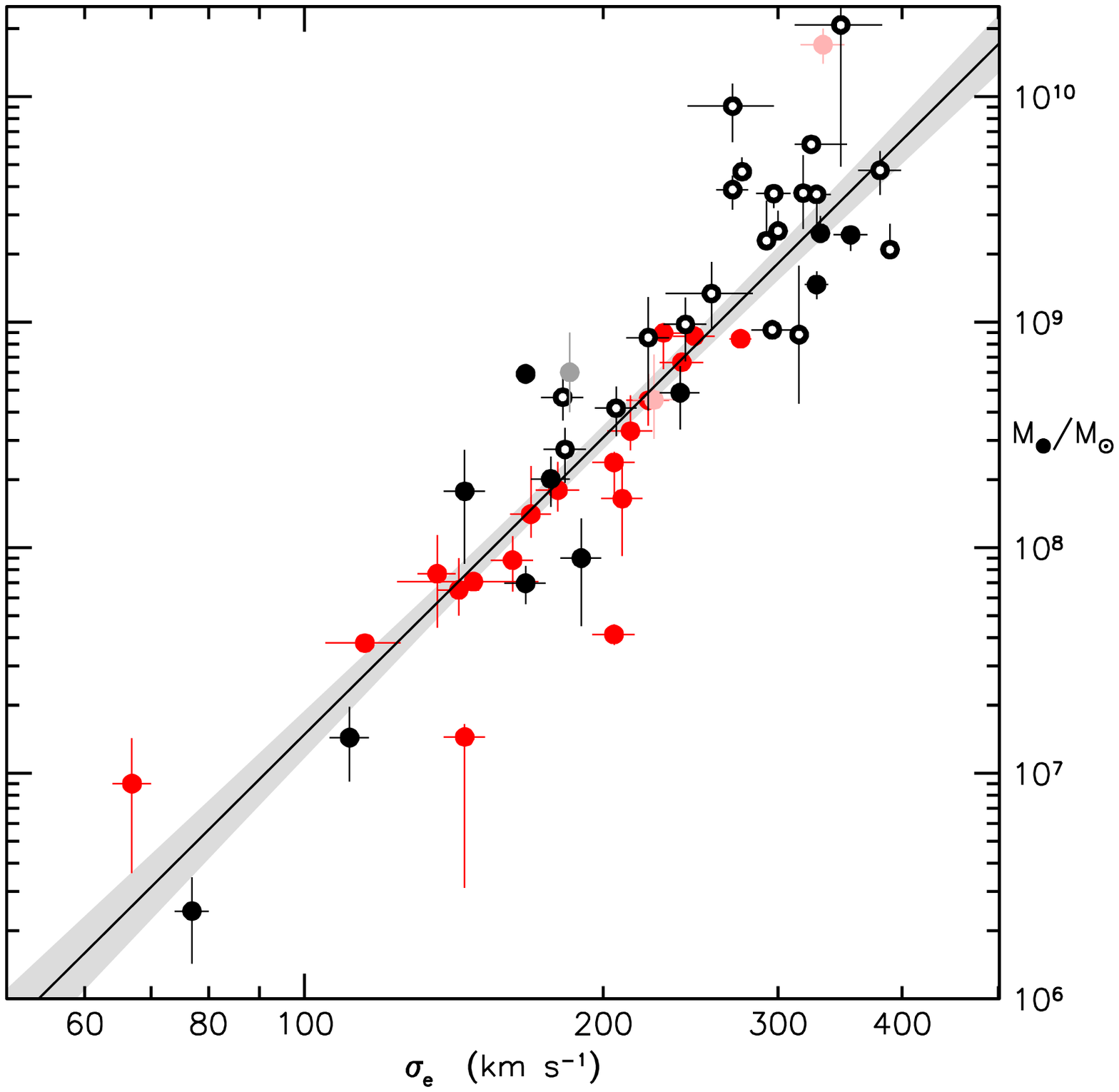}

\ni {\bf \textBlue Figure 19}\textBlack 

\vskip 2pt
\hrule width \hsize
\vskip 3pt
\ni BH correlations and fits from {\bf Figure\ts17} with core Es identified.  The $M_\bullet$\ts--\ts$L_{K,\rm bulge}$
relation remains log-linear at high $M_\bullet$, but the biggest BHs have masses that become essentially independent of $\sigma_e$.
    
\vs

\noindent \ARRed {\bf 6.7.2 The M\lower0.5ex\hbox{$\bullet$}\ts--\ts$\sigma\lower0.5ex\hbox{\kern -0.8pt e}$ relation saturates at high 
                                                                    $\sigma\lower0.5ex\hbox{\kern -0.8pt e}$}\textBlack~ {\bf Figure 19}
suggests \hbox{that the $M_\bullet$\ts--\ts$\sigma_e$} relation becomes vertical at $\sigma$\ts\gapprox 270 km s$^{-1}$.  The sample is small, but 
indications are that core ellipticals are responsible.   That is, for core ellipticals, $M_\bullet$ becomes almost independent of $\sigma_e$ 
even while the correlation with bulge luminosity remains unchanged to the highest luminosities.  

      Hints of this effect were already seen as a result of the upward revision of the BH~mass~for~M{\ts}87 (Gebhardt \etal 2011).
It has the third-highest $M_\bullet$ in {\bf Figure 19}.  At that time, M{\ts}87 stood out more than it does now, because the similar-$M_\bullet$
Rusli \etal (2013) BH points were not available and because BH masses based on ionized gas rotation curves were still being included. 
The $\sigma_e$ saturation was more convincingly found for the two highest-$M_\bullet$ BHs in NGC 3842 and NGC 4889 by their discoverers 
(McConnell \etal 2011a, 2012).  Then, Kormendy (2013) and McConnell \& Ma (2013) note that the upward kink in 
$M_\bullet$\ts--\ts$\sigma_e$ at high $\sigma_e$ involves core galaxies.  Kormendy (2013) and {\bf Figure\ts19} further show that the three 
coreless galaxies at $\sigma_e > 300$ km s$^{-1}$ -- a tiny sample, to be sure -- do not participate in this effect.

      Most high-$\sigma_e$ core galaxies scatter to the high-$M_\bullet$ side of the fit in {\bf Figure 19} and in this sense contribute to 
the impression that $\sigma_e$ saturates.  However, in fact, only NGC 3842 and NGC 4889 are significantly outside the upper envelope of the 
scatter, and only they were omitted~from~our~fits.  We checked that the other high-$M_\bullet$ BHs do not significantly bias our correlation 
slope by also fitting only the galaxies that have $\sigma_e < 270$ km s$^{-1}$.~These remaining 29 galaxies give
\vskip -28pt
\null
$$
\log\biggl(\kern -2pt{{M_\bullet} \over {10^9{\ts}M_\odot}}\kern -2pt\biggr) 
         = -(0.54 \pm 0.07) + (4.26 \pm 0.44) \log \biggl({{\sigma} \over {200~{\rm km~s}^{-1}}}\biggr);~{\rm intrinsic~scatter} = 0.30. \eqno{(12)}
$$ 
\vskip -5pt
\noindent The zeropoint and slope are closely similar to those, $-0.51 \pm 0.05$ amd $4.38 \pm 0.29$, in Equation 3.

      The saturation of $M_\bullet$\ts--\ts$\sigma_e$ is only now becoming apparent directly via the detection of \hbox{higher-$M_\bullet$}
BHs than fits predict, but this effect has been known indirectly for a long~time.  The ``contradiction between the
$M_\bullet$--$\sigma$ and $M_\bullet$--$L$ relationships'' is illustrated for a large sample of galaxies in Figure 2 of Lauer \etal (2007a).
It plots $M_\bullet$ predictions from the two relations against each other and shows that  $M_\bullet$--$L$ predicts $M_\bullet$ up
to $\sim$\ts$10^{10}$\ts$M_\odot$, whereas $M_\bullet$--$\sigma$ predicts~almost~no $M_\bullet$\ts$>$\ts2\ts$\times$\ts$10^9$\ts$M_\odot$.
The reason is well known:~the Faber-Jackson (1976) $L \propto \sigma^m$ correlation, $m \sim 4$, ``level[s] off for large~$L$; indeed, 
there appears to be little relationship between $\sigma$ and $L$ for galaxies with $M_V < -22$'' (Lauer \etal 2007a).  
Not coincidentally, this is where ellipticals switch from coreless galaxies to ones with cores.~The saturation of the 
Faber-Jackson relation in the biggest galaxies has been noted before
(Davies \etal 1983; 
Oegerle \& Hoessel 1991;
Matkovi\'c\ \& Guzm\'an 2005).
It is directly responsible for the saturation in $M_\bullet$\ts--\ts$\sigma_e$ that is now seen in {\bf Figure 19}, as Lauer \etal (2007a)
predicted.

     High-$L$ saturation in the Faber-Jackson relation is seen in merger simulations by Boylan-Kolchin, Ma, \& Quataert (2006).
They found that merger orbits become more radial for the most massive galaxies.  Then, dry mergers feed more orbital energy into internal energy
and help to puff up the remnant and thereby keep $\sigma$ from growing.~They, too, predicted that $M_\bullet$\ts--\ts$\sigma_e$ 
saturates at high~$L$.

     {\bf Figure 20} updates the Faber-Jackson relation for core and coreless ellipticals (from Kormendy \& Bender 2013a).
It is similar to Figure 3 in Lauer \etal (2007a) with a few core classifications and Hubble types corrected (KFCB) and a 
few low-luminosity ellipticals added.~The~coreless~galaxies show the familiar relation $\sigma$\ts$\propto$\ts$L^{1/4}$.
But the relation is very shallow for core Es, $\sigma \propto L^{0.12 \pm 0.02}$.  Also, Kormendy \& Bender (2013a)
show that coreless galaxies have the above, steep~slope~and core galaxies have the above, shallow slope even in the absolute
magnitude range where~they~overlap.  The difference in slope in {\bf Figure 20} and hence the $M_\bullet$--$\sigma_e$ saturation 
in {\bf Figure 19} result from different formation physics for core and coreless ellipticals.

      A shallow slope for core ellipticals arises naturally if they form by dry major (not minor) mergers.
Hilz \etal (2012) numerically simulate equal-mass mergers and remergers of bulges embedded in dark matter halos.
They show that successive mergers lead to a slow increase in the stellar velocity dispersion, because violent relaxation
broadens the energy distribution, so some bound particles become more tightly bound while some weakly 
bound particles escape.  They find that the effective line-of-sight velocity dispersion increases as mergers increase
the galaxy mass as $\sigma \propto M^{0.15}$.  They also present results for growth by minor mergers.  Such growth happens 
mostly by adding a low-$\sigma$ halo at large radii, so it leads to a decrease in the total projected velocity dispersion, 
$\sigma \propto M^{-0.05}$.  

      If we convert $\sigma$\ts$\propto$\ts$L^{0.12 \pm 0.02}$ to a relation in mass $M$ via $M/L$\ts$\propto$\ts$L_I^{0.32 \pm 0.06}$ 
(Cappellari{\ts}et{\ts}al.{\ts}2006), we get $\sigma \propto M^{+0.09 (+0.03, -0.02)}$, similar to the Hilz and Boylan-Kolchin results.
Kormendy \& Bender (2013a) conclude that the shallow $\sigma$\ts--\ts$L$ correlation observed for core galaxies is consistent with their 
formation in major (not minor) mergers.  This is consistent with KFCB.

      We noted previously that core ellipticals are so massive that plausible immediate progenitors must have been bulge-dominated.
Then we expect that each progenitor brings a BH to the merger and the galaxy and BH both grow together during the merger.  But
those mergers have relatively little effect on $\sigma_e$ (see also Volonteri \& Ciotti 2013).
Thus we suggest that $\sigma_e$ saturation becomes noticeable once a significant number of dry mergers have happened in a galaxy's
formation history.  Thus it seems reasonable that the faintest core ellipticals do not show the effect.

      Another argument for dry major mergers is made in Section 6.13, where we note that BH binary mass ratios must be $\sim 1$
in order to lift a sufficient mass in stars to account for cores.  And the mergers must be dry in order not to be swamped by
a ULIRG-like starburst.

\vsss


\includegraphics{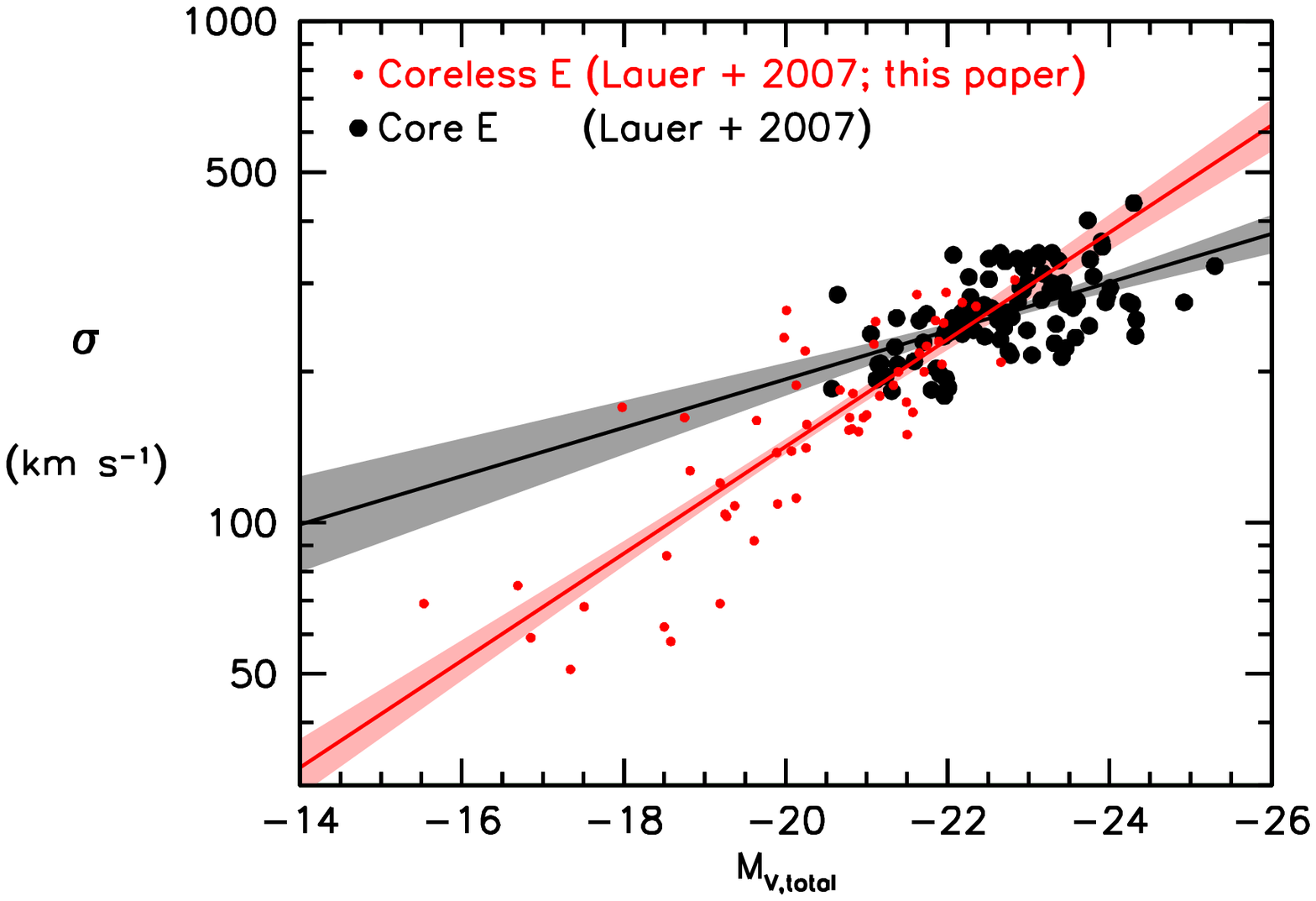}

\def\figindent{\indent \hangindent=3.0truein}

{\parindent=3.0truein
\figindent \ni {\bf \textBlue Figure 20}\textBlack 

\vskip 3pt
\nointerlineskip \moveright 3.0truein\vbox{\hrule width 2.9truein}\textBlack\nointerlineskip
\vskip 3pt

\figindent  Faber-Jackson (1976) relations for core and coreless ellipticals (Kormendy{\ts}\&{\ts}Bender\ts2013a).   
    Total $V$-band absolute magnitudes $M_{V, {\rm total}}$, velocity dispersions $\sigma$, and profile types are 
    mostly from Lauer{\ts}et{\ts}al.{\ts}(2007b).  The lines are symmetric least-squares fits to core ({\it black line\/}) and 
    coreless galaxies ({\it red line\/}).  One-sigma fit uncertainties are shaded.  The coreless galaxies show the 
    familiar relation, $\sigma \propto L_V^{0.27 \pm 0.02}$.  But velocity dispersions in core ellipticals increase
    very slowly with luminosity, $\sigma \propto L_V^{0.12 \pm 0.02}$. 

}
 
\eject

\vs
\ni {\big\ARRed 6.8 BHs do not correlate with pseudobulges of disk galaxies}\textBlack
\vs

      Section 4 discusses the observational distinction between classical bulges, which are essentially indistinguishable 
from ellipticals, and pseudobulges, which usually have properties that are more disk-like than those of classical bulges.
We believe that we understand why: Classical bulges are made by major galaxy mergers, whereas pseudobulges are grown secularly 
out of disks.  But for this section, an explanation is not essential.  The Supplementary Information expands on Kormendy \& Kennicutt (2004) 
by listing observational criteria that can be used to distinguish between classical and pseudo bulges, independent of interpretation.  
Do they correlate similarly with BHs?

      Papers on this issue are divided.  Kormendy \& Gebhardt (2001) were the first to ask the
question.  They concluded that classical and pseudo bulges correlate in the same~way~with~BHs.  They were wrong,
because, with data available then, the classical bulges in NGC 4258 and NGC 7457 were mistakenly called
pseudobulges and because their sample included only six (after correction: four) pseudobulges.  G\"ultekin 
\etal (2009c) also did not see a significant difference, but their sample also was small, and 14 of their 23
disk galaxies had no available (pseudo)bulge parameters. 

      Other papers find different BH correlations for bulges and pseudobulges.  This was first suggested by Hu (2008).  
He examined only the \hbox{$M_\bullet$ -- $\sigma$} correlation, which is relatively safe, because it does not require 
bulge-disk decomposition.  He used the pseudobulge classification criteria from Kormendy \& Kennicutt (2004).  His sample 
Tables 1 and 2 do not completely overlap with our {\bf Tables 2}~and~{\bf 3}, because a few different judgments were made 
about reliable BH detections.  But the sample was large enough and accurate enough to suggest that pseudobulges have
smaller $M_\bullet$ at a given $\sigma$ than do classical bulges and ellipticals.  This proves to be a robust result.

      A possible indirect detection of a difference is by Graham (2008a,\ts2008b) and Graham\ts\&{\ts}Li{\ts}(2009).
They argued that the correlations are different for barred galaxies than they are for unbarred disk galaxies
and ellipticals.  The sense of the difference is similar to that found by Hu (2008): $M_\bullet$ is anomalously
small or $\sigma$ is anomalously large in barred galaxies.  Graham adopted the latter explanation and suggested that 
velocity dispersions are enhanced in barred galaxies.  This could be a detection of the effect that was seen by Hu, 
because barred galaxies preferably have pseudobulges (Kormendy \& Kennicutt 2004).  However, of Graham's (2008a) 
seven barred galaxies, three have classical bulges, and of his unbarred disk galaxies, at least one (NGC 1068) 
contains a prominent pseudobulge.  Similar comments apply to the 2008b paper and to Graham \& Scott (2013).  Graham \& Li (2009) 
derived BH masses from emission-line widths in AGNs; less is known about these more distant objects, but they are also offset 
from $M_\bullet$\ts--\ts$\sigma$ in the direction of low $M_\bullet$ or high $\sigma$.

      Seven BH discoveries based on H$_2$O maser dynamics were recently published by Kuo~et~al.~(2011).  The host galaxy velocity 
dispersions were measured and their brightness distributions analyzed by Greene \etal (2010).  They decisively concluded that the 
pseudobulges in their sample have smaller BH masses than predicted by the $M_\bullet$ -- $\sigma$ correlation for classical bulges 
and ellipticals.  Our bulge-disk decompositions do not completely agree with theirs, but their sample is well suited to the problem, 
because maser galaxies tend to be late in type and therefore to contain pseudobulges. 

      Meanwhile, Kormendy, Bender, \& Cornell (2011) measured (pseudo)bulge parameters for the full G\"ultekin \etal (2009c) sample 
and used them to conclude that BHs do not correlate significantly with either the luminosities or the velocity dispersions 
of their host pseudobulges.  ``Pseuodbulges at best show a much larger scatter [than do classical bulges and ellipticals].''
``Whether pseudobulges correlate with $M_\bullet$ with large scatter or not at all, the weakness of any correlation \dots~makes no 
compelling case that pseudobulges and BHs coevolve, beyond the obvious expectation that it is easier to grow bigger BHs and bigger
pseudobulges in bigger galaxies that contain more fuel.''

       {\bf Figure~21} confirms these results with the new sample from {\bf Table 3}.  The total amount of BH growth is not extremely
different in classical and pseudo bulges.  But the scatter in all panels~is~large.

\vfill\eject

\cl{\null}

\vskip 2.4truein

 \includegraphics{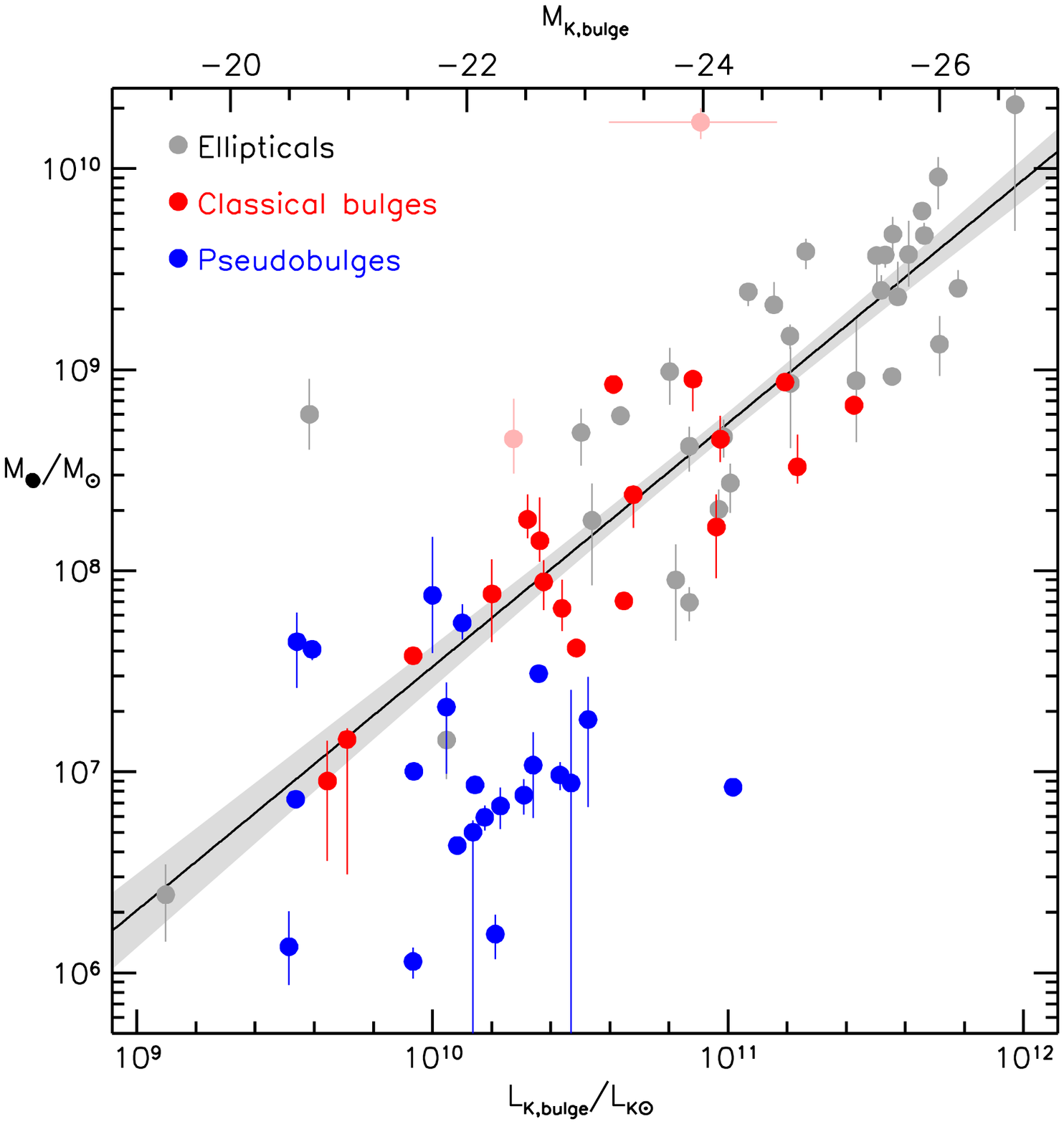}
 \includegraphics{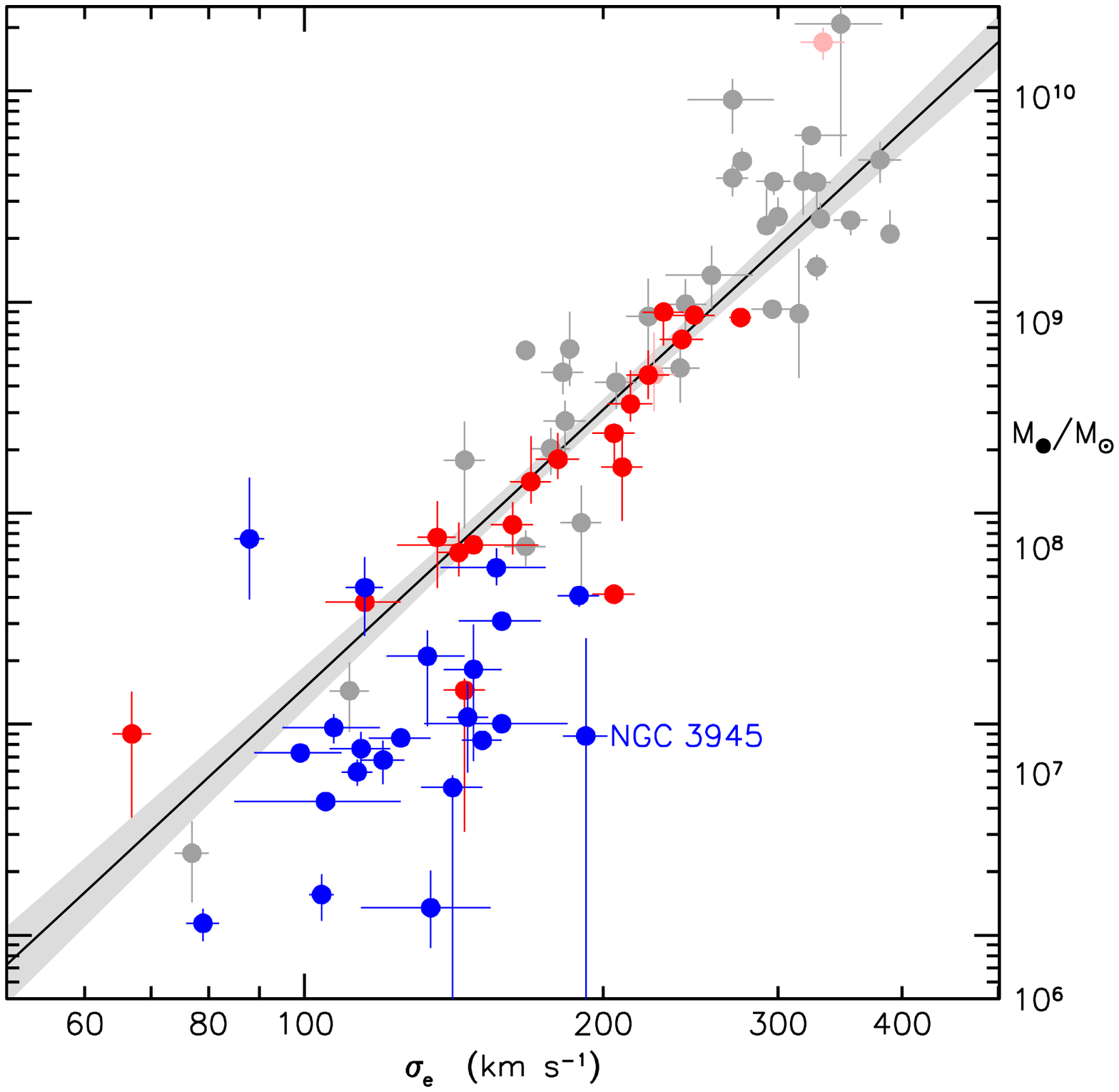}

\vskip 0.05truein

 \includegraphics{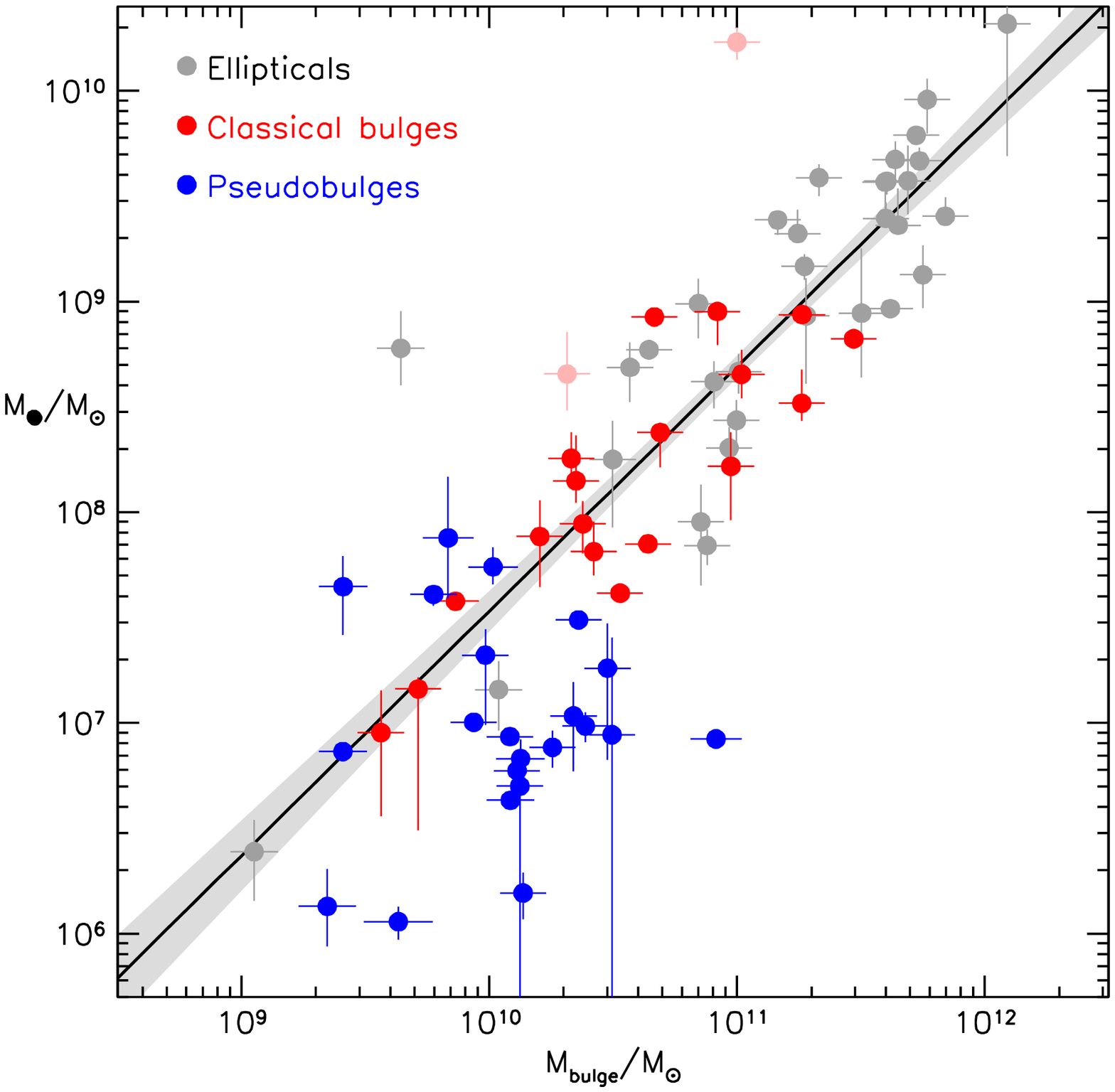}

\def\figindent{\indent \hangindent=3.0truein}

{\parindent=3.0truein
\figindent \ni {\bf \textBlue Figure 21}\textBlack 

\vskip 3pt
\nointerlineskip \moveright 3.0truein\vbox{\hrule width 2.9truein}\textBlack\nointerlineskip
\vskip 3pt

\figindent  BH correlations ({\bf Figures 17} and {\bf 18}) including ({\it blue points}) pseudobulges~with~dynamical BH masses. 
            Ellipticals are plotted in gray because  we want to emphasize the contrast between classical and pseudo bulges.
            In all panels, pseudobulge BHs are offset toward smaller $M_\bullet$ from the correlations for
            classical bulges and ellipticals.  Because star formation is common in pseudobulges, we worry that the offset in
            the top-left panel is caused by young stars.  But the same offset is seen against bulge mass ({\it bottom panel\/}).
            Recall that we designed our $M/L_K$ estimator to work as well as possible for a wide range of stellar populations.  In any case,
            seven~of~the~22 pseudobulge galaxies are S0s or Sas and do not have large amounts of star formation.  

}

\vfill

\noindent  The same result is seen by Sani \etal (2011) and indirectly (for late-type galaxies) by McConnell \& Ma (2013).
           Also, BH searches in pseudobulges appear to fail more often than those in classical bulges and ellipticals (e.{\ts}g., 
           NGC 3945: G\"ultekin \etal 2009b).  For these galaxies, it is still possible that we detect only the largest-mass BHs and therefore 
see the upper envelopes of $M_\bullet$ distributions that extend down to smaller BHs than we can currently find (Barth, Greene, \& Ho 2005).  

      Section 7.2 extends the $M_\bullet$ dynamic range by considering BH masses determined for AGNs.  Taken together, AGN BHs of all masses do
show some correlation with host properties even down to $M_\bullet \sim 10^5$ $M_\odot$.    But the scatter at low $M_\bullet$ is very large.  
No coevolution is implied.  Nevertheless, we emphasize: We do not know whether the decrease in BH correlation scatter as we move from pseudobulges 
to classical bulges to ellipticals and as we look at larger $M_\bullet$ and $M_{\rm bulge}$ is due only to merger averaging (Section 8.5) or
whether additional coevolution physics -- presumably in wet mergers (Section 8.6) -- is required.  What we need most is more BH detections in 
small classical bulges and ellipticals to see whether and how much the scatter in their $M_\bullet$\ts--{\ts}host  correlations is smaller than 
that for pseudobulges in the BH mass range where they overlap.

\eject

\vs
\ni {\big\ARRed 6.9 BHs do not correlate with galaxy disks}\textBlack
\vs

      {\bf Figure 22} confirms the results of Kormendy \& Gebhardt (2001) and Kormendy \etal (2011)
that BHs do not correlate with galaxy disks.  The Kormendy \& Gebhardt (2001) conclusion was based on showing
that the good $M_\bullet$ -- $M_{B,\rm bulge}$ correlation essentially disappears when the total absolute
magnitude is used instead of the bulge absolute magnitude.  The Kormendy \etal (2011) result is more direct:
it is based on a version of {\bf Figure 22} for the G\"ultekin \etal (2009c) sample.  The full sample
of {\bf Table\ts3} strengthens this result.~Many pseudobulges are in late-type, star-forming disks;
if $M_{K,\rm disk}$ were converted to disk mass, the corresponding blue points would move leftward with 
respect to the red points and the weak anticorrelation of $M_\bullet$ with disk luminosity would get weaker.  
Thus BHs do not ``know about'' disks at all.  This extends
previous results: BHs somehow coevolve with bulges but not with disks or with the total baryonic content of their
host galaxies.

      And yet, we now know that BHs can exist even in pure-disk galaxies.  Some pure disks 
show AGN activity, although it is rare in bulgeless galaxies (Ho, Filippenko \& Sargent 1997b;~Ho 2004b,~2008).  And BHs 
with $M_\bullet$ = 10$^4$\ts--\ts10$^6${\ts}$M_\odot$ have confidently been discovered in (pseudo)bulge-less 
galaxies.  The most extreme case is NGC\ts4395, an Sm galaxy with a tiny, \hbox{globular-cluster-like} nucleus that has an 
AGN powered by a BH (Filippenko \& Ho 2003) with $M_\bullet$ = (3.6 $\pm$ 1.1) $\times$ 10$^5$~$M_\odot$ measured by 
reverberation mapping (Peterson \etal 2005).  It is shown in {\bf Figure 22}.  Similar objects include NGC 1042 (Scd; 
Shields \etal 2008), NGC 3621 (Sd; Barth \etal 2009), NGC 4178 (Sdm; Secrest \etal 2012).  Also, the Sph galaxy POX 52
contains an AGN powered by a BH of mass $M_\bullet$\ts$\simeq$\ts10$^5$\ts$M_\odot$ 
(Barth \etal 2004; Thornton \etal 2008).  See Section 7.1 for details.

      The lack of correlation of BHs with disks and pseudobulges plus the discovery of BHs in 
(pseudo)bulgeless galaxies are critical clues to BH evolution.  They motivate 
the hypothesis in Section 8 that there are two feeding mechanisms for BHs, (1) a global feeding mechanism
that is connected with bulge formation and therefore presumably with galaxy mergers and that engineers
BH -- bulge coevolution and (2) one or more local feeding mechanisms that operate even in pure-disk galaxies
but that result in no BH coevolution with any part of the host galaxy.

\vsss


\includegraphics{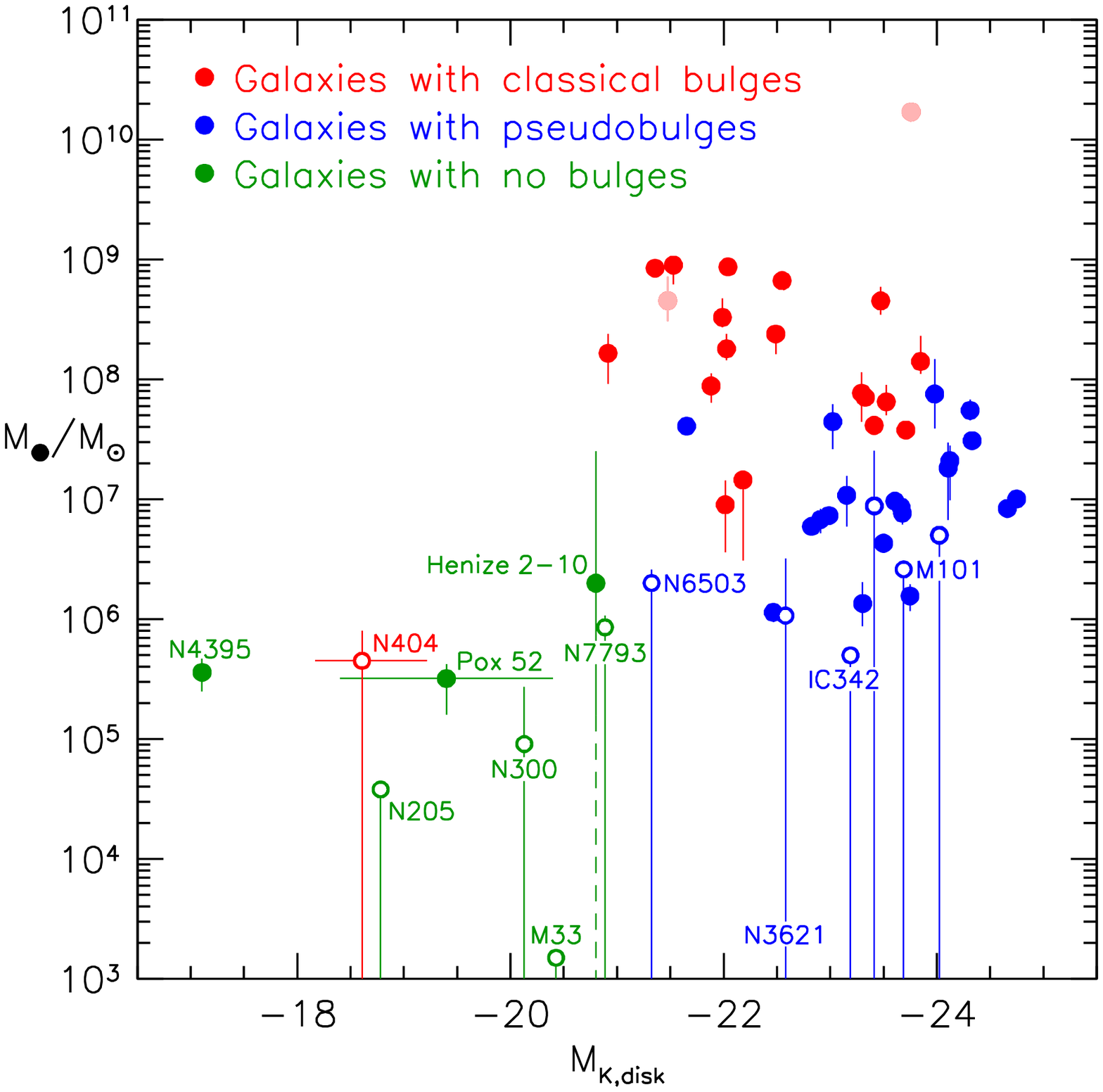}

{\parindent=2.95truein
\figindent \ni {\bf \textBlue Figure 22}\textBlack 

\vskip 3pt
\nointerlineskip \moveright 2.95truein\vbox{\hrule width 2.95truein}\textBlack\nointerlineskip
\vskip 3pt

\figindent  BH mass versus $K$-band absolute magnitude of the disk of the host galaxy.~Red and blue~filled circles
            are from {\bf Table 3}.  Open circles show $M_\bullet$ upper limits; the strongest is 
            $M_\bullet$ \lapprox \ts1500\ts$M_\odot$ for M{\ts}33 (Gebhardt \etal 2001).  M{\ts}101 and NGC 6503
            are from Kormendy \etal (2010).  IC~342 is from B\"oker \etal (1999).  NGC 3621 is from Barth \etal (2009).
            Green points are for galaxies with no classical bulge and almost~no or certainly no pseudobulge but only a nuclear star cluster.
            NGC\ts4395 (Peterson \etal 2005) and Pox\ts52 (Barth \etal 2004;~Thornton \etal 2008) are discussed in Section 7.1.  Henize 2-10 
            (Reines \etal 2011) is discussed in Section 7.3.  Limits: NGC 205 is from Valluri \etal (2005); NGC 300 and NGC 7793 are
            from Neumayer \& Walcher (2012; several other limits in that paper are broadly similar, but the galaxies~are~much farther 
            away).~NGC\ts404 is from Seth{\ts}et{\ts}al.{\ts}(2010).

}
\eject

\vs
\ni {\big\ARRed 6.10 BHs do not correlate with dark matter halos~in~a~way~that~is}
\vskip 1pt
\ni {\big\phantom{6.10}~\hbox{more fundamental than the M\lower2pt\hbox{$\bullet$}\ts--{\ts}M\lower2pt\hbox{\almostbig bulge} and 
                                                  M\lower2pt\hbox{$\bullet$}\ts--\ts$\sigma$\lower2pt\hbox{\almostbig\kern -1.2pt e} correlations}
\vskip 1pt
\ni {\big\phantom{6.10}~for classical bulges and ellipticals}}\textBlack
\vsss

\hsize=15.0truecm  \hoffset=0.0truecm  \vsize=21.4truecm  \voffset=1.2truecm

      Ferrarese (2002) suggested that $M_\bullet$ correlates as closely with DM halos of galaxies as it does~with bulges and ellipticals. 
In fact, she suggested that the DM correlation is the more fundamental~one.  This was not based on galaxies with BH detections.
Rather, she used proxy parameters: $\sigma$~for~$M_\bullet$ and the asymptotic outer rotation velocity $V_{\rm circ}$ of galaxy disks for DM.  
Ferrarese's conclusion was based on observing a tight correlation between $\sigma$ and $V_{\rm circ}$.  Baes \etal (2003)~lent further support, 
and the idea quickly became popular in galaxy formation theory, because it provided a natural way for AGN feedback to control BH growth
(Booth \& Schaye 2010), and because it led to a simple prescription for including feedback in semianalytic models of galaxy formation. 
But the implications were more profound than this.  The conclusion -- if correct -- implied 
that the unknown, exotic physics of nonbaryonic DM might be necessary to engineer BH--galaxy coevolution.

      Ferrarese's conclusion was surprising and counterintuitive.  It was known that BHs do not 
correlate with galaxy disks (Kormendy \& Gebhardt 2001), whereas disks correlate closely with DM
(van Albada \& Sancisi 1986; Sancisi \& van Albada 1987).  How could BHs and disks separately correlate 
with DM without also correlating with each other?

      Kormendy \& Bender (2011) re-examined the $V_{\rm circ}$--$\sigma$ correlation and concluded that BHs do not correlate with DM 
in a way that goes beyond the known correlations with bulges and ellipticals.  Section 6.10.1 updates their arguments.~Section 6.10.2
then presents a new, more direct~argument that is based not on $V_{\rm circ}$ but rather on new results about global halo parameters. 

\vsss
\noindent \ARRed {\bf 6.10.1 The V\lower0.5ex\hbox{circ}\ts--\ts$\sigma$ correlation}\textBlack~is shown in {\bf Figure 23}.
The left panel shows the correlation as Ferrarese derived it except that incorrect $\sigma$ measurements 
are omitted or corrected (see below) and that new measurements of bulgeless galaxies are added.  
The right panel includes more galaxies.  \phantom{000000000000}

\vfill

\includegraphics{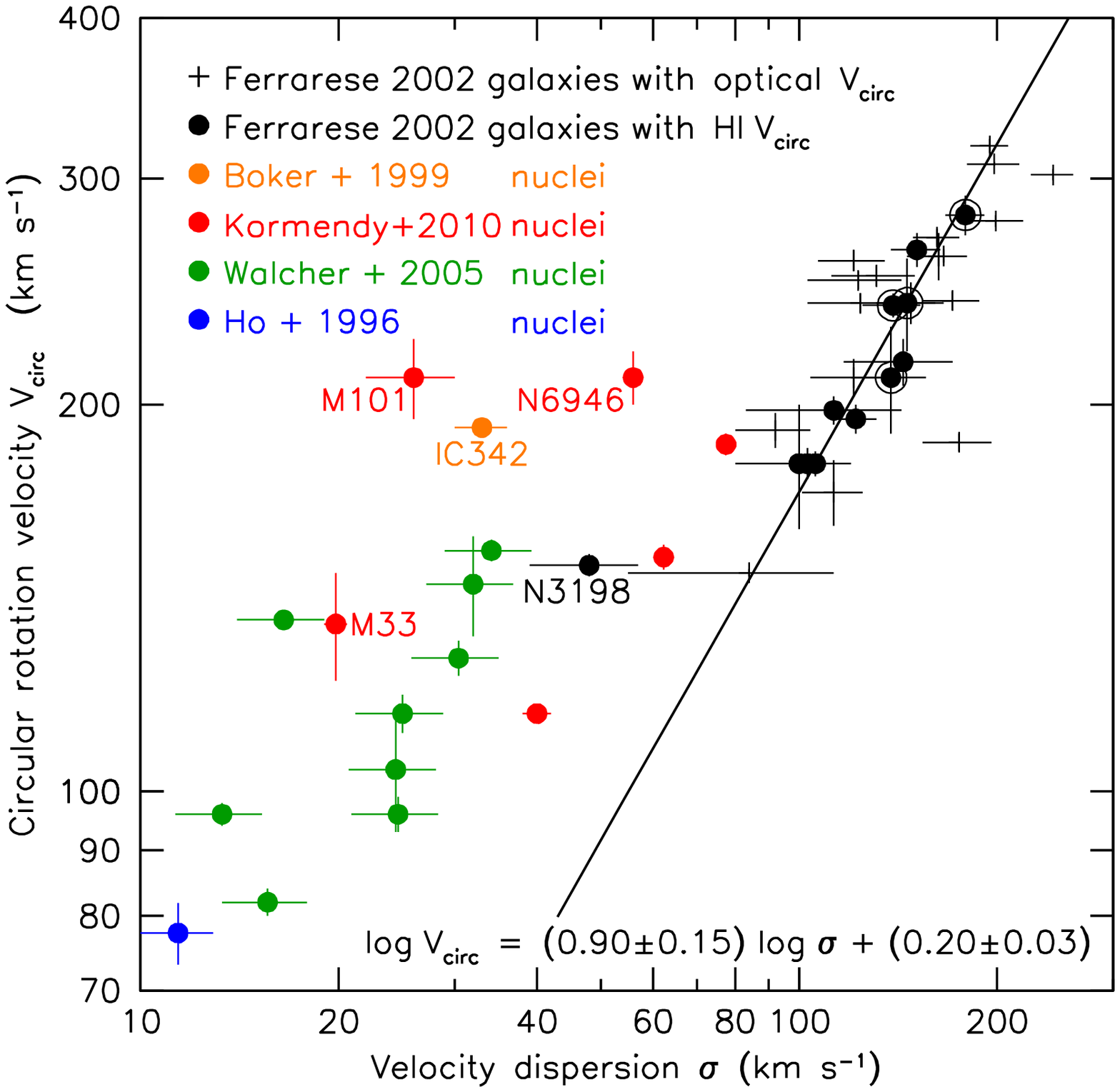}
\includegraphics{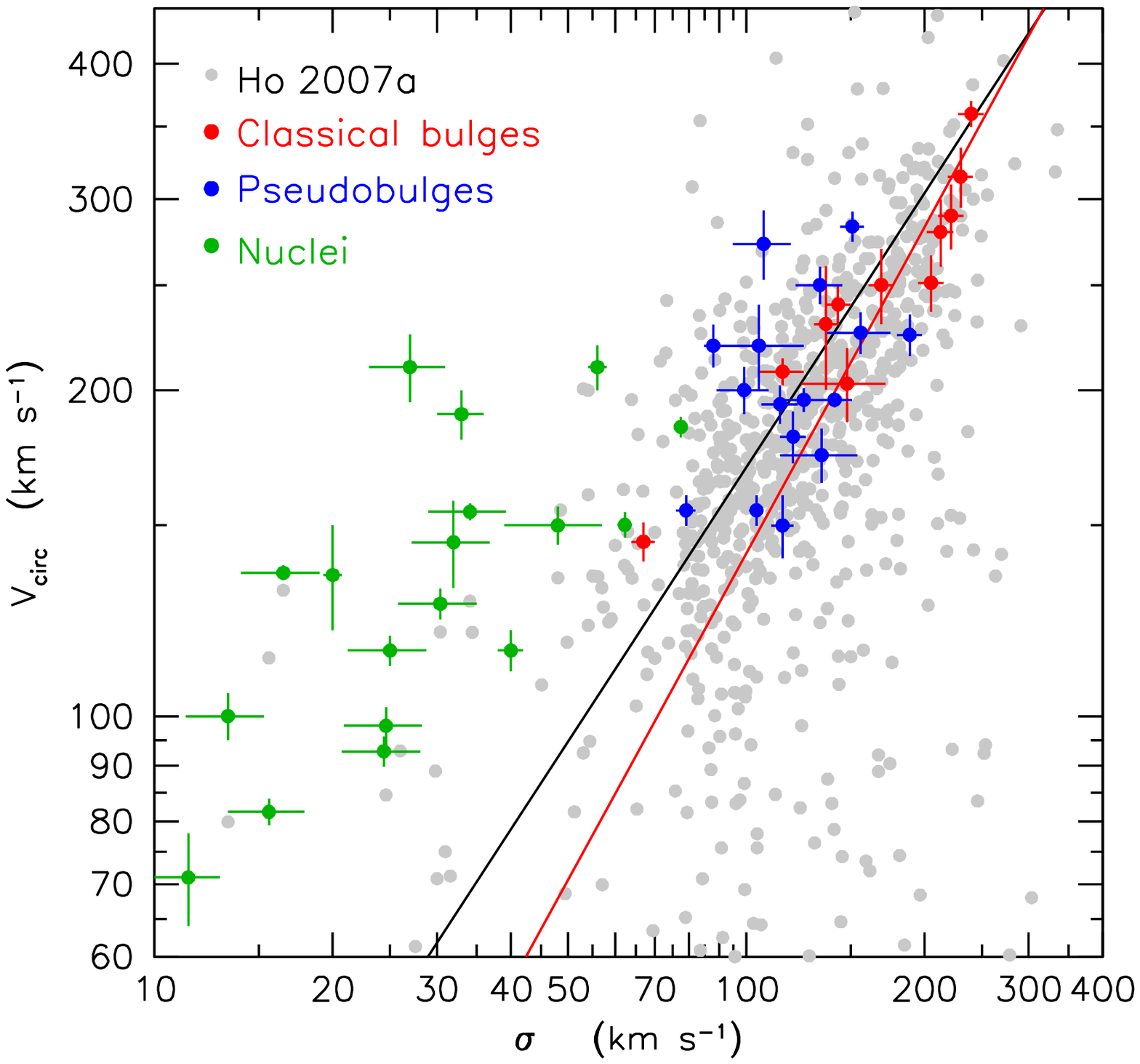}

\ni {\bf \textBlue Figure 23}\textBlack 

\vskip 2pt
\hrule width \hsize
\vskip 3pt
\ni Outer rotation velocities $V_{\rm circ}$ of spiral galaxy disks vs.~central velocity dispersions~$\sigma$ (Fig.\ts1~and~S3
in Kormendy \& Bender 2011).~Left:~Ferrarese~(2002) correlation (black points, circled if the galaxy has a classical bulge). 
Ferrarese points are omitted if the $\sigma$ measurement had insufficient velocity resolution or corrected if higher-resolution data 
are available.  Added in color are points for galaxies that have no classical bulge and essentially no pseudobulge but that are 
measured with instrumental resolution (expressed as a dispersion) $\sigma_{\rm instr}  <  10$ km s$^{-1}$ high enough to resolve the 
smallest $\sigma$ in galactic nuclei.  The line (equation at bottom, velocities in units of 200 km s$^{-1}$) is a least-squares 
fit to the black circles omitting NGC 3198.~Right:~$V_{\rm circ}$--$\sigma$ relation for the large sample in Ho (2007).  
It includes nuclei ({\it left panel\/}) and classical bulges and pseudobulges with BH detections ({\bf Table 2}).  The black line is 
Ferrarese's fit to her correlation.  The red line is not a fit; it shows $V_{\rm circ} = \sqrt{\rm 2}\ts\sigma$.

\eject

\hsize=15.0truecm  \hoffset=0.0truecm  \vsize=21.0truecm  \voffset=1.4truecm

\noindent The Kormendy \& Bender (2011) arguments were based on the left panel of {\bf Figure 23} and on the right panel using the
G\"ultekin \etal (2009c) subsample of the present, {\bf Table 3} disk galaxies with dynamical BH detections.  Ellipticals are not included;
we cannot directly measure $V_{\rm circ}$ and do not have a sufficiently accurate way to measure DM that is not intimately connected with 
the stellar mass distribution.  We already know that BHs correlate tightly with that stellar mass distribution.

      Before we proceed, we need to be clear about what we are testing.  

      First, how do we measure DM?  Here in Section 6.10.1, we assume that $V_{\rm circ}$ measures the inner parts of DM halos.  Provided 
that the measurements reach out beyond most of the visible galaxy, rotation curve decompositions suggest 
that this is a good approximation.  Ferrarese argues and we agree that H{\ts}I rotation curves reach farther out into DM halos than optical 
ones and provide more reliable asymptotic $V_{\rm circ} \simeq$ constant measurements of halos.  This is why the filled circles in 
{\bf Figure 23} ({\it left\/}) show smaller scatter than the error bars withough central points.  A variety of recent studies
further support our assumption that $V_{\rm circ}$ is a good measure of the inner parts of DM halos; i.{\ts}e., approximately the maximum 
rotation velocity of the halo if it has a Navarro, Frenk, \& White (1997) density distribution (e.{\ts}g.,
Dutton \etal 2010, 2011). 
Thus, in the rest of this section, we follow Ferrarese (2002) and use $V_{\rm circ}$ to measure DM.  Then, in Section 6.10.2, we broaden 
the discussion to include estimates of the total dark matter content of galaxies.  This will provide an additional point that is independent 
of the arguments in this section.

      Second, what are we testing?  The issue is not just whether a plot of $V_{\rm circ}$ versus $\sigma$ shows a tight correlation.  At stake
are implications for BH growth and galaxy formation.  We test two competing ideas.  The known $M_\bullet$\ts--\ts$\sigma$ relation for bulges
suggests that BHs and bulges coevolve.  An equally tight $M_\bullet$\ts--\ts$V_{\rm circ}$ correlation would raise the possibility that baryons 
are irrelevant and that the physics of nonbaryonic DM controls BH growth.  But we emphasize: Intermediate interpretations are part of the first 
alternative, because we know that DM gravity (but not its exotic physics) drives hierarchical clustering.  So the issue is:~{\it Do BHs correlate 
with DM in a way that goes beyond the BH\ts--{\ts}bulge relation?  Does this BH{\ts}--{\ts}DM correlation demand the conceptual leap that it is
not bulge growth but rather DM properties that engineer BH{\ts}--{\ts}galaxy coevolution?}

\def\vs{\vskip 3pt}

\vs
We begin using Ferrarese's (2002) correlation ({\bf Figure 23}, {\it left\/}) and the assumption that $\sigma$ serves as a surrogate for 
$M_\bullet$.  The Kormendy \& Bender (2011) arguments are:
\vs

\def\nhi1{\indent \hangindent=0.8truecm}

\nhi1 1.~Some $\sigma$ values used by Ferrarese (2002) were measured with insufficient velocity 
         resolution and are known to be incorrect.  The bulgeless Scd galaxy IC\ts342 (orange point 
in {\bf Figure 23} {\it left}) was shown at $\sigma$ = 77 $\pm$ 12 km s$^{-1}$, consistent with the black points. 
But the measurement (Terlevich, D\'\i az, \& Terlevich 1990) had low resolution:~the instrumental velocity dispersion 
$\sigma_{\rm instr}$\ts$\equiv$\ts(resolution~FWHM)/2.35 was 77 km s$^{-1}$, similar to $\sigma$ measured in IC 342.  
Low resolution often results in overestimated $\sigma$.  Terlevich \etal (1990) also got \hbox{$\sigma$ = 77 km s$^{-1}$} 
for the nucleus of M{\ts}33, which has $\sigma$ = 20 $\pm$ 1 km s$^{-1}$ as measured at high resolution (Kormendy \& McClure 1993; 
Gebhardt \etal 2001; Kormendy \etal 2010).  In fact, a high-resolution measurement of IC 342 was available: at 
$\sigma_{\rm instr} = 5.5$ km s$^{-1}$, B\"oker, van der Marel \& Vacca (1999) got \hbox{$\sigma$ = 33 $\pm$ 3 km s$^{-1}$} 
(orange point).  So {\bf Figure 23} omits black points if  $\sigma_{\rm instr}$ \gapprox \ts$\sigma$.  The range over which 
the remaining black points show a correlation is much reduced.
\vs

\nhi1 2.~{\it If DM halos and not bulges control $M_\bullet$, then the galaxies that should demonstrate
              this are the biggest ones that do not contain bulges.}  Thus motivated, Kormendy \etal
(2010) measured~$\sigma$ in six Sc -- Scd galaxies that contain nuclear star clusters (``nuclei'') but no 
classical bulges.  Even pseudobulges make up only a few percent of these galaxies.   The observations were 
made with the 9.2 m Hobby-Eberly Telescope High Resolution Spectrograph at $\sigma_{\rm instr}$\ts=\ts8~km~s$^{-1}$.  
Results are shown by the red points in {\bf Figure 23} together with other high-dispersion measurements of (pseudo)bulgeless 
galaxies (color points).   Kormendy and Bender concluded that (pseudo)bulgeless galaxies show only a weak correlation
between $V_{\rm circ}$ and $\sigma$.  A weak correlation is expected, because bigger galaxies tend to have bigger nuclei 
(B\"oker \etal 2004; Rossa \etal 2006).  But the scatter is much larger than the measurement errors:~$\chi^2$\ts=\ts15.7.  
No tight correlation is suggestive of any more compelling formation physics than the expectation that 
galaxies grow bigger nuclei when they contain more cold gas.  The black and color points overlap for 
180 km s$^{-1}$ \lapprox $V_{\rm circ}$ \lapprox \ts220 km s$^{-1}$.  In this DM range, galaxies participate
in a tight $V_{\rm circ}$ -- $\sigma$ correlation only if they have bulges.  That is, baryons and BH growth are closely
connected only if they are in a bulge.  Baryons in a disk are not enough.  DM by itself is not enough.
M{\ts}101 ({\it top-left red point\/}) has a halo  that is similar to those of half of the galaxies in the remaining tight correlation,
but that halo did not manufacture a canonical BH in the absence of a bulge.  This suggests that bulges, 
not halos, coevolve~with~BHs. \vs

\nhi1 3.~The large, unbiased sample of galaxies studied by Ho (2007a) leads to the same conclusions.  In
         {\bf Figure 23} ({\it left\/}), the galaxy sample shown by the color points is intentionally biased 
against galaxies that contain bulges.  There, we want to know whether DM correlates with BHs 
in the absence of the component that we know correlates with BHs.  However, Ho (2007a) made a similar 
study of a large galaxy sample that is not biased against bulges.  His results are shown by the gray points
in {\bf Figure 23} ({\it right\/}).  He concluded that $V_{\rm circ}$ correlates weakly with $\sigma$, especially
in galaxies that contain bulges, but the scatter is large and ``these results render questionable any attempt 
to supplant the bulge with the halo as the fundamental determinant of the central black hole mass in galaxies.'' 
\vs

\hfuzz=20pt

\def\vs{\vskip 3pt}

\nhi1 4.~Kormendy \& Bender (2011) argued that the tight correlation of black points in {\bf Figure 23} 
is a result of the conspiracy between baryons and DM to make featureless, nearly flat rotation curves with 
no distinction between radii that are dominated by baryonic and nonbaryonic matter 
(van Albada \& Sancisi 1986;
Sancisi \& van Albada 1987).
This is a natural consequence of the observation that baryons make up 16\ts\% of the matter in the universe
(Hinshaw \etal 2013) and that, to make stars, they
need to dissipate inside their halos until they self-gravitate.  This is enough to engineer 
that $V_{\rm circ}$ is approximately the same for DM halos and for the disks that are embedded in them
(Gunn 1987;
Ryden \& Gunn 1987).
That bulges participate in the conspiracy is less well known and not implied by the above arguments.
However, in their Figures S1 and S2, Kormendy \& Bender (2011) demonstrate that, when $V_{\rm circ}
\simeq 200$ km s$^{-1}$, the bulge, disk, and halo all have approximately the same velocity scales.
All galaxies that participate in the tight correlation in {\bf Figure 23} have bulges or pseudobulges.
The suggestion then is that that correlation is nothing more nor less than a restatement of the rotation
curve conspiracy for bulges and DM.
This means that the correlation of black points in {\bf Figure 23} is a consequence of 
DM-mediated baryonic galaxy formation.  \vs

\hsize=15.0truecm  \hoffset=0.0truecm  \vsize=21.4truecm  \voffset=1.2truecm

\nhi1 5.~So far, we have discussed BH correlations indirectly using the assumption that $\sigma$ is a 
surrogate for BH mass $M_\bullet$.  However, Section 6.8 establishes that $\sigma$ is not a valid
surrogate for $M_\bullet$ in pseudobulges.  Only 4 galaxies plotted with black points in {\bf Figure 23}
({\it left\/}) contain classical bulges, M{\ts}31, NGC 2841, NGC 4258, and NGC 7331.  Their points are circled.  
The other black filled circles represent pseudobulges.  {\it For pseudobulges, the demonstration of a tight 
$V_{\rm circ}$\ts--\ts$\sigma$ correlation is not a demonstration that DM and BHs correlate.  Moreover,
the circled points for classical bulges agree with the correlation for pseudobulges.  It is 
implausible to suggest that the correlation for the four circled points is caused by \hbox{BH{\ts}--{\ts}DM}
coevolution whereas the identical correlation for the other points has nothing to do with BHs.  Ferrarese's
observation that classical and pseudo bulges show the same $V_{\rm circ}$\ts--\ts$\sigma$ correlation
in {\bf Figure 23} is by itself a strong argument against the \hbox{hypothesis that either correlation reflects BH{\ts}--{\ts}DM
coevolution.}}   Our BH galaxies show a tighter $V_{\rm circ}$\ts--\ts$\sigma$ correlation when they contain classical bulges~than 
when they contain pseudobulges ({\bf Figure\ts23},\ts{\it right}).~This is consistent with our picture that pseudobulges grow by 
local processes that are not closely connected with the depth of the gravitational potential well, whereas (point 4, above) classical 
bulges form in a way that engineers the DM{\ts}--visible-matter conspiracy revealed by rotation curve decomposition.

\vfill\eject

\cl{\null}

\vskip 2.55truein

\includegraphics{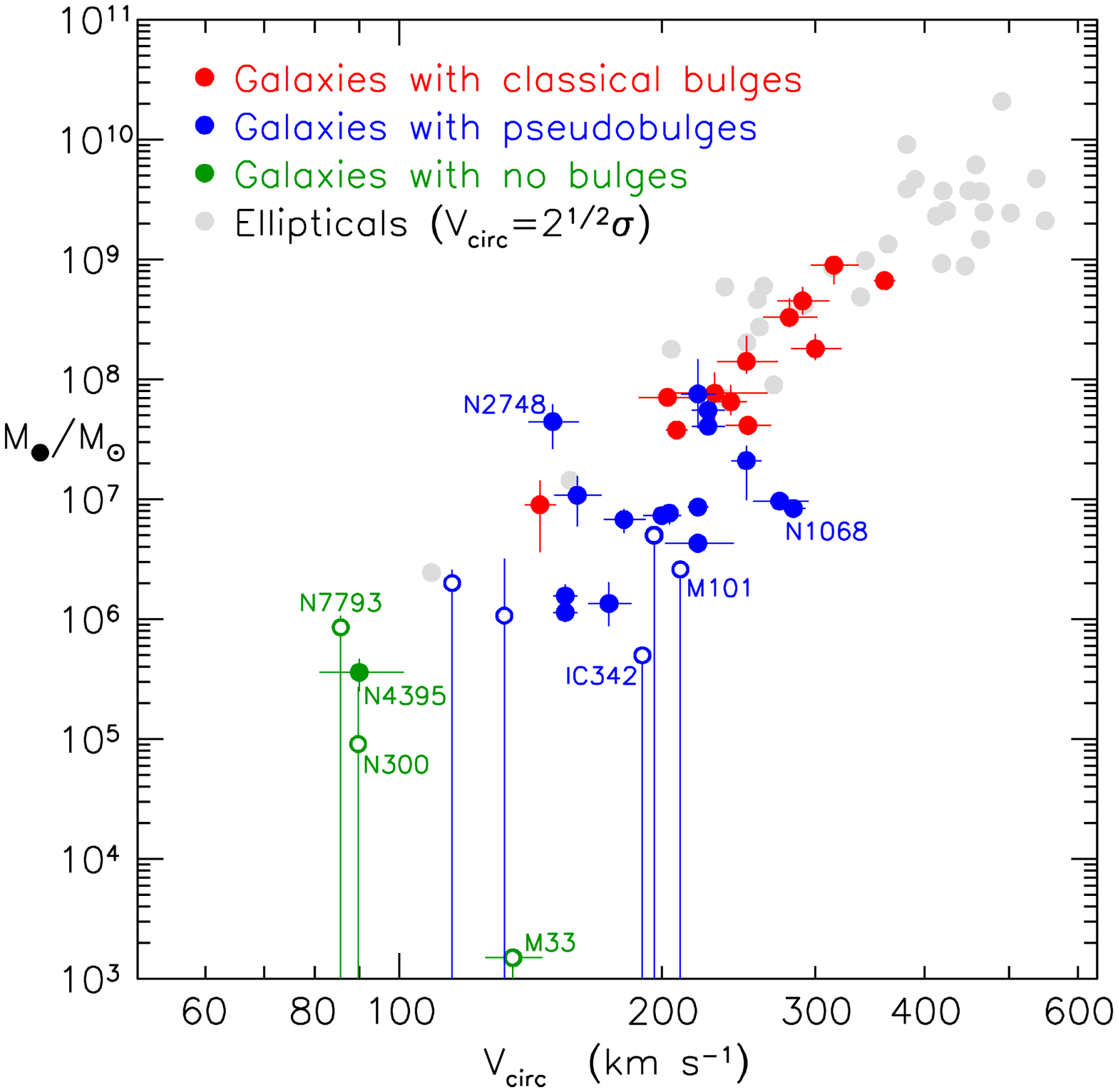}

\ni {\bf \textBlue Figure 24}\textBlack 

\vskip 2pt
\hrule width \hsize
\vskip 3pt
\ni Dynamically detected BH mass versus asymptotic, outer circular rotation velocity of the disk~of~the 
    host galaxy.~Kormendy \& Bender (2011) show this plot for the G\"ultekin \etal (2009c) sample~in their Figure 2(d).~The present version 
    is for the {\bf Table 3} sample.~Open circles show $M_\bullet$ upper limits.   Red points are for galaxies 
    with classical bulges; blue points are for galaxies with pseudobulges, and green points are for galaxies with neither a classical 
    nor a pseudo bulge but only a nuclear star cluster.  Gray points for elliptical galaxies are plotted assuming that 
    $V_{\rm circ} = \sqrt{2}\sigma$ (cf.~{\bf Figure\ts23}, {\it right}); they remind us that classical bulges and ellipticals have 
    similar parameter correlations.  Unlike arguments 1{\ts}--{\ts}5, this is a direct test for a BH{\ts}--{\ts}DM correlation.  But it 
    is a difficult one, because $V_{\rm circ}$ is not known for most bulge-dominated galaxies (hence our assumption for ellipticals).

\vs

\nhi1 6.~{\bf Figure 24} shows that dynamically measured BH masses do not correlate with $V_{\rm circ}$ and hence with DM for pseudobulges.  
    For classical bulges and possibly for ellipticals, there is a good correlation, but it remains likely that this is a consequence
    of the rotation curve conspiracy.  This confirms directly what we deduced indirectly above.  No conceptual leap to a BH\ts--{\ts}DM 
    correlation is compelled by~the~data.
\vs

\nhi1 7.~A final point from Kormendy \& Bender (2011) is reproduced here from their Supplementary Information.
         The DM halos of galaxy clusters predict giant BHs that~are~not observed.  This is important in the context of
         Ferrarese's (2002) conclusion that more massive halos are more efficient in growing BHs.~Her
         $V_{\rm circ}$ -- $\sigma$ correlation, reproduced to within errors in {\bf Figure\ts23},~is
\vskip -10pt
$$
\log{V_{\rm circ}} = (0.84 \pm 0.49)\ts\log{\sigma} + (0.55 \pm 0.19)~,                                    \eqno{(13)}
$$
\vskip -5pt
Substituting for $\sigma$ in her $M_\bullet$ -- $\sigma$ relation,
\vskip -5pt
$$
M_\bullet = 
(1.66 \pm 0.32) \times 10^8~M_\odot~\biggl({{\sigma} \over {200~{\rm km~s}^{-1}}}\biggr)^{4.58 \pm 0.52}~,  \eqno{(14)}
$$
\vskip -5pt
yields
\vskip -10pt
$$
{{M_\bullet} \over {10^8~M_\odot}} = 0.168~\biggl({V_{\rm circ} \over {200~{\rm km~s}^{-1}}}\biggr)^{5.45}~. \eqno{(15)}
$$
\vskip -5pt
As a check on previous arguments, $V_{\rm circ} = 210$ km s$^{-1}$ for M{\ts}101 predicts 
$M_\bullet \simeq 2.2 \times 10^7$ $M_\odot$.  This conflicts with the observed upper limit, 
$M_\bullet$ \lapprox \ts$(2.6 \pm 0.5) \times 10^6$ $M_\odot$ (Kormendy~et al.~2010).  At the high end of the 
range of $V_{\rm circ}$ values for DM halos, rich galaxy clusters typically have velocity dispersions 
$\sigma \sim 1000$ km s$^{-1}$ and can have dispersions as high as $\sim 2000$ km s$^{-1}$.  
Whether we can use these cluster halos in our argument depends on whether the DM is already 
distributed in the cluster or whether it is still attached only to the galaxies, with the result that 
the total mass is large but that individual halos are not.  Large-scale simulations of hierarchical 
clustering show that, while substructure certainly exists, much of the DM in rich, relaxed clusters is 
distributed ``at large'' in the cluster (Springel \etal 2005b).  In fact, DM hierarchical clustering is 
so nearly scale-free that ``it is virtually impossible to distinguish [the halo of an individual galaxy 
from that of a cluster of galaxies] even though the cluster halo is nearly a thousand times more massive''
(Moore \etal 1999).  Therefore we can treat cluster halos like galaxy halos in predicting~$M_\bullet$.
They provide an especially important test of ``baryon-free'' BH{\ts}--{\ts}DM coevolution, because 
Ferrarese (2002) concluded that higher-mass DM halos are more efficient at growing large BHs.

      DM halos of $10^{15}$ $M_\odot$ are not rare (Faltenbacher, Finoguenov \& Drory 2010).
A cluster like Coma has $\sigma \sim 1000$  km s$^{-1}$ and therefore presumably has 
$V_{\rm circ} \sim \sqrt{2}\ts\sigma \sim 1400$ km s$^{-1}$ (Kent \& Gunn 1982).  Equation (15) then predicts that
\vskip -9pt
$$M_\bullet \sim 1 \times 10^{11}~M_\odot~.                             \eqno(16)$$
\vskip -1pt
\noindent Sunk to the center of NGC 4874 or NGC 4889, such a BH would be hard to hide.  Both galaxies have 
$\sigma \simeq 300$ km{\ts}s$^{-1}$ and normal, shallow $\sigma(r)$ profiles
(Fisher, IIllingworth \& Franx 1995).  But an $10^{11}$-$M_\odot$ BH would have a sphere-of-influence radius 
of 2\sd2.  We would have found such a BH.  In fact, we now have a measurement of $M_\bullet = 2.1(0.5-3.7) \times 10^{10}$
$M_\odot$ in NGC 4889 (McConnell \etal 2012).  But this mass is more appropriate for the host galaxy than it is for the
cluster halo.

      Therefore, if baryons do not matter---if the hypothesis is that DM makes BHs independent of how 
baryons are involved (as a galaxy, as a group of galaxies, or not at all)---then Equation (15) predicts 
unrealistically large BH masses for rich clusters of galaxies, {\it especially when there is not a giant 
elliptical at the cluster center.}  

      Combining 1\ts--\ts7: Over the whole range of $V_{\rm circ}$ values associated with DM halos,
i.{\ts}e., at least
\vskip -8pt
$$
50~{\rm km~s^{-1}} \lesssim V_{\rm circ} \lesssim 2000~{\rm km~s^{-1}}~, \eqno{(17)} 
$$
\vskip -2pt
\noindent the only part of the range in which {\bf Figure 23} shows a tight correlation between $\sigma$ and 
$V_{\rm circ}$ is \hbox{150\ts--\ts300} km s$^{-1}$ and only if the galaxy 
contains a classical bulge.  At most $V_{\rm circ}$ values even in this range, there are DM halos
that do not contain classical bulges, and then $V_{\rm circ}$ does not correlate with $M_\bullet$.  Outside
this $V_{\rm circ}$ range, Equation (15) gets into trouble with observations.
This situation argues against the hypothesis that baryons are irrelevant and that DM controls BH growth.
We conclude that BHs do not correlate causally with DM halos. 

      The consequences for galaxy formation are simple and profound.  There is no reason to expect that 
the unknown, exotic physics of nonbaryonic DM directly influences BH growth.   Even the presence 
of the halo's gravitational potential well appears not to be solely and directly responsible for BH\ts--{\ts}galaxy 
coevolution.  Rather, coevolution physics appears to be as simple as it could be: it happens only in the context
of the major galaxy mergers that make classical bulges and ellipticals.

\vsss
\noindent \ARRed {\bf 6.10.2 The relationship between DM and stellar masses of galaxies and consequences 
                  \null\quad\quad\quad for our understanding of BH{\ts}--{\ts}galaxy coevolution}\textBlack~

      Our discussion so far has been confined to the inner parts of DM halos, where rotation curve decomposition and other
observations support the assumption that $V_{\rm circ}$ is a relatively~secure measure of DM properties.~However, rotation curves
sample small fractions of galaxy DM halos.  What new things do we learn if we can measure total halo mass $M_{\rm DM}$? Behroozi \etal (2010,\ts2012) 
review and extend recent work that provides a statistical connection between baryonic~and~DM masses of galaxies.  A variety of techniques 
are involved, ranging from relatively direct observations of satellite dynamics, clustering properties, and weak lensing, to indirect but
powerful methods such as abundance matching of DM halo masses in cosmological simulations with galaxy mass distributions
observed in large volumes in the universe.  References are given in the above papers and in {\bf Figure 25}.
From this work, we cannot correlate $M_\bullet$ with $M_{\rm DM}$ for individual galaxies, but we can{\ts}--{\ts}at least 
statistically{\ts}--{\ts}combine the $M_\bullet$\ts--\ts$M_{\rm bulge}$ correlation with the $M_*/M_{\rm DM}$\ts--\ts$M_{\rm DM}$ 
correlation to infer how $M_\bullet$ is related to $M_{\rm DM}$.  For the purposes of this section, the difference between $M_{\rm bulge}$ 
and the total stellar mass $M_*$ is unimportant; we assume that they are equal.  The results provide a further argument
that BHs coevolve with bulges and not with DM halos.

\vfill\eject

      {\bf Figure 25} shows the relationship between visible matter (mainly stellar) mass and DM mass.  The left panel shows the 
mass function of cold dark matter halos ({\it upper dashed curve\/}) and~the expected baryonic mass function if all galaxies contained
the cosmological baryon fraction of $\sim$\ts1/6.  It is well known that the baryon content of galaxies falls short of the cosmological 
value at all masses.  This is shown more directly at right, where Behroozi, Conroy, \& Wechsler (2012) summarize results from many 
techniques on the stellar-to-DM mass ratios in galaxies, this time as a function of DM halo mass.  At a ``sweet-spot'' DM mass of 
$\sim$\ts$1 \times 10^{12}$\ts$M_\odot$, $M_*/M_{\rm DM}$ reaches a maximum of 1/5 of the cosmological value.  But the stellar 
mass fraction is smaller at both lower and higher DM masses.  The sweet-spot DM mass corresponds to $M_* \approx M_{\rm bulge} \simeq 
3 \times 10^{10}$\ts$M_\odot$.  This is in the middle of the $M_\bullet$\ts--\ts$M_{\rm bulge}$ correlation in {\bf Figure 18}.
So $M_\bullet$ shows a simple, log-linear correlation with $M_{\rm bulge}$ but a complicated relation with $M_{\rm DM}$ that has a 
kink at the center of the $M_\bullet$ range.  This argues that the more fundamental relation is the one between $M_\bullet$ and $M_{\rm bulge}$.

\vfill


 \includegraphics{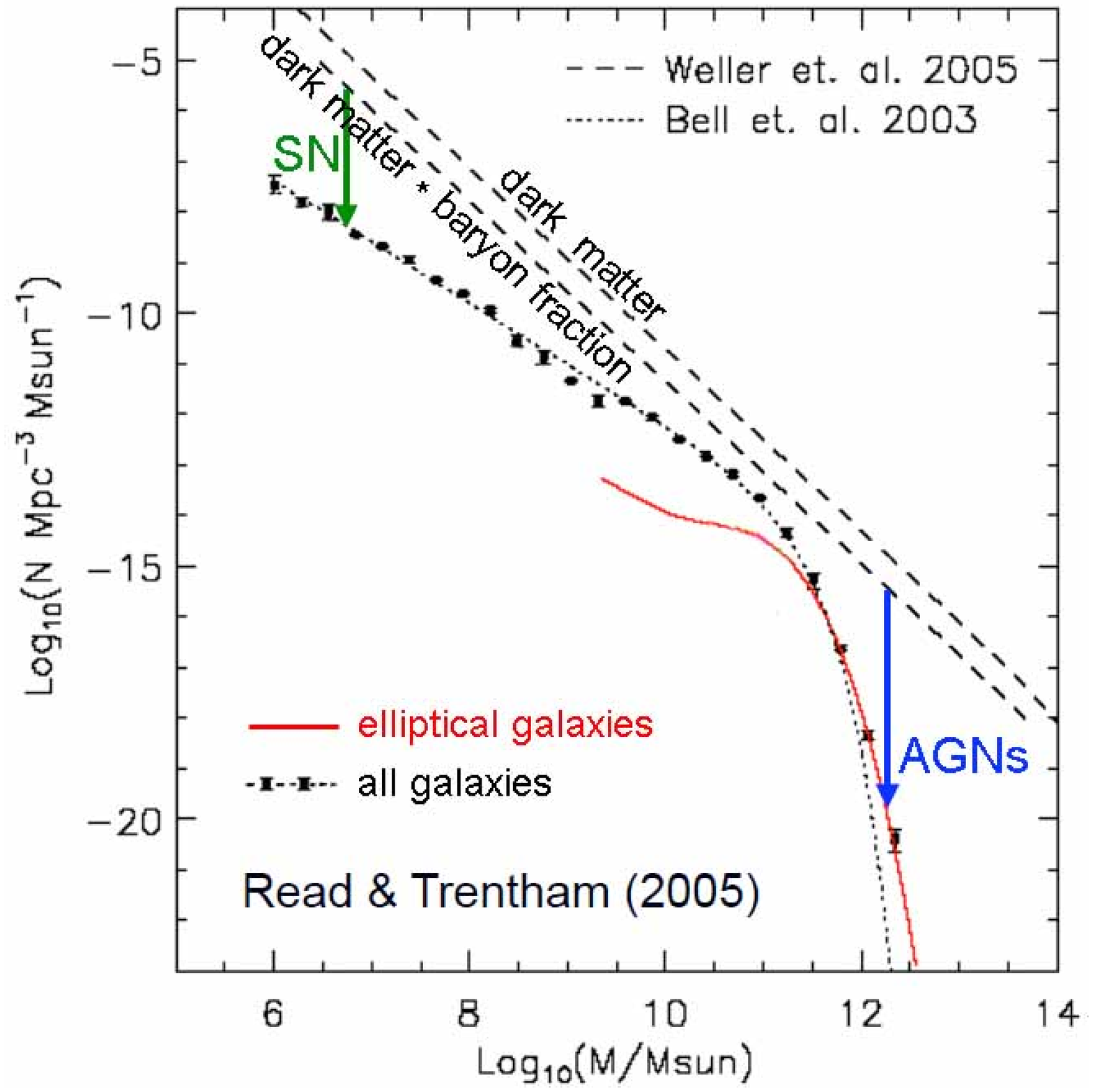}

\includegraphics{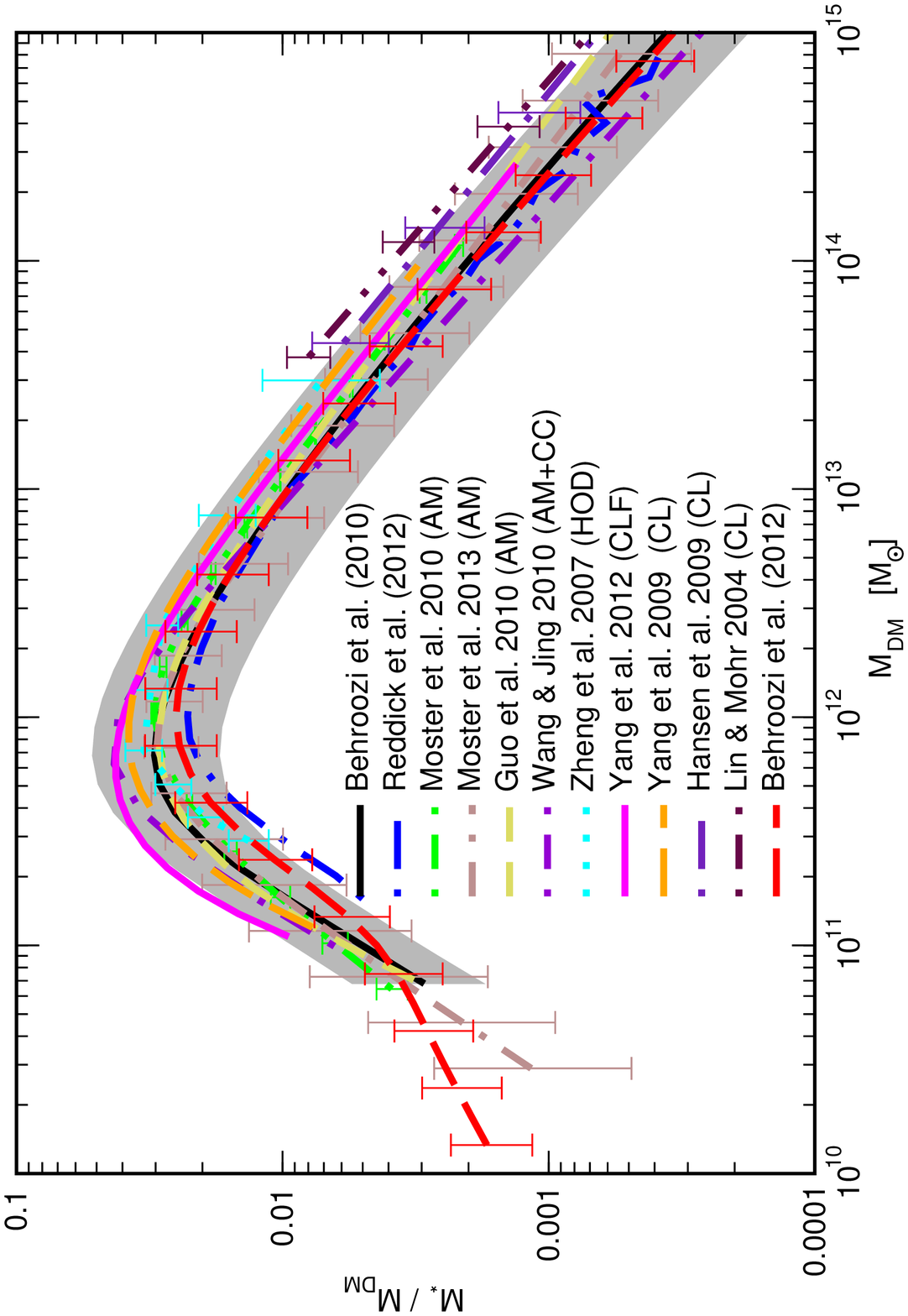}

\ni {\bf \textBlue Figure 25}\textBlack 

\vskip 2pt
\hrule width \hsize
\vskip 3pt
\ni ({\it left}) From Read \& Trentham \& (2005), the total field galaxy baryonic mass function ({\it black~points}) and Schechter (1976)
                 function fit (Bell \etal 2003:~{\it dotted curve)} compared to the mass spectrum of cold DM halos from numerical simulations 
                 by Weller{\ts}et{\ts}al.{\ts}(2005) and to that DM mass function multiplied by the universal baryon fraction of 0.163 (Hinshaw \etal 2013) 
                 ({\it lower dashed curve}).  
     ({\it right}) From Behroozi, Wechsler, \& Conroy (2012), comparison of their abundance matching of DM halos and visible galaxies with published results.
                 Abbreviations:~AM = abundance matching; 
                 CC = clustering constraints; 
                 HOD = modeling of halo occupation distributions;
                 CLF = conditional luminosity function; 
                 CL = various (e.{\ts}g., X-ray) results on galaxy clusters; see Behroozi \etal (2012) for details.
                 Gray shading shows 68\ts\% confidence limits from a similar analysis in Behroozi~et al.~(2010).~Both panels show that 
      galaxies have baryon-to-DM mass fractions that are less than the cosmic value.  
      The largest baryon mass fractions are seen in halos of mass $M_{\rm DM} \sim 10^{12}$\ts$M_\odot$; the remaining
      shortfall there is believed to be in a Warm-Hot Intergalactic Medium (WHIM: Dav\'e \etal 2001) and in cooler gas that has not yet 
      accreted onto galaxies.  Smaller galaxies are thought to miss progressively more baryons because they were ejected by (e.{\ts}g., 
      supernova-driven) winds (Dekel \& Silk 1986).  Halos that are much more massive than $M_{\rm crit} \equiv M_{\rm DM} \sim 10^{12}$ $M_\odot$
      are missing progressively more stars because baryons are kept suspended in hot gas by a combination of AGN feedback and cosmological gas infall 
      (see Section 8.4).  The important point here is this: Because $M_\bullet \propto M_*^{(1.16 \pm 0.08)}$ and because the ratio of stellar mass $M_*$
      to dark mass $M_{\rm DM}$ is not monotonic, therefore the relationship between BH mass and DM mass is complicated and not monotonic.  This suggests 
      that BH growth is controlled by stellar mass, not DM mass. \lineskip=-30pt \lineskiplimit=-30pt
\eject


\hsize=15.0truecm  \hoffset=0.0truecm  \vsize=20.1truecm  \voffset=1.5truecm

     More specifically, Behroozi \etal (2010) find that $M_* \propto M_{\rm DM}^{2.3}$  when $M_{\rm DM} \ll 10^{12}$\ts$M_\odot$ and that 
                                                        $M_* \propto M_{\rm DM}^{0.29}$ when $M_{\rm DM} \gg 10^{12}$\ts$M_\odot$.
Broadly similar results are obtained in many papers, some included in the above summary (e.{\ts}g., 
Yang \etal 2009, 2012;  
Moster \etal 2010;
Guo \etal 2010;
Leauthaud \etal 2012).
The lowest DM mass in {\bf Figure 25} at which the above correlation is well defined (the low-mass end of the 68-\%-probability
gray confidence band) corresponds to a baryonic mass of $M_* \approx 2 \times 10^8$ $M_\odot$ and the highest DM mass corresponds to
a baryonic mass of $M_* \approx 4 \times 10^{11}$ $M_\odot$.  That is, the multivalued correlation of $M_*/M_{\rm DM}$ with $M_{\rm DM}$ 
in {\bf Figure 25} holds over almost the same mass range as the $M_\bullet$\ts--\ts$M_{\rm bulge}$ correlation in {\bf Figure 18}.
Substituting $M_{\rm bulge} \approx M_*$ and $M_\bullet \propto M_{\rm bulge}^{1.16}$ (Equation 10) in the above, we conclude that
\vskip -7pt
$$ M_\bullet \propto M_{\rm DM}^{2.7}~~~{\rm at}~~M_{\rm DM}  \ll 10^{12}{\ts}M_\odot   \eqno{(18)}$$ 
\vskip -9pt
\noindent and that
\vskip -9pt
$$ M_\bullet \propto M_{\rm DM}^{0.34}~~~{\rm at}~~M_{\rm DM} \gg 10^{12}{\ts}M_\odot~, \eqno{(19)}$$
\vskip -1pt
\noindent with a kink in the correlation at $M_{\rm DM} \simeq 10^{11.8 \pm 0.5}$\ts$M_\odot$. 

 Moreover, we think we understand why the
$M_*/M_{\rm DM}$\ts--\ts$M_{\rm DM}$ correlation in {\bf Figure 25} changes slope in the middle; the explanation (figure caption)
involves BHs only very indirectly.  Meanwhile, the $M_\bullet$\ts--\ts$M_{\rm bulge}$ correlation
is log linear with small scatter from the lowest to the highest bulge masses in {\bf Figure 18}.  That correlation shows no kink
at $M_{\rm DM} \sim 10^{12}$ $M_\odot$.  The simplicity of the $M_\bullet$\ts--\ts$M_{\rm bulge}$ correlation versus the complexity
of the $M_\bullet$\ts--\ts$M_{\rm DM}$ correlation is another argument in favor of the hypothesis that BHs coevolve with bulges and 
ellipticals and not with DM halos.

\vs
\noindent \hbox{\ARRed {\bf 6.10.3 Revised strategies for including BHs in semianalytic models of galaxy~formation}}\textBlack\par
\noindent are suggested by the above results.  For studies that focus on the mass inventory~of~BHs~and~stars, it is reasonable 
to ignore or simplify the distinction between classical and pseudo bulges.  A way to do the latter would be to decrease the 
$M_\bullet/M_{\rm bulge}$ ratio by factors of several at (say) $M_\bullet < 10^8$\ts$M_\odot$.  Small BHs and small (pseudo)bulges 
do not dominate either mass budget.  At high BH masses, one simple strategy continues to be to use the $M_\bullet${\ts}--{\ts}$\sigma$ 
correlation but with our revised slope and especially with the larger $M_\bullet$ zeropoint in Equation 7.  The shortcoming of 
this approach is the observation that $\sigma$ saturates at high BH masses, and these are not negligible to the mass budget.
So a more attractive approach is to start with $M_{\rm DM}$ from a cosmological simulation, then use the Behroozi \etal (2010, 2012) 
or similar results ({\bf Figure 25} here) to estimate the associate $M_{\rm bulge}$, and then use our $M_\bullet$\ts--\ts$M_{\rm bulge}$
relation (Equation 10, including cosmic scatter) to get the BH mass.  
 
\vs\vs
\ni {\big\ARRed 6.11 Do galaxies contain either BHs or nuclear star clusters~and
                    \phantom{0\ts~\ts00}do these correlate in the same way with host galaxies?}\textBlack
\vs

      Wehner \& Harris (2006), Ferrarese \etal (2006), and Rossa \etal (2006) were the first~to~find~that the masses $M_\bullet$ of BHs 
and the masses $M_{\rm nuc}$ of nuclei show similar-slope correlations with the masses of the bulge (in the case of BHs) or whole galaxy 
(in the case of Virgo spheroidals).  C\^ot\'e{\ts}et{\ts}al.\ts(2006) showed explicitly that BHs and nuclei (collectively ``central massive objects'' 
or ``CMOs'") have the same distributions of the ratio of their mass to the mass of the host galaxy, in agreement with the Wehner and Ferrarese papers. 
In contrast, Rossa \etal (2006) found a zeropoint offset betweeen nuclei and BHs corresponding to a typical mass ratio $M_\bullet/M_{\rm nuc}$\ts$\simeq$\ts0.30.
Ferrarese \etal (2006) further found that $M_\bullet$ and $M_{\rm nuc}$ both correlate with galaxy velocity dispersion; the slopes of the
correlations are the same, but nuclei are 10 times more massive than BHs at a given $\sigma$. However, they found that CMO masses have a single, 
continuous correlation with the virial mass $M_{\rm gal} = 5 r_e \sigma^2 / G$ (Cappellari \etal 2006) of the host. 
Finally, Graham \& Driver (2007) found a continuous correlation between CMO mass and host S\'ersic index.
These papers propose that BHs and nuclei form by similar processes that favor BHs in large galaxies and nuclei in small ones.  
Some papers suggest further that galaxies contain either a BH or a nucleus but generally not both.  The change from high-mass BHs 
to lower-mass nuclei happens at a CMO mass of $\sim 10^{7.5}$ $M_\odot$.

      We partly agree and we partly disagree.  AGNs in late-type, bulgeless galaxies show that BHs exist 
even at the absolute magnitudes of dwarf galaxies that usually have nuclei (Ho 2008; Section\ts7 here).  
Also, we observe nuclei in bulgeless galaxies at high luminosities at which BHs predominate when galaxies have
bulges (e.{\ts}g., M\ts101 and NGC 6946).  Most importantly, many galaxies 
contain both nuclei and BHs.  So BHs and nuclei are not almost mutually exclusive.  In contrast, they routinely coexist.  
However, the more important point of the above papers is well supported by the data: BHs and nuclei do 
show some remarkable similarities in masses, as discussed in this section.

      We begin by expanding on the coexistence of BHs and nuclei in similar (often the same)~galaxies. 
AGN-based $M_\bullet$ measurements discussed in Section 7 and illustrated in 
{\bf Figure~32} show that many late-type, bulgeless galaxies contain BHs with 
$M_\bullet \simeq 10^5$\ts$M_\odot$ to at least $10^{6.5}$\ts$M_\odot$.   Their central velocity dispersions 
range from $\sigma \sim 30$ km s$^{-1}$ to at least 100 km s$^{-1}$.  These $M_\bullet$ and $\sigma$
values overlap with galaxy nuclei, but they show little correlation between $M_\bullet$ 
and $\sigma$, consistent with the result that BHs do not correlate with pseudobulges or disks 
(Sections 6.8 and 6.9).  Also, {\bf Figure\ts24} includes nuclei in some of the most massive pure-disk 
galaxies (blue open circles from Kormendy \etal 2010); these objects overlap about half of the $V_{\rm circ}$ range 
of BH galaxies.  Lauer \etal (2005) show that nuclei also exist in brighter ellipticals.  Therefore there
is no segregation of galaxies into giants that only contain BHs and dwarfs that only contain nuclei.

      Our second point is that BHs and nuclei coexist in many galaxies, and the ratio of BH mass to nuclear mass
varies widely from $\gg 1$ to $\ll 1$ 
(Seth \etal 2008;
Graham \& Spitler 2009;
KFCB).  
The most robustly determined comparisons of BH and nuclear masses are summarized in {\bf Table~4},
ordered by $M_\bullet/M_{\rm nuc}$.  We include upper limits on $M_\bullet$ when they constrain our discussion.

%
\def\m{~~~}
\hfuzz=20pt
\table
\scbaselines
\hskip -10pt
\tablewidth{15.3truecm}
\tablewidth{16.9truecm} 
\tablewidth{15.0truecm}
\tablespec{\l\c\c\c\l\l}
\body{
\header{\bf \Blue{Table 4 \quad Masses of coexisting nuclear star clusters and supermassive black holes}}
\skip{5pt}
\hline
\skip{2pt}
& Galaxy   &   $D$   &  $M_\bullet$ ($M_{\rm low},M_{\rm high}$) & $M_{\rm nuc}$          & ~~~$M_\bullet/M_{\rm nuc}$ &  Ref.   \end
&          & {(Mpc)} &  {($M_\odot$)}                            & {($M_\odot$)}          &                          &         \end
&(1)       &   (2)   &  (3)                                      & (4)                    & ~~~(5)                      & (6)     \end
\skip{2pt}
\hline
\skip{4pt}
& NGC 4026 & \013.35  &  1.80 (1.45\ts\--\ts2.40) $\times\, 10^8$   &$  1.44$ $\times\, 10^7$            & \ts\ts\ts12.4             & 1     \end \skip{2pt}
& NGC 3115 &\0\09.54  &  8.97 (6.20\ts\--\ts9.54) $\times\, 10^8$   &$ (1.04 \pm 0.29)$ $\times\, 10^8$  &$\m8.6^{+5.0}_{-2.9}      $& 2,3   \end \skip{2pt}
& M{\ts}31 &\0\0\00.774& 1.43 (1.12\ts\--\ts2.34) $\times\, 10^8$   &$ (3.5 \pm 0.8)$ $\times\, 10^7$    &$\m4.1^{+2.6}_{-1.2}      $& 4,5,6 \end \skip{2pt}
& NGC 1023 & \010.81  &  4.13 (3.71\ts\--\ts4.56) $\times\, 10^7$   &$  0.99$ $\times\, 10^7$            & \m4.1                     & 1     \end \skip{2pt}
& NGC 3384 & \011.49  &  1.08 (0.59\ts\--\ts1.57) $\times\, 10^7$   &$  2.3$ $\times\, 10^7$             & \m0.48                    & 7     \end \skip{2pt}
& NGC 7457 & \012.53  &  8.95 (3.60\ts\--\ts14.3) $\times\, 10^6$   &$  2.7$ $\times\, 10^7$             & \m0.33                    & 7     \end \skip{2pt}
& Galaxy&\0\0\0\00.0083& 4.30 (3.94\ts\--\ts4.66) $\times\, 10^6$   &$ (2.9 \pm 1.5)$ $\times\, 10^7$    &$\m0.15^{+0.075}_{-0.075}$ & 8     \end \skip{2pt}
& NGC 4395 & \04.3    &  3.6 (2.5\ts--\ts4.7) $\times\, 10^5$       &$ (3.5 \pm 2.4)$ $\times\, 10^6$    &$\m0.10^{+0.077}_{-0.077}$ & 7,9   \end \skip{2pt}
& $\omega$ Cen & \0\0\0\00.0048 & 4.7 (3.7\ts--\ts5.7) $\times\, 10^4$ &$ (2.6 \pm 0.1)$ $\times\, 10^6$ &$\m0.018 \pm 0.004$        & 10,11 \end \skip{2pt}
& NGC 205  &\0\00.82  &  $\lesssim 2.4$ $\times\, 10^4$             &$ (1.4 \pm 0.1)$ $\times\, 10^6$    &$\lesssim$\ts0.017         & 12,13 \end \skip{2pt}
& G1       &\0\00.77  &  1.8 (1.3\ts--\ts2.3) $\times\, 10^4$       &$ (8 \pm 1)$ $\times\, 10^6$        &$\m0.0023 \pm 0.0007$      & 14    \end \skip{2pt}
& M{\ts}33 &\0\00.82  &  $\lesssim$ 1540                            &$ (1.0 \pm 0.2)$ $\times\, 10^6$     &$\lesssim$\ts0.0015       & 15    \end \skip{2pt}
\hline
}
\endtable
\scbaselines
\null
\vskip -12pt
\null
{\sc\noindent
Column 2 is the assumed distance (e.{\ts}g., {\sbf Table 3}).  
Column 3 is the BH mass from {\bf Tables 2} and {\bf 3} or from sources given in the text. 
Column 4 is the nuclear mass.
Column 5 is the ratio of BH mass to nuclear mass.
Column 6 lists the references for the nuclear mass: 1 = Lauer \etal (2005);
                                                    2 = Kormendy \etal (1996b);
                                                    3 = Emsellem, Dejonghe \& Bacon (1999);
                                                    4 = Kormendy \& Bender (1999);
                                                    5 = Kormendy (1988a);
                                                    6 = Bacon \etal (1994);
                                                    7 = Seth \etal (2008);
                                                    8 = Launhardt, Zylka \& Mezger (2002);
                                                    9 = Filippenko \& Ho (2003);
                                             \hbox{10 = van de Ven \etal (2006);}
                                                   11~=~Jalali \etal (2011);
                                                   12 = Jones \etal (1996);
                                                   13 = De Rijcke \etal (2006);
                                                   14 = Baumgardt \etal (2003);
                                             \hbox{15 = Kormendy \& McClure (1993);} Kormendy \etal (2010).
    
\lineskip=-20pt \lineskiplimit=-20pt
}

\dblbaselines

\eject

      Curiously absent from most discussions of this subject are the nuclei of M{\ts}31 and NGC 3115.  
They are examples of galaxies in which the BH is substantially more massive than the nucleus.

      In contrast, our Galaxy is a typical example of galaxies in which $M_\bullet < M_{\rm nuc}$.
It contains a normal nucleus composed of a mixture of old and young stars.
The mass determined by Launhardt, Zylka, \& Mezger (2002) is
$M_{\rm nuc} = (2.9 \pm 1.5) \times 10^7$ $M_\odot$
(see also Sch\"odel, Merritt \& Eckart 2009, who emphasize uncertainties).  This gives 
$M_\bullet/M_{\rm nuc} = 0.15 \pm 0.075$.  

      NGC 4395 has the smallest $M_\bullet$ determined
from reverberation mapping (Peterson \etal 2005).  The nuclear mass is uncertain; the value in {\bf Table 4}
is the mean of one obtained from the effective radius and velocity dispersion of the nuclear cluster
(Filippenko \& Ho 2003 corrected for $M_\bullet$) and a value obtained using a typical $I$-band
mass-to-light ratio of 0.5 (Seth \etal 2008).

      We treat the Galactic globular cluster $\omega$ Cen and the M{\ts}31 globular cluster G1 as defunct 
galactic nuclei (see Section 7.4).  Then, for $\omega$ Cen, the BH mass is from Noyola \etal (2010; see also Noyola, Gebhardt \& Bergmann 2008)
and the cluster mass is from Jalali \etal (2011, see also van de Ven \etal 2006 and D'Souza \& Rix 2013).  
For G1, the BH mass is
from Gebhardt, Rich \& Ho (2002, 2005) and the cluster mass is from Baumgardt \etal (2003).  The BH-to-nuclear
mass ratios of $\omega$ Cen and G1 are the smallest that have so far been measured: $M_\bullet/M_{\rm nuc} =
0.018 \pm 0.004$ and $0.0023 \pm 0.0007$, respectively.

      Finally, two upper limits on BHs in bulgeless galaxies have a strong impact on our argument.  
They are $M_\bullet \lesssim 1500$ $M_\odot$ in M{\ts}33 (Gebhardt \etal 2001; see also Merritt, Ferrarese \& Joseph 2001) 
and  $M_\bullet \lesssim 2.4 \times 10^4$ $M_\odot$ in NGC 205.  For M{\ts}33, $M_\bullet/M_{\rm nuc} 
\lesssim 0.00015$.

      Therefore $M_\bullet/M_{\rm nuc}$ ranges over factors of $>$ 5000.  If ellipticals were included 
(Lauer{\ts}et{\ts}al.\ts2005), the biggest $M_\bullet/M_{\rm nuc}$ ratios would be even bigger.
So BHs and nuclei are not always similar~in~mass.  

      On the other hand, the more important point raised by the C\^ot\'e, Ferrarese, Wehner, and Graham
papers is the hint that $M_\bullet$ and $M_{\rm nuc}$ are similarly related to properties of their
host galaxies.  We confirm this result.  In fact, the relationship is even more intriguing than
the above papers suggest:

      {\bf Figure 26} ({\it top\/}) plots the ratio of the CMO mass to the mass of the host (pseudo)bulge
against (pseudo)bulge mass.  
The BH mass fraction $M_\bullet/M_{\rm bulge}$ for classical bulges and ellipticals shows the small scatter and near-independence
from $M_{\rm bulge}$ that we noted in Section 6.6.1.  The Equation\ts11 relation,
$M_\bullet/M_{\rm bulge} \propto M_{\rm bulge}^{0.14 \pm 0.08}$, is shown.   In contrast, BHs in pseudobulges 
show the larger scatter that we found in Section 6.8.  The remarkable new result in {\bf Figure\ts26}\ts({\it top\/}) 
emerges when we include the nuclei that coexist with BHs in seven galaxies.  Consistent with published
work, they have $M_{\rm nuc}/M_{\rm bulge}$ in the same range as the BHs.  The intriguing new result is
that, with one exception (NGC\ts1023), when the BH mass is slightly high with respect to the scatter in
{\bf Figure\ts26} ({\it top\/}), the nuclear mass is slightly low, and vice versa.  To put it another way,
$(M_\bullet + M_{\rm nuc})/M_{\rm bulge}$ shows less scatter than do either BHs or nuclei by themselves.
{\bf Figure 26} ({\it top\/}) therefore hints that the building of nuclei and the growth of BHs are somehow related. 
We do not understand either growth well enough to speculate about what the relation might be. \lineskip=-1pt


      {\bf Figure 26} ({\it bottom}) compares CMOs to the total mass of the host galaxy.  We can therefore
include nuclei in bulgeless galaxies (e.{\ts}g., NGC 4395).  The points for ellipticals do not change.  
The points for classical bulges move to slightly smaller BH-to-host-galaxy mass ratios.  The pseudobulge BHs move 
the most from the top panel, because PB/T is generally small.   The differences between BHs in bulges and BHs in 
pseudobulges are accentuated.

      From Seth \etal (2008), {\bf Figure 26} ({\it bottom}) shows $M_{\rm nuc}/M_{\rm gal}$ for nuclei
in early-type galaxies (red crosses), late-type galaxies (blue plus signs), and spheroidal galaxies
(light green triangles).  Nuclei from Table 1 and from Kormendy \etal (2010: large blue plus signs)
are included.  Relative 
masses of nuclei are slightly larger in Sphs than in late-type galaxies,
consistent with the idea that
Sphs are defunct late-type galaxies that lost some of their baryons (Kormendy 1985, 1987; KFCB;
Kormendy \& Bender 2012).   Minus spheroidals, {\bf Figure 26} ({\it bottom})
strengthens previous results (e.{\ts}g., 
Rossa \etal 2006; 
Seth \etal 2008) 
that nuclei in early-type galaxies have larger mass fractions
than nuclei in late-tpe galaxies.  It further confirms that BHs and nuclei in early-type galaxies 
have similar, somewhat large mass fractions, with nuclei showing more scatter than BHs.  Nuclei in
high-mass, late-type galaxies havesmaller mass fractions, like BHs in pseudobulges, which also occur 
predominately in late-type galaxies.  There is a hint that nuclear mass fractions for
late-type galaxies decrease toward the right approximately as $M_{\rm gal}$ increases.
That is, much of the decrease in $M_{\rm nuc}/M_{\rm gal}$ is due to the increase in $M_{\rm gal}$.

      {\it These again are hints that BHs and nuclei are related.  In fact, nuclei in early-type
galaxies look more closely related to BHs in early-tpe galaxies than they are to nuclei in late-type
galaxies.  Similarly, late-type galaxy BHs and nuclei look more closely related to each other than
either are to their counterparts in early-type galaxies. }

     These results appear at least superficially consistent with the suggestion in Sections 8
that different feeding mechanisms grow large BHs in bulges\ts$+${\ts}ellipticals and small BHs in galaxies
with pseudobulges or no bulges.  Evidently, this is also true of nuclei.    Nuclei in early-type galaxies
are older than nuclei in late-type galaxies.  It is possible (K.~C.~Freeman, private communication) that
their higher masses are a consequence of the general downsizing with decreasing formation redshifts of the
masses of the galaxy components that are produced by nearly all processes of galaxy formation.  This clearly
includes BHs.  It also seems inevitable, if the largest BHs grow as much as is inherent in the So\l tan (1982)
argument, that they will swallow any nuclei that~are~present.  That is, the largest BHs may include
any relevant nuclear mass.  Beyond this, we do not speculate about the physics that underlies the
BH{\ts--{\ts}nucleus relationship.

      The rest of this section compares our conclusions with recent published results.  
      Graham (2012) and Scott \& Graham (2013) find, as did Ferrarese \etal (2006), that, at a given CMO mass,
nuclei are associated with smaller central velocity dispersions than are BHs.  They also find that nuclei have
shallower correlations with host luminosity, $\sigma$, and stellar mass.  And they find that $M_{\rm nuc}/M_{\rm gal}$
decreases with increasing $M_{\rm gal}$; we do, too, but the decrease~that~they~see (Figure\ts3~in Scott \& Graham 2013) 
is steeper than the one that we find.  They conclude that ``nuclear stellar clusters and [BHs] do not form a single family of CMOs''.
Leigh, B\"oker, \& Knigge (2012) agree.  Their fit $M_{\rm nuc} \propto M_{\rm bulge}^{1.18 \pm 0.16}$ gives a
mean $M_{\rm nuc}/M_{\rm bulge} \simeq 0.0023$ at $M_{\rm bulge} \simeq 10^{9.7}$ $M_\odot$, roughly similar to 
the results in Scott \& Graham (2013) and also to those in {\bf Figure 26}.  Like Graham and Scott, they conclude that 
nuclei and BHs ``do not share a common origin''.  We do not necessarily disagree.  However, we cannot easily interpret the
above and many other papers on galactic nuclear star clusters in the context of the picture of galaxy structure and evolution
that we advocate in this paper for the following reasons.  (1) Most papers treat Sph galaxies as faint ellipticals
(e.{\ts}g., Ferrarese \etal 2006; C\^ot\'e \etal 2006; Turner \etal 2012).  We argue that elliptical and Sph galaxies are unrelated,
so the relevant comparison in {\bf Figure 26} is not between green triangles and red crosses.  Rather, Sphs are transformed dwarf 
spiral and irregular galaxies, so the relevant comparison (discussed above) is between green triangles and blue crosses.  (2) The above papers treat all
central extra light above the inward extrapolation of an outer S\'ersic function fit as nuclei.  In contrast, Hopkins \etal (2009b,
see Figure 45) show that extra light in low-luminosity ellipticals and nuclei in Sph and disk galaxies are physically different.
We agree with Turner \etal (2012) that (faint) nuclei and (brighter) extra light components form differently.  But they
have different parameter correlations, and confusing them creates mixed correlations that are difficult to interpret.
Finally, (3) the above papers do not differentiate between classical bulges and pseudobulges, both of which can contain
nuclei.  So they do not find the ~~~~~~~
\includegraphics{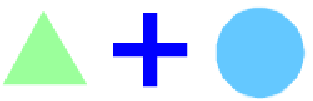} or the ~~~~~~ 
\includegraphics{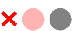} continuities, 
nor do they see the difference between the ~~~~~~~
\includegraphics{gBb.eps} correlation and the much flatter ~~~~~~ 
\includegraphics{Rrb.eps} correlation in {\bf Figure\ts26}.  

\vfill\eject

\cl{\null} \vfill
\includegraphics{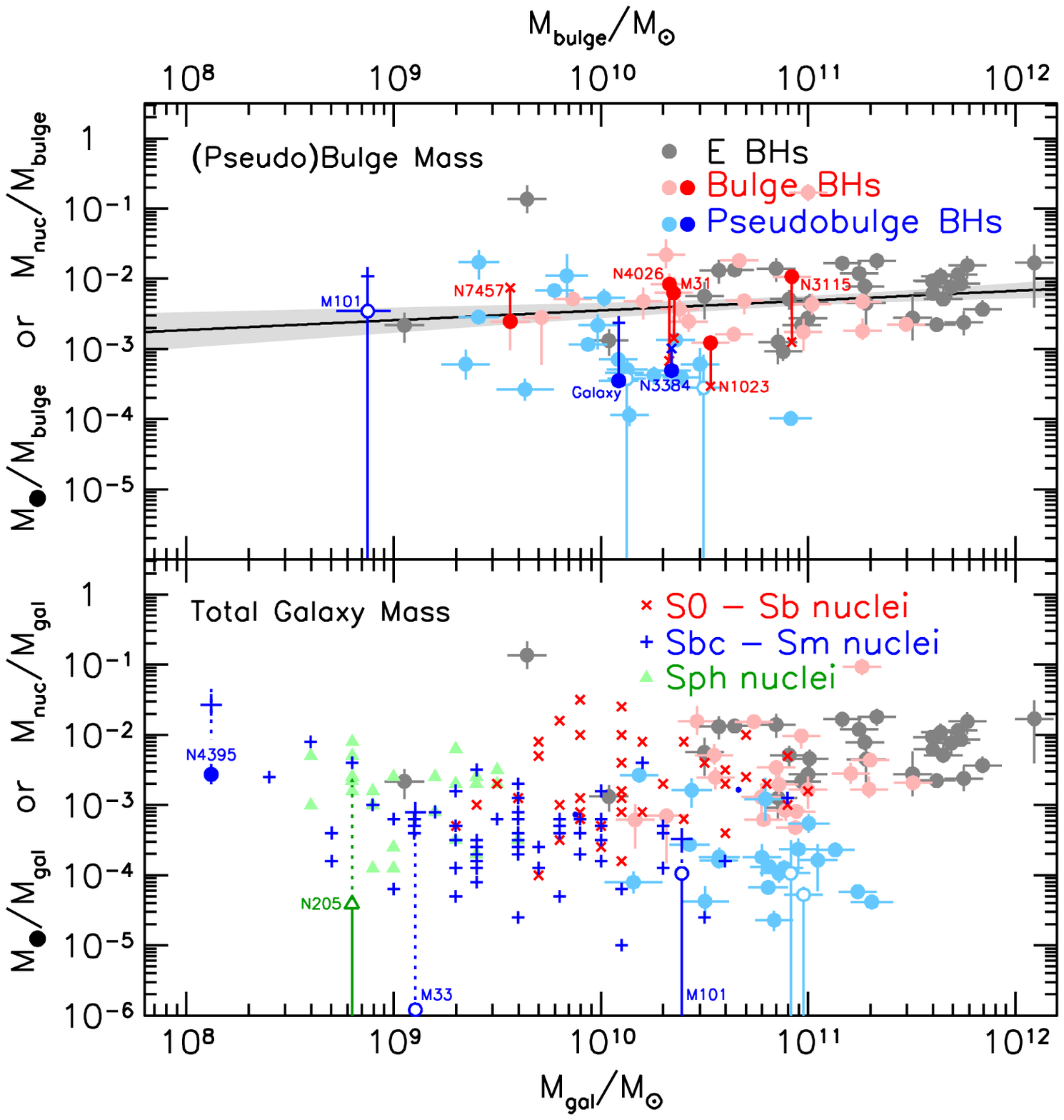}

\ni {\bf \textBlue Figure 26}\textBlack 

\vskip 2pt
\hrule width \hsize
\vskip 3pt
\ni ({\it top\/}) Ratio of CMO mass ($M_\bullet$ or the mass $M_{\rm nuc}$ of the nuclear star cluster) to
the mass $M_{\rm bulge}$ of the host (pseudo)bulge or elliptical galaxy plotted against $M_{\rm bulge}$.
\hbox{The black line with gray-shaded} 1-sigma error is the least-squares fit to the classical bulges and ellipticals 
(Section\ts6.6.1, {\bf Figure\ts18} and Equation 11).~({\it bottom\/}){\ts}Ratio of CMO mass to total host galaxy mass 
$M_{\rm gal}$ plotted against~$M_{\rm gal}$.  In both panels,
filled circles denote galaxies with BH mass measurements (see the key).
Open symbols denote galaxies with $M_\bullet$ upper limits.  When BHs and nuclei coexist, the points for 
$M_\bullet$ and $M_{\rm nuc}$ are joined by a vertical line and the name of the galaxy is given.  In the 
upper panel, NGC 3384 is an S0 with a pseudobulge, so $M_\bullet$ is plotted in blue; $M_{\rm nuc}/M_{\rm bulge}$ is 
also plotted blue for clarity.  In the lower panel, crosses, plus signs, and filled light-green triangles
denote nuclei in early- and late-type spirals~and~Sphs, respectively, all from Seth \etal (2008).

\eject

\vs
\hfuzz=150pt
\ni \hbox{\big\ARRed 6.12 BHs correlate with globular cluster systems in bulges and ellipticals}\textBlack
\vs

      Burkert \& Tremaine (2010) discovered that BH masses correlate with the total numbers $N_{\rm GC}$
of globular clusters in galaxies.  This was startling, because it connects the most compact 
objects at galaxy centers with one of the outermost galaxy components.  The sample contained
only 13 galaxies.  But the rms scatter in $M_\bullet$ was remarkably small, only 0.2 dex.  That is, 
the $M_\bullet$\ts--\ts$N_{\rm GC}$ correlation looks tighter than the $M_\bullet$\ts--\ts$\sigma$ correlation.  
To a good approximation, the total mass in globular clusters is the same as the mass in the central BH.

      Harris \& Harris (2011) confirmed these results for 33 galaxies.  They, too, found that $N_{\rm GC} \propto M_\bullet$, 
with a cosmic scatter (in addition to measurement uncertainties) of 0.2 dex in either parameter.  They got this result for 
ellipticals.  Three spirals agree with it; one (our~Galaxy)~deviates, and S0s by themselves show no correlation.  
Our Galaxy, NGC 4382, NGC 5128, and NGC 7457 all deviate from the main trend in having BHs that are at least a factor of 
10 undermassive.  For the galaxies that satisfy the $M_{\rm GC}$\ts--\ts$M_\bullet$ correlation,
$M_\bullet \simeq 1.5$ times the total mass of globulars.

      {\bf Figure 27} ({\it left}) updates the Harris \& Harris (2011) sample with $M_\bullet$ from {\bf Tables\ts2}~and~{\bf 3}. 
\bf Figure 27} ({\it right}) shows $N_{\rm GC}$ versus galaxy velocity dispersion for BH hosts in {\bf Tables\ts2}~and~{\bf 3}
and for other galaxies with published globular cluster counts.  Correcting several host-galaxy morphological 
types allows us to understand the puzzling results in the Harris \& Harris paper in the context of other results presented in this paper.
   
\vfill

\includegraphics{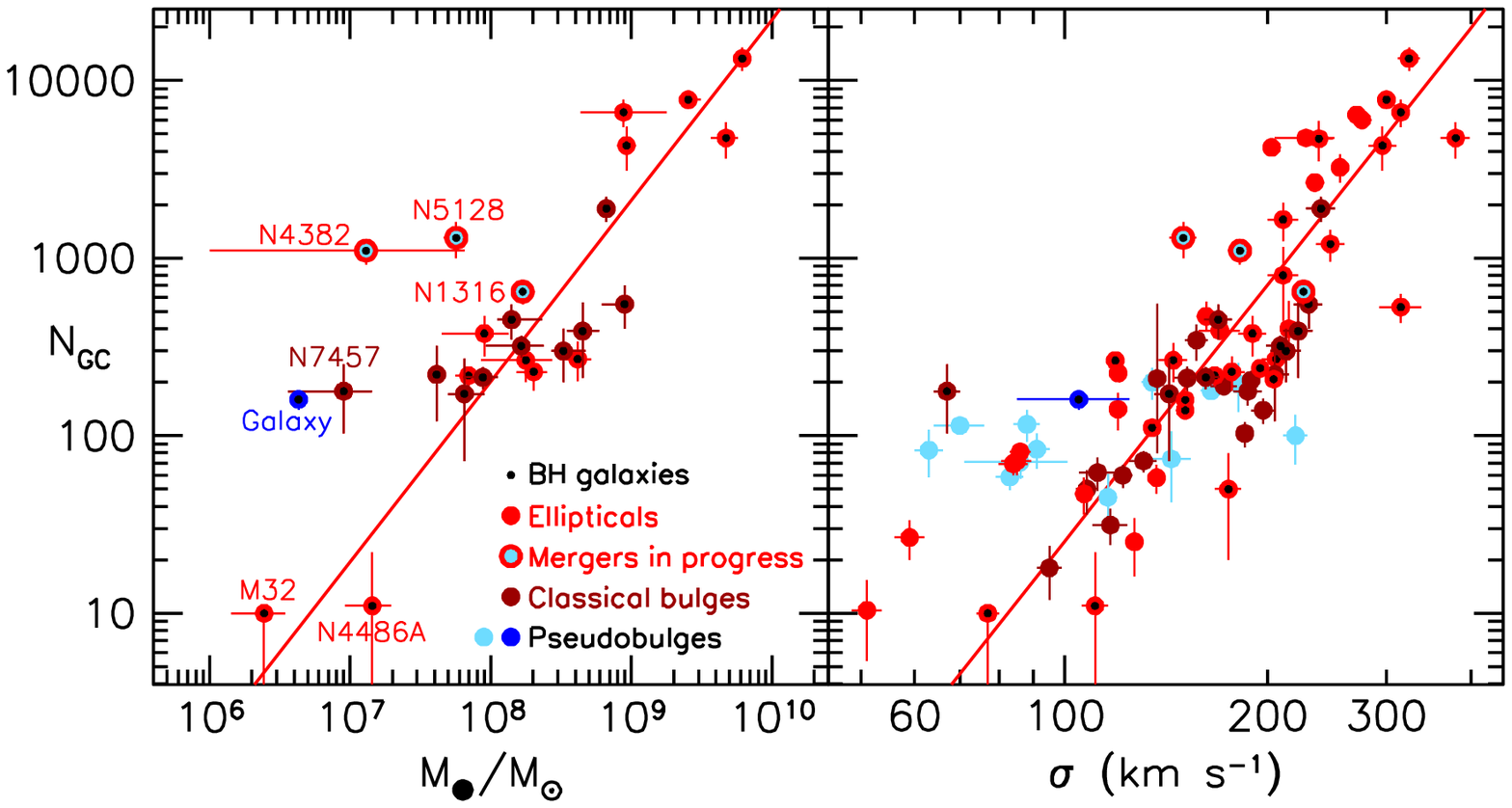}

\ni {\bf \textBlue Figure 27}\textBlack 

\vskip 1pt
\hrule width \hsize
\vskip 2pt
\ni Correlation of the total number of globular clusters with ({\it left}) dynamically measured~BH~mass 
and ({\it right}) the velocity dispersion of the (pseudo)bulge or elliptical host.  Galaxy types 
are given in the key.  The left panel omits galaxies with reliable BH detections but uncertain BH masses (turquoise lines 
in {\bf Tables 2} and {\bf 3}); these are included in the right panel.
The right panel~also~shows additional galaxies with $N_{\rm GC}$ values from 
Peng \etal (2008), 
Spitler \etal (2008),
Harris \& Harris (2011), and
Cho \etal (2012) that have velocity dispersions tabulated in Ho \etal (2009) or in Hyperleda.  The lines are symmetrical
least-squares fits to the classical bulges and ellipticals omitting mergers in progress, points that involve limits,
and NGC\ts7457.  We confirm that classical bulges and ellipticals show the same correlation between $N_{\rm GC}$ and 
BH mass or its proxy $\sigma$, whereas pseudobulges show little correlation. Pseudobulges and mergers in progress
frequently contain undermassive BHs.

\eject

      Elliptical galaxies and classical bulges satisfy the same linear correlations~in~both~panels.  
The left panel is similar to the $M_\bullet$\ts--\ts$N_{\rm GC}$ correlations shown by Burkert \& Tremaine
(2010) and by Harris \& Harris (2011), and the right panel is similar the $N_{\rm GC}$\ts--\ts$\sigma$ correlation 
shown by Snyder, Hopkins \& Hernquist (2011).  However, outliers are easier to interpret here. 

      In the left panel of {\bf Figure 27}, a symmetric least-squares fit to the classical bulges and ellipticals 
omitting mergers in progress, points that involve limits, and NGC 7457 gives
\vskip -2pt
$$
\log\ts\biggl(  {{N_{\rm GC}}\over{500}}  \biggr) 
       = (1.017 \pm 0.132)\ \log\ts\biggl( {{M_\bullet} \over {10^8~M_\odot}}  \biggr) - (0.393 \pm 0.106), \eqno{(20)}
$$
\vskip -2pt
\noindent similar to the results in the Burkert and Harris papers.  Harris finds that there are $\sim 250$
globulars per $10^8$ $M_\odot$ of BH mass; we find $202^{+56}_{-44}$.

      The total rms scatter of all points with respect to the fit is 0.32 in both $\log{N_{\rm GC}}$ and 
$\log {M_\bullet}$.  Ellipticals alone have an rms scatter of 0.34 dex in $\log{N_{\rm GC}}$.
Ellipticals and spirals together have a scatter of 0.30 dex, whereas S0s have a scatter of 0.36 dex.  
These differences are not significant.

      In constructing the above fit, we do not use the individual parameter errors but rather assume that all
galaxies have the mean estimated error of $\pm 0.12$ in $\log{M_\bullet}$ and $\pm 0.11$ in $\log{N_{\rm GC}}$.
If we assume that the intrinsic scatter is the same in both parameters, then that intrinsic scatter is
$\epsilon(\log{M_\bullet}) = \epsilon(\log{N_{\rm GC}}) = 0.20$.

      In the right panel of {\bf Figure 27}, a symmetric least-squares fit, again omitting pseudobulges, 
mergers in progress, points that involve limits, and NGC 7457 gives
\vskip -4pt
$$
\log\ts\biggl(  {{N_{\rm GC}}\over{500}}  \biggr) 
       = (4.80 \pm 0.39)\ \log\ts\biggl( {{\sigma} \over {200~\rm km~s^{-1}}}  \biggr) + (0.150 \pm 0.064), \eqno{(21)}
$$
\vskip -2pt
\noindent similar to the results in Snyder \etal (2011).  
The rms scatter for all points with respect to the above fit is 0.45 in $\log{N_{\rm GC}}$ and 
0.095 in $\log {\sigma}$.  The ellipticals alone have an rms scatter of 0.49 dex in $\log{N_{\rm GC}}$, and 
the classical bulges have a scatter of 0.29 dex.  S0s do not have larger scatter than other galaxy types.
In the above, we  again do not use the individual parameter errors but rather assume that all
galaxies have the mean estimated error of $\pm 0.019$ in $\log{\sigma}$ and $\pm 0.10$ in $\log{N_{\rm GC}}$.
If we assume that the intrinsic scatter is the same in both parameters, then that intrinsic scatter is
$\epsilon(\log{\sigma}) = \epsilon(\log{N_{\rm GC}}) = 0.089$. 

     Pseudobulges and mergers in progress deviate from the above correlations:

      In Harris \& Harris (2011), the deviant elliptical is NGC 5128 (Cen A).  However, this is a
merger in progress; i.{\ts}e., an elliptical in formation.  One deviant ``S0'' in the Harris paper is NGC 4382.
KFCB show that this is also a merger in progress.  An additional ``S0'', NGC 1316, is reclassified 
in {\bf Figure 27} as a merger.  The three mergers in progress behave the same way 
in {\bf Figure 27} as they do in the $M_\bullet$\ts--\ts$M_{\rm K,bulge}$ and $M_\bullet$\ts--\ts$M_\sigma$ 
correlations in {\bf Figure\ts14} (Section\ts6.4).  They deviate from the correlations for ellipticals and 
classical bulges in the sense that $M_\bullet$ is unusually low for the galaxy $K$-band luminosity and for the
number of globular clusters.  The latter deviation confirms what we concluded in Section 6.4: the
deviation there is not due to any temporary enhancement of the luminosity due to merger-induced star formation.
The $K$-band luminosity is insensitive to modest amounts of star formation, and the main stellar populations of
the mergers in progress are in any case old.  The globular cluster systems are also old, and NGC 4382 and NGC 5128
have \gapprox 10 times more globular clusters than ``normal'' for their BH masses.  Even NGC 1316 lies near
the top of the scatter in {\bf Figure 27}.  As in Section 6.4, we conclude that the BH
masses are unusually low in these mergers in progress. 
The correlation with $\sigma$ is more normal, as it was in {\bf Figure 14}.    All in all, {\bf Figure 27}
provides compelling independent corroboration for the BH results found in Section 6.4 for mergers in progress.
Of course, the sample is very small.  The importance of further BH and globular cluster measurements in ongoing mergers 
is correspondingly high.

      In Harris \& Harris (2011), NGC 7457 is the other S0 and our Galaxy is the spiral that deviates from the
$N_{\rm GC}$\ts--\ts$M_\bullet$ correlation by having very small $M_\bullet$.  But our Galaxy contains a boxy pseudobulge
(Weiland \etal 1994;
Dwek \etal 1995),
now understood to be an almost-end-on bar 
(Combes \& Sanders 1981;
Blitz \& Spergel 1991;
see Kormendy \& Kennicutt 2004 for a review).
There is no photometric or kinematic sign of a merger-built classical bulge
(Freeman 2008;
Howard \etal 2009; 
Shen \etal 2010).
It is therefore plotted in blue in {\bf Figure 27}.  For NGC 7457, bulge classification criteria provide ``mixed signals'' -- 
the relatively round shape, high S\'ersic index, and moderate rotation ($V/\sigma^* \simeq 1$, Kormendy \etal 2011) led us 
to classify the bulge as classical in {\bf Table 3}, but its remarkably low velocity dispersion of $\sigma = 67 \pm 3$ km s$^{-1}$ 
has long suggested that it is pseudo (Kormendy 1993b).  We know that classical and pseudo bulges coexist in some galaxies.
This may be one of them, and NGC 7457 may deviate from the {\bf Figure 27} correlations for this reason.  We know that pseudobulge 
properties correlate little with $M_\bullet$ and particularly that some pseudobulges contain smaller BHs than do classical bulges.  
This is how our Galaxy and NGC 7457 deviate in {\bf Figure 27}.   Burkert \& Tremaine (2010) already note this result 
for our Galaxy.  For NGC 7457, the unusually small $\sigma$ in the right panel suggests that our classification in
{\bf Table 3} may be wrong and that the pseudobulge may be dominant.  Classifying mixed cases is difficult.

      In the right panel of {\bf Figure 27}, pseudobulges are plotted in light blue.  Bulges from Peng \etal (2008) 
and Spitler \etal (2008) are classified in Kormendy \& Bender (2011) or here.
The classical bulges form part of the correlation discussed above.  Overall, the pseudobulges here are consistent with 
results from previous sections:~they do not show the same correlations as classical bulges.  However, it is still possible
that the scatter in {\bf Figure\ts27} ({\it right}) just increases toward lower $N_{\rm GC}$ and  $\sigma$.
More measurements would be welcome.

      For completeness, we note that six Harris \& Harris (2011) galaxies are omitted in {\bf Figure 27}.  NGC 4552, NGC 4621, NGC 5813, and NGC 5846
had BH masses from Graham (2008b) that were read from an unlabeled plot in Cappellari \etal (2008); see our discussion in Section 5.1.
They agree with the {\bf Figure 27} correlations.  The NGC 4350 BH mass is based on observations with low spatial resolution 
(Section 6.5).  And NGC 4486B is one of the ``monsters'' that deviate toward high $M_\bullet$ (Section 6.5).  By and large, BHs
that deviate from the correlations described in earlier sections also deviate in the same way here.

      Returning to the main result of this section, what does the tight $M_\bullet$\ts--\ts$N_{\rm GC}$ correlation 
for bulges and ellipticals tell us about galaxy formation?  The formation of globular cluster systems is an 
large subject with a long history (see Harris 1991 and Brodie \& Strader 2006 for reviews).  We confine 
ourselves to a few remarks:

      Based on their conclusion that the $M_\bullet$ correlates more tightly with globular clusters than with~$\sigma$, 
Burkert \& Tremaine (2010) suggest that major mergers may provide the connection between the 
smallest and largest scales in galaxies.  They suggest that the most rapid BH growth may happen in
the most dissipative mergers that manufacture the most globular clusters (Ashman \& Zepf 1992;
Zepf \& Ashman 1993).  That suggestion is not invalidated by our results on mergers in progress, because mergers at 
low redshifts are almost certainly poorer in gas than mergers at high redshifts.  We suggested in 
Section 6.4 that these mergers convert disk mass that does not correlate with $M_\bullet$ into bulge mass with 
relatively little new star formation; the resulting $M_\bullet$ is smaller than normal.  Our mergers
are not ultraluminous infrared galaxies; they are among the gas-poorest ones~at~low~$z$.  We do not 
expect much late formation of globular clusters.  Even at $z \sim 1$, our understanding 
of cores requires that the most recent mergers that made the biggest elliptical galaxies were dry 
(Section 6.13).  Any correlations between $M_\bullet$, $\sigma$, and $N_{\rm GC}$ likely were
put in place very early.  Late BH growth by ``maintenance-mode accretion'' from hot gas in clusters 
(Section 8.4) is unlikely to increase $M_\bullet$ by much, and it is unlikely to form globular clusters
(C\^ot\'e \etal 2001).  

      Harris \& Harris (2011) note that most globulars have ages of $\sim 10$ to 13 Gyr, corresponding to redshifts 
$z \simeq 2$ to 7.  They suggest that ``$N_{\rm GC}$ and $M_\bullet$ should be closely correlated simply because 
they are both byproducts of similarly extreme conditions in high-density locations during the main period of galaxy 
formation'' (cf.~the Burkert \& Tremaine 2010 picture).  Both they and Burkert \& Tremaine (2010) point out that the 
present similarity in the masses of BHs and globular cluster systems is a coincidence, because BHs have been growing 
and globular cluster systems have been suffering attrition for many billions of years.   

       Snyder \etal (2011) also argue that the tight $M_\bullet$\ts--\ts$N_{\rm GC}$ correlation
does not imply any direct connection between globular clusters and BHs.  Rather, they argue that $M_\bullet$
and $N_{\rm GC}$ both correlate primarily with galaxy binding energy $\propto M_* \sigma^2$, where $M_*$ is the
stellar mass of the galaxy.  The $M_\bullet$\ts--\ts$M_* \sigma^2$ correlation is called the ``black hole
fundamental plane''
(Hopkins \etal 2007a, b;
Aller \& Richstone 2007;
Section 6.14 here).
Rhode (2012) is another study of correlations between $M_\bullet$, $N_{\rm GC}$, and other host galaxy properties
based on a sample that mostly overlaps with the above studies and with ours.  She, too, concludes that BH masses and numbers
of globular clusters are not directly linked but are correlated because both depend on the depths of galaxy potential wells.

      It is well known that globular cluster systems are bimodal in color and metallicity.  A metal-rich,
red population is relatively concentrated to galaxy centers, whereas a metal-poor, blue population forms a radially more
extended cloud; the red globulars are more nearly coeval with the main galaxy, whereas the blue globulars likely
predate the main galaxy (e.{\ts}g, 
Forbes, Brodie \& Grillmair 1997;
Larsen \etal 2001;
Peng \etal 2006c;
Harris \etal 2006;
Strader \etal 2006;
see Brodie \& Strader 2006 for a review).
It is important to emphasize: We do not yet know whether $M_\bullet$ correlates best with the red globulars,
the blue globulars, or their sum.  It will be important to make this test.  But as Burkert \& Tremaine (2010)
note, this will be difficult, because the relative numbers of red and blue clusters do not vary by much from 
galaxy to galaxy.  

      Sadoun \& Colin (2012) take a first step in this direction by showing that the correlation between $M_\bullet$ 
and the velocity dispersions of globular cluster systems (not the numbers of globular clusters) is tighter for red
than for blue globular clusters.  This is perhaps not surprising, because red globular cluster systems are more 
closely associated with the main bodies of their hosts and we already know that $M_\bullet$ correlates tightly with
the velocity dispersions of those hosts.  Nevertheless, it seems reasonable to suggest that, if larger samples that 
include cluster counts and perhaps other parameters confirm that BHs correlate best with red globulars, then this points
to coevolution processes such as mergers and the origin of the BH fundamental plane.  In contrast, if BHs correlate best 
with blue globular clusters, then this points to unknown initial conditions in the environments that later turn into giant ellipticals. 
E.{\ts}g., M{\ts}87 now has one of the highest numbers of globular clusters and one of the highest BH masses.  But we know 
almost nothing about the (probably~many) progenitor pieces in which its stars were formed, and we know still less about the 
(probably earlier) times when its blue globular clusters formed.  For further discussion, see, e.{\ts}g., 
Ashman \& Zepf (1992), 
Forbes, Brodie \& Grillmair (1997), 
Kundu \etal (1999), 
Brodie \& Strader (2006), 
Hopkins \etal (2007a, b), and 
Yoon \etal (2011).

\vfill\eject

\vs
\hfuzz=150pt
\ni \hbox{\big\ARRed 6.13 BHs correlate with the ``missing light'' that defines galaxy cores}\textBlack
\vs

      BHs modify the structures of their host galaxies out to several times the sphere-of-influence radius 
$r_{\rm infl} = G M_\bullet/\sigma^2$ inside which they dominate the gravitational potential.  E.{\ts}g., many stars which visit 
$r < r_{\rm infl}$ live on orbits that take them out to radii
$r > r_{\rm infl}$ where they~spend~most~of~their~time.  Scattering off of the BH results in gradual changes to the orbital 
structure, a slow evolution that we do not have space to discuss (see Merritt 1999 for a review). 
More relevant here is this: We expect that galaxy mergers generally result in BH binaries, and their
interactions with host galaxies have observable consequences.  This {\it local\/} BH{\ts}--{\ts}galaxy 
coevolution driven by gravity is different from the {\it global\/} coevolution driven by AGN energy feedback 
that is the subject of Section\ts8.  Local evolution driven by BH binaries provides a natural explanation{\ts}--{\ts}albeit one 
that is difficult to prove{\ts}--{\ts}of an almost ubiquitous feature of giant elliptical galaxies, namely galaxy cores.

      The term ``core'' has a specific meaning, different from its colloquial use to refer to a galaxy center.  
It is illustrated in {\bf Figure 28}.  Observations over many years have shown that surface brightness profiles $I(r)$ of 
elliptical galaxies are well described by S\'ersic (1968) functions $\log {I(r)} \propto r^{1/n}$, where $n \simeq 2$ to 12 is 
the ``S\'ersic index'' (see KFCB for a review).  For historical reasons (early data suggested that $n = 4$; de Vaucouleurs 1948), 
it is convenient to plot profiles against $r^{1/4}$; then profiles that are concave-up have $n > 4$ 
and profiles that are concave-down have $n < 4$.  {\bf Figure\ts28} shows that an $n = 6$ S\'ersic function
fits NGC 4472, the brightest elliptical in the Virgo cluster and one of our BH galaxies, over a range of a factor of 
more than 200 in radius and 6000 in surface brightness.  Then, at small radii, $I(r)$ breaks downward from the 
inward extrapolation of the outer S\'ersic profile.  This region of lower surface brightness where $I(r)$ is a shallow 
power-law cusp is called the core.  In contrast, the smaller BH elliptical NGC 4459 shows an inner break in the opposite direction; 
instead of a core, it has  extra light at small radii with respect to the outer $n = 3.2$ S\'ersic profile. 

\vfill


 \includegraphics{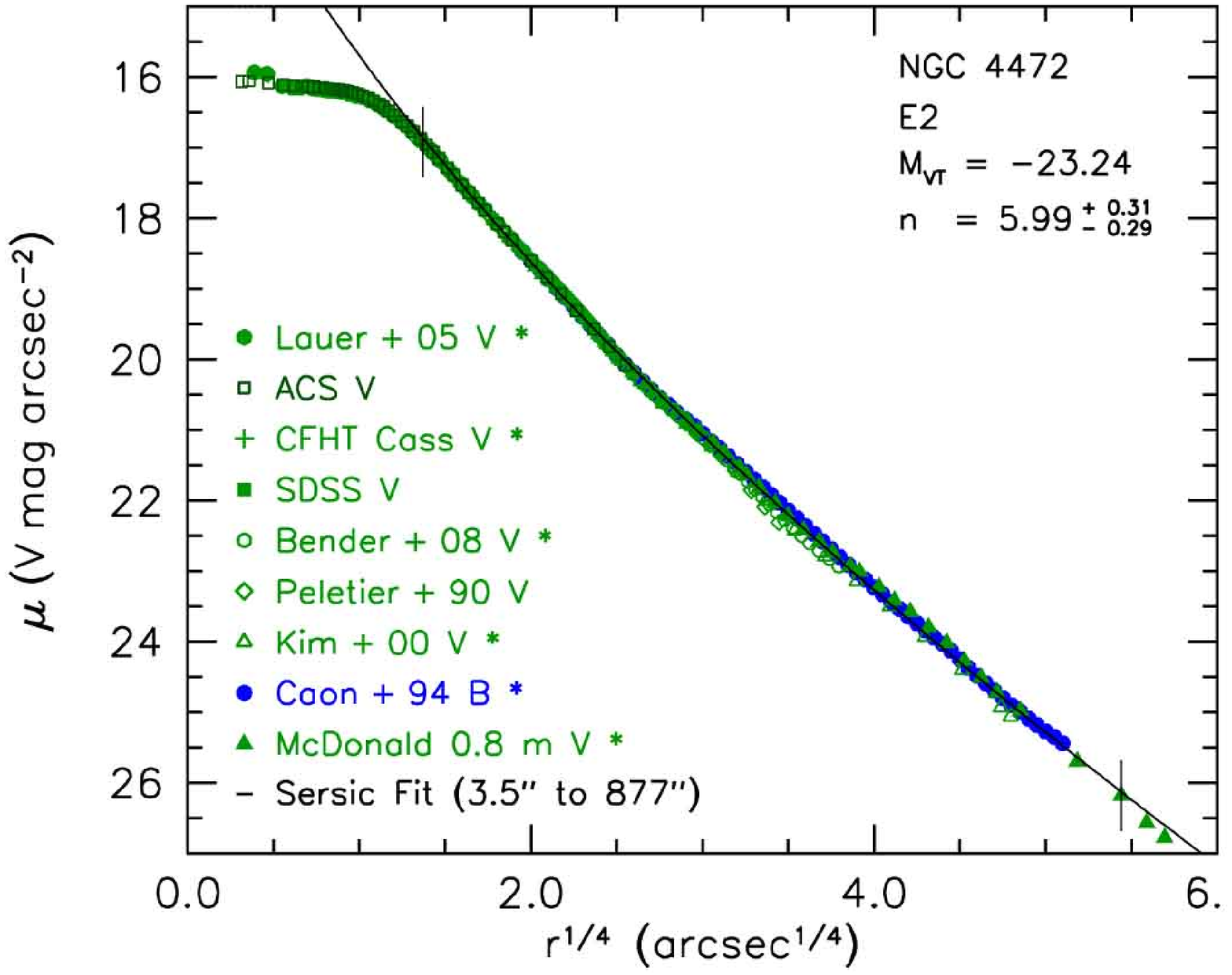}
 \includegraphics{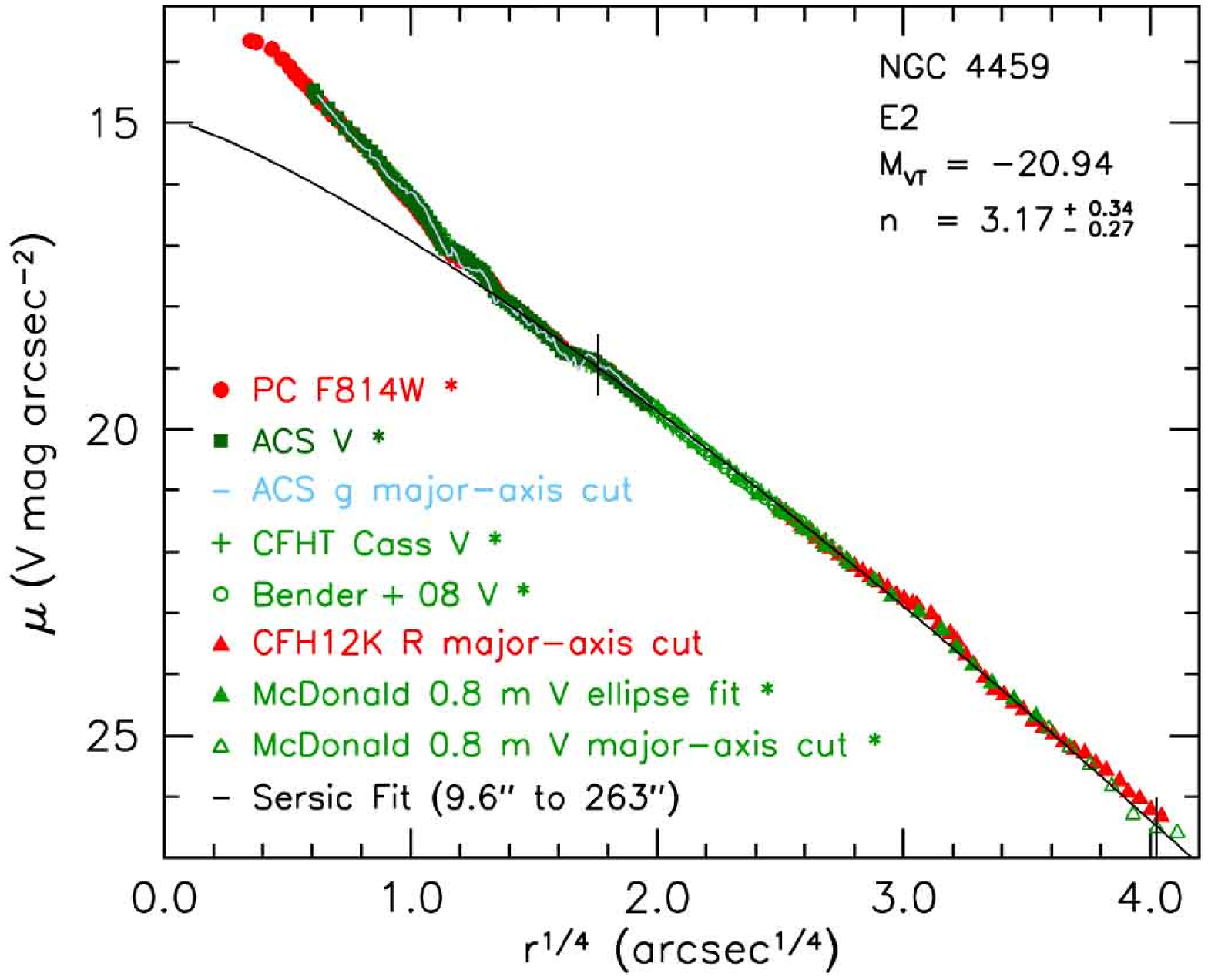}

\ni {\bf \textBlue Figure 28}\textBlack 

\vskip 2pt
\hrule width \hsize
\vskip 2pt
\ni Observed $V$-band ($\sim$\ts5500\ts\AA) surface brightness profiles of ({\it left\/}) NGC 4472, a prototypical giant 
elliptical with a core, and ({\it right\/}) NGC 4459, a lower-luminosity elliptical with no core but instead with extra 
light at small radii above the inward extrapolation of the outer brightness profile.  The photometry is from 
KFCB.  References to data sources are in the key.

\eject

      Section 6.7 introduced the difference between ellipticals with cores and ones with extra light as part of an ``E{\ts}--{\ts}E dichotomy''
that also involves dynamical properties (slow versus fast rotation), the presence or absence of X-ray-emitting gas, 
and the presence or absence of AGNs.  The change from extra light to core happens in the Virgo cluster at an
absolute magnitude of $M_V$\ts$\simeq$\ts$-21.5$,  
a luminosity of $\sim$\ts3\ts$\times$\null$10^{10}$\ts$L_{V\odot}$, 
a stellar mass of $\sim$\ts2\ts$\times$\null$10^{11}$\ts$M_\odot$, and
a dark matter mass of $\sim$\ts5\ts$\times$\null$10^{12}$\ts$M_\odot$ ({\bf Figure\ts25}).
KFCB review these observations~and suggest that the most recent mergers that formed core ellipticals were
``dry'' (no cold gas and no significant star formation), whereas the most recent mergers that formed extra-light 
ellipticals were ``wet'' (cold gas dissipation fed a starburst that built the extra-light component).
The idea is not new; e.{\ts}g., Faber \etal (1997) already suggested it as the explanation for the difference
between core and coreless ellipticals.  Sections 8.4 and 8.6 discuss the E{\ts}--{\ts}E dichotomy in the context of 
BH--host-galaxy coevolution.

      Core formation is hard to understand in our standard picture in which ellipticals form by
\hbox{major} galaxy mergers, i.{\ts}e.,  \hbox{ones in which the progenitors were equal in mass to within a factor~of~a~few.}
{\bf Figure 29} shows why.  The surface brightness of the extra-light elliptical NGC\ts4459 is fainter than that of
NGC\ts4472 at most radii, but it is brighter than NGC\ts4472 inside
the latter's~core.  Absent BHs, dry mergers preserve the highest densities in the progenitor galaxies  
(Kormendy 1984;
Barnes\ts1992;
Barnes\ts\&{\ts}Hernquist\ts1992;
Holley-Bockelmann{\ts}\&{\ts}Richstone{\ts}1999,{\ts}2000;
Merritt \& Cruz 2001;
Boylan-Kolchin, Ma \& Quataert 2004,{\ts}2005;
Hopkins \etal 2009a,{\ts}b).  
NGC\ts4472 is so bright that the only plausible progenitors are smaller ellipticals like NGC\ts4459.   How do
progenitors with dense centers merge to form remnant ellipticals with fluffy centers (Faber{\ts}et{\ts}al.{\ts}1997)?

      The suggested solution is this:~Cores may be scoured by the orbital decay of BH binaries that form 
in galaxy mergers
(Begelman, Blandford \& Rees 1980; 
Ebisuzaki, Makino \& Okamura 1991; 
Makino \& Ebisuzaki 1996; 
Quinlan \& Hernquist 1997; 
Faber \etal 1997; 
Milosavljevi\'c \& Merritt 2001; 
Milosavljevi\'c et al.~2002; 
Merritt 2006).  
When the BHs are far apart, they decay separately by dynamical friction.  When they get close together, they form a binary, 
and then the tendency toward energy equipartition between light and heavy particles drives the very generic result that 
the BH binary decays by flinging stars away (Szebehely \& Peters 1967 show a beautiful example).  \phantom{0000000000}

\vfill

\includegraphics{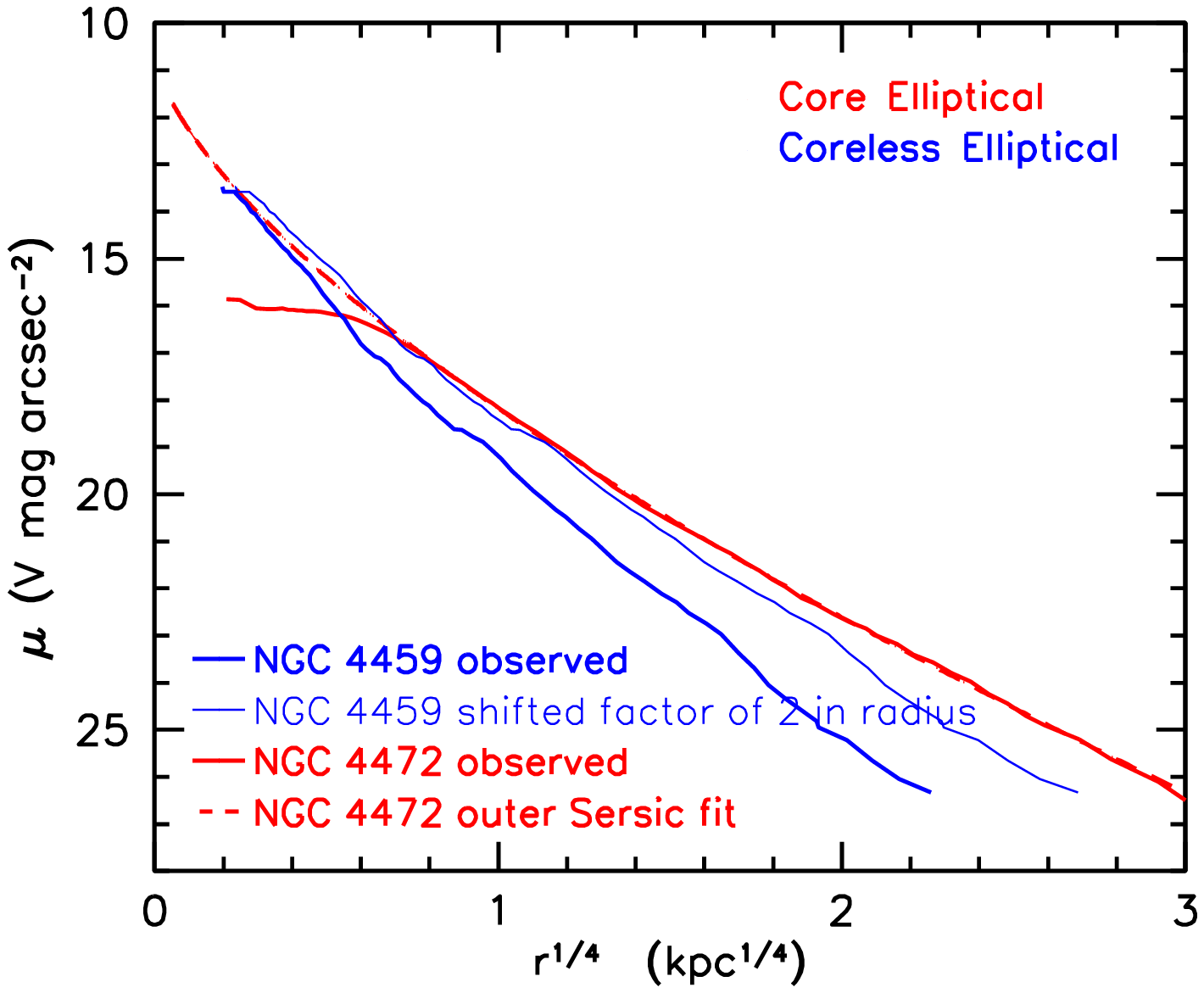}

\includegraphics{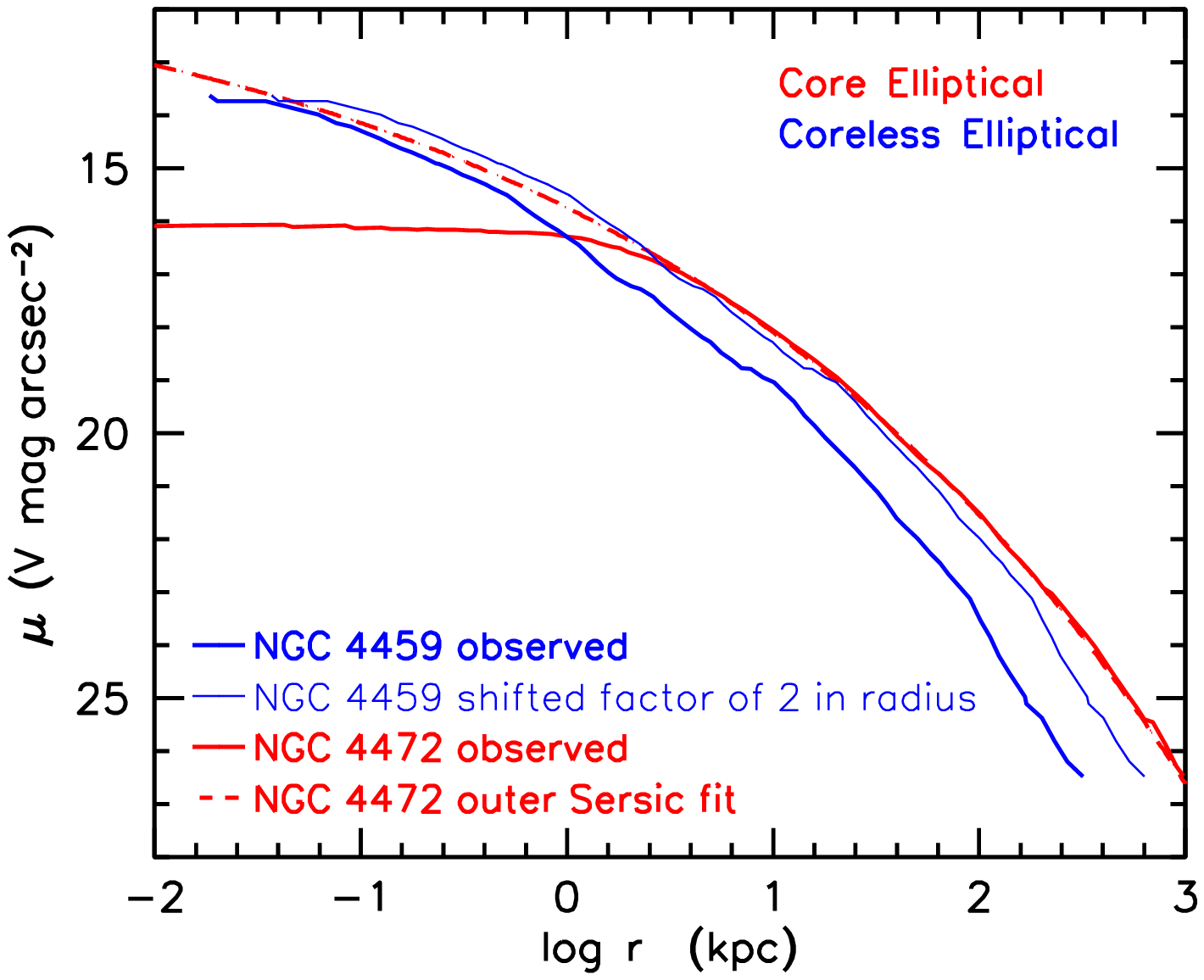}

\ni {\bf \textBlue Figure 29}\textBlack 

\vskip 2pt
\hrule width \hsize
\vskip 2pt
\ni Major-axis brightness profiles of NGC 4472 and NGC 4459 corrected for Galactic absorption and plotted against $r^{1/4}$ ({\it left\/}) and
$\log {r}$ ({\it right\/}).  In each panel, the thin blue line is the observed profile of NGC 4459 shifted to larger
radii by a factor of 2.  Near the center, this shifted profile is similar to the inward extrapolation of the outer
S\'ersic fit to the core elliptical NGC 4472 ({\it red dashed line}).

\eject

\noindent 
This decreases the surface brightness and excavates a core.  The effect of a series of mergers should be 
cumulative; if the mass deficit after one merger is $f M_\bullet$, then the mass deficit after $N$ 
mergers should be $\sim N f M_\bullet$.  The above papers predict that $f \sim 0.5$ to 2.  Observations of mass
deficits $M_{\rm def}$ are consistent with this picture; $M_{\rm def} \propto M_\bullet$ and $Nf \sim 1${\ts}--{\ts}5, 
consistent with formation by several dry mergers 
(e.{\ts}g., Milosavljevi\'c \& Merritt 2001;
Milosavljevi\'c et al.~2002; 
Ravindranath, Ho \& Filippenko 2002;
Graham 2004; 
Ferrarese \etal 2006; 
Merritt 2006;
Hopkins \& Hernquist~2010). 

     As an explanation of cores, scouring by binary BHs is ``the only game in town''. Still, how~can~we be sure that any mass is missing?
How can we know what the density profile would be in the absence of any proposed mechanism such as core scouring?  Differences between 
different authors' estimates of mass deficits result mainly from different assumptions about this unmodified profile.  

      Following Kormendy \& Bender (2009), we argue here that the best way to calculate $M_{\rm def}$ is by comparing the
observed core profile with the inward extrapolation of a S\'ersic function fitted to the profile at larger radii 
(Graham 2004; 
Ferrarese \etal 2006;
KFCB;
Kormendy \& Bender 2009).  
The arguments for calculating $M_{\rm def}$ in this way are: \vs

\nhi1 1.~Some authors assume that the central profile before scouring was an inward extrappolation of a power law
         fitted just outside the core.  {\bf Figure 29} ({\it right\/}) shows why this is not ideal.  Plotted
         against $\log {r}$, the outer profile is curved at all radii.  Over large radius ranges, elliptical galaxy profiles 
         are not power laws.  Picking a small radius range near the center in which to fit a power law requires
         a choice that is essentially arbitrary.  The figure also shows that a power-law extrapolation lies
         above the S\'ersic extrapolation~at~all~$r$.  It is therefore a less conservative choice than a S\'ersic
         extrapolation. \vs

\nhi1 2.~We can, in fact, know something about merger progenitors.  Core ellipticals are more luminous than
         any other galaxies in the nearby universe.  NGC 4472 is $\sim$\ts8 times more luminous than NGC 4459.  
         Faber \etal (2007) and KFCB argue that core galaxies form by major mergers that are dry because they are so massive
         that even their progenitors generally~have~DM~masses $M_{\rm DM}$ \gapprox \ts$M_{\rm crit} \simeq 10^{12}$ $M_\odot$ large enough 
         to retain X-ray-emitting gas that heats any~cold~gas and prevents star formation ($M_{\rm crit}$ quenching).
         Further evidence for dry mergers is discussed in Sections 6.7.2 and 8.4.  And mergers with mass ratios that are
         not too different from 1:1 are preferred in order to life enough stars to explain observed cores (Merritt 2006).
         Then, over a large range in luminosity and mass, the only plausible present-day progenitors 
         whose mergers could produce the global properties of core ellipticals are slightly fainter ellipticals and disk galaxies
         with large bulges.  Coreless ellipticals are like NGC 4459: their profiles have extra light near the 
         center with respect to the inward extrapolation of outer S\'ersic profiles.
         It is conceivable that the progenitors of core ellipticals were nothing like any galaxy now in the universe.  If so,
         it would be necessary to use them all up.  This seems unlikely.  So NGC 4459 is representative
         of known galaxies that could plausible merge to make core ellipticals. \vs
 
\nhi1 3.~{\bf Figure 29} shows that, over the radius range of the core of NGC\ts4472, the profile of NGC\ts4459
         is similar to the inward extrapolation of the outer S\'ersic fit to NGC 4472.  In addition: \vs

\nhi1 4.~The profile of a dry-merger remnant is similar to that of the (say:~equal-mass) progenitors, with little change in 
         density but a shift of the profile to larger radii (Hopkins \etal 2009c).  {\bf Figure\ts29} includes the profile of NGC\ts4459 
         shifted outward in radius by a factor of 2 (a fourfold increase in luminosity).  Except at large $r$, this resembles the profile 
         of NGC\ts4472, including the inward extrapolation of the outer S\'ersic fit.  So numerical simulations support our assumption 
         that remergers of realistic progenitors produce remnant profiles such that we can estimate the missing light that defines cores 
         by comparing observed profiles (we assume: after BH scouring) with the inward extrapolation of S\'ersic functions fitted to the 
         outer profiles.\vs

\nhi1 5a.{\ts}More specifically, $n$-body simulations show that the remnants of dry major mergers{\ts}--{\ts}that is, remnants of 
         mergers of two, approximately equal-mass galaxies{\ts}--{\ts}are approximately S\'ersic at all radii that are not
         compromised by resolution effects (e.{\ts}g., Hopkins \etal 2009b). \vs

\nhi1 5b.{\ts}Moreover, simulations of wet mergers
          (Mihos \& Hernquist 1994; 
          Hopkins \etal 2009a) 
          produce remnants that closely resemble extra-light ellipticals, with a S\'ersic outer component that is
          the product of violent relaxation of the pre-existing stars and a separate, dense central component that is
          the result of the starburst.  This has been emphasized by 
          Kormendy (1999),
          C\^ot\'e \etal (2007),
          Hopkins \etal (2009a), and
          KFCB. \vs

      Thus, within our picture of the merger formation of ellipticals, it is reasonable to expect that the near-central profile of a giant 
elliptical would not show a core.  Rather, the central profile should be a steep continuation of the outer profile if it formed by violent
relaxation in dry mergers and in the absence of BHs.  This is the motivation behind our procedure to estimate $L_{\rm def}$.

     Hopkins \& Hernquist (2010) suggest a nonparametric way to estimate $L_{\rm def}$ that is is similar~in spirit to what we do.~They 
use lower-mass galaxies that are plausible $z$\ts$\simeq$\ts0 progenitors to provide template profiles against which to measure missing 
light (not) in the cores of higher-mass ellipticals.  We prefer our approach, because it relies on the robustness (KFCB) of 
S\'ersic fits over large dynamic ranges in the core galaxies under study.~Using template profiles is inherently more uncertain, 
(1) because relatively few profiles get averaged into the templates, as Hokpkins \& Hernquist emphasize, and 
(2) because progenitors at high $z$ may differ from those at low $z$.  So we use low-$z$ galaxies as a guide in our argument, but we depend
on well determined S\'ersic fits to the actual galaxies under study to make what is, after all, a tricky differential measurement of $L_{\rm def}$.
Nevertheless, the Hopkins \& Hernquist (2010) approach should give results that are similar to ours, and it does so.

      Using the above prescription, KFCB and Kormendy\ts\&{\ts}Bender\ts(2009) measure central light and mass
excesses and deficits as illustrated in {\bf Figure 30}.  BH masses for core galaxies are updated to include DM in
the dynamical models.  We confirm previous results
(Milosavljevi\'c\ts\&{\ts}Merritt\ts2001;
Milosavljevi\'c et al.~2002; 
Ravindranath, Ho \& Filippenko 2002;
Graham 2004; 
Ferrarese~et~al.~2006; 
Merritt 2006)
that $M_{\rm def} \propto M_\bullet$.  
Averaging in the $\log$, $<$\null${M_{\rm def}/M_\bullet}$\null$>$ = $4.1^{+0.8}_{-0.7}$.  Our  $M_{\rm def}$ measures
are larger than previous published values.  However, BH masses are revised upward also ({\bf Table\ts2}).  As a result,
the ratio  $<$\null${M_{\rm def}/M_\bullet}$\null$>$ is in the range given by previous papers.  

\vfill




\includegraphics{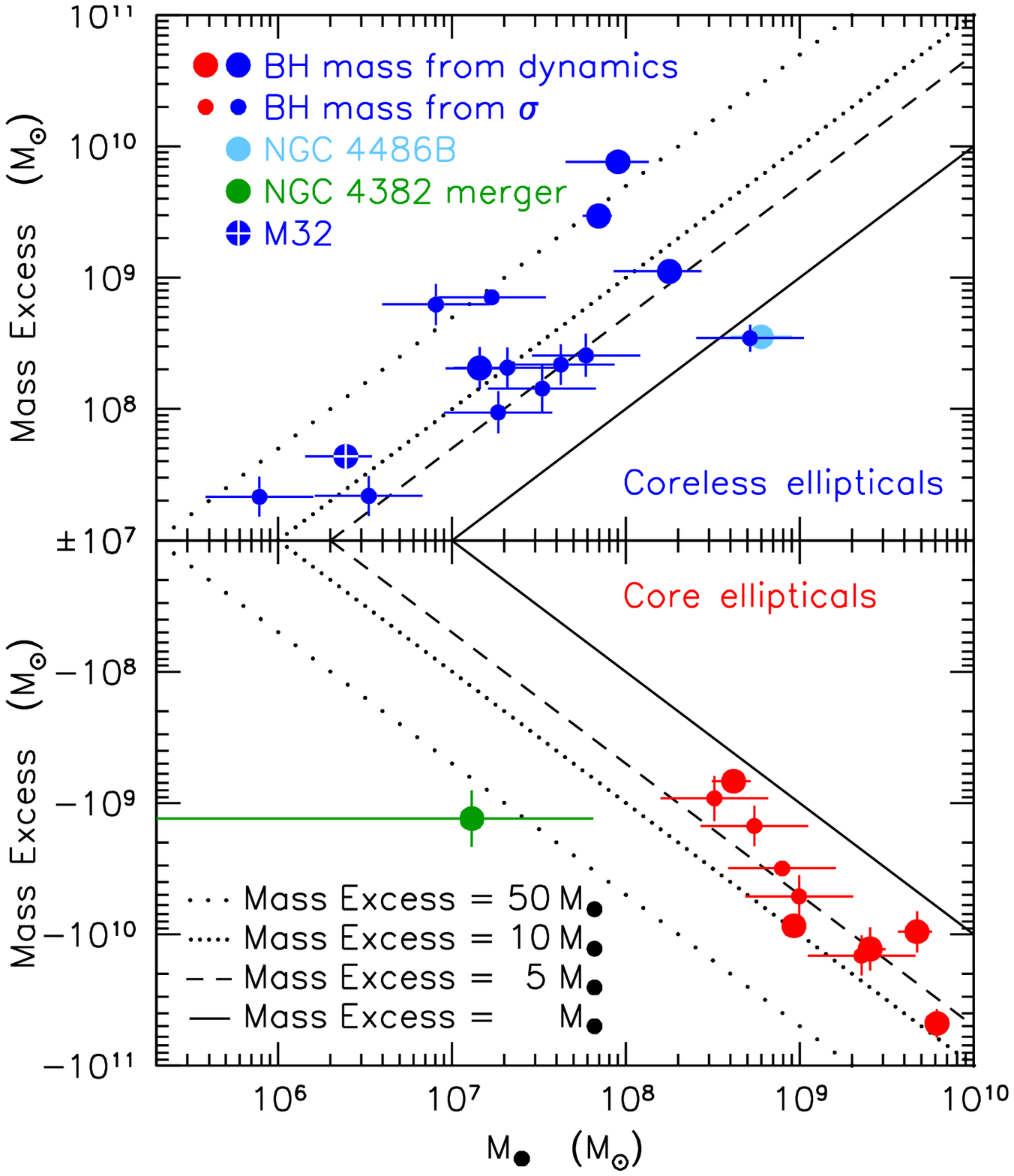}

\def\figindent{\indent \hangindent=2.95truein}

{\parindent=2.95truein
\figindent \ni {\bf \textBlue Figure 30}\textBlack 

\vskip 3pt
\nointerlineskip \moveright 2.95truein\vbox{\hrule width 2.75truein}\textBlack\nointerlineskip
\vskip 3pt

\widebaselines

\figindent Stellar mass ``missing''~in~cores~({\it lower~panel\/}) or ``extra'' in coreless galaxies ({\it upper~panel\/}) 
vs.~BH~mass. Large symbols are for galaxies with dynamical $M_\bullet$ measurements; small ones use $M_\bullet$ 
given~by~the \hbox{$M_\bullet$\ts--\ts$\sigma$} relation.  Core~mass~deficits correlate tightly with~BH~mass: \hbox{$M_{\rm def} \simeq 4{\ts}M_\bullet$}.  Mass excesses tend to be larger than mass deficits and to show more scatter with $M_\bullet$. 
It is imporant to note that the merger in progress NGC 4382 has a normal core for its luminosity but deviates
to small ${M_{\rm def}/M_\bullet}$.   If cores are excavated by BH binaries, this suggests that the galaxy
does not lack a big BH (or BH binary) but rather that this BH (or BH binary) is not resident at the center.
Updated from Kormendy \& Bender (2009).

}

\eject

       The mean ratio $<$\null${M_{\rm def}/M_\bullet}$\null$>$ = $4.1^{+0.8}_{-0.7}$ is larger than the Merritt (2006) prediction 
that $M_{\rm def}/M_\bullet \simeq 0.5$ per major merger.   However, there are reasons to believe that $M_{\rm def}/M_\bullet$ can be 
larger than 0.5 per merger event.  First, more accurate simulations of the late stages of BH mergers suggest that $M_{\rm def}/M_\bullet$ 
can be as large as $\sim 4$ per merger (Merritt, Mikkola \& Szell 2007).  Second, an additional process has been found that is likely to
make large-$M_{\rm def}$ cores 
(Merritt \etal 2004; 
Boylan-Kolchin, Ma \& Quataert 2004; 
Gualandris \& Merritt 2008).  
Coalescing binary BHs emit gravitational radiation anisotropically and therefore recoil at velocities that are comparable 
to galaxy escape velocities.  If they do not escape, they decay back to the center by dynamical friction.  In the process, 
they further heat the core.  Gualandris \& Merritt (2008) estimate that they can excavate as much as 
$M_{\rm def}/M_\bullet \sim 5$ in addition to the mass that was already scoured by the pre-coalescence binary.   
{\bf Figure 30} appears at least roughly consistent with the suggestion that cores are made by a combination of BH scouring 
mechanisms acting over the course of one or more successive dry mergers.  Our picture of core formation by BH binary scouring 
is thereby strengthened.

      In particular, Kormendy \& Bender (2009) discuss two correlations that provide a ``smoking gun'' connection between BHs and
cores.  Their Figure 2 shows that observed BH mass fractions scatter substantially around their mean value (this mean has since been 
revised upward~in~our~{\bf Figure\ts18}).  But the scatter is not random: it correlates tightly with the ratio of $L_{\rm def}$
to galaxy luminosity.  Also, their Figure 4 shows that $L_{\rm def}$ (expressed as a magnitude) corrrelates tightly with
$\sigma$, which we know is an $M_\bullet$ surrogate for elliptical galaxies.  This correlation has the advantage that no 
dynamical modeling is involved.  It is purely a correlation between observables.  Both figures support the idea that
BHs and cores are closely connected.

      At the same time, {\bf Figures 28 -- 30} remind us that smaller ellipticals have extra, not missing, light near
their centers.  Why?  KFCB emphasize that BHs are detected in extra-light ellipticals; examples in {\bf Table 2\/} include 
M{\ts}32, NGC 3377, and NGC 4459.  Extra-light ellipticals satisfy the $M_\bullet$ -- $\sigma$ correlation.  We believe 
that they formed in mergers, and these mergers cannot all have involved only pure-disk, BH-less galaxies.  Why did 
core scouring by BH binaries fail?

      KFCB  and Kormendy \& Bender (2009) suggest that this is a consequence of the formation of extra-light 
galaxies in wet mergers.  Like Kormendy (1999) and C\^ot\'e \etal (2007), they suggest that the extra light was 
formed by the starburst that results naturally from the dissipation and central concentration of cold gas in a wet
merger.  This is universally seen in $n$-body simulations (e.{\ts}g.,
Mihos \& Hernquist 1994;
Springel \& Hernquist 2005;
Cox \etal 2006;
Hopkins \etal 2009a)
and observed in young merger remnants
(Rothberg \& Joseph 2004, 2006).
We do not suggest that scouring did not happen, although gas may help BH binaries to merge quickly
(Gould \& Rix 2000;
Armitage \& Natarajan 2002, 2005).
Rather, the observation that extra mass generally is larger than missing mass ({\bf Figure 30})
suggests that the starburst swamped any core scouring.  Thus, extra-light ellipticals present no problem for
our picture of core formation by BH binary scouring.

\vfill\eject

\hfuzz=150pt
\ni \hbox{\big\ARRed 6.14 BH fundamental plane:~the correlation with galaxy binding energy}\textBlack
\vs\vs

      Correlations between $M_\bullet$ and two of $r_e$, $I_e$, and $\sigma$ are slightly tighter than correlations between $M_\bullet$ 
and any one of these variables
(Aller  \& Richstone 2007;
Hopkins \etal 2007a, b;
Snyder  \etal 2011). 
These correlations are commonly referred to as the BH ``fundamental plane'' (not to be confused with a correlation
between $M_\bullet$ and the X-ray and radio luminosities~of~AGNs [Merloni \etal 2003] which has the same name).
The existence of a fundamental plane (not a line) for ellipticals
(Djorgovski \& Davis 1987; 
Faber      \etal 1987;
Dressler   \etal 1987; 
Djorgovski \etal 1988; 
Djorgovski 1992; 
Bender \etal 1992, 1993;
Saglia \etal 1993; 
J\"orgensen \etal 1996)
means that the correlation of $M_\bullet$ with the E-galaxy fundamental plane variables is also multivariate.

      This result is related to a demonstration by Kormendy \& Gebhardt (2001) that galaxies which have abnormally
large $M_\bullet$ for their stellar masses tend to have abnormally large $\sigma$ and small $r_e$; i.{\ts}e., to have
dissipated and collapsed~to~smaller radii.  And indeed, the above papers find that BH mass correlates well with the
binding energy of the spheroidal component.  This is yet another connection between BH growth and the details of the
formation of classical bulges and ellipticals.  In particular, it involves the physics of wet mergers and not just the 
overall depth of the potential well (which can be large even for pure-disk galaxies that contain no or only small BHs).  
We do not pursue these correlations further because of length constraints.

\vs\vs\vs
\ni {\big\ARRed 6.15 Other correlations?}\textBlack
\vs\vs
      
       We also do not pursue a number of additional correlations that have been proposed between~$M_\bullet$ and
various structural parameters of host galaxies.~Graham \etal (2001) discuss a correlation with a concentration
index that is a mixture of galaxy stellar mass and structural properties such as S\'ersic index $n$.  It could be related
to the result in the previous section that more compact ellipticals have higher BH masses at the same $M_*$.  However,
we prefer to correlate $M_\bullet$ with $M_*$, $n$, and other parameters separately.  

        Graham \& Driver (2007) find a tight correlation between $M_\bullet$ and $n$.  Beifiori \etal (2012) and Vika \etal (2012)
do not confirm a tight correlation, and neither do we.

        Erwin, Graham \& Caon (2004) review the subjects of this subsection.

\vs\vs\vs
\ni {\big\ARRed 6.16 Summary -- Which components co-evolve with BHs?}\textBlack
\vs\vs

      BH masses correlate tightly enough to imply coevolution with the properties of classical bulges and elliptical galaxies 
and with no other galaxy components.

\vfill

\ni \hbox{{\big\ARRed 7.~CENTRAL BLACK HOLES IN BULGELESS GALAXIES}}
\vskip 1pt

\ni {\big\ARRed\phantom{7.~}AND GLOBULAR CLUSTERS}\textBlack
\vs

      In the early days of work on this subject, it appeared that $10^6$ to $10^{9.5}$ $M_\odot$
BHs are almost universally associated with bulges and elliptical galaxies (e.{\ts}g., KR95; 
Magorrian \etal 1998).  Then the exclusion of a BH with $M_\bullet > 1500$ $M_\odot$  in M{\ts}33 (Gebhardt \etal 2001; 
Merritt, Ferrarese \& Joseph 2001), which has neither a classical nor a pseudo bulge (Kormendy \& Kennicutt
2004), led to the impression that BHs do not exist in pure-disk galaxies.  However, the hints from
AGNs always were that some pure-disk galaxies do contain BHs (see Ho 2008 for a recent review).  Since~then,
it has been convincingly demonstrated that bulges are not necessary equipment for the formation of BHs.
Here we review these and related results.

\vfill\eject

\vs
\ni {\big\ARRed 7.1 A new population of AGNs in late-type galaxies}\textBlack
\vs

      The most secure example of an AGN BH in a pure-disk galaxy is NGC 4395.
The galaxy is classified Sd III-IV in the Revised Shapley-Ames Catalog (Sandage \& Tammann 1981)
and SA(s)m? in the RC3.  At a distance of $D = 4.61$ Mpc (Kennicutt \etal 2008), 
it has absolute~magnitudes~of $M_B = -17.75$ and $M_V = -18.20$.  {\bf Figure\ts31} emphasizes how 
enormously different NGC 4395 is from a bulge-dominated spiral or elliptical.  It is a dwarf spiral 
with no classical bulge, no significant pseudobulge, and only a nuclear star cluster with an
absolute magnitude of $M_B \simeq -11.0$ and a velocity dispersion of 
\hbox{$\sigma$ \lapprox \ts30 $\pm$ 5 km s$^{-1}$} (Filippenko \& Ho 2003; Ho \etal 2009).

\includegraphics{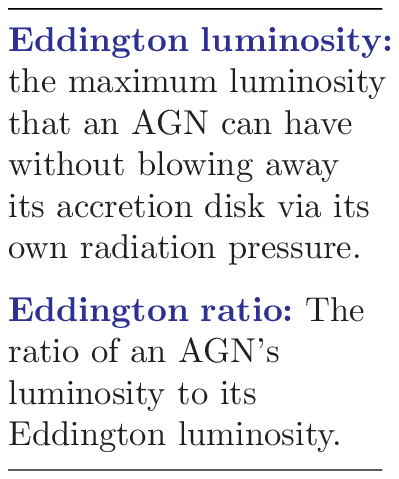}

      NGC 4395 is the nearest Seyfert 1 galaxy known and one of the least luminous (Filippenko \& Ho 2003).
It has all the hallmark signatures of BH accretion, including prominent optical and UV broad emission lines
(Filippenko, Ho \& Sargent 1993), highly variable X-ray emission~(Shih, Iwasawa, \& Fabian 2003), and a compact,
flat-spectrum radio core (Wrobel \& Ho 2006).  Peterson \etal (2005) measured $M_\bullet = (3.6 \pm 1.1) \times 10^5 \, M_\odot$ 
by reverberation mapping of C {\sc IV} $\lambda$1549 using HST STIS.  This is the smallest $M_\bullet$ measured by 
reverberation mapping.  The low AGN luminosity corresponds to a small Eddington ratio of $\sim$\ts$1.2 \times 10^{-3}$.  
Filippenko \& Ho argue that NGC 4395's BH mass may be as low as $M_\bullet \approx 10^4 \, M_\odot$, consistent 
with the estimate of $M_\bullet = (4.9 \pm 2.6) \times 10^4 \, M_\odot$ by Edri \etal (2012) based on a novel approach using
broad-band photometric reverberation mapping.  All of these masses are larger than $M_\bullet$ \lapprox \ts1500
$M_\odot$ in the brighter pure-disk galaxy M{\ts}33 ($M_V = -19.0$).

\vs


\includegraphics{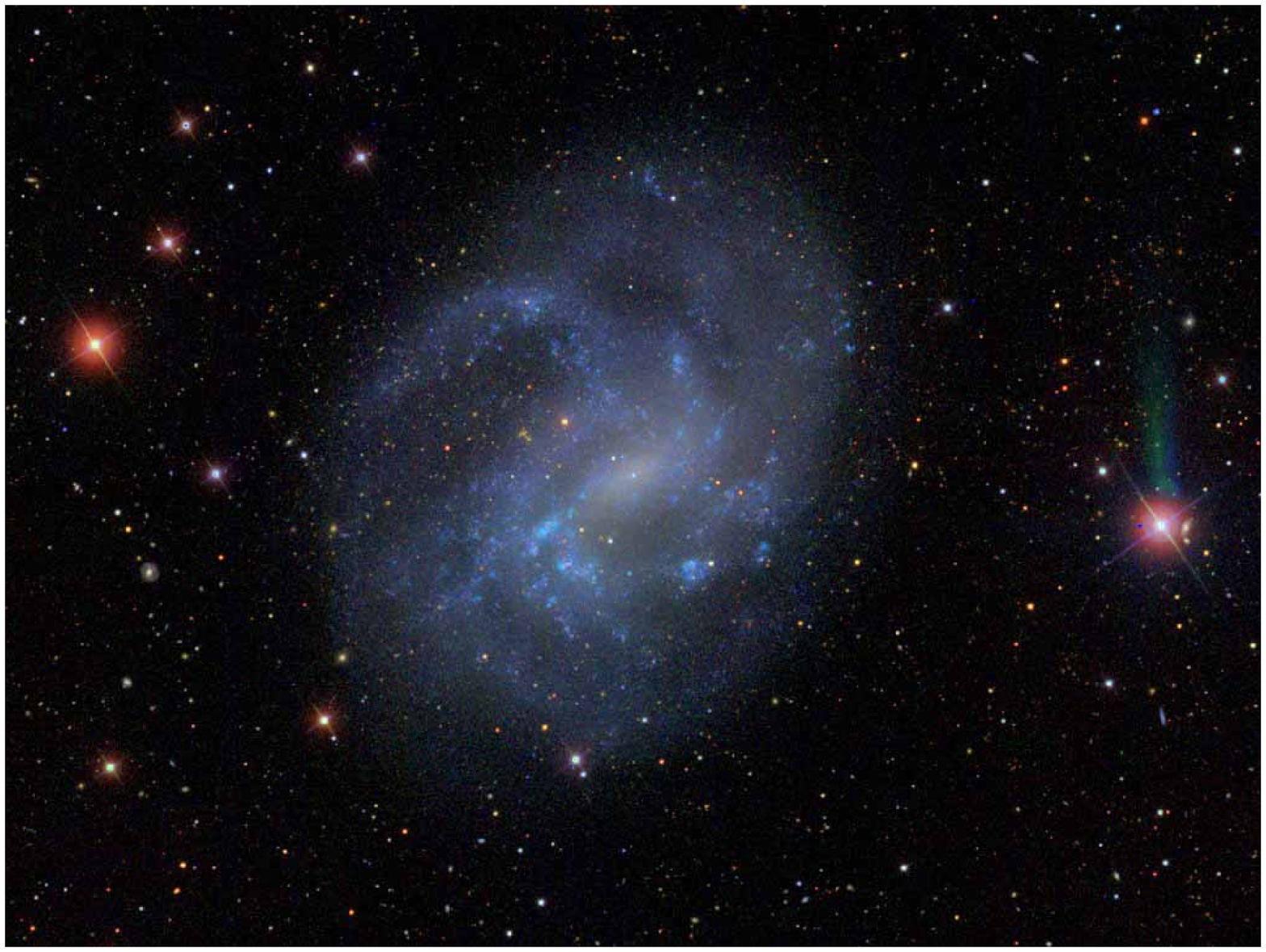}

\vskip 282pt

\vfill

\ni {\bf \textBlue Figure 31}\textBlack 

\vskip 2pt
\hrule width \hsize
\vskip 2pt

\ni The Sd -- Sm galaxy NGC 4395 (SDSS $gri$ image from NED).  The low surface brightness, the lack of a 
bulge, and the presence of a nuclear star cluster are characteristic of dwarf, late-type galaxies.
M{\ts}33 and M{\ts}101 are more luminous and higher-surface-brightness analogs (Kormendy \etal 2010).

\eject

      No less remarkable is POX 52 (Barth \etal 2004; Thornton \etal 2008).  It is an almost identical twin 
to NGC 4395 in terms of its AGN properties and BH mass, except that it radiates at a higher fraction 0.2\ts--\ts0.5 of its 
Eddington rate.  The galaxy is very dwarfish ($M_V = -17.6$) but completely devoid of a disk.  
Barth \etal (2004) measured a central stellar velocity dispersion of $36 \pm 5${\ts}km{\ts}s$^{-1}$, which, together with its
structural parameters, led them tentatively to classify this as a ``dE{\ts}galaxy.''  That is, they concluded 
that its parameters are consistent with the fundamental plane sequence for faint Sph galaxies like Draco, UMi, and 
Sculptor and bright Sph galaxies like NGC 147, NGC 185, and NGC 205 (see Kormendy 1985, 1987; KFCB).
Kormendy \& Bender (2011) call these ``Sph galaxies'' to differentiate them from the disjoint fundamental 
plane parameter sequence of ellipticals shown in the above papers and in Bender, Burstein, \& Faber (1992, 1993).  

\includegraphics{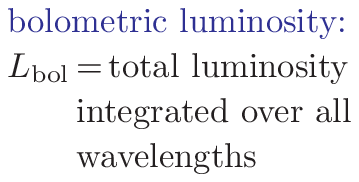}

Galaxies like NGC~4395 and POX~52 are very rare but not unique.  The large spectroscopic database from SDSS provides 
a perfect resource to uncover statistically significant numbers of such hard-to-find objects.  Greene \& Ho (2004, 2007c) 
performed a systematic search for AGNs with low-mass BHs from a detailed characterization of the broad-line (type 1) AGN 
population at $z < 0.35$ (Greene \& Ho 2007b).  They used the telltale broad H$\alpha$ line as a signpost for AGN activity 
and estimated BH masses using the virial mass estimator for this feature calibrated by Greene \& Ho (2005b).  From a parent 
sample of nearly 600,000 galaxies, Greene \& Ho found $\sim 200$ type 1 AGNs with $M_\bullet < 2 \times 10^6 \,M_\odot$.  This is rare
indeed but still a hundred-fold increase from the previous sample of two!  The sensitivity threshold of the selection 
method strongly biases the sample toward luminous objects.  As a consequence, the majority of the sample has relatively 
high accretion rates, with a median $L_{\rm bol}/L_{\rm Edd} = 0.4$.  Dong \etal (2012)  revisited the SDSS 
database with slightly different selection criteria and increased the sample by another $\sim 30$\%.  Consistent with their 
AGN nature, most of the objects emit X-rays (Greene \& Ho 2007a; Desroches, Greene, \& Ho 2009; Ai \etal 2011; Dong, Greene, \&
Ho 2012), but, interestingly, the vast majority are extremely radio-quiet. Greene, Ho, \& Ulvestad (2006) attribute this
to their high accretion rates.

In a complementary effort, Barth, Greene, \& Ho (2008) used SDSS to identify candidate low-mass BHs in type 2 Seyferts.
Although BH virial masses cannot be estimated for these narrow-line sources, one can maximize the probability of discovering 
low-mass BHs by selecting AGNs with low-luminosity host galaxies, to the extent that BH mass scales roughly with total 
galaxy mass.  Follow-up Keck spectroscopy confirms that the hosts indeed have low masses and likely very late morphological 
types: a significant number have $\sigma < 60$ \kms, some as low as $\sigma \approx 40$ \kms.  If these objects obeyed the 
$M_\bullet - \sigma$ relation, they would have masses and Eddington ratios comparable with those of the Greene-Ho type 1 
objects.  Consistent with the type 2 classification, their X-ray properties suggest moderately high obscuration (Thornton \etal 2009).

Apart from the systematic optical searches, there have been a number of reports of AGN activity in late-type galaxies based on 
detections with Spitzer of the mid-infrared [Ne~V] 14.3 $\mu$m and/or 24.3 $\mu$m line 
(Satyapal \etal 2007, 2008).  With an ionization potential of 97 eV, Ne$^{+4}$ is usually considered to be an unambiguous 
indicator of nonstellar processes, and the mid-infrared fine-structure transitions are especially insensitive to dust 
extinction.  Follow-up X-ray observations, where available, support the AGN interpretation; heavy obscuration is implied 
(Gliozzi \etal 2009; McAlpine \etal 2011; Secrest \etal 2012).  Despite initial suspicions that a large population of AGNs 
may lie undiscovered in late-type galaxies, a dedicated Spitzer survey of 18 nearby Sd/Sdm spirals detected [Ne~V] in only 
one galaxy (NGC~4178: Satyapal \etal 2009; Secrest \etal 2012).  Thus, while some bulgeless spirals definitely harbor AGNs, 
most do not.  Interestingly, all four secure cases of AGNs in pure-disk spirals contain a nuclear star cluster (NGC 4395: 
Filippenko \& Ho 2003; NGC 1042: Shields \etal 2008; NGC 3621: Barth \etal 2009; NGC 4178: Satyapal \etal 2009).  But not 
all nuclear star clusters host AGNs (e.{\ts}g., M{\ts}33; see also Shields \etal 2012).

X-ray variability is another way to find dwarf AGNs.~Kamizasa, Terashima, \& Awaki (2012) used the XMM-Newton Serendipitous 
Source Catalogue to identify 15 AGNs with candidate low-mass BHs on the basis of strong X-ray variability whose amplitude
is known to correlate inversely with BH mass.  NGC 4395 and the Greene-Ho objects, for example, are among the 
most highly variable X-ray AGNs known (Miniutti \etal 2009).  The empirical correlation between X-ray variability amplitude 
and BH mass implies $M_\bullet \approx (1-7) \times 10^6 \,M_\odot$ for the new candidates, and the most extreme case 
may be as tiny as $M_\bullet$ \lax\ $10^5 \,M_\odot$ (Ho, Kim, \& Terashima 2012; Terashima \etal 2012).

Most of the low-mass BHs discovered to date may be strongly biased toward high Eddington ratios.  Are we seeing just the tip 
of an iceberg?  Or are there many more low-mass BHs with low accretion rates yet to be found?  X-ray observations again 
provide tantalizing clues.  Taking advantage of Chandra's high angular resolution, which is crucial for isolating faint 
sources in crowded fields, Desroches \& Ho (2009; see also Zhang \etal 2009) report that $\sim$\ts25\ts\% of nearby Scd--Sm spirals 
contain a central X-ray core consistent with low-level AGN emission at the level of $L_{\rm bol}/L_{\rm Edd} \approx 10^{-6}$ to
$10^{-3}$, comparable to the lowest accretion rates seen in nearby supermassive BHs (Ho 2009a, b, c, 2002).  Some of the 
sources have 2~--\ts10 keV luminosities no larger than $L_{\rm X} \approx 10^{37}-10^{38}$ erg~s$^{-1}$, but the authors 
argue that stellar contamination by \hbox{X-ray} binaries or supernova remnants is unlikely.  If the nonstellar origin of these 
sources can be confirmed, then this would imply that a sizable fraction of bulgeless spirals host central BHs.

\vs\vs
\ni {\big\ARRed 7.2 Host galaxy properties}\textBlack
\vs

The observations described in Section 7.1 serve as a proof of concept that Nature can and does manufacture BHs in the mass 
range $M_\bullet \approx 10^4-10^6 \,M_\odot$.  This extends the dynamic range of BH masses below the threshold of 
$M_\bullet \approx 10^6 \,M_\odot$ of most spatially resolved dynamical searches.  Importantly, these low-mass BHs 
apparently can form without the aid of a bulge of any kind.  Truly bulgeless hosts such NGC~4395 or spheroidal hosts like 
POX~52 are, however, relatively rare.  The majority of the Greene-Ho sample has been imaged with HST (Greene, Ho \& Barth 
2008; Jiang \etal 2011b).  Fewer than $\sim 10$\% of these objects are qualitatively consistent with being spheroidals. 
Among the galaxies that have extended, late-type disks, only $\sim 5$\% appear to be truly bulgeless.  The vast majority 
contain resolved central components that resemble pseudobulges on the photometric projections of the fundamental plane.  That 
is, low-mass BHs have not been found in genuine ellipticals or classical bulges.  Given their late-type morphologies and low 
luminosities ($\sim 1$ mag below $L^*$; Greene \& Ho 2004, 2007b), it is not surprising that these host galaxies tend to have 
relatively low gas-phase metallicities (Ludwig \etal 2012).

      Dwarf hosts and their low-mass BHs let us probe the BH--host-galaxy scaling relations 
in a regime that previously was poorly constrained by dynamically detected objects.   A weak $M_\bullet-\sigma$ relation for AGNs 
continues to hold down to $M_\bullet \approx 10^5\,M_\odot$ and $\sigma \approx 30$ \kms\ (Barth, Greene \& Ho 2005; Greene \& Ho 2006b; 
Xiao \etal 2011; {\bf Figure\ts32}, {\it left\/}).  There is some tension between these results and {\bf Figure 21} ({\it upper-right\/}).  
Here, the scatter is larger than in inactive galaxies or more massive AGNs, but it is difficult to disentangle the intrinsic scatter
from the contribution due to the inherently larger systematic uncertainty of the virial mass estimators extrapolated~to~low~masses.  
The correlation between BH mass and (pseudo)bulge luminosity is more complicated to interpret.  The Greene-Ho objects lie significantly 
below the fiducial $M_\bullet - L_{\rm bulge}$ relation for inactive galaxies: (pseudo)bulges are overluminous at a fixed BH mass 
(Greene \etal 2008; Jiang \etal 2011a) or the BHs are undermassive at fixed (pseudo)bulge mass (Section 6.9).
Part of this offset can be ascribed to younger stellar populations in pseudobulges.  But that is not the whole 
story.  A significant offset remains after applying an $M/L$ correction and even after replacing the 
(pseudo)bulge luminosity with its dynamical mass computed using the measured velocity dispersion and effective radius 
(Jiang \etal 2011a; {\bf Figure 32}, right).  Therefore {\bf Figure 32} ({\it right\/}) is consistent with Section 6.9.

\eject

\cl{\null}

\vfill

\includegraphics{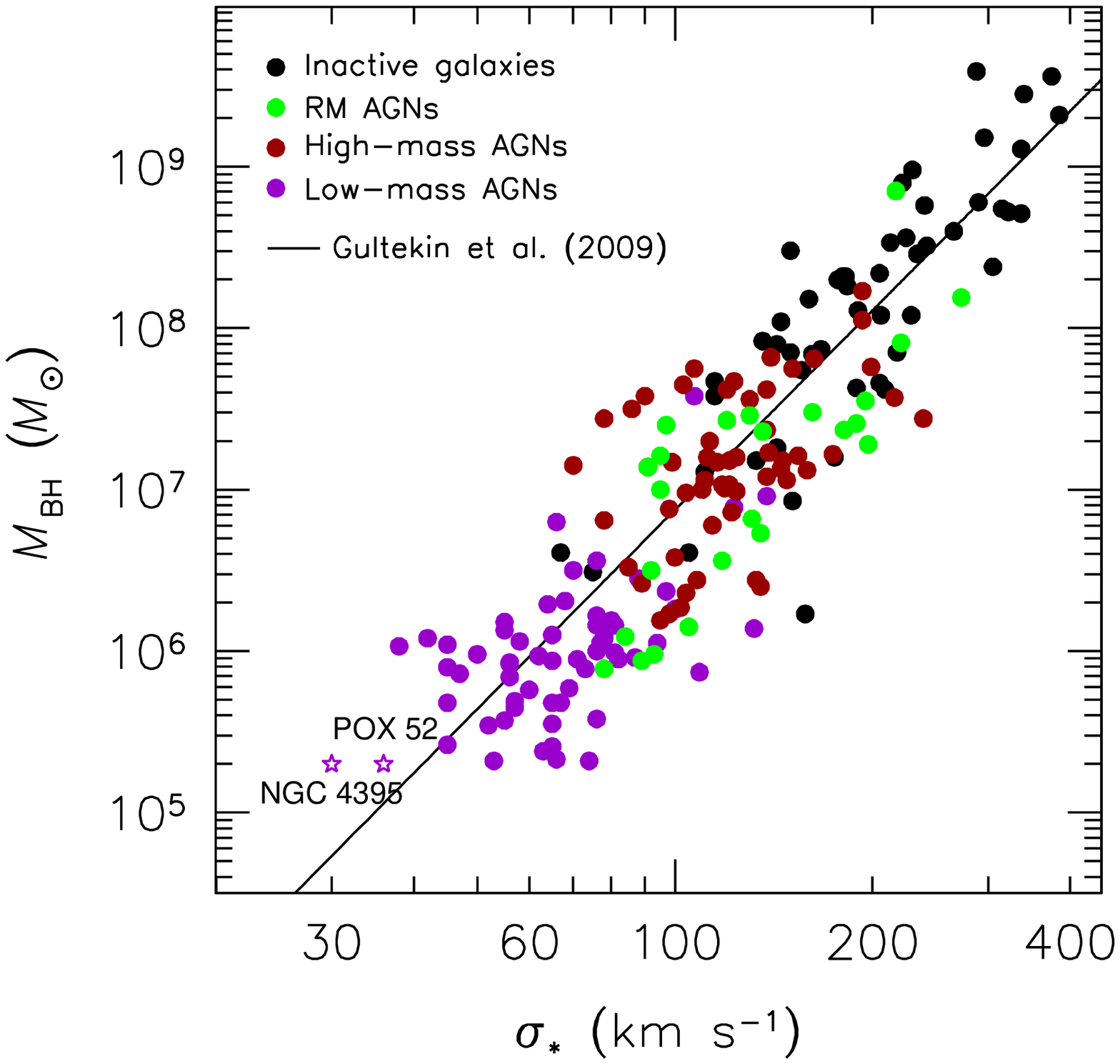}

\includegraphics{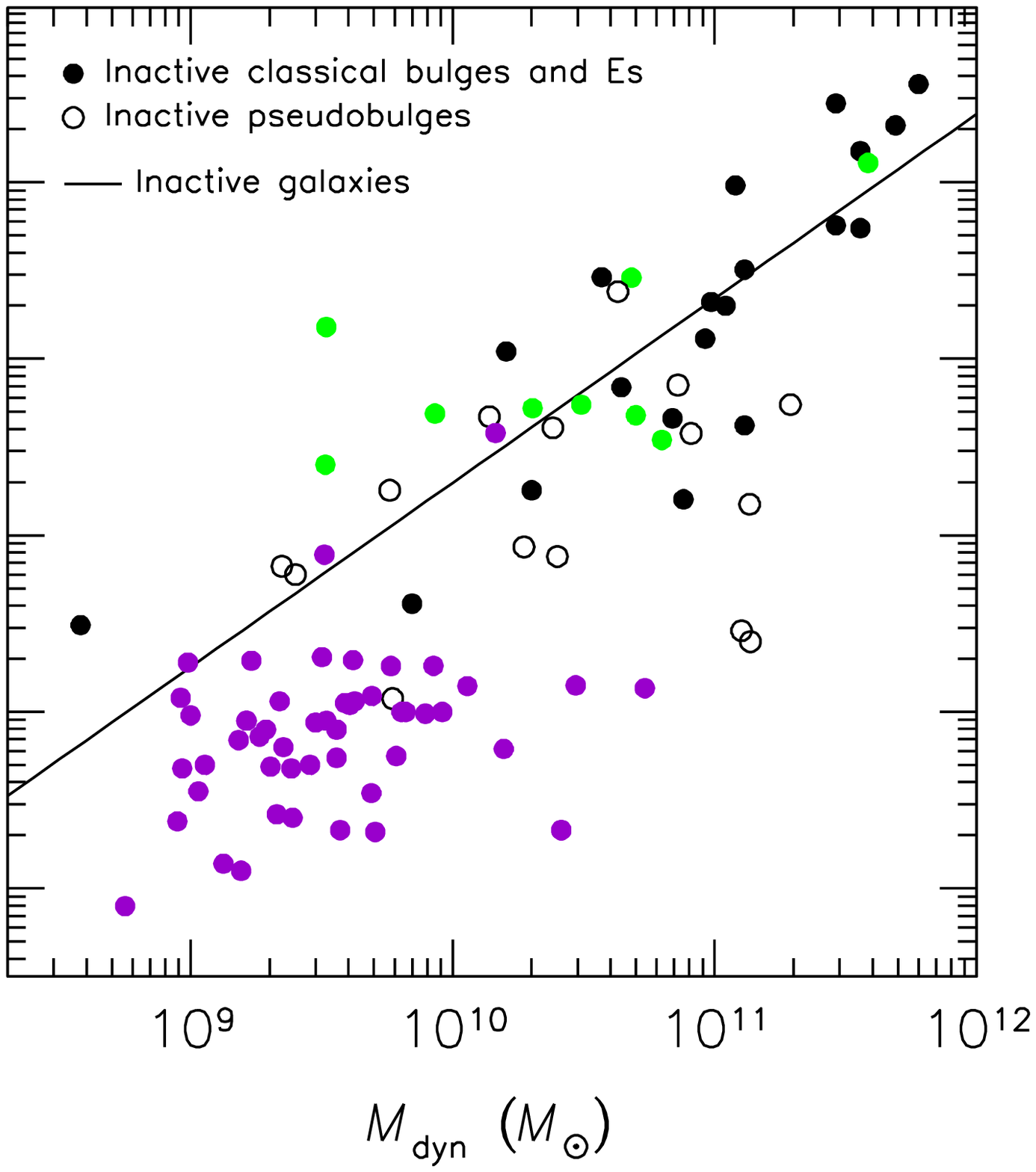}

\vskip 235pt
\ni {\bf \textBlue Figure 32}\textBlack

\vskip 1pt
\hrule width \hsize
\vskip 2pt

\ni BH--host-galaxy correlations for AGNs with BH masses derived from 
reverberation~mapping~(RM) and from single-epoch spectroscopy of broad AGN emission 
lines.  ({\it left}) AGNs ({\it colored points\/}) show considerably larger 
scatter in the $M_\bullet - \sigma$ relation than inactive galaxies ({\it black 
points\/}).  Moreover, the scatter increases toward lower BH masses; most of 
these galaxies contain pseudobulges.  Adapted from Xiao \etal (2011). 
({\it right}) Inactive BHs in classical bulges and ellipticals obey a 
well-defined relation between $M_\bullet$ and dynamical mass  $M_{\rm dyn}$; 
the same holds for reverberation-mapped AGNs with $M_\bullet \gtrsim 10^7\,M_\odot$
({\it green points\/}).  Low-mass AGNs, on the other hand, along 
with inactive BHs in pseudobulges, fall notably below the correlation for 
classical bulges and ellipticals.  Adapted from Jiang \etal (2011a).

\vfill

\eject

\vs
\ni {\big\ARRed 7.3 A BH in the starbursting dwarf galaxy Henize 2-10}\textBlack
\vs

     The ``poster child'' for BH discovery using the radio--X-ray--$M_\bullet$ fundamental plane (Merloni, Heinz, \& Di Matteo 2003)
is Henize 2-10, illustrated in {\bf Figure 33}.  Reines \etal (2011) present 
a good case that the galaxy contains a BH, based on the observation of an 
\hbox{X-ray} point source with 2\ts--\ts10 keV luminosity ${L_X} \sim 10^{39.4}$ erg s$^{-1}$
and a radio core with 4.9 GHz and 8.5 GHz luminosities of $L_R \approx 10^{35.9}$ erg s$^{-1}$. 
Careful astrometry implies that they are the same source.  It lies far from the distribution of points for X-ray binary
stars and inside the distribution of AGN points in the X-ray--radio-luminosity correlation. 
The compactness of the radio source (\lax\ 3 pc $\times$ 1 pc) and its high
brightness temperature ($> 3 \times 10^5$ K) point to a nonthermal origin (Reines \& Deller 2012).
Therefore the case for an intermediate-mass or supermassive BH is strong.  

\vfill

\cl{\null} 

 \includegraphics{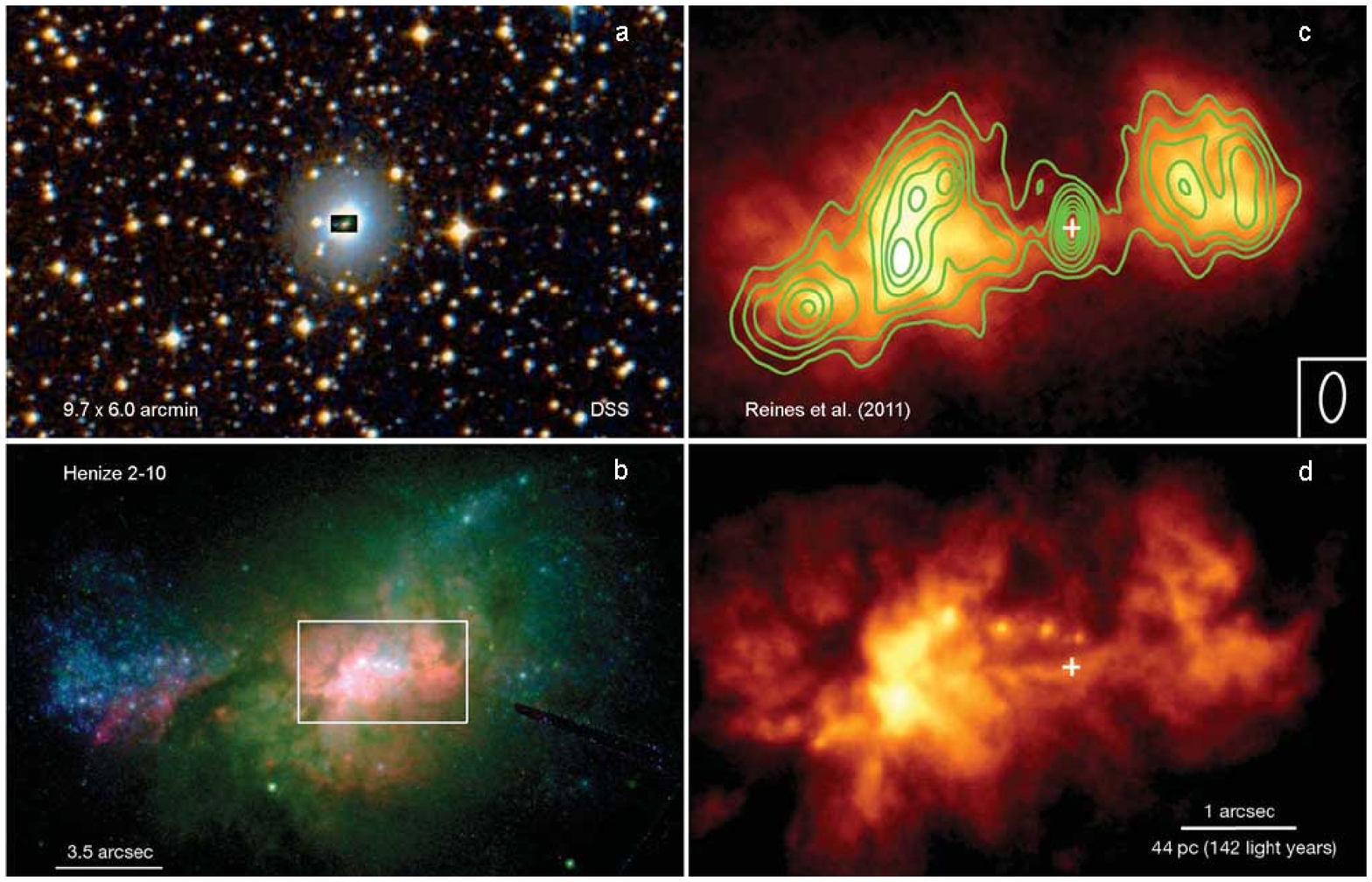}

\ni {\bf \textBlue Figure 33}\textBlack 

\vskip 2pt
\hrule width \hsize
\vskip 3pt

\ni The starbursting blue compact dwarf galaxy Henize 2-10.  {\it (a)} An optical color image from the 
Digital Sky Survey 
(DSS, courtesy {\bf http://www.wikisky.org}) shows that the outer parts of the galaxy are blue but have a smooth light 
distribution.  This proves to have an exponential profile.  The burned-out central parts contain the ongoing starburst.
The tiny rectangle in {\it (a)} shows a scaled version of panel {\it (b)} from Reines \etal (2011).  It is an HST image
showing H$\alpha$ emission~in~red, the $I$-band stellar continuum in green, and $U$ band in blue.  The white box shows the
field of panels {\it (c)} and {\it (d)}, also from Reines \etal (2011).  {\it (c)} Shows radio emission contours ({\it green})
superposed on a continuum-subtracted, HST H$\alpha$ image.  Radio and H$\alpha$ emission from ionized gas regions~match.
The radio point source (beam size in the inset) is coincident with the X-ray source (not shown).  Panel {\it (d)} is an
HST H$\alpha$ image that does not have continuum subtracted; it shows that the AGN~($+$) is not coincident with 
any of the super star clusters that are part of the ongoing starburst.

\eject

     However, determining $M_\bullet$ is tricky.  Reines \etal (2011) estimate $M_\bullet$ using the Merloni, Heinz \& Di Matteo (2003) 
correlation between BH mass, radio luminosity, and X-ray luminosity.  They get $\log {(M_\bullet/M_\odot)} = 6.3 \pm 1.1$ for $D = 9$ Mpc.  
But the error estimate comes from the uncertainty in the coefficients of a linear fit to the above correlation, not from the scatter 
of BHs around that correlation.  Also, the radio and X-ray observations were made two years apart in time, and variations 
in radio luminosity can easily be factors of several over this time (G\"ultekin, private communication). 
For example, a variation in radio luminosity of a factor of 3 would change the derived $\log {M_\bullet}$ by $\pm 0.41$.  
Overall, $M_\bullet$ derivations from the above correlation are at best uncertain.

      Still, the case for an AGN-mass BH in Henize 2-10 is strong.  A mass $M_\bullet \sim 10^6$ $M_\odot$ is surprisingly large
for a dwarf galaxy.  Reines \etal (2011) estimate that the accretion rate is $\sim 5 \times 10^{-6}$ $M_\odot$~yr$^{-1}$ and 
that the Eddington ratio is $L/L_{\rm Edd} \simeq 10^{-4}$.  Given the 
uncertainties, $M_\bullet$ could be 10 times smaller.  But this is unlikely to be a stellar-mass~BH.

      What do we learn from Henize 2-10?  The answer depends on the mass of Henize 2-10 and on what kind of galaxy it is 
evolving into.  Both are uncertain:  

      The mass and BH mass fraction are easier.  Noeske \etal (2003) find a total $K_s$ magnitude~of~8.89 and 
2MASS gets $K_s = 9.004 \pm 0.024$.  We adopt the mean, $K_s \simeq 8.95$.  With $D = 9$ Mpc and a Galactic absorption
of $A_K = 0.041$, the total absolute magnitude is $M_K \simeq -20.86$.  This corresponds to a luminosity of 
$L_K \simeq 5 \times 10^9$ $L_{K\odot}$.  The galaxy is undergoing a starburst, so the mass-to-light ratio is uncertain,
and it probably varies with radius.  But it seems safe to assume that $M/L_K \simeq 0.3$ with an uncertainty of a factor of 3. 
Then the stellar mass is $M_* \sim 1.4 \times 10^9$~$M_\odot$ with an uncertainty of a factor of 3.  The mass in atomic and
molecular gas is $M_{\rm gas} \sim 0.9 \times 10^9$~$M_\odot$ (Kobulnicky \etal 1995; cf.~Meier \etal 2001).  This gives a total mass 
of $M \sim 2.3 \times 10^9$ $M_\odot$ to within a factor of $\sim$\ts3.  Consistent with the above, the H{\ts}I rotation curve 
implies a dynamical mass $M \sim (2.7 \times 10^9~M_\odot)/\sin^2 {i}$, where $i$ is the inclination (Kobulnicky \etal 1995).  
So the BH mass fraction is $M_\bullet/M \sim 0.0006$ to within a combined uncertainty of a factor of 10.  This is an order of magnitude 
smaller than our revised canonical mass fraction of $\sim 0.005$ (\S\ts6.6.1).  Henize 2-10 is consistent with the pattern that is emerging
for other low-mass BHs in late-type galaxies (Greene \etal 2008; Jiang \etal 2011a).

\vs\vs
\ni {\big\ARRed 7.4 Intermediate-mass black holes in globular clusters?}\textBlack
\vs

      The subject of BHs in globular clusters is important, interesting, and controversial.  Reviews are provided by
van der Marel (2004) and by
Miller \& Colbert (2004).
The evidence is not yet compelling that BHs have been found in any ``real'' globular clusters.  A BH has probably been 
found in G1 and may have been found in $\omega$ Cen.  But these may be defunct galactic nuclei. If this is correct, then 
these BHs are like the ones in NGC 4395 and in other bulgeless galaxies. 

      Based on HST STIS measurements of an inward-rising $\sigma$ profile, it has been suggested that 
the post-core-collapse globular cluster M{\ts}15 (Gebhardt \etal 2000c; Gerssen \etal 2002; van der Marel \etal 2002)
and the M{\ts}31 globular cluster G1 (Gebhardt, Rich \& Ho 2002, 2005) contain IMBHs with
masses of $\sim 4000$ and 20000 $M_\odot$, respectively.  Similarly, Gemini GMOS integral-field spectroscopy
of $\omega$ Cen leads to the conclusion that it contains an IMBH of mass $4.0^{+0.75}_{-1,0} \times 10^4$ 
$M_\odot$ (Noyola, Gebhardt \& Bergmann 2008).  These conclusions have been -- and remain -- controversial:

      First, have central dark masses reliably been detected?  Sollima \etal (2009) and van der Marel
\& Anderson (2010) argue against the detection in $\omega$ Cen.  Noyola \etal (2010) obtain further kinematic
data with VLT-FLAMES and confirm their earlier result.  But their dispersion profile and van der Marel's
disagree, and the reason is not known.  For M{\ts}15, Gerssen \etal (2003) correct an error in their
2002 modeling paper; the result is that a BH-less model fits the data to $\sim 1\ts\sigma$.  

      Second, if the dark mass detections are correct, do they see single IMBHs or central clusters 
of dark stellar remnants?  All the above globulars
are very old; any stars that initially were massive enough to leave behind stellar-mass BHs
or neutron stars have long since died.  Dynamical evolution times for globular clusters are
moderately short; any remnants that get retained by the clusters should
to sink to the center.  In defense of the BH detections, two arguments are cited: (1) the relaxation
time of M{\ts}15 is short enough to have allowed core collapse, so mass segregation of remnants
is plausible (this test was involved in the above error).  But the central relaxation 
time in G1 and $\omega$ Cen as given by the observed stellar density and velocity dispersion profiles
is longer than the age of the Universe.  (2) The supernovae that leave behind stellar mass BHs
and neutron stars are frequently asymmetric~(Wang~\&~Wheeler~2008).   Remnants recoil. 
Free-flying neutron stars are observed in the Galaxy with peculiar velocities of hundreds of km s$^{-1}$.  
If heavy remnants are not retained, then they cannot mimic an IMBH.  In contrast, white dwarf remnants --
which are now being manufactured in all three clusters -- are lighter than the highest-mass main sequence stars that
remain; they actually migrate out of the core (albeit slowly).

      Both arguments are only somewhat persuasive.  Argument (1) is weaker than it sounds.  When there is a large 
mass range, mass segregation  is rapid compared to relaxation times calculated for single-mass systems.  
Lee (1995, 1996, 1998) emphasizes that a cluster of remnants separates dynamically from lower-mass, visible stars 
and quickly ends up with a short relaxation time.  Then the apparent relaxation time of the
visible stars is not a good estimate of the evolution timescale for a central cluster of massive remnants.
Thus remnants can core-collapse even when the visible stars imply a long relaxation time.  
Argument 2's uncertainty is that we don't know the distribution of recoil velocities.
But the detection of two stellar-mass BHs in the globular cluster M{\ts}22 (Strader \etal 2012a) is
evidence that BH stellar remnants can be retained in globular clusters.  And it is well known that
globular clusters retain many neutron stars in the form of millisecond pulsars (e.{\ts}g.,
Kulkarni, Hut, \& McMillan 1993;
Phinney \& Kulkarni 1994;
Phinney 1996;
D'Amico \etal 2001)
including at least as many as 20 in 47{\ts}Tuc
(Camilo \etal 2000;
Bogdanov \etal 2006)
and~21 in Terzan\ts4
(Ransom \etal 2005). 

      We are uncomfortable with the fact that this subject has not shown steady progress toward becoming
more convincing.  In particular, it would be very compelling to find independent evidence for IMBHs
in the same way that the detection of AGNs in dwarf, bulgeless galaxies points to BHs
even when they cannot be found dynamically.  Finding AGN-like sources in globular clusters would be
persuasive.  Kong \etal (2010) discuss a central X-ray source in G1 but cannot differentiate
between IMBH accretion and a stellar X-ray binary.  More generally, X-ray detections of globular clusters 
appear most consistent with stellar-mass BHs (Maccarone \etal 2007, 2010).

      This uncomfortably uncertain situation has not improved.  The radio source in G1 reported by 
Ulvestad, Greene \& Ho (2007) was welcome news.  Together with an X-ray source detection, 
it helped to lend confidence to the IMBH interpretation.  Unfortunately, a recent observation with 
much improved sensitivity failed to confirm the earlier radio measurement (Miller-Jones \etal 2012). 
This does not invalidate the dynamical constraint, but it certainly does not help to
build a case that this is an IMBH.  Additional deep X-ray and radio searches of globular 
clusters -- particularly bright ones that might be defunct galactic nuclei -- would be helpful.  
So far the news has not been good (Ho, Terashima \& Okajima 2003; Wrobel, Greene \& Ho 2011; 
Strader \etal 2012b).

      If the IMBHs in $\omega$ Cen and G1 prove to be real, there remains the complication that both
hosts may be nuclei of defunct dwarf spheroidal galaxies and not globular clusters.  We already know that 
some bulgeless galaxies contain nuclear star clusters that host BHs whereas others~(M{\ts}33)~do~not.  Work
on this subject is continuing (e.{\ts}g., L\"utzgendorf \etal 2013).

      Whether or not these possible IMBHs lie on the extrapolation of the $M_\bullet$ -- $\sigma$ relation, 
it is unlikely that the same physics is responsible as at high BH masses.  This is especially true since 
supermassive BHs with $M_\bullet \sim 10^6$ to $10^7$ $M_\odot$ do not correlate tightly with disks or pseudobulges.

\vs
\ni {\big\ARRed 7.5 BHs in ultraluminous X-ray sources?}\textBlack
\vs

      Ultraluminous X-ray sources (ULXs) in starbursting galaxies are often suggested to be IMBHs.  Their interpretation as 
accreting BHs is not much in doubt, but the BH masses are controversial.  Zhang \etal (2009) provide an excellent review 
and arguments that ``the majority of [these sources] are nuclear BHs, rather than X-ray binaries.''  In contrast, Roberts (2007) 
concludes that ``New observational evidence is now pointing~away from the interpretation of ULXs as $\sim$\ts1000\ts$M_\odot$ 
black holes.''  These differences are symptomatic of the fact that application of the $M_\bullet$--X-ray-luminosity--radio-luminosity 
correlation and other indirect techniques such as variability timescales can fail.  We find it suspicious that most ULXs 
are associated with starbursts and not with galactic nuclei (e.{\ts}g., Kong \etal 2007).  The closeness of that association argues for short lifetimes 
and stellar-mass BHs (Roberts 2007).   This subject deserves further~work.  High-resolution observations to look for compact 
radio sources are the biggest need.  K\"ording, Colbert, \& Falcke (2005) is a step in this direction and further illustrates 
the difficulty in distinguishing AGN BHs from stellar-mass BHs and BH alternatives.

      The solution is important for an accurate census of low-mass BHs, but it is not crucial to the conclusions of this paper.
We already know that bulgeless galaxies can contain AGN BHs with $M_\bullet \sim 10^{5 \pm 1}$ $M_\odot$ and that these can serve
as seeds for higher-mass BHs grown in galaxy mergers.

\vfill\eject

\ni {\big\ARRed 8. COEVOLUTION (OR NOT) OF BHs AND HOST GALAXIES}\textBlack
\vs\vs

      The discovery of the $M_\bullet$\ts--\ts$\sigma$ relation enormously 
energized work on AGN feedback and its effects on galaxy evolution.  Two 
circumstances together make a case that is widely regarded as compelling 
(Section\ts8.2).  First, many observations and astrophysical arguments provide 
motivation: they can be tied together into a multiply connected tapestry of 
the sort that we associate with successful~and mature ideas.  Second, if a 
tunable amount of energy is available -- source unspecified~-- several 
longstanding problems of galaxy formation can be ``solved''.  Hundreds of 
papers attempt to combine these circumstances to suggest that galaxy formation 
and BH growth regulate each other.

      However, the observations reviewed in this paper lead us to a 
qualitatively different picture.  The evidence for close coevolution -- for 
mutual growth of bulges and BHs in lockstep -- is less compelling than we 
thought.  In the present universe, any high-$z$ era of coevolution is over.  
At $z \sim 0$, no process significantly engineers tight $M_\bullet$--host 
correlations.  Indeed, while some processes help to maintain the correlations, 
others erode them.  Even out to $z \sim 1.5-2$, the observations 
suggest~that~most~BHs do not grow in lockstep with their host bulges.  To be 
sure, some observations do show that AGN activity affects galaxy formation.  
This convincingly solves some problems of galaxy formation, such as heating 
hot gas to prevent cooling flows.  Nevertheless, the physics that seems most 
responsible for establishing the $M_\bullet$--host-galaxy correlations is not 
some magic aspect of AGN feedback.  Rather, it is at least partly and perhaps 
mostly the averaging of BH masses that is inherent in galaxy and BH mergers.  
These ideas are heretical, so we develop them in Sections 8.3 -- 8.6 as 
carefully as space allows.

\vs\vs\vs
\ni {\big\ARRed 8.1 Four Regimes of AGN Feedback: An Introduction}\textBlack
\vs\vs

      The conclusion that BHs correlate differently with different galaxy 
components allows~us~to~refine our ideas on coevolution.~Section 8 reviews 
evidence for the ``punch line'' conclusions~of~this~paper. We present 
the case that there are four regimes of AGN feedback in three different kinds 
of galaxies.  \vsss

\nhii (1) {\bf\ARRed Galaxies that lack dominant classical bulges}
\textBlack~can contain BHs, but these grow by low-level AGN activity that 
involves too little energy to affect the host galaxy (Section\ts8.3).  
Decisively at $z \sim 0$ and probably out to $z \sim 1.5-2$, most AGNs are of 
this kind. \vsss

\nhii (2) Feedback may help to establish $M_\bullet$--host relations 
during dissipative (``wet'') major mergers that make {\bf\ARRed classical 
bulges and low- to moderate-luminosity elliptical galaxies}\textBlack.  The 
jury is still out and the physics remains obscure.  But if this is to work, 
it must work mostly at high $z$ (Section\ts8.6).  
\pretolerance=10000   \tolerance=10000 \vsss

\nhii (3) {\bf\ARRed The highest-mass ellipticals have cores}\textBlack~and 
otherwise are recognizably different from their lower-mass counterparts.  We 
review evidence that they form in dissipationless (``dry'') mergers.  These 
giant ellipticals inherit any feedback effects from (2).  In them, AGN feedback 
plays the essentially negative role of keeping galaxy formation from ``going 
to completion'' by keeping baryons locked up in hot gas.  Here the controlling 
point is that these galaxies are massive enough to hold onto hot, 
X-ray-emitting gas.  ``Maintenance-mode AGN feedback'' helps to keep the hot 
gas hot and to prevent late star formation and BH accretion (Section\ts8.4). \vsss

\nhii (4) {\bf\ARRed Averaging inherent in galaxy and BH mergers may be the 
most important effect that leads to BH--host-galaxy correlations.}\textBlack~ 
Then the central limit theorem ensures that the scatter in BH correlations 
with their hosts decreases as $M_\bullet$ increases (Section\ts8.5). \vs

      The AGN population is dominated by (1) at all redshifts for 
which we have data.  However, the highest-mass BHs and the total mass in BHs 
are dominated by processes (2)\ts--\ts(4).

\vfill\eject

\vs
\ni {\big\ARRed 8.2 Feedback from AGNs (and Star Formation): 
Motivation}\textBlack
\vs

Before we indulge in heresy, we review the motivation for the idea that 
BHs and bulges coevolve. \vs

\nhii (1) The tightness of the $M_\bullet$ correlations with bulge velocity 
dispersion, mass, numbers~of~globular clusters, and core properties suggests 
that BH growth and galaxy formation are connected. Especially compelling is 
the conclusion that the scatter in the $M_\bullet$\ts--\ts$\sigma$ correlation 
is consistent with (Ferrarese \& Merritt 2000; Gebhardt \etal 2000a) or only 
moderately larger than (Tremaine \etal 2002; G\"ultekin \etal 2009c; this paper)
the measurement errors.  However, seeing a tight correlation is not enough to 
ensure that coevolution happens because of AGN feedback.  For example, we argued in 
Section\ts6.13 that the correlations with core properties are produced by 
dynamical processes that are purely gravitational.  \vs 

\nhii (2) The binding energy $\sim$\ts$0.1 M_\bullet c^2$ of a BH (assuming a radiative 
efficiency of 10\ts\%) is much larger than the binding energy  $\sim M_{\rm bulge} \sigma^2$
of its host bulge,.  For $M_\bullet \approx 5.3 \times
10^{-3} M_{\rm bulge}$ (Section\ts6.6), the ratio of binding energies is $5.3 
\times 10^{-4} (c/\sigma)^2$, or $\gtrsim 500$ for $\sigma \lesssim 300$ \kms.
If only one percent of the AGN energy output couples to gas in the forming 
galaxy, then all of the gas can be blown away (Silk \& Rees 1998; Ostriker \& 
Ciotti 2005).  AGN feedback can be radiative (acting via photons) or 
mechanical (acting via energetic particles or a wind or a jet).  Thus, BH 
growth may be self-limiting, and AGNs may quench star formation. \vs

\nhii (3) The histories of BH growth and star formation in the universe are 
similar ({\bf Figure 34}).  Quasars and starbursts appear, at least 
superficially, to be closely linked. The most luminous starbursts always
show signs of buried AGNs, even if these do not dominate the bolometric output 
(Genzel \etal 1998), and the host galaxies of AGNs often show concurrent or 
recent star formation (e.{\ts}g., Kauffmann et al. 2003; Shi et al. 2009).  Heckman 
\etal (2004) note that the volume-averaged ratio of BH accretion rate to star 
formation rate today is $\sim 10^{-3}$, eerily close to $M_\bullet/M_{\rm bulge}$.  
Is this just a remarkable coincidence, or does it signify a profound causal 
connection between BH and galaxy growth? \vs

\vfill

\includegraphics{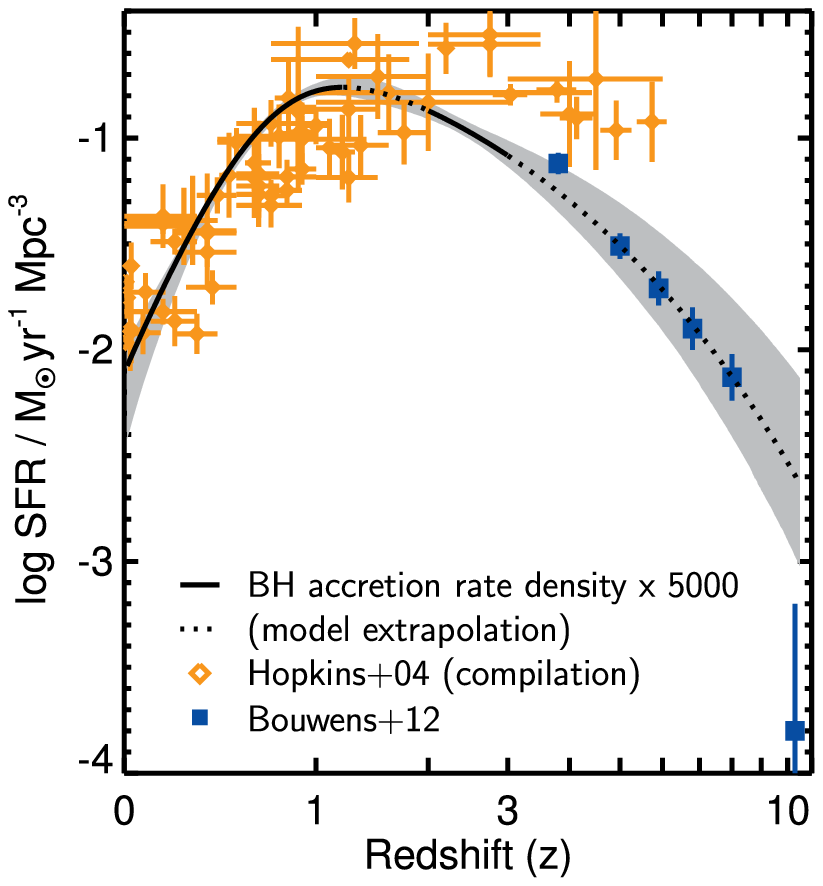}

\ni {\bf \textBlue Figure 35}\textBlack

\vskip 1pt
\hrule width \hsize
\vskip 2pt

\ni  Evolution with redshift of the volume density of black hole 
accretion rate ({\it black line with~grey~band}), scaled up by a factor of 
5000.  This closely tracks the evolution of cosmic star formation rate compiled 
by Hopkins (2004; {\it orange points}) and by Bouwens \etal (2012; {\it blue
squares}).  Adapted and updated from Aird \etal (2010). 

\eject

\def\vs{\vskip 10pt} 

\nhii (4) The observed, zero-redshift volume density of BH mass agrees with 
the density of the fuel that is required to power quasars for plausible values 
of the radiative efficiency of BH accretion (So\l tan 1982; Yu \& Tremaine 2002; 
Marconi \etal 2004).~The uncertainty is a factor~of~$\sim$\ts2.   Unless this 
agreement is a coincidence, most BH mass must have been acquired through radiatively 
efficient gas accretion while the galaxy hosted an AGN.   BH growth by radiatively 
inefficient accretion and by swallowing of stars without disruption (at 
$M_\bullet \gg 10^8$ $M_\odot$) must play a small enough role to not 
invalidate So\l tan's argument.  This applies to intermediate masses 
$M_\bullet \sim 10^{8 \pm 1}$\ts$M_\odot$.  No argument based on energetics
constrains the behavior of the smallest BHs, because they contribute 
negligibly to the mass density.~And no dark mergers of BHs of any mass can 
affect the So\l tan argument, as long as their progenitors grew by radiatively 
efficient gas accretion.  The point is that a large amount of AGN energy is 
available as input to arguments~(1)\ts--\ts(3). \vs

      Points (1) to (4) are suggestive but indirect.  Direct observations that are
consistent with feedback in action (see Alexander \& Hickox 2012 for a general review)
include the following: \vs

\nhii (5) Individual giant ellipticals and groups and clusters of 
galaxies contain large amounts of hot, X-ray-emitting gas.  This gas often 
shows X-ray cavities or ``bubbles'' that are believed to be inflated by AGN 
jets.  Plausibly, this is AGN feedback in action (Section\ts8.4). This
subject is reviewed by 
McNamara \& Nulsen (2007, 2012),
Cattaneo \etal (2009), and 
Fabian (2012). \vs

\nhii (6) Some bright quasars show blue-shifted X-ray spectral absorption 
lines interpreted as coming from winds with velocities $v$\ts\gapprox\ts$0.1c$ 
and with mass loss rates of one to tens of $M_\odot$ yr$^{-1}$ (e.{\ts}g., 
Pounds \& Page 2006; Reeves \etal 2009; Tombesi \etal 2012).
Nonrelativistic outflows are also seen (Cano-D\'\i az \etal 2012 and 
references therein).  Maiolino \etal (2012) report on a quasar at $z = 6.4$ 
with an inferred outflow rate of $>$\ts3500 $M_\odot$ yr$^{-1}$.  Such an 
outflow can clean a galaxy of cold gas in a single AGN episode. \vs

\nhii (7) Some powerful radio galaxies at $z$\ts$\sim$\ts2 show ionized gas 
outflows with velocities of $\sim$\ts$10^3${\ts}km{\ts}s$^{-1}$ and gas masses 
of $\sim 10^{10}$ $M_\odot$.  Kinetic energies of $\sim$\ts0.2\ts\% of BH rest 
masses are one argument among several that the outflows are driven by the 
radio sources and not by star formation (Nesvadba \etal 2006, 2008).  
Molecular gas outflows are also seen (Nesvadba \etal 2010). \vs

\nhii (8) We focus on AGN energy feedback in this section, but we emphasize 
that it may very generally cooperate with starburst-driven feedback.  A number 
of nearby ultraluminous infrared galaxies (ULIRGs), all composite 
starburst/AGN systems, show strong neutral and molecular outflows with 
velocities $\gtrsim 10^3$ \kms\ and outflow rates of several hundred to more than 
a thousand  $M_\odot$ yr$^{-1}$ (Rupke \& Veilleux 2011; Sturm \etal 2011).  After 
the fireworks subside and the dust clears, these systems transform to 
post-starbust systems still enveloped in outflowing gas detectable as
$10^3$ \kms\ Mg~II absorbers (Tremonti, Moustakas \& Diamond-Stanic 2007). 
However, winds are not restricted to the most extreme starbursts.  Rather, they
seem to be generic features of star-forming galaxies at $z \sim 1-2$ (Weiner 
\etal 2009; Genzel \etal 2011; Newman \etal 2012).  Nevertheless, their relation to AGN 
feedback in such objects is unclear.

\vfill

\eject

\def\vs{\vskip 1.5pt} 

AGN feedback is also popular because it may resolve longstanding problems in 
galaxy formation: \vs

\nhii (9) Episodic AGN feedback is believed to solve the ``cooling flow'' 
problem (Fabian 1994) that, in the absence of energy input, X-ray halos in 
giant galaxies and in clusters of galaxies would cool quickly, but cool gas 
and star formation are not seen in the predicted large amounts (e.{\ts}g., 
Ostriker \& Ciotti 2005).  Related issues are the origin of the 
entropy floor and the steeper-than-expected temperature scaling of the 
X-ray luminosity of the intracluster and intragroup medium (see Cattaneo et 
al. 2009 for a review).  This ``maintenance-mode AGN feedback'' is discussed
in Section\ts8.4.\vs

\nhij (10) At high masses, the galaxy mass function drops more steeply than 
the mass function of dark halos that is predicted by our standard cosmology.  
The proposed solution is that~higher-$M_\bullet$ AGNs are more efficient at  
preventing late galaxy growth, again through the action of radio jets that keep
baryons suspended in hot gas (e.{\ts}g., Bower~et~al.~2006; Croton \etal 2006). \vs

\nhij (11) The star formation histories of the biggest ellipticals are not trivially 
consistent with galaxy formation by hierarchical clustering (Faber, Worthey, \& Gonzalez 1992). 
Stellar populations in the biggest ellipticals are very old and very enhanced in 
$\alpha$ elements compared to the Sun.  This implies that essentially all star formation 
was completed, respectively, at high $z$ and in \lapprox\ts$10^9$ yr 
(e.{\ts}g., 
Worthey,{\ts}Faber,\ts\&{\ts}Gonzalez{\ts}1992;
Matteucci{\ts}1994;
Bender{\ts}1996).~In~general, 
more massive galaxies formed their stars earlier and more rapidly 
(Thomas{\ts}et{\ts}al.\ts1999,\ts2005).  In contrast, hierarchical clustering implies that the biggest 
ellipticals were assembled late via the longest histories  of successive mergers.
If each merger involved star formation, then neither the old ages nor the $\alpha$-element 
enhancement could be preserved.  The solution is the realization that star formation and 
galaxy assembly could easily have happened separately and at different times.  
In particular, the above and other properties of giant ellipticals require
that the last mergers that formed them were mostly dry (Faber \etal 2007; KFCB;
Sections 6.7, 6.13, and 8.4 here).  AGN feedback is one way to quench late star formation 
by maintaining hot gas halos (points 9,\ts10) (Bower \etal 2006; Croton \etal 2006). \vs

\nhij (12) Mergers convert spiral galaxies into classical bulges and 
ellipticals (Toomre~\&~Toomre~1972; Toomre 1977; Schweizer 1990).  Since the 
former are gas-rich and star-forming whereas the latter are gas-poor and 
``red and dead'', something connected with mergers presumably removes gas and 
quenches star formation.  The observed bimodality in the color-magnitude 
correlation (Strateva \etal 2001; Baldry \etal 2004) can be explained, at 
least in part, by rapid quenching of star formation (cf.~point 11).  Expulsion 
or heating of residual cold gas may be accomplished by AGN feedback (Springel, 
Di Matteo \& Hernquist 2005a; Schawinski \etal 2007).  One sign of this may be 
the close association of high-luminosity AGNs with post-starburst stellar 
populations (Kauffmann \etal 2003). Another is the detection of fast 
($\gtrsim 10^3$ \kms) outflows in post-starburst systems (Tremonti, Moustakas 
\& Diamond-Stanic 2007).  However, the jury is still out as to whether AGNs drive 
these winds (Diamond-Stanic \etal 2012).  Also, feedback from starbursts
may play an integral and necessary part in the removal of (most) cold gas from
moderate-luminosity elliptical galaxies (point 8). \vs

      Galaxy formation studies that incorporate AGN feedback include
semi-analytic models 
(Kauffmann \& Haehnelt 2000; 
Granato \etal 2004; 
Bower \etal 2006; 
Croton \etal 2006; 
Sijacki~et~al.~2007;
Somerville \etal 2008),
hydrodynamic models 
(Quilis, Bower, \& Balogh 2001; 
Dalla Vecchia \etal 2004; 
Br\"uggen \& Scannapieco 2009; 
Ciotti, Ostriker, \& Proga 2010), 
and analytical models (Cattaneo \etal 2011).  They explain some observations 
(luminosity functions of galaxies; see Silk \& Mamon 2012 for a review) 
but fail to explain others (bulgeless galaxies;
see Abadi \etal 2003 for a model discussion and
Kormendy \etal 2010 for the observations).
An advantage of 
such models is that they efficiently explore how various physical effects are
helpful.  A disadvantage is that they do not uncover new physics.  And, 
even when they seem to work (e.{\ts}g., solving the cooling flow problem in 
clusters), they do not tell us which of several competing mechanisms is 
responsible.  Are hot gas halos that protect giant ellipticals
from late star formation kept hot by AGN feedback, by cosmological gas infall, or by 
gas recycled from dying stars? 

We therefore turn to observations for guidance.  The following sections 
summarize what the observations tell us about the four regimes of coevolution 
(or not) between BHs and host galaxies.  We start with the $z \sim 0$ 
universe, where answers are more clearcut, and then move out to higher 
redshifts, where conclusions are more tentative.  Evidence for coevolution 
will not be overwhelming. 

\vs\vsss\vskip 1pt
\ni {\big\ARRed 8.3 BH Growth in Disk Galaxies: No Coevolution at Low  M\lower.3ex\hbox{$\bullet$}}\textBlack
\vsss

The observation that $M_\bullet$ correlates closely with classical bulges and 
ellipticals~but~not~with pseudobulges and disks motivated Kormendy \etal (2011)
to suggest that there~are two different~BH feeding mechanisms.  (1) BHs in bulges
and ellipticals grow rapidly when mergers drive gas infall that feeds quasar-like events 
(e.{\ts}g., Hopkins \etal 2006).  (2) In contrast, small BHs in largely disk-dominated galaxies, 
most of which host pseudobulges, grow as low-level Seyferts; their feeding is driven locally and 
stochastically, and they do not coevolve~with~any~part of their host galaxy.  
This suggestion was made independently by Greene \etal (2008) based on 
the observation that BH masses in low-luminosity Seyfert\ts1s (Greene \& Ho 
2004) are smaller than predicted by the $M_\bullet$\ts--\ts$L_{\rm bulge}$ 
relation.  Greene \etal (2008; updates in Jiang \etal 2011a, b)
had less information about the bulges of their galaxies but argued 
that most of them are pseudo.  Related pictures are suggested by Hopkins \& 
Hernquist (2009) and by Kauffmann \& Heckman (2009).  BH growth without host 
coevolution is the subject of this section.  It applies to most AGNs at 
$z$\ts$\sim$\ts0, and possibly also to many out to $z \sim 1.5$ (Section\ts8.6).

     Most AGNs in $z$\ts$\sim$\ts0 galaxies are weak (Ho 2008).  BHs that accrete at significant fractions of
their Eddington rates are rare (Heckman \etal2004; Ho\ts2009a).  Almost all are low-mass, $10^5$\ts--$10^7$-$M_\odot$ BHs that live in 
low-mass, Sbc\ts--{\ts}Sd spirals (Greene~\&~Ho~2004, 2007c) that contain pseudobulges (Greene{\ts}et{\ts}al.\ts2008).  
{\bf Figure 35} illustrates how, among $z \lesssim 0.35$ Seyfert 1 galaxies and quasars \phantom{0000000000}


 \includegraphics{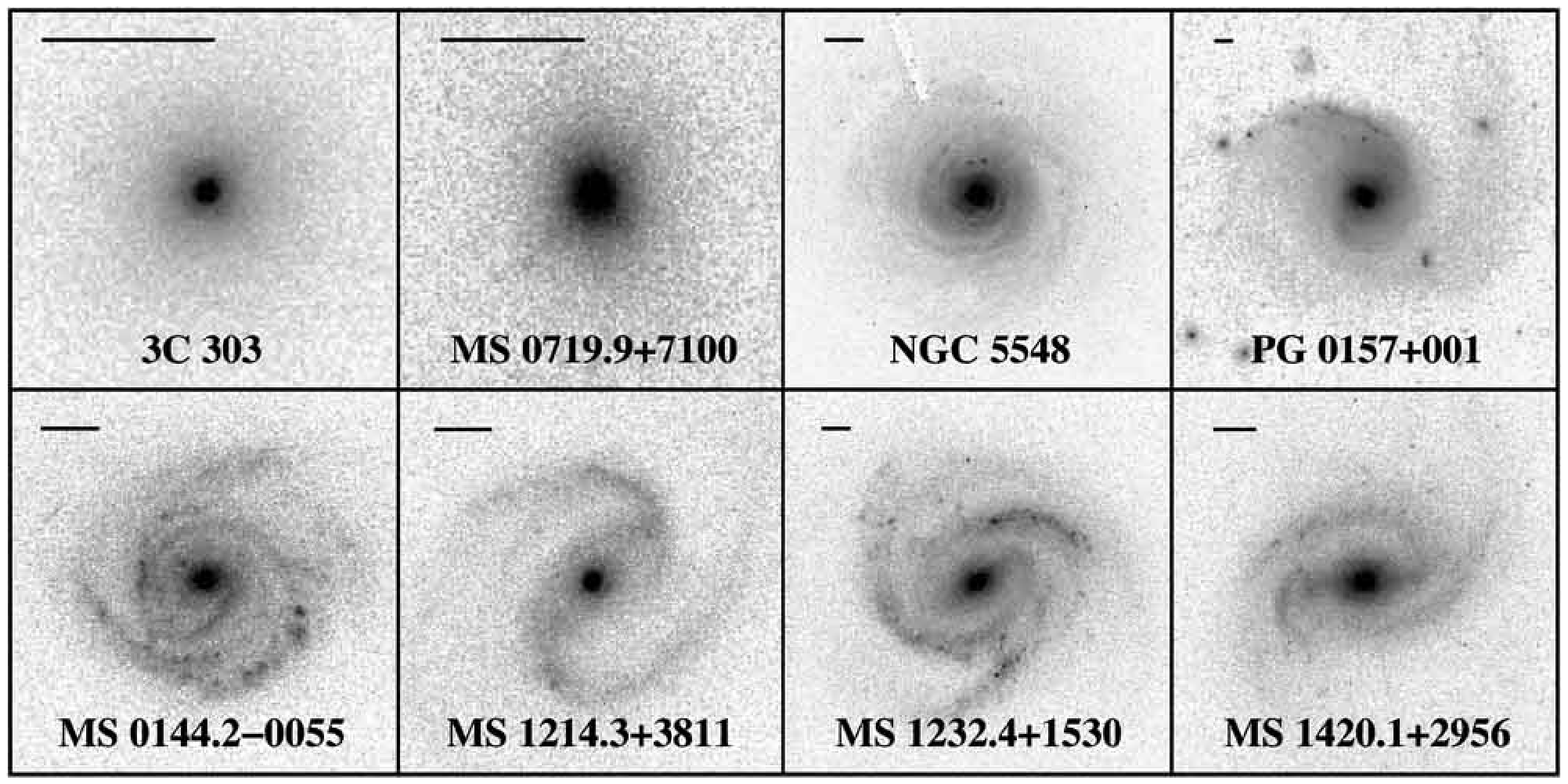}

\vfill

\ni {\bf \textBlue Figure 35}\textBlack

\vskip 1pt
\hrule width \hsize
\vskip 2pt

\ni Optical $R$-band HST images of $\langle z \rangle \sim 0.1$ Seyfert 1 galaxies and quasars.  
Some of these nearby AGNs have elliptical hosts (the two leftmost objects on the top row), but most
have hosts that are spirals with smallish bulge-to-disk luminosity ratios.  At least 1/3 are barred
(bottom row).  The horizontal bar in each panel represents a scale of $2^{\prime\prime}$.  
Adapted from Kim \etal (2013).

\eject

\noindent that have been studied with HST, 
$\sim 1/3$ of the hosts contain bars (Jiang \etal 2011b; Kim \etal 2013; {\bf Figure 35} here).
This fraction is not higher than the already-large bar fraction in local spirals (Eskridge \etal 2000).  
Bars, oval disks, and tidal torques help to transport gas to 
the central $\sim$\ts0.1\ts--\ts1{\ts}kpc.  Most gas that is accumulated there forms stars 
and builds pseudobulges, but some can -- at least in principle -- feed the BH. 
Numerous observational papers are devoted to this popular idea, but evidence for any
enhancement of AGN activity by bars or environmental perturbations is murky (see 
Ho 2008 for review), in part because most papers on this subject ignore the effects
of oval disks in unbarred galaxies ({\bf Figure 10}).  Also, the global supply 
of H{\ts}I gas in the hosts seems to have no impact on the level of AGN activity 
(Ho, Filippenko \& Sargent 2003; Ho, Darling \& Greene 2008; Fabello \etal 2011). 
The apparent insensitivity~of~$z \sim 0$ AGN activity to galaxy gas content and to the
mechanisms that can deliver it to the center reflects~two~facts: (1) BH feeding 
and AGN activity are episodic and not always switched on, and (2) feeding 
rates of low-mass BHs are very modest.  Even if accretion occurs at the maximum Eddington 
rate, $\dot M_{\rm Edd} = 2.2\times 10^{-8} \left(\eta/0.1\right)\left(
M_\bullet/M_{\odot}\right)\,M_\odot\, {\rm yr}^{-1}$, where $\eta$ is the radiative efficiency, 
the accretion rate for a $10^6$-$M_\odot$ BH is only $0.02\, M_\odot\, {\rm yr}^{-1}$. 
This is tiny.  A variety of stochastic processes driven by local, circumnuclear ``weather'' 
can supply the necessary fuel. In fact, the paradox for local BHs is not whether there is 
enough fuel to light them up.  Rather, the puzzle is how to keep them so dim despite the 
ready abundance of {\it in situ}\ gas (Kormendy \& Richstone 1995; Ho 2008, 2009a). 

      BHs are pervasive across much of the Hubble sequence, but is there any evidence that AGNs directly 
affect galaxy properties?  For small BHs in nearby disk galaxies, the answer is:~``Not~much.''  The most 
powerful AGNs do seem to pump up the velocities in associated ionized gas (Greene \& Ho 2005a; Ho 2009b), 
but it is unclear whether this has an impact beyond the relatively compact narrow-line region.  Empirical 
evidence for a connection between AGN activity and star formation is, admittedly, compelling.  Local Seyferts
often show ongoing or recent star formation (e.{\ts}g., Cid Fernandes \etal 2001), and stronger starbursts 
are associated with more powerful AGNs (Kauffmann \etal 2003; Wild \etal 2007).  Nevertheless, this evidence 
is largely circumstantial and does not prove that BHs {\it cause}\ the starbursts.  A link between BH accretion 
and star formation arises naturally because both depend on gas from the same reservoir and on the same, largely 
secular processes that drive gas inward (Kormendy \& Kennicutt 2004).  Wild, Heckman, \& Charlot (2010)
find that the peak accretion rate onto the BH typically occurs $\sim 250$ Myr {\it after}\ the onset of the starburst. 
The two processes are not strictly coeval, as Ho (2005) had suggested independently based on the apparent reduction 
of star formation efficiency in type~1 AGNs.  

      The tendency for AGN galaxies to inhabit the ``green valley'' of the color-magnitude relation
(Martin \etal 2007; Nandra \etal 2007; Schawinski \etal 2007, 2009, 2010; but see Xue~et~al.~2010 
for complications due to selection effects) has helped to promote the idea that AGN feedback quenches 
star formation and drives the color transformation of galaxies from the blue cloud to the red sequence.  
However, as in the case of star formation, it is dangerous to infer a direct, physical connection 
between AGN activity and galaxy color.  We are unaware of any concrete proof that AGNs orchestrate the 
star formation histories of disk galaxies at any epoch from 
$z \sim 0$ (Wild \etal 2010) to 
$z \sim 1$ (Aird \etal 2012) or 
higher (Bongiorno \etal 2012). 

      In summary, most local AGNs accrete at very sub-Eddington rates.  Very few galaxies are still growing their BHs 
at a significant level.  Rapid BH growth by radiatively efficient accretion took place mostly in more massive galaxies 
that are largely quenched today.~That is, the~era~of~BH~growth by radiatively efficient accretion is now mostly over.  
The era of major mergers is mostly over, too. We suggest that today's BHs grow mainly by secular processes that involve 
too little energy to result in structural coevolution with their host galaxies.  BHs and host galaxies have stopped coevolving.  
If coevolution happened at all, it happened at high $z$.  Section\ts8.6 reviews this subject.

\vs
\ni {\big\ARRed 8.4 Maintenance-Mode AGN Feedback at z{\ts}$\sim${\ts}0} \textBlack
\vs

\def\vs{\vskip 10pt}

      Still focusing on AGN feedback at $z$\ts$\sim$\ts0, we turn next to the opposite extreme from Section\ts8.3, 
i.{\ts}e., the highest-mass BHs in giant elliptical galaxies with cores.  Then Section\ts8.5 reviews how \hbox{BH{\ts}--{\ts}BH} 
mergers affect the $M_\bullet$ correlations.  In these sections, the big-picture physics is moderately well understood, although
it remains a challenge to engineer the details.  More uncertain is the situation at high $z$ (Section 8.6), 
where AGN feedback may be substantially different.  

      The differences between core and coreless ellipticals are reviewed in Sections 6.7 and 6.13; they are part of an ``E{\ts}--{\ts}E
dichotomy'' of many physical properties that are diagnostic~of~galaxy~evolution.  The physics that creates these differences is summarized 
there, but the motivating observations are the subject of this section and are illustrated~in~{\bf Figure\ts36}.~The suggestion is that 
coreless ellipticals form via wet mergers in which cold-gas dissipation feeds a central starburst that builds a dense ``extra light''
component that fills in any core ({\bf Figure 28}).  {\bf Figure 36} shows that~these~coreless or ``power-law'', disky ellipticals generally 
lack hot gas. In contrast, giant, core-boxy ellipticals contain hot, X-ray-emitting gas.  It is the essential agent that evaporates any 
cold gas and prevents the dissipation that leads to star formation (Nipoti \& Binney 2007).  The dividing line between galaxies with and 
without hot gas is at $\sim$\ts1/2 of the luminosity of the faintest core galaxies.  Above this luminosity and corresponding mass, even 
merger progenitors should have had enough hot gas to keep mergers dry and allow core scouring by BH binaries.~Moreover ({\bf Figure~36}),
core ellipticals contain radio sources that help, via ``maintenance-mode AGN feedback'', to keep the hot gas hot.


 \includegraphics{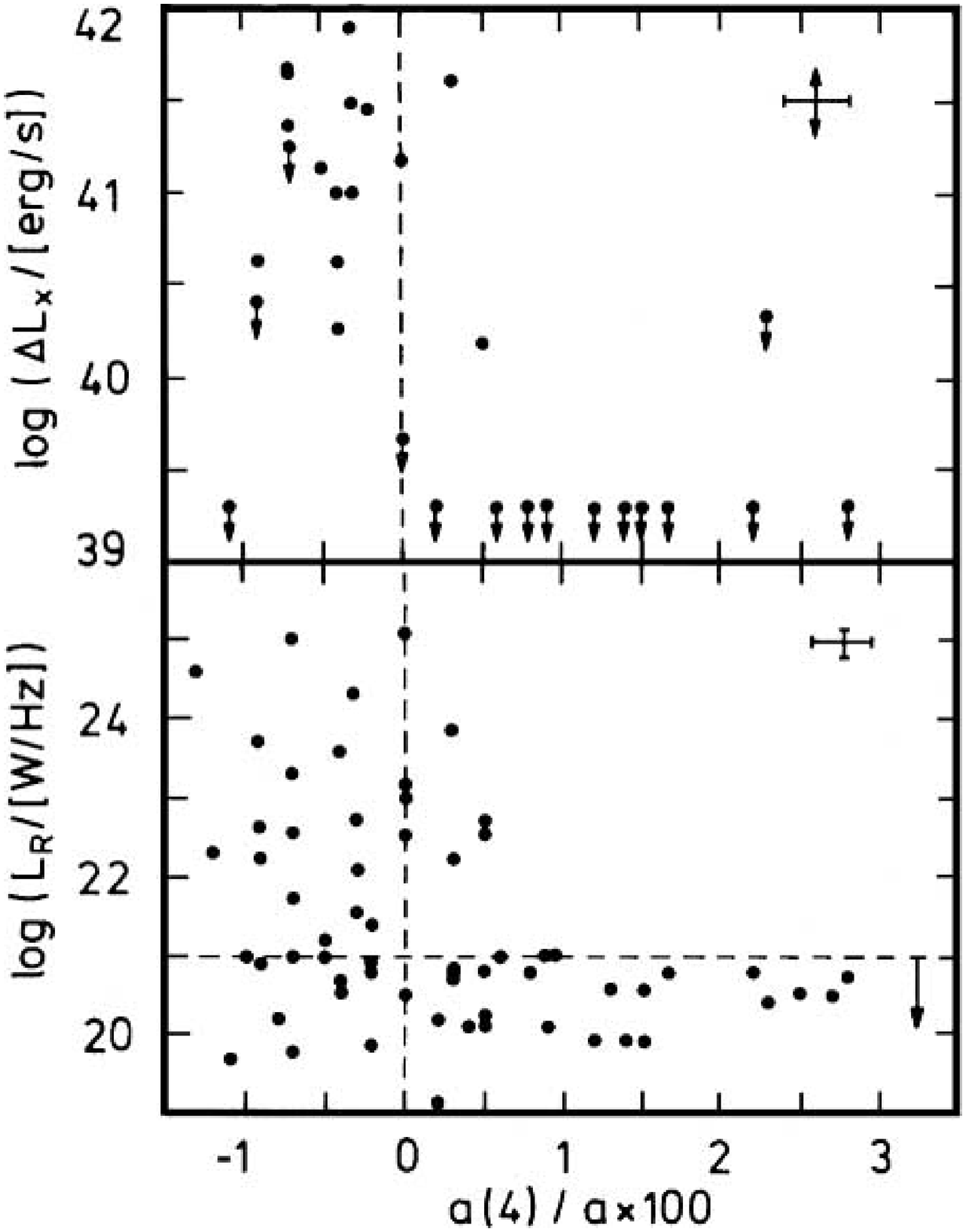}


\includegraphics{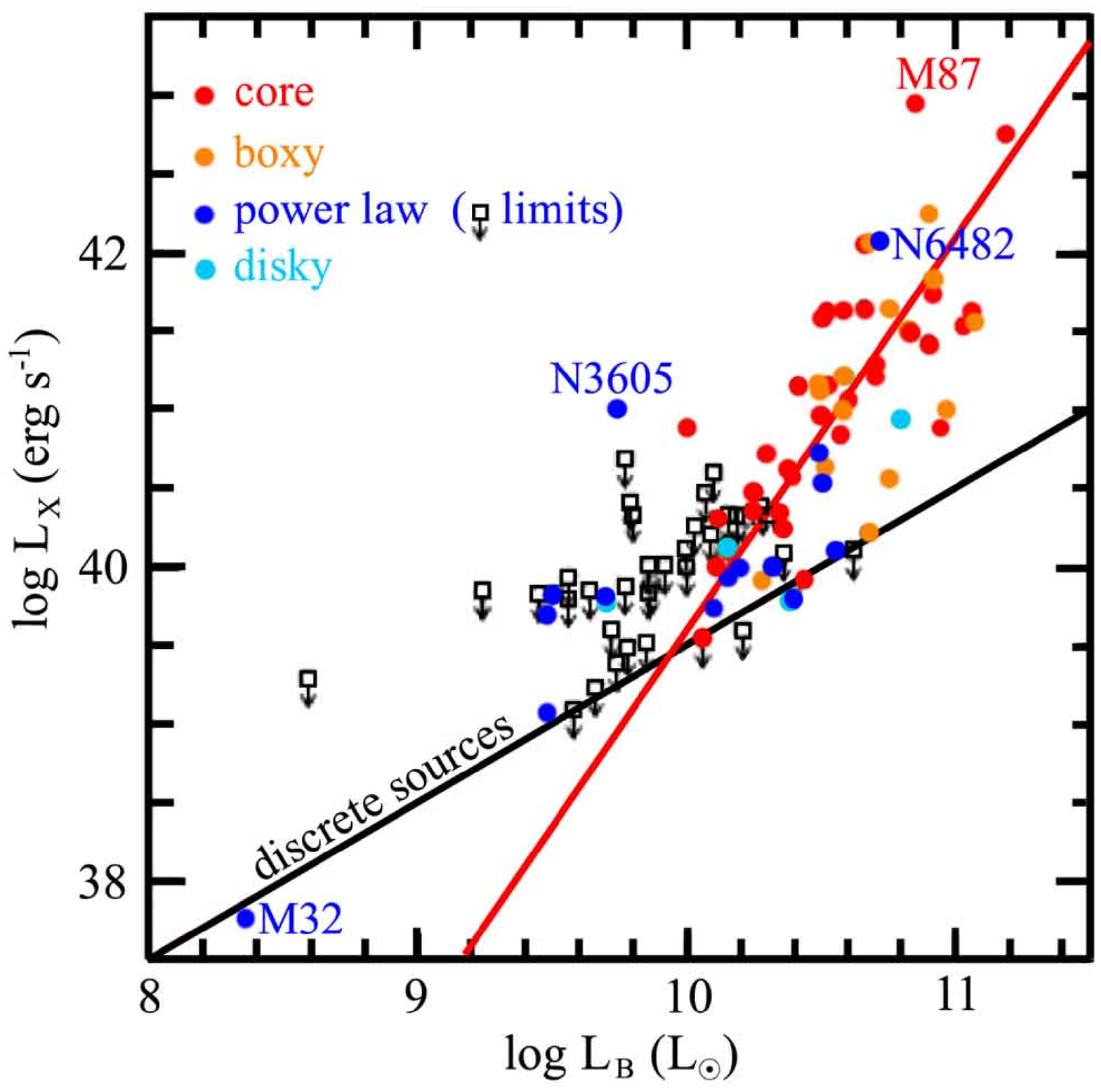}

\vfill

\ni {\bf \textBlue Figure 36}\textBlack

\vskip 1pt
\hrule width \hsize
\vskip 2pt

\ni ({\it left}) From Bender{\ts}et{\ts}al.{\ts}(1989), the correlation with isophote shape parameter~$a_4$ of ({\it top\/}) 
\hbox{X-ray} emission from hot gas and ({\it bottom\/}) radio emission.  Boxy ellipticals ($a_4 < 0$) contain \hbox{X-ray-emitting} gas
and strong radio sources; disky ellipticals ($a_4 > 0$) do not.~({\it right\/}) KFCB update~of~the~\hbox{X-ray} correlation:~This
plot shows total X-ray emission versus galaxy $B$-band luminosity, adapted from Ellis \& O\kern -0.55pt'\kern 0.4ptSullivan (2006).~Detections 
are color-coded according~to~the~E\ts--{\ts}E dichotomy.  The emission from discrete sources (X-ray binary stars) is estimated by the 
black line (O\kern -0.55pt'\kern 0.4ptSullivan \etal 2001) and was already subtracted before constructing the left figure.  The red line is 
a bisector fit to the core-boxy ellipticals.  They statistically reach $L_X = 0$ from hot gas at log\ts$L_B \simeq 9.94$.  This corresponds to 
$M_V \simeq -20.4$, 1 magnitude fainter than the luminosity that divides the two kinds of ellipticals.  Similar results were derived
by Pellegrini (1999, 2005) and by Ellis \& O\kern -0.55pt'\kern 0.4ptSullivan (2006).

\eject

      ``Radio\kern 1pt-mode'' or ``maintenance-mode'' AGN feedback requires a working surface against which the feedback can act (Begelman\ts2004; 
Best 2006; McNamara \& Nulsen 2007).  This is provided by the X-ray gas.  It also serves as an energy storage medium that smooths out the episodic
energy input from the AGN and ensures that hot gas is always available whenever~cold~gas~is~accreted.  A galaxy's ability to hold onto this gas 
depends on the depth of its gravitational potential well, so in this case, the velocity dispersion~$\sigma$ is the fundamental parameter that 
controls the physics.  The transition between galaxies without and with hot gas occurs roughly at a critical halo mass 
$M_{\rm crit} \simeq 10^{12}$\ts$M_\odot$ at which the gas cooling time is equal to the collapse time
(Rees \& Ostriker 1977;
Dekel \& Birnboim~2006;
Kere\v s~et~al.~2005;
Cattaneo \etal 2006, 2008;
Dekel \& Birnboim 2006, 2008;
Faber \etal 2007).
Faber \etal (2007) and KFCB get a corresponding stellar mass of $(1\null-\null2) \times 10^{11}$ $M_\odot$ or $M_{V,\rm crit} \simeq -21.3$. 
This disagrees \hbox{slightly with the DM-to-visible-matter calibration in} {\bf Figure\ts25}\ts({\it right\/}).  If that calibration is wrong by 
1/2 dex, the argument made there is unaffected.  Here, we note that hot gas does not disappear suddenly and completely at $M_{\rm crit}$.  
Some lower-mass halos contain smaller amounts of hot gas (e.{\ts}g., Bogd\'an \etal 2012, 2013), and still lower-mass halos likely contain 
warm-hot gas (Dav\'e \etal 2001).  KFCB emphasize the big-picture conclusion that there is remarkably good agreement between (1) the observations 
of which galaxies contain enough hot gas to be seen in {\bf Figure 36} and which do not and (2) the absolute magnitude $M_V \simeq -21.6$ that 
divides core and coreless ellipticals.  At the more detailed level that takes into account the gradual decrease in amount and temperature of gas 
as $M_{\rm DM}$ decreases, we suggest that the practical way to approach our engineering problem is this: Formation~of~extra-light components in 
dissipative mergers is easily switched off by even a modest amount of feedback (e.{\ts}g., Cox \etal 2006).  So the change at $M_V \sim -21.6$ 
from extra light to cores tells us how much hot gas and how much AGN activity is needed for feedback to be effective. The answer agrees with what 
{\bf Figure 35} shows.  In summary, KFCB suggest that $M_{\rm crit}$ quenching is the origin of the E\ts--{\ts}E dichotomy.  The corollary  is that 
maintenance-mode radio AGN feedback plausibly can operate in core ellipticals but not in coreless ellipticals or in most classical bulges.

      McNamara \& Nulsen (2007, 2012) and Fabian (2012) review observations of feedback in action.  X-ray gas often 
shows cavities or bubbles that are connected to and{\ts}--{\ts}we believe{\ts}--{\ts}inflated by jets.  The story 
is most convincing in galaxy clusters such as Perseus (Fabian \etal 2003,~2006, ~2011).  X-ray cavities are also seen
in individual galaxies such as M{\ts}87 (Forman \etal 2005, 2007).  The observations make a compelling case that heating
is relatively isotropic even though jets are strongly collimated.  However, we emphasize:~The physics that makes this happen
is not well understood.   The biggest puzzle is how to confine at least some effects of well-collimated jets within their galaxies.  
Firing a rifle in a room does not much heat the air in the room.  Maybe{\ts}--{\ts}as in protostellar jets (Shu \etal 1994, 1995)\ts--{\ts}the 
AGN jets are not as one-dimensional as they look.  After all, a BH can more easily accrete gas if an outflow carries away some angular momentum. 
Maybe most of the impact comes not from the infrequent, episodic, well-collimated, extended jets, but from the accumulated effects of steady, 
slower outflows associated with the pervasive compact cores.  Diehl \& Statler (2008a, 2008b) find widespread evidence that even weak AGNs 
can create significant disturbances in the spatial distribution and thermal structure of X-ray-emitting gas in normal ellipticals.  
Ho (2009b) saw hints of this, too, based on analysis of the kinematics of the warm ($10^4$ K) gas.  Is this level of ``stirring'' 
by compact radio sources enough to keep even field ellipticals quenched?

      Also, it is not certain that every giant elliptical can host a powerful radio jet, either periodically, as
required to keep hot gas hot, or even occasionally.  While $\sim 30\%$ of galaxies with $M_\star \sim 10^{11.5}\,M_\odot$ host a 
radio source with power above $P_{\rm 1.4~GHz} = 10^{23}$ W Hz$^{-1}$ (cf.~{\bf Figure 36}, {\it lower-left}); this could
be consistent either with the hypothesis that jets are switched on $\sim$\ts1/3 of the time or with the (for present purposes)
less benign hypothesis that only $\sim$\ts1/3 of giant ellipticals can make jets.  Could jet production require special conditions 
such as rapid BH spin (Sikora, Stawarz \& Lasota 2007; Fabian 2013) that exist in only a subset of giant ellipticals?  If so, then it 
gets harder to understand how hot gas is kept hot.

      However the detailed physics turns out, AGN feedback has enthusiastically been embraced by the galaxy formation community.
In one fell swoop, maintenance-mode feedback and $M_{\rm crit}$ quenching seem to offer a solution to many thorny problems 
that have plagued our picture of galaxy evolution: 
(1) how to prevent cooling flows, 
(2) how to shape the upper end of the galaxy luminosity function, 
(3) how to quench star formation in massive galaxies and keep them red, dead, and $\alpha$-element-enhanced, and 
(4) how to explain the dichotomy into core and extra light elliptical galaxies.
The effect of maintenance-mode feedback is largely negative.  It stops the completion of galaxy formation by keeping baryons 
suspended in hot gas.  Observations and theory lead to a picture in which heating and cooling are, on average, finely balanced 
but in which relaxation oscillations occur: 
accretion lights up the AGN; 
its feedback heats gas and drives it away;
the BH is starved of further fuel and switches off; and 
cooling resumes until new cool gas revives the AGN.
Supporting observations include 
(1) that the estimated AGN energy output is similar to the radiative losses in the hot gas, 
(2) that weak AGNs are found in the majority of clusters that have short cooling times, and 
(3) that low-density bubbles are younger than the cooling time of the hot gas through which they are rising, 
    so AGNs recur quickly enough to prevent runaway cooling.
This picture is developed or reviewed in Binney \& Tabor (1995), Ostriker \& Ciotti (2005), Croton \etal (2006), 
Rafferty \etal (2006, 2008), Sijacki \& Springel (2006), Ciotti \& Ostriker (2007), McNamara \& Nulsen (2007, 2012),
Cattaneo \etal (2009), Fabian (2012), and many other papers.

      Notwithstanding the popularity of radio-mode AGN feedback, we should not forget that there are other sources of heating.  
Ostriker \& Ciotti (2005) discuss the role of radiative heating.  Gravitational heating from cosmological accretion (Dekel \& 
Birnboim 2006, 2008) and satellite infall in dense environments (Khochfar \& Ostriker 2008) should also be important.
Even internal stellar processes such as stellar mass loss and Type 1a supernovae may contribute.  The important
point is this:~The ingredient that is necessary to solve the above problems is hot gas in giant galaxies.  Observations
of those galaxies tell us that the gas is present, and temperature measurements tell us that the gas is not cooling 
catastrophically.  The rest is engineering.  We include the subject because AGN feedback is relevant to this review.
But, as KFCB emphasize, any combination of the above and possibly other heating mechanisms suffices provided that it keeps 
the gas in the state in which we observe it.

\def\vs{\vskip 6pt}

\vs\vsss
\ni \hbox{{\big\ARRed 8.5 Making the M\lower.3ex\hbox{$\bullet$}{\ts}--{\ts}M\lower.3ex\hbox{\sbf bulge} Correlation by Mass Averaging in Mergers}}\textBlack
\vs

A radical alternative to making the $M_\bullet$\ts--\ts$M_{\rm bulge}$ relation by AGN feedback emerges from the statistics of 
mass averaging in galaxy mergers.  This idea was first articulated clearly and explicitly by Peng (2007), although elements of it 
were implicit in previous work (e.{\ts}g., Croton \etal 2006), and it has been emphasized independently by Gaskell (2010, 2011).  The original 
Monte Carlo experiments of Peng have now been elaborated using more realistic semi-analytic models of galaxy formation properly embedded
in $\Lambda$CDM merger trees (Hirschmann \etal 2010; Jahnke \& Macci\`o 2011).   As illustrated in {\bf Figure 37} ({\it left}), 
if the $i^{\rm th}$ progenitor galaxy of mass $M_{*,i}$ contains one BH of mass $M_{\bullet,i}$ 
that may or (as in the figure) may not correlate with $M_{*,i}$, then building bigger galaxies by a succession of $N$ major mergers in 
which the galaxy masses and BH masses separately add (``$\Sigma$'') builds a correlation $\Sigma M_{\bullet,i} 
\propto (\Sigma M_{*,i})^\beta$ with $\beta \simeq 1$ and a fractional scatter 
that decreases with increasing mass as $1/\sqrt{N}$.  This is a consequence of 
the Law of Large Numbers and arises completely independently of any input 
astrophysics that couples the BH and the galaxy. 

\vfill\eject

\cl{\null}

 \includegraphics{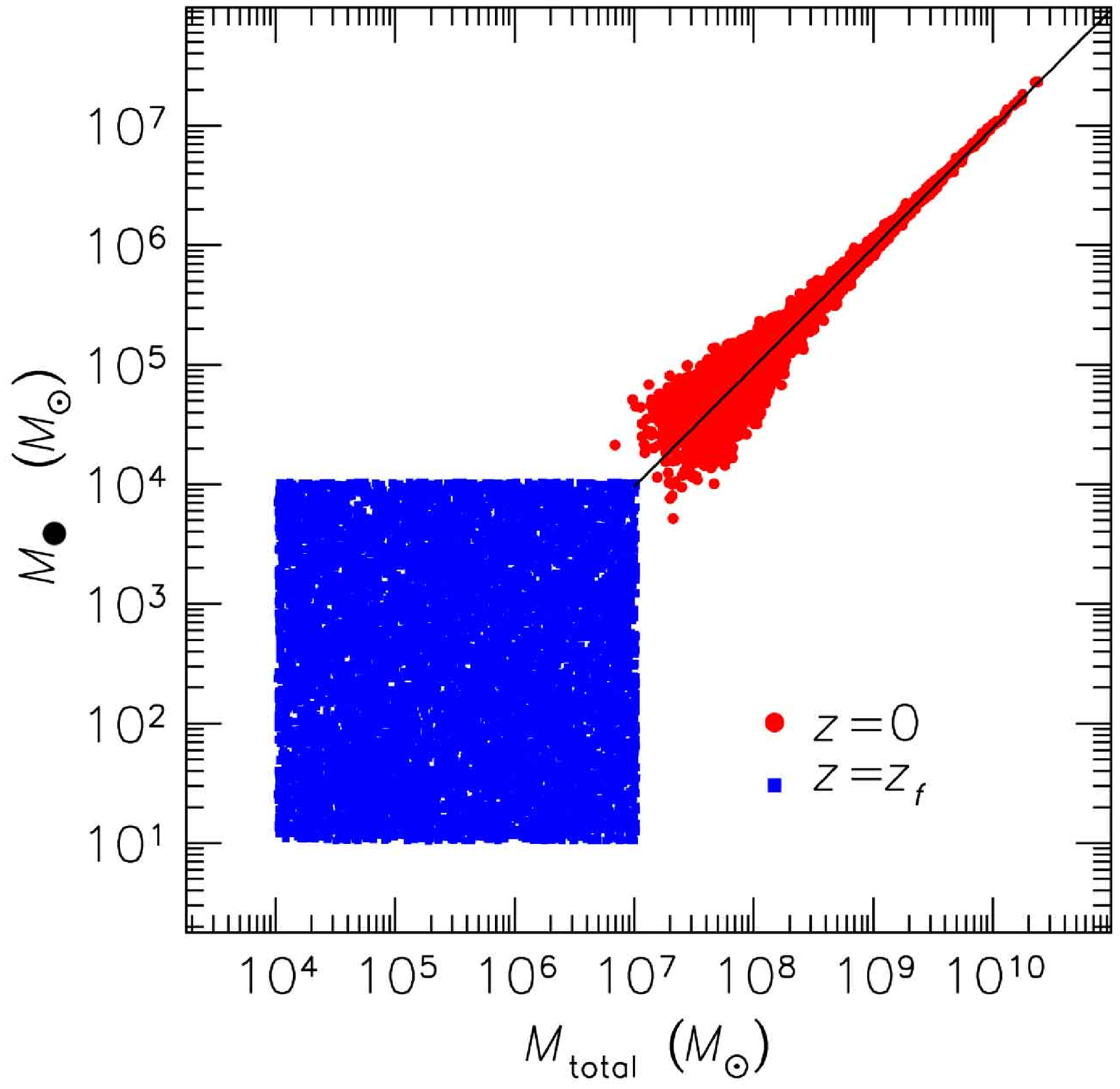}

 \includegraphics{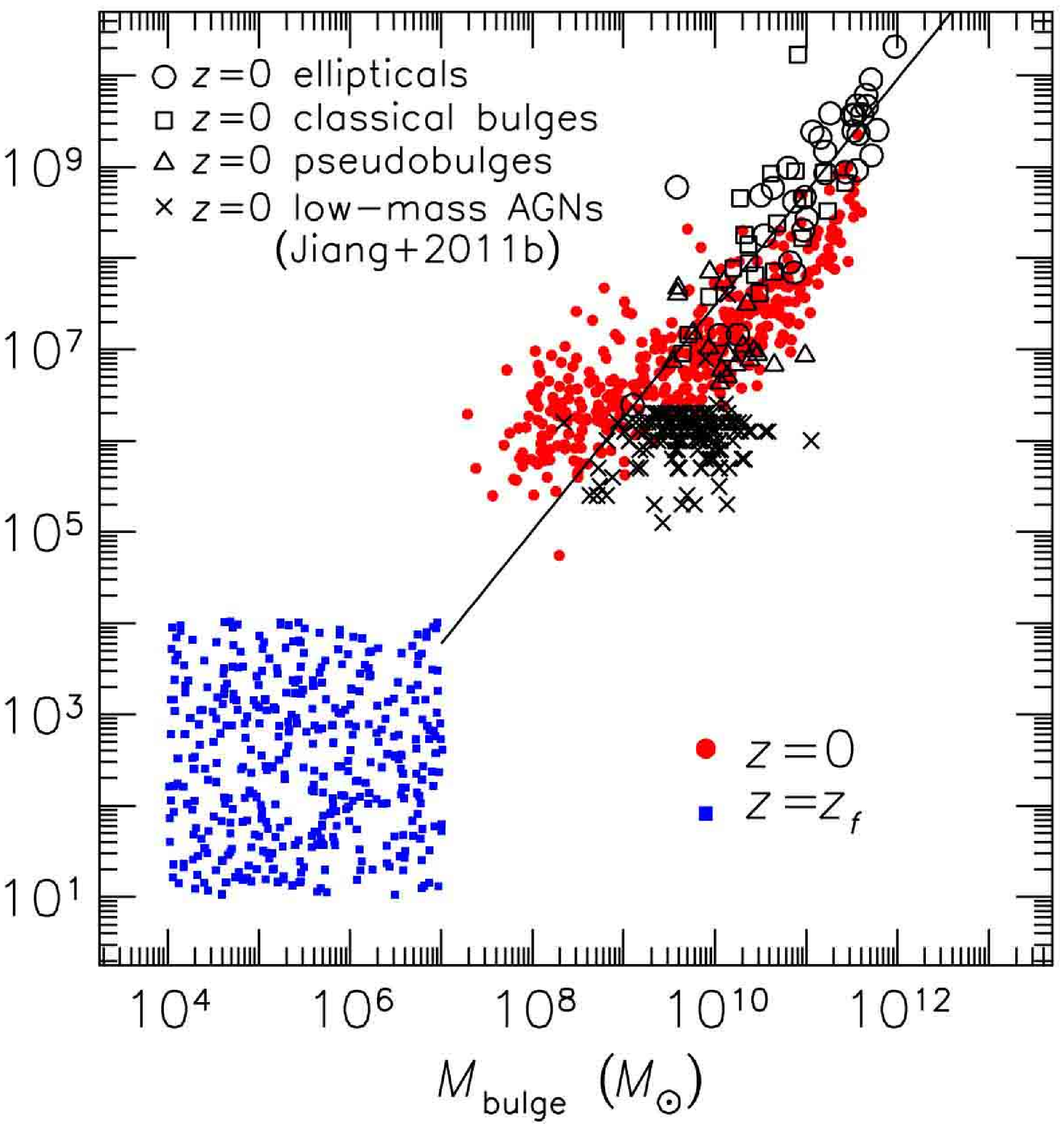}

\vskip 200pt

\ni {\bf \textBlue Figure 37}\textBlack

\vskip 1pt
\hrule width \hsize
\vskip 2pt

\ni ({\it left}) Numerical experiment to show how a correlation ({\it red points\/}) 
between BH mass and bulge stellar mass is produced from an initially uncorrelated distribution 
({\it blue points\/}) by a succession of dry mergers.  A tight linear correlation emerges, and
the scatter decreases toward higher masses.  The solid line has a slope of 1. 
({\it right\/}) Here, a subset of 400 randomly selected merger trees include simple prescriptions
for star formation, BH accretion, population of dark matter~halos~with~stars, and conversion of disks 
to bulges during major mergers.~Overplotted~in~black~is~the Section\ts5 database of dynamically 
measured BH masses supplemented by low-mass BHs from Greene\ts\&{\ts}Ho (2004,\ts2007c) that have bulge 
photometry in Jiang \etal (2011b).  The line is the best-fit $M_\bullet-M_{\rm bulge}$
relation from Section 6.6.1 (Equation 10).  Simulations adapted from Jahnke \& Macci\`o (2011).  

\vskip 10pt

      If the mean and dispersion $\equiv$ $\sqrt{\rm variance}$ of the mass distribution of the 
progenitors galaxies are $\overline{m_g}$ and $s_g$, and the mean and 
dispersion of the mass distribution of seed BHs are $\overline{m_\bullet}$ 
and $s_\bullet$, then, for a large number $N$ of mergers, the galaxy mass
tends toward $N \overline{m_g}$ with dispersion $N s_g / \sqrt{N}$ and the BH 
mass independently tends toward $N \overline{m_\bullet}$ with dispersion 
$N s_\bullet / \sqrt{N}$.  Thus the fractional dispersions in 
$N \overline{m_g}$ and $N \overline{m_\bullet}$ both decrease as 
$1/\sqrt{N}$.  Since $d\kern0.12ex(\log_{10}{x}) \simeq 
0.434{\ts}d{\kern.10ex}x/x$ for small $dx/x$, the scatter in a 
log$_{10}$-log$_{10}$ plot decreases as $1/\sqrt{N}$ in both coordinates, 
again for large $N$.  The Central Limit Theorem further~tells~us~that, 
independent of the forms of the initial distributions, the distributions of 
final galaxy and BH masses converge to Gaussians.  

     This result is broadly consistent with the observation that the scatter 
in the $M_\bullet$\ts--{\ts}host-galaxy correlations decreases with increasing 
mass.  We say ``broadly consistent" because $N$ is not extremely large.   To 
better reproduce the normalization, slope, and scatter seen in local 
galaxies, the simulations shown in {\bf Figure 37} ({\it right}) incorporate 
physically motivated recipes which describe how DM halo mass 
relates to stellar mass, how star formation rates vary with redshift, and how 
AGN accretion rates vary with redshift.  Jahnke \& Macci\`o further 
adopt a simple recipe guided by simulations to convert disks to bulges 
during major mergers.  The numerical results roughly reproduce our BH database,
including the decrease in scatter toward higher masses.  There is a minor offset 
in normalization, but this is not surprising, because the simulations were not 
\hbox{fine-tuned} to our data.~In particular, they were made prior to the change in 
$M_\bullet$ zeropoint derived in this review.  
Gaskell (2010) reports a similar trend for AGNs; it is further amplified in
Section\ts8.6 ({\bf Figure 38}).  

      {\it It is possible that merger averaging is at least as important as feedback-driven coevolution in creating the 
observed BH{\ts}--{\ts}host correlations.}  The observation that points to this possibility is the result that $M_\bullet$
correlates tightly with the properties of bulges and ellipticals but not with those of pseudobulges and disks.~These are 
not remnants of major mergers.~Merger averaging~is~not~relevant for them.  In contrast, we concluded 
that at least several dry, major mergers are required to excavate realistic cores.  And these can easily have been 
preceeded by one or more wet mergers.

      We do not know the relative importance of merger averaging and AGN feedback in setting up the BH--host
correlations.  If feedback plays a major role, then -- we suggest in Section 8.6 -- it is quasar-mode feedback that 
happens at large $z$.  It is poorly understood.  However, as summarized in the next section, observations of high-$z$ quasars 
and submillimeter galaxies (SMGs) suggest that a nascent correlation between $M_\bullet$ and $M_{\rm host}$ was already
in place at $z \approx 2-6$.  Given the prodigious accretion and star formation rates of these gas-rich systems, it seems likely 
that some form of energy feeback{\ts}--{\ts}from the AGN, from the starburst, or (most likely) from both{\ts}--{\ts}{\it did\/}~help 
to establish the BH{\ts}--{\ts}host correlations during the quasar era.  The intrinsic scatter of the correlations, 
on the other hand, appears to be larger than at $z \sim 0$.   Moreover, the observations suggest that 
BH accretion and star formation were not precisely coeval.  Thus the BH{\ts}--{\ts}host correlations seen in nearby galaxies may
have been tightened by hierarchical merging at intermediate-to-late times.

      One caveat is inherent in the downsizing of both AGNs and star formation as $z \rightarrow 0$.  As discussed
in Section 6.4, it involves giant, pure-disk galaxies at $z \sim 0$.  Prototypes are galaxies like M{\ts}101,
NGC\ts6946, and IC\ts342 (Kormendy \etal 2010).  Their gas content and BH masses are small.  When two such pure-disk galaxies merge, 
their stars get scrambled up into a bulge or elliptical, and their cold gas presumably feeds a modest starburst that helps to make
the resulting bulge dense enough to satisfy the fundamental plane structural parameter correlations.  But the sum of the BH masses 
remains too small for the newly converted bulge mass, and there is uncomfortably little gas in $z \sim 0$ progenitors to feed up the 
BH mass in proportion to how much disk mass got converted into bulge mass.  Mergers of pure-disk galaxies have become somewhat rare at 
$z \sim 0$, but the observation in Section 6.4 of undermassive BHs in mergers in progress may be a sign that merger averaging
can actually erode the BH{\ts}--{\ts}host correlations in the nearby universe.

      Clearly we need more work to determing the evolving relative importance of AGN feedback and merger averaging as functions of cosmic time.

\vs
\ni {\big\ARRed 8.6 Quasar-Mode AGN Feedback in Wet Mergers}
\vs

\noindent \ARRed {\bf 8.6.1 The formation of coreless ellipticals in wet 
mergers} \textBlack is relatively well understood.  It is reviewed in detail 
in KFCB; some of the present discussion is abstracted from that review.  
Galaxy collisions lead to mergers that scramble disks into ellipticals 
(Toomre \& Toomre 1972; Toomre 1977); numerical simulations (see Barnes \& 
Hernquist 1992; Barnes 1998 for reviews and Steinmetz \& Navarro 2002 for 
a case study) and observations of mergers-in-progress (see Schweizer 1990, 
1998 for reviews) make a definitive~case.  When the progenitors contain gas, 
gravitational torques drive it to the center 
and feed starbursts. These build a 
distinct,~extra~stellar component that is recognizably smaller and denser 
than the S\'ersic-function main body of the remnant (Mihos \& Hernquist 1994; 
Springel \& Hernquist 2005, Fig.~43 in KFCB; Cox \etal 2006; Hopkins \etal 
2008, 2009a).  Kormendy (1999) was the first to observe and recognize the 
extra component; he interpreted it as the ``smoking gun'' result which shows 
that coreless ellipticals formed in wet mergers.  Further observational 
confirmation followed, both for mergers-in-progress (Rothberg \& Joseph 2004, 
2006) and for old, relaxed ellipticals (C\^ot\'e \etal 2007; KFCB; {\bf 
Figure 28} here).  KFCB emphasize that the extra light often has 
properties such as disky structure and rapid rotation that further point to 
dissipative formation.  

Classical bulges and coreless ellipticals make up \gapprox 2/3 of the dynamic 
range~in~the~BH correlations of Section\ts6.  Knowing how they got their tight 
scatter is the key to understanding coevolution.  They bequeath their tight scatter 
to core ellipticals via dry mergers; no further coevolution is needed for core galaxies 
beyond the need to preserve the tight correlations.

\vs
\noindent \ARRed {\bf 8.6.2 ULIRGs are prototypes of the formation of coreless ellipticals.} \textBlack 
ULIRGs are enormous starbursts that are almost completely shrouded by dust.  By definition, they have 
infrared luminosities greater than $10^{12}$ $L_\odot$.  In the nearby universe, almost all of them 
are observed to be mergers in progress 
(Joseph \& Wright 1985; 
Sanders \etal 1988a, 1988b; 
Sanders \& Mirabel~1996; 
Rigopoulou \etal 1999; 
Dasyra \etal 2006a).~They are local prototypes of the formation of ellipticals by very wet mergers.   

What kind of ellipticals are they forming?  Their structural parameters are consistent with the fundamental plane
(Kormendy \& Sanders 1992;
Doyon    \etal 1994; 
Genzel   \etal 2001; 
Tacconi  \etal 2002; 
Veilleux \etal 2006; 
Dasyra   \etal 2006a, 2006b), 
so they are making normal ellipticals.  The crucial 
observation is that ULIRGs have stellar velocity dispersions of $\sigma 
\simeq 100$\ts--\ts230~km~s$^{-1}$ (Genzel \etal 2001; Tacconi \etal 2002; 
Dasyra \etal 2006b, 2006c).  Therefore they make moderate-mass ellipticals;
i.{\ts}e., the disky-coreless-rotating side of the E{\ts}--{\ts}E dichotomy 
discussed in Section\ts6.7.

\vs
\noindent \ARRed {\bf 8.6.3 ULIRGs are prototypes of BH--host coevolution driven by quasar-mode feedback.} \textBlack 
Remnants of major mergers show tight $M_\bullet$ correlations whereas disks and pseudobulges~do~not, even over a 
substantial range where they overlap in $\sigma$ ({\bf Figure 21}).~\hbox{Therefore $\sigma$ as a measure} of the 
depth of the gravitational potential well does not control whether coevolution happens or not.  That is, the 
lowest-mass objects that coevolve with BHs (small ellipticals like M{\ts}32 and NGC\ts4486A and small classical 
bulges like the one in NGC 4258) are much~lower in mass than~the~\hbox{highest-mass} disks and pseudobulges 
that do not coevolve (e.{\ts}g., M{\ts}101, IC\ts342, and NGC\ts1068).  It is important to increase the numbers of
objects on which these conclusions are based.  Meanwhile, the observations imply that coevolution is controlled by 
the process that makes coreless ellipticals and not just by galaxy mass. That process is dissipative mergers with 
starbursts, as exemplified by ULIRGs.  We now enlarge on these suggestions, starting with observed connections 
between ULIRGs and AGNs.

\vs

\noindent \ARRed {\bf 8.6.4 The ULIRG -- AGN connection: What is the energy 
source that powers ULIRGs?} \textBlack A connection between ULIRGs and AGNs 
emerged very early.  Sanders \etal (1988a,{\ts}b) suggested that 
``ultraluminous infrared galaxies represent the initial, dust-enshrouded 
stages of quasars''.  ULIRGs and quasars have similar number densities at 
$z$\ts\lapprox\ts0.08 (ULIRGs are slightly more common) and luminosities (the 
ULIRG Mrk 231 is as luminous as the brightest nearby quasars).  Both show 
frequent signs of mergers (essentially always, in the case of ULIRGs).  And
ULIRG spectra range from completely starburst-dominated (Arp 220) through 
many Seyfert 1 objects to quasars.  Sanders \etal (1988a) state, ``The 
discovery of several more distant (and presumably rarer) optical quasars with 
substantial infrared excess, as illustrated by the energy distributions for 
3C\ts48 and Mrk\ts1014, strengthens the case for an orderly progression from 
ultraluminous infrared galaxy to optical quasar'' (3C 273:~their 
Figure 17).  Similar  suggestions were made by Sanders \etal (1988b) and 
Mirabel \etal (1989).  Recent papers along the same line
include{\ts}--{\ts}among many others\ts-- Urrutia \etal (2008), Wang \etal 
(2010), Simpson \etal (2012), and Xia \etal (2012).  They find large amounts 
of molecular gas in quasars of all redshifts and support the case that the 
transition from starburst-dominated to AGN-dominated evolution is rapid.  

These suggestions led to an enduring debate about whether starbursts or AGNs 
power ULIRGs (Joseph 1999; Sanders 1999).  It is now clear that both are 
important but that starbursts usually dominate energetically (e.{\ts}g., 
Genzel \etal 1998; see KFCB for a review).  ULIRGs~are~rare~locally, but they 
get more common rapidly with increasing redshift (Sanders \& Mirabel 1996; 
Le{\ts}Floc'h \etal 2005) as do quasars.  Many theoretical papers pursue and 
develop this picture; a review is beyond the scope of this review.  We 
mention only one paper: Hopkins \etal (2006) develop ``a~unified, 
merger-driven model of the origin of starbursts, quasars, the cosmic X-ray 
background, supermassive black holes, and galaxy spheroids''.  Importantly, 
while the early starburst and associated, dust-obscured quasar phase last 
longer, ``the total mass growth and radiated energy are dominated by the 
final blowout stage visible as a bright optical quasar''.  This is necessary 
by the So\l tan (1982) argument.

At the same time, nothing in the preceding discussion tells us whether AGN feedback 
or starburst-driven feedback engineers the BH{\ts}--{\ts}host-galaxy correlations.  
Both may be important.

\vs
\noindent \ARRed {\bf 8.6.5 Morphologies of AGN hosts at high redshifts.}~ \textBlack 
We saw in Section\ts8.3 that, in the nearby universe, most AGN host galaxies are not mergers in progress.  
This is consistent~with~what we know about mass functions of different kinds of galaxies:~the most 
numerous galaxies~are~small ones, and overwhelmingly, they are disk galaxies.  The fraction of galaxies 
or galaxy components that are merger remnants increases with increasing mass.  The highest-mass 
galaxies are core-boxy-nonrotating ellipticals, and are all remnants of major mergers.  This is mirrored 
in the distribution of BH 
masses: the lowest-mass BHs are in secularly evolving disks, whereas the 
highest-mass BHs are all in merger remnants.  Observations suggest that this 
same picture holds at least out to $z \sim 2$.

Classifying quasar hosts at high $z$ is difficult, but a consensus is 
emerging.  Most~AGNs~of moderate luminosity (X-ray luminosity 
$L_X$\ts$\simeq$\ts$10^{43 \pm 1}$ erg s$^{-1}$) live in disk galaxies, 
even when they are optically obscured 
(Cisternas \etal 2011a, b;
Schawinski \etal 2011, 2012;
Kocevski~et~al.~2012; 
Treister \etal 2012; 
B\"ohm \etal 2013; 
Schramm \& Silverman 2013). 
On the other hand, the highest-bolometric-luminosity AGNs, true quasars with $L_X > 10^{44}$ erg s$^{-1}$ and 
$L_{\rm bol} > 10^{45}$ erg s$^{-1}$, commonly are in mergers-in-progress 
and specifically in dust-shrouded, merger-induced starbursts (e.{\ts}g.,
Urrutia \etal 2008; Treister \etal 2010, 2012). As noted earlier,
observations indicate that a substantial fraction ($\sim$\ts1/2) of the 
growth of the largest BHs happens in this dust-shrouded phase. 

We emphasize that these different results are not inconsistent.  They address different questions: \vsss\vskip 1pt

\nhii (1) How do typical AGNs behave?  For any realistic mass or energy 
function, typical objects are the smallest ones that can be found by the 
detection limits. Observations indicate -- as many authors
\hbox{concluded~--~that} BH growth by secular evolution of disk galaxies is 
more important at large $z$ than we thought.  In contrast: \vsss\vskip 1pt

\nhii (2) How do ellipticals form, how do big BHs grow, and how do the 
answers conspire to produce tight BH--host correlations?  Our answer is the 
subject of this section: We suggest that ellipticals form by major mergers of 
the above disks and their small BHs, coevolution happens in the context of 
dissipative events and dust-shrouded starbursts, and statistical averaging 
via mergers finishes the job by tightening the BH-host correlations.
At $z < 0.3$, the few remaining optically luminous quasars are already 
predominantly in ellipticals or in objects that are about to become 
ellipticals (e.{\ts}g., Bahcall \etal 1997). \vsss\vskip 1pt

\noindent A similar distinction between high-luminosity AGNs fed by wet 
mergers and lower-luminosity AGNs fed by long-duration secular processes is 
discussed by Hopkins \& Hernquist (2009).  It is important to note that 
classical bulges and ellipticals are a moderately small fraction of all 
galaxy components.  Giant ellipticals are particularly rare.  Thus the 
results on AGN host galaxies at high redshifts appear to be consistent with 
the picture of coevolution that we suggest.

The new result that we did not anticipate is that well-formed disk galaxies 
already exist at $z \sim 2$ and that these have had time and opportunity to 
evolve secularly and to feed their BHs by local processes.  We have no 
concrete reason to believe that these low-luminosity AGN BHs correlate 
tightly with any property of their host galaxy.

\vs\vsss
\noindent \ARRed {\bf 8.6.6 Evolution of $M_\bullet/M_{\rm bulge}$ at higher redshifts}~ \textBlack 

      The complexity of the BH-host correlations at $z\sim 0$ and the uncertain role of AGN feedback~in 
shaping them compel us to turn to observations at high $z$ for guidance.  Here we briefly summarize this 
very active field, concentrating on the aspects that most affect our preceding discussion.  

      Any attempt to map out the origin and evolution of the BH-host correlations beyond the local universe 
faces two major obstacles.~BH masses cannot yet be measured directly for distant~galaxies; this forces us 
to resort to less-accurate, single-epoch virial mass estimators based on broad AGN emission lines (Supplemental Information).   
Moreover, the glare of the bright nucleus compounded by the large distances 
of the sources severely limits our ability to measure host galaxies, especially bulges and especially in
the systems for which we must rely on broad emission lines to get $M_\bullet$.  

      Progress on this subject has therefore depended on high-resolution images to separate~the~host from the AGN
and the bulge from the rest of the galaxy.~Most of this work uses HST, with important recent contributions from 8-meter-class, 
\hbox{ground-based} telescopes (e.{\ts}g., 
McLeod \& Bechtold 2009; 
Targett, Dunlop \& McLure 2012).  
Deblending the host from the AGN is feasible for relatively nearby sources 
with modest AGN-to-host brightness constrasts (e.{\ts}g., Kim \etal 2008; 
Bennert \etal 2010), but systematic errors due to PSF mismatch become 
increasingly problematic for more luminous, distant quasars (Kukula \etal
2001; Ridgway \etal 2001).  At the highest redshifts, no detection is 
possible at all (Mechtley \etal 2012), except for rare cases in which strong 
gravitational lenses stretch the host into an Einstein ring (Peng \etal
2006b).  To circumvent blinding by the AGN, some 
authors have resorted to sources where line-of-sight obscuration creates
a natural coronograph (e.{\ts}g., McLure \etal 2006; Sarria \etal 2010).  
However, even when the host can be detected, we still need to convert
measurements of light into estimates of stellar~mass.  This requires at least
some color information or assumptions about the star formation history.  
Only a handful of studies have attempted to do this 
(Jahnke \etal 2009; 
Bennert \etal 2010, 2011; 
Decarli \etal 2010; 
Merloni \etal 2010; 
Sarria \etal 2010; 
Cisternas \etal 2011b; 
Schramm \& Silverman 2013).

       Photometry is hard, but kinematic measurements are harder.  The AGN not only dilutes 
the stellar features but corrupts them with a plethora of emission lines that can result in 
systematic errors in the inferred central stellar velocity dispersions $\sigma$ (Greene \& Ho 2006a).  
Several studies have used AGN samples that were chosen to minimuze these complications, 
but so far, efforts to quantify the $M_\bullet$\ts--\ts$\sigma$ relation have been restricted to 
$z \lesssim 0.3-0.4$ 
(Treu, Malkan \& Blandford 2004; 
Barth, Greene \& Ho 2005; 
Greene \& Ho 2006b; 
Woo \etal 2006; 
Shen \etal 2008; 
Xiao \etal 2011; 
Canalizo \etal 2012; 
Harris \etal 2012; 
Hiner \etal 2012), 
with just one venturing out to $z \sim 0.6$ (Woo \etal 2008).  AO assisted by laser guide 
stars appears well poised to contribute to this subject soon (Inskip \etal 2011).  
In the meantime, some have used the widths of narrow emission lines to estimate $\sigma$ (e.{\ts}g., 
Shields \etal 2003; 
Salviander \etal 2007), 
while others advocate the use of radio emission lines to probe the rotation velocities of the hosts on larger scales 
(e.{\ts}g., 
Walter \etal 2004; 
Shields \etal 2006; 
Ho 2007b; 
Ho \etal 2008; 
Riechers \etal 2008; 
Wang \etal 2010).
 
      Broad-brush summary: Provided that the above caveats concerning $M_\bullet$ and $M_{\star, \rm bulge}$ 
are not too serious, the zero point and possibly the intrinsic scatter of the $M_\bullet-$host relations appear 
to evolve strongly with redshift.  At higher redshifts, BHs appear to be more massive with respect to their hosts
than they are in $z \sim 0$ galaxies of the same mass or gravitational potential.  This suggests that BHs grew faster than
their hosts formed stars or were assembled.

      This conclusion is illustrated in {\bf Figure 38}.  It summarizes all available observations of AGNs at
$z \approx 0.1-7.1$ that have reliable estimates of $M_\bullet$, of the stellar masses of the host galaxies, and
of bulge-to-disk ratios determined from photometric component decompositions (low-$z$ objects only).  We compare
these to the $z \sim 0$, $M_\bullet$\ts--\ts$M_{\rm bulge}$ relation from  H\"aring \& Rix (2004), because the AGN 
BH masses are zeropointed to that relation.  This is discussed further below.

\cl{\null}
\vfill

\cl{\null} 

\includegraphics{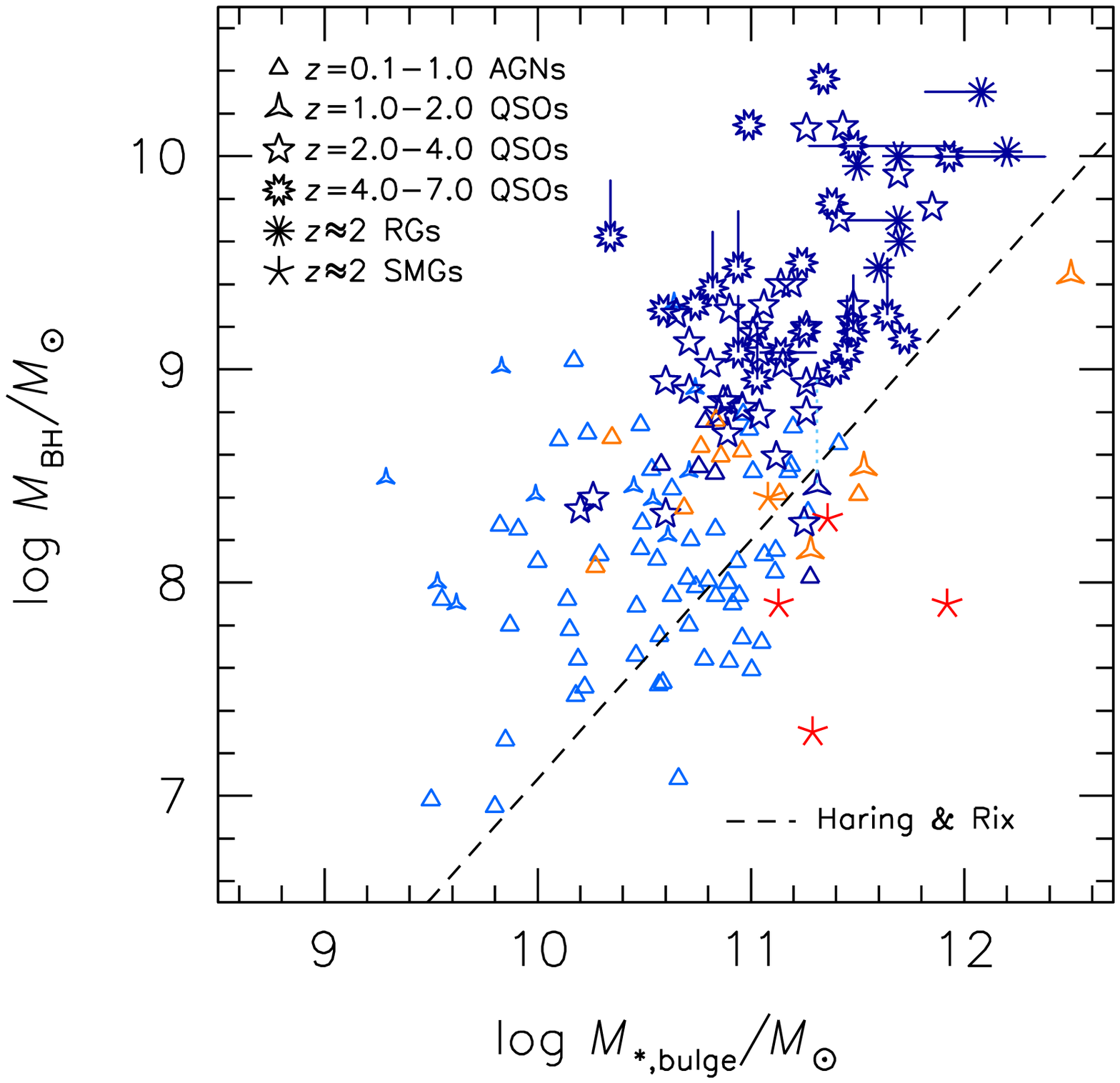}
\includegraphics{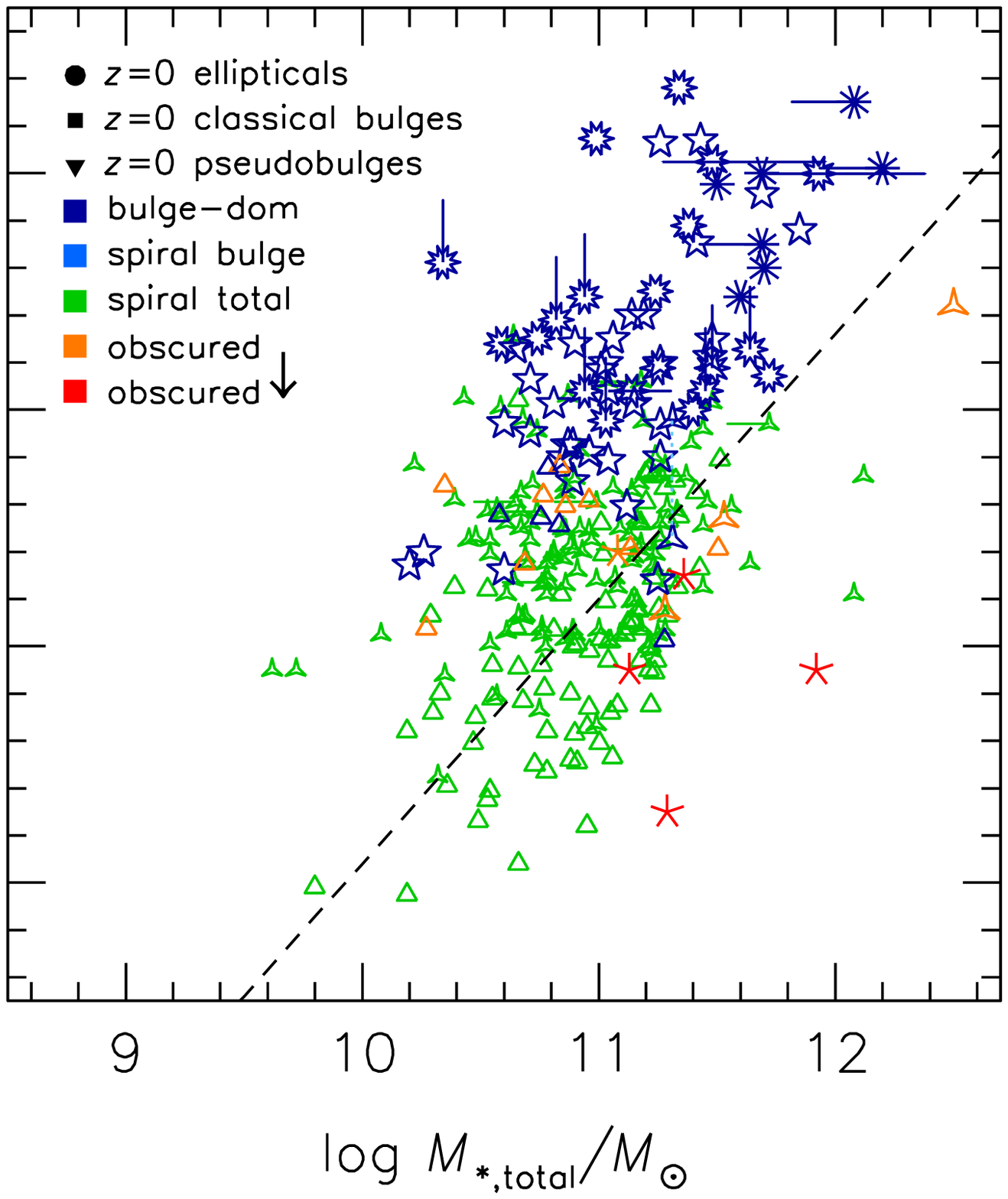}
\includegraphics{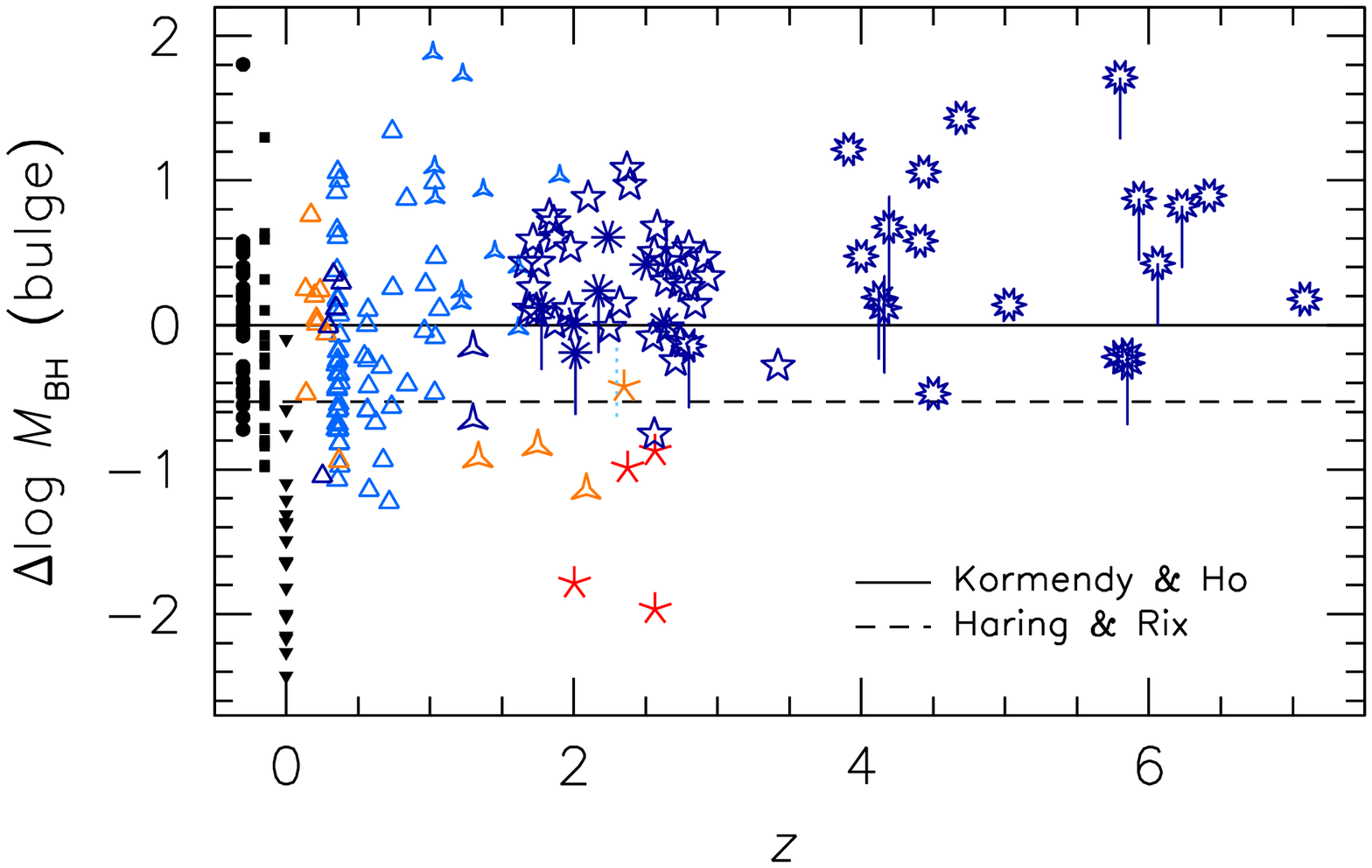}

\vskip 280pt
\ni {\bf \textBlue Figure 38}\textBlack

\vskip 1pt
\hrule width \hsize
\vskip 2pt

\ni Correlation between $M_\bullet$ and host galaxy stellar mass \hbox{(explicitly:~$M_*$)
of AGNs from $z\sim 0.1$ to 7.1,} shown separately ({\it upper left}) for the bulge 
only and ({\it upper right}) for the entire galaxy.  The dashed line is the 
H\"aring \& Rix (2004) correlation between $M_\bullet$ and $M_{\star,{\rm bulge}}$ 
at $z \sim 0$ for inactive galaxies, which is roughly consistent with the zeropoint 
calibration of AGN BH masses.  Dark blue points denote hosts that are known to be ellipticals 
or that contain classical bulges or that are massive enough so that they must be bulge-dominated
by $z \sim 0$.  They obey a moderately strong correlation.  
Dust-reddened quasars and other obscured AGNs appear to have preferentially undermassive BHs, 
the most extreme being the submm galaxies (SMGs).  Less massive hosts are mostly disk-like, 
spiral galaxies at $z \lesssim 2$; they show a larger scatter in $M_\bullet-M_{\star,{\rm bulge}}$
like that of pseudobulges at $z \sim 0$.  
({\it bottom}) 
Offset of $\log M_\bullet$ with respect to the local $M_\bullet-M_{\star,{\rm bulge}}$ relation 
derived by H\"aring \& Rix (2004; {\it dashed line}) and here in Section 6.6 ({\it solid line}).  
The black points at $z \simeq 0$ are our sample of ({\it left to right\/}) classical bulges, 
ellipticals, and pseudobulges with dynamically detected BHs ({\bf Tables 2} 
and {\bf 3}; objects are slightly offset from $z = 0$ for clarity).  Adapted from Ho (2013).

\eject

      {\bf Figure 38} show that, on average, $M_\bullet/M_{\star, \rm bulge}$ is larger than the local value 
from H\"aring \& Rix (2004) by factors of 
$\sim 2$ at $z \approx 0.2-0.6$ (Treu \etal 2004; Woo \etal 2006, 2008; Canalizo \etal 2012) to factors of 
$\sim 4$ at $z \approx 2$ (e.{\ts}g., Peng \etal 2006a, 2006b; Shields \etal 2006; Ho 2007b), or more at 
$z \approx 4-6$ (Walter \etal 2004; Wang \etal 2010; Targett \etal 2012).  
Parameterizing the evolution as $M_\bullet/M_{\star, \rm bulge} \propto (1+z)^\beta$, published values span 
the range $\beta \approx 0.7-2$ 
(McLure \etal 2006; 
Bennert \etal 2010, 2011; 
Decarli \etal 2010; 
Merloni \etal 2010).  
Taken at face value, these results imply that BH growth preceded or outpaced the growth of the bulge.   
However, it is tricky to interpret the observed distribution of points.  Selection efffects are difficult to quantify 
(Lauer \etal 2007c; 
Shen \& Kelly 2010; 
Schulze \& Wisotzki 2011), 
and current samples of high-$z$ AGNs may be biased toward the most luminous quasars.  We cannot exclude the possibility
that the points in {\bf Figure 38}\ are the upper envelope of a distribution that extends to smaller BH masses.  Still, 
two conclusions seem robust.  First, the population of actively growing BHs at high $z$ contains members with higher
$M_\bullet/M_{\star, \rm bulge}$ ratios than are seen in most present-day massive galaxies.  {\bf Section 6.5}\ and
{\bf Figure 18} illustrate a few local examples of BH monsters that are much more massive with respect to their host bulges
than the well-defined scatter shown by other galaxies.  As noted there, these may be fossils of high-$z$ quasars with
higher-than-normal $M_\bullet/M_{\star, \rm bulge}$.  And second, whatever its true distribution, the observed scatter in 
$M_\bullet/M_{\star, \rm bulge}$ at high $z$ is almost certainly larger than it is at low $z$.  Coevolution may take place,
but it does not look finely coordinated.  

      The most undermassive BHs are seen in $z \sim 2$ SMGs ({\it red points\/}).~SMGs are the \hbox{high-$z$}~analogs of ULIRGs 
(Sections 8.6.2\ts--\ts8.6.4).  Like ULIRGs, SMGs are dusty, gas-rich mergers powered simultaneously by a massive starburst and 
often by a heavily buried AGN (Tacconi \etal 2008).  Importantly, SMGs are not just run-of-the-mill, high-$z$ 
star-forming galaxies.  Borys \etal (2005) find that $z \sim 2$ SMGs have a median stellar mass of $\sim 2\times 10^{11}\,M_\odot$, 
fully compatible with being the precursors of most $z \sim 2$ quasars and perhaps of present-day coreless ellipticals (the masses 
are too large to be entirely comfortable).  Within this backdrop, it is noteworthy that SMGs have $M_\bullet/M_{\star, \rm bulge}$ 
values that are at least an order of magnitude smaller than the fiducial local value 
(Borys    \etal 2005; 
Alexander \etal 2008; 
Coppin    \etal 2008; 
Carrera   \etal 2011).  
A weaker but similar effect is seen in less extreme obscured systems (orange points).  Given their already substantial stellar masses, 
SMGs clearly have been forming stars already for some time.  And they are still at it.  Their central BH, then, has {\it a lot}\ 
of catching up to do.  How does it know how to control itself so that it ends up with the ``right'' value of 
$M_\bullet/M_{\star, \rm bulge}$ at $z \simeq 0$?

      It may be a coincidence with a different explanation, but we note that our mergers in progress (Section 6.4), 
while not ULIRG-like starbursts, similarly contain undermassive BHs.

      Curiously, any sign that BH-to-host mass ratios evolve with $z$ weakens or disappears at $z \lesssim 2$~if~we consider not the bulge 
component but the whole of the host galaxy ({\it green points\/}, {\bf Figure 38}, {\it right\/}).  At intermediate redshifts,
the samples shift to moderate-luminosity, Seyfert-like AGNs, which tend to have lower-mass BHs ($M_\bullet \approx 10^7-10^8\,M_\odot$) 
that live in spiral disk galaxies (Section\ts8.6.6).  The vast majority show no signs of morphological peculiarity 
(Cisternas \etal 2011a; 
Schawinski \etal 2011, 2012; 
Kocevski   \etal 2012; 
Treister   \etal 2012; 
B\"ohm     \etal 2013; 
Schramm \& Silverman 2013).  
Internal processes, not mergers, govern the evolution of these galaxies.  If, at these intermediate redshifts, the bulge lies off the 
local correlation but the galaxy as a whole does not, then, over the next $7-10$ Gyr, the galaxy must find a way to redistribute the 
stars from its disk to its bulge.  The obvious possibility is major mergers, but a serious caveat is that present data suggest
that too many mergers are required and that they make too many big bulges and ellipticals.  Alternatively, recent studies frequently 
invoke minor mergers and bulge formation through disk instabilities 
(Jahnke   \etal 2009;
Merloni   \etal 2010;
Bennert   \etal 2011; 
Cisternas \etal 2011b; 
Schramm \& Silverman 2013).  
Here, it is important to note that violent disk instabilities that make $10^8$\ts--\ts$10^9$-$M_\odot$, kpc-size star-forming clumps 
which then sink to the center and merge 
(Elmegreen et{\ts}al.~2005,\ts2007,\ts2008a,\ts2009a,{\ts}b;
Bournaud \etal 2007;
Genzel \etal 2008;
F\"orster Schreiber \etal 2009;
Tacconi{\ts}et{\ts}al.\ts2010)
are not secular processes (Kormendy \& Kennicutt 2004; Kormendy 2012).  Rather, this is a variant of the merger picture.
It makes classical bulges (Elmegreen \etal 2008b).
Connections between these processes and AGN BH growth remain to be explored, though it is reasonable to expect that they more nearly 
resembles major mergers than they do the secular growth of pseudobulges.  Also, more accurate observations -- especially 
to determine mass-to-light rations -- and more rigorous calculations are needed to check whether the results seen in 
{\bf Figure\ts38} are robust and to see what physical processes link high-$z$ progenitors to $z \simeq 0$ descendants.  

      Our inference above that $M_\bullet/M_{\star, \rm bulge}$ evolves with $z$ is based on single-epoch 
BH mass estimates for AGNs zeropointed to $z \simeq 0$ $M_\bullet$--host-galaxy relations published before the present work (e.{\ts}g.,
the {\it dashed line\/} in {\bf Figure 38}).   A major, unanticipated development in this review has been that the zeropoint and, 
to a lesser extent, the slope of the BH-host correlations need to be revised (Section 6.6).  Therefore the zeropoint for AGN $M_\bullet$
estimates must be revised, too.  This work is in progress but has not yet been completed.  If we provisionally use our present
$M_\bullet$\ts--\ts$M_{\rm bulge}$ relation ({\it solid line\/} in the bottom panel of {\bf Figure 38}), then the amount of evolution 
inferred for the $z > 0$ AGN population is significantly reduced.  The $\sim 0.3-0.5$ dex evolution reported above for the $z < 1$ 
objects essentially vanishes, and the offset inferred for the $z \gtrsim 2$ quasars reduces to a comfortable factor of $2-3$.  

      It seems entirely reasonable that BHs in $z \gtrsim 2$ quasars were overmassive by the above factor compared to their
remnants today.  In fact, this is expected.  Our understanding of the growth and assembly histories of galaxies is that
high-mass galaxies at  $z \approx 2-3$ -- i.{\ts}e., candidate quasar hosts when their BHs were deactivated --
roughly double in stellar mass and quadruple in size by $z = 0$ (e.{\ts}g.,
Buitrago   \etal 2008; 
van~Dokkum \etal 2010;
Szomoru    \etal2012).  
If the stellar mass of these systems grows mainly by dry, minor mergers (e.{\ts}g., 
Oser  \etal 2010, 2012; 
Huang \etal 2013b; 
Naab        2013), 
we expect that $M_\bullet$ was more-or-less ``locked in'' at the value that it had at $z \approx 2-3$.  
Then, by the time massive ellipticals arrive at $z = 0$, the factor-of-two difference in $M_\bullet/M_{\star, \rm bulge}$ is erased.  
In support of this scenario, Targett \etal (2012) find that the host galaxies of $z \sim 4$ quasars were not only undermassive but also 
too compact by a factor of $\sim$\ts5 with respect to local ellipticals.  In other words, the hosts of high-$z$ quasars are closely 
related to red nuggets.  And perhaps not coincidentally, to the hosts of BH monsters today (Section 6.5).
 
      How much our recalibration of the $M_\bullet/M_{\rm bulge}$ distribution affects the above conclusions remains to be seen.
Many objects used to calibrate AGN BH masses (Onken \etal 2004; Park \etal 2012) have low-mass BHs that live in pseudobulges.  
For these, $M_\bullet$ revisions are modest.  Stay tuned.

\def\vs{\vskip 4pt}

      While the absolute masses are still unsettled, the relative masses of the AGN BHs and those of their hosts should be 
more secure.  As a summary of the above discussion, the take-away points on what we learn from {\bf Figure 38} are as follows: \vs

\nhii (1) Quasars at $z$ \gapprox \ts2 ({\it dark blue points\/}) populate a reasonably well-defined locus that lies above and parallels the 
          sequence of BHs at $z = 0$.  These quasars generally have $M_\bullet$ \gapprox \ts$ 10^{8.5}\,M_\odot$ and 
          $M_{\star, \rm total}$ \gapprox \ts$10^{10.5}\,M_\odot$. \vs

\nhii (2) The observed distribution of points has a maximum perpendicular spread of \gapprox \ts1 dex.  While the measurements of 
          $M_\bullet$ and $M_{\star,\rm total}$ are both uncertain, it seems likely that the 
          intrinsic scatter for the high-$z$ points is larger than the local value of 0.29 dex (Section\ts6.6). \vs

\nhii (3) If SMGs are any guide, it appears that high-$z$ BHs initially grow more slowly than their galaxies. The BHs quickly 
          ``catch up,'' so that, by the time the dust is cleared away, they achieve $M_\bullet/M_{\star, \rm total}$ ratios
          seen in optically selected quasars.  If quasar-mode feedback moderates star formation and accretion, it presumably happens
          in this context.  The details, however, are murky.  Who is in control? \vs

\nhii (4) Whatever the initial distribution of $M_\bullet/M_{\star,\rm total}$, it evidently settles to a relatively restricted range 
          of values quite early on.  A log-linear trend between $M_\bullet$ and $M_{\star,\rm total}$ is already recognizable in the 
          quasar population between $z \sim 2$ and 6 with no obvious change in scatter across this redshift range ({\bf Figure 38}, 
          {\it lower panel\/}).  To attain their enormous masses by these early times, we know that quasars had to grow by a combination 
          of gas-rich mergers and near-Eddington-limited accretion (e.{\ts}g., Li \etal 2007).  The nascent $M_\bullet- M_{\star,\rm total}$ 
          relation at these high redshifts could not have been established entirely through merger averaging. \vs

\nhii (5) In Section\ts6.7, we suggested that coreless ellipticals are made in wet mergers in which quasar-mode feedback takes place.  
          We now identify these events with the \hbox{high-$z$} SMG population.  SMGs are analogs of nearby ULIRGs, and like ULIRGs,
          most SMGs are gas-rich mergers (e.{\ts}g., Tacconi \etal 2008). These highly dissipative systems naturally produce the high central
          densities and globally disky structure that are characteristic of coreless ellipticals.  Because major, wet 
          mergers are quasi-spherical train wrecks, their gas and dust distributions may achieve a high enough covering factor to 
          effectively couple the AGN radiation field with the interstellar medium of the host galaxy.  This is a generic requirement 
          for any AGN feedback mechanism to work (Silk \& Rees 1998; Fabian 1999; King 2003; Murray, Quataert \& Thompson 2005).  
          Later, the remnants of these early SMGs presumably get converted to core ellipticals by dry mergers. \vs

\nhii (6) Evolution in the $M_\bullet$\ts--\ts$M_{\star, \rm bulge}$ correlation for high-$z$ quasars slows down dramatically at
          $z$\ts\lapprox\ts2 when star formation and AGN activity cease.  AGNs persist in massive, quiescent galaxies 
          (Olsen \etal 2013), but at low levels that do not much grow $M_\bullet$.  By contrast, we believe that high-mass galaxies roughly 
          double their stellar mass from $z \sim 2$ to 0 through a series of dry, mostly minor mergers that are necessary to add their
          low-S\'ersic-index halos (e.{\ts}g., van~Dokkum \etal 2010).  By $z \sim 0$, this presumably lowers the zero point in the
          $M_\bullet - M_{\star, \rm bulge}$ correlation as stellar mass is driven from disks into bulges. \vs

\nhii (7) Given that little $M_\bullet$ growth happens in dry mergers beyond the merging of BH binaries, core ellipticals inherit the
          $M_\bullet$--host correlations from smaller (and ultimately coreless) bulges and ellipticals.  Then merger averaging is essentially
          the only effect that tights the correlations (Section\ts8.5).  Radio-mode AGN feedback protects the galaxies from late star formation. 
          These BHs coast on into the future quite unrelated to the BHs in point (8).\vs

\nhii (8) At  $z \lesssim 1.5$, BHs with $M_\bullet$\ts\lapprox \ts$10^{8.5}\,M_\odot$ are generally hosted by disk galaxies
          ({\it light blue points\/} in {\bf Figure 38}).  The scatter of these points in the $M_\bullet - M_{\star, \rm bulge}$ plane 
          flares up, and they show little correlation.  Ironically, for these objects,
          the total stellar mass of the galaxy correlates better with $M_\bullet$ than the bulge mass ({\it green points\/} in {\bf Figure 38},
          {\it upper right}).  This is not entirely unexpected, in view of the complex mass growth history of spiral galaxies, which is 
          governed by a combination of internal, secular processes and external factors such as gas accretion and minor mergers (e.{\ts}g., 
          Martig \etal 2012; Sales \etal 2012).   Disks settle (Kassin \etal 2012), bars and spiral arms develop (Kraljic, Bournaud \& 
          Martig 2012), and material gradually drains toward the center to build pseudobulges (Kormendy \& Kennicutt 2004).  Some minute 
          fraction of the gas gets accreted by the BH.  This is enough to sustain modest AGN activity and continued BH growth that persists 
          to the present day.  Given the essentially stochastic and local nature of these processes, it would be remarkable if BH accretion and star
          formation had much to do with each other.  We suggest that this accounts for the poor BH-host correlations seen in pseudobulges. \vs

\vfill\eject

\vs
\ni {\big\ARRed 9. CONCLUSIONS}\textBlack
\vs

\vs
\ni {\big\ARRed 9.1 Summary of our Picture of BH and Host Galaxy Coevolution}\textBlack
\vs

      We have reviewed observations which tell us that BH masses correlate tightly only with classical bulges and ellipticals.  
In contrast, they correlate weakly enough with pseudobulges and dark halos to imply no relationship closer than the fact that it is 
easier to grow bigger BHs in bigger galaxies that contain more fuel.~And BHs do not correlate with disk properties at all.~We 
conclude that the physics that engineers tight $M_\bullet$\ts--\ts$\sigma$ and other correlations happens in the context 
of dissipative mergers that build disky-coreless-rotating bulges and elliptical galaxies.  We use these constraints to navigate 
through a variety of sometimes-conflicting theory and observational results to construct a consensus picture of the relationships 
between BHs and host galaxies.  

      This picture is most uncertain at $z \simeq 2$\ts--\ts4 in the quasar era.  Most growth of large BHs happens then 
by radiatively efficient gas accretion.  Any coevolution that engineers the $M_\bullet$\ts--\ts$\sigma$ relation mostly
happens then, too.  Observations of the host galaxies of high-$z$ quasars are fraught~with challenges because of the 
enormous distances of these sources and the bright glare of their AGNs.  Nevertheless, the evidence suggests that, even 
at these early times, the BHs already ``knows about'' the gravitational potential wells of the host galaxies, although the $M_\bullet$
zeropoint of the correlation may be a factor of \hbox{2\ts--\ts3} higher and the scatter may be larger than at $z \sim 0$. 
The progenitors of high-$z$ quasars appear to be gas-rich systems possibly involved in major mergers and possibly related to
submillimeter galaxies.  In light of the substantial stellar masses and star formation rates of submillimeter galaxies and the
evidence for subdominant AGN activity and moderate BH masses, this scenario implies that gas-rich major mergers convert most 
gas into stars before they much build their BHs.  When the BHs reach a critical threshold, ``quasar-mode energy feedback'' 
balances outward radiation or mechanical pressure against gravity.  Then the AGNs and starbursts together limit galaxy growth by 
blowing away (or preventing the accretion of) a larger fraction of baryon mass from bigger galaxies.  This quenches star 
formation and leaves the galaxies red and dead soon after their AGNs become visible.  

      The $z \sim 0$ situation is clearer.  ``Downsizing'' means that most star formation~and~most~AGN~$M_\bullet$ growth 
now happen in relatively low-mass galaxies. By and large, these galaxies lack classical bulges.  NGC 1068 and NGC 4151 are 
prototypical Seyfert (1943) galaxies that contain \hbox{high-mass} pseudobulges (Kormendy 2012).  We suggest that BHs in such galaxies 
are fed by local processes and that the resulting AGNs produce too little energy to affect their hosts.  The AGN activity is episodic but 
intrinsically secular; not much changes over long time periods.  BHs that now live in classical bulges and ellipticals continue
to correlate tightly with their hosts, but the physics that engineered this has mostly stopped happening.  The highest-mass 
BHs at $z \sim 0$ live in giant, boxy-core-nonrotating elliptical galaxies that, at late times, grow mostly by dry mergers.  They 
inherit coevolution magic from smaller galaxies.  AGN feedback is different from that at high~$z$: it is ``maintenance-mode feedback'' 
in which radio jets help to heat hot, X-ray-emitting gas halos in the biggest galaxies and in clusters of galaxies.  This solves 
the ``cooling flow problem'' by keeping baryons locked up in hot gas.  The result is to arrest star formation in the biggest galaxies.  
BH mass growth in this phase is relatively small. 

      At all $z$, major mergers concurrently grow larger BHs and larger elliptical hosts.  The averaging that is inherent in mergers
helps to decrease the scatter in BH-host correlations.  Judging by the observed large scatter of the $M_\bullet-M_{\star, \rm host}$ 
relation for high-$z$ quasars, any viable model for BH-galaxy coevolution must explain how it converges to the tight local relation. 
The relative importance of merger averaging and feedback magic in engineering tight correlations is not known.  Merger averaging probably 
dominates the largest $M_\bullet$ and may be significant at all BH masses.  

      The era in which major mergers help to build BH correlations is already ending.  \hbox{Low-$M_\bullet$}~outliers to the 
$M_\bullet$\ts--{\ts}host-galaxy correlations are\ts--{\ts}we suggest\ts--{\ts}being created at $z \sim 0$~by~mergers that convert 
disk mass into bulge mass without much BH growth beyond the mergers of BH binaries.  Then mergers-in-progress have undermassive BHs if 
their progenitor disks had low-mass BHs.   Major mergers are becoming rarer.  They must largely have stopped happening in clusters,
because these have velocity dispersions that are too large compared to galaxy velocity dispersions.  In sparser environments, major 
mergers can make exceptions to the BH correlations at $z \sim 0$ whereas they help to create the correlations at large $z$ because gas fractions 
in progenitor galaxies are small in the nearby universe but large at large $z$.  ULIRGs that resemble our picture of large-$z$ BH growth 
still occur today, but a growing fraction of major mergers in the nearby universe involve only modest star formation (Schweizer 1990)
and necessarily only modest BH growth.  Thus the coevolution era has largely ended and we have entered a time when the correlations slowly
erode as mergers transform stellar disks into bulges without concurrent BH feeding.   

     In contrast, a few $z \sim 0$ exceptions to $M_\bullet$\ts--\ts$\sigma$ have unusually high BH masses in unusually compact host galaxies. 
These may be the rare remnants of galaxies in which BH growth spurted ahead of galaxy growth at high $z$.  I.{\ts}e., they may be remnants 
of a time when an $M_\bullet$\ts--\ts$\sigma$ relation either was not yet established or was offset from today's relation in the direction
of high BH masses.  They are most likely to survive in rich, dynamically relaxed clusters of galaxies, where remnant gas is too hot and 
galaxies move too rapidly to grow.  In such environments, the rich get richer and the poor get poorer -- giant ellipticals tend to grow
via minor mergers, whereas small galaxies get starved of cold gas infall and whittled by dynamical harrassment.  It seems no accident that
the BH monsters are found in small galaxies that live near the centers of rich clusters

\vs
\ni {\big\ARRed 9.2 Burning Issues Regarding AGN Feedback}\textBlack
\vs

     Remaining tensions involve a recurring theme: Supporting observations{\ts}--{\ts}even ``smoking~gun'' observations such as
relativistic outflows in quasars{\ts}--{\ts}are essentially circumstantial.  We need to develop a more rigorous theoretical understanding 
that coevolution works as we suggest.  Or not.  Burning issues include the following: \vsss

\nhij (1) What processes clean cold gas out of early-type galaxies?  For coreless-disky-rotating elliptical galaxies and classical bulges, 
          we suggest that the fundamental initial cleaning process is an unknown combination of starburst-driven and quasar-mode AGN feedback.  
          Is this correct?  Which is more important?  And what keeps these galaxies clean?
          Is it AGN activity that recurs when gas builds up? \vsss

\nhij (2) For core-boxy-nonrotating ellipticals, we suggest that radio, maintenance-mode feedback is
          responsible for keeping hot gas hot in individual giant ellipticals and in clusters of galaxies.  The microphysics
          of this process is not well understood.  Can radio jet energy be confined within individual galaxies and
          can it be thermalized? \vsss

\nhij (3) A broader issue about X-ray gas is whether AGN feedback is the main energy source~that keeps it hot.~Alternatives include 
           (i) continued infall of gas from the cosmological~web,~which feeds a shock at the boundary to the quasistatic
               halo and keeps hot gas hot
               (Dekel \& Birnboim 2006, 2008), and
           (ii) gas recycled from dying stars; this is injected into~the~gas~halo at the kinetic temperature of stellar
                motions, which, by the virial theorem, is necessarily just right to maintain hot gas temperatures.
           It is not critical for our coevolution picture~to know the relative importance of these (and possibly other)
           heating mechanisms; it is sufficient to observe that the gas is hot enough in all objects observed.
           But the balance between heating mechanisms affects conclusions about AGN duty cycles and about the amount of 
           BH mass growth by gas accretion that happens during the maintenance-mode feedback phase. 

\vs\vs
\ni {\big\ARRed 9.3 Open Issues That We Have Not Reviewed}\textBlack

\vs\vsss
\noindent {\ARRed\bf 9.3.1. How do BHs grow so quickly at high redshifts?}~\textBlack  The highest-redshift~($z$\ts$=$\ts7.085)
                  quasar known already had $M_\bullet \sim (2.0^{+1.5}_{-0.7}) \times 10^9$ $M_\odot$ only 770 million years after the
                  Big Bang (Mortlock \etal 2011).  The BH mass is based on the quasar's luminosity and on its Mg II $\lambda$2798\ts\AA\ line 
                  width.  It is uncertain.  But this is only the latest and most extreme 
                  of a growing number of known giant BHs at early times whose rapid growth, within the (somewhat squishy) 
                  constraint of the Eddington limit, is difficult to understand.  The best bet is that these BHs get a 
                  head~start on radiatively efficient growth by merging many small seed BHs, possibly Population III 
                  remnants.   The point worth making is this: Such objects are so rare that any attempt to find a ``natural'' 
                  explanation is probably wrong.  If the suggested process that makes these objects is not extremely unusual, 
                  it is probably the wrong process.  This subject is reviewed by Volonteri (2010). 

\vs
\noindent {\ARRed\bf 9.3.2 Why do we not see many BH binaries near galaxy centers?}~\textBlack BH binaries formed in galaxy merges shrink
                  in separation by several processes.  At moderate separations, the tendency toward energy equipartition causes binaries
                  to fling stars away, thereby -- we believe -- excavating cores.  At small separations, they emit gravitational 
                  radiation.  In between, there can be a bottleneck at separations of $\sim 1$ pc where decay processes are slow.  
                  This ``final parsec problem'' is discussed or reviewed in
                  Begelman, Blandford, \& Rees (1980);
                  Yu (2002);
                  Milosavljevi\'c \& Merritt (2003);
                  Makino \& Funato (2004), and
                  Merritt \& Milosavljevi\'c (2005).
                  The subject is complicated; we have neither the space nor the expertise to review it.  Komossa (2006) reviews the
                  observations.  The ``bottom line'' is that
                  BH binaries with separations $\sim$ 1 pc are surprisingly rare, especially in big classical bulges and
                  ellipticals.  Additional decay processes are discussed in the above papers.  We bring
                  this subject up because it leads to interesting expectations:

\vs
\noindent {\ARRed\bf 9.3.3 Why do we not see BHs that are not at galaxy centers?}~\textBlack If a second merger supplies a third BH to a 
                  BH binary, the resulting three-BH interactions generally fling all BHs away from the center.  Even if the most massive 
                  BHs make a binary that ejects the third BH, the binary recoils.  This leads to expectations that we have not observed: 
                  Where are the BH-less bulges and ellipticals?  And where are the free-flying BHs and their very compact cloaks of stars? 

          An important corollary is this: BHs that have been evicted from the centers of their
          galaxies by multiple-BH interactions should decay back to the center by dynamical friction, but
          not instantly.  So we expect that some BHs are located near but not at the centers
          of their host galaxies.  Except in M{\ts}31 (Kormendy \& Bender 1999), it is not realistic to expect that we see this in 
          galaxies with dynamical BH detections.  But radio observations can measure AGN positions very accurately.  An astrometric 
          survey to look for offcenter AGNs could pay interesting dividends.~The~problem~is hard,
          because radio and optical observations are made at different times, and the stars whose astrometry 
          is needed to tie radio frames of reference to optical ones have poorly known proper motions. 

An even more important corollary involves observations of early-type galaxies at \hbox{$z$\ts$\sim$\ts$2\pm1$} that reveal high-mass
red~nuggets that are
           factors of $\sim$\ts4 smaller than similar-mass ellipticals at $z$\ts$\sim$\ts0 
(e.{\ts}g.,~Daddi~et~al.~2005;
           Trujillo \etal 2006,~2007;
           van Dokkum \etal 2008;
           Buitrago \etal 2008;~Damjanov~et~al.~2009;
           Szomoru \etal 2012).
           They are sometimes also described as denser and higher in velocity dispersion than $z$\ts$\sim$\ts 0 ellipticals, 
           but this is misleading.  These quantities look high when measured at or averaged inside small half-light radii $r_e$.  But 
           their central brightness profiles are very similar to those of nearby giant ellipticals when corrected for cosmological
           dimming and stellar population age (Hopkins{\ts}et{\ts}al.{\ts}2009;
           Bezanson{\ts}et{\ts}al.{\ts}2009;
           van Dokkum{\ts}et{\ts}al.{\ts}2010).
           So puffing up is not needed.  Instead, giant ellipticals at $z \sim 0$ and $z \sim 2$ are similar near their centers; 
           they differ at large radii because nearby core-boxy-nonrotating~ellipticals have \hbox{high-S\'ersic-index} outer halos (KFCB)
           that are not present in their high-$z$ counterparts.  These red nuggets are usually interpreted as the
           ancestors of today's core-boxy-nonrotating ellipticals.  The favored evolution
           scenario is inside-out growth in which outer halos are added onto already-formed centers mostly by minor mergers
          (Khochfar \& Silk 2006;
           Naab \etal 2009;
           Hopkins \etal 2010;
           Feldmann \etal 2010;
           van Dokkum \etal 2010; 
           Oser \etal 2010, 2012;
           Hilz, Naab \& Ostriker 2013; see
           Naab 2013 for a review of this subject). 

How does this story affect our picture of BH{\ts}--{\ts}host coevolution?  The whole answer is not known.
           However, one suggestion follows immediately from looking at typical merger trees.  In most minor mergers, the
           small galaxies are not vanishingly small.  Many should bring their own BHs -- probably not ones that coevolved,
           but nevertheless ones with $M_\bullet$ \gapprox \ts$10^5$ $M_\odot$~-- to the burgeoning large galaxy.  Cold gas
           is not available to speed mergers, and dynamical friction is slower for smaller objects.  Many minor mergers occur.
           We suggest that the result could often be a swarm of smallish but still supermassive BHs surrounding the central 
           giant BH.  They should reveal themselves as a cluster of AGNs during phases when the central BH is most active.  
           The devil is in the details.  But it is worth considering whether sensitive radio searches should already have
           found AGN clusters in the many radio galaxies that have been surveyed.  This may constrain the histories of minor
           mergers.

      Finally, it seems inevitable that some BHs with masses $M_\bullet \sim 10^4$ to $10^7$ $M_\odot$ in bulgeless galaxies
will be accreted by much bigger spirals.  For example, imagine that a giant galaxy like M{\ts}31 inhales a fluffy dwarf like
NGC\ts4395 ({\bf Figure\ts31}).  Its stars and gas will be deposited at large radii in the bigger galaxy.  After that, the BH, 
probably still cloaked in its nucleus, should fly free in the halo potential.  Dynamical friction is slow.  This suggests that 
there should be low-level AGNs in the disks and (more often) the halos of galaxies like M{\ts}31 and our own.  Can we find them?

\vs\vsss
\ni {\big\ARRed 9.4 Conclusion}\textBlack
\vs

      Our inward journey to observe closer to the Schwarzschild radius and to observe relativistic effects such as 
BH spin is progressing rapidly.  We are optimistic that spectacular advances are just around the corner.  Our optimism is also 
based on a concurrent outward 
journey -- the increasingly broad and convincing connections between the $M_\bullet$ demographics that are the subjects 
of this paper and a variety of aspects of galaxy physics.  Robust connections between disparate research fields are a sign of the 
developing maturity of this subject.  The observation that BH masses correlate differently with different galaxy components increases 
the richness of this subject and opens the door to a more nuanced and reliable understanding of BH--host-galaxy coevolution.

\vs\vsss
\ni {\big\ARRed DISCLOSURE STATEMENT}\textBlack
\vs

The authors are not aware of any affiliations, memberships, funding, or financial holdings that 
might be perceived as affecting the objectivity of this review.  Indeed, they are conscious of a
deplorable lack of financial holdings that might be perceived as affecting anything at all.

\vs\vsss
\ni{\big\ARRed ACKNOWLEDGEMENTS}\textBlack
\vs

      Scott Tremaine has been associated with this paper since the beginning, and we have
benefited enormously from his advice and constructive criticism.  His comments on most sections 
of this paper have been invaluable and have influenced both the science and its presentation.
We do not mean to imply that he agrees with all of our conclusions; any results that prove to be 
wrong are, of course, our responsibility.  We sincerely thank Scott for all his help. 

      We acknowledge with pleasure our many years of productive collaboration with the 
Nuker~team (Sandy Faber and Doug Richstone, past and present PIs).  Some of our ideas
on BHs and BH-galaxy coevolution matured in the intense and enjoyable environment of this
collaboration.  For conversations or emails that were specifically helpful to this paper, we thank
Ralf Bender,
Andrew Benson,
Andi Burkert,
Michele Cappellari,
Ken Freeman,
Reinhard Genzel,
Kayhan G\"ultekin,
John Mulchaey,
Doug Richstone, and
Mark Sarzi.~We are also grateful to Ralf Bender for letting~us use his symmetric least-squares fitting code
and the sm macro that makes 1$\sigma$ shading around plots of those fits.
And JK thanks
Martin Gaskell
for helpful conversations and comments on AGNs.

This paper would be much weaker without the generosity of people who provided 
data before publication.~We are most sincerely grateful to Karl Gebhardt, 
Stephanie Rusli, and Roberto Saglia for allowing us to use their BH 
discoveries and $M_\bullet$ measurements.  We thank Michele Cappellari for 
providing ATLAS3D $\sigma_e$ measurements of some of our BH host galaxies.
The following colleagues provided figures or data that contributed to some of the 
figures: James Aird, Reinhard Genzel, Knud Jahnke, Yan-Fei Jiang, Minjin Kim, Andrea Macci\`o, 
and Ting Xiao.

      It is a particular pleasure to thank our Scientific Editor, Sandy Faber, for her thorough reading of the paper and
for detailed comments that led to significant improvements both in science and in presentation.  We thank Production
Editor Roselyn Lowe-Webb and the production team at Annual Reviews for their masterful handling of a difficult paper.
Also, we note that~we~have~been greatly influenced by {\bf http://www.annualreviews.org/page/infooverload},
an Annual Reviews White paper about ``The Role of the Critical Review Article in Alleviating Information Overload''.
We~recommend this paper to all ARA\&A authors.

      JK warmly thanks Mary Kormendy for her editorial help and for her support and understanding during the 
many years when this paper dominated his thinking and controlled our schedule.  

      This work would not have been practical without extensive use of NASA's Astrophysics 
Data System bibliographic services and the NASA/IPAC Extragalactic Database (NED).
NED is operated by the Jet Propulsion Laboratory and the California Institute of Technology 
under contract with NASA.  We also used the HyperLeda electronic database (Paturel \etal 2003)
at {\bf http://leda.univ-lyon1.fr} and the WIKISKY image database ({\bf www.wikisky.org}).
And we made extensive use of images from the {\it Hubble Space Telescope\/} archive 
and from the Two Micron All Sky Survey (2MASS: Skrutskie \etal 2006).  2MASS is a joint project
of the University of Massachusetts and the Infrared Processing and Analysis Center/California 
Institute of Technology, funded by NASA and the NSF.  

      JK's research was supported in part by NSF grant AST-0607490.  Also, this multi-year 
research project would not have been possible without the long-term support provided to JK 
by the Curtis T.~Vaughan, Jr.~Centennial Chair in Astronomy.  We are most sincerely grateful to
Mr.~and Mrs.~Curtis T.~Vaughan, Jr.~for their continuing support of Texas 
astronomy.  LCH's work is supported by the Carnegie Institution for Science and 
by NASA grants from the Space Telescope Science Institute (operated by AURA, 
Inc., under NASA contract NAS5-26555).   LCH thanks the Chinese Academy of
Sciences and the hospitality of the National Astronomical Observatories, where
part of this review was written.

\hsize=15.0truecm  \hoffset=0.0truecm  \vsize=20.5truecm  \voffset=1.5truecm

\vs\vsss
\ni {\big\ARRed LITERATURE CITED}\textBlack
\vs

\def\nhi{\noindent \hangindent=1.0truecm}

{\frenchspacing \ninerm \refbaselines

\nhi Abadi, M. G., Navarro, J. F., Steinmetz, M., \& Eke, V. R. 2003, ApJ, 591, 499

\nhi Ai, Y. L., Yuan, W., Zhou, H. Y., Wang, T. G., \& Zhang, S. H. 2011, ApJ, 727, 31

\nhi Aird, J., Coil, A. L., Moustakas, J., \etal 2012, ApJ, 746, 90

\nhi Aird, J., Nandra, K., Laird, E. S., \etal 2010, MNRAS, 401, 2531

\nhi Alexander, D.~M., Brandt, W.~N., Smail, I., \etal  2008, AJ, 135, 1968

\nhi Alexander, D. M., \& Hickox, R. C. 2012, NewAR, 56(4), 93

\nhi Aller, M. C., \& Richstone, D. O. 2007, ApJ, 665, 120                                             

\nhi Argon,{\ts}A.{\ts}L., Greenhill,{\ts}L.{\ts}J., Reid,{\ts}M.{\ts}J., Moran,{\ts}J.{\ts}M., \& Humphreys,{\ts}E.{\ts}M.{\ts}L.\ts2007, ApJ, 659, 1040

\nhi Arkhipova, V. P., \& Saveleva, M. V. 1984, Trudy Gosudarstvennogo Astronomicheskogo Instituta P. K. Sternberga, 54, 33

\nhi Armitage, P. J., \& Natarajan, P. 2002, ApJ, 567, L9

\nhi Armitage, P. J., \& Natarajan, P. 2005, ApJ, 634, 921

\nhi Ashman, K. M., \& Zepf, S. E. 1992, ApJ, 384, 50  

\nhi Atkinson, J. W., Collett, J. L., Marconi, A., \etal 2005, MNRAS, 359, 504

\nhi Bacon, R., Emsellem, E., Combes, F., \etal 2001, A\&A, 371, 409

\nhi Bacon, R., Emsellem, E., Monnet, G., \& Nieto, J.-L. 1994, A\&A, 281, 691 

\nhi Baes, M., Buyle, P., Hau, G. K. T., \& Dejonghe, H. 2003, MNRAS, 341, L44

\nhi Bahcall, J. N., Kirhakos, S., Saxe, D. H., \& Schneider, D. P. 1997, ApJ, 479, 642  

\nhi Bahcall, J. N., \& Wolf, R. A. 1976, ApJ, 209, 214

\nhi Bajaja, E., van der Burg, G., Faber, S. M., \etal 1984, A\&A, 141, 309

\nhi Baldry, I. K., Glazebrook, K., Brinkmann, J., \etal 2004, ApJ, 600, 681

\nhi Barbosa, F.~K.~B., Storchi-Bergmann, T., Cid Fernandes, R., Winge, C., \& Schmitt, J.~2006, MNRAS, 371, 170

\nhi Barnes, J. E. 1989, Nature, 338, 123

\nhi Barnes, J. E. 1992, ApJ, 393, 484                                          

\nhi Barnes, J. E. 1998, in 26$^{\rm th}$ Advanced Course of the Swiss Society of Astronomy and Astrophysics, Galaxies: Interactions and Induced Star Formation, ed. D. Friedli, L. Martinet, \& D. Pfenniger (New York: Springer), 275

\nhi Barnes, J. E., \& Hernquist, L. 1992, ARA\&A, 30, 705       

\nhi Barth, A. J., Greene, J. E., \& Ho, L. C. 2005, ApJ, 619, L151

\nhi Barth, A. J., Greene, J. E., \& Ho, L. C. 2008, AJ,  136, 1179

\nhi Barth, A. J., Ho, L. C., Rutledge, R. E., \& Sargent, W. L. W. 2004, ApJ, 607, 90                   

\nhi Barth, A. J., Sarzi, M., Rix, H.-W., \etal 2001, ApJ, 555, 685  

\nhi Barth, A. J., Strigari, L. E., Bentz, M. C., Greene, J. E., \& Ho, L. C. 2009, ApJ, 690, 1031

\nhi Bastian, N., Covey, K. R., \& Meyer, M. R. 2010, ARA\&A, 48, 339

\nhi Batcheldor, D., Axon, D., Merritt, D., \etal 2005, ApJS, 160, 76

\nhi Baumgardt, H., Makino, J., Hut, P., McMillan, S., \& Portegies Zwart, S. 2003, ApJ, 589, L25                                 

\nhi Beckers, J. M. 1993, ARA\&A, 31, 13

\nhi Begelman, M.~C. 2004, in Carnegie Observatories Astrophysics Series, Vol. 1: Coevolution of Black Holes and Galaxies, ed. L. C. Ho (Cambridge: Cambridge Univ. Press), 375

\nhi Begelman, M. C., Blandford, R. D., \& Rees, M. J. 1980, Nature, 287, 307  

\nhi Begelman, M. C., Blandford, R. D., \& Rees, M. J. 1984, Rev. Mod. Phys., 56, 255

\nhi Behroozi, P. S., Conroy, C., \& Wechsler, R. H. 2010, ApJ, 717, 379

\nhi Behroozi, P. S., Wechsler, R. H., \& Conroy, C. 2012, arXiv:1207.6105

\nhi Beifiori, A., Courteau, S., Corsini, E. M., \& Zhu Y. 2012, MNRAS, 419, 2497

\nhi Beifiori, A., Sarzi, M., Corsini, E. M., \etal 2009, ApJ, 692, 856

\nhi Bell, E. F., \& de Jong, R. S. 2001, ApJ, 550, 212

\nhi Bell E. F., McIntosh D. H., Katz N., \& Weinberg M. D., 2003, ApJ, 585, L117

\nhi Bender, R. 1996, in IAU Symposium 171, New Light on Galaxy Formation, ed. R. Bender \& R. L. Davies (Dordrecht: Kluwer), 181

\nhi Bender, R., Burstein, D., \& Faber, S. M. 1992, ApJ, 399, 462

\nhi Bender, R., Burstein, D., \& Faber, S. M. 1993, ApJ, 411, 153

\nhi Bender, R., Kormendy, J., \etal 2013, in preparation

\nhi Bender, R., Kormendy, J., Bower, G., \etal 2005, ApJ, 631, 280

\nhi Bender, R., Kormendy, J., \& Dehnen, W. 1996, ApJ, 464, L123   

\nhi Bender, R., Pierce, M. J., Tully, R. B., \& Kormendy, J. 2008, unpublished

\nhi Bender, R., Surma, P., D\"obereiner, S., M\"ollenhoff, C., \& Madejsky, R. 1989, A\&A, 217, 35

\nhi Benedict, G. F. 1976, AJ, 81, 799

\nhi Bennert, N., Schulz, H., \& Henkel, C. 2004, A\&A, 419, 127

\nhi Bennert, V. N., Auger, M. W., Treu, T., Woo, J.-H., \& Malkan, M. A. 2011, ApJ, 742, 107

\nhi Bennert, V. N., Treu, T., Woo, J.-H., \etal 2010, ApJ, 708, 1507

\nhi Bessell, M. S. 2005, ARA\&A, 43, 293

\nhi Best, P. N. 2006, Paper Presented at the Workshop on The Role of Black Holes in Galaxy Formation and Evolution, 
     Potsdam, Germany, 2006 September 10{\ts}--{\ts}13 (see Cattaneo \etal 2009)

\nhi Bezanson, R., van Dokkum, P. G., Tal, T., \etal 2009, ApJ, 697, 1290

\nhi Binggeli, B., \& Cameron, L. M.~1991, A\&A, 252, 27

\nhi Binggeli, B., Sandage, A., \& Tammann, G. A.~1985, AJ, 90, 1681

\nhi Binggeli, B., Sandage, A., \& Tammann, G. A.~1988, ARA\&A, 26, 509

\nhi Binggeli, B., Sandage, A., \& Tarenghi, M.~1984, AJ, 89, 64

\nhi Binggeli, B., Tammann, G. A., \& Sandage, A.~1987, AJ, 94, 251

\nhi Binney, J. 1978a, Comments Astrophys., 8, 27

\nhi Binney, J. 1978b, MNRAS, 183, 501

\nhi Binney, J., \& de Vaucouleurs, G. 1981, MNRAS, 194, 679

\nhi Binney, J., \& Mamon, G. A.~1982, MNRAS, 200, 361

\nhi Binney, J., \& Tabor, G. 1995, MNRAS, 276, 663

\nhi Binney, J., \& Tremaine, S. 1987, Galactic Dynamics (Princeton:~Princeton~Univ.~Press)

\nhi Binney,{\ts}J., \& Tremaine,{\ts}S.{\ts}2008, Galactic Dynamics, 2nd Edition (Princeton:{\ts}Princeton{\ts}Univ.{\ts}Press)

\nhi Blakeslee, J. P., Cantiello, M., Mei, S., \etal 2010, ApJ, 724, 657

\nhi Blakeslee, J. P., Jord\'an, A., Mei, S., \etal 2009, ApJ, 694, 556

\nhi Blitz, L., \& Spergel, D. N. 1991, ApJ, 379, 631

\nhi Bogd\'an, A., David, L. P., Jones, C., Forman, W. R., \& Kraft, R. P.  2012, ApJ, 758, 65

\nhi Bogd\'an, A., Forman, W. R., Vogelsberger, M., \etal 2013, ApJ, in press (arXiv:1212.0541)

\nhi Bogdanov, S., Grindlay, J. E., Heinke, C. O., Camilo, F., Freire, P. C. C., \& Becker, W. 2006, ApJ, 646, 1104

\nhi B\"ohm, A., Wisotzki, L., Bell, E. F., \etal 2013, A\&A, 549, A46

\nhi B\"oker, T. 2010, in IAU Symposium 266, Star Clusters: Basic Galaxy Building Blocks, ed. R. de Grijs, \& J. R. D. L\'epine (Cambridge: Cambridge Univ. Press), 58

\nhi B\"oker, T., Laine, S., van der Marel, R. P., \etal 2002, AJ, 123, 1389

\nhi B\"oker, T., Sarzi, M., McLaughlin, D. E., \etal 2004, AJ, 127, 105

\nhi B\"oker, T., van der Marel, R. P., \& Vacca, W. D.~1999, AJ, 118, 831

\nhi Bond, J. R., Arnett, W. D., \& Carr, B. J. 1984, ApJ, 280, 825  

\nhi Bongiorno, A., Merloni, A., Brusa, M., \etal 2012, MNRAS, 427, 3103

\nhi Booth, C. M., \& Schaye, J. 2010, MNRAS, 405, L1

\nhi Borys, C., Smail, I., Chapman, S.~C., \etal 2005, ApJ, 635, 853

\nhi Bosma, A. 1981, AJ, 86, 1825   

\nhi Bournaud, F., Elmegreen, B. G., \& Elmegreen, D. M. 2007, ApJ, 670, 237

\nhi Bouwens, R.~J., llingworth, G. D., Oesch, P. A., \etal 2012, ApJ, 754, 83

\nhi Bower, G. A., Green, R. F., Bender, R., \etal 2001, ApJ, 550, 75    

\nhi Bower, G. A., Green, R. F., Danks, A., \etal  1998, ApJ, 492, L111  

\nhi Bower, G. A., Heckman, T. M., Wilson, A. S., \& Richstone, D. O. 1997, ApJ, 483, L33   

\nhi Bower, G. A., Wilson, A. S., Heckman, T. M., \etal 2000, BAAS, 32, 1566  

\nhi Bower, R. G., Benson, A. J., Malbon, R., \etal 2006, MNRAS, 370, 645

\nhi Boylan-Kolchin, M., Ma, C.-P., \& Quataert, E.~2004, ApJ, 613, L37

\nhi Boylan-Kolchin, M., Ma, C.-P., \& Quataert, E.~2005, MNRAS, 362, 1184                          

\nhi Boylan-Kolchin, M., Ma, C.-P., \& Quataert, E.~2006, MNRAS, 369, 1081                          

\nhi Braatz, J. A., Reid, M. J., Humphreys, E. M. L, \etal 2010, ApJ, 718, 657    

\nhi Brodie, J. P., \& Strader, J. 2006, ARA\&A, 44, 193

\nhi Br\"uggen, M., \& Scannapieco, E. 2009, MNRAS, 398, 548

\nhi Buitrago, F., Trujillo, I., Conselice, C. J., \etal 2008, ApJ, 687, L61

\nhi Burkert, A., \& Tremaine, S. 2010, ApJ, 720, 516

\nhi Burkhead, M. S. 1986, AJ, 91, 777

\nhi Buta, R., Corwin, H. G., \& Odewahn, S. C. 2007, The de Vaucouleurs Atlas of Galaxies (Cambridge: Cambridge Univ.~Press)

\nhi Camilo, F., Lorimer, D. R., Freire, P., Lyne, A. G., \& Manchester, R. N. 2000, ApJ, 535, 975

\nhi Canalizo, G., Wold, M., Hiner, K. D., \etal 2012, ApJ, 760, 38

\nhi Cano-D\'\i az, M., Maiolino, R., Marconi, A., \etal 2012, A\&A, 537, L8

\nhi Caon, N., Capaccioli, M., \& D'Onofrio, M. 1994, A\&AS, 106, 199

\nhi Cappellari, M., Bacon, R., Bureau, M., \etal 2006, MNRAS, 366, 1126                         

\nhi Cappellari, M., Bacon, R., Davies, R. L., \etal 2008, in IAU Symposium 245,  Formation and Evolution of Galaxy Bulges,
     ed. M. Bureau, E. Athanassoula, \& B. Barbuy (Cambridge: Cambridge Univ.~Press), 215

\nhi Cappellari, M., Emsellem, E., Bacon, R., \etal 2007, MNRAS, 379, 418                        

\nhi Cappellari, M., Emsellem, E., Krajnovi\'c, D., \etal 2011, MNRAS, 416, 1680                 

\nhi Cappellari, M., Neumayer, N., Reunanen, J., \etal 2009, MNRAS, 394, 660                     

\nhi Cappellari, M., Scott, N., Alatalo, K., \etal 2013, MNRAS, in press (arXiv:1208.3522)       

\nhi Cappellari, M., Verolme, E. K., van der Marel, R. P., \etal 2002, ApJ, 578, 787             

\nhi Carollo, C. M. 1999, ApJ, 523, 566

\nhi Carpenter, J. M. 2001, AJ, 121, 2851

\nhi Carrera, F.~J., Page, M. J., Stevens, J. A., \etal 2011, MNRAS, 413, 2791

\nhi Cattaneo, A., Dekel, A., Devriendt, J., Guiderdoni, B., \& Blaizot, J. 2006, MNRAS, 370, 1651

\nhi Cattaneo, A., Dekel, A., Faber, S. M., \& Guiderdoni, B. 2008, MNRAS, 389, 567

\nhi Cattaneo, A., Faber, S. M., Binney, J., \etal 2009, Nature, 460, 213

\nhi Cattaneo, A., Mamon, G. A., Warnick, K., \& Knebe, A. 2011, A\&A, 533, A5

\nhi Cecil, G., Greenhill, L. J., De Pree, C. G., \etal 2000, ApJ, 536, 675

\nhi Cecil, G., Wilson, A. S., \& De Pree, C. 1995, ApJ, 440, 181

\nhi Centrella, J., Baker, J. G., Kelly, B. J., \& van Meter, J. R.~2010, Rev. Mod. Phys., 82, 3069

\nhi Chakrabarty, D., \& Saha, P. 2001, AJ, 122, 232  

\nhi Chemin, L., Carignan, C., \& Foster, T. 2009, ApJ, 705, 1395

\nhi Cherepashchuk, A. M., Afanas'ev, V. L., Zasov, A. V., \& Katkov, I. Yu. 2010, Astr. Reports, 54, 578

\nhi Cho, J., Sharples, R. M., Blakeslee, J. P., \etal 2012, MNRAS, 422, 3591

\nhi Cid Fernandes, R., Heckman, T. M., Schmitt, H. R., \etal 2001, ApJ, 558, 81

\nhi Ciotti, L., \& Ostriker, J. P. 2007, ApJ, 665, 1038

\nhi Ciotti, L., Ostriker, J. P., \& Proga, D. 2010, ApJ, 718, 708

\nhi Cisternas, M., Jahnke, K., Bongiorno, A., \etal 2011a, ApJ, 741, L11

\nhi Cisternas, M., Jahnke, K., Inskip, K. J.,  \etal 2011b, ApJ, 726, 57

\nhi Coccato, L., Sarzi, M., Pizzella, A., \etal  2006, MNRAS, 366, 1050

\nhi Combes, F., \& Sanders, R. H. 1981, A\&A, 96, 164

\nhi Conroy, C., \& van Dokkum, P. G. 2012, ApJ, 760, 71

\nhi Cooke, A. J., Baldwin, J. A., Ferland, G. J., Netzer, H., \& Wilson, A. S. 2000, ApJS, 129, 517  

\nhi Coppin, K. E. K., Swinbank, A. M., Neri, A., \etal 2008, MNRAS, 389, 45

\nhi C\^ot\'e, P., Ferrarese, L., Jord\'an, A., \etal 2007, ApJ, 671, 1456    

\nhi C\^ot\'e, P., McLaughlin, D. E., Hanes, D. A., \etal 2001, ApJ, 559, 828     

\nhi C\^ot\'e, P., Piatek, S., Ferrarese, L., \etal 2006, ApJS, 165, 57     

\nhi Courteau, S., Widrow, L. M., McDonald, M., \etal 2011, ApJ, 739, 20

\nhi Cox, T. J., Jonsson, P., Primack, J. R., \& Somerville, R. S. 2006, MNRAS, 373, 1013

\nhi Crane, P., Stiavelli, M., King, I. R., \etal 1993, AJ, 106, 1371

\nhi Cretton, N., de Zeeuw, P. T., van der Marel, R. P., \& Rix, H.-W. 1999a, ApJS, 124, 383   

\nhi Cretton, N., \& van den Bosch, F. C. 1999b, ApJ, 514, 704   

\nhi Croton, D. J., Springel, V., White, S. D. M., \etal 2006, MNRAS, 365, 11  

\nhi Daddi, E., Renzini, A., Pirzkal, N.,\etal 2005, ApJ, 626, 680

\nhi Dalla Bont\`a, E., Ferrarese, L., Corsini, E. M., \etal 2009, ApJ, 690, 537

\nhi Dalla Vecchia, C., Bower, R. G., Theuns, T., \etal 2004, MNRAS, 355, 995

\nhi D'Amico, N., Lyne, A. G., Manchester, R. N., Possenti, A., \& Camilo, F. 2001, ApJ, 548, L171

\nhi Damjanov, I., McCarthy, P. J., Abraham, R. G., \etal 2009, ApJ, 695, 101

\nhi Dasyra, K. M., Tacconi, L. J., Davies, R. I., \etal 2006a, ApJ, 638, 745

\nhi Dasyra, K. M., Tacconi, L. J., Davies, R. I., \etal 2006b, ApJ, 651, 835

\nhi Dasyra, K. M., Tacconi, L. J., Davies, R. I., \etal 2006c, New Astron. Rev., 50, 720

\nhi Dav\'e, R., Cen, R., Ostriker, J. P., \etal 2001, ApJ, 552, 473

\nhi Davies, R., \& Kasper, M. 2012, ARA\&A, 50, 305  

\nhi Davies, R. I., Thomas, J., Genzel, R., \etal 2006, ApJ, 646, 754

\nhi Davies, R. L., Efstathiou, G., Fall, S. M., Illingworth, G., \& Schechter, P. L. 1983, ApJ, 266, 41

\nhi Davis, T. A., Bureau, M., Cappellari, M., Sarzi, M., \& Blitz, L. 2013, Nature, 494, 328

\nhi Decarli, R., Falomo, R., Treves, A., \etal 2010, MNRAS, 402, 2453

\nhi De Francesco, G., Capetti, A., \& Marconi A. 2006, A\&A, 460, 439  

\nhi De Francesco, G., Capetti, A., \& Marconi A. 2008, A\&A, 479, 355  

\nhi Dehnen, W. 1993, MNRAS, 265, 250

\nhi Dehnen, W. 1995, MNRAS, 274, 919  

\nhi Dekel, A., \& Birnboim, Y. 2006, MNRAS, 368, 2

\nhi Dekel, A., \& Birnboim, Y. 2008, MNRAS, 383, 119

\nhi Dekel, A., \& Silk, J. 1986, ApJ, 303, 39

\nhi De Rijcke, S., Prugniel, P., Simien, F., \& Dejonghe, H. 2006, MNRAS, 369, 1321

\nhi Desroches, L.-B., \& Ho, L. C. 2009, ApJ, 690, 267

\nhi Desroches, L.-B., Greene, J. E., \& Ho, L. C. 2009, ApJ, 698, 1515

\nhi de Vaucouleurs, G. 1948, Ann.~Astrophys., 11, 247

\nhi de Vaucouleurs, G. 1961, ApJS, 6, 213

\nhi de Vaucouleurs, G., de Vaucouleurs, A., Corwin, H. G., \etal 1991, Third Reference Catalogue of Bright Galaxies (Berlin: Springer) (RC3)

\nhi Devereux, N., Ford, H., Tsvetanov, Z., \& Jacoby, G. 2003, AJ, 125, 1226  

\nhi Diamond-Stanic, A. M., Moustakas, J., Tremonti, C. A., \etal 2012, ApJ, 755, L26

\nhi Diehl, S., \& Statler, T. S. 2008a, ApJ, 680, 897

\nhi Diehl, S., \& Statler, T. S. 2008b, ApJ, 687, 986

\nhi Djorgovski, S. 1992, in Morphological and Physical Classification of Galaxies, ed. G. Longo, M. Capaccioli, \& G. Busarello (Dordrecht: Kluwer), 337

\nhi Djorgovski, S., \& Davis, M. 1987, ApJ, 313, 59

\nhi Djorgovski, S., de Carvalho, R., \& Han, M.-S. 1988, in The Extragalactic Distance Scale, ed. S. van den Bergh \& C. J. Pritchet (San Francisco: ASP), 329

\nhi Dong, R., Greene, J. E., \&  Ho, L. C. 2012, ApJ, 761, 73

\nhi Dong, X.-B., Ho, L. C., Yuan, W., \etal 2012, ApJ, 755, 167

\nhi Doyon, R., Wells, M., Wright, G. S., Joseph, R. D., Nadeau, D., \& James, P. A. 1994, ApJ, 437, L23

\nhi Dressler, A. 1989, in IAU Symposium 134, Active Galactic Nuclei, ed. D. E. Osterbrock \& J. S. Miller (Dordrecht: Kluwer), 217

\nhi Dressler, A., Lynden-Bell, D., Burstein, D., \etal 1987, ApJ, 313, 42

\nhi Dressler, A., \& Richstone, D. O. 1988, ApJ, 324, 701

\nhi Dressler, A., \& Richstone, D. O. 1990, ApJ, 348, 120

\nhi Dressler, A., \& Sandage, A. 1983, ApJ, 265, 664

\nhi D'Souza, R., \& Rix, H.-W.~2013, MNRAS, 429, 1887

\nhi Duncan, M. J., \& Wheeler, J. C.~1980, ApJ, 237, L27

\nhi Dutton, A. A., Conroy, C., van den Bosch, F. C., Prada, F., \& More, S. 2010, MNRAS, 407, 2  

\nhi Dutton, A. A., Conroy, C., van den Bosch, F. C., \etal 2011, MNRAS, 416, 322

\nhi Dwek, E., Arendt, R. G., Hauser, M. G., \etal 1995, ApJ, 445, 716

\nhi Ebisuzaki, T., Makino, J., \& Okamura, S. K. 1991, Nature, 354, 212

\nhi Eckart, A., \& Genzel, R. 1997, MNRAS, 284, 576  

\nhi Edri, H., Rafter, S. E., Chelouche, D., Kaspi, S., \& Behar, E. 2012, ApJ, 756, 73

\nhi Ellis, S. C., \& O'Sullivan, E. 2006, MNRAS, 367, 627

\nhi Elmegreen, B. G., Bournaud, F., \& Elmegreen, D. M. 2008a, ApJ, 684, 829                      

\nhi Elmegreen, B. G., Bournaud, F., \& Elmegreen, D. M. 2008b, ApJ, 688, 67                       

\nhi Elmegreen, B. G., \& Elmegreen, D. M. 2005, ApJ, 627, 632                                     

\nhi Elmegreen, B. G., Elmegreen, D. M, Fernandez, M. X., \& Lemonias, J. J. 2009a, ApJ, 692, 12   

\nhi Elmegreen, D. M., Elmegreen, B. G, Marcus, M. T., \etal 2009b, ApJ, 701, 306                  

\nhi Elmegreen, D. M., Elmegreen, B. G, Ravindranath, S., \& Coe, D. A. 2007, ApJ, 658, 763        

\nhi Emsellem, E., Bacon, R., \& Monnet, G. 1995, in IAU Colloquium 149, Tridimensional Optical Spectroscopic Methods in Astrophysics, 
     ed. G. Comte \& M. Marcelin (San Francisco: ASP), 282

\nhi Emsellem, E., Cappellari, M., Krajnovi\'c, D., \etal 2007, MNRAS, 379, 401  

\nhi Emsellem, E., Cappellari, M., Krajnovi\'c, D., \etal 2011, MNRAS, 414, 888  

\nhi Emsellem, E., Cappellari, M., Peletier, R. F., \etal 2004, MNRAS, 352, 721

\nhi Emsellem, E., \& Combes, F. 1997, A\&A, 323, 674 

\nhi Emsellem, E., Dejonghe, H., \& Bacon, R. 1999, MNRAS, 303, 495   

\nhi Emsellem, E., Monnet, G., Bacon, R., \& Nieto, J.-L. 1994, A\&A, 285, 739  

\nhi Erwin, P., Graham, A. W., \& Caon, N. 2004, in Carnegie Observatories Astrophysics Series, Vol.\ts1: Coevolution of Black Holes and Galaxies, ed. L. C. Ho (Cambridge: Cambridge Univ. Press), 12

\nhi Erwin, P., Vega Beltr\'an, J. C., Graham, A. W., \& Beckman, J. E.~2003, ApJ, 597, 929  

\nhi Eskridge, P. B., Frogel, J. A., Pogge, R. W., \etal 2000, AJ, 119, 536

\nhi Evans, N. W., \& de Zeeuw, P. T. 1994, MNRAS, 271, 202  

\nhi Fabello, S., Kauffmann, G., Catinella, B., \etal 2011, MNRAS, 416, 1739

\nhi Faber, S. M., Balick, B., Gallagher, J. S., \& Knapp, G. R. 1977, ApJ, 214, 383

\nhi Faber, S. M., Dressler, A., Davies, R. L., \etal 1987, in Nearly Normal Galaxies: 
     From the Planck Time to the Present, ed. S. M. Faber (New York: Springer), 175

\nhi Faber, S. M., \& Jackson, R. E. 1976, ApJ, 204, 668
 
\nhi Faber, S. M., Tremaine, S., Ajhar, E. A., \etal ~1997, AJ, 114, 1771 

\nhi Faber, S. M., Willmer, C. N. A., Wolf, C., \etal 2007, ApJ, 665, 265  

\nhi Faber, S. M., Worthey, G., \& Gonzalez, J. J. 1992, in IAU Symposium 149, The Stellar Populations
     of Galaxies, ed. B. Barbuy \& A. Renzini (Dordrecht: Kluwer), 255

\nhi Fabian, A. C. 1994, ARA\&A, 32, 277  

\nhi Fabian, A. C. 1999, MNRAS, 308, L39

\nhi Fabian, A. C. 2012, ARA\&A, 50, 455  

\nhi Fabian, A. C. 2013, in IAU Symposium 290, Feeding Compact Objects: Accretion on All Scales, 
     ed. C. M. Zhang, T. Belloni, M. Mendez \& S. N. Zhang (Cambridge: Cambridge Univ. Press), 3 (arXiv:1211.2146)

\nhi Fabian, A. C., Sanders, J. S., Allen, S. W., \etal 2003, MNRAS, 344, L43    

\nhi Fabian, A. C., Sanders, J. S., Allen, S. W., \etal 2011, MNRAS, 418, 2154   

\nhi Fabian, A. C., Sanders, J. S., Taylor, G. B., \etal 2006, MNRAS, 366, 417    

\nhi Faltenbacher, A., Finoguenov, A., \& Drory, N. 2010, ApJ, 712, 484

\nhi Feldmann, R., Carollo, C. M., Mayer, L., \etal 2010, ApJ, 709, 218

\nhi Ferrarese, L. 2002, ApJ, 578, 90

\nhi Ferrarese, L., C\^ot\'e, P., Dalla Bont\`a, E., \etal 2006, ApJ, 644, L21  

\nhi Ferrarese, L., \& Ford, H. 2005, SSRev, 116, 523

\nhi Ferrarese, L., \& Ford, H. C. 1999, ApJ, 515, 583            

\nhi Ferrarese, L., Ford, H. C., \& Jaffe, W. 1996, ApJ, 470, 444   

\nhi Ferrarese, L., \& Merritt, D. 2000, ApJ, 539, L9

\nhi Filippenko, A. V., \& Ho, L. C. 2003, ApJ, 588, L13   

\nhi Filippenko, A. V., Ho, L. C., \& Sargent W. L. W. 1993, ApJ, 410, L75

\nhi Fisher, D. 1997, AJ, 113, 950

\nhi Fisher, D. , Illingworth, G., \& Franx, M. 1995, ApJ, 438, 539

\nhi Fisher, D. B., \& Drory, N. 2008, AJ, 136, 773

\nhi Fisher, D. B., \& Drory, N. 2010, ApJ, 716, 942

\nhi Forbes, D. A., Brodie, J. P., \& Grillmair, C. J. 1997, AJ, 113, 1652  

\nhi Ford, H. C., Harms, R. J., Tsvetanov, Z. I., \etal 1994, ApJ, 435, L27

\nhi Forman, W., Jones, C., Churazov, E., \etal 2007, ApJ, 665, 1057

\nhi Forman, W., Nulsen, P., Heinz, S., \etal 2005, ApJ, 635, 894

\nhi F\"orster Schreiber, N.~M., Genzel, R., Bouch\'e, N., \etal 2009, ApJ, 706, 1364

\nhi Freeman, K. C. 1970, ApJ, 160, 811

\nhi Freeman, K. C. 2008, in Formation and Evolution of Galaxy Disks, ed. J. G. Funes, \& E. M. Corsini (San Francisco, CA: ASP), 3

\nhi Fryer, C. L., Woosley, S. E., \& Heger, A. 2001, ApJ, 550, 372  

\nhi Gallimore, J. F., Baum, S. A., O'Dea, C. P., Brinks, E., \& Pedlar, A. 1996, ApJ, 462, 740

\nhi Gaskell, C. M. 2010, in The First Stars and Galaxies: Challenges for the Next Decade, ed. D. J. Whalen, V. Bromm, \& N. Yoshida (Melville, NY: AIP), 261

\nhi Gaskell, C. M. 2011, in SF2A-2011: Proceedings of the Annual Meeting of the French Society of Astronomy and Astrophysics, 
     ed. G. Alecian, K. Belkacem, R. Samadi, \& D. Valls-Gabaud (Paris: SF2A), 577

\nhi Gebhardt, K. 2004, in Carnegie Observatories Astrophysics Series, Vol. 1: Coevolution of Black Holes and Galaxies, 
     ed. L. C. Ho (Cambridge: Cambridge Univ. Press), 37

\nhi Gebhardt, K., Adams, J., Richstone, D., \etal 2011, ApJ, 729, 119    

\nhi Gebhardt, K., Bender, R., Bower, G.,  \etal 2000a, ApJ, 539, L13

\nhi Gebhardt, K., Kormendy, J., Ho, L. C., \etal 2000b, ApJ, 543, L5    

\nhi Gebhardt, K., Lauer, T. R., Kormendy, J., \etal 2001,  AJ, 122, 2469   

\nhi Gebhardt, K., Lauer, T. R., Pinkney, J., \etal 2007,  ApJ, 671, 1321  

\nhi Gebhardt, K., Pryor, C., O'Connell, R. D., Williams, T. B., \& Hesser, J. E. 2000c, AJ, 119, 1268   

\nhi Gebhardt, K., Rich, R. M., \& Ho, L. C. 2002,  ApJ, 578, L41   

\nhi Gebhardt, K., Rich, R. M., \& Ho, L. C. 2005,  ApJ, 634, 1093  

\nhi Gebhardt, K., Richstone, D., Ajhar, E. A., \etal 1996,  AJ, 112, 105    

\nhi Gebhardt, K., Richstone, D., Kormendy, J., \etal 2000d, AJ, 119, 1157   

\nhi Gebhardt, K., Richstone, D., Tremaine, S., \etal 2003,  ApJ, 583, 92    

\nhi Gebhardt, K., \& Thomas, J. 2009, ApJ, 700, 1690 
  
\nhi Gebhardt, K., \etal 2013, in preparation  

\nhi Genzel, R., Burkert, A., Bouch\'e, N., \etal 2008, ApJ, 687, 59

\nhi Genzel, R., Eckart, A., Ott, T., \& Eisenhauer, F. 1997, MNRAS, 291, 219

\nhi Genzel, R., Eisenhauer, F., \& Gillessen, S. 2010, Rev. Mod. Phys., 82, 3121 (GEG10)

\nhi Genzel, R., Hollenbach, D., \& Townes, C. H. 1994, Rep. Prog. Phys., 57, 417

\nhi Genzel, R., Lutz, D., Sturm, E., \etal 1998, ApJ, 498, 579

\nhi Genzel, R., Newman, S., Jones, T., \etal 2011, ApJ, 733, 101 

\nhi Genzel, R., Pichon, C., Eckart, A., Gerhard, O. E, \& Ott T. 2000, MNRAS, 317, 348

\nhi Genzel, R., Tacconi, L. J., Rigopoulou, D., Lutz, D., \& Tecza, M. 2001, ApJ, 563, 527

\nhi Genzel, R., Thatte, N., Krabbe, A., Kroker, H., \& Tacconi-Garman, L. E. 1996, ApJ, 472, 153

\nhi Genzel, R., \& Townes, C. H. 1987, ARA\&A, 25, 377

\nhi Gerhard, O. 2013, in IAU Symposium 295, The Intriguing Live of Massive Galaxies, ed.
     D. Thomas, A. Pasquali, \& I. Ferreras (Cambridge: Cambridge University Press), in press (arXiv:1212.2768)

\nhi Gerhard, O. E., \& Binney, J. 1985, MNRAS, 216, 467

\nhi Gerssen, J., van der Marel, R. P., Gebhardt, K., \etal 2002, AJ, 124, 3270. 

\nhi Gerssen, J., van der Marel, R. P., Gebhardt, K., \etal 2003, AJ, 125, 376   

\nhi Ghez, A. M., Becklin, E., Duch\^ene, G., \etal 2003, AN Suppl.~1, 324, 527          

\nhi Ghez, A. M., Klein, B. L., Morris, M., \& Becklin, E. E. 1998, ApJ, 509, 678                                                      

\nhi Ghez, A., Morris, M., Lu, J., \etal 2009, Astro2010, 89 (arXiv:0903.0383)

\nhi Ghez, A. M., Salim, S., Hornstein, S. D., \etal 2005, ApJ, 620, 744  

\nhi Ghez, A. M., Salim, S., Weinberg, N. N, \etal 2008, ApJ, 689, 1044                                             

\nhi Gillessen, S., Eisenhauer, F., Fritz, T. K., \etal 2009a, ApJ, 707, L114     

\nhi Gillessen, S., Eisenhauer, F., Trippe, S., \etal 2009b, ApJ, 692, 1075     

\nhi Gliozzi, M., Satyapal, S., Eracleous, M., Titarchuk, L., \& Cheung, C. C. 2009, ApJ, 700, 1759

\nhi Goodman, J., \& Binney, J. 1984, MNRAS, 207, 511

\nhi Gould, A., \& Rix, H.-W. 2000, ApJ, 532, L29

\nhi Graham, A. W. 2004, ApJ, 613, L33

\nhi Graham, A. W. 2007, MNRAS, 379, 711

\nhi Graham, A. W. 2008a, ApJ, 6580, 143  

\nhi Graham, A. W. 2008b, PAS Australia, 25, 167  

\nhi Graham, A. W. 2012, MNRAS, 422, 1586         

\nhi Graham, A. W., Colless, M.M., Busarello, G., Zaggia, S., \& Longo, G. 1998, A\&AS, 133, 325

\nhi Graham, A. W., \& Driver, S. P. 2007, ApJ, 655, 77  

\nhi Graham, A. W., Erwin, P., Caon, N., \& Trijillo, I. 2001, ApJ, 563, L11  

\nhi Graham, A. W., \& Li, I.-H. 2009, ApJ, 698, 812

\nhi Graham, A. W., Onken, C. A., Athanassoula, E., \& Combes, F. 2011, MNRAS, 412, 2211

\nhi Graham, A. W., \& Scott, N. 2013, ApJ, 764, 151

\nhi Graham, A. W., \& Spitler, R. L. 2009, MNRAS, 397, 2148

\nhi Granato, G. L., De Zotti, G., Silva, L., Bressan, A., \& Danese, L. 2004, ApJ, 600, 580

\nhi Graves, G. J., \& Faber, S. M. 2010, ApJ, 717, 803

\nhi Greene, J. E. 2012, Nature Communications, 3, 1304

\nhi Greene, J. E., \& Ho, L. C. 2004, ApJ, 610, 722

\nhi Greene, J. E., \& Ho, L. C. 2005a, ApJ, 627, 721

\nhi Greene, J. E., \& Ho, L. C. 2005b, ApJ, 630, 122

\nhi Greene, J. E., \& Ho, L. C. 2006a, ApJ, 641, 117

\nhi Greene, J. E., \& Ho, L. C. 2006b, ApJ, 641, L21

\nhi Greene, J. E., \& Ho, L. C. 2007a, ApJ, 656, 84

\nhi Greene, J. E., \& Ho, L. C. 2007b, ApJ, 667, 131

\nhi Greene, J. E., \& Ho, L. C. 2007c, ApJ, 670, 92   

\nhi Greene, J. E., Ho, L. C, \& Barth, A. J. 2008, ApJ, 688, 159

\nhi Greene, J. E., Ho, L. C, \& Ulvestad, J. S. 2006, ApJ, 636, 56

\nhi Greene, J. E., Peng, C. Y., Kim, M., \etal 2010, ApJ, 721, 26  

\nhi Greenhill, L. J. 2007, in IAU Symposium 242, Astrophysical Masers and their Environments, ed. J. M. Chapman, \& W. A. Baan (Cambridge: Cambridge Univ. Press), 381

\nhi Greenhill, L. J., Booth, R. S., Ellingsen, S. P., \etal 2003, ApJ, 590, 162  

\nhi Greenhill, L. J., \& Gwinn, C. R. 1997, ApSS, 248, 261           

\nhi Greenhill, L. J., Gwinn, C. R, Antonucci, R., \& Barvainis, R. 1996, ApJ, 472, L21 

\nhi Greenhill, L. J., Kondratko, P. T., Moran, J. M., \& Tilak, A. 2009, ApJ, 707, 787    

\nhi Greenhill, L. J., Moran, J. M., \& Herrnstein, J. R. 1997, ApJ, 481, L23           

\nhi Gualandris, A., \& Merritt, D. 2008, ApJ, 678, 780

\nhi G\"ultekin, K., Cackett, E. M., Miller, J. M., \etal 2009a, ApJ, 706, 404   

\nhi G\"ultekin, K., Richstone, D. O., Gebhardt, K., \etal 2009b, ApJ, 695, 1577  

\nhi G\"ultekin, K., Richstone, D. O., Gebhardt, K., \etal 2009c, ApJ, 698, 198   

\nhi G\"ultekin, K., Richstone, D. O., Gebhardt, K., \etal 2011,  ApJ, 741, 38    

\nhi Gunn, J. E. 1987, in IAU Symposium 117, Dark Matter in the Universe, ed. J. Kormendy \& G. R. Knapp (Dordrecht: Reidel), 537

\nhi  Guo, Q., White, S., Li, C., \& Boylan-Kolchin, M. 2010, MNRAS, 404, 1111

\nhi Haller, J. W., Rieke, M. J., Rieke, G. H., \etal 1996, ApJ, 456, 194   

\nhi Hansen, S. M., Sheldon, E. S., Wechsler, R. H., \& Koester, B. P. 2009, ApJ, 699, 1333

\nhi H\"aring, N., \& Rix, H.-W. 2004, ApJ, 604, L89  

\nhi H\"aring-Neumayer, N., Cappellari, M., Rix, H.-W., \etal 2006, ApJ, 643, 226  

\nhi Harms, R. J., Ford, H. C., Tsvetanov, Z. I., \etal 1994, ApJ, 435, L35  

\nhi Harris, C. E., Bennert, V. N., Auger, M. W., \etal 2012, ApJS, 201, 29

\nhi Harris, G. L. H., \& Harris, W. E. 2011, MNRAS, 410, 2347  

\nhi Harris, W. E. 1991, ARA\&A, 29, 543

\nhi Harris, W. E, Whitmore, B. C., Karakla, D., \etal 2006, ApJ, 626, 90 

\nhi Haschick, A. D., Baan, W. A., \& Peng, E. W. 1994, ApJ, 437, L35

\nhi Heckman, T.~M., Kauffmann, G., Brinchmann, J., \etal 2004, ApJ, 613, 109

\nhi Heger, A., Fryer, C. L., Woosley, S. E., Langer, N., \& Hartmann, D. H. 2003, ApJ, 591, 288  

\nhi Henkel, C., Braatz, J. A., Greenhill, L. J., \& Wilson, A. S. 2002, A\&A, 394, L23

\nhi H\'eraudeau, Ph., Simien, F., Maubon, G., \& Prugniel, Ph. 1999. A\&AS, 136, 509

\nhi Herrnstein, J. R., Moran, J. M., Greenhill, L. J., \etal 1999,Nature, 400, 539  

\nhi Herrnstein, J. R., Moran, J. M., Greenhill, L. J., \& Trotter, A. S. 2005, ApJ, 629, 719  

\nhi Hicks, E. K. S., \& Malkan, M. A. 2008, ApJS, 174, 31

\nhi Hilz, M., Naab, T., \& Ostriker, J. P. 2013, MNRAS, 429, 2924

\nhi Hilz, M., Naab, T., Ostriker, J.~P., \etal 2012, MNRAS, 425, 3119

\nhi Hiner, K. D., Canalizo, G., Wold, M., Brotherton, M. S., \& Cales, S. L. 2012, ApJ, 756, 162

\nhi Hinshaw, G., Larson, D., Komatsu, E., \etal 2013, ApJS, in press (arXiv:1212.5226)  

\nhi Hirschmann, M., Khochfar, S., Burkert, A., \etal 2010, MNRAS, 407, 1016 

\nhi Ho, L. C. 1999a, in Observational Evidence for Black Holes in the Universe, ed. S. K. Chakrabarti (Dordrecht: Kluwer), 157

\nhi Ho, L. C. 1999b, ApJ, 510, 631

\nhi Ho, L. C. 1999c, ApJ, 516, 672

\nhi Ho, L. C. 2002, ApJ, 564, 120

\nhi Ho, L. C, ed. 2004a, Carnegie Observatories Astrophysics Series, Vol. 1: Coevolution of Black Holes and Galaxies (Cambridge: Cambridge Univ. Press)

\nhi Ho, L. C. 2004b, in Carnegie Observatories Astrophysics Series, Vol. 1: Coevolution of Black Holes and Galaxies, ed. L. C. Ho (Cambridge: Cambridge Univ. Press), 293

\nhi Ho, L. C. 2005, ApJ, 629, 680

\nhi Ho, L. C. 2007a, ApJ, 668, 94   

\nhi Ho, L. C. 2007b, ApJ, 669, 821

\nhi Ho, L. C. 2008, ARA\&A, 46, 475

\nhi Ho, L. C. 2009a, ApJ, 699, 626

\nhi Ho, L. C. 2009b, ApJ, 699, 638

\nhi Ho, L. C. 2013, ApJ, in preparation

\nhi Ho, L. C., Darling, J., \& Greene, J. E. 2008, ApJ, 681, 128

\nhi Ho, L. C., \& Filippenko, A. V. 1996, ApJ, 472, 600

\nhi Ho, L. C., Filippenko, A. V., \& Sargent, W. L. W. 1997a, ApJS, 112, 315

\nhi Ho, L. C., Filippenko, A. V., \& Sargent, W. L. W. 1997b, ApJ, 487, 568

\nhi Ho, L. C., Filippenko, A. V., \& Sargent, W. L. W. 2003, ApJ, 583, 159

\nhi Ho, L. C., Greene, J. E., Filippenko, A. V., \& Sargent, W. L. W. 2009, ApJS, 183, 1  

\nhi Ho, L. C., Kim, M., \& Terashima, Y. 2012, ApJ, 759, L16

\nhi Ho, L. C., Sarzi, M., Rix, H.-W., \etal 2002, PASP, 114, 137

\nhi Ho, L. C., Terashima, Y., \& Okajima, T. 2003, ApJ, 587, L35

\nhi Holley-Bockelmann, K., \& Richstone, D. 1999, ApJ, 517, 92

\nhi Holley-Bockelmann, K., \& Richstone, D. 2000, ApJ, 531, 232

\nhi Hopkins, A.~M. 2004, ApJ, 615, 209

\nhi Hopkins, P. F., Bundy, K., Hernquist, L., Wuyts, S., \& Cox, T. J.~2010, MNRAS, 401, 1099                    

\nhi Hopkins, P. F., Bundy, K., Murray, N., \etal 2009a, MNRAS, 398, 898       

\nhi Hopkins, P. F., Cox, T. J., Dutta, S. N., \etal 2009b, ApJS, 181, 135 

\nhi Hopkins, P. F., \& Hernquist, L. 2009, ApJ, 694, 599

\nhi Hopkins, P. F., \& Hernquist, L. 2010, MNRAS, 407, 447

\nhi Hopkins, P. F., Hernquist, L., Cox, T. J., Dutta, S. N., \& Rothberg, B. 2008, ApJ, 679, 156                 

\nhi Hopkins, P. F., Hernquist, L., Cox, T. J., Robertson, B., \& Krause, E. 2007a, ApJ, 669, 45                  

\nhi Hopkins, P. F., Hernquist, L., Cox, T. J., Robertson, B., \& Krause, E. 2007b, ApJ, 669, 67                  

\nhi Hopkins, P. F., Hernquist, L., Cox, T. J., \etal 2006, ApJS, 163, 1  

\nhi Hopkins, P. F., Lauer, T. R., Cox, T. J., Hernquist, L., \& Kormendy, J. 2009c, ApJS, 181, 486               


\nhi Houghton, R. C. W., Magorrian, J., Sarzi, M., \etal 2006, MNRAS, 367, 2  

\nhi Howard, C. D., Rich, R. M., Clarkson, W., \etal 2009, ApJ, 702, L153

\nhi Hoyle, F., \& Fowler, W. A. 1963, Nature, 197, 533   

\nhi Hu, J. 2008, MNRAS, 386, 2242                     

\nhi Huang, S., Ho, L. C., Peng, C. Y., Li, Z.-Y., \& Barth, A. J. 2013a, ApJ, 766, 47

\nhi Huang, S., Ho, L. C., Peng, C. Y., Li, Z.-Y., \& Barth, A. J. 2013b, ApJL, in press (arXiv:1304.2299)

\nhi Hubble, E. 1930, ApJ, 71, 231 

\nhi Hughes, M. A., Axon, D., Atkinson, J., \etal 2005, AJ, 130, 73

\nhi Humphreys,{\ts}E.{\ts}M.{\ts}L., Reid,{\ts}M.{\ts}J., Greenhill,{\ts}L.{\ts}J., Moran,{\ts}L.{\ts}M., \& Argon,{\ts}A.{\ts}L. 2008, ApJ, 672, 800

\nhi Hur\'e, J.-M. 2002, A\&A, 395, L21

\nhi Hur\'e, J.-M., Hersant, F., Surville, C., Nakai, N., \& Jacq, T. 2011, A\&A, 530, 145

\nhi Illingworth, G. 1977, ApJ, 218, L43

\nhi Inskip, K.~J., Jahnke, K., Rix, H.-W., \& van de Ven, G. 2011, ApJ, 739, 90

\nhi Into, T., \& Portinari, L. 2013, MNRAS, 430, 2715

\nhi Ishihara, Y., Nakai, N., Iyomoto, N., \etal 2001, PASJ, 53, 215

\nhi Jahnke, K., Bongiorno, A., Brusa, M., \etal 2009, ApJ, 706, L215

\nhi Jahnke, K., \& Macci\`o, A. V. 2011, ApJ, 734, 92  

\nhi Jalali, B., Baumgardt, H., Kissler-Patig, M., \etal 2011, A\&A, 538, A19  

\nhi Jardel J., Gebhardt, K., Shen, J., \etal 2011, ApJ, 739, 21

\nhi Jarrett, T. H., Chester, T., Cutri, R., Schneider, S. E., \& Huchra, J. P. 2003, AJ, 125, 525

\nhi Jarvis, B. J., \& Freeman, K. C., 1985, ApJ, 295, 324


\nhi Jiang, Y.-F., Greene, J. E., \& Ho, L. C. 2011a, ApJ, 737, L45

\nhi Jiang, Y.-F., Greene, J. E., Ho, L. C., Xiao, T., \& Barth, A. J. 2011b, ApJ, 742, 68

\nhi Johnson, H. L. 1962, ApJ, 135, 69

\nhi Jones, D. H., Mould, J. R., Watson, A. M., \etal 1996, ApJ, 466, 742

\nhi J\o rgensen, I., Franx, M., \& Kj\ae rgaard, P. 1996, MNRAS, 280, 167

\nhi Joseph, C. L., Merritt, D., Olling, R., \etal 2001, ApJ, 550, 668

\nhi Joseph, R. D. 1999, Ap\&SS, 266, 321

\nhi Joseph, R. D., \& Wright, G. S.~1985, MNRAS, 214, 87

\nhi Kamizasa, N., Terashima, Y., \& Awaki, H. 2012, ApJ, 751, 39

\nhi Karachentsev,{\ts}I.{\ts}D., Karachentseva,{\ts}V.{\ts}E., Huchtmeier,{\ts}W.{\ts}K., \& Makarov,{\ts}D.{\ts}I.~2004, AJ, 127, 2031 

\nhi Kassin, S.~A., Weiner, B. J., Faber, S. M., \etal 2012, ApJ, 758, 106

\nhi Kauffmann, G., \& Haehnelt, M. 2000, MNRAS, 311, 576

\nhi Kauffmann, G.,\&  Heckman, T. M. 2009, MNRAS, 397, 135

\nhi Kauffmann, G., Heckman, T. M., Tremonti, C., \etal 2003, MNRAS, 346, 1055

\nhi Kennicutt, R. C. 1989, ApJ, 344, 685

\nhi Kennicutt, R. C. 1998a, ARA\&A, 36, 189

\nhi Kennicutt, R. C. 1998b, in 26$^{\rm th}$ Advanced Course of the Swiss Society of Astronomy and Astrophysics, 
     Galaxies: Interactions and Induced Star Formation, ed. D. Friedli, L. Martinet, \& D. Pfenniger (New York: Springer), 1

\nhi Kennicutt, R. C, Lee, J. C., Funes, J. G., Sakai, S., \& Akiyama, S. 2008, ApJS, 178, 247  

\nhi Kent, S. M. 1990, AJ, 100, 377

\nhi Kent, S. M.~1992, ApJ, 387, 181

\nhi Kent, S. M, Dame, T. M., \& Fazio, G. 1991, ApJ, 378, 131

\nhi Kent, S. M, \& Gunn, J. E. 1982, AJ, 87, 945

\nhi Kere\v s, D., Katz, N., Weinberg, D. H., \& Dav\'e, R. 2005, MNRAS, 363, 2

\nhi Khochfar, S., \& Ostriker, J. P. 2008, ApJ, 680, 54

\nhi Khochfar, S., \& Silk, J., 2006, ApJ, 648, L21

\nhi Kim, E., Lee, M. G., \& Geisler, D. 2000, MNRAS, 314, 307

\nhi Kim, M., Ho, L. C., Peng, C. Y., Barth, A. J., \& Im, M. 2008, ApJS, 179, 283

\nhi Kim, M., Ho, L. C., Peng, C. Y., \& Im, M. 2013, ApJ, in preparation

\nhi King, A. 2003, ApJ, 596, L27

\nhi Kinney, A. L., Schmitt, H. R., Clarke, C. J., \etal 2000, ApJ, 537, 152 

\nhi Kobulnicky, H. A., Dickey, J. M., Sargent, A. I., Hogg, D. E., \& Conti, P. S. 1995, AJ, 110, 116

\nhi Kocevski, D. D., Faber, S. M., Mozena, M., \etal 2012, ApJ, 744, 148

\nhi Komatsu, E., Dunkley, J., Nolta, M. R., \etal 2009, ApJS, 180, 330  

\nhi Komossa, S. 2006, MemSAI, 77, 733

\nhi Kondratko, P. T., Greenhill, L. J., \& Moran, J. M. 2005, ApJ, 618, 618  

\nhi Kondratko, P. T., Greenhill, L. J., \& Moran, J. M. 2006, ApJ, 652, 136  

\nhi Kondratko, P. T., Greenhill, L. J., \& Moran, J. M. 2008, ApJ, 678, 87   

\nhi Kong, A. K. H., Heinke, C. O., di Stefano, R., \etal 2010, MNRAS, 407, L84

\nhi Kong, A. K. H., Yang, Y. J., Hsieh, P.-Y., Mak, D. S. Y., \& Pun, C. S. J. 2007, ApJ, 671, 349

\nhi K\"ording, E., Colbert, E., \& Falcke, H. 2005, A\&A, 436, 427

\nhi Kormendy, J. 1977, ApJ, 217, 406

\nhi Kormendy, J. 1982, in Morphology and Dynamics of Galaxies, 12th Advanced Course of the Swiss Society of Astronomy and Astrophysics, ed. L. Martinet \& M. Mayor (Sauverny: Geneva Obs.), 113

\nhi Kormendy, J. 1984, ApJ, 287, 577

\nhi Kormendy, J. 1985, ApJ, 295, 73

\nhi Kormendy, J. 1987, in Nearly Normal Galaxies: From the Planck Time to the Present, ed. S. M. Faber (Berlin: Springer), 163

\nhi Kormendy, J. 1988a, ApJ, 325, 128

\nhi Kormendy, J. 1988b, ApJ, 335, 40

\nhi Kormendy, J. 1992a, Testing the AGN Paradigm, ed. S. S. Holt, S. G. Neff, \& C. M. Urry (New York: AIP), 23

\nhi Kormendy, J. 1992b, in High Energy Neutrino Astrophysics, ed. V. J. Stenger, J. G. Learned, S. Pakvasa, \& X. Tata (Singapore: World Scientific), 196

\nhi Kormendy, J. 1993a, in The Nearest Active Galaxies, ed. J. Beckman, L. Colina, \& H. Netzer (Madrid: Consejo Superior de Investigaciones Cient\'\i ficas), 197

\nhi Kormendy, J. 1993b, in IAU Symposium 153, Galactic Bulges, ed. H. Dejonghe \& H. J. Habing (Dordrecht: Kluwer), 209

\nhi Kormendy, J. 1999, in Galaxy Dynamics: A Rutgers Symposium, ed. D. Merritt, J. A. Sellwood, \& M. Valluri (San Francisco: ASP), 124

\nhi Kormendy, J. 2004, in Carnegie Observatories Astrophysics Series, Vol. 1: Coevolution of Black Holes and Galaxies, ed. L. C. Ho (Cambridge: Cambridge Univ. Press), 1

\nhi Kormendy, J. 2009, in Galaxy Evolution: Emerging Insights and Future Challenges, ed. S. Jogee, I. Marinova, L. Hao, \& G. A. Blanc (San Francisco: ASP), 87

\nhi Kormendy, J. 2012, in XXIII Canary Islands Winter School of Astrophysics, Secular Evolution~of Galaxies, ed.{\ts}J.{\ts}Falc\'on-Barroso\ts\&{\ts}J.{\ts}H.{\ts}Knapen (Cambridge:~Cambridge Univ.~Press), in press

\nhi Kormendy, J. 2013, in IAU Symposium 295, The Intriguing Life of Massive Galaxies, ed.~D. Thomas, A. Pasquali, \& I. Ferreras 
     (Cambridge: Cambridge University Press), in press

\nhi Kormendy, J., \& Bender, R. 1999, ApJ, 522, 772

\nhi Kormendy, J., \& Bender, R. 2009, ApJ, 691, L142

\nhi Kormendy, J., \& Bender, R. 2011, Nature, 469, 377

\nhi Kormendy, J., \& Bender, R. 2012, ApJS, 198, 2

\nhi Kormendy, J., \& Bender, R. 2013a, ApJL, submitted       

\nhi Kormendy, J., \& Bender, R. 2013b, ApJS, in preparation  

\nhi Kormendy, J., Bender, R., Ajhar, E. A., \etal 1996a, ApJ, 473, L91     

\nhi Kormendy, J., Bender, R., \& Cornell, M. E. 2011, Nature, 469, 374

\nhi Kormendy, J., Bender, R., Evans, A. S., \& Richstone, D. 1998, AJ, 115, 1823                  

\nhi Kormendy, J., Bender, R., Magorrian, J., \etal 1997, ApJ, 482, L139  

\nhi Kormendy, J., Bender, R., Richstone, D., \etal 1996b, ApJ, 459, L57   

\nhi Kormendy, J., Byun, Y.-I., Ajhar, E. A., \etal 1996c, in IAU Symposium 171, New Light on Galaxy Evolution, 
     ed. R. Bender \& R. L. Davies (Dordrecht: Kluwer), 105

\nhi Kormendy, J., Drory, N., Bender, R., \& Cornell, M. E. 2010, ApJ, 723, 54 

\nhi Kormendy, J., Fisher, D. B., Cornell, M. E., \& Bender, R. 2009, ApJS, 182, 216 (KFCB)

\nhi Kormendy, J., \& Gebhardt, K. 2001, in 20$^{\rm th}$ Texas Symposium on Relativistic Astrophysics, ed. J. C. Wheeler \& H. Martel (Melville, NY: AIP), 363

\nhi Kormendy, J., Gebhardt, K., Fisher, D. B., \etal 2005, AJ, 129, 2636                                     

\nhi Kormendy, J., Gebhardt, K., \& Richstone, D. 2000, BAAS, 32, 702

\nhi Kormendy, J., \& Illingworth, G. 1982, ApJ, 256, 460

\nhi Kormendy, J., \& Kennicutt, R. C. 2004, ARA\&A, 42, 603

\nhi Kormendy, J., \& McClure, R. D. 1993, AJ, 105, 1793

\nhi Kormendy, J., \& Richstone, D. 1992, ApJ, 393, 559 

\nhi Kormendy, J., \& Richstone, D. O. 1995, ARA\&A, 33, 581 (KR95)

\nhi Kormendy, J., \& Sanders, D. B. 1992, ApJ, 390, L53

\nhi Krabbe, A., Genzel, R., Eckart, A., \etal 1995, ApJ, 447, L95

\nhi Krajnovi\'c, D., McDermid, R. M., Cappellari, M., \& Davies, R. L. 2009, MNRAS, 399, 1839           

\nhi Krajnovi\'c, D., Sharp, R., \& Thatte, N. 2007, MNRAS, 374, 385  

\nhi Kraljic, K., Bournaud, F., \& Martig, M. 2012, ApJ, 757, 60

\nhi Kroupa, P. 2001, MNRAS, 322, 231

\nhi Kukula, M.~J., Dunlop, J.~S., McClure, R.~J., \etal 2001, MNRAS, 326, 1533

\nhi Kulkarni, S. R., Hut, P., \& McMillan, S. 1993, Nature, 364, 421

\nhi Kumar, P.~1999, ApJ, 519, 599

\nhi Kundu, A., Whitmore, B. C., Sparks, W. B., \etal 1999, ApJ, 513, 733  

\nhi Kuo, C. Y., Braatz, J. A., Condon, J. J., \etal 2011, ApJ, 727, 20 

\nhi Laor, A. 2001, ApJ, 553, 677

\nhi Larsen, S. S., Brodie, J. P., Huchra, J. P., Forbes, D. A., \& Grillmair, C. J. 2001, AJ, 121, 2974              

\nhi Larson, R. B. 2000, in ESLAB Symposium 33, Star Formation from the Small to the Large Scale, ed. F. Favata, A. Kaas, \& A. Wilson (Noordwijk: European Space Agency), 13


\nhi Lauer, T. R. 1999, PASP, 111, 227

\nhi Lauer, T. R. 2012, ApJ, 759, 64

\nhi Lauer, T. R., Ajhar, E. A., Byun, Y.-I., \etal 1995, AJ,   110, 2622    

\nhi Lauer, T. R., Bender, R., Kormendy, J., Rosenfield, P., \& Green, R. F. 2012, ApJ,  745, 121     

\nhi Lauer, T. R., Faber, S. M., Ajhar, E. A., Grillmair, C. J., \& Scowen, P. A. 1998, AJ,   116, 2263    

\nhi Lauer, T. R., Faber, S. M., Currie, D. G., \etal 1992, ApJ,  104, 522     

\nhi Lauer, T. R., Faber, S. M., Gebhardt, K., \etal 2005, AJ,   129, 2138    

\nhi Lauer, T. R., Faber, S. M., Groth, E. J., \etal 1993, AJ,   106, 1436    

\nhi Lauer, T. R., Faber, S. M., Lynds, D. R., \etal 1992, AJ,   103, 703     

\nhi Lauer, T. R., Faber, S. M., Richstone, D., \etal 2007a, ApJ, 662, 808     

\nhi Lauer, T. R., Gebhardt, K., Faber, S. M., \etal 2007b, ApJ, 664, 226     

\nhi Lauer, T. R., Gebhardt, K., Richstone, D.,\etal 2002,  AJ,  124, 1975    

\nhi Lauer, T.~R., Tremaine, S., Richstone, D., \& Faber, S. M. 2007c, ApJ, 670, 249 

\nhi Launhardt, R., Zylka, R., \& Mezger, P. G. 2002, A\&A, 384, 112  

\nhi Leauthaud, A., Tinker, J., Bundy, K., \etal 2012, ApJ 744, 159


\nhi Lee, H. M. 1995, MNRAS, 272, 605

\nhi Lee, H. M. 1996, in IAU Symposium 174, Dynamical Evolution of Star Clusters: Confrontation of Theory and Observations,
     ed. P. Hut \& J. Makino (Dordrecht: Kluwer), 293

\nhi Lee, H. M. 1998, J. Korean Physical Society, 33, S549

\nhi Le Floc'h, E., Papovich, C., Dole, H., \etal 2005, ApJ, 632, 169

\nhi Lema\^\i tre, G. E. 1931, quoted in IAU Symposium 92, Objects of High Redshift, ed. G. O. Abell \& P. J. E. Peebles (Dordrecht: Reidel),  Frontispiece

\nhi Li, Y., Hernquist, L., Robertson, B., \etal 2007, ApJ, 665, 187

\nhi Liebling, S. L., \& Palenzuela, C. 2012, Living Reviews in Relativity, 15, No.~6, cited 2012 October~4, {\ssbf http://relativity.livingreviews.org/Articles/lrr-2012-6/}

\nhi Light, E. S., Danielson, R. E., \& Schwarzschild, M. 1974, ApJ, 194, 257

\nhi Lin, Y.-T., \& Mohr, J. J. 2004, ApJ, 617, 879

\nhi Lodato, G., \&  Bertin, G. 2003, A\&A, 398, 517

\nhi Ludwig, R. R., Greene, J. E., Barth, A. J., \& Ho, L. C. 2012, ApJ, 756, 51

\nhi L\"utzgendorf, N., Kissler-Patig, M., Gebhardt, K., \etal 2013, A\&A, 552, A49

\nhi Lynden-Bell, D. 1969, Nature, 223, 690

\nhi Lynden-Bell, D. 1978, Physica Scripta, 17, 185

\nhi Lynden-Bell, D., \& Rees, M. J. 1971, MNRAS, 152, 461

\nhi Maccarone, T. J., Bergond, G., Kundu, A., \etal 2007, in IAU Symposium~246, Dynamical
Evolution of{\ts}Dense{\ts}Stellar{\ts}Systems,{\ts}ed.{\ts}E.{\ts}Vesperini,{\ts}M.{\ts}Giersz,\ts\&{\ts}A.{\ts}Sills{\ts}(Cambridge:{\ts}Cambridge{\ts}Univ.{\ts}Press),\ts336

\nhi Maccarone, T. J., Kundu, A., Zepf, S. E., \& Rhode, K. L. 2010, MNRAS, 409, L84

\nhi Macchetto, F., Marconi, A., Axon, D. J., \etal 1997, ApJ, 489, 579

\nhi Maciejewski, W., \& Binney, J. 2001, MNRAS, 323, 831   

\nhi Madau, P., \& Rees, M. J. 2001, ApJ, 551, L27  

\nhi Magorrian, J., Tremaine, S., Richstone, D., \etal 1998, AJ, 115, 2285

\nhi Maiolino, R., Gallerani, S., Neri, R., \etal 2012, MNRAS, 425, L66

\nhi Maiolino, R., Krabbe, A., Thatte, N., \& Genzel, R. 1998, ApJ, 493, 650

\nhi Makino, J., \& Ebisuzaki, T. 1996, ApJ, 465, 527

\nhi Makino, J., \& Funato, Y. 2004, ApJ, 602, 93

\nhi Mamyoda, K., Nakai, N., Yamauchi, A., Diamond, P., \& Hur\'e, J.-M. 2009, PASJ, 61, 1143  

\nhi Maoz, E.~1995, ApJ, 447, L91

\nhi Maoz, E.~1998, ApJ, 494, L181


\nhi Marconi, A., Axon, D. J., Capetti, A., Maciejewski, W., \& Atkinson, J. 2003, ApJ,   586, 868 

\nhi Marconi, A., Capetti, A., Axon, D. J., \etal 2001, ApJ,   549, 915  

\nhi Marconi, A., \& Hunt, L. K.  2003, ApJ, 589, L21  

\nhi Marconi, A., Pastorini, G., Pacini, F., \etal 2006, A\&A,  448, 921  

\nhi Marconi, A., Risaliti, G., Gilli, R., \etal 2004, MNRAS, 351, 169  


\nhi Martig, M., Bournaud, F., Croton, D. J., Dekel, A., \& Teyssier, R. 2012, ApJ, 756, 26

\nhi Martin, D. C., Wyder, T. K., Schiminovich, D., \etal 2007, ApJ, 173, 342

\nhi Marziani, P., \& Sulentic, J. 2012, NewAR, 56, 49

\nhi Matkovi\'c, A., \& Guzm\'an, R.~2005, MNRAS, 362, 289

\nhi Matteucci, F. 1994, A\&A, 288, 57

\nhi McAlpine, W., Satyapal, S., Gliozzi, M., \etal 2011, ApJ, 728, 25

\nhi McConnell, N. J., \& Ma, C.-P. 2013, ApJ, 764, 184

\nhi McConnell, N. J., Ma, C.-P., Gebhardt, K., \etal 2011a, Nature, 480, 215

\nhi McConnell, N. J., Ma, C.-P., Graham, J. R., \etal 2011b, ApJ, 728, 100

\nhi McConnell, N. J., Ma, C.-P., Murphy, J. D., \etal 2012, ApJ, 756, 179

\nhi McDermid, R. M., Emsellem, E., Shapiro, K. L. \etal 2006, MNRAS, 373, 906

\nhi McLeod, K.~K., \& Bechtold, J. 2009, ApJ, 704, 415

\nhi McLure, R. J., \& Dunlop, J. S. 2002, MNRAS, 331, 795     

\nhi McLure, R.~J., Jarvis, M.~J., Targett, T. A., Dunlop, J. S., \& Best, P. N. 2006, MNRAS, 368, 1395

\nhi McNamara, B. R., \& Nulsen, P. E. J. 2007, ARA\&A, 45, 117

\nhi McNamara, B. R., \& Nulsen, P. E. J. 2012, New J. Phys., 14, 055023

\nhi Mechtley, M., Windhorst, R. A., Ryan, R. E., \etal 2012, ApJ, 756, L38

\nhi Mei, S., Blakeslee, J. P., C\^ot\'e, P., \etal 2007, ApJ, 655, 144

\nhi Meier, D. S., Turner, J. L., Crosthwaite, L. P., \& Beck, S. C. 2001, AJ, 121, 740

\nhi Melia, F. 2007, The Galactic Supermassive Black Hole (Princeton: Princeton Univ. Press)

\nhi Melia, F., \& Falcke, H. 2001, ARA\&A, 39, 309

\nhi M\'endez-Abreu, J., Aguerri, J. A. L., Corsini, E. M., \& Simonneau, E. 2008, A\&A, 478, 353  

\nhi Merloni, A., Bongiorno, A., Bolzonella, M., \etal 2010, ApJ, 708, 137 

\nhi Merloni, A., Heinz, S., \& Di Matteo, T. 2003, MNRAS, 345, 1057  

\nhi Merritt, D. 1999, PASP, 111, 129  

\nhi Merritt, D. 2006, ApJ, 648, 976

\nhi Merritt, D., \& Cruz, F. 2001, ApJ, 551, L41

\nhi Merritt, D., \& Ferrarese, L. 2001, MNRAS, 320, L30  

\nhi Merritt, D., Ferrarese, L., \& Joseph, C. J. 2001, Science, 293, 1116  

\nhi Merritt, D., Mikkola, S., \& Szell, A. 2007, ApJ, 671, 53

\nhi Merritt, D., \& Milosavljevi\'c, M.~2005, Living Reviews in Relativity, 8, No.~8, cited 2012 October~23, \par {\ssbf http://www.livingreviews.org/lrr-2005-8}

\nhi Merritt, D., Milosavljevi\'c, M., Favata, M., Hughes, S. A., \& Holz, D. E. 2004, ApJ, 607, L9

\nhi Merritt, D., \& Quinlan, G. D. 1998, ApJ, 498, 625

\nhi Meyer, L., Ghez, A. M., Sch\"odel, R., \etal 2012, Science, 338, 84

\nhi Mihalas, D., \& Routly, P. M. 1968, Galactic Astronomy (San Francisco: Freeman)

\nhi Mihos, J. C., \& Hernquist, L. 1994, ApJ, 437, L47

\nhi Miller, M. C., \& Colbert, E. J. M. 2004, International J. Mod. Phys. D, 13, 1

\nhi Miller-Jones, J. C. A., Wrobel, J. M., Sivakoff, G. R., \etal 2012, ApJ, 755, L1

\nhi Milosavljevi\'c, M., \& Merritt, D. 2001, ApJ, 563, 34

\nhi Milosavljevi\'c, M., \& Merritt, D. 2003, in AIP Conference Proceedings No. 686, The Astrophysics of Gravitational Wave Sources, ed. J. M. Centrella (Melville, NY: AIP), 201
     
\nhi Milosavljevi\'c, M., Merritt, D., Rest, A., \& van den Bosch, F. C. 2002, MNRAS, 331, L51

\nhi Miniutti, G., Ponti, G., Greene, J. E., \etal 2009, MNRAS, 394, 443

\nhi Mirabel, I. F., Sanders, D. B., \& Kaz\`es, I. 1989, ApJ, 340, L9

\nhi Miyoshi, M., Moran, J., Herrnstein, J., \etal 1995, Nature, 373, 127  

\nhi M\"ollenhoff, C., Matthias, M., \& Gerhard, O. E. 1995, A\&A, 301, 359

\nhi Monachesi, A., Trager, S. C., Lauer, T. R., \etal 2011, ApJ, 727, 55

\nhi Moore, B., Ghigna, S., Governato, F., \etal 1999, ApJ, 524, L19

\nhi Moran, J. M. 2008, in Frontiers of Astrophysics: A Celebration of NRAO's 50$^{\rm th}$ Anniversary, ed. A. H. Bridle, J. J. Condon, \& G. C. Hunt (San Francisco: ASP), 87

\nhi Mortlock, D. J., Warren, S. J., Venemans, B. P., \etal 2011, Nature, 474, 616

\nhi Moster, B. P., Naab, T., \& White, S. D. M. 2013, MNRAS, 428, 3121

\nhi Moster, B. P., Somerville, R. S., Maulbetsch, C., van den Bosch, F. C., Macci\`o, A. V., Naab, T., \& Oser, L. 2010, ApJ, 710, 903

\nhi Mouhcine, M., Ferguson, H. C., Rich, R. M., Brown, T. M., \& Smith, T. E. 2005, ApJ, 633, 810


\nhi Mould, J., \& Sakai, S. 2008, ApJ, 686, L75

\nhi M\"uller-S\'anchez, F., Davies, R. I., Eisenhauer, F., \etal 2006, A\&A, 454, 481

\nhi Mundell, C. G., Pedlar, A., Axon, D. J., Meaburn, J., \& Unger, S. W. 1995, MNRAS, 277, 641

\nhi Murray, N., Quataert, E., \& Thompson, T. A. 2005, ApJ, 618, 569

\nhi Naab, T., 2013, in  IAU Symposium 295, The Intriguing Life of Massive Galaxies, ed. D. Thomas, A. Pasquali, \& I. Ferreras 
     (Cambridge: Cambridge University Press), in press (arXiv:1211.6892)

\nhi Naab, T., Johansson, P. H., \& Ostriker, J. P.~2009, ApJ, 699, L178 

\nhi Nagino, R., \& Matsushita, K. 2009, A\&A, 501, 157

\nhi Nandra, K., Georgakakis, A., Willmer, C. N. A., \etal 2007, ApJ, 660, L11

\nhi Navarro, J. F., Frenk, C. S., \& White, S. D. M. 1997, ApJ, 490, 493

\nhi Nesvadba, N. P. H., Boulanger, F., Salom\'e, P., \etal 2010, A\&A, 521, A65

\nhi Nesvadba,{\ts}N.{\ts}P.{\ts}H., Lehnert, M.{\ts}D., De Breuck, C., Gilbert, A.{\ts}M., \& van Breugel, W. 2008, A\&A, 491, 407

\nhi Nesvadba, N. P. H., Lehnert, M. D., Eisenhauer, F., \etal 2006, ApJ, 650, 693


\nhi Neumayer, N., Cappellari, M., Reunanen, J., \etal 2007, ApJ, 671, 1329  

\nhi Neumayer, N., \& Walcher, C. J. 2012, Adv. Astron., 2012, 709038

\nhi Newman, S. F., Genzel, R., F\"orster-Schreiber, N. M., \etal 2012, ApJ, 761, 43

\nhi Nipoti, C., \& Binney, J. 2007, MNRAS, 382, 1481

\nhi Noel-Storr, J., Baum, S. A., \& O'Dea, C. P. 2007, ApJ, 663, 71

\nhi Noel-Storr, J., Baum, S. A., Verdoes Kleijn, G. A., \etal 2003, ApJS, 148, 419

\nhi Noeske, K. G., Papaderos, P., Cair\'os, L. M., \& Fricke, K. J. 2003, A\&A, 410, 481  

\nhi Noordermeer, E., van der Hulst, J.~M., Sancisi, R., Swaters, R.~S., \& van Albada, T.~S.~2007, MNRAS, 276, 1513

\nhi Norman, C. A., May, A., \& van Albada, T. S. 1985, ApJ, 296, 20

\nhi Nowak, N., Saglia, R. P., Thomas, J., \etal 2008, MNRAS, 391, 1629    

\nhi Nowak, N., Saglia, R. P., Thomas, J., \etal 2007, MNRAS, 379, 909     

\nhi Nowak, N., Thomas, J., Erwin, P., \etal 2010, MNRAS, 403, 646 

\nhi Noyola, E., Gebhardt, K., \& Bergmann, M. 2008, ApJ, 676, 1008      

\nhi Noyola, E., Gebhardt, K., Kissler-Patig, M., \etal  2010, ApJ, 719, L60 

\nhi Oegerle, W. R., \& Hoessel, J. G. 1991, ApJ, 375, 15

\nhi Oliva, E., Origlia, L., Kotilainen, J. K., \& Moorwood, A. F. W. 1995, A\&A, 301, 55  

\nhi Olsen, K. P., Rasmussen, J., Toft, S., \& Zirm, A. W. 2013, ApJ, 764, 4

\nhi Onken, C. A., Ferrarese, L., Merritt, D., \etal 2004, ApJ, 615, 645  

\nhi Onken, C. A., Valluri, M., Peterson, B. M., \etal 2007, ApJ, 670, 105

\nhi Oser, L., Naab, T., Ostriker, J. P., \& Johansson, P. H.~2012, ApJ, 744, 63

\nhi Oser, L., Ostriker, J. P., Naab, T., Johansson, P. H., \& Burkert, A.~2010, ApJ, 725, 2312 

\nhi Ostriker, J. P., \& Ciotti, L. 2005, Phil. Trans. R. Soc. London, 363, A667

\nhi O'Sullivan, E., Forbes, D. A., \& Ponman, T. J. 2001, MNRAS, 328, 461

\nhi Park, D., Woo, J.-H., Treu, T., \etal 2012, ApJ, 747, 30

\nhi Pastorini, G., Marconi, A., Capetti, A., \etal 2007, A\&A, 469, 405 

\nhi Paturel, G., Fang, Y., Petit, C., Garnier, R., \& Rousseau, J. 2000, A\&AS, 146, 19

\nhi Paturel, G., Petit, C., Prugniel, Ph., \etal 2003, A\&A, 412, 45

\nhi Peebles, P. J. E. 1972, ApJ, 178, 371

\nhi Peiris, H. V., \& Tremaine, S. 2003, ApJ, 599, 237

\nhi Peletier, R. F., Davies, R. L., Illingworth, G. D., Davis, L. E., \& Cawson, M. 1990, AJ, 100, 1091

\nhi Pellegrini, S. 1999, A\&A, 351, 487

\nhi Pellegrini, S. 2005, MNRAS, 364, 169


\nhi Peng, C. Y. 2007, ApJ, 671, 1098  

\nhi Peng, C.~Y., Impey, C.~D., Ho, L.~C., Barton, E. J., \& Rix, H.-W. 2006a, ApJ, 640, 114

\nhi Peng, C.~Y., Impey, C.~D., Rix, H.-W., \etal                       2006b, ApJ, 649, 616

\nhi Peng, E. W., Jord\'an, A., C\^ot\'e, P., \etal                     2006c, ApJ, 639, 95

\nhi Peng, E. W., Jord\'an, A., C\^ot\'e, P., \etal                     2008, ApJ, 681, 197  

\nhi Peterson, B. M. 2008, NewAR, 52, 240

\nhi Peterson, B. M., Bentz, M. C., Desroches, L.-B., \etal 2005, ApJ, 632, 799; Erratum. 2005, ApJ, 641, 638 

\nhi Phinney, E. S. 1996, in ASP Conference Series, Vol. 90, The Origins, Evolution, and Destinies of Binary Stars in Clusters,
     ed. E. F. Milone \& J.-C. Mermilliod (San Francisco: ASP), 163

\nhi Phinney, E. S., \& Kulkarni, S. R. 1994, ARA\&A, 32, 591

\nhi Pignatelli, E., Salucci, P., \& Danese, L. 2001, MNRAS, 320, 124

\nhi Pinkney, J., Gebhardt, K., Bender, R., \etal 2003, ApJ, 596, 903

\nhi Portinari, L., \& Into, T. 2011, in ASP Conference Series, Vol.~445, Why Galaxies Care About AGB Stars II: 
     Shining Examples and Common Inhabitants, ed. F. Kerschbaum, T. Lebzelter, \& R.F. Wing (San Francisco: ASP), 403 

\nhi Pounds, K.~A., \& Page, K.~L. 2006, MNRAS, 372, 1275

\nhi Qian, E. E., de Zeeuw, P. T., van der Marel, R. P., \& Hunter, C. 1995, MNRAS, 274, 602

\nhi Quilis, V., Bower, R. G., \& Balogh, M. L. 2001, MNRAS, 328, 1091

\nhi Quinlan, G. D., \& Hernquist, L. 1997, NewA, 2, 533

\nhi Rafferty, D. A., McNamara, B. R., \& Nulsen, P. E. J. 2008, ApJ, 687, 899

\nhi Rafferty, D. A., McNamara, B. R., Nulsen, P. E. J., \& Wise, M. W. 2006, ApJ, 652, 216

\nhi Ransom, S. M., Hessels, J. W. T., Stairs, I. H., \etal 2005, Science, 307, 892

\nhi Ravindranath, S., Ho, L. C., \& Filippenko, A. V. 2002, ApJ, 566, 801

\nhi Read, J. I., \& Trentham, N. 2005, Phil. Trans. R. Soc. London, A363, 2693

\nhi Reddick, R. M., Wechsler, R. H., Tinker, J. L., \& Behroozi, P. S. 2012, arXiv:1207.2160

\nhi Rees, M. J. 1984, ARA\&A, 22, 471

\nhi Rees, M. J., \& Ostriker, J. P. 1977, MNRAS, 179, 541

\nhi Reeves, J. N., O'Brien, P. T., Braito, V., \etal 2009, ApJ, 701, 493     

\nhi Reid, M. J., Braatz, J. A., Condon, J. J., \etal 2009, ApJ, 695, 287 

\nhi Reines, A. E, \& Deller, A. T. 2012, ApJ, 750, L24                                      

\nhi Reines, A. E., Sivakoff, G. R., Johnson, K. E., \& Brogan, C. L. 2011, Nature, 470, 66  

\nhi Renzini, A. 1999, in The Formation of Galactic Bulges, ed. C. M. Carollo, H. C. Ferguson, \& R. F. G. Wyse (Cambridge: Cambridge Univ. Press), 9

\nhi Rhode, K. L. 2012, AJ, 144, 154

\nhi Richstone, D., Ajhar, E. A., Bender, R., \etal 1998, Nature, 395, A14  

\nhi Richstone, D., Bower, G., \& Dressler, A.~1990, ApJ, 353, 118

\nhi Richstone, D., Gebhardt, K., Aller, M., \etal 2004, arXiv:astro-ph/0403257

\nhi Richstone, D. O., \& Tremaine, S.~1984, ApJ, 286, 27  

\nhi Richstone, D. O., \& Tremaine, S.~1985, ApJ, 296, 370 

\nhi Richstone, D. O., \& Tremaine, S.~1988, ApJ, 327, 82  

\nhi Ridgway, S.~E., Heckman, T.~M., Calzetti, D., \& Lehnert, M. 2001, ApJ, 550, 122

\nhi Riechers, D. A., Walter, F., Brewer, B. J., \etal 2008, ApJ, 686, 851

\nhi Rigopoulou, D., Spoon, H. W. W., Genzel, R., Lutz, D., Moorwood, A. F. M., \& Tran, Q. D. 1999, AJ, 118, 2625

\nhi Rix, H.-W. 1993, in IAU Symposium 153, Galactic Bulges, ed. H. Dejonghe, \& H. J. Habing (Dordrecht: Kluwer), 423

\nhi Rix, H.-W., Kennicutt, R. C., Braun, R., \& Walterbos, R. A. M. 1995, ApJ, 438, 155

\nhi Roberts, T. P. 2007, ApSS, 311, 203

\nhi Rossa, J., van der Marel, R. P., B\"oker, T., \etal 2006, AJ, 132, 1074  

\nhi Rothberg, B., \& Joseph, R. D.~2004, AJ, 128, 2098

\nhi Rothberg, B., \& Joseph, R. D.~2006, AJ, 131, 185

\nhi Rupke, D. S., \& Veilleux, S. 2011, ApJ, 729, L27

\nhi Rusli, S. P., Thomas, J., Erwin, P., \etal 2011, MNRAS, 410, 1223   

\nhi Rusli, S. P., Thomas, J., Saglia, R. P., \etal 2013, preprint

\nhi Ryden, B. S., \& Gunn, J. E. 1987, ApJ, 318, 15

\nhi Sadoun, R., \& Colin, J. 2012, MNRAS, 426, L51

\nhi Saglia, R. P, Bender, R., \& Dressler, A. 1993, A\&A, 279, 75


\nhi Sales, L. V., Navarro, J. F., Theuns, T., \etal 2012, MNRA, 423, 1544 

\nhi Salpeter, E. E. 1964, ApJ, 140, 796

\nhi Salviander, S., Shields, G. A., Gebhardt, K., \& Bonning, E. W. 2007, ApJ, 662, 131

\nhi Samurovi\'c, S., \& Danziger, I. J. 2005, MNRAS, 363, 769

\nhi S\'anchez-Portal, M., D\'\i az, `'A., Terlevich, E., \& Terlevich, R. 2004, MNRAS, 350, 1087

\nhi Sancisi, R., \& van Albada T. S. 1987, in IAU Symposium 117, Dark Matter in the Universe, ed. J. Kormendy \& G. R. Knapp (Dordrecht: Reidel), 67

\nhi Sandage, A. 1961, The Hubble Atlas of Galaxies (Washington: Carnegie Institution of Washington)

\nhi Sandage, A., \& Bedke, J. 1994, The Carnegie Atlas of Galaxies (Washington, DC: Carnegie Institution of Washington)

\nhi Sandage, A., \& Binggeli, B. 1984, AJ, 89, 919

\nhi Sandage, A., Binggeli, B., \& Tammann, G. A. 1985, AJ, 90, 1759

\nhi Sandage, A., Freeman, K. C., \& Stokes, N. R. 1970, ApJ, 160, 831

\nhi Sandage, A., \& Tammann, G. A. 1981, A Revised Shapley-Ames Catalog of Bright Galaxies (Washington: Carnegie Institution of Washington)

\nhi Sanders, D. B. 1999, Ap\&SS, 266, 331

\nhi Sanders, D. B., \& Mirabel, I. F. 1996, ARA\&A, 34, 749

\nhi Sanders, D. B., Soifer, B. T., Elias, J. H., \etal 1988a, ApJ, 325, 74

\nhi Sanders, D. B., Soifer, B. T., Elias, J. H., Neugebauer, G., \& Matthews, K. 1988b, ApJ, 328, L35

\nhi Sani, E., Marconi, A., Hunt, L. K., \& Risaliti, G. 2011, MNRAS, 413, 1479

\nhi Sargent, W. L. W., Young, P. J., Boksenberg, A., Shortridge, K., Lynds, C. R., \& Hartwick, F. D. A. 1978, ApJ, 221, 731 

\nhi Sarria, J. E., Maiolino, R., La Franca, F., \etal 2010, A\&A, 522, L3

\nhi Sarzi, M., Rix, H.-W., Shields, J. C., \etal 2001, ApJ, 550, 65  

\nhi Sarzi, M., Rix, H.-W., Shields, J. C., \etal 2002, ApJ, 567, 237

\nhi Satyapal, S., B\"oker, T., Mcalpine, W., \etal 2009, ApJ, 704, 439

\nhi Satyapal, S., Vega, D., Dudik, R. P., Abel, N. P., \& Heckman, T. 2008, ApJ, 677, 926

\nhi Satyapal, S., Vega, D., Heckman, T., O'Halloran, B., \& Dudik, R. 2007, ApJ, 663, L9

\nhi Schawinski, K. 2012, in 2011 Frank N. Bash Symposium, New Horizons in Astronomy, ed.  
     S. Salviander, J. Green, \& A. Pawlik, Proc. Sci., {\ssbf http://pos.sissa.it/cgi-bin/reader/conf.cgi?confid=149}

\nhi Schawinski, K., Dowlin, N., Thomas, D., Urry, C. M., \& Edmondson, E. 2010, ApJ, 714, L108

\nhi Schawinski, K., Simmons, B. D., Urry, C. M., Treister, E., \& Glikman, E. 2012, MNRAS, 425, L61

\nhi Schawinski, K., Thomas, D., Sarzi, M., \etal 2007, MNRAS, 382, 1415

\nhi Schawinski, K., Treister, E., Urry, C. M., \etal 2011, ApJ, 727, L31

\nhi Schawinski, K., Virani, S., Simmons, B., \etal 2009, ApJ, 692, L19

\nhi Schechter, P. 1976, ApJ, 203, 297

\nhi Schlafly, E. F., \& Finkbeiner, D. P. 2011, ApJ, 737, 103

\nhi Schlegel, D. J., Finkbeiner, D. P., \& Davis, M. 1998, ApJ, 500, 525

\nhi Schmidt, M. 1959, ApJ, 129, 243

\nhi Schmidt, M. 1963, Nature, 197, 1040

\nhi Sch\"odel, R., Merritt, D., \& Eckart, A. 2009, A\&A, 502, 91  

\nhi Sch\"odel, R., Ott, T., Genzel, R., \etal 2002, Nature, 419, 694

\nhi Schramm, M., \& Silverman, J. D. 2013, ApJ, 767, 13

\nhi Schulze, A., \& Gebhardt, K. 2011, ApJ, 729, 21

\nhi Schulze, A., \& Wisotzki, L. 2011, A\&A, 535, A87

\nhi Schunck, F. E., \& Mielke, E. W. 2003, Classical and Quantum Gravity, 20, 301

\nhi Schwarzschild, M.~1979, ApJ, 232, 236 

\nhi Schwarzschild, M.~1993, ApJ, 409, 563

\nhi Schweizer,{\ts}F.{\ts}1990, in Dynamics and Interactions of Galaxies, ed.{\ts}R.{\ts}Wielen{\ts}(New{\ts}York:~Springer),{\ts}60

\nhi Schweizer, F. 1998, in 26$^{\rm th}$ Advanced Course of the Swiss Society of Astronomy and Astrophysics, Galaxies: Interactions and Induced Star Formation, ed. D. Friedli, L. Martinet, \& D. Pfenniger (New York: Springer), 105

\nhi Scorza, C., \& Bender, R. 1995, A\&A, 293, 20

\nhi Scott, N., \& Graham, A. W. 2013, ApJ, 763, 76  

\nhi Secrest, N., Satyapal, S., Gliozzi, M., \etal 2012, ApJ, 753, 38

\nhi Seigar, M. S., Barth, A. J., \& Bullock, J. S. 2008, MNRAS, 389, 1911

\nhi Sellgren, K., McGinn, M. T., Becklin, E. E., \& Hall, D. N. B. 1990, ApJ, 359, 112  

\nhi S\'ersic, J. L. 1968, Atlas de Galaxias Australes (C\'ordoba: Obs. Astron. Univ. Nacional de C\'ordoba)

\nhi Seth, A. C., Ag\"ueros, M., Lee, D., \& Basu-Zych, A. 2008, ApJ, 678, 116   

\nhi Seth, A. C., Cappellari, M., Neumayer, N., \etal 2010, ApJ, 714, 713

\nhi Seyfert, C. K. 1943, ApJ, 97, 28

\nhi Shapiro, K. L., Cappellari, M., de Zeeuw, T., \etal 2006, MNRAS, 370, 559  

\nhi Shen, J., \& Gebhardt, K. 2010, ApJ, 711, 484

\nhi Shen, J., Rich, R. M., Kormendy, J., \etal 2010, ApJ, 720, L72

\nhi Shen, J., Vanden Berk, D. E., Schneider, D. P., \& Hall, P. B. 2008, AJ, 135, 928

\nhi Shen, Y., \& Kelly, B. C. 2010, ApJ, 713, 41

\nhi Shi, Y., Rieke, G. H., Ogle, P., Jiang, L., \& Diamond-Stanic, A. M. 2009, ApJ, 703, 1107

\nhi Shields, G.~A., Gebhardt, K., Salviander, S., \etal 2003, ApJ, 583, 124

\nhi Shields, G.~A., Menezes, K. L., Massart, C. A., \& Vanden Bout, P. 2006, ApJ, 641, 683

\nhi Shields, J. C., B\"oker, T., Ho, L. C., \etal 2012, AJ, 144, 12

\nhi Shields, J. C., Rix, H.-W., Sarzi, M., \etal 2007, ApJ, 654, 125

\nhi Shields, J. C., Walcher, C. J., B\"oker, T., \etal 2008, ApJ, 682, 104

\nhi Shih, D. C., Iwasawa, K., \& Fabian, A. C. 2003, MNRAS, 341, 973

\nhi Shostak, G. S. 1987, A\&A, 175, 4

\nhi Shu, F., Najita, J., Ostriker, E., \etal 1994, ApJ, 429, 781

\nhi Shu, F. H., Najita, J., Ostriker, E. C., \& Shang, H. 1995, ApJ, 455, L155

\nhi Sijacki, D., \& Springel, V. 2006, MNRAS, 366, 397

\nhi Sijacki, D., Springel, V., Di Matteo, T., \& Hernquist, L. 2007, MNRAS, 380, 877

\nhi Sikora, M., Stawarz, L., \& Lasota, J.-P. 2007, ApJ, 658, 815

\nhi Silge, J. D., Gebhardt, K., Bergmann, M., \& Richstone, D. 2005, AJ, 130, 406  

\nhi Silk, J., \& Mamon, G. A. 2012, Res. Astron. Ap., 12, 917

\nhi Silk, J., \& Rees, M. J. 1998, A\&A, 331, L1  

\nhi Simien, F., \& Prugniel, Ph. 2000, A\&AS, 145, 263

\nhi Simien, F., \& Prugniel, P. 2002, \aa, 384, 371

\nhi Simpson, J. M., Smail, I., Swinbank, A. M., \etal 2012, MNRAS, 426, 3201

\nhi Siopis, C., Gebhardt, K., Lauer, T. R., \etal 2009, ApJ, 693, 946

\nhi Skrutskie, M. F., Cutrie, R. M., Stiening, R., \etal 2006, AJ, 131, 1163

\nhi Snyder, G. F., Hopkins, P. F., \& Hernquist, L. 2011, ApJ, 728, L24  

\nhi Sollima, A., Bellazzini, M., Smart, R. L., \etal 2009, MNRAS, 396, 2183  

\nhi So\l tan, A. 1982, MNRAS, 200, 115

\nhi Somerville, R. S., Hopkins, P. F., Cox, T. J., Robertson, R. E., \& Hernquist, L. 2008, MNRAS, 391, 481

\nhi Spitler, L. R., Forbes, D. A., Strader, J., Brodie, J. P., \& Gallagher, J. S. 2008, MNRAS, 385, 361   

\nhi Springel, V., Di Matteo, T., \& Hernquist, L. 2005a, MNRAS, 361, 776  

\nhi Springel, V., \& Hernquist, L. 2005, ApJ, 622, L9 (see arXiv:astro-ph/041379)

\nhi Springel, V., White, S. D. M., Jenkins, A., \etal 2005b, Nature, 435, 629    

\nhi Steinmetz, M., \& Navarro, J. F. 2002, NewA, 7, 155

\nhi Strader, J., Brodie, J. P. , Spitler, L., \& Beasley, M. A. 2006, AJ, 132, 2333  

\nhi Strader, J., Chomiuk, L., Maccarone, T. J., Miller-Jones, J. C. A., \& Seth, A. C. 2012a, Nature, 490, 71

\nhi Strader, J., Chomiuk, L., Maccarone, T. J., \etal 2012b, ApJ, 750, L27

\nhi Strateva, I. V., Ivezi\'c,\ts\v Z., Knapp, G. R., \etal 2001, AJ, 122, 1861

\nhi Sturm, E., Gonz\'alez-Alfonso, E., Veilleux, S., \etal 2011, ApJ, 733, L16

\nhi Szebehely, V., \& Peters, C. F. 1967, AJ, 72, 876

\nhi Szomoru, D., Franx, M., \& van Dokkum, P. G. 2012, ApJ, 749, 121

\nhi Tacconi, L. J., Genzel, R., Lutz, D., \etal 2002, ApJ, 580, 73

\nhi Tacconi, L.~J., Genzel, R., Neri, R., \etal 2010, Nature, 463, 781

\nhi Tacconi, L.~J., Genzel, R., Smail, I., \etal 2008, ApJ, 680, 246

\nhi Tadhunter, C., Marconi, A., Axon, D., \etal 2003, MNRAS, 342, 861                             

\nhi Tanaka, Y., Nandra, K., Fabian, A. C., \etal 1995, Nature, 375, 659

\nhi Targett, T. A., Dunlop, J. S., \& McLure, R. J. 2012, MNRAS, 420, 3621

\nhi Tempel, E., Tamm, A., \& Tenjes, P. 2010, A\&A, 509, A91

\nhi Terashima, Y., Kamizasa, N., Awaki, H., Kubota, A., \& Ueda, Y. 2012, ApJ, 752, 154

\nhi Terlevich, E., D\'\i az, A. I., \& Terlevich, R. 1990, MNRAS, 242, 271

\nhi Thomas, D., Greggio, L., \& Bender , R. 1999, MNRAS, 302, 537

\nhi Thomas, D., Maraston, C., Bender, R., \& Mendes de Oliveira, C. 2005, ApJ, 621, 673

\nhi Thomas, J. 2010, arXiv:1007.3591

\nhi Thomas, J., Saglia, R. P., Bender, R., \etal 2004, MNRAS, 353, 391

\nhi Thomsen, B., Baum, W. A., Hammergren, M., \& Worthey, G. 1997, ApJ, 483, L37  

\nhi Thornton, C. E., Barth, A. J., Ho, L. C., \& Greene, J. E. 2009, ApJ, 705, 1196

\nhi Thornton, C. E., Barth, A. J., Ho, L. C., Rutledge, R. E., \& Greene, J. E. 2008, ApJ, 686, 892

\nhi Thornton, R. J., Stockton, A., \& Ridgway, S. E. 1999, AJ, 118, 1461

\nhi Tombesi, F., Cappi, M., Reeves, J. N., \& Braito, V. 2012, MNRAS, 422, L1

\nhi Tonry, J. L.~1984, ApJ, 283, L27

\nhi Tonry, J. L.~1987, ApJ, 322, 632

\nhi Tonry, J. L., Dressler, A., Blakeslee, J. P., \etal 2001, ApJ, 546, 681

\nhi Toomre, A. 1977, in The Evolution of Galaxies and Stellar Populations, ed. B. M. Tinsley \& R. B. Larson (New Haven: Yale University Observatory), 401

\nhi Toomre, A., \& Toomre, J. 1972, ApJ, 178, 623

\nhi Torres, D. F., Capozziello, S., \& Lambiase, G. 2000, Phys. Rev. D, 62, 10012  

\nhi Treister, E., Natarajan, P., Sanders, D. B., \etal 2010, Science, 328, 600

\nhi Treister, E., Schawinski, K., Urry, C. M., \& Simmons, B. D. 2012, ApJ, 758, L39

\nhi Tremaine, S. 1995, AJ, 110, 628

\nhi Tremaine, S. 1997, in Unsolved Problems in Astrophysics, ed. J. N. Bahcall \& J. P. Ostriker (Princeton: Princeton University Press), 137

\nhi Tremaine, S, Gebhardt, K., Bender, R., \etal 2002, ApJ, 574, 740

\nhi Tremaine, S., Richstone, D. O., Byun, Y.-I., \etal 1994, AJ, 107, 634

\nhi Tremblay, B., \& Merritt, D. 1996, AJ, 111, 2243

\nhi Tremonti, C. A., Moustakas, J., \& Diamond-Stanic, A. M. 2007, ApJ, 663, L77

\nhi Treu, T., Malkan, M. A., \& Blandford, R. D. 2004, ApJ, 615, L97

\nhi Trujillo, I., Conselice, C. J., Bundy, K., \etal 2007, MNRAS, 382, 109

\nhi Trujillo, I., F\"orster Schreiber, N. M., Rudnick, G., \etal 2006, ApJ, 650, 18


\nhi Turner, M. L., C\^ot\'e, P., Ferrarese, L., \etal 2012, ApJS, 203, 5

\nhi Ulvestad, J. S., Greene, J. E., \& Ho, L. C. 2007, ApJ, 661, L151

\nhi Urrutia, T., Lacy, M., \& Becker, R. H. 2008, ApJ, 674, 80

\nhi Valluri, M., Ferrarese, L., Merritt, D., \& Joseph, C. L. 2005, ApJ, 628, 137   

\nhi Valluri, M., Merritt, D., \& Emsellem, E. 2004, ApJ, 602, 66

\nhi van Albada, G. D. 1980, A\&A, 90, 123  

\nhi van Albada, T. S., \& Sancisi, R. 1986, Phil.~Trans.~R.~Soc.~London A, 320, 447

\nhi van den Bergh, S., Li, W., \& Filippenko, A. V. 2002, PASP, 113, 820

\nhi van den Bosch, R. C. E., \& de Zeeuw, P. T. 2010, MNRAS, 401, 1770

\nhi van den Bosch, R. C. E., Gebhardt, K., G\"ultekin, K., \etal 2012, Nature, 491, 729

\nhi van den Bosch,{\ts}R.{\ts}C.{\ts}E., van de Ven,{\ts}G., Verolme,{\ts}E.{\ts}K., \& de Zeeuw,{\ts}P.{\ts}T. 2008, MNRAS, 385, 647 

\nhi van der Kruit, P. C., \& Freeman, K. C. 2010, ARA\&A, 49, 301

\nhi van der Kruit, P. C., Oort, J. H., \& Mathewson, D. S. 1972, A\&A, 21, 169


\nhi van der Marel, R. P. 1994a, MNRAS, 270, 271

\nhi van der Marel, R. P. 1994b, ApJ, 432, L91  

\nhi van der Marel, R. P. 1995, Highlights of Astronomy, 10, 527 

\nhi van der Marel, R. P. 2004, in Carnegie Observatories Astrophysics Series, Vol. 1: 
     Coevolution of Black Holes and Galaxies, ed. L. C. Ho (Cambridge: Cambridge Univ. Press), 37

\nhi van der Marel, R. P., \& Anderson, J.                                                          2010,  ApJ,    710, 1063  

\nhi van der Marel, R. P., Cretton, N., de Zeeuw, P. T., \& Rix, H.-W.                              1998,  ApJ,    493, 613   

\nhi van der Marel, R. P., de Zeeuw, P. T., \& Rix, H.-W.                                           1997a, ApJ,    488, 119   

\nhi van der Marel, R. P., de Zeeuw, P. T., Rix, H.-W., \& Quinlan, G. D.                           1997b, Nature, 385, 610   

\nhi van der Marel, R. P., Evans, N. W., Rix, H.-W., White, S. D. M., \& de Zeeuw, T.               1994a, MNRAS,  271, 99    

\nhi van der Marel, R. P., Gerssen, J., Guhathakurta, P., Peterson, R. C., \& Gebhardt, K.          2002,  AJ,     124, 3255  

\nhi van der Marel, R. P., Rix, H.-W., Carter, D., \etal.                                           1994b, MNRAS,  268, 521   

\nhi van der Marel, R. P., \& van den Bosch, F. C.                                                  1998, AJ,     116, 2220  

\nhi van de Ven, G., van den Bosch, R. C. E., Verolme, E. K., \& de Zeeuw, P. T. 2006, A\&A, 445, 513                  

\nhi van Dokkum, P. G., Franx, M., Kriek, M., \etal 2008, ApJ, 677, L5

\nhi van Dokkum,     P.     G., Whitaker,     K.     E., Brammer,     G., \etal 2010, ApJ, 709, 1018   

\nhi van Driel, W., \& van Woerden, H. 1991, A\&A, 243, 71

\nhi Veilleux, S., Kim, D.-C., Peng, C. Y., \etal 2006, ApJ, 643, 707

\nhi Verdoes Kleijn, G. A., van der Marel, R. P., Carollo, C. M., \& de Zeeuw, P. T. 2000, AJ, 120, 1221

\nhi Verdoes Kleijn, G. A., van der Marel, R. P., de Zeeuw, P. T., Noel-Storr, J., \& Baum, S. A. 2002, AJ, 124, 2524

\nhi Verdoes Kleijn, G. A., van der Marel, R. P., \& Noel-Storr, J. 2006, AJ, 131, 1961

\nhi Verolme E. K., Cappellari M., Copin Y., \etal 2002, MNRAS, 335, 517

\nhi Vika, M., Driver, S. P., Cameron, E., Kelvin, L., \& Robotham, A. 2012, MNRAS, 419, 2264

\nhi Volonteri, M. 2010, A\&AR, 18, 279

\nhi Volonteri, M., \& Ciotti, L. 2013, ApJ, 768, 29

\nhi Volonteri, M., Haardt, F., \& Madau, P. 2003, ApJ, 582, 559     

\nhi Volonteri, M., \& Natarajan, P. 2009, MNRAS, 400, 1911          

\nhi Volonteri, M., \& Rees, M. J. 2005, ApJ, 633, 624               

\nhi Walcher, C. J., van der Marel, R. P., McLaughlin, D., \etal 2005, ApJ, 618, 237

\nhi Walsh, J. L., Barth, A. J., \& Sarzi, M. 2010, ApJ, 721, 762  

\nhi Walsh, J. L., van den Bosch, R. C. E., Barth, A. J., \& Sarzi, M. 2012, ApJ, 753, 79

\nhi Walter, F., Carilli, C., Bertoldi, F., \etal 2004, ApJ, 615, L17

\nhi Walterbos, R. A. M., \& Kennicutt, R. C. 1987, A\&AS, 69, 311

\nhi Wang, L., \& Jing, Y. P. 2010, MNRAS, 402, 1796

\nhi Wang, L., \& Wheeler, J. C. 2008, ARA\&A, 46, 433

\nhi Wang, R., Carilli, C. L., Neri, R., \etal 2010, ApJ, 714, 699

\nhi Watson, W. D., \& Wallin, B. K. 1994, ApJ, 432, L35

\nhi Wehner, E. H., \& Harris, W. E. 2006, ApJ, 644, L17  

\nhi Weiland, J. L., Arendt, R. G., Berriman, G. B., \etal 1994, ApJ, 425, L81

\nhi Weiner, B. J., Coil, A. L., Prochaska, J. X., \etal 2009, AJ, 692, 187

\nhi Weinzirl, T., Jogee, S., Khochfar, S., Burkert, A., \& Kormendy, J. 2009, ApJ, 696, 411

\nhi Weller, J., Ostriker, J. P., Bode, P., \& Shaw, L. 2005, MNRAS, 364, 823

\nhi Wild, V., Heckman, T., \& Charlot, S. 2010, MNRAS, 405, 933

\nhi Wild, W., Kauffmann, G., Heckman, T., \etal 2007, MNRAS, 668, 543

\nhi Williams, M. J., Bureau, M., \& Cappellari, M. 2009, MNRAS, 400, 1665

\nhi Wilson, A. S., Braatz, J. A., \& Henkel, C. 1995, ApJ, 455, L127

\nhi Wirth, A., \& Gallagher, J. S.~1984, ApJ, 282, 85

\nhi Wold, M., Lacy, M., K\"aufl, H. U., \& Siebenmorgen, R. 2006, A\&A, 460, 449

\nhi Wolf, J., Martinez, G. D., Bullock, J. S., \etal 2010, MNRAS, 406, 1220

\nhi Woo, J.-H., Treu, T., Malkan, M. A., \& Blandford, R. D. 2006, ApJ, 645, 900

\nhi Woo, J.-H., Treu, T., Malkan, M. A., \& Blandford, R. D. 2008, ApJ, 681, 925

\nhi Worthey, G., Faber, S. M., \& Gonzalez, J. J. 1992, ApJ, 398, 69

\nhi Wrobel, J. M., Greene, J. E., \& Ho, L. C. 2011, AJ, 142, 113

\nhi Wrobel, J. M., \& Ho, L. C. 2006, ApJ, 646, L95

\nhi Wyithe, J. S. B. 2006a, MNRAS, 365, 1082

\nhi Wyithe, J. S. B. 2006b, MNRAS, 371, 1536

\nhi Xia, X. Y., Gao, Y., Hao, C.-N., \etal 2012, ApJ, 750, 92

\nhi Xiao, T., Barth, A. J., Greene, J. E., \etal 2011, ApJ, 739, 28

\nhi Xue, Y. Q., Brandt, W. N., Luo, B., \etal 2010, ApJ, 720, 368

\nhi Yamauchi, A., Nakai, N., Ishihara, Y., Diamond, P., \& Sato, N. 2012, PASJ, 64, 103

\nhi Yang, X., Mo, H. J., \& van den Bosch, F. C. 2009, ApJ, 695, 900

\nhi Yang, X., Mo, H. J., van den Bosch, F. C., Zhang, Y., \& Han, J. 2012, ApJ, 752, 41

\nhi Yang, Y., Li, B., Wilson, A. S., \& Reynolds, C. S.~2007, ApJ, 660, 1106  

\nhi Yelda, S., Ghez, A. M., Lu, J. R., \etal 2011, in The Galactic Center: A Window to the Nuclear Environment of Disk Galaxies, ed. M. R. Morris, Q. D. Wang, \& F. Yuan (San Francisco: ASP), 167

\nhi Yoon, S.-J., Lee, S.-Y., Blakeslee, J. P., \etal 2011, ApJ, 743, 150   

\nhi Young, P. J. 1980, ApJ, 242, 1232

\nhi Young, P. J., Westphal, J. A., Kristian, J., Wilson, C. P., \& Landauer, F. P. 1978, ApJ, 221, 721

\nhi Yu, Q. 2002, MNRAS, 331, 935

\nhi Yu, Q, \& Tremaine, S. 2002, MNRAS, 335, 965  

\nhi Zel'dovich, Ya. B. 1964, Soviet Physics -- Doklady, 9, 195

\nhi Zepf, S. E., \& Ashman, K. M. 1993, MNRAS, 264, 611  

\nhi Zhang W. M., Soria R., Zhang S. N., Swartz D. A, \& Liu J. 2009, ApJ, 699, 281

\nhi Zheng, Z., Coil, A. L., \& Zehavi, I. 2007, ApJ, 667, 760

}  

\hsize=15.0truecm  \hoffset=0.0truecm  \vsize=20.1truecm  \voffset=1.5truecm

\vfill\eject

\end